\documentclass[11pt, twoside]{Thesis} % The default font size and one-sided printing (no margin offsets)

\usepackage[square, numbers, comma, sort&compress]{natbib} 
\usepackage{siunitx}
\usepackage{mathtools}
\usepackage{enumitem}
\usepackage{comment}
\usepackage{bbold}
\usepackage{caption}
\def\bra#1{\mathinner{\langle{#1}|}}
\def\ket#1{\mathinner{|{#1}\rangle}}
\def\braket#1{\mathinner{\langle{#1}\rangle}}

\DeclarePairedDelimiter\kket{\lvert}{\rangle}
\DeclarePairedDelimiterX\bbrakket[2]{\langle}{\rangle}{#1 \delimsize\vert #2}

\title{\ttitle}

\begin{document}

\frontmatter
\setstretch{1.3}
\fancyhead{}
\rhead{\thepage}
\lhead{}
\pagestyle{fancy}

\newcommand{\virgolette}[1]{``#1''}
\newcommand{\capo}{\\\hspace*{6mm}}
\newcommand{\spazio}{\hspace*{6mm}}
\newcommand{\HRule}{\rule{\linewidth}{0.5mm}}

\newcommand{\approptoinn}[2]{\mathrel{\vcenter{
			\offinterlineskip\halign{\hfil$##$\cr
				#1\propto\cr\noalign{\kern2pt}#1\sim\cr\noalign{\kern-2pt}}}}}
\newcommand{\appropto}{\mathpalette\approptoinn\relax}

\makeatletter
\newcommand\footnoteref[1]{\protected@xdef\@thefnmark{\ref{#1}}\@footnotemark}
\makeatother

%----------------------------------------------------------------------------------------
%	TITLE PAGE
%----------------------------------------------------------------------------------------

\begin{titlepage}
\begin{center}

\includegraphics[scale=.12]{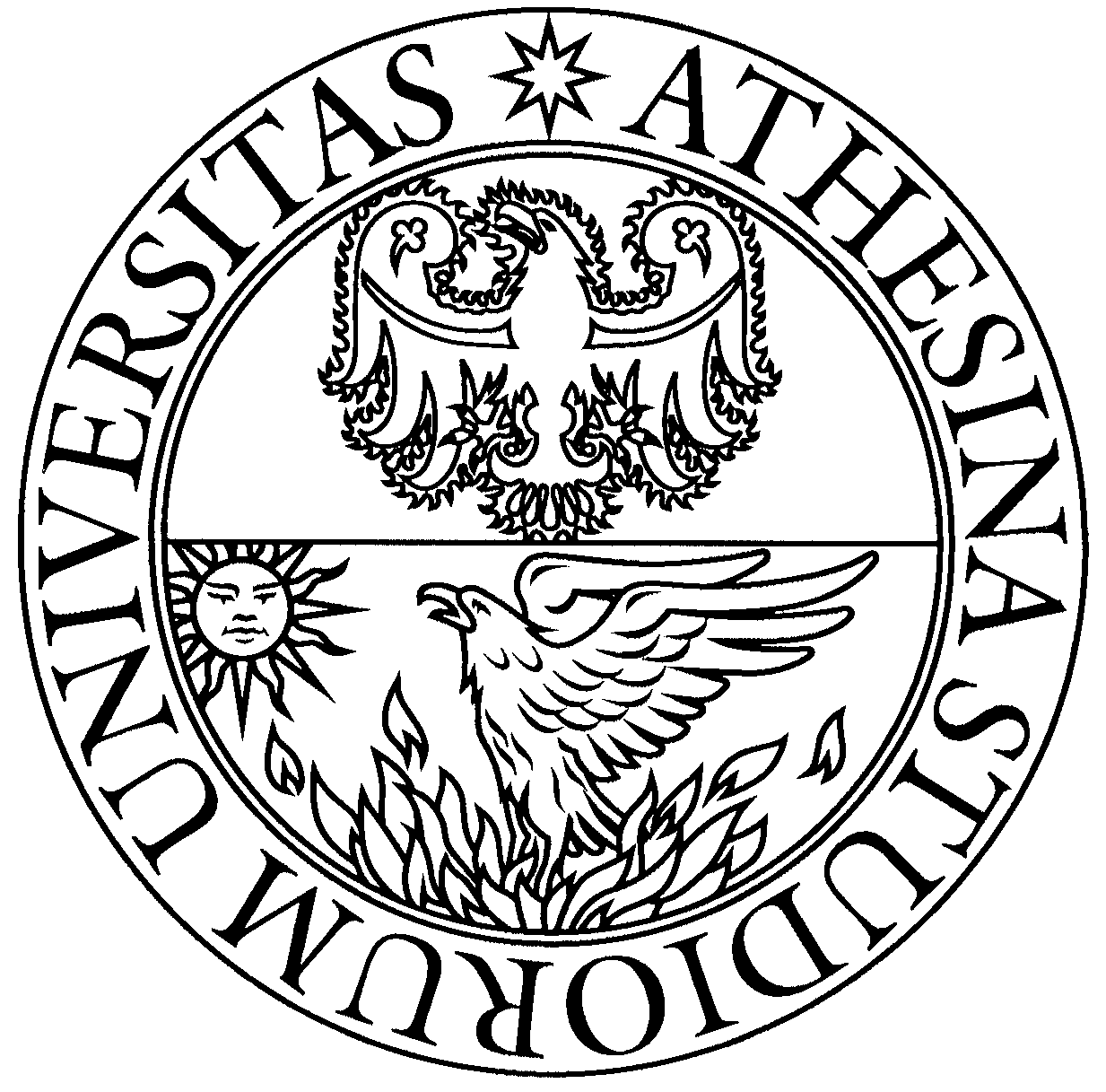}
\\[1cm]
\textsc{\LARGE UNIVERSITY OF TRENTO}\\[.5cm] % University name
\textsc{\LARGE DEPARTMENT OF PHYSICS}\\[.5cm]
\textsc{\Large Master Degree Course in Physics}\\[0.5cm]
\textsc{\Large Final Thesis}\\[0.5cm] % Thesis type

\HRule \\[0.4cm] % Horizontal line
{\huge \bfseries \ttitle}\\[0.4cm] % Thesis title
\HRule \\[4cm] % Horizontal line

\begin{minipage}{0.4\textwidth}
\begin{flushleft} \large
\emph{Supervisor:}\\
{\supname} % Author name - remove the \href bracket to remove the link
\end{flushleft}
\end{minipage}
\begin{minipage}{0.4\textwidth}
\begin{flushright} \large
\emph{Graduand:} \\
{\authornames} % Supervisor name - remove the \href bracket to remove the link  
\end{flushright}
\end{minipage}\\[3cm]

{\large Academic Year 2018/2019}\\[2cm] % Date
 % University/department logo - uncomment to place it

\vfill
\end{center}

\end{titlepage}

\newpage
\null
\thispagestyle{empty}
\newpage

%----------------------------------------------------------------------------------------
%	QUOTATION PAGE
%----------------------------------------------------------------------------------------

% \pagestyle{empty} % No headers or footers for the following pages

% \null\vfill % Add some space to move the quote down the page a bit

% \textit{``Thanks to my solid academic training, today I can write hundreds of words on virtually any topic without possessing a shred of information, which is how I got a good job in journalism."}

% \begin{flushright}
% Dave Barry
% \end{flushright}

% \vfill\vfill\vfill\vfill\vfill\vfill\null % Add some space at the bottom to position the quote just right

% \clearpage % Start a new page

%----------------------------------------------------------------------------------------
%	ABSTRACT PAGE
%----------------------------------------------------------------------------------------

%\addtotoc{Abstract} % Add the "Abstract" page entry to the Contents

%\abstract{\addtocontents{toc}{\vspace{1em}} % Add a gap in the Contents, for aesthetics
%...
%}

%\cleardoublepage % Start a new page
% \newpage

%----------------------------------------------------------------------------------------
%	LIST OF CONTENTS/FIGURES/TABLES PAGES
%----------------------------------------------------------------------------------------

\pagestyle{fancy} % The page style headers have been "empty" all this time, now use the "fancy" headers as defined before to bring them back

\lhead{\emph{Contents}} % Set the left side page header to "Contents"
\tableofcontents % Write out the Table of Contents

\mainmatter % Begin numeric (1,2,3...) page numbering

\pagestyle{fancy} % Return the page headers back to the "fancy" style

% Include the chapters of the thesis as separate files from the Chapters folder
% Uncomment the lines as you write the chapters

%!TEX root = ../main.tex
% Chapter Template

\graphicspath{{./pic0/}}

\chapter{Introduction}\label{ch0} % Change X to a consecutive number; for referencing this chapter elsewhere, use \ref{ChapterX}

\lhead{Chapter 1. \emph{Introduction}}

The quantum Hall effect is surely one of the most surprising condensed matter phenomena observed in the last century. 
The setup is surprisingly simple, a system of electrons at very low temperature constricted to move in a two-dimensional plane pierced by a strong perpendicular magnetic field, the consequences astonishing.
\begin{figure}[htp!]
	\centering
	\includegraphics[width=.7\textwidth]{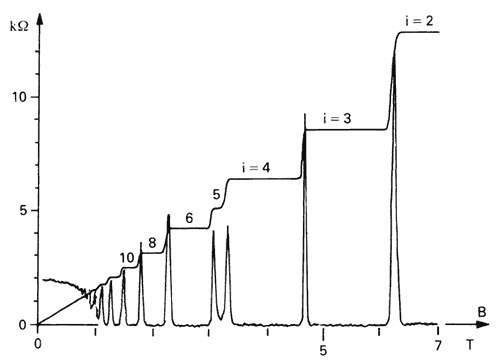}
	\caption[The LOF caption]{The Hall resistance and the transverse resistance are measured as a function of the applied magnetic field. The Hall resistance varies stepwise, the transverse one shows a bump whenever the transverse resistance transition from one plateau to another (Kosmos, 1986).}
\end{figure}

First of all, when in 1980 von Klitzing measured the Hall (transverse) conductivity of such a system (\cite{VonKlitzing1980}, \cite{VonKlitzing1986}), he realised that it does not change smoothly as the magnetic field is varied; rather it exhibits robust plateaus at integer multiples of $\frac{e^2}{h}$
\[
\sigma_\text{Hall} = \frac{e^2}{h}\,\nu
\]
regardless of the system details. For the discovery, von Klitzing was awarded the 1985 Nobel prize in physics.
This is per-se a remarkable fact, since experimental samples are far from ideal: what he observed is an emergent property in a macroscopic, \virgolette{dirty} system and not a single atom property.
The effect is now known as the \textit{Integer Quantum Hall Effect} (IQHE).
The phenomenon can be explained without taking into account the interactions between the electrons: von Klitzing immediately recognised that the strong magnetic field makes the electron levels \virgolette{coalesce} into highly degenerate discretely spaced Landau levels. 
He related the integer number $\nu$ appearing in the Hall conductance formula to the number of filled Landau levels. 
The robustness and universality of the transverse conductivity quantization was then recognised by Laughlin to be a more general fact; by a clever argument he was able to show the Hall quantization to be a manifestation of gauge invariance (\cite{Laughlin1981}); this seminal work gave a hint in the direction of understanding the deep connection between the quantum Hall effect and topology: the filling factor $\nu$ was indeed later recognised as a topological invariant of the system (\cite{NiuThoulessWu1985}).
\newline The importance of the system edges soon emerged; their prominent role can be guessed even classically.
A classical electron in a magnetic field moves in a circular path (a so called cyclotron orbit), without changing its average position. Near the system edges however the same electron gets scattered by the edge potential and thus performs a skipping motion. Particles near the boundary will thus move in a preferential direction.
This last point remains true at the quantum level, the quantum counterpart of the skipping orbits being gapless current-carrying electron states localised near the sample boundary but extended over the whole edge of the system.
As pointed out by Halperin (\cite{Halperin1982}), MacDonald and Středa (\cite{MacDonaldAndStreda1984}) these edge states play a prominent role in the quantization of the Hall conductivity, since the Hall current can be viewed as arising from the edge response:
the filling factor $\nu$ can be as well interpreted as being the number of current-carrying edge modes.
These edge states help moreover explaining the robustness of the Hall conductance quantization: 
as pointed out by Büttiker (\cite{Buttiker1988}) the chiral edge modes are immune to both elastic and inelastic backscattering as an effect of the presence of the strong magnetic field, and thus continue to provide current carrying channels even in the presence of modest disorder.
\newline That the number of edge modes equals the filling factor $\nu$ (a topological invariant) may seem surprising but turned out not to be a coincidence; rather it is a consequence of a general theorem known as the bulk-edge correspondence (\cite{QiWuZhang2006}), which establishes a deep connection between the bulk and edge properties of the system.

%\newline The fact that the Hall conductance quantization is so robust has been recognised as being the consequence of these chiral edge modes being immune to both elastic and inelastic backscattering (and thus continue to provide current carrying channels) as an effect of the presence of the strong magnetic field, as pointed out by Büttiker (\cite{Buttiker1988}), and to disorder not disrupting them.

%\newline For a bulk system the robustness of the transverse conductivity quantization can be understood to be related to topology, as pointed out by Niu, Thouless and Wu (\cite{NiuThoulessWu1985}), the filling factor $\nu$ being recognised as a topological invariant (the so called Chern number). That the number of edge modes equals a bulk topological invariant is not a coincidence, rather it is a consequence of the bulk-edge correspondence (\cite{QiWuZhang2006}).
%The interplay between bulk properties and edge ones is thus deeply connected to topology.

It was later found that the so called \virgolette{filling factor} $\nu$ could assume a whole plethora of fractional values too.
The first observation by Störmer and Tsui (\cite{StormerTsui}) in 1982 reported plateaus with $\nu=\frac{1}{3}$ and $\nu=\frac{2}{3}$.
The phenomenon is now known as the \textit{Fractional Quantum Hall Effect} (FQHE), and it is explained by the formation of a highly correlated quantum state whose excitations are fractionally charged and have fractional particle statistics.

In this thesis work the quantum dynamics of an integer Quantum Hall state is studied, and special importance is given to the edge states. 
Given the insensitivity of the Hall conductance to the particular system details, a very simple semi-infinite Hall bar has been considered: along one direction we imposed periodic conditions at the boundary, in the other one a simple analytical confining potential has been used in order to model the edges.
\newline The $\nu=1$ Landau level has been considered, and correspondingly the Fermi level has been chosen to lie between the first and second Landau levels. 
\newline The \virgolette{static} properties of the system have been studied, both via analytical methods and numerically (by diagonalising the single particle Hamiltonian). 
\newline To investigate the dynamics the system has been subjected to an external perturbation, with different spatial profiles.
Single particle time-dependent perturbation theory beyond the leading order and numerical methods have been used; concerning the latter one an alternate direction implicit algorithm has been implemented. Both linear and non-linear phenomena are discussed and studied.

We decided to tackle this very simple non-interacting system attracted by the possibility of studying from a microscopic point of view and without approximations the linear edge dynamics beyond the chiral Luttinger liquid paradigm and more interestingly the non-linear density dynamics at the edges, since there is a whole flourishing research area which aims at studying the boundary dynamics beyond the linear response theory, especially in the direction of the fractional quantum Hall effect, where electron-electron correlations play a major role (e.g. \cite{Wiegmann2012}, \cite{BettelheimAbanovWiegmann2006}).

The thesis' structure is the following.
\begin{itemize}
	\item \textbf{Chapter 2.} The system which has been considered is described, and its static properties are reviewed. The classical dynamics is tackled first, and used as a starting point to analyse the Landau quantisation of the classical orbits, both in the bulk and at the system edges. 
	We then move to a full quantum approach. Before discussing the diagonalization method, some considerations regarding the system edge are made. The particle-hole spectrum is discussed, as well as the Hall conductivity.
	\item \textbf{Chapter 3.} The dynamics of the many-body Fermion system is discussed, and it is reduced to the study of the single electron one, thanks to the absence of interactions among the constituents. 
	The numerical machinery is then briefly discussed, and the algorithm is tested in some simple situations, such as the motion of a Gaussian wavepacket in a magnetic field.
	\item \textbf{Chapter 4.} Before diving into the numerical analysis, an effective edge theory (which de facto is a non-interacting chiral Luttinger liquid) of the edge excitations is discussed starting from a second quantisation approach and integrating out the spatial direction orthogonal to both the magnetic field and the system edge, under some special assumption on the structure of the Landau levels.	
	\item \textbf{Chapter 5.} From this chapter we begin the discussion of the new results obtained during this thesis work. First the dynamics generated by a spatially periodic excitation is studied. The bulk is explicitly shown to behave as an incompressible medium as long as excitations to higher Landau levels can be neglected; the edge dynamics is studied up to second perturbative order, and the importance of the Fermi velocity and effective mass at the Fermi point highlighted. 
	The currents flowing in the system are also discussed and related to the density variation induced by the excitation.
	Numerical results are analysed during the whole discussion.
	
	\item \textbf{Chapter 6.} In this chapter a Gaussian shaped potential is used to study a localised density excitation of the system; perturbation theory is again used to give it some simple characterisation. Important non-linear corrections are discussed starting from the numerical results; in particular big differences emerge between the behaviour of the system edge response to a dip and bump like potential.
	
	\item \textbf{Chapter 7.} In order to generate the propagation of an initially bell-shaped packet, the response of the system to a sigmoid shaped excitation is studied in this chapter. The results of the numerical simulations are analysed and used to qualitatively understand the non-linear edge dynamics.
	
	\item \textbf{Chapter 8.} The work conclusions are made as well as some future perspectives.
	
\end{itemize}
%!TEX root = ../main.tex
% Chapter 1

\graphicspath{{./pic1/}}

\chapter{The Landau Levels and the Integer Quantum Hall Effect}\label{ch1}
\lhead{Chapter 2. \emph{The Landau Levels and the Integer Quantum Hall Effect}}

In this chapter the general problem is very briefly introduced, starting with a classical discussion and undertaking a full quantum route afterwards. The static properties of the system are discussed both theoretically and numerically, and some of the general ideas which are found almost everywhere in this work will be exposed too.

\section{The problem}
Consider a quantum gas of non-interacting free electrons moving in a thin slab with dimensions $L_x, L_y\gg L_z $, which is pierced by a uniform magnetic field $\textbf{B}=B\hat{z}$.
\newline The system will be considered to be infinite along $\hat{y}$, and periodic boundary conditions will be used along such a direction; 
\newline The energy levels for the confinement along $\hat{z}$ will roughly be of order $\epsilon_z\sim\frac{\hbar^2}{2m\,L_z^2}$, which we will assume to be much larger than any other relevant energy scale of the system so that the degrees of freedom along this axis can be safely considered to be frozen and the problem may be regarded as being effectively 2D.

The one-electron Hamiltonian reads
\begin{equation}
\label{eq:problem_hamiltonitan}
\mathcal{H}=\frac{\boldsymbol{\pi}^2}{2m}+V_c(x)=\frac{1}{2m}\left(\mathbf{p}+e \mathbf{A}\right)^2+V_{c}(x)
\end{equation}
where $\boldsymbol{\pi}=\mathbf{p}+e\mathbf{A}$ is the mechanical momentum and $V_{c}(x)$ is some confining potential along the $\hat{x}$ direction, very steep at the edges of the slab but nearly constant within the bulk. We approximate it to be $y$-independent. %As a side comment, one can notice that the Hamiltonian is not gauge invariant, but the Schrödinger is.
\newline Electrons do have spin, which in a magnetic field would result in a twofold splitting of each Landau level. For simplicity though the electron gas has been assumed to be perfectly polarised and scattering between spin up and down electrons neglected.

\section{The bulk}
It is convenient to discuss the bulk case first, since the calculations can be carried out analytically and the results are simple.
We take the classical route first, discussing the cyclotron orbits, also in the presence of an electric field, since this will turn out to be useful when the edges will be discussed. Semiclassical quantization of the orbits is also briefly treated.

The more complicated Hamiltonian in eq. \ref{eq:problem_hamiltonitan} greatly simplifies in a bulk system, where the edge potential can be safely neglected. In particular, the Hamiltonian is quadratic and thus the problem very simple
\begin{equation}
\label{eq:bulk_hamiltonian}
\mathcal{H}_\text{bulk}=\frac{\boldsymbol{\pi}^2}{2m}.
\end{equation}

\subsection{Semiclassical treatment}\label{section:semiclassical}
As already stated, it is insightful to consider the classical case first. The electron dynamics is described by Hamilton's equations of motion
\begin{equation}
\begin{cases}
\dot{p}_i=-\frac{\partial \mathcal{H}}{\partial q_i}
\\
\dot{q}_i=+\frac{\partial \mathcal{H}}{\partial p_i}
\end{cases}
\end{equation}
which easily reduce to Newton's second equation $\frac{d\boldsymbol{\pi}}{dt}=-\frac{e}{m} \boldsymbol{\pi}\times\mathbf{B}$.
The solution in a uniform magnetic field is
\begin{equation}
\begin{cases}
\pi_x = \pi_0 \cos\left(\omega_c t+\phi_0\right)
\\
\pi_y = \pi_0 \sin\left(\omega_c t+\phi_0\right)
\end{cases}
\end{equation}
where $\omega_c=\frac{eB}{m}$ is the cyclotron frequency. 
\newline The kinetic energy is a conserved quantity, since magnetic forces do no work
\begin{equation}
E = \frac{\pi_x^2+\pi_y^2}{2m}=\frac{\pi_0^2}{2m}.
\end{equation}

A further integration gives the classical orbits
\begin{equation}
\label{eq:cyclotron}
\begin{cases}
x = x_0+\frac{\pi_0}{m\omega_c} \sin\left(\omega_c t+\phi_0\right) = x_0 + \frac{\pi_y}{m \omega_c}
\\
y = y_0-\frac{\pi_0}{m\omega_c} \cos\left(\omega_c t+\phi_0\right) = y_0 - \frac{\pi_x}{m \omega_c}
\end{cases}
\end{equation}
where $\mathcal{C}_0=(x_0,y_0)$ is the classical centre of the orbit, whose radius is $r_0=\frac{\pi_0}{m \omega_c}$.
\newline Going to the quantum scenario, we expect a quantization of the classical orbits; according to the Bohr-Sommerfeld semiclassical quantization rule (as derived from the semiclassical approximation) we indeed have
\begin{equation}
\begin{split}
h\left(n+\frac{1}{2}\right)
=&\oint_{H(\mathbf{q},\mathbf{p})=E} \boldsymbol{p}\cdot d\mathbf{q}
=\oint_{H(\mathbf{q},\mathbf{p})=E} \left(\boldsymbol{\pi}-e\mathbf{A}\right)\cdot d\mathbf{q}=\\
=&\int_{0}^{\frac{2\pi}{\omega_c}} \boldsymbol{\pi}\cdot \frac{d\mathbf{q}}{dt}\,dt-e\int_{H(\mathbf{q},\mathbf{p})\leq E} \nabla\times \mathbf{A}\cdot d\mathbf{S}=\\
=&2E\int_{0}^{\frac{2\pi}{\omega_c}} dt-eB\int_{H(\mathbf{q},\mathbf{p})\leq E} dx dy
=E \,\frac{2\pi}{\omega_c}.
\end{split}
\end{equation}
In the last line the fact that the orbits with energy $\leq E$ are enclosed within a circle of radius $r_0=\frac{\pi_0}{m\omega_c}$ has been used, as well as the relation between $\pi_0$ and the energy $E$ of the orbit. We thus obtain
\begin{equation}
\label{eq:semiclassical_LL}
E=\hbar \omega_c\left(n+\frac{1}{2}\right)
\end{equation}
which are the quantized energy levels of a quantum harmonic oscillator, and it will be shown that a full quantum treatment yields exactly the same result.
\newline From $\frac{\hbar\omega_c}{2}=\frac{\pi_0^2}{2m}$ we can get $\pi_0$ and thus we see that the semiclassical orbit takes up a minimum amount of space with radius $r_0=\sqrt{\frac{\hbar}{m\omega_c}}$, which is the harmonic oscillator typical length.

\subsubsection{In the presence of a uniform electric field}\label{section:classical_electric_field}
Suppose we subject the system to a uniform electric field in the $x$ direction. The new Hamiltonian reads
\begin{equation}
\mathcal{H}_\text{bulk}=\frac{\boldsymbol{\pi}^2}{2m}+eEx.
\end{equation}
Newton's second equation in this case becomes  $\frac{d\boldsymbol{\pi}}{dt}=-\frac{e}{m} \boldsymbol{\pi}\times\mathbf{B}+eE\hat{x}$, or
\begin{equation}
\begin{cases}
\dot{\pi}_x=-\omega_c\pi_y-eE\\
\dot{\pi}_y=+\omega_c\pi_x
\end{cases}
\end{equation}
or, defining $\pi_y'=\pi_y+\frac{eE}{\omega_c}$
\begin{equation}
\begin{cases}
\dot{\pi}_x=-\omega_c\pi_y'\\
\dot{\pi}_y'=+\omega_c\pi_x.
\end{cases}
\end{equation}
It is readily checked that $\widetilde{\pi}_0^2=\pi_y'^2+\pi_x^2$ is a conserved quantity. Integrating twice this equation yields
\begin{equation}
\begin{cases}
\begin{alignedat}{2}
x &= x_0&&+\frac{\widetilde{\pi}_0}{m\omega_c} \sin\left(\omega_c t+\phi_0\right)
\\
y &= y_0-\frac{eE}{m\omega_c}\,t &&-\frac{\widetilde{\pi}_0}{m\omega_c} \cos\left(\omega_c t+\phi_0\right)
\end{alignedat}
\end{cases}
\end{equation}
i.e. the centre of the cyclotron orbit $\mathcal{C}_0=(x_0,y_0-\frac{eE}{m\omega_c}\,t)$  drifts in time with speed $-\frac{eE}{m\omega_c}=-\frac{E}{B}$ along the $y$ direction as a consequence of the presence of an $x$ directed electric field. 
We moreover see that the rotation radius is $\widetilde{r}_0=\frac{\widetilde{\pi}_0}{m\omega_c}$.
\newline If during a cyclotron period the displacement of the centre of the orbit is much smaller than the orbit radius, i.e. if we have $\omega_c \gg \frac{eE}{m\omega_c}\frac{1}{\widetilde{r}_0}$, the trajectory will practically correspond to a slowly drifting circular orbit (the effect of the magnetic field is much larger than that of the electric field); vice-versa when $\omega_c \ll \frac{eE}{m\omega_c}\frac{1}{\widetilde{r}_0}$ we have a deeply stretched-out trajectory.

\subsection{Quantum-mechanical treatment}
Let's now take the full quantum-mechanical route. A gauge invariant treatment is employed first, to highlight the gauge independence of quantities which will be used throughout the whole thesis work.

The mechanical momentum along $x$ and $y$ fail to commute, but their commutator is a c-number
\begin{equation}
[\pi_x,\pi_y]=-i \hbar e \left(\partial_x A_y - \partial_y A_x\right) = -i \hbar e B.
\end{equation}
It is possible to define operators obeying harmonic oscillator commutation relations $[a,a^\dagger]=1$
\begin{equation}
\begin{cases}
a = \frac{\pi_x-i \pi_y}{\sqrt{2\hbar e B}}
\\
a^\dagger = \frac{\pi_x+i \pi_y}{\sqrt{2\hbar e B}}
\end{cases}
\end{equation}
in terms of which the bulk Hamiltonian reads
\begin{equation}
\label{eq:Landau_level_algebraic}
\mathcal{H}_\text{bulk}=\hbar \omega_c\,\left(a^\dagger a+\frac{1}{2}\right)
\end{equation}
The bulk Hamiltonian therefore is a one-dimensional quantum harmonic oscillator, whose (gauge independent) spectrum can be constructed in a standard way
\begin{equation}
E_n = \hbar \omega_c\left(n+\frac{1}{2}\right).
\end{equation}
The spectrum is discrete, and depends on a single quantum number. These states are called Landau levels.
\newline We expect that there will be many degenerate states having this same energy eigenvalue. 
Heuristically the degeneracy can be obtained by a simple reasoning: since the total number of states is conserved when the magnetic field is switched on, we will have (in the thermodynamic limit and using periodic boundary conditions both along $x$ and $y$)
\begin{equation}
\label{eq:conservation_of_number_of_states}
\begin{alignedat}{1}
\begin{cases}
\substack{{\text{\# of states}}\\{\text{without a B field}}}
&=\sum_{k_x^2+k_y^2\leq k_F^2}\rightarrow\frac{L_x}{2\pi} \frac{L_y}{2\pi}\, \int_{k \leq k_F}dk_x dk_y
\\
\substack{{\text{\# of states}}\\{\text{with a B field}}}&=\mathcal{N} \,\times\,\substack{{\text{\# of filled}}\\{\text{Landau levels}}}
\end{cases}
\end{alignedat}
\end{equation}
where $\mathcal{N}$ is the degeneracy of each Landau level (assumed to be Landau level independent), which can be  understood as being caused by the magnetic field making the free-particle states \virgolette{collapse} onto the discretely spaced Landau levels.
\newline The number of filled Landau levels can be obtained by equating the Fermi energy to the energy of the highest occupied Landau level, or $E_F=\frac{\hbar^2k_F^2}{2m}=\hbar\omega_c\left(M+\frac{1}{2}\right)\approx \hbar\omega_c M$.
From \ref{eq:conservation_of_number_of_states} we therefore obtain
\begin{equation}
\label{eq:ll_degeneracy}
\mathcal{N}=(B L_x L_y)\,\frac{e}{h}=\frac{\Phi_B}{\Phi_0}
\end{equation}
where $\Phi_B$ is the flux of $B$ through the sample area $L_xL_y$ and $\Phi_0$ is the flux quantum $\frac{h}{e}$.
Though heuristically derived, this is indeed the correct result. A more rigorous route will will be discussed below in this section.
\newline Since the eigenvalues of $H_0$ are degenerate, there must be some other observable commuting with the Hamiltonian which can label these states. Since the system is uniform, the energy cannot depend on \textit{where} the cyclotron orbit centre is. Inspired by the semiclassical treatment (see eq. \ref{eq:cyclotron}) we then define self-adjoint operators corresponding to the classical centre of the cyclotron orbit
\begin{equation}
\label{eq:py_gauge_invariance}
\begin{cases}
X_0=x-\frac{\pi_y}{m \omega_c}
\\
Y_0=y+\frac{\pi_x}{m \omega_c}
\end{cases}
\end{equation}
in which only the observable mechanical momentum appears (they are thus gauge invariant self-adjoint operators). It is easy to shown that both commute with the bulk Hamiltonian, $[X_0,\mathcal{H}_\text{bulk}]=[Y_0,\mathcal{H}_\text{bulk}]=0$, but they fail to commute with one other, $[X_0,Y_0]=i l_B^2$, which means that we cannot perfectly localize the centre of the classical orbit. (Here $l_B=\sqrt{\frac{\hbar}{eB}}$ is the magnetic length).
We can then simultaneously diagonalize $\{\mathcal{H}_\text{bulk}, X_0\}$ or $\{\mathcal{H}_\text{bulk}, Y_0\}$. Whichever pair we choose to diagonalise, an important point is that the associated eigenvalue equations and eigenvalues will be independent of the choice of gauge.

Using these operators it is possible to more rigorously derive eq. \ref{eq:ll_degeneracy}. 
Suppose we simultaneously diagonalised $\{\mathcal{H}_\text{bulk}, X_0\}$, and let $\ket{n,x_0}$ be the common eigenvectors. Since the system is finite in the $x$-direction, we require $-\frac{L_x}{2}\leq x_0\leq \frac{L_x}{2}$.
We now need to find an operator which commutes with the Hamiltonian but allows us to move within the degenerate subspace corresponding to a given Landau level. Such a operator can easily be identified with $U_\alpha=e^{-i\alpha Y_0}$. 
\newline Not every value of $\alpha$ is admitted: since we are using periodic boundary conditions along $y$, the operators $Y_0$ and $Y_0+\mathbb{1} L_y$ are perfectly equivalent. This requires that $\alpha$ is quantised in units of $\frac{2\pi}{L_y}$.
\newline It is easy to show by induction that $[X_0, Y_0^n]=il_B^2 \,n\, Y_0^{n-1}$. It follows that $[X_0, U_\alpha]=\alpha l_B^2  U_\alpha$, and thus
\begin{equation}
X_0\,U_\alpha\,\ket{n,x_0}= (x_0+\alpha l_B^2)\, U_\alpha\,\ket{n,x_0}
\end{equation}
i.e. if $\ket{n,x_0}$ is an eigenket of $X_0$ with eigenvalue $x_0$, then also $U_\alpha\,\ket{n,x_0}$ is an eigenstate of the same operator but with a shifted eigenvalue $\alpha l_B^2  + x_0$. 
\newline The idea is rather simple; by acting with $U_{\alpha=\frac{2\pi}{L_y}}$ on $\ket{n,-\frac{L_x}{2}}$ a number $\mathcal{N}$ of times we eventually will span all the degenerate subspace, arriving at $\ket{n,+\frac{L_x}{2}}$. The degeneracy of a Landau level is then easily found by solving $-\frac{L_x}{2}+\mathcal{N}\,\frac{2\pi}{L_y}\,l_B^2=\frac{L_x}{2}$, and we find precisely eq. \ref{eq:ll_degeneracy}.

As a final comment, notice that since $[X_0,Y_0]=i l_B^2$ the Heisenberg uncertainty theorem gives
\begin{equation}
\Delta X_0 \Delta Y_0\geq\frac{1}{2}\,l_B^2
\end{equation}
i.e. the classical orbit takes a minimum amount of space due to quantum uncertainty of order $\sim l_B^2$; this agrees with the semiclassical conclusions derived in section \ref{section:semiclassical}. 
\newline The zero-temperature non-interacting ground state of the system is obtained by filling all the lowest lying energy states, starting from the least energetic and going on until there are no more electron, respecting the Pauli exclusion principle. Heuristically we expect the electronic density associated to a single Landau level to roughly be of the order $\sim \frac{1}{\Delta X_0 \Delta Y_0}\sim l_B^{-2}$, i.e. one electron occupying a quantized cyclotron orbit of area $\sim l_B^2$.
More quantitatively, knowing the Landau level degeneracy (eq. \ref{eq:ll_degeneracy}), if we have $\nu$ completely filled Landau levels the bulk electronic density becomes
\begin{equation}
\label{eq:bulk_denisty}
\rho_\text{bulk}=\nu\frac{\mathcal{N}}{L_xL_y}=\frac{\nu}{2\pi l_B^2}
\end{equation}
which is indeed of the expected order of magnitude.

\subsubsection{Choosing a gauge}\label{section:choosing_a_gauge}
Having established some result which does not depend on the choice of the gauge, we are now in the position of picking up one which preserves the most the symmetries of our system.
\newline The Landau gauge $\mathbf{A}=Bx\,\hat{y}$ is a convenient choice for our problem, so that translational invariance along $\hat{y}$ is preserved.
The bulk Hamiltonian in this case reads
\begin{equation}
\mathcal{H}_\text{bulk} = \frac{p_x^2}{2m}+\frac{1}{2}m\omega_c^2\left(\frac{p_y}{eB}+x\right)^2
\end{equation}
and it is apparent that commutes with $p_y$, i.e. the generator of translations along the $\mathbf{y}$ direction. The fact that such a operator seems to have a gauge dependent spectrum is just an artifact of our gauge choice, indeed from eq. \ref{eq:py_gauge_invariance}
\begin{equation}
X_0=x-\frac{\pi_y}{m \omega_c}=-\frac{p_y}{eB}.
\end{equation}
The eigenvalue of $p_y$ should then be though of as being proportional to the centre of the classical orbit along $\hat{x}$.

Eigenfunctions of $\mathcal{H}_\text{bulk}$ are separable
\begin{equation}
\label{eq:eigenfunctions_tranlational_symmetry}
\psi_{n,k}(x,y)=\frac{e^{i k y}}{\sqrt{L_y}}\,\Phi_{n,k}(x)
\end{equation}
which give
\begin{equation}
\label{eq:y_dof_decoupled}
\underbrace{\left(e^{-i k y} \mathcal{H}_\text{bulk} e^{i k y}\right)\,\,\,\Phi_{n,k}(x)}_{\widetilde{\mathcal{H}}_\text{bulk}(k)\,\Phi_{n,k}(x)}=E \Phi_{n,k}(x)
\end{equation}
with
\begin{equation}
\widetilde{\mathcal{H}}_\text{bulk}(k) = \frac{p_x^2}{2m} + \frac{1}{2}m \omega_c^2 \left(x+\frac{\hbar k}{eB}\right)^2.
\end{equation}
This is just a harmonic oscillator with a shifted minimum, so its spectrum is $k$-independent and equals the usual harmonic oscillator one
\begin{equation}
\label{eq:LandauSpectrum}
E_n=\hbar \omega_c \left(n+\frac{1}{2}\right)
\end{equation}
in agreement with the previous results eq. \ref{eq:Landau_level_algebraic} and eq. \ref{eq:semiclassical_LL}. 
The eigenfunctions are centred at $x_0=-\frac{\hbar k}{eB}$, and localised within a region of whose spatial extension is of the order of the magnetic length $l_B=\sqrt{\frac{\hbar}{eB}}$
\begin{equation}
\label{eq:HO_WF}
\psi_{n,k}(x,y)=\sqrt{\frac{1}{l_B\sqrt{\pi}\,2^n\,n!}}\,H_{n}\left(\frac{x}{l_B}+k\,l_B\right)e^{-\frac{1}{2}\left(\frac{x}{l_B}+k\,l_B\right)^2}\,\frac{e^{i k y}}{\sqrt{L_y}}
\end{equation}
where $H_n(x)$ are Hermite polynomials and the allowed values the wavevector $k$ can assume are quantised in units of $\frac{2\pi}{L_y}$ due to the presence of periodic boundary conditions along the $y$ direction.
\newline Notice that the result \ref{eq:bulk_denisty} for the density of an infinitely extended bulk can be recovered from this discussion; indeed we have
\begin{equation}
\begin{split}
\rho_\text{bulk}=&\sum_{n,k} |\psi_{n,k}(x,y)|^2 \\=& \frac{1}{2\pi}\sum_{n}\,\frac{2\pi}{L_y}\sum_k \frac{1}{l_B\sqrt{\pi}\,2^n\,n!}\,H_{n}^2\left(\frac{x}{l_B}+k\,l_B\right)e^{-\left(\frac{x}{l_B}+k\,l_B\right)^2}.
\end{split}
\end{equation}
In the thermodynamic limit $\frac{2\pi}{L_y}\sum_k\rightarrow\int dk$; shifting the integration variable
\begin{equation}
\rho_\text{bulk}= \frac{1}{2\pi\,l_B^2}\sum_{n}\,\int \frac{1}{\sqrt{\pi}\,2^n\,n!}\,H_{n}^2\left(\mu\right)e^{-\mu^2}\,d\mu.
\end{equation}
Since the harmonic oscillator wavefunctions are normalized the last integral is exactly one; the summation over $n$ just yields the number of occupied Landau levels, and thus as expected
\begin{equation}
\rho_\text{bulk}= \frac{\nu}{2\pi\,l_B^2}.
\end{equation}

Finally it is interesting to compute again the degeneracy of the levels. The semiclassical orbit associated to a Landau level must lie between the x-edges of the sample, i.e. \mbox{$-\frac{L_x}{2}\leq x_0 \leq \frac{L_x}{2}$;} periodic boundary conditions along $y$ however require $k=\frac{2\pi}{L_y}n$. Thus $-\frac{1}{2}\frac{\Phi_B}{\Phi_0}\leq n \leq \frac{1}{2}\frac{\Phi_B}{\Phi_0}$, and the number of states within a single Landau level is
$\mathcal{N}=\frac{\Phi_B}{\Phi_0}$, as obtained above.

\subsubsection{The integer quantum Hall effect}
We can now give a naïve explanation of the integer quantum Hall effect .
\newline If we switch on a constant electric field along $\hat{x}$, the Hamiltonian will include an electrostatic potential term as well
\begin{equation}
\Delta\mathcal{H}=-e\phi=+eEx
\end{equation}
which causes a current to flow in the $\hat{y}$ direction, as also seen classically in section \ref{section:classical_electric_field}. 
\newline I will take a semiclassical route here. The classical current density is described by means of the usual expression
\begin{equation}
J_y(\mathbf{r})=-e\sum_i v_{y,i}\,\delta^{(2)}(\mathbf{r}-\mathbf{r}_i)=-\frac{e}{m}\sum_i \pi_{y,i}\,\delta^{(2)}(\mathbf{r}-\mathbf{r}_i)
\end{equation}
the summation being over all the system electrons.
We define the quantum current density operator by symmetrizing such an expression 
\begin{equation}
J_y(\mathbf{r})=-\frac{e}{2m}\sum_i \left(\pi_{y,i}\,\delta^{(2)}(\mathbf{r}-\mathbf{r}_i)+\delta^{(2)}(\mathbf{r}-\mathbf{r}_i)\pi_{y,i}\right)
\end{equation}
which is a one-body operator. If we deal with a system of non-interacting particles, its expectation value simplifies considerably
\begin{equation}
\label{eq:semiclassical_derivation_current}
\begin{split}
\braket{J_y(\mathbf{r})}=&-\frac{e}{2m}\sum_\alpha\braket{ \pi_{y,1}\,\delta^{(2)}(\mathbf{r}-\mathbf{r}_1)+\delta^{(2)}(\mathbf{r}-\mathbf{r}_1)\pi_{y,1}}_\alpha=\\
=&
-\frac{e}{2m}\sum_\alpha\int dx_1\,dy_1 \,\Psi_\alpha^*\left( \pi_{y,1}\,\delta^{(2)}(\mathbf{r}-\mathbf{r}_1)+\delta^{(2)}(\mathbf{r}-\mathbf{r}_1)\pi_{y,1}\right)\Psi_\alpha=\\
=&
-\frac{e}{2m}\sum_\alpha\left(\Psi_\alpha^*\left(-i \hbar \frac{\partial}{\partial y}+e B x\right)\Psi_\alpha +
\Psi_\alpha\left(i \hbar \frac{\partial}{\partial y}+e B x\right)\Psi_\alpha^*\right)
\end{split}
\end{equation}
where the summation runs over all the filled states. The last equation can be recognized to be the usual Schroedinger probability current (in the presence of a magnetic field the canonical momenta are replaced by the kinetic ones) multiplied by the charge of the electron, evaluated for each non-interacting constituent of the system and then summed over, as one could have written down from scratch. (A more rigorous derivation is briefly discussed in the following chapter.)
\newline Such a current operator can be averaged over the size of the sample
\begin{equation}
\label{eq:JyResponse}
\begin{split}
\overline{J}_y=&\int\frac{dx}{L_x}\,\frac{dy}{L_y}\,\braket{J_y(\mathbf{r})} = -\frac{1}{L_xL_y}\,\frac{e}{m}\,\sum_\alpha \int\Psi_\alpha^*\left(-i \hbar \frac{\partial}{\partial y}+e B x\right)\Psi_\alpha\,dx\,dy=\\
=&
-\frac{1}{L_xL_y}\,\frac{e}{m}\,\sum_\alpha \braket{\pi_y}_\alpha.
\end{split}
\end{equation}
This can be easily computed if all the electron wavefunctions are eigenfunctions of $\mathcal{H}+\Delta\mathcal{H}$: notice that the expectation value of the commutator
\begin{equation}
\left[\mathcal{H}_\text{bulk}+eEx,p_x\right]=i\hbar\left[\frac{eB}{m}\left(p_y+eBx\right)+eE\right]
\end{equation}
vanishes identically on these eigenstates. Then we have (as expected on purely classical grounds)
\begin{equation}
\braket{\pi_y} = -m\,\frac{E}{B}
\end{equation}
which gives, for $\nu$ fully filled Landau levels,
\begin{equation}
\begin{split}
\overline{J}_y=
\frac{1}{L_xL_y}\,e\frac{E}{B}\,\sum_\alpha = E\,\frac{e^2}{h}\nu.
\end{split}
\end{equation}
The Hall conductivity is defined to be the linear response coefficient 
\begin{equation}
\label{eq:HallConductivity0}
\sigma_\text{Hall} = \frac{\overline{J}_y}{E} = \nu\, \frac{e^2}{h}
\end{equation}
and it evidently appears to be an integer multiple of $\frac{e^2}{h}$.

This simple explanation however is still lacking some important detail: we basically \virgolette{fine-tuned} the number of electrons in the system in order to have exactly $\nu$ filled Landau levels. This will not in general be the case, since as the magnetic field is varied the degeneracy of each Landau level changes continuously.
\newline The system edges provide a mechanism to remedy to such a drawback\footnote{Actually, system disorder is another fundamental ingredient in the discussion; however I will not discuss about it in this thesis work.}, as will be discussed below when the Hall conductivity will be computed again by including the edges from scratch.

\section{The edges}
As before, we begin with a classical discussion, since it helps developing an intuitive understanding of the physics of the problem. 
\newline We then move to the full quantum treatment of the problem; the associated eigenvalue problem is numerically solved, and some discussion is made. The existence of peculiar symmetry points which admit an analytical (and easy) solution is highlighted; this allows some simple considerations about the behaviour of the energy spectrum to be made, as well as some re-discussion of the Landau level degeneracy $\mathcal{N}$ in the presence of edges (previously obtained using periodic boundary conditions).
\newline Finally, a perturbative analysis is discussed, which leads to substantially the same results obtained from the classical discussion.

\subsection{Classical approach}
We can now understand what will happen near the system edges. Suppose these are included in the treatment by means of a confining potential $V_c(x)$, and that we can linearise $V_c(x)$ neglecting quadratic terms over the whole extension $\sim\widetilde{r}_0$ of the cyclotron orbit along the x direction. Then
\begin{equation}
V_c(x)\simeq V_c(x_0)+(x-x_0)V'_c(x_0)+\mathcal{O}(x-x_0)^2=\text{constant} + V'_c(x_0) x.
\end{equation}
If we drop the constant term (being just an irrelevant energy shift) we see that the Hamiltonian is perfectly analogous to the one in the presence of a uniform electric field along the $x$ direction, and the results above can be extended to this case too just by substituting $eE\rightarrow V'_c(x_0)$. 
\newline In the presence of the edges, we see that the electron orbit acquires a $y$ velocity which does not vanish if we average over the cyclotron period $\omega_c^{-1}$, and that it is proportional to the gradient of the confining potential
\begin{equation}
\label{eq:semiclassical_drifting_speed}
v=-\frac{1}{m\omega_c}\,\left.\frac{\partial V_c}{\partial x}\right|_{x_0}=-\frac{1}{eB}\,\left.\frac{\partial V_c}{\partial x}\right|_{x_0}.
\end{equation}
This result can as well be heuristically understood in terms of skipping orbits, i.e. electrons elastically bouncing on a hard-wall confinement and chirally moving along the one-dimensional boundary of the system.

\subsection{Quantum mechanical viewpoint}\label{section:quantum_edges}
\begin{figure}[htp!]
	\centering
	\includegraphics[width=1.\textwidth]{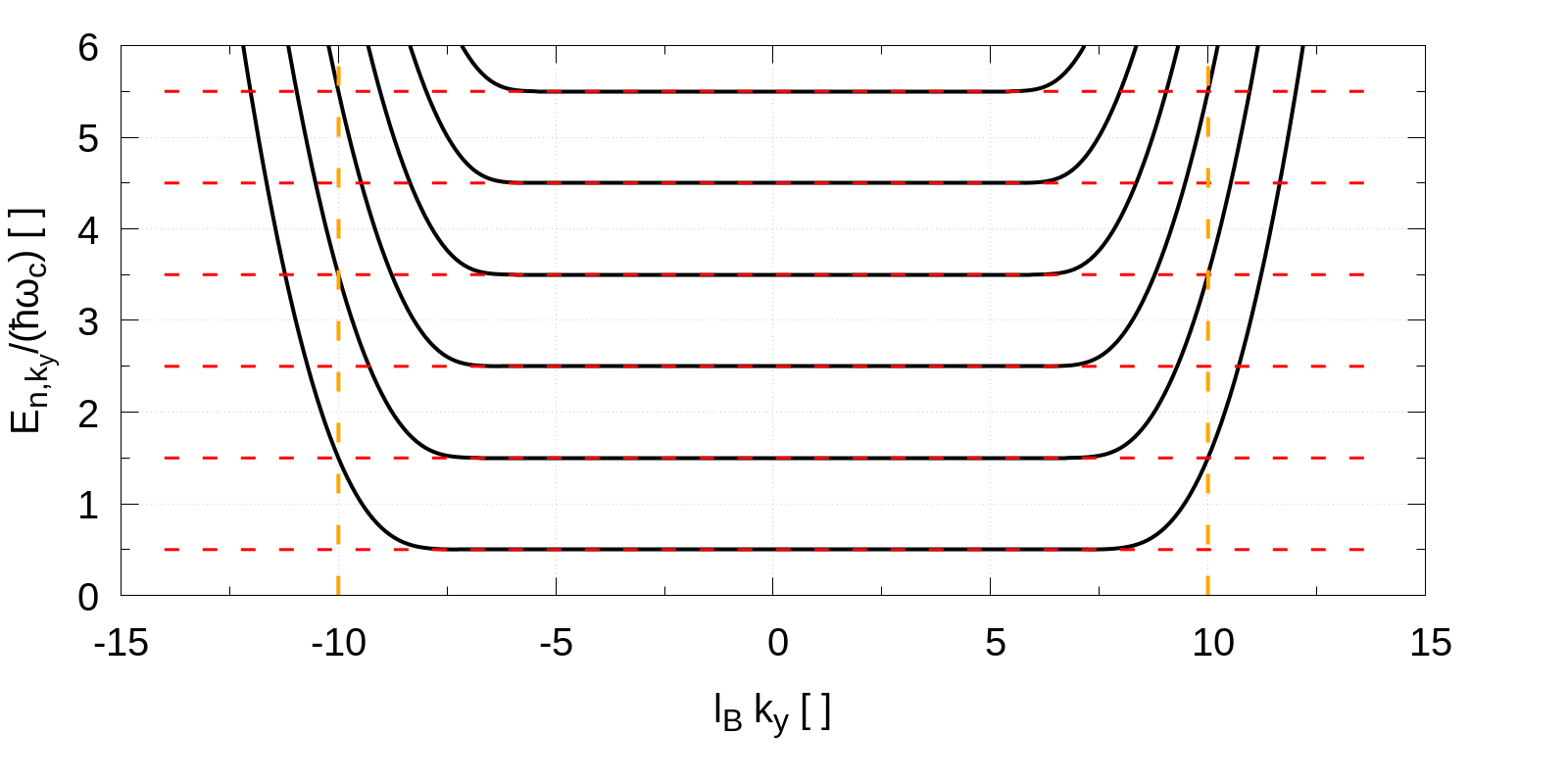}
	\caption[The LOF caption]{The numerically computed spectrum of the bulk Hamiltonian (eq. \ref{eq:bulk_hamiltonian}) in the presence of an hard-wall confining potential is shown. Dashed orange lines at $k=\pm\frac{L_x}{2l_B^2}$ are drawn, together with red dashed lines at the bulk Landau level energies (eq. \ref{eq:LandauSpectrum}). 
	\newline The system length has been chosen to be $L_x=20l_B$.}.
	\label{fig:DBCspectrum}
\end{figure}

\begin{figure}[htp!]
	\begin{minipage}{.5\textwidth}
		\centering
		\includegraphics[width=1.\textwidth]{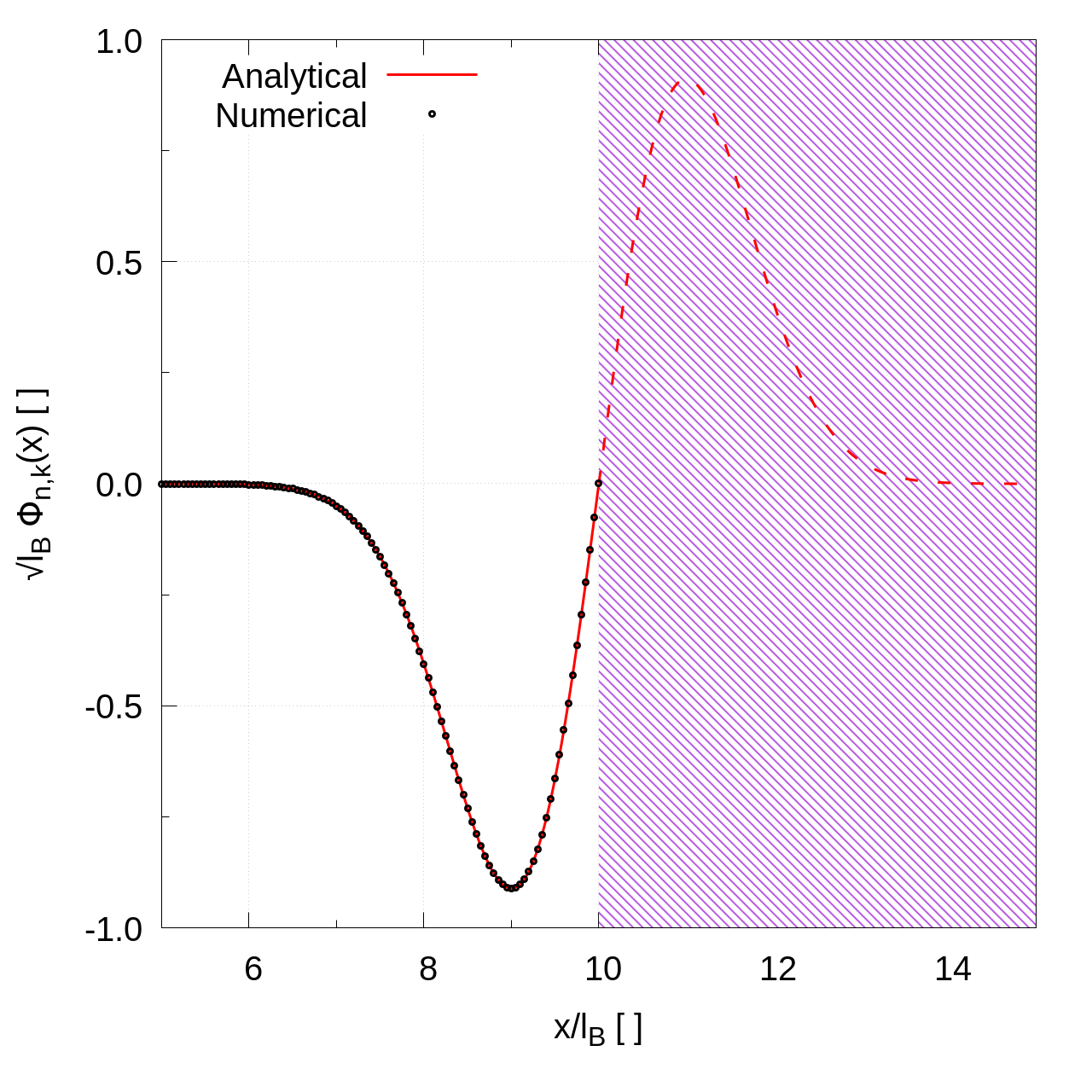}
	\end{minipage}%
	\begin{minipage}{0.5\textwidth}
		\centering
		\includegraphics[width=1.\textwidth]{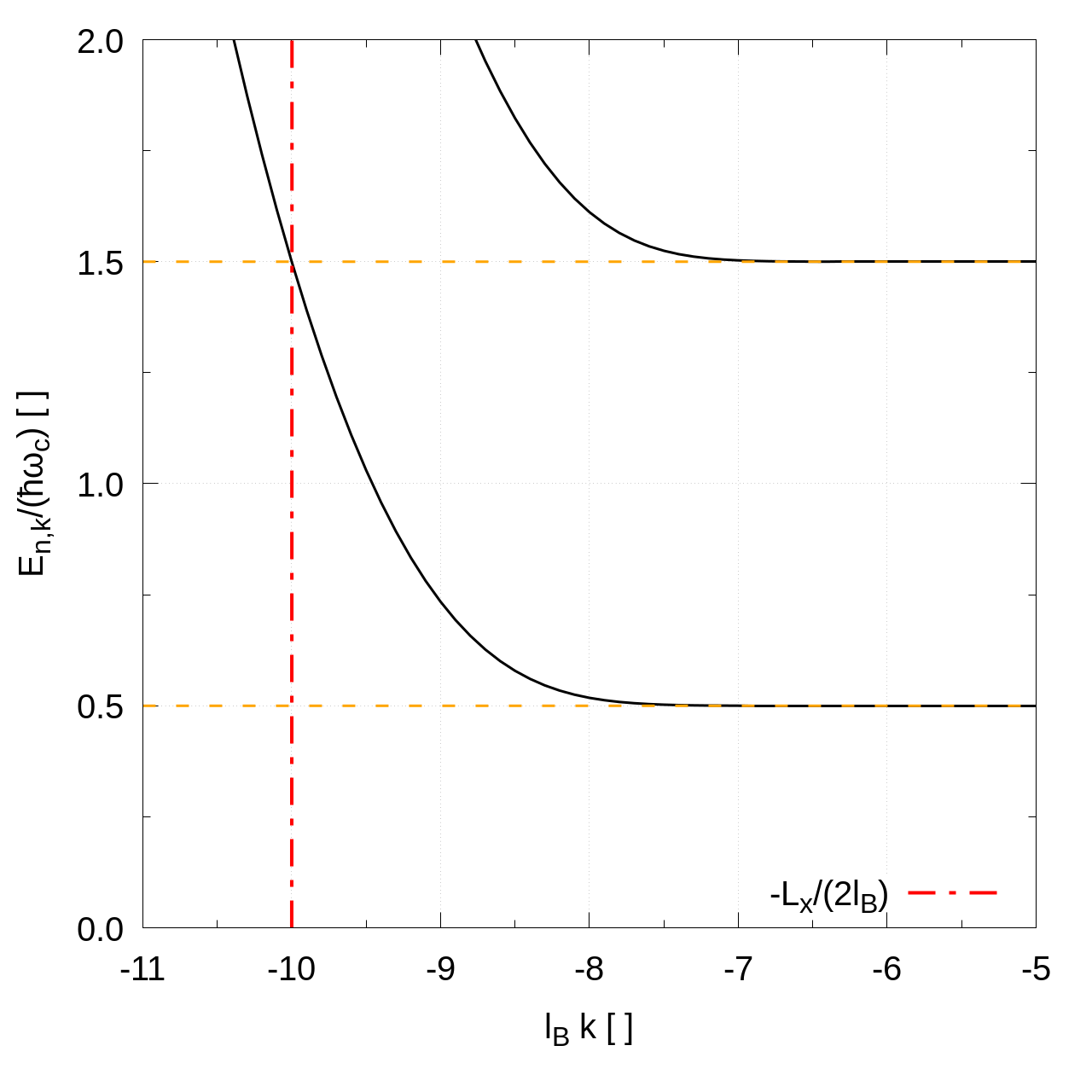}
	\end{minipage}    
	\caption[The LOF caption]{On the left hand side, the numerically computed eigenstate of the bulk Hamiltonian (eq. \ref{eq:bulk_hamiltonian}) in the presence of an hard-wall confining potential in the case $n=0$, $l_B k=-\frac{L_x}{2l_B}$ is compared with properly shifted and normalized $n=1$ bulk eigenstates (eq. \ref{eq:HO_WF}). On the right hand side the spectrum is plotted instead, and dashed lines at the relevant wavevectors are drawn. 
		\newline The system length has been chosen to be $L_x=20l_B$.}.
	\label{fig:EnergySpectrum0}
\end{figure}

The inclusion of a confining potential $V_c(x)$ makes an analytical closed form solution for the problem (to the best of my knowledge) unfeasible.
However some understanding can be gained by enclosing the system between $\pm\frac{Lx}{2}$ with the simplest possible model potential, a hard-wall potential.
\newline In Fig. \ref{fig:DBCspectrum} the spectrum of eq. \ref{eq:bulk_hamiltonian} with Dirichlet boundary conditions\footnote{The spectrum, as well as the wavefunctions, have been obtained using a numerical procedure substantially identical to the one discussed below in sec. \ref{sec:NumericalProcedureStaticProperties}. All the numerical details can be found in such a section.} is shown. 
Some comments are mandatory. In the bulk one recognises the structure of the quantised Landau levels; near the system edge however these levels \virgolette{bend upwards} due to the increased kinetic energy (this point will be discussed in greater detail below).
One immediately notices at $k=\pm\frac{L_x}{2l_B^2}$ the energy of the $n$-th level seems to be remarkably close to the bulk energy of an excited $2n+1$ level. This is not a coincidence indeed.
\newline Diagonalising the bulk Hamiltonian (eq. \ref{eq:bulk_hamiltonian}) in a box is indeed equivalent to look for a solution of the time-independent Schrödinger equation with Dirichlet boundary conditions, i.e. with the requirements that the solutions have nodes at the two edges and vanish outside of the domain. 
\newline We assume that $Lx\gg l_B$, so that a wavefunction localized at one edge is exponentially small at the other one (due to the harmonic confinement).
If we choose the wavevector to lie exactly at the edge, i.e. $k=-\frac{Lx}{2l_B^2}$, the solution is easy to find since the minimum of the Harmonic potential coincides with one of the walls. 
Since every odd solution of the quantum harmonic oscillator has (by symmetry) a node at the bottom of the well which confines the particle, we can conclude that these functions when properly shifted (and set to zero outside of the domain) are exact solution of the problem
\begin{equation}
\psi_{n, \pm\frac{Lx}{2l_B^2}}\propto\psi_{2n+1}^{\text{ho}}\left(x\pm\frac{L_x}{2}\right)\theta\left(\frac{L_x}{2}\pm x\right)
\end{equation}
where $\theta$ is the Heaviside step function. The proportionality constant being just a numerical factor (the wavefunction needs to be properly normalised). That the $(2n+1)$-th bulk level is the one related to the $n$-th edge level can easily be understood by some simple considerations on the number of nodes of the two wavefunctions. The $n$-th edge level has exactly $n-1$ nodes in the sample and $1$ at the edge; on the other hand the bulk state related to it has the same nodes within the sample but $n-1$ additional ones lying outside of the boundary, for a total of $2n+1$ nodes.
\newline The spectrum at this symmetry point reads
\begin{equation}
\label{eq:steep_edge_edge_energy}
E_{n, \pm\frac{Lx}{2l_B^2}}=\hbar \omega_c\Bigl((2n+1)+	\frac{1}{2}\Bigr)=E^{(\text{ho})}_{2n+1}
\end{equation}
as indeed expected when we first looked at Fig. \ref{fig:DBCspectrum}.
\newline In Fig. \ref{fig:EnergySpectrum0} numerically computed eigenvectors and energies of $\ref{eq:bulk_hamiltonian}$ with Dirichlet boundary conditions modelling the edges are compared with the results that emerged from the previous discussion. (More details on the numerical procedure will be given below; the images here are shown only with clarifying purposes).

This fact has a nice consequence. Without the presence of periodic boundary conditions along $y$, we could theoretically put as many electrons as we wish into the $n=0$ level, but $\mathcal{N}=\frac{L_xL_y}{2\pi l_B^2}$ (equal to the Landau level degeneracy \ref{eq:ll_degeneracy}) is the amount which keeps the highest energetic occupied state below the first excited Landau level.

\begin{figure}[htp!]
	\begin{minipage}{.5\textwidth}
		\centering
		\includegraphics[width=1.\textwidth]{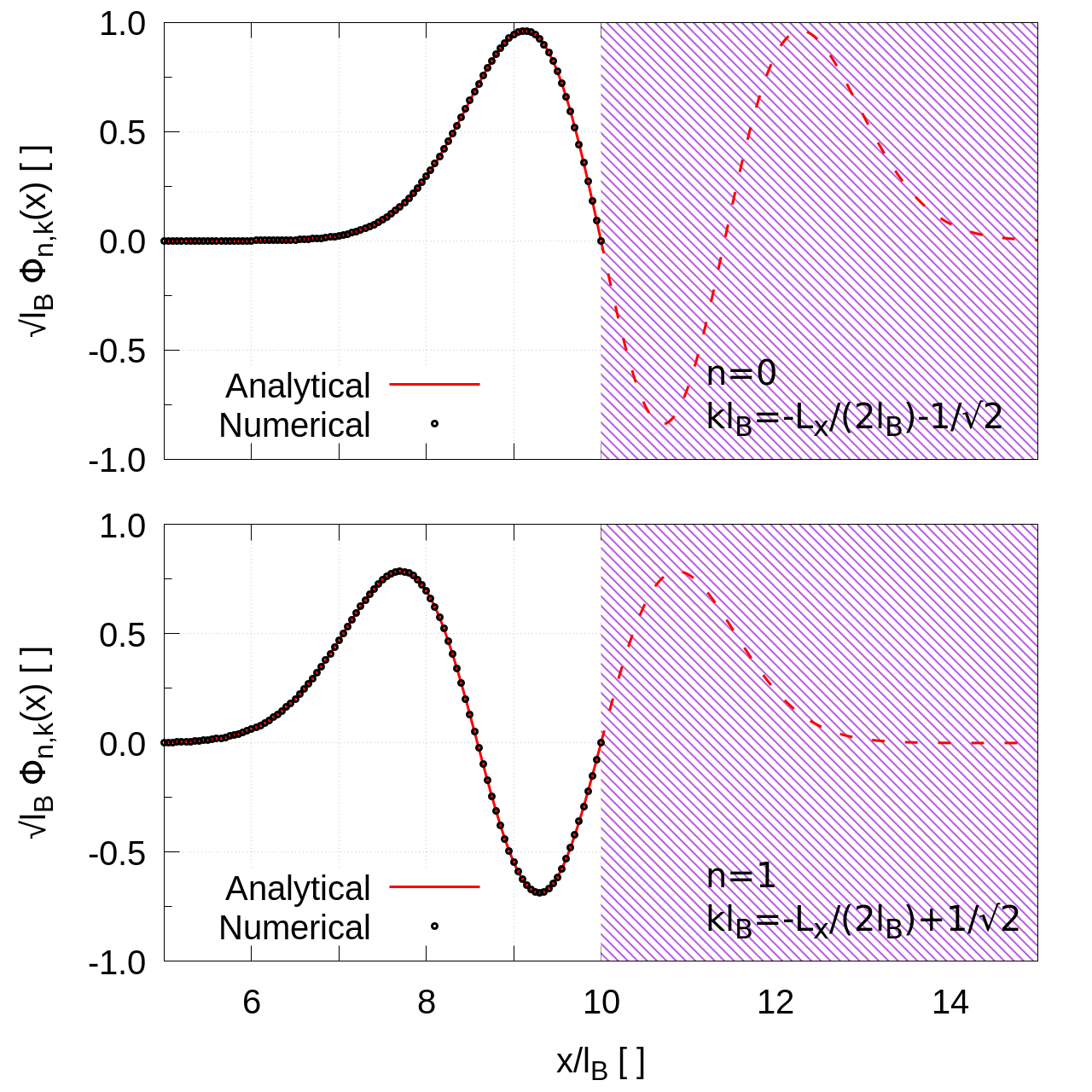}
	\end{minipage}%
	\begin{minipage}{0.5\textwidth}
		\centering
		\includegraphics[width=1.\textwidth]{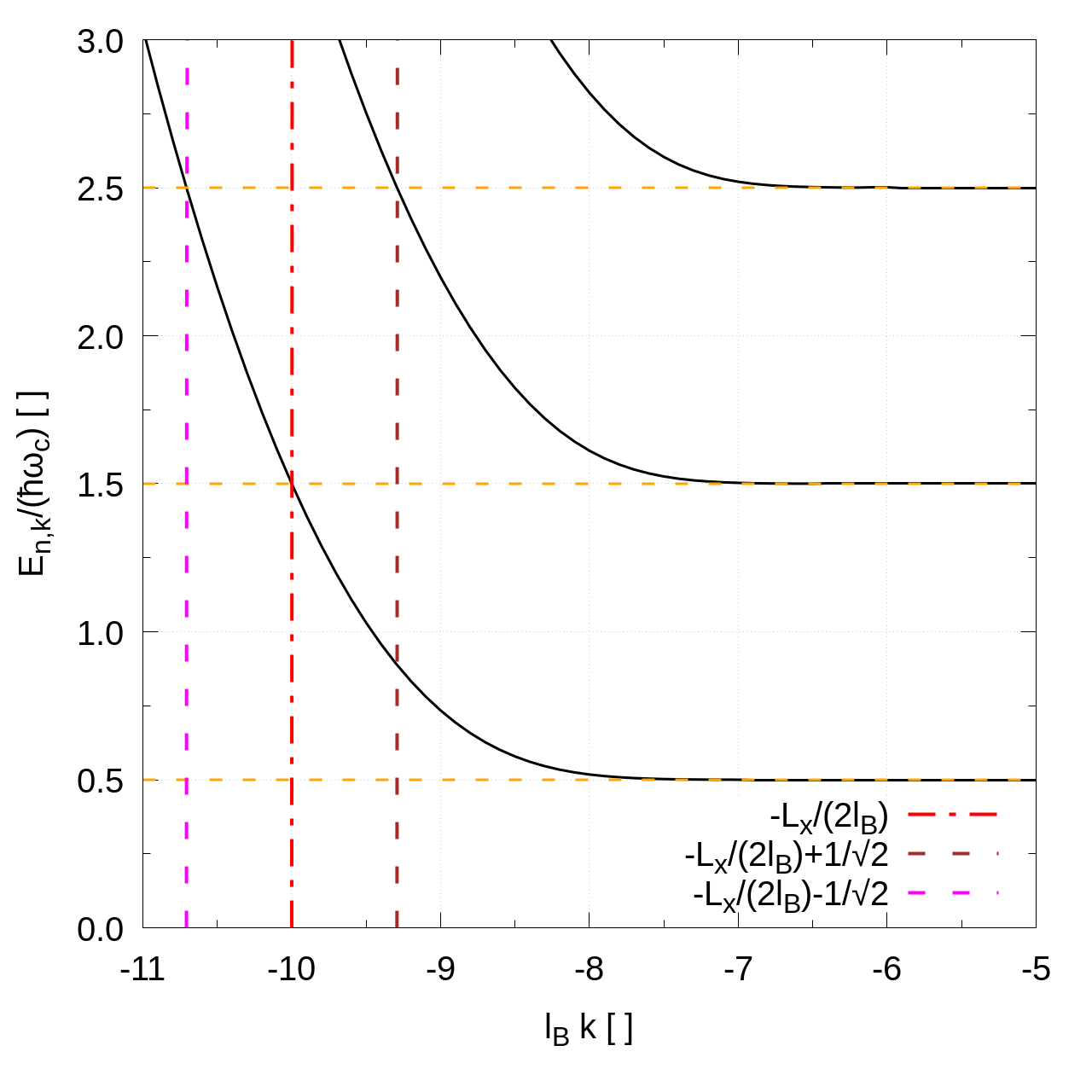}
	\end{minipage}    
	\caption[The LOF caption]{On the left hand side, the numerically computed eigenstates of the bulk Hamiltonian (eq. \ref{eq:bulk_hamiltonian}) in the presence of an hard-wall confining potential in the cases $n=0$, $l_B k=-\frac{L_x}{2l_B}-\frac{1}{\sqrt{2}}$ (top) and $n=1$, $l_B k=-\frac{L_x}{2l_B}+\frac{1}{\sqrt{2}}$ (bottom) are shown and compared with properly shifted and normalized\footnote{The new normalization coefficients can be analytically computed in terms of error functions, but are not reported here since not much insightful.} $n=2$ bulk eigenstates (eq. \ref{eq:HO_WF}). On the right hand side the spectrum is plotted instead, and dashed lines at the relevant wavevectors are drawn. 
	\newline The system length has been chosen to be $L_x=20l_B$.}.
	\label{fig:EnergySpectrum1}
\end{figure}

One could hope to generalise this fact to include higher Landau levels as well; the reasoning made above concerning the special point in momentum space $k=\pm\frac{L_x}{2l_B^2}$ can indeed be generalised with little effort and extended to include some other special values of $k$. 
\newline An edge eigenstate belonging to the $n$-th Landau level will equal a truncated harmonic oscillator wavefunction whenever the latter (eq. \ref{eq:HO_WF}) has a node at the edge and the correct number of nodes in the sample.
\newline As an example, consider the $n=2$ harmonic oscillator wavefunction (eq. \ref{eq:HO_WF}). 
\newline For $k=-l_B^{-1}\left(\frac{L_x}{2l_B}+\frac{1}{\sqrt{2}}\right)$, we have a node at the positive sample edge $x=\frac{L_x}{2}$ and no other node for $x\leq\frac{L_x}{2}$; the restriction of such a function to the inner part of the sample\footnote{The function needs to be normalised accordingly.} provides therefore the proper edge $n=0$, $k=-l_B^{-1}\left(\frac{L_x}{2l_B}+\frac{1}{\sqrt{2}}\right)$ Landau level, with an energy $\frac{5}{2}\hbar\omega_c$. 
The situation is graphically represented in the top left panel of Fig. \ref{fig:EnergySpectrum1}; in the right hand side image the energy levels are shown instead.
\newline Analogously, for $k=-l_B^{-1}\left(\frac{L_x}{2l_B}-\frac{1}{\sqrt{2}}\right)$, the $n=2$ harmonic oscillator wavefunction has a node at the positive sample edge $x=\frac{L_x}{2}$ and just one node for $x\leq\frac{L_x}{2}$; 
the restriction of this function to the inner part of the sample is therefore the proper $n=1$, $k=-l_B^{-1}\left(\frac{L_x}{2l_B}-\frac{1}{\sqrt{2}}\right)$ Landau level, again with energy $\frac{5}{2}\hbar\omega_c$.
The situation is graphically represented in the bottom left image of Fig. \ref{fig:EnergySpectrum1}. The energy levels are shown instead in the right hand side panel.

\noindent The maximum number of electrons which can be accommodated in the lowest two ($n=0$ and $n=1$) Landau levels to build a proper ground state is $2\mathcal{N}$, as in the case analysed before with periodic boundary conditions; 
however, not $\mathcal{N}$ per Landau level, rather the number of electrons which can be accommodated in the lowest level is slightly increased ($\mathcal{N}+\frac{2}{l_B\Delta k\sqrt{2}}$) while the occupation number of the $n=1$ level correspondingly decreased.

This example makes clear that the special values of $k$ we talked about above are zeroes of the Hermite polynomials.

\begin{figure}[htp!]
	\begin{minipage}{.5\textwidth}
		\centering
		\includegraphics[width=1.\textwidth]{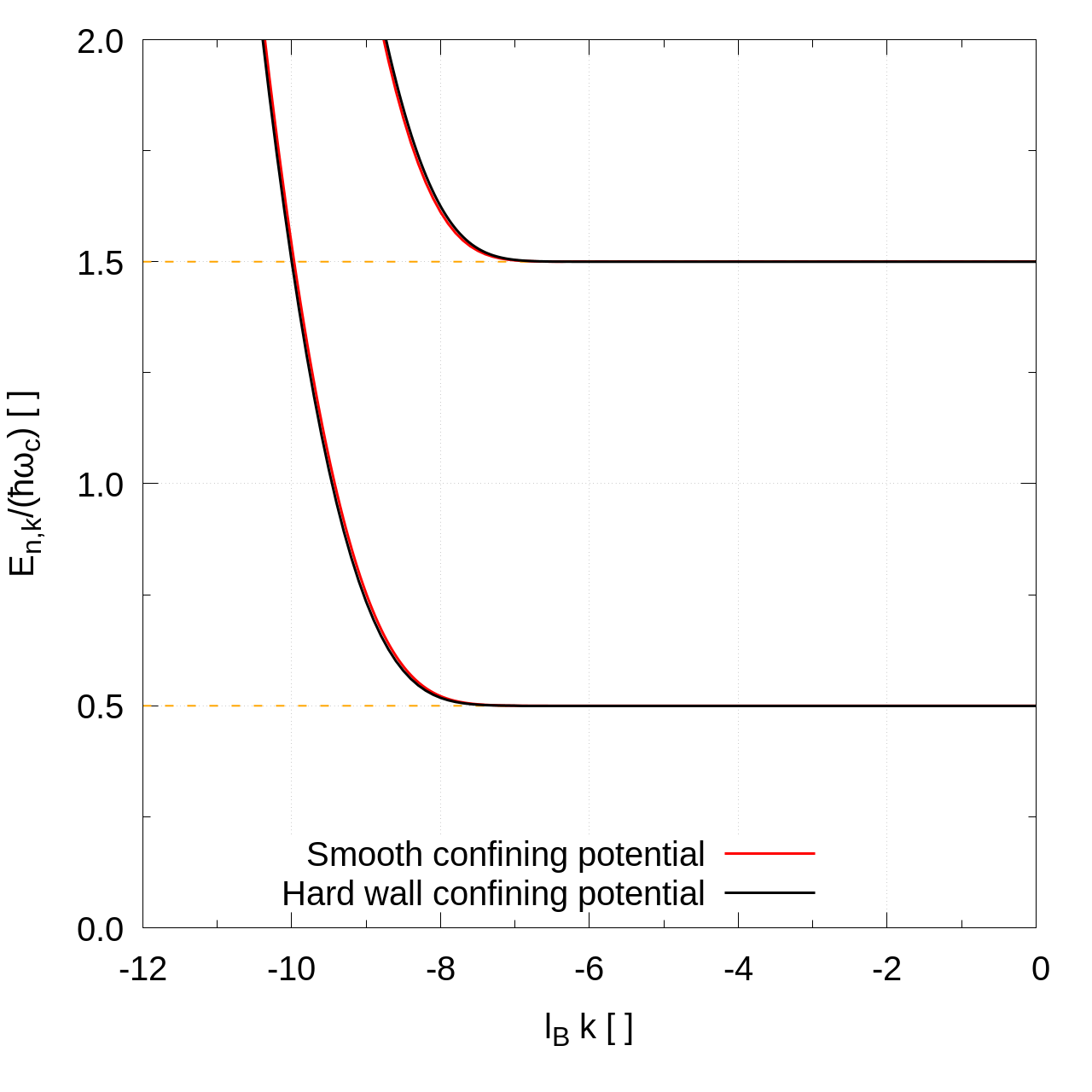}
	\end{minipage}%
	\begin{minipage}{0.5\textwidth}
		\centering
		\includegraphics[width=1.\textwidth]{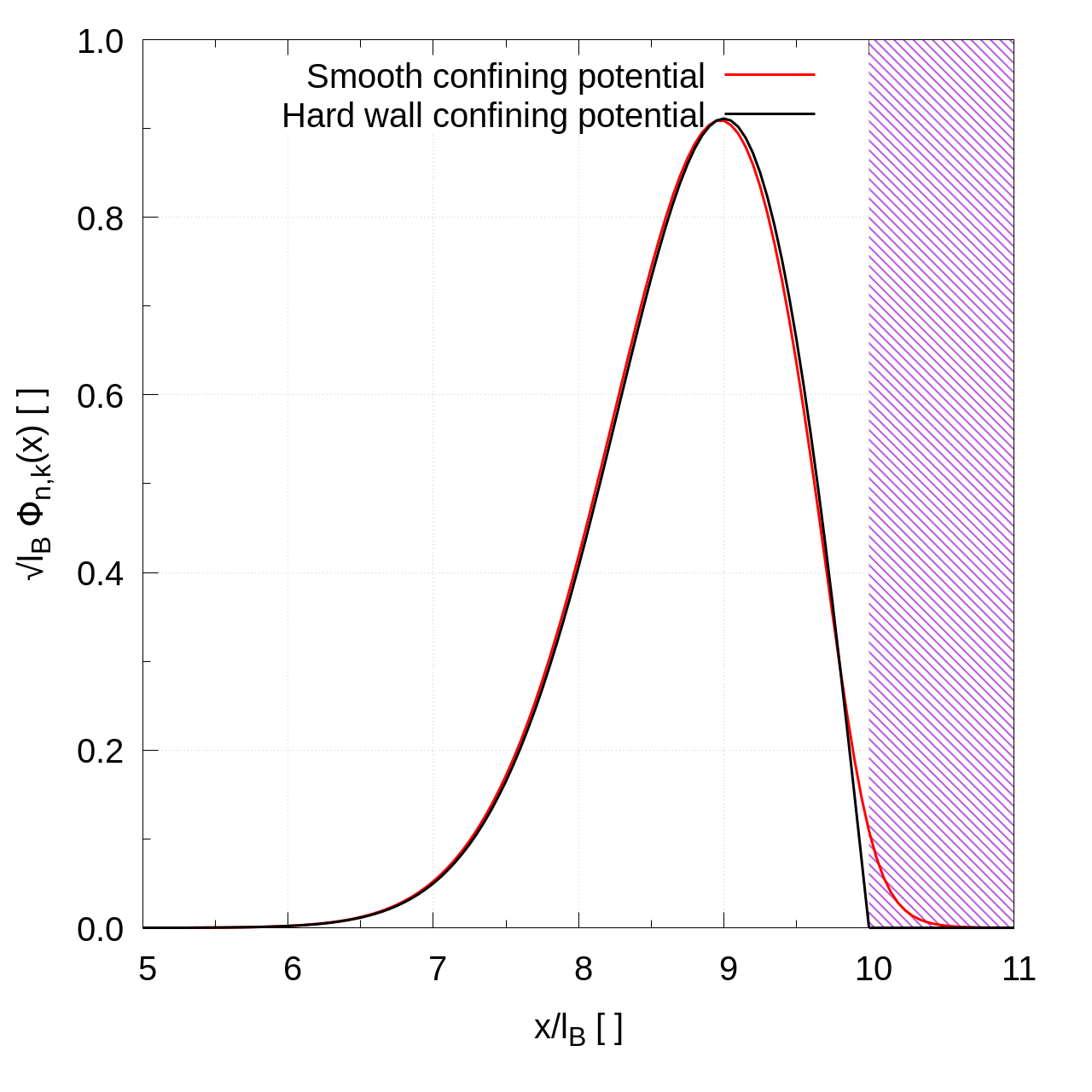}
	\end{minipage}    
	\caption[The LOF caption]{The image on the left compares numerical results for the energy spectrum computed using hard-wall boundary conditions at $L_x=20l_B$ with those obtained by using the smooth confining potential that will be introduced below (eq. \ref{eq:confining_potential}, using $\sigma_c=0.1l_B$ and $V_0=30\hbar\omega_c$).
		\newline On the right hand side the $n=0$, $l_B k=-\frac{L_x}{2l_B}$ wavefunctions in the same cases are directly compared.}.
	\label{fig:BCcomparison}
\end{figure}

If the confining potential is not hard-wall like, this argument evidently ceases to be true; however, if the potential rises steeply on a scale $\ll l_B$and its typical magnitude $V_c\gg\hbar\omega_c$ these conclusions will still be good approximations, as can be seen from the images in Fig. \ref{fig:BCcomparison}.

\noindent More generally we expect the energy associated to a given electronic state to increase as it moves from the bulk towards the edges, since it will get \virgolette{squeezed} between the hard wall and the harmonic potential; the confinement becoming tighter, the kinetic energy will henceforth increase, as can be seen in the next-section Figures \ref{fig:EnergySpectrum} (right hand side image) and \ref{fig:real_and_momentum_space_widths} (left hand side image).

There is still one more point which is worth discussing about the nature of the edge states.
\newline Consider a bulk system first; suppose we have a completely filled Landau level. The lowest lying excitations of such a system are gapped, the gap being the Landau level splitting $\hbar\omega_c$. The consequence is that the bulk of the system substantially behaves as an insulator.
On the other hand, near the edges the Landau levels do tilt upwards. Near the Fermi points there will be unoccupied states to which an edge electron can jump to with vanishingly small energy cost $\approx v q$, $q$ being the \virgolette{momentum kick} associated to the transition and $v$ the slope of the dispersion at the Fermi point. In this respect the edges behave thus as a conductor.
\newline We will come back on this point later on in this chapter, in section \ref{section:ParticleHole}.

\subsubsection*{Perturbative viewpoint}
Some further insight on the edge physics can be gained via approximation methods.
If the lengthscale over which the confining potential varies appreciably is much larger than the magnetic length $l_B$, the electron labelled by the quantum numbers $(n,k)$ will only feel a constant energy shift set by
\begin{equation}
E_{n,k} \simeq \hbar \omega_c\left(n+\frac{1}{2}\right)+\bra{n,k}V_c\ket{n,k}.
\end{equation}
which under the same approximations can be replaced with
\begin{equation}
E_{n,k} \simeq \omega_c\left(n+\frac{1}{2}\right)+V_c\left(-\frac{\hbar k}{eB}\right).
\end{equation}
The energy bands become $k$-dependent, or, equivalently, do depend on $x_0$.
\newline The average velocity for a particle with momentum $k$ in a Landau level $n$ will be given by
\begin{equation}
v_{n,k}=\frac{1}{\hbar}\,\frac{\partial E_{n,k}}{\partial k}\simeq -\frac{1}{eB}\,\frac{\partial V_c}{\partial x_0}
\end{equation}
where $x_0=-\frac{\hbar k}{eB}$. This is the same result obtained above, from purely classical Newtonian dynamics (eq. \ref{eq:semiclassical_drifting_speed}). 
The speed has different signs on opposite sides of the sample; this implies that along the one-dimensional boundary particles will move in a preferential direction, i.e. they are chiral.
\newline These approximations give an heuristic idea of what should happen at the edge of the system, but they are inevitably going to quantitatively fail, since one expects the confining potential to rise steeply.
However the main conclusions will hold even in this case; as already stated the effect of a \virgolette{high and steep} edge is indeed to asymmetrically \virgolette{squeeze} the harmonic oscillator parabola, leading to an increased energy for the states with higher $|k|$, i.e. those which are localized nearer to the edge, thus giving rise to a non-zero velocity $\frac{1}{\hbar}\frac{\partial E_{n,k}}{\partial k}$ which has opposite sign on opposite sides of the sample.

\section{Diagonalization of the semi-infinite free system single particle Hamiltonian}
We now discuss the numerical solution of the eigenvalue problem $\mathcal{H}\psi_{n,k}=E_{n,k}\psi_{n,k}$ with the Hamiltonian \ref{eq:problem_hamiltonitan}.
\newline The $y$ degree of freedom can be decoupled easily; thanks to translational symmetry along the $y$ direction, the eigenfunctions have the structure in eq. \ref{eq:eigenfunctions_tranlational_symmetry}, and transforming the Hamiltonian as in \ref{eq:y_dof_decoupled} we get
\begin{equation}
\label{eq:problem_hamiltonitan_transformed}
\widetilde{\mathcal{H}}_k = \frac{p_x^2}{2m} + \frac{1}{2}m \omega_c^2 \left(x+\frac{\hbar k}{eB}\right)^2 + V_c(x).
\end{equation}
The problem is purely one-dimensional, and can be easily solved numerically once $V_c$ is known.
If the lengthscale over which $V_{c}$ rises steeply is much smaller than the magnetic length $l_B$, and if the magnitude of $V_{c}\gg \hbar \omega_c$, the exact shape of the confining potential is to any extent irrelevant. A symmetric form has been used
\begin{equation}
\label{eq:confining_potential}
V_c(x) = \frac{V_0}{\exp\left[-\frac{1}{L_x\sigma_\text{c}}\left(x^2-\left(\frac{L_x}{2}\right)^2\right)\right]+1}
\end{equation}
where $\sigma_c$ characterizes the lengthscale over which $V_c$ rises from $V\sim0$ to $V\sim V_0$ near $x=\pm\frac{L_x}{2}$; indeed in a neighbourhood of say $x=\frac{L_x}{2}$ we can approximate $x^2-\left(\frac{L_x}{2}\right)^2\approx L_x(x-\frac{L_x}{2})$, obtaining
\begin{equation}
V_c\left(x-\frac{L_x}{2}\ll L_x\right) \simeq \frac{V_0}{\exp\left[-\frac{1}{\sigma_\text{c}}\left(x-\frac{L_x}{2}\right)\right]+1}
\end{equation}
and we see that $\sigma_c$ regulates as stated the steepness of the edge potential.
\begin{figure}[htp!]
	\begin{minipage}{.5\textwidth}
		\centering
		\includegraphics[width=1.\textwidth]{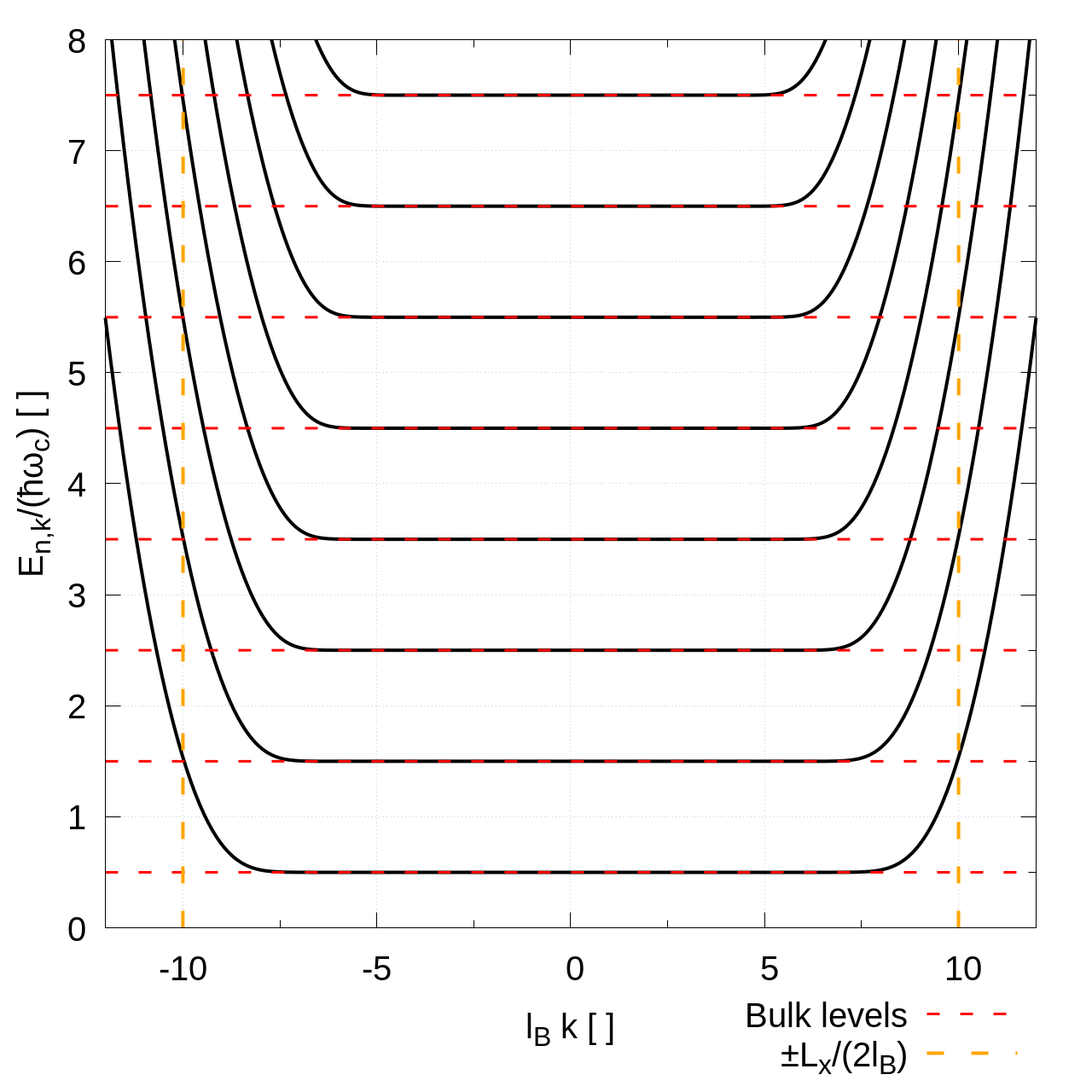}
	\end{minipage}%
	\begin{minipage}{0.5\textwidth}
		\centering
		\includegraphics[width=1.\textwidth]{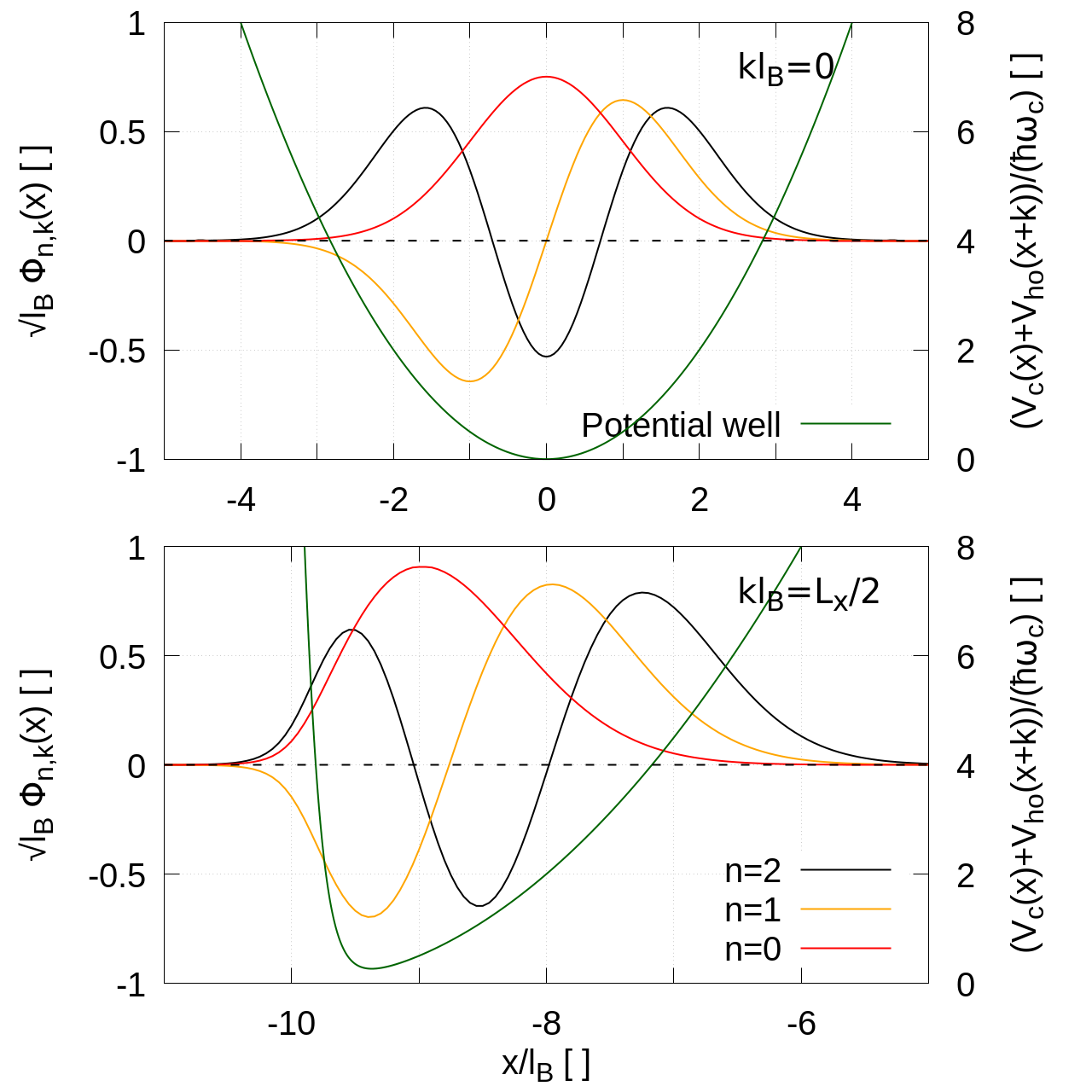}
	\end{minipage}    
	\caption[The LOF caption]{The image on the left shows the first eight Landau levels for a system of length $L_x=20l_B$, and confining parameters $V_0=30\hbar\omega_c$, $\sigma_c=0.1l_B$. The right hand side panel shows some bulk (top panel) versus edge (bottom) wavefunctions. On the second y-axis the total effective potential well in which the electron is trapped (magnetic field plus confining potential) is shown.}
	\label{fig:EnergySpectrum}
\end{figure}

The potential is symmetric with respect to $x=0$; the spectrum will then be symmetric about $k=0$, since the Hamiltonian is left invariant by the joint transformation\footnote{Let $\mathcal{T}$ be the operator implementing such a transformation. Then
	\begin{equation}
	\widetilde{\mathcal{H}}_{k}\mathcal{T}\Phi_{n,k}(x)
	=\left(\mathcal{T}\widetilde{\mathcal{H}}_{k}\mathcal{T}^{-1}\right)\mathcal{T}\Phi_{n,k}(x)
	=\mathcal{T}\widetilde{\mathcal{H}}_k\Phi_{n,k}(x)=E_{n,k} \mathcal{T}\Phi_{n,k}(x)
	\end{equation}
	which tells us that $\mathcal{T}\Phi_{n,k}(x)=\Phi_{n,-k}(-x)$ is an eigenvalue of $\widetilde{\mathcal{H}}_k$ with the same energy of $\Phi_{n,k}(x)$, i.e. $E_{n,k}=E_{n,-k}$.
	Moreover, $\Phi_{n,-k}(-x)$ and $\Phi_{n,k}(x)$ must be equal up to an irrelevant phase.} 
$x\rightarrow-x$, $k\rightarrow-k$. 
\newline The one-dimensional Hamiltonian written above (eq. \ref{eq:problem_hamiltonitan_transformed}) can easily be diagonalized by standard methods (e.g. Numerov algorithm or diagonalization of the matrix problem emerging when replacing the derivatives by finite differences). The one I used is briefly explained in the following section, as well as some detail on the boundary conditions.
\newline From the diagonalization procedure one obtains both the spectrum $E_{n,k}$ and the mutually orthogonal stationary states, which can be normalized to give
\begin{equation}
\int_{-\infty}^{\infty}\Phi^*_{m,k}(x)\Phi_{n,k}(x)\,dx =\delta_{n,m}.
\end{equation}
Notice that such a orthogonality relation holds only if the two states have the same wavevector $k$ (i.e. the two harmonic oscillators need not to be shifted one with respect to the other). However, since
\begin{equation}
\int_{0}^{L_y} \frac{e^{-iqy}}{\sqrt{L_y}}\frac{e^{iky}}{\sqrt{L_y}}\,dy=\delta_{k,q}
\end{equation}
we have
\begin{equation}
\int_{-\infty}^{\infty}dx\int_{0}^{L_y}dy\,\, \psi^*_{m,q}(x,y)\psi_{n,k}(x,y) =\delta_{n,m}\delta_{k,q}.
\end{equation}

\begin{figure}[htp!]
	\begin{minipage}{.5\textwidth}
		\centering
		\includegraphics[width=1.\textwidth]{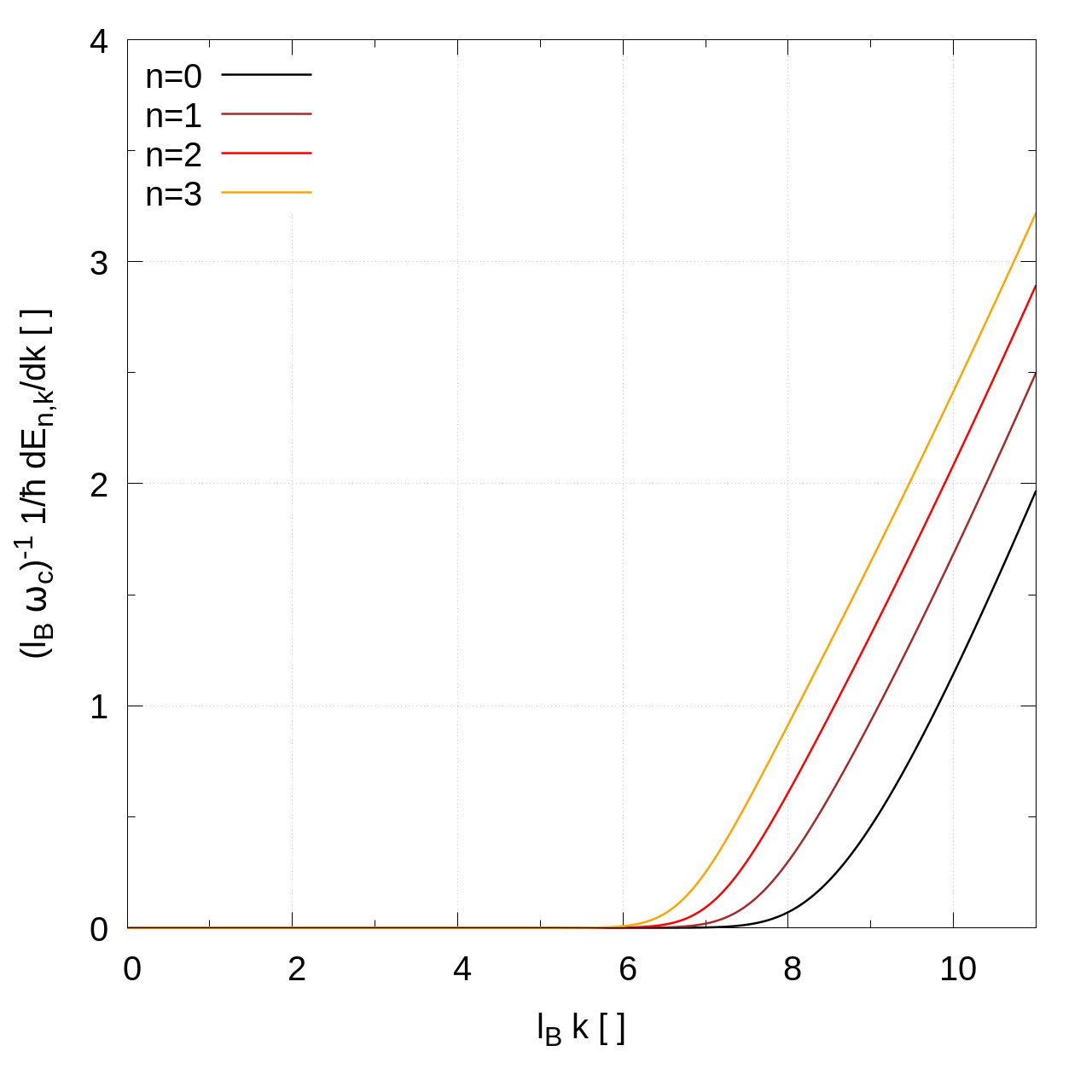}
	\end{minipage}%
	\begin{minipage}{0.5\textwidth}
		\centering
		\includegraphics[width=1.\textwidth]{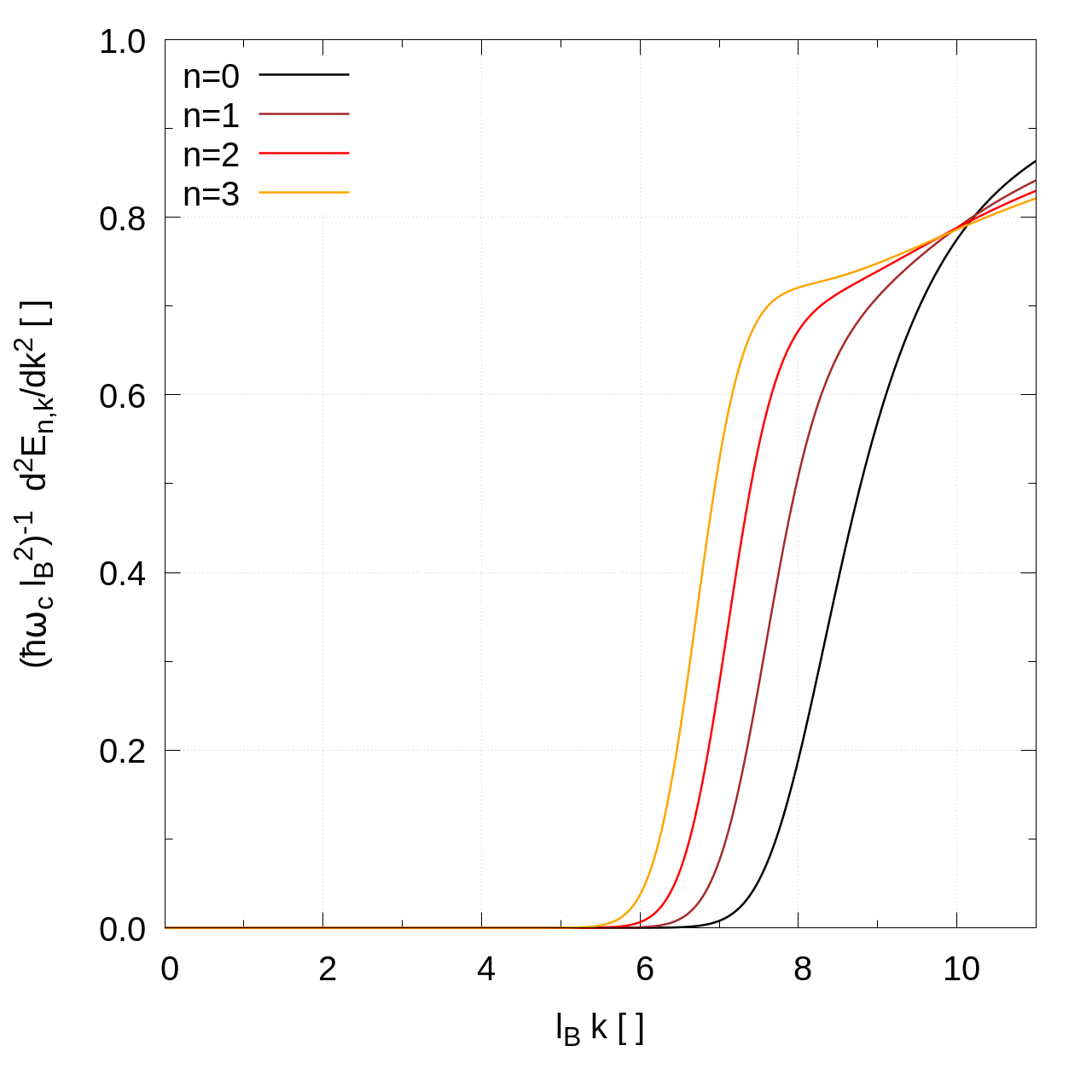}
	\end{minipage}    
	\caption[The LOF caption]{The image on the left shows the group velocity $\frac{1}{\hbar}\,\partial_k\,E_{n,k}$ as computed directly by differentiating the numerically computed energy spectrum. The one on the right on the other hand shows the curvature of the band $\partial_k^2 E_{n,k}$, for different Landau levels. Again the system length was chosen to be $L_x=20l_B$ and the confining parameters $V_0=30\hbar\omega_c$, $\sigma_c=0.1l_B$.}.
	\label{fig:GroupVelocity_Curvature}
\end{figure}
The spectrum is shown in the left panel of Fig. \ref{fig:EnergySpectrum}. The panel on the right shows some of the (normalized) wavefunctions together with the total potential well (harmonic potential plus confining one).
It can be immediately noticed that in the bulk we essentially have pure Landau levels; at the edges of the system however the oscillator is \virgolette{pushed} toward the walls of the system; since the confining region becomes smaller, the energy of the system will increase accordingly. This behaviour can be recognized when looking at the right hand side image of Fig. \ref{fig:EnergySpectrum}.

It is probably evident that the first and second derivatives of the Landau level will have a prominent role in this thesis work, the first one being related to the velocity of the chiral edge electrons and the second one to dispersion phenomena (from a classical viewpoint, when an electron makes a cyclotron orbit near the edge, its effective velocity is slightly larger when it is closer to the edge as opposed to when it dives into the bulk\footnote{For our choice of a \virgolette{steep and heigh} potential at least.}).
The numerical results are shown in Fig. \ref{fig:GroupVelocity_Curvature}; the behaviour qualitatively matches what we expected from the discussion in section \ref{section:quantum_edges}.

\begin{figure}[htp!]
	\begin{minipage}{.5\textwidth}
		\centering
		\includegraphics[width=1.\textwidth]{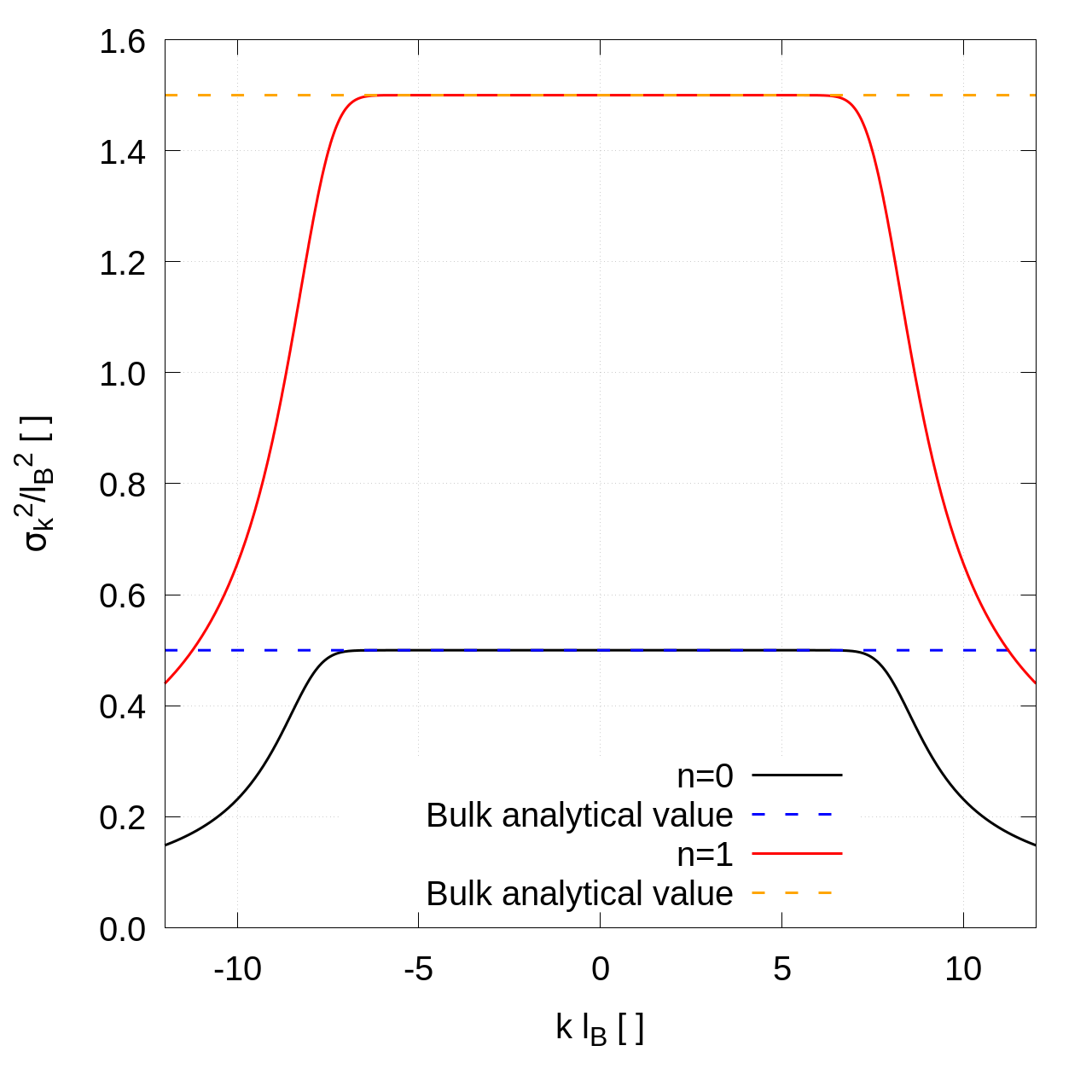}
	\end{minipage}%
	\begin{minipage}{0.5\textwidth}
		\centering
		\includegraphics[width=1.\textwidth]{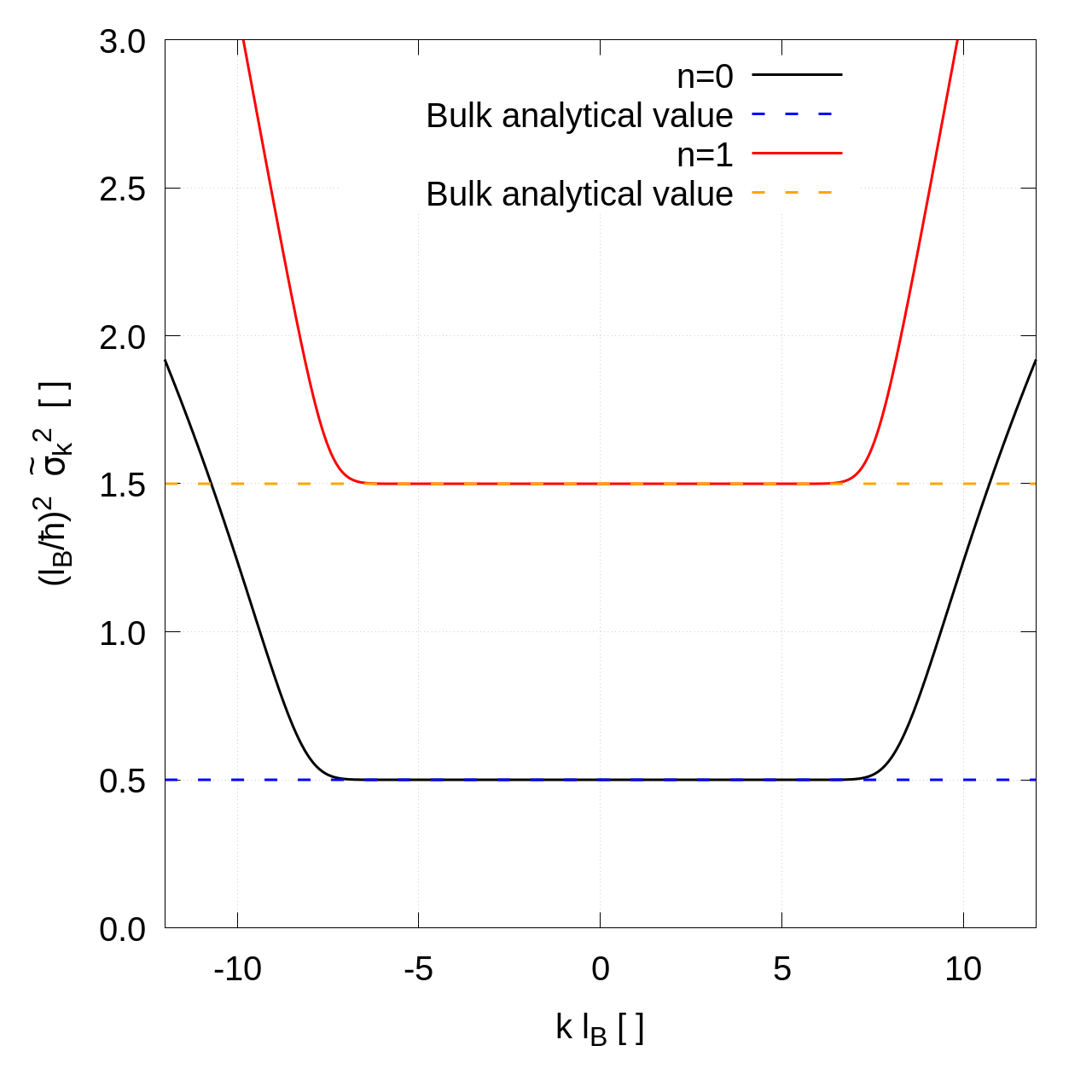}
	\end{minipage}    
	\caption[The LOF caption]{The left hand side image shows the real space (squared) width $\braket{(x-\braket{x}_{n,k})^2}_{n,k}$ of the $(n,k)$ eigenstate as a function of $k$ in the $n=0$ and $n=1$ cases. 
		\newline On the right hand side we plot the numerically evaluated momentum space (squared) width $\braket{(p_x-\braket{p_x}_{n,k})^2}_{n,k}$. To generate the images I chose $L_x=20l_B$, $V_0=30\hbar\omega_c$ and $\sigma_c=0.1l_B$.}
	\label{fig:real_and_momentum_space_widths}
\end{figure}
In the left panel of Fig. \ref{fig:real_and_momentum_space_widths} the squared real space width $\sigma_{n,k}^2=\braket{(x-\braket{x}_{n,k})^2}_{n,k}$ is plotted as a function of $k$, for the two lowest lying Landau levels. In the right panel, the squared momentum space width $\widetilde{\sigma}_{n,k}^2=\braket{(p_x-\braket{p_x}_{n,k})^2}_{n,k}=\braket{p_x^2}_{n,k}$ is shown instead.
As heuristically expected, as the eigenstates move closer to one edge of the system (i.e. $k l_B$ approaches $\pm \frac{L_x}{2l_B}$) they get \virgolette{trapped} between the harmonic oscillator parabola on one side and the confining potential on the other one, resulting in a tighter confinement and thus a reduced $\sigma_k^2$. In momentum space on the other hand the state must broaden due to quantum uncertainty, $\widetilde{\sigma}_{k}^2\appropto\frac{1}{\sigma_{k}^2}$.

\begin{figure}[htp!]
	\begin{minipage}{.5\textwidth}
		\centering
		\includegraphics[width=1.\textwidth]{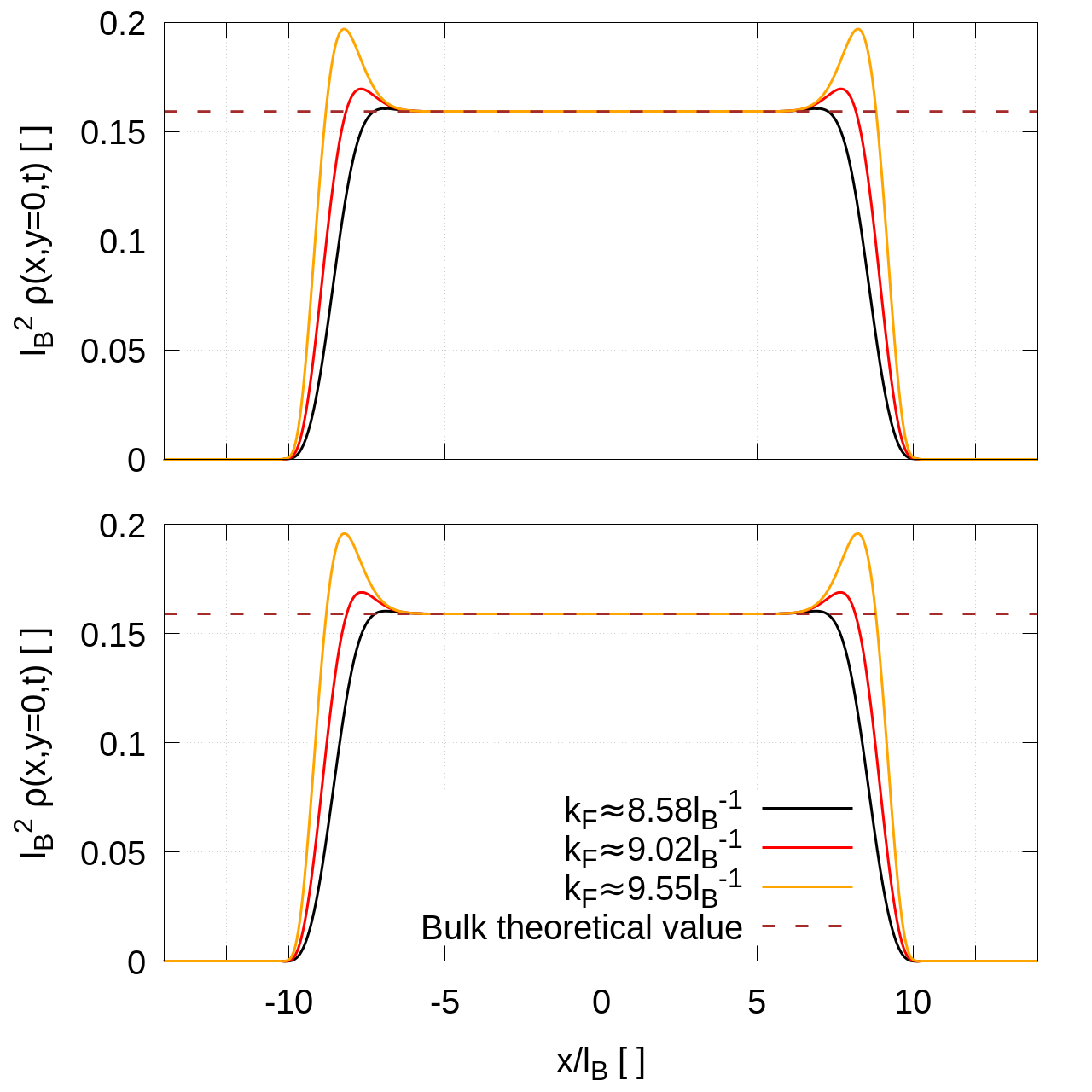}
	\end{minipage}%
	\begin{minipage}{0.5\textwidth}
		\centering
		\includegraphics[width=1.\textwidth]{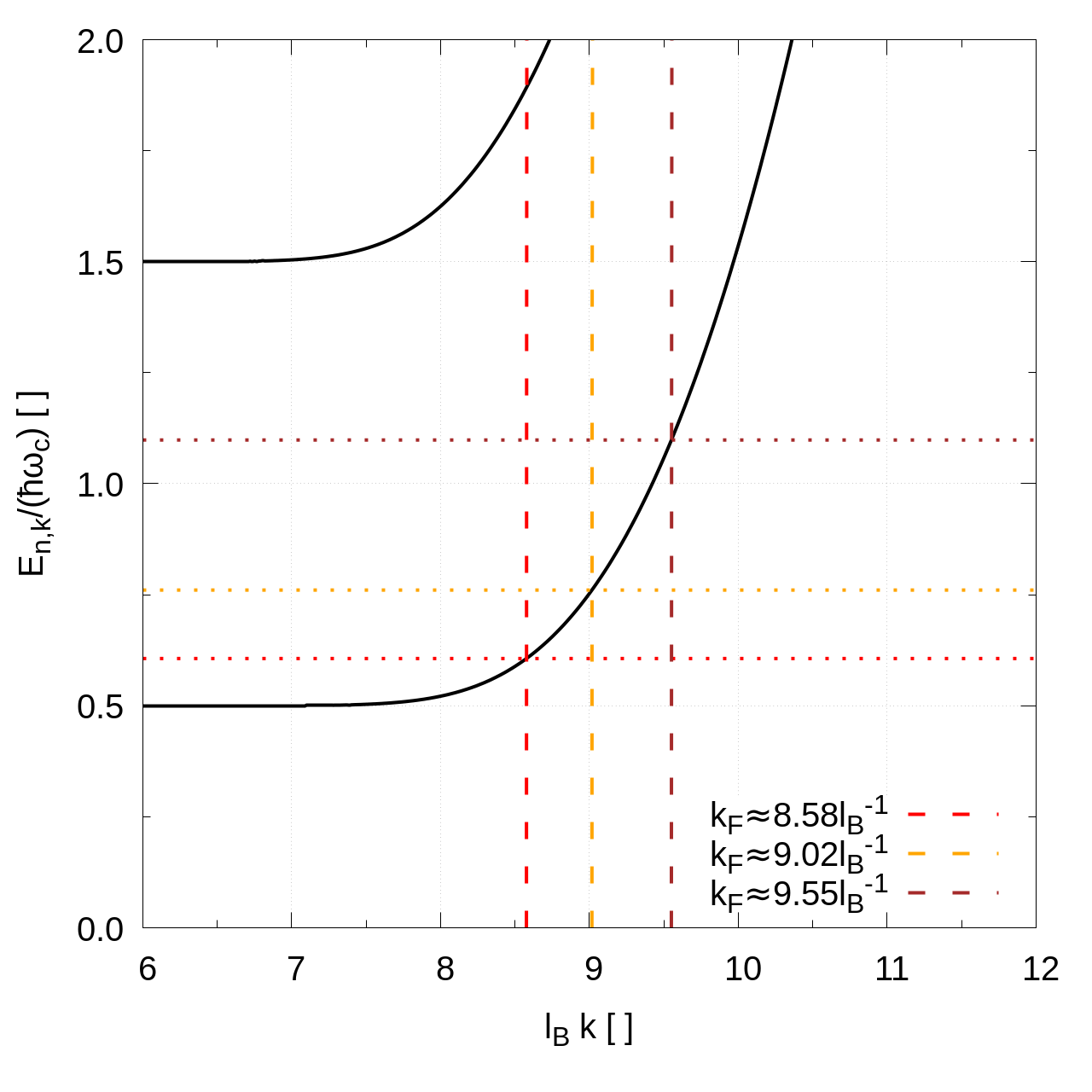}
	\end{minipage}    
	\caption[The LOF caption]{
		In the left panel the numerically computed density of a $n=0$ ground state is plotted for different values of the Fermi momentum, in the case $L_x=20l_B$, $V_0=30\hbar\omega_c$, $\sigma_c=0.1l_B$. $L_y=200l_B$ is plotted in the top panel, $L_y=2000l_B$ in the bottom one.
		Since the ground state density is $y$ independent, it is plotted at fixed $y$ as a function of $x$ alone. It can be seen that the two figures are qualitatively very similar.
		\newline In the right hand side panel the Fermi momenta and Fermi energies which have been used in the left hand side image are highlighted on top of a portion of the system energy spectrum, respectively as dashed and dotted lines.
	}
	\label{fig:ground_state_density}
\end{figure}

Finally, in the left hand side image in Fig. \ref{fig:ground_state_density} the ground state density is plotted for $L_y=200l_B$ (left hand side) and $L_y=2000l_B$ (right hand side), for different values of the Fermi momentum $k_F$. 
It can be seen that in the bulk it approaches the expected value given in eq. \ref{eq:bulk_denisty}, and that it quickly decays to zero at the system edges. As the number of electrons is increased, \virgolette{bumps} appear; the phenomenon can be heuristically understood as follows: as the Fermi level moves between two bulk Landau levels, the edge modes get populated by more and more electrons. 
Since these states are localised near the system edge, the system density in this small region will start to increase. 
It however becomes highly energetically unfavourable to stack electrons in a way which would make the system density everywhere uniform and equal to $\rho_\text{bulk}$ (as for a system with periodic conditions at the boundaries); rather the system density rises above the bulk level before dropping to zero when approaching the edges.
That the system density must be higher than its bulk value can also be understood by taking a slightly different path. From the discussion made in section \ref{section:quantum_edges} we know that the $\nu=1$ Landau level can at most accommodate $\mathcal{N}$ electrons, even when the edges are included\footnote{At least when these are modelled by a hard-wall potential.}. Thus for a saturated $\nu=1$ level we have
\begin{equation}
\mathcal N=\int\rho(x)\, dx\,dy=\rho_\text{bulk}\int\,dx\,dy.
\end{equation}
If the edges are now included the density $\rho(x)$ is not uniform, but will approach zero when getting closer to them due to the energetic considerations above. The two integrals being equal implies that $\rho(x)$ must increase near the edges. 

\subsection{Numerical procedure}\label{sec:NumericalProcedureStaticProperties}
In this section some more detail on the numerical diagonalization of the single particle one dimensional Hamiltonian \ref{eq:problem_hamiltonitan_transformed} is given.

Expressing lengths in units of the magnetic length $l_B$, times in units of the inverse cyclotron frequency $\omega_C$ and energies accordingly in units of $\hbar\omega_C$, the time-dependent Schrödinger equation for a single electron in a magnetic field in the Landau gauge becomes
\begin{equation}
i\,\frac{\partial}{\partial\widetilde{t}}\,\,\widetilde{\psi}\left(\tilde{x},\tilde{y};\,\tilde{t}\,\right)
= \left[-\frac{1}{2}\,\frac{\partial^2}{\partial \tilde{x}^2} + \frac{1}{2}\left(\tilde{x}-i\,\frac{\partial}{\partial \tilde{y}}\right)^2 + \frac{V_c\left(\tilde{x}\right)}{\hbar\omega_C} \right]\,\widetilde{\psi}\left(\tilde{x},\tilde{y};\,\tilde{t}\,\right)
\end{equation}
where $\widetilde{\psi}=l_B\,\psi$ is normalized
\begin{equation}
\int \,\bigl|\tilde{\psi}\,\bigr|^2 d\tilde{x}\,d\tilde{y}=1.
\end{equation}
The stationary states $\widetilde{\psi}\left(\tilde{x},\tilde{y};\,\tilde{t}\,\right)=\widetilde{\Phi}_{\tilde{k}}\left(\tilde{x}\right) \frac{e^{i \tilde{k}\,\tilde{y}}}{\sqrt{\widetilde{L}_y}}\,e^{-i\tilde{E}\,\tilde{t}}$ obey
\begin{equation}
\Biggl[-\frac{1}{2}\,\frac{\partial^2}{\partial \tilde{x}^2} + \underbrace{\frac{1}{2}\left(\tilde{x}+\tilde{k}\right)^2 + \frac{V_c\left(\tilde{x}\right)}{\hbar\omega_C}}_{\widetilde{V}(\tilde{x})} \,\Biggr]\,\widetilde{\Phi}_{\tilde{k}}\left(\tilde{x}\,\right)=\widetilde{E}\,\,\widetilde{\Phi}_{\tilde{k}}\left(\tilde{x}\,\right)
\end{equation}
The tildes will be dropped for notational convenience in the rest of this section, and in the following chapter too.
\newline We introduced an initial grid point $x_0$ far enough from the edges so that at such a point and beyond the wavefunction can be considered to any extent to vanish, and a regularly spaced mesh of positions $x_j = x_0 + h_x j$. 
The step $h_x$ will set the discretization error of the computation; after some quick convergence tests it has been chosen to be $h_x=\frac{l_B}{20}$ (unless otherwise specified).
We define the discrete versions of the wavefunction $\Phi_{j}=\Phi_k(x_j)$ and analogously of the potential.
The second derivative can be approximated as
\begin{equation}
\partial_x^2 \Phi_{k}\Bigl.\Bigr|_{x_j} = \frac{\Phi_{j+1}-2\Phi_{j}+\Phi_{j-1}}{h_x^2}+\mathcal{O}(h_x^2).
\end{equation}
Points farther away than the nearest neighbours can be included to achieve smaller discretization error; this has indeed been done, but to keep the discussion simple I will not write them down here. 
The same five-point discretization rule which has been used for solving the time-dependent Schrödinger equation has been used though, and can be found in the next chapter eq. \ref{eq:five_point_rules}.
\newline With this approximation the continuous eigenvalue problem becomes a finite dimensional one
\begin{equation}
\label{eq:discretized_1d_problem}
-\frac{1}{2}\,\frac{\Phi_{j+1}-2\Phi_{j}+\Phi_{j-1}}{h_x^2} + V_j \Phi_j =E\, \Phi_j
\end{equation}
which however must be further supplemented with some boundary condition, since we need to truncate the system somewhere.
To this purpose we notice that the system is enclosed within $\sim \pm\frac{L_x}{2}$ by the presence of the confining potential $V_c$; 
for an edge electron the harmonic oscillator potential will locally be much smaller than the confining one, so we can for a moment forget about the presence of the $\propto x^2$ term in the Hamiltonian. 
Since the confining potential rises steeply, as a lowest order heuristic discussion we can set $\sigma_c\approx 0$. In this limit the confining potential becomes a \virgolette{step}, and thus once the wavefunction crosses the edge it gets exponentially killed with a typical lengthscale of order $\lambda^*\sim l_B\,\sqrt{\frac{\hbar\omega_c}{2V_0}}$ (see the images in the left hand side panels of Fig. \ref{fig:BCcomparison} and \ref{fig:EnergySpectrum}). The presence of the harmonic confinement will actually make the decay faster (Gaussian) further away, but the important point is that a low energy electron is confined within the system edges apart for a small leaking tail.
In this work $V_0=30\hbar\omega_c$ has been used throughout, which gives $\lambda^*\approx0.13l_B$. (compare with the bottom right image in Fig. \ref{fig:EnergySpectrum}). 
This allows to safely \virgolette{chop} the discretized equation \ref{eq:discretized_1d_problem} a few magnetic lengths beyond the edges, which de facto is equivalent to put hard-wall boundary conditions to the system. The key point however is that the low-energy physics will be totally insensitive to these additional boundary conditions (necessary from the numerical viewpoint): only the confining potential $V_c$ will matter.

\begin{figure}[htp!]
	\begin{minipage}{.5\textwidth}
		\centering
		\includegraphics[width=1.\textwidth]{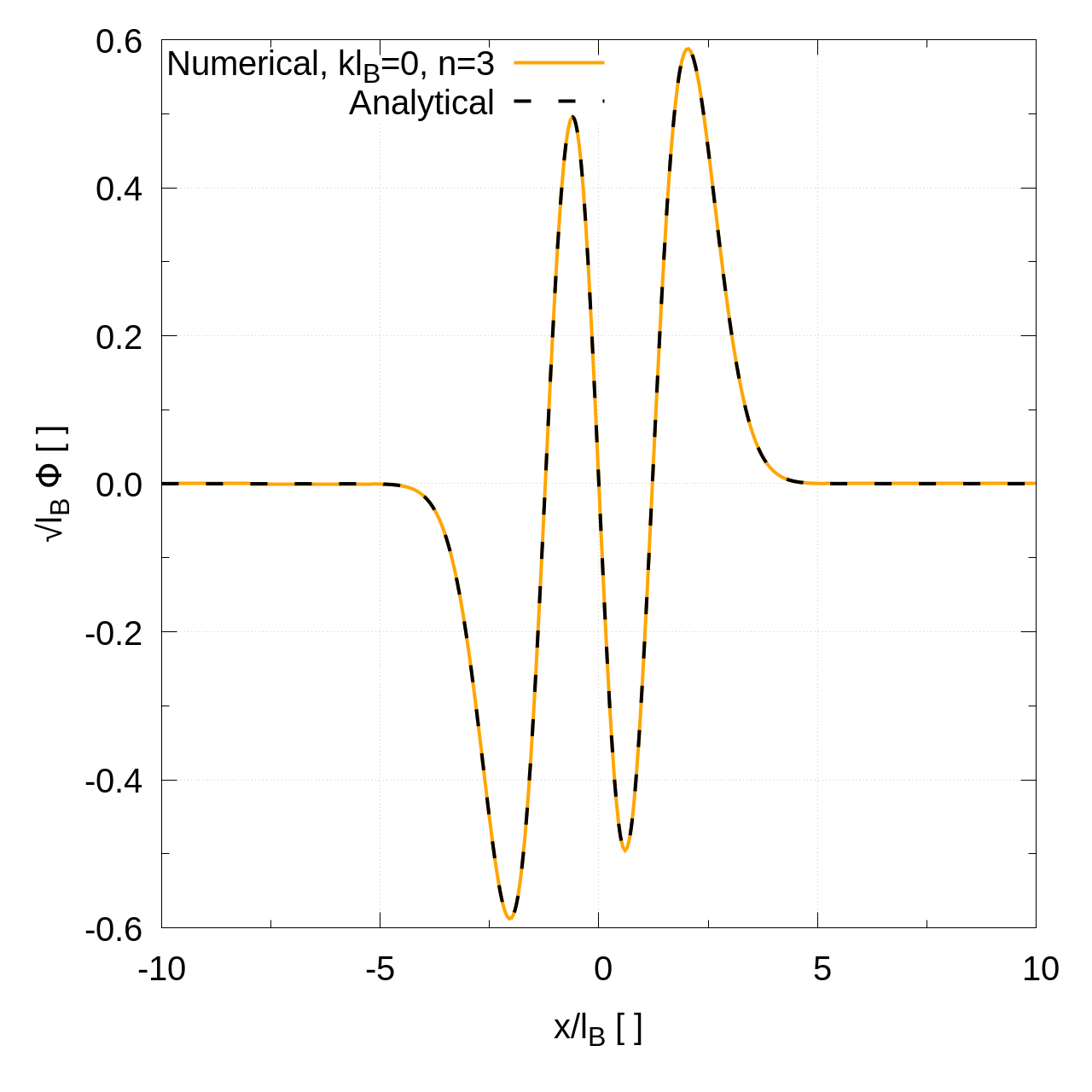}
	\end{minipage}%
	\begin{minipage}{0.5\textwidth}
		\centering
		\includegraphics[width=1.\textwidth]{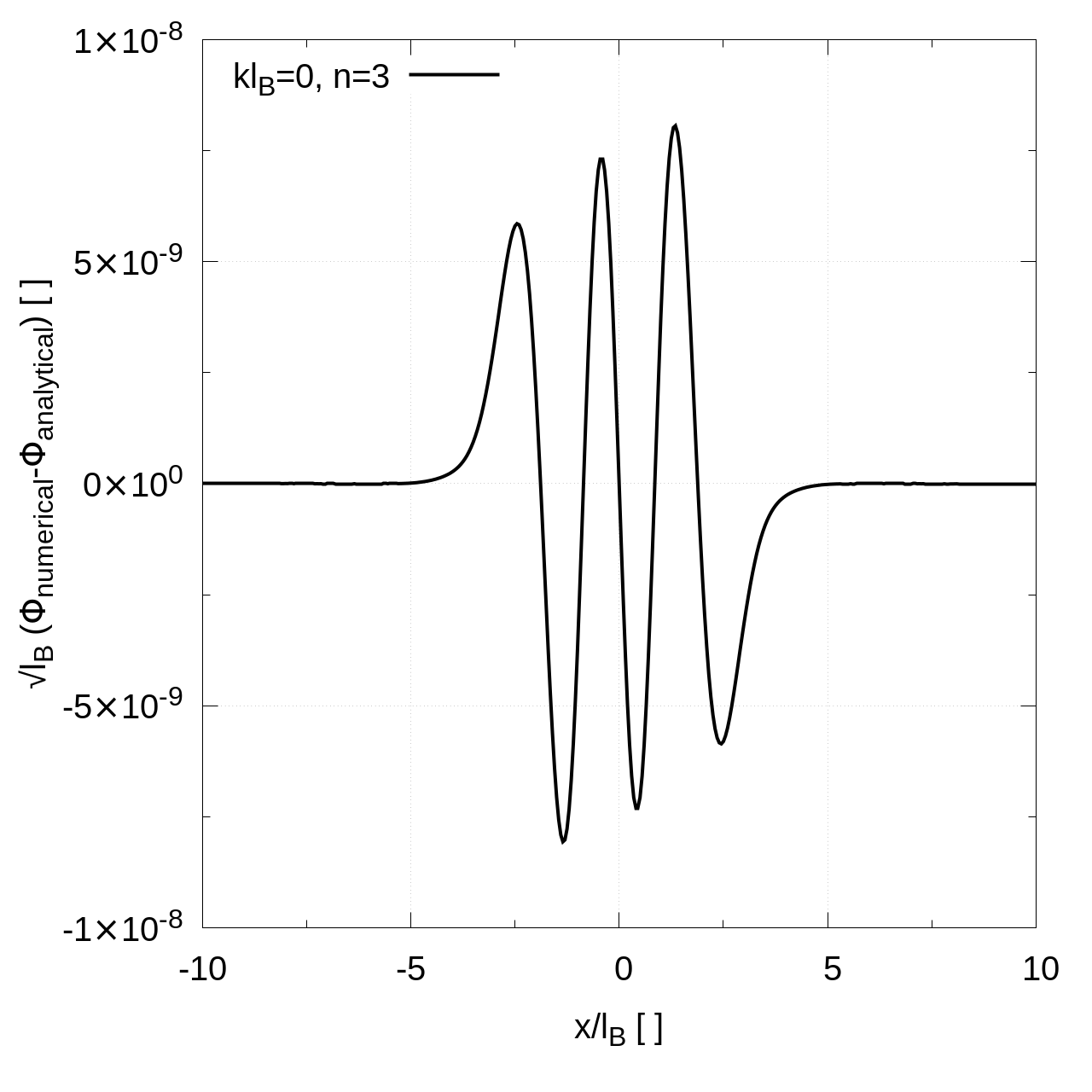}
	\end{minipage}    
	\caption[The LOF caption]{The left hand side image qualitatively compares a bulk wavefunction with the exact Harmonic oscillator one ($k=0$, $n=3$). At its right the difference between the numerically computed and the analytical function is shown.}
	\label{fig:NumericalWFcomparison}
\end{figure}

Equation \ref{eq:discretized_1d_problem} together with the box-like boundary conditions just discussed can finally be compactly rewritten using a matrix notation
\begin{equation}
M=\left( 
\begin{array}{ccccccc} 
\frac{1}{h_x^2}+V_1 & -\frac{1}{2h_x^2} & 0  & & \dots & &\\
&   & \ddots	& \ddots  &	   &   &   \\
\dots  & 0 	 & -\frac{1}{2h_x^2} 	& \frac{1}{h_x^2}+V_j  & -\frac{1}{2h_x^2} & 0 		  & \dots \\ 
& 	     & 						& \ddots			   & \ddots			   &    & 	  \\
& 	     & 			\dots			& 			   & 0			   &  -\frac{1}{2h_x^2}   & \frac{1}{h_x^2}+V_{N_x}	  \\
\end{array} 
\right)
\end{equation}
and
\begin{equation}
\mathbf{\Phi}=
\left( 
\begin{array}{c} 
\vdots  	\\ 
\Phi_{j-1}  \\
\Phi_{j}    \\
\Phi_{j+1}  \\ 
\vdots		\\
\end{array} 
\right).
\end{equation}
we have
\begin{equation}
M\mathbf{\Phi} = E\,\mathbf{\Phi}
\end{equation}
which can easily be handled numerically with one of the many linear-algebra libraries freely available (GSL routines were used). Once the diagonalization has been carried out, the eigenfunctions can be normalized by approximating $\int |\Phi|^2\,dx$ with a finite summation. Simpson's rule has been used to such a purpose.
\newline Since in the system's bulk the problem's solution is analytical, numerical results can be quantitatively checked. Fig. \ref{fig:NumericalWFcomparison} shows a comparison between the numerical and analytical solution; from the panel on the right hand side we see the error to be absolutely negligible.

\begin{figure}[htp!]
	\begin{minipage}{.5\textwidth}
		\centering
		\includegraphics[width=1.\textwidth]{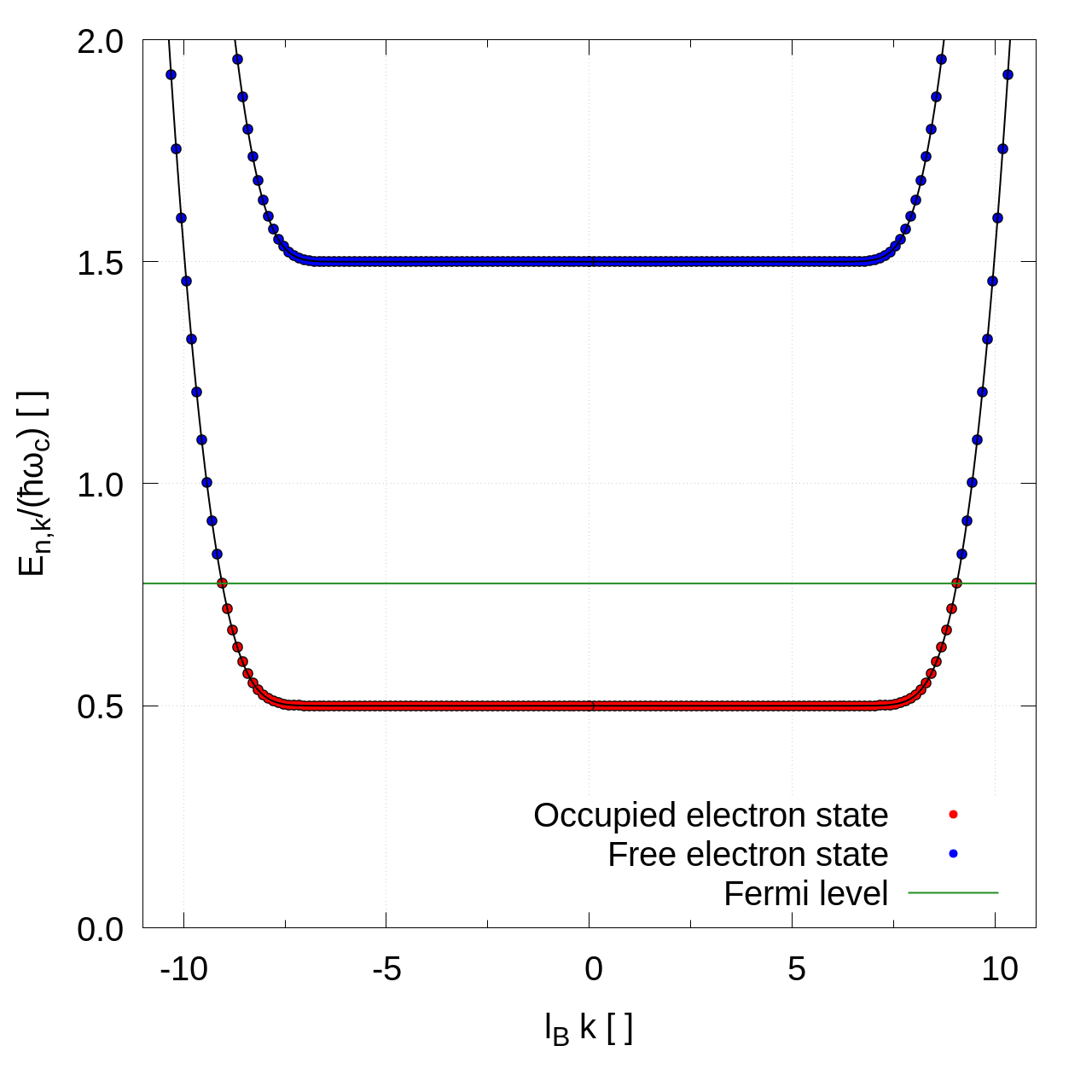}
	\end{minipage}%
	\begin{minipage}{0.5\textwidth}
		\centering
		\includegraphics[width=1.\textwidth]{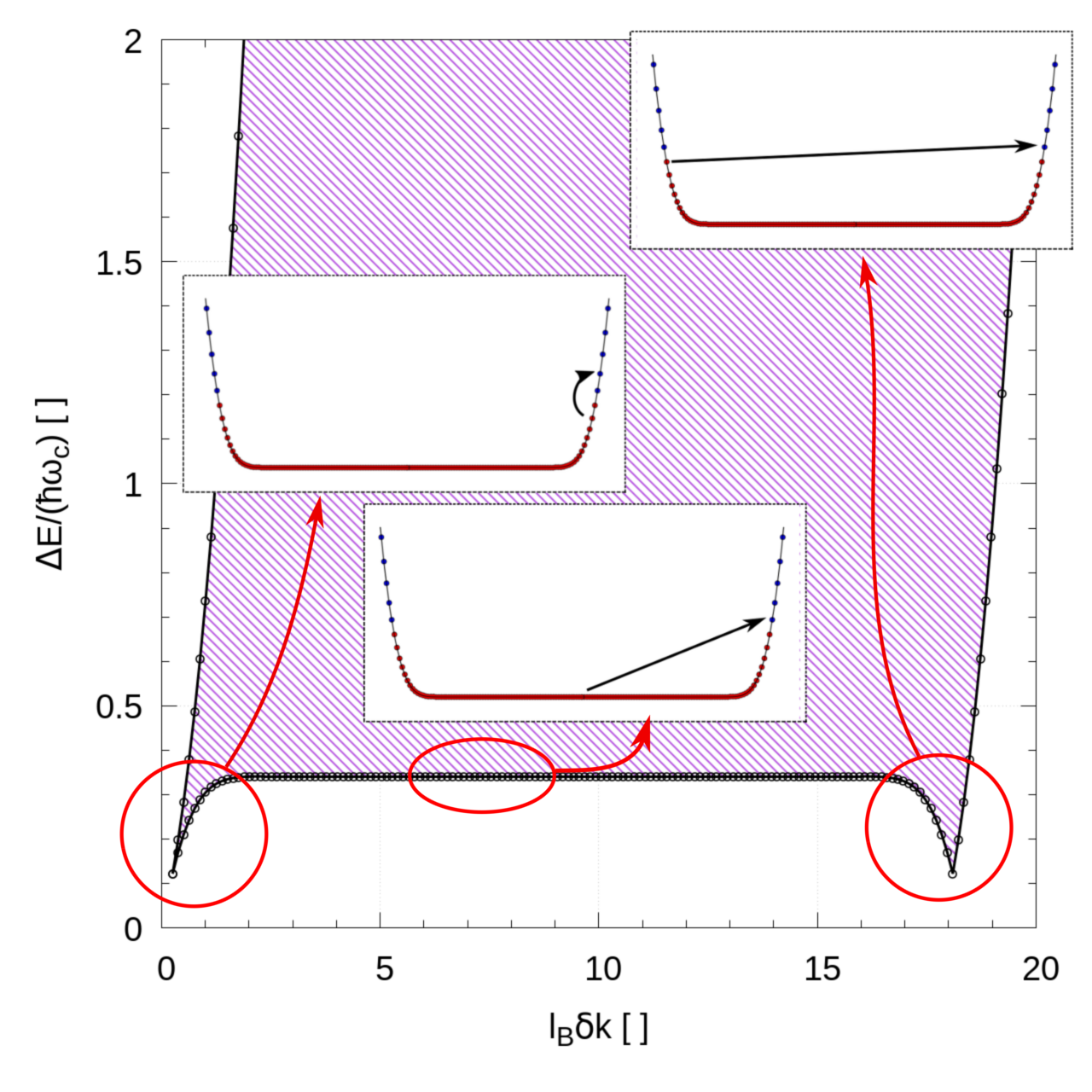}
	\end{minipage}    
	\caption[The LOF caption]{The image on the left shows the two lowest ($n=0$ and $n=1$) Landau levels for a system of size $L_x=20l_B$, together with the allowed $k$ states (depicted as circles) for $L_y=50l_B$ (the size has been chosen only for plotting purposes so that the circles were not too closely spaced). I chose $k_F=72\Delta k\simeq 9.05l_B^{-1}$. The occupied states are coloured in red, the unoccupied ones in blue.
		\newline On the right, the single particle-hole excitation spectrum from $n=0$ to $n=0$ is contoured for the same system parameters (the image should look like a bunch of discretely spaced points within the shaded area, but it looked awful and does not provide much more information; only the allowed points lying on the boundary of the shaded area have been highlighted).
		The two kind of transitions (bulk to edge, edge to edge) are schematically shown in small panels connected with red arrows to the associated region of the particle-hole spectrum.}
	\label{fig:systemGS}
\end{figure}

\section{Particle-Hole excitations}\label{section:ParticleHole}
We are now going to discuss the particle-hole excitations of such a system, i.e. excitations obtained by promoting an electron in the ground state of the system to an unoccupied level above the Fermi surface.

Consider a given number of electrons occupying the lowest lying Landau level states, in the configuration minimising the total energy, aka the ground state of the system. 
As already stated above the electronic states are filled with the available electrons starting from the least energetic ones, and appealing to the Pauli exclusion principle (no two electrons with the same quantum numbers, since the many-body wavefunction needs to be antisymmetric under the exchange of the coordinates of two electrons).
The energy of the highest occupied electron is dubbed \virgolette{Fermi energy}.
\newline The situation is graphically illustrated in Fig. \ref{fig:systemGS}, where the boundary condition allowed states are shown as small circles, differently coloured according to their occupation number (which at zero temperature is just a step function).
The key assumption here is that the bulk levels are completely filled, and some of the system electrons are edge ones. The Fermi energy has moreover been taken well below the first Landau level.

The single particle-hole excitation spectrum at $T=0$ can be easily numerically plotted as a function of the wavevector variation; its behaviour will qualitatively match the characteristics outlined at the end of section \ref{section:quantum_edges}. This in turn gives a good understanding of the response of the system under a small perturbation, and will be an useful tool for the following chapters.
\begin{figure}[htp!]
	\begin{minipage}{.5\textwidth}
		\centering
		\includegraphics[width=1.\textwidth]{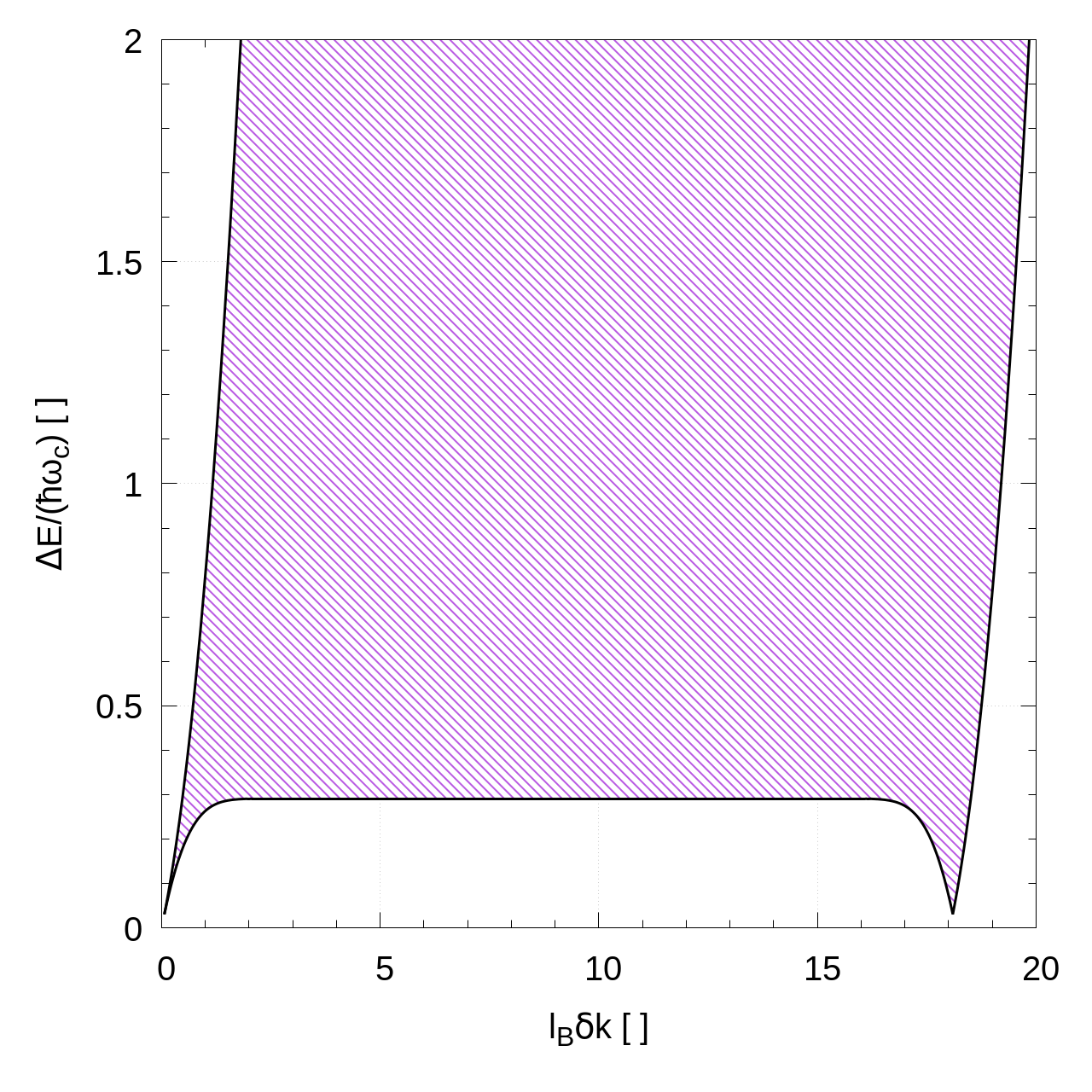}
	\end{minipage}%
	\begin{minipage}{0.5\textwidth}
		\centering
		\includegraphics[width=1.\textwidth]{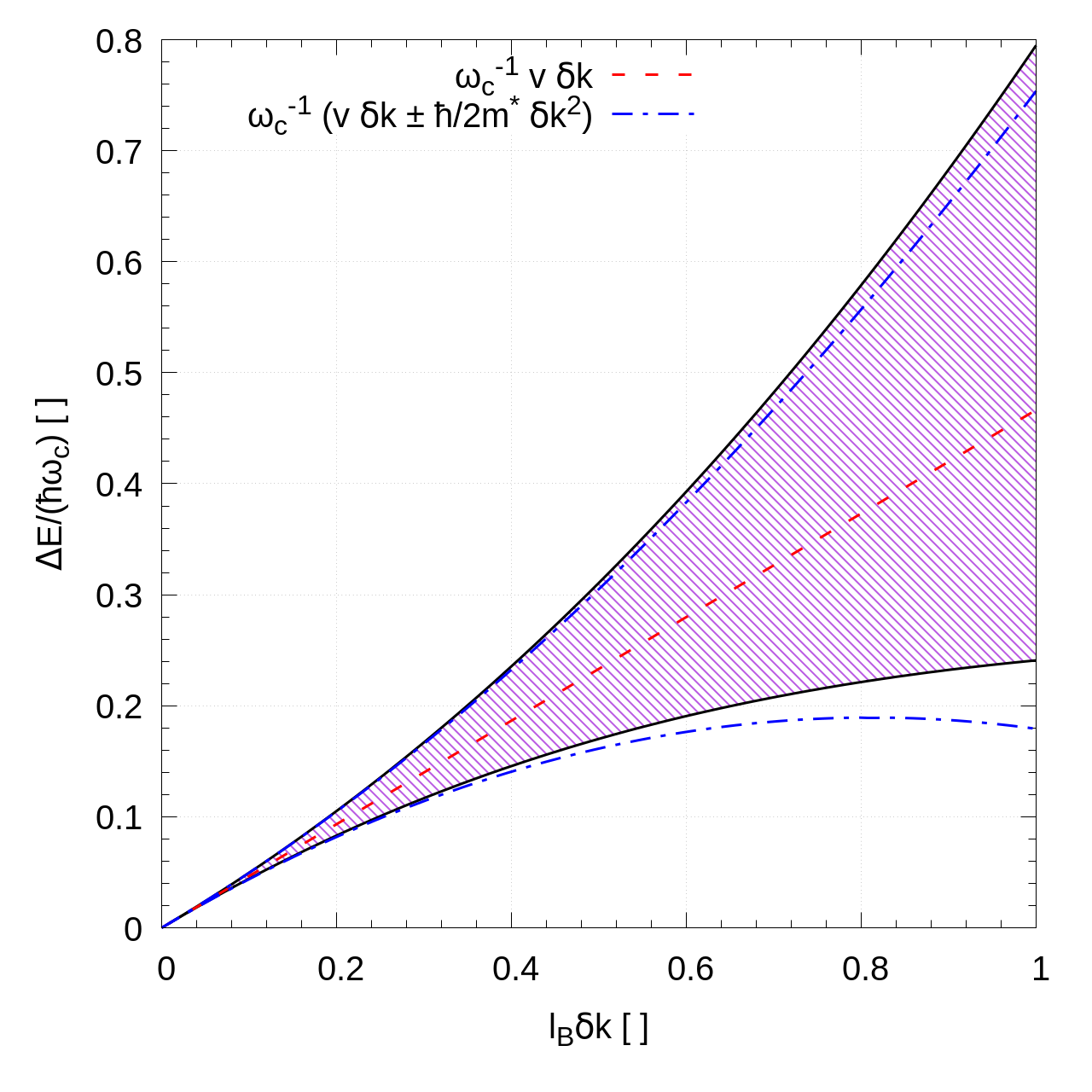}
	\end{minipage}    
	\caption[The LOF caption]{
The image on the left shows the single particle-hole excitation spectrum from the $n=0$ to the $n=0$ Landau level (only the $\delta k\geq0$ part is shown).
The system dimensions were chosen to be $L_x=20l_B$, $L_y=200l_B$ (i.e. four times as large as the system y-length $L_y$ used in Fig. \ref{fig:systemGS}). The Fermi wavevector is the same as the one used to generate Fig. \ref{fig:systemGS}, $k_F=288\Delta k\simeq 9.05l_B^{-1}$. 
\newline On the right hand side an even larger-system version ($L_y=800l_B$) is plotted in the region of $\delta k\sim 0$, and compared with the curves of eq. \ref{eq:ph_spectrum_curves} plotted as dotted dashed blue lines. The dashed red curve is the linear order truncation of eq. \ref{eq:ph_spectrum0}.  
The same Fermi wavevector has been used.}
\label{fig:gapless_excitations}
\end{figure}

Neglect for the moment inter-band transitions (consider only particle-hole excitations promoting an electron from the $n=0$ level to the same one).

\noindent A bulk electron has $k\ll k_F$; under the above conditions there are no available states to which the electron can be excited with a \virgolette{small} energetic cost: the closest unoccupied states in the same Landau level are edge states; these transitions correspond to a momentum kick $\delta k$ of the order of many $l_B^{-1}$, and the energetic gap is $\sim E_F-\frac{\hbar\omega_c}{2}$, $E_F$ being the Fermi energy (the energy of the highest occupied level). 
This behaviour can be recognized in the right hand side image of Fig. \ref{fig:systemGS} (bulk to edge transitions, in the central part of the particle-hole spectrum).

\noindent For edge electrons the situation is different, since with a momentum kick $\delta k\approx 0$ or $\delta k\approx 2k_F$
they can \virgolette{jump} to another empty edge state with a small energy cost $\epsilon$. This behaviour can be recognized in Fig. \ref{fig:systemGS}  as well (edge to edge transitions).
In the $L_y\rightarrow\infty$ limit, the spacing $\Delta k=\frac{2\pi}{L_y}$ between the allowed states in a given Landau level shrinks and eventually approaches zero in the limit, so that we have available energy states infinitesimally above the Fermi points and thus $\epsilon\rightarrow0$ as well. 
This is clearly seen when comparing the right hand side panel of Fig. \ref{fig:systemGS} with \ref{fig:gapless_excitations}. 
We can conclude then that long wavelength excitations in the thermodynamic limit ($L_y\rightarrow\infty$) are gapless (this is not perfectly the case for finite $L_y$, since the allowed values of $k$ are discretely spaced), or, equivalently stated, that the excitation spectrum looks like a sound wave. 
\newline The long wavelength/low energy properties of the system excitations can be explained by Taylor expanding the dispersion relation at quadratic order. Considering only the positive momentum Fermi point
\begin{equation}
\label{eq:ph_spectrum0}
\omega_k\simeq \omega_{k_F}+v_F(k-k_F)+\frac{\hbar}{2m^*}(k-k_F)^2.
\end{equation}
Consider now $k_1\leq k_F$ and $k_2>k_F$ (i.e. an occupied electron state below the Fermi surface and a free one above it) and fix $\delta k=k_2-k_1$. If an electron transitions from $k_1$ to $k_2$, the energy change expressed as a function of the momentum kick $\delta k$ can be bounded by the following inequalities\footnote{Consider $k_1$ and $k_2$ both to be positive and near $k_F$ so that the Taylor expansion of the dispersion relation does hold.}
\begin{equation}
\label{eq:ph_spectrum_curves}
\begin{cases}
\begin{alignedat}{1}
\delta\omega_k =& v_{F} \delta k+\frac{\hbar}{2m^*}\delta k^2+\frac{\hbar}{m^*}\delta k(k_1-k_F)\leq v_{k_F} \delta k+\frac{\hbar}{2m^*}\delta k^2\\
\delta\omega_k=&
v_{F} \delta k-\frac{\hbar}{2m^*}\delta k^2+\frac{\hbar}{m^*}\delta k(k_2-k_F)>v_{k_F} \delta k-\frac{\hbar}{2m^*}\delta k^2.
\end{alignedat}
\end{cases}
\end{equation}
These curves, together with the linear order expression, are plotted as dotted-dashed line over a numerically computed particle-hole spectrum in the right image of Fig. \ref{fig:gapless_excitations}.
\begin{figure}[htp!]
	\centering
	\includegraphics[width=1.\textwidth]{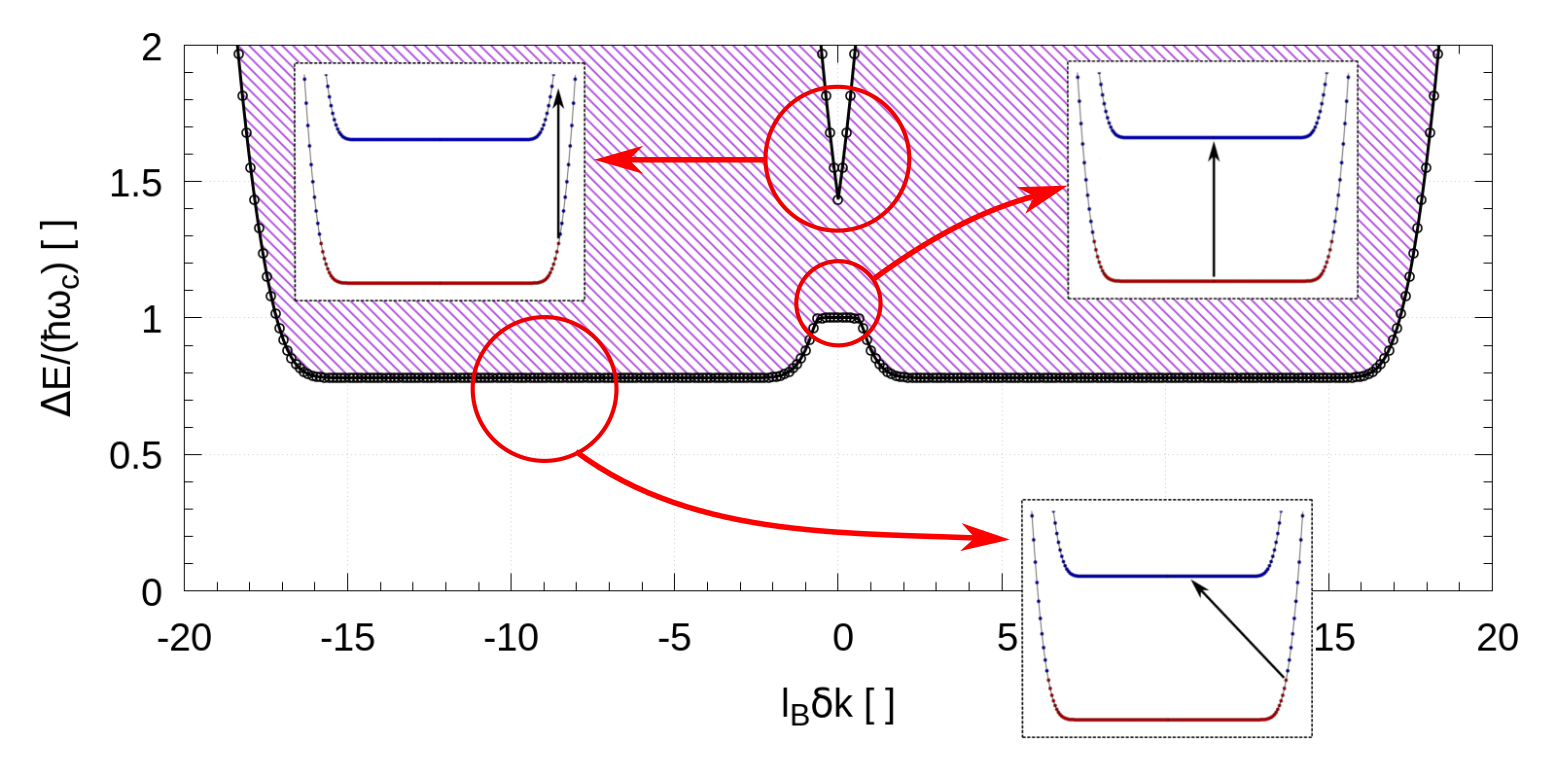}
	\caption[The LOF caption]{The image shows the single particle-hole excitation spectrum from the $n=0$ to the $n=1$ level.
		The system dimensions were chosen to be $L_x=20l_B$, $L_y=50l_B$. The Fermi wavevector has been fixed at $k_F=72\Delta k\simeq 9.05l_B^{-1}$.
		\newline Again, as in Fig. \ref{fig:systemGS} the region within which transitions can occur has been greyed out, even though for a finite sized system it should like as a bunch of discretely spaced points. Only those lying on the edge of the dashed area have been explicitly plotted as empty points.
		\newline The three kind of transitions (bulk to bulk, edge to edge, edge to bulk) which explain the main features of the spectrum are shown as insets connected with red arrows to the associated region of the particle-hole spectrum.}
	\label{fig:TransitionsTo1stLL}
\end{figure}

Finally, a comment on transitions from the $n=0$ to the $n=1$ Landau level. 
If the excitation occurs between two bulk states, the required energy is precisely the Landau level spacing $\hbar\omega_c$, as can indeed be seen in \ref{fig:TransitionsTo1stLL} (bulk to bulk transitions): at $\delta k\sim0$ these kind of transitions are the least energetic ones, indeed the spacing between two neighbouring Landau levels is greater than $\hbar \omega_c$ near the edges (thus edge to edge transitions are more energetic).
\newline For bigger $\delta k$ we have transitions ranging from an edge state and ending in the $n=1$ bulk (edge to bulk transitions). Notice that if the number of electrons is increased the edge states will become almost degenerate with the excited Landau levels\footnote{Filling edge states with energy greater than the one of the closest Landau levels in the bulk is not meaningful, since such a state can not represent a stable ground state.}, so that low energy transitions with non-vanishing $\delta k$ become possible.

\section{The Hall conductivity revisited}
We are now in the position to include the edges in the computation of the Hall conductivity.

The formula derived above (eq. \ref{eq:HallConductivity0} together with eq. \ref{eq:JyResponse}) can conveniently be rewritten in terms of dimensionless quantities
\begin{equation}
\label{eq:HallConductivity1}
\tilde{\sigma}_\text{Hall}=\sigma_\text{Hall}\frac{h}{e^2}=-\frac{2\pi}{\widetilde{L}_x\widetilde{L}_y}\,\frac{1}{\widetilde{E}}\,\sum_{n,k} \braket{-i \partial_{\tilde{y}}+\tilde{x}}_{n,k}.
\end{equation}
where $\widetilde{E}=\frac{eEl_B}{\hbar \omega_C}$ is the transverse electric field magnitude expressed in \virgolette{natural} units. 
The tildes will be dropped for convenience, every quantity is therefore measured in \virgolette{natural} units.
\newline Even in the presence of a uniform electric field directed in the $\hat{x}$ direction we have
\begin{equation}
\braket{-i \partial_{y}+x}_{n,k}=\braket{k+x}_{n,k}=\braket{\partial_k \mathcal{H}_k}_{n,k}
\end{equation}
If the states over which the expectation value is taken are eigenstates of the system Hamiltonian in the presence of an electric field directed along the $x$ axis, the last term equals $\partial_k \braket{\mathcal{H}_k}_{n,k}=\partial_k E_{n,k}$ by the Hellman-Feynman theorem.
\newline In the thermodynamic limit ($L_y\rightarrow\infty$)
\begin{equation}
\sigma=-\frac{2\pi}{L_xL_y}\,\frac{1}{E}\sum_{n,k}\partial_k E_{n,k}\rightarrow
-\frac{1}{L_x}\,\frac{1}{E}\sum_{n}\int dk\,\partial_k E_{n,k}
\end{equation}
where the integral ranges between the two Fermi points $\int dk\,\partial_k E_{n,k}=\Delta E_{n}$. 
When the electric field is \virgolette{small}, the energy difference between the two edges is given by 
\begin{equation}
\label{eq:ChemicalPotentialDifference}
\Delta E_{n}=\braket{E x}_{n,{k_{F,n}^{(+)}}}-\braket{E x}_{n,{k_{F,n}^{(-)}}}\simeq - E L_x
\end{equation}
which is just the classical potential energy difference between the two edges. 
Again we get
\begin{equation}
\sigma=\sum_{n}=\nu.
\end{equation}
\begin{figure}[htp!]
	\begin{minipage}{.5\textwidth}
		\centering
		\includegraphics[width=1.\textwidth]{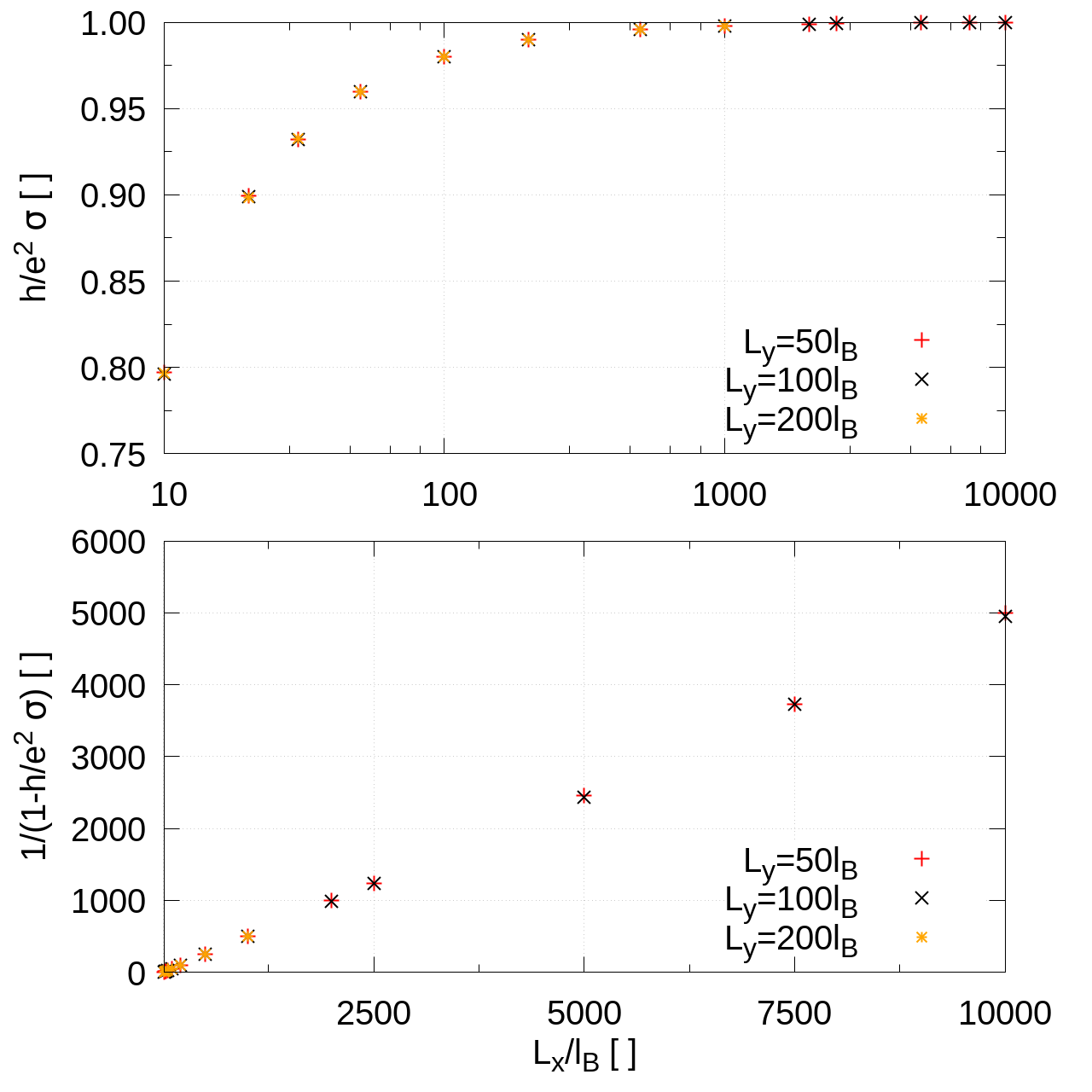}
	\end{minipage}%
	\begin{minipage}{0.5\textwidth}
		\centering
		\includegraphics[width=1.\textwidth]{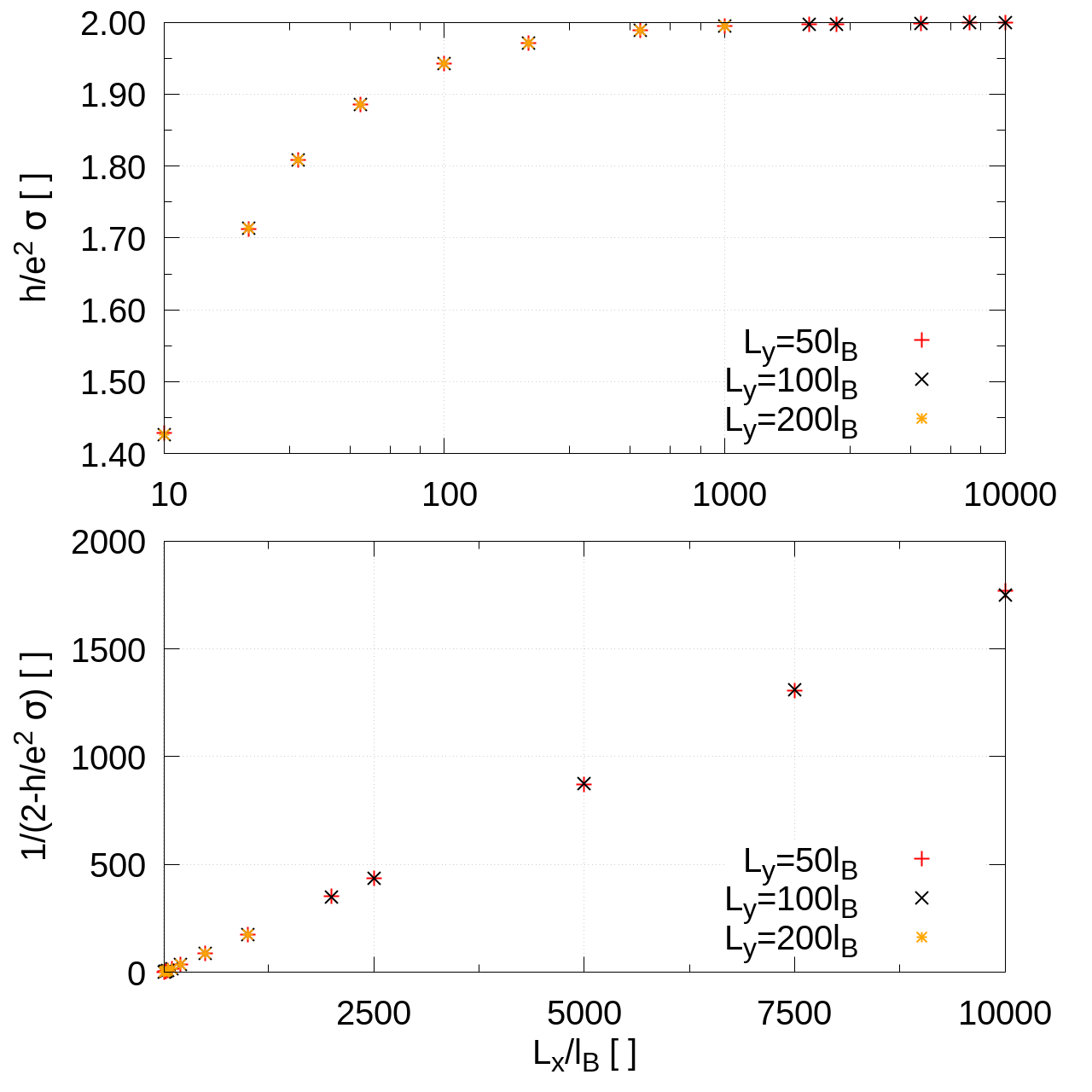}
	\end{minipage}    
	\caption[The LOF caption]{In the left panel the Hall conductivity $\sigma$ for the filled $n=0$ Landau level is plotted as a function $L_x$, for different values of $L_y$ on top. The bottom image instead shows $\frac{1}{\nu-\widetilde{\sigma}_\text{Hall}}$, which is expected to increase linearly with $L_x$ is plotted instead.
		\newline The images in the right hand side panel show the same quantities, but computed by filling both the $n=0$ and $n=1$ levels.}
	\label{fig:HallConductivity}
\end{figure}

\noindent Notice that the result is not sensible to the number of electrons in the Landau level as long as the Fermi energy lies in the gap between two such levels.
\newline This completes our simple discussion of the Hall conductivity: since the number of electron states is constant, as the magnetic field is varied electrons are transferred from the bulk to the system edges (the Fermi energy moves through the bulk gap); the value of the measured conductivity can not change though, since only the potential energy difference between the two sides of the sample matters.
It is only when the Fermi level crosses a Landau level that the conductivity can change, jumping from a plateau to the other one.

With the computational methods discussed above, an electric field along the $\hat{x}$ direction can be included in the problem and the new perturbed Hamiltonian can be diagonalized. The eigenstates can be used to compute the Hall conductivity using the relation in eq. \ref{eq:HallConductivity1}. 
The results for the first two Landau levels are shown in the top panels of Fig. \ref{fig:HallConductivity}.
We find a non-perfect quantisation, with an error decreasing $\propto \frac{l_B}{L_x}$, with practically no $L_y$ dependence. 
This is probably an artifact of the highly idealised system studied; that the committed error is of this order of magnitude could have also been expected from eq. \ref{eq:ChemicalPotentialDifference}: 
since the Fermi energy lies in the gap between two Landau levels, the average electron position is localised at the system edge near $\pm\frac{L_x}{2}$ within a length of the order of the magnetic length, so that the error committed when approximating \ref{eq:ChemicalPotentialDifference} will be of order $\mathcal{O}\left(\frac{l_B}{L_x}\right)$.

\graphicspath{{./pic2/}}

\chapter{System dynamics}\label{ch2}
\lhead{Chapter 3. \emph{System dynamics}}
Now that the system Hamiltonian has been diagonalized and the initial state of the system determined (in the independent particle model the task is easy since it is sufficient to fill the lowest lying states and thus building the Fermi sphere for the system), we would like to study what happens when the two-dimensional electron gas in a integer quantum Hall state is perturbed.
\newline I will briefly outline how the many-body system is formally described in terms of density matrices and how these can be computed from single-electron wavefunctions.
\newline The numerical apparatus which has been used to study the system dynamics in the upcoming chapters is then outlined, and the algorithm reliability has been tested in many different ways comparing its outputs with some simple cases which admit analytical solutions or which can be studied by means of different numerical techniques.
\newline In the last section some of the numerical expedients which have been used are outlined.

\section{State of the many-body quantum system}
First of all the density matrix description of a many-body system is outlined. 
In the independent particle case the one-body density matrix can be easily computed by the knowledge of single electron states, and contains all the information one needs; this is evidently much more convenient than dealing with some giant Slater determinant. Through such a device the expectation value of physical observables can be computed. 
\newline The time evolution of the density matrix is considered next

\subsection{Density matrices}
The properly symmetrized state of a system of non-interacting electrons is described by means of a Slater determinant, which, in coordinate space is written
\begin{equation}
\begin{split}
\Psi(\mathbf{r}_1,..,\mathbf{r}_N)=&\frac{1}{\sqrt{N!}}\,
\left|
\begin{array}{cccc} 
\psi_{\alpha_1}(\mathbf{r}_1) & \psi_{\alpha_1}(\mathbf{r}_2)  & \dots   & \psi_{\alpha_1}(\mathbf{r}_N) \\
\psi_{\alpha_2}(\mathbf{r}_1) & \psi_{\alpha_2}(\mathbf{r}_2)  & \dots   & \psi_{\alpha_2}(\mathbf{r}_N) \\ 
\vdots 					      & \vdots						   &          & \vdots 						  \\
\psi_{\alpha_N}(\mathbf{r}_1) & \psi_{\alpha_N}(\mathbf{r}_2)  & \dots   & \psi_{\alpha_N}(\mathbf{r}_N) \\ 
\end{array} 
\right|=\\
=& \frac{1}{\sqrt{N!}}\,\sum_\pi\,(-1)^{\pi}\,\psi_{\alpha_{\pi_1}}(\mathbf{r}_1)\dots\psi_{\alpha_{\pi_N}}(\mathbf{r}_N)
\end{split}
\end{equation}
where the summation runs over all the permutations of $N$ numbers and $(-1)^\pi$ is the sign of the permutation.
\newline The single particle properties of the system are encoded in the one-body density matrix, which in the non-interacting case has a particularly simple form
\begin{equation}
\label{eq:1BD}
\begin{split}
\rho^{(1)}(\mathbf{r},\mathbf{r}')=&N\int d^2\mathbf{r}_2\dots d^2\mathbf{r}_N\, \Psi^*(\mathbf{r}, \mathbf{r}_2\dots\mathbf{r}_N)\Psi(\mathbf{r}', \mathbf{r}_2\dots\mathbf{r}_N)=\\
=&\frac{1}{(N-1)!}
\sum_{\pi,\pi'}\,(-1)^{\pi+\pi'}\,\psi_{\alpha_{\pi_1}}^*(\mathbf{r})\,\psi_{\alpha_{\pi'_1}}(\mathbf{r}')\,
\delta_{\pi_2,\pi'_2}\,\dots \delta_{\pi_N,\pi'_N}=\\
=&\sum_\alpha \psi_{\alpha}^*(\mathbf{r})\psi_{\alpha}(\mathbf{r}').
\end{split}
\end{equation}
Here, congruently with the problem dealt, two spatial dimensions have been assumed. The generalization is however straightforward.
\newline The diagonal part represents the system's number density
\begin{equation}
\begin{split}
\rho^{(1)}(\mathbf{r},\mathbf{r})=\sum_\alpha \left|\psi_{\alpha}(\mathbf{r})\right|^2
\end{split}
\end{equation}
properly normalized to the number of particles in the system
\begin{equation}
\int d^2\mathbf{r}\,\rho^{(1)}(\mathbf{r},\mathbf{r})=N.
\end{equation}

The one-body density matrix can be used to compute the expectation value of one-body operators $O(\mathbf{r}_1...\mathbf{r}_N)=\sum_i O_i(\mathbf{r}_i)$
\begin{equation}
\begin{split}
\braket{O}=&\sum_i \int d^2\mathbf{r}_1\dots d^2\mathbf{r}_N\, \Psi^*(\mathbf{r}_1\dots\mathbf{r}_N)O_i\Psi(\mathbf{r}_1\dots\mathbf{r}_N)=\\
=&
N \int d^2\mathbf{r}_1\dots d^2\mathbf{r}_N\, \Psi^*(\mathbf{r}_1\dots\mathbf{r}_N)O_1\Psi(\mathbf{r}_1\dots\mathbf{r}_N)=\\
%=&
%N \int d^2\mathbf{r}_1\,d^2\mathbf{r}_1' \,\delta^{(2)}(\mathbf{r}_1-\mathbf{r}_1') \,O_1\int d^2\mathbf{r}_2\dots %d^2\mathbf{r}_N\, \Psi^*(\mathbf{r}_1'\dots\mathbf{r}_N)\Psi(\mathbf{r}_1\dots\mathbf{r}_N)=\\
=& 
\int d^2\mathbf{r}_1\,d^2\mathbf{r}_1' \,\delta^{(2)}(\mathbf{r}_1-\mathbf{r}_1') \,O_1\rho^{(1)}(\mathbf{r}_1',\mathbf{r}_1).
\end{split}
\end{equation}
In the second line the symmetry of the wavefunction under the exchange of two particle's labels has been used. Notice that the arguments of the density matrix are exchanged with respect to those in eq. \ref{eq:1BD}. 
In the independent particle model the expression simplifies to
\begin{equation}
\braket{O}= \sum_\alpha \int \, \psi_{\alpha}^*(\mathbf{r})O_1\psi_{\alpha}(\mathbf{r}) \, d^2\mathbf{r}.
\end{equation}
Higher order densities can be defined too, by straightforward generalization of the above discussion. These can be used to express the expectation value of many-body quantum operators, and thus they encode the many-particle correlations of the system. 
However since we are dealing with non-interacting particles all the physically relevant information is encoded in the one-body density; the higher order densities can indeed be expressed in terms of the one-body one alone (Wick's theorem). For example, it is easy to show that
\begin{equation}
\begin{split}
\rho^{(2)}(\mathbf{r}_1,\mathbf{r}_2;\mathbf{r}_1',\mathbf{r}_2')=&N(N-1)\int d^2\mathbf{r}_3\dots d^2\mathbf{r}_N\, \Psi^*(\mathbf{r}_1, \mathbf{r}_2\dots\mathbf{r}_N)\Psi(\mathbf{r}'_1, \mathbf{r}'_2\dots\mathbf{r}_N)=\\
=&
\rho^{(1)}(\mathbf{r}_1,\mathbf{r}'_1)\rho^{(1)}(\mathbf{r}_2,\mathbf{r}'_2)
-\rho^{(1)}(\mathbf{r}_1,\mathbf{r}'_2)\,\rho^{(1)}(\mathbf{r}_2,\mathbf{r}'_1).
\end{split}
\end{equation}

Finally we can define a probability current field $\mathbf{J}$. Taking the time derivative of the diagonal part of the one-body density matrix and using the Schrödinger continuity equation
\begin{equation}
\frac{\partial\rho^{(1)}}{\partial t}=\sum_\alpha -\boldsymbol{\nabla} \cdot \mathbf{J}_\alpha \equiv -\boldsymbol{\nabla}\cdot\mathbf{J}
\end{equation}
where $\mathbf{J}_\alpha = \Re\left(\psi_\alpha^*(\mathbf{r})\frac{\boldsymbol{\pi}}{m}\psi_\alpha(\mathbf{r})\right)$ is the usual probability current\footnote{The presence of a magnetic field is taken into account by replacing the canonical momentum with the gauge-invariant kinetic one, which is physically intuitive but can easily be proved by carrying out some simple algebra.}. For the many-body system, the one-body counterpart $\mathbf{J}(\mathbf{r};t)$ is thus easily computed by summing over all the occupied electron states, $\mathbf{J}(\mathbf{r};t)=\sum_\alpha \mathbf{J}_\alpha$, as indeed heuristically done in eq. $\ref{eq:semiclassical_derivation_current}$. 

\subsection{Time evolution}
If the N-particle Hamiltonian $\sum_i\mathcal{H}_i$ is time independent, the time-evolution operator factorizes $\mathcal{U}=\exp\left(-\frac{i}{\hbar}\mathcal{H}t\right) = \prod_j \exp\left(-\frac{i}{\hbar}\mathcal{H}_j t\right)=\prod_j \mathcal{U}_j$ since $\left[\mathcal{H}_i,\mathcal{H}_j\right]=0$. Then each component of the Slater determinant evolves in time independently from one other, i.e.
\begin{equation}
\begin{split}
\Psi(\mathbf{r}_1&,..,\mathbf{r}_N; t)=\,\mathcal{U}(t,t_0)\,\Psi(\mathbf{r}_1,..,\mathbf{r}_N; t_0)
=\\
=& \frac{1}{\sqrt{N!}}\,\sum_\pi\,(-1)^{\pi}\,e^{-\frac{i}{\hbar}\mathcal{H}_1 (t-t_0)}\psi_{\alpha_{\pi_1}}(\mathbf{r}_1; t_0)\dots e^{-\frac{i}{\hbar}\mathcal{H}_N (t-t_0)}\psi_{\alpha_{\pi_N}}(\mathbf{r}_N; t_0)=\\
=& \frac{1}{\sqrt{N!}}\,\sum_\pi\,(-1)^{\pi}\,\psi_{\alpha_{\pi_1}}(\mathbf{r}_1; t)\dots \psi_{\alpha_{\pi_N}}(\mathbf{r}_N; t).
\end{split}
\end{equation}
and thus the dynamics of the non-interacting many-body system can be studied by studying the evolution of the single particle states.

Such a property will hold even if the Hamiltonian does depend on time, owing to the absence of interactions among the system constituents.
\newline Suppose we solve the single electron time-evolution equation
\begin{equation}
i \hbar \frac{\partial\mathcal{U}_j}{\partial t}=\mathcal{H}_j\,\mathcal{U}_j
\end{equation}
for every electron (since no interaction is present and the Hamiltonian is just a sum of single-particle Hamiltonians). Evidently $\left[\mathcal{U}_i,\mathcal{U}_j\right]=\left[\mathcal{H}_i,\mathcal{U}_j\right]=0$. Thus, operating on both sides of every equation with $\prod_{j'\neq j}\mathcal{U}_{j'}$ and adding them up one gets
\begin{equation}
\begin{split}
i \hbar& \frac{\partial\mathcal{U}_1}{\partial t}\mathcal{U}_2\dots\mathcal{U}_N+
\dots+i \hbar \,\mathcal{U}_1\mathcal{U}_2\dots\frac{\partial\mathcal{U}_N}{\partial t}=
i \hbar \frac{\partial}{\partial t} \prod_j\mathcal{U}_j=
\\
=&\mathcal{H}_1\,\mathcal{U}_1\mathcal{U}_2\dots\mathcal{U}_N+\dots+\mathcal{H}_N\,\mathcal{U}_1\mathcal{U}_2\dots\mathcal{U}_N=\sum_i \mathcal{H}_i \prod_j\mathcal{U}_j
\end{split}
\end{equation}
which is the evolution equation for the many-body (non-interacting) system \mbox{$i \hbar \frac{\partial\mathcal{U}}{\partial t}=\mathcal{H}\,\mathcal{U}$.}

\subsection{Excitations to higher Landau levels}
One can keep track of the number of electrons getting excited to an higher Landau level by simply computing the expectation value of the projector onto the $n$-th Landau level. This results in
\begin{equation}
\mathcal{P}_n(t)=\sum_{i}\sum_{k}||\braket{\psi_{n,k}|\Psi_{i}(t)}||^2
\end{equation}
where the first summation runs over all the electrons of the system, while the second one over all the discretely spaced $k$ values allowed by the periodic boundary conditions (i.e. over all the elements belonging to the eigensubspace of a given Landau level).

\section{Time-evolution of single-particle states}\label{sec:time_evo_alg}
An algorithm for solving the single-particle time-dependent Schrödinger equation has been implemented in order to study the non-interacting many-body dynamics.
\newline  The algorithm (and a possible alternative) will now be presented.
\newline The set of dimensionless variables introduced in the first chapter will be used throughout the whole section. 

\subsection{Alternate Direction Implicit Crank Nicolson method}
Given some initial state $\psi\left(x,y;\,t_0\,\right)$, we want to study how the wavefunction $\psi\left(x,y;\,t\,\right)$ evolves in time under the general time-dependent Hamiltonian
\begin{equation}
\begin{cases}
\mathcal{H}=-\frac{1}{2}\,\frac{\partial^2}{\partial x^2} + \frac{1}{2}\left(x-i\,\frac{\partial}{\partial y}\right)^2 + V_c(x) + V_\text{ext}(x,y;t)
\\
i\,\frac{\partial}{\partial t}\,\,\psi\left(x,y;\,t\,\right)
= \mathcal{H}\,\psi\left(x,y;\,t\,\right).
\end{cases}
\end{equation}
The problem has been tackled by means of an Alternate Direction Implicit (ADI) Crank-Nicolson algorithm, which can be motivated as follows.
\newline We begin by introducing a discrete grid for the time variable; integrating the Schrödinger equation between two such grid-points one gets
\begin{equation}
\begin{split}
i\,\int_{t_{n}}^{t_{n+1}} \frac{\partial\,\psi}{\partial t'}\,dt'=&\,
i\,\,\left(\psi^{(n+1)}-\psi^{(n)}\right)
= \int_{t_{n}}^{t_{n+1}} \mathcal{H}\,\psi \,dt'\\
=& \,\Delta t\,\frac{\left(\mathcal{H}\,\psi\right)^{(n)}+\left(\mathcal{H}\,\psi\right)^{(n+1)}}{2}+\mathcal{O}(\Delta t^3)
\end{split}
\end{equation}
where the time integral of $\mathcal{H}\,\psi$ in the second line has been approximated using the trapezoidal rule (Crank-Nicolson method). The previous expression can be rewritten as
\begin{equation}
\label{eq:CrankNicolson}
\left(\mathbb{1}-\frac{\Delta t}{2i}\,\mathcal{H}^{(n+1)}\right)\psi^{(n+1)}=\left(\mathbb{1}+\frac{\Delta t}{2i}\,\mathcal{H}^{(n)}\right)\psi^{(n)}
\end{equation}
which is nothing but the Crank-Nicolson algorithm.
%\left(\mathcal{H}\,\psi\right)^{\left(n+\frac{1}{2}\right)}
The Hamiltonian is split into two parts as $\mathcal{H}=\mathcal{H}_x+\mathcal{H}_y$, the first containing the differential operators along $x$ and the second those along $y$. To order $\Delta t^2$ the operators acting on the $\psi$ in the above equation can be split as the product of two operators, one acting along $x$ and the other along $y$
\begin{equation}
\begin{split}
\left(\mathbb{1}-\frac{\Delta t}{2i}\,\mathcal{H}_x^{(n+1)}\right)&\left(\mathbb{1}-\frac{\Delta t}{2i}\,\mathcal{H}_y^{(n+1)}\right)\psi^{(n+1)}
=\\
=&\left(\mathbb{1}+\frac{\Delta t}{2i}\,\mathcal{H}_x^{(n)}\right)\left(\mathbb{1}+\frac{\Delta t}{2i}\,\mathcal{H}_y^{(n)}\right)\psi^{(n)}.
\end{split}
\end{equation}
A new variable $\psi^{\left(n+\frac{1}{2}\right)}$ is introduced, which obeys by construction the following equation
\begin{equation}
\label{eq:ADI1}
\left(\mathbb{1}-\frac{\Delta t}{2i}\,\mathcal{H}_x^{(n+1)}\right)\psi^{(n+\frac{1}{2})}=\left(\mathbb{1}+\frac{\Delta t}{2i}\,\mathcal{H}_y^{(n)}\right)\psi^{(n)}.
\end{equation}
Plugging this equation in the previous one
\begin{equation}
\begin{split}
\left(\mathbb{1}-\frac{\Delta t}{2i}\,\mathcal{H}_x^{(n+1)}\right)&\left(\mathbb{1}-\frac{\Delta t}{2i}\,\mathcal{H}_y^{(n+1)}\right)\psi^{(n+1)}
=\\
=&\left(\mathbb{1}+\frac{\Delta t}{2i}\,\mathcal{H}_x^{(n)}\right)\left(\mathbb{1}-\frac{\Delta t}{2i}\,\mathcal{H}_x^{(n+1)}\right)\psi^{(n+\frac{1}{2})}=\\
=&\left(\mathbb{1}-\frac{\Delta t}{2i}\,\mathcal{H}_x^{(n+1)}\right)\left(\mathbb{1}+\frac{\Delta t}{2i}\,\mathcal{H}_x^{(n)}\right)\psi^{(n+\frac{1}{2})} + \mathcal{O}(\Delta t^2)
\end{split}
\end{equation}
and operating with $\left(\mathbb{1}-\frac{\Delta t}{2i}\,\mathcal{H}_x^{(n+1)}\right)^{-1}$ on the left we get
\begin{equation}
\label{eq:ADI2}
\left(\mathbb{1}-\frac{\Delta t}{2i}\,\mathcal{H}_y^{(n+1)}\right)\psi^{(n+1)}
=\left(\mathbb{1}+\frac{\Delta t}{2i}\,\mathcal{H}_x^{(n)}\right)\psi^{(n+\frac{1}{2})}.
\end{equation}
We thus have a system of two coupled equations (eq. \ref{eq:ADI1} and \ref{eq:ADI2}) that needs to be solved at each time-step
\begin{equation}
\begin{alignedat}{2}
\begin{cases}
\left(\mathbb{1}-\frac{\Delta t}{2i}\mathcal{H}_x^{\left(n+1\right)}\right)\,\psi^{\left(n+\frac{1}{2}\right)} &= \left(\mathbb{1}+\frac{\Delta t}{2i}\mathcal{H}_y^{\left(n\right)}\right)\,\psi^{(n)}
\\
\left(\mathbb{1}-\frac{\Delta t}{2i}\mathcal{H}_y^{\left(n+1\right)}\right)\,\psi^{(n+1)} &=  \left(\mathbb{1} + \frac{\Delta t}{2i}\mathcal{H}_x^{\left(n\right)}\right)\,\psi^{\left(n+\frac{1}{2}\right)}.
\end{cases}
\end{alignedat}
\end{equation}
Notice that the times at which the Hamiltonian is evaluated can all be symmetrically shifted at $t_{n+\frac{1}{2}}$, since the error committed is still higher order.
\begin{equation}
\label{eq:ADI}
\begin{alignedat}{2}
\begin{cases}
\left(\mathbb{1}-\frac{\Delta t}{2i}\mathcal{H}_x^{\left(n+\frac{1}{2}\right)}\right)\,\psi^{\left(n+\frac{1}{2}\right)} &= \left(\mathbb{1}+\frac{\Delta t}{2i}\mathcal{H}_y^{\left(n+\frac{1}{2}\right)}\right)\,\psi^{(n)}
\\
\left(\mathbb{1}-\frac{\Delta t}{2i}\mathcal{H}_y^{\left(n+\frac{1}{2}\right)}\right)\,\psi^{(n+1)} &=  \left(\mathbb{1} + \frac{\Delta t}{2i}\mathcal{H}_x^{\left(n+\frac{1}{2}\right)}\right)\,\psi^{\left(n+\frac{1}{2}\right)}.
\end{cases}
\end{alignedat}
\end{equation}
\newline This pair of equations is equivalent up to $\mathcal{O}(\Delta t^2)$ to the Schrödinger equation we started with;
such a formulation is however very convenient since, given $\psi^{(n)}$, the first equation can be solved for $\psi^{\left(n+\frac{1}{2}\right)}$ at fixed $y$. Once $\psi^{\left(n+\frac{1}{2}\right)}$ has been obtained for every $y$, the second equation can be solved for $\psi^{(n+1)}$ at fixed $x$, and the procedure can be iterated to obtain the full time evolution.
The main advantage of this formulation of the problem is apparent: the spatial differential operators are treated in a \virgolette{staggered way}. This is an enormous computational advantage actually, since at every step the linear system of equations in the above algorithm can be solved very efficiently in both the $x$ and $y$ directions, as will become apparent in a while.

We introduce a spatial grid with spacings $\Delta x$, $\Delta y$. The wavefunction is sampled at these mesh points only, and denoted $\psi(x_i,y_j;t_n)=\psi_{i,j}^{(n)}$. The spatial derivatives are then approximated with finite differences, and the corresponding operator denoted with a $\delta$; the ADI equations eq. \ref{eq:ADI} (multiplied by the imaginary unit) can then be written as
\begin{equation}
\begin{alignedat}{2}
\begin{cases}
\Biggl(i-\frac{\Delta t}{2}\left(-\frac{1}{2}\delta_x^2 + V_{ij}^{\left(n+\frac{1}{2}\right)}\right)\Biggr)\,\psi_{ij}^{\left(n+\frac{1}{2}\right)} &= u^{(n)}_{ij}
\\
\Biggl(i-\frac{\Delta t}{2}\left(\frac{1}{2}\left(x_i-i \delta_y\right)^2+U_{ij}^{\left(n+\frac{1}{2}\right)}\right)\Biggr)\,\psi^{(n+1)}_{ij} &= v^{\left(n+\frac{1}{2}\right)}_{ij}
\end{cases}
\end{alignedat}
\end{equation}
where $V$ is the potential energy part included in $\mathcal{H}_x$, $U$ the one in $\mathcal{H}_y$ and
\begin{equation}
\begin{alignedat}{2}
\begin{cases}
u_{ij}^{(n)} &=\left(i+\frac{\Delta t}{2}\left(\frac{1}{2}\left(x_i-i \delta_y\right)^2+U_{ij}^{\left(n+\frac{1}{2}\right)}\right)\right)\,\psi^{(n)}_{ij}
\\
v^{\left(n+\frac{1}{2}\right)}_{ij} &=
\left(i + \frac{\Delta t}{2}\left(-\frac{1}{2}\delta_x^2 + V_{ij}^{\left(n+\frac{1}{2}\right)}\right)\right)\,\psi^{\left(n+\frac{1}{2}\right)}
\end{cases}
\end{alignedat}
\end{equation}
are the known terms.
\newline The five-point rule has been used to get a $\mathcal{O}(\Delta x^4,\Delta y^4)$ accuracy; for example, along $x$ we approximate
\begin{equation}
\label{eq:five_point_rules}
\begin{cases}
\partial_x\psi^{(n)} \Bigl.\Bigr|_{x_i,y_j}= \delta_x\psi_{ij}=\frac{4}{3}\,\frac{\psi_{i+1,j}^{(n)} -  \psi_{i-1,j}^{(n)}}{2\Delta x} - \frac{1}{3}\,\frac{\psi_{i+2,j}^{(n)} -  \psi_{i-2,j}^{(n)}}{4\Delta x}
\\
\partial_x^2\psi^{(n)} \Bigl.\Bigr|_{x_i,y_j}= \delta_x^2\psi_{ij}= \frac{4}{3}\,\frac{\psi_{i+1,j}^{(n)} - 2\psi_{i,j}^{(n)} + \psi_{i-1,j}^{(n)}}{\Delta x^2} - \frac{1}{3}\,\frac{\psi_{i+2,j}^{(n)} - 2\psi_{i,j}^{(n)} + \psi_{i-2,j}^{(n)}}{(2\Delta x)^2}.
\end{cases}
\end{equation}
Identical expressions do obviously hold for the $y$ derivatives.
\newline Along the $x$ direction, due to the presence of a smooth confining potential, the wavefunction is smooth and vanishingly small after some fraction of a magnetic length from the system's edge. A few magnetic lengths away from $L_x$, the wavefunction can be safely considered to be vanishing at all the times (for reasonable external potentials at least, which are much smaller in magnitude than the confining potential).
With this boundary condition, the first equation for $\psi_{ij}^{\left(n+\frac{1}{2}\right)}$ can be conveniently written in matrix form as
\begin{equation}
\label{eq:pentadiagonal}
\left(
\begin{array}{ccccccccc} 
	D_{1,j} & \Delta_1				& \Delta_2		   				& 0 & 0 & 0 & 0 & 0 &\dots \\
	%\Delta_1 & D_2 &  \Delta_1				& \Delta_2		   				& 0 & 0 & 0 & \dots \\
	%\Delta_2 & \Delta_1 & D_2 &  \Delta_1				& \Delta_2		   				& 0 & 0 & \dots \\		
	%0 & \Delta_2 & \Delta_1 & D_2 &  \Delta_1				& \Delta_2		   				& 0 & \dots \\		
	 &  & \ddots & \ddots &  \ddots	& \ddots &  &  \\
	\dots & 0 & \Delta_2 & \Delta_1 & D_{ij} &  \Delta_1				& \Delta_2 & 0 &\dots   \\
		 &  & \ddots & \ddots &  \ddots	& \ddots &  &  \\	
	\dots & 0 & 0 & 0 & 0 & 0 & \Delta_2 & \Delta_1 & D_{N_x,j} \\ 
\end{array} 
\right)
\,
\boldsymbol{\psi}_{j}^{\left(n+\frac{1}{2}\right)}	
=
\mathbf{u}_j^{(n)}
\end{equation}
where
\begin{equation}
\boldsymbol{\psi}_{j}^{\left(n+\frac{1}{2}\right)}	=
\left( 
\begin{array}{c} 
\vdots  									\\ 
\psi_{i-1,j}^{\left(n+\frac{1}{2}\right)}   \\
\psi_{i,j}^{\left(n+\frac{1}{2}\right)}     \\ 
\psi_{i+1,j}^{\left(n+\frac{1}{2}\right)}   \\	
\vdots										\\
\end{array}
\right)
,\qquad\qquad
\mathbf{u}_j^{(n)}=
\left( 
\begin{array}{c} 
\vdots  		    \\
u_{i-1,j}^{(n)}	    \\
u_{i,j}^{(n)}	    \\ 
u_{i+1,j}^{(n)}	    \\
\vdots				\\	
\end{array} 
\right)
\end{equation}
and
\begin{equation}
\begin{cases}
D_{i,j} = i-\frac{\Delta t}{2}\left(\frac{5}{4\Delta x^2} + V_{ij}^{\left(n+\frac{1}{2}\right)}\right)
\\
\Delta_1 =-\frac{\Delta t}{2}\left(-\frac{2}{3\Delta x^2}\right)
\\
\Delta_2 =-\frac{\Delta t}{2}\left(+\frac{1}{24\Delta x^2}\right).
\end{cases}
\end{equation}
It is apparent that a purely pentadiagonal linear system of equations needs to be solved at each step, for every $y$ value. This operation can be efficiently performed in $\mathcal{O}(N_x)$ operations for fixed $y$.
An algorithm (\cite{EngelnMuellgesUhlig1996}) based on the LU decomposition of the pentadiagonal matrix
eq. \ref{eq:pentadiagonal} has been implemented for the purpose.
\newline Along the $y$ direction we have periodic boundary conditions instead. The second equation for $\psi^{(n+1)}$ then reads
\begin{equation}
\label{eq:cyclicpentadiagonal}
\left(
\begin{array}{ccccccccc} 
D'_{i,1} & \sigma_i				& \theta_i		   				& 0 					    & \dots & & 0 & \theta_i^* & \sigma_i^* \\
\sigma_i^* & D'_{i,2} & \sigma_i				& \theta_i		   				& 0 					    & \dots  & & 0 & \theta_i^* \\	
\theta_i^* & \sigma_i^* & D'_{i,3} & \sigma_i				& \theta_i		   				& 0 					    & \dots & & 0  \\
		 &  & \ddots & \ddots &  \ddots	& \ddots & & \dots &  \\
\dots & 0 & \theta_i^* & \sigma_i^* & D_{i,j} & \sigma_i & \theta_i	& 0 & \dots \\
& \dots & & \ddots & \ddots & \ddots & \ddots  & & \\
0 & & \dots & 0 & \theta_i^* & \sigma_i^* & D'_{i,N_y-2} & \sigma_i & \theta_i \\
\theta_i & 0 & & \dots & 0 & \theta_i^* & \sigma_i^* & D'_{i,N_y-1} & \sigma_i \\
\sigma_i & \theta_i & 0 & & \dots & 0 & \theta_i^* & \sigma_i^* & D'_{i,N_y} \\
\end{array} 
\right)
\,
\boldsymbol{\psi}_{i}^{\left(n+1\right)}	
=
\mathbf{v}_i^{(n)}
\end{equation}
where
 \begin{equation}
\begin{cases}
D'_{i,j} = i-\frac{\Delta t}{2}\left(\frac{5}{4\Delta y^2} +\frac{1}{2}x_i^2+ U_{ij}^{\left(n+\frac{1}{2}\right)}\right)
\\
\sigma_i =-\frac{\Delta t}{2}\left(-\frac{2}{3\Delta y^2} - i x_i\frac{2}{3\Delta y}\right)
\\
\theta_i =-\frac{\Delta t}{2}\left(+\frac{1}{24\Delta x^2} - i x_i \left(-\frac{1}{12\Delta y}\right)\right).
\end{cases}
\end{equation}
Along $y$ we thus have to solve a cyclic pentadiagonal linear system of equations for every $x$. Such a task can efficiently be done in $\mathcal{O}(N_y)$ operations (for fixed $x$).
An algorithm (\cite{NguetchueAbelman2008}) based on the LU decomposition of the cyclically pentadiagonal matrix in eq. \ref{eq:cyclicpentadiagonal} has been implemented in order to solve the associated system of linear equations.

Analogous methods were initially used with the more usual but less accurate three-point rules for the finite-difference derivatives. Higher order methods were however implemented to obtain more accurate results with practically the same effort; in particular the truncation error is $\mathcal{O}(\Delta t^2, \Delta x^4, \Delta y^4)$ instead of $\mathcal{O}(\Delta t^2, \Delta x^2, \Delta y^2)$. 
This scaling will be checked in the upcoming section.

\subsection{Unitarity}
The unitarity of the algorithm will now be briefly discussed.

\noindent It is convenient to use the bra-ket notation. Formally we can write
\begin{equation}
\label{eq:ADI_braket}
\begin{alignedat}{2}
\begin{cases}
\kket*{\psi^{\left(n+\frac{1}{2}\right)}} &= \Bigl(\mathbb{1}-\frac{\Delta t}{2i}\mathcal{H}_x^{\left(n+\frac{1}{2}\right)}\Bigr)^{-1}
\Bigl(\mathbb{1}+\frac{\Delta t}{2i}\mathcal{H}_y^{\left(n+\frac{1}{2}\right)}\Bigr)\,\kket*{\psi^{(n)}}
\\
\kket*{\psi^{(n+1)}} &=  \Bigl(\mathbb{1}-\frac{\Delta t}{2i}\mathcal{H}_y^{\left(n+\frac{1}{2}\right)}\Bigr)^{-1}
\Bigl(\mathbb{1} + \frac{\Delta t}{2i}\mathcal{H}_x^{\left(n+\frac{1}{2}\right)}\Bigr)\,\kket*{\psi^{\left(n+\frac{1}{2}\right)}}.
\end{cases}
\end{alignedat}
\end{equation}
The norm at time $t_{n+1}$ can then be expressed as a function of the one at the previous time-step. 
\newline Since $\mathcal{H}_x$ and $\mathcal{H}_y$ will not in general commute, we expect the algorithm not to be unitary. After some calculations (taking the second eq. in \ref{eq:ADI_braket}, computing $\bbrakket*{\psi^{\left(n+1\right)}}{\psi^{\left(n+1\right)}}$, using the first one and then commuting all the remaining operators) one sees that the term $\propto \Delta t^2$ vanishes identically, but the next order term does not
\begin{equation}
\bbrakket*{\psi^{\left(n+1\right)}}{\psi^{\left(n+1\right)}}= \,\bbrakket*{\psi^{(n)}}{\psi^{(n)}}+
%+\,&\frac{\Delta t^3}{8i}\bbra*{\psi^{(n)}}2\left[\mathcal{H}_x,\mathcal{H}_y^2\right]
%+ \left[\mathcal{H}_x^2,\mathcal{H}_y\right]\kket*{\psi^{(n)}}
\mathcal{O}(\Delta t^3).
\end{equation}
Since the error committed at each step is $\mathcal{O}(\Delta t^3)$, %\footnote{The error term reads \[\frac{\Delta t^3}{8i}\bbra*{\psi^{(n)}}2\left[\mathcal{H}_x,\mathcal{H}_y^2\right]+ \left[\mathcal{H}_x^2,\mathcal{H}_y\right]\kket*{\psi^{(n)}}.\]}. 
this kind of algorithm is not norm-preserving, and thus in principle we would need to normalize the wavefunction at every time-step.
However, since the error is higher order with respect to the one committed on the wavefunction, which is $\mathcal{O}(\Delta t^2)$, neglecting such an issue will not modify any of the physically relevant expectation values.

A unitary algorithm can be easily devised as well. Considering a time-independent Hamiltonian for simplicity and splitting the time evolution operator symmetrically we get
\begin{equation}
\psi(t+\Delta t)=e^{-i\mathcal{H} \Delta t}\psi(t)=e^{-i\mathcal{H}_x \frac{\Delta t}{2}}e^{-i\mathcal{H}_y \Delta t}e^{-i\mathcal{H}_x \frac{\Delta t}{2}}\psi(t)+\mathcal{O}(\Delta t)^3
\end{equation}
where $\mathcal{H}_x$ and $\mathcal{H}_y$ as before contain the differential operators along $x$ and $y$ respectively. Notice that the approximated time-evolution operator is unitary, and we still have the advantages of the Alternate Direction Implicit algorithm. Introducing intermediate functions, the action of such an operator on $\psi(t)$ can be written as
\begin{equation}
\begin{alignedat}{2}
\begin{cases}
u_1&=e^{-i \mathcal{H}_x \frac{\Delta t}{2}}\psi(t)
\\
u_2&=e^{-i \mathcal{H}_y \Delta t}u_1
\\
\psi(t+\Delta t)&=e^{-i \mathcal{H}_x \frac{\Delta t}{2}}u_2
\end{cases}
\end{alignedat}
\end{equation}
each of which can be solved through the unitary Crank-Nicholson algorithm with error $\mathcal{O}(\Delta t^2)$
\begin{equation}
\label{eq:unitary_ADI}
\begin{alignedat}{2}
\begin{cases}
\left(\mathbb{1}+i \frac{\Delta t}{4}\mathcal{H}_x\right)u_1&=\left(\mathbb{1}-i \frac{\Delta t}{4}\mathcal{H}_x\right)\psi(t)
\\
\left(\mathbb{1}+i \frac{\Delta t}{2}\mathcal{H}_y\right)u_2&=\left(\mathbb{1}-i \frac{\Delta t}{2}\mathcal{H}_y\right)u_1
\\
\left(\mathbb{1}+i \frac{\Delta t}{4}\mathcal{H}_x\right)\psi(t+\Delta t)&=\left(\mathbb{1}-i \frac{\Delta t}{4}\mathcal{H}_y\right)u_2.
\end{cases}
\end{alignedat}
\end{equation}
It is apparent that the two spatial directions are treated in the same \virgolette{staggered} way as the previous algorithm eq. \ref{eq:ADI} did, so that solving eq. \ref{eq:unitary_ADI} instead of eq. \ref{eq:CrankNicolson} carries substantially the same advantages as the ones introduced by the pair of equations \ref{eq:ADI}, yet it is mildly more computationally expensive (one linear system of equations more to solve per time-step), so it has only been used as a toy to check whether the non-unitarity of the algorithm used was actually relevant.
With no surprise (for the reasons explained above) it turned out that this was not the case.

\section{Testing the algorithm} 
The algorithm has been tested in several cases which either admit analytical solution or can be treated via different substantially simpler numerical methods.

\subsection{Gaussian wavepacket in a uniform magnetic field}
Suppose we localize an electron in the system's bulk, and assume for the moment that finite system effects can be neglected. This second assumption is well motivated if the wavepacket is vanishingly small at the system edges; since in the classical limit the electron will undergo cyclotron motion, the initial state can be chosen so that the electron packet actually never feels the edge effects along its circular orbit.
\newline The initial state has chosen to be Gaussian
\begin{equation}
\label{eq:InitialStateGaussian}
\psi(x,y; t_0) = \left(\frac{1}{4\pi^2 \sigma_x^2\sigma_y^2}\right)^\frac{1}{4}\exp\left(-\frac{(x-x_0)^2}{4\sigma_x^2}\right)\exp\left(-\frac{(y-y_0)^2}{4\sigma_y^2}\right)
\end{equation}
and under the assumptions above it will evolve in time under the bulk Hamiltonian
\begin{equation}
\mathcal{H}= \frac{\pi_x^2+\pi_y^2}{2}=-\frac{1}{2}\frac{\partial^2}{\partial x^2}+\frac{1}{2}\,\left(x-i\frac{\partial}{\partial y}\right)^2 
\end{equation}
which is quadratic; it is interesting that an electron prepared in the above state will have a Gaussian wavefunction at any time. The state is thus, apart for a phase factor, completely characterized by the first and second order moments.
This fact can be straightforwardly shown by plugging a Gaussian with time-dependent means and variances and a space-time dependent phase into the Schrödinger equation, as pointed out in \cite{KimWeiner1973}. 
\newline Equations of motion for the first and second order moments are derived in Appendix \ref{app:GaussianMoments}, together with their solution. Analytical expressions are compared to numerical data. The mesh should be chosen so that  $\Delta x\ll\sigma_x$, $\Delta y\ll\sigma_y$ and $\Delta t\ll\frac{1}{\braket{\mathcal{H}}}$. In practice, each one has been chosen to be approximatively a hundredth of the associated typical variation scale, i.e. $\Delta x=0.05l_B$, $\Delta y=0.05l_B$ and $\omega_c\Delta t=\frac{\pi}{5000}$.
\begin{figure}[htp!]
	\begin{minipage}{.5\textwidth}
		\centering
		\includegraphics[width=1.\textwidth]{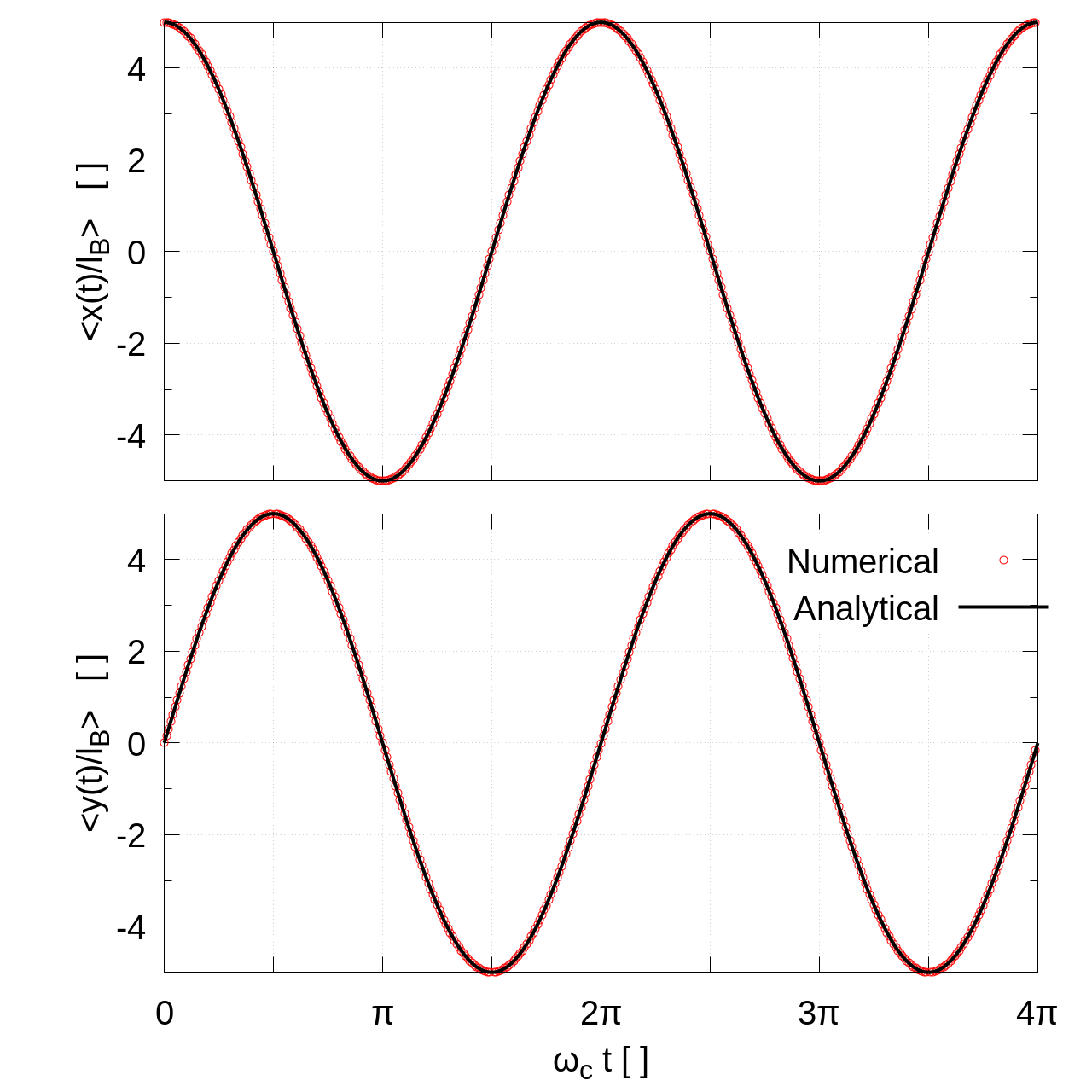}
	\end{minipage}%
	\begin{minipage}{0.5\textwidth}
		\centering
		\includegraphics[width=1.\textwidth]{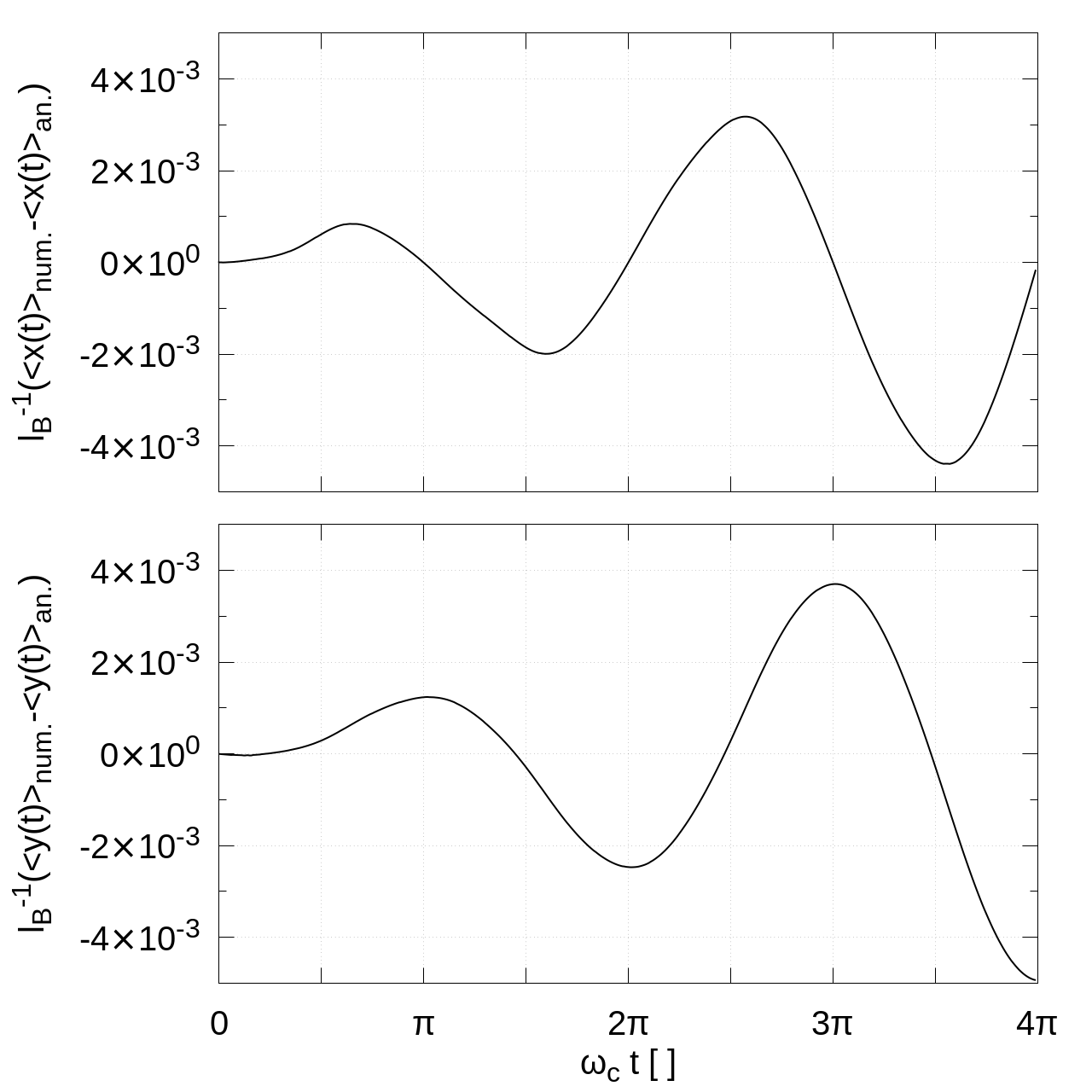}
	\end{minipage}    
	\caption[The LOF caption]{The first order spatial moments have been numerically computed using eq. \ref{eq:ADI}. The results are directly compared to the analytical expressions eq. \ref{fig:FirstOrderMoments} in the left hand side panel.
	The difference between the analytical and the numerical result is shown in the right panel.
	\newline The parameters of the Gaussian initial state from eq. \ref{eq:InitialStateGaussian} have been chosen to be $x_0=5l_B$, $y_0=0$, $\sigma_x=1.1l_B$ and $\sigma_y=1.8l_B$.}
	\label{fig:FirstOrderMoments}
\end{figure}

\noindent The average position follows a classical cyclotron motion eq. \ref{eq:first_order_moments}
\begin{equation}
\begin{cases}
\braket{x}_t = x_0 \cos(t)
\\
\braket{y}_t = y_0+x_0 \cos(t)
\end{cases}
\end{equation}
On the left hand side of Figure \ref{fig:FirstOrderMoments} these equations are compared to the numerically computed expectation values $\braket{x}_t, \braket{y}_t$. 
On the right hand side the difference between the exact solution and the numerical one is shown. The committed error exhibits an oscillatory behaviour, without blowing up exponentially, which gives an hint on the numerical stability of the algorithm. The oscillating growth is due to truncation error\footnote{A simple example. Consider $y'(t)=i \omega y(y)$ with $y(0)=1$. The solution is evidently $y(t)=e^{i\omega t}$. The Crank-Nicolson algorithm gives the solution as a discrete sequence which at a given time $t=n \Delta t$ can be found by simple iteration
	\small{\[
	y_{\Delta t}(t)=\frac{1+i\frac{\omega \Delta t}{2}}{1-i\frac{\omega \Delta t}{2}}y((n-1)\Delta t)=\left(\frac{1+i\frac{\omega \Delta t}{2}}{1-i\frac{\omega \Delta t}{2}}\right)^\frac{t}{\Delta t}y(0)=\exp\left(i\omega t\,\frac{\arctan\left(\frac{\omega \Delta t}{2}\right)}{\frac{\omega \Delta t}{2}}\right).
	\]}
	We can see that the solution is indeed an oscillating exponential, but with a frequency shift owing to the truncation error committed in writing down the Crank-Nicolson formula; the error $y(t)-y_{\Delta t}(t)$ is evidently bounded, the real and imaginary parts being modulated oscillations; the error will initially appear to be an oscillation (at frequency $\omega$) with increasing amplitude.}
and can be controlled by an adequate choice of the time-step $\Delta t$ and the final integration time. The simulations which have been performed show both qualitative and quantitative agreement with the analytical expectations.

\begin{figure}[htp!]
	\begin{minipage}{.5\textwidth}
		\centering
		\includegraphics[width=1.\textwidth]{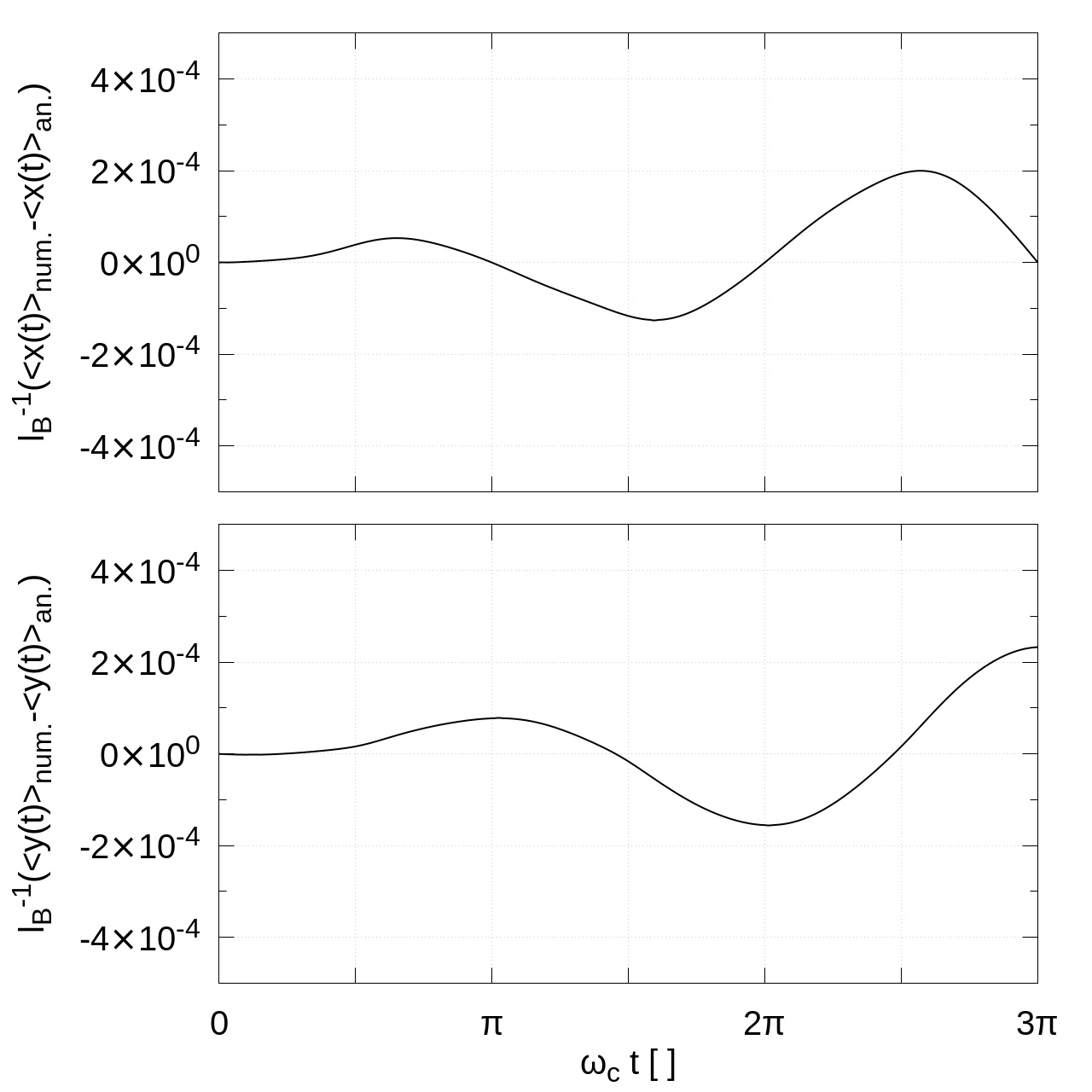}
	\end{minipage}%
	\begin{minipage}{0.5\textwidth}
		\centering
		\includegraphics[width=1.\textwidth]{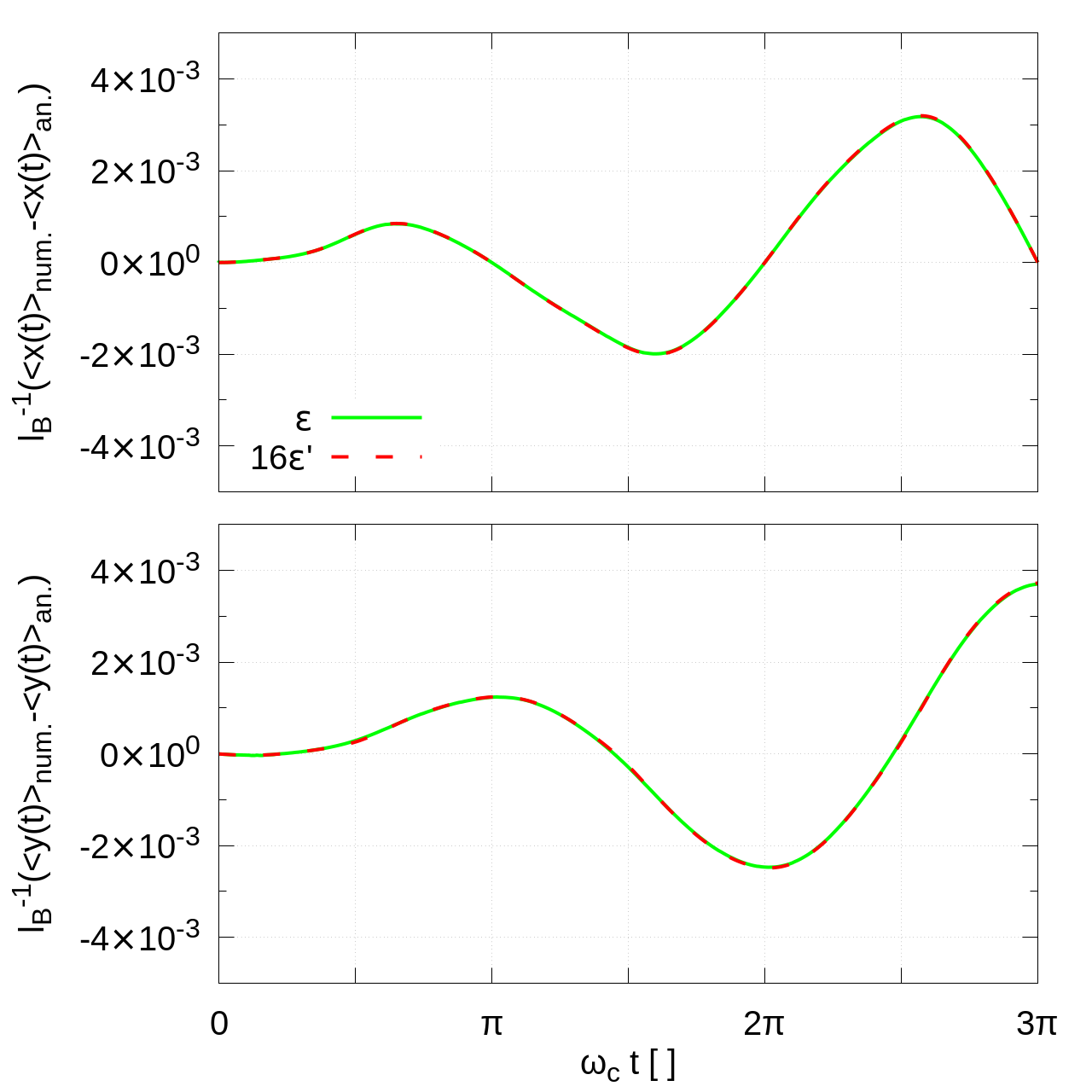}
	\end{minipage}    
	\caption[The LOF caption]{On the left hand side the computed error on the average positions $\braket{x(t)}$ and $\braket{y(t)}$ is shown when the spatial mesh resolution is doubled and the temporal one increased by a factor of four (i.e. $dx=dy=0.025l_B$ and $\omega_c\Delta t=\frac{\pi}{20000}$). The error is consequently expected to decrease by a factor $\sim16$ (see eq. \ref{eq:error_scaling2}) with respect to the one shown in Fig. \ref{fig:FirstOrderMoments}.
	\newline In the left hand side image we directly compare $\epsilon$ with $16\epsilon'$ (see equations \ref{eq:error_scaling} and \ref{eq:error_scaling2}).}
	\label{fig:ErrorScaling}
\end{figure}
These results can be used to check whether the error scales as predicted.
As already stated the numerical error is expected to scale $\mathcal{O}(dx^4,dy^4,dt^2)$. At some given time-step $n$ it can hence be expressed as
\begin{equation}
\label{eq:error_scaling}
\epsilon_n=a_n dx^4+b_n dy^4+c_n dt^2
\end{equation}
where $a_n$, $b_n$ and $c_n$ are probably very complicated numerical coefficients which however should be independent on the steps of the meshes. If, say, we halve the spatial mesh step and divide by four the temporal one we get
\begin{equation}
\label{eq:error_scaling2}
\epsilon_n'=a_n \frac{dx^4}{16}+b_n \frac{dy^4}{16}+c_n \frac{dt^2}{16}=\frac{1}{16}\,\epsilon_n.
\end{equation}
Notice that \textit{every} step needs to be reduced in order for the scaling to be observed, otherwise one of the other terms may dominate the total error and the reduction be useless.
By comparing the right hand side image of Fig. \ref{fig:FirstOrderMoments} with the left hand side one in Fig. \ref{fig:ErrorScaling} one can see that the numerical error committed scales as predicted by eq. \ref{eq:error_scaling2}. 
The image in the right hand side panel of Fig. \ref{fig:ErrorScaling} directly compares $\epsilon_n$ with $16\epsilon'_n$; a good overlap between the two curves can be seen at a glance.

\begin{figure}[htp!]
	\begin{minipage}{.5\textwidth}
		\centering
		\includegraphics[width=1.\textwidth]{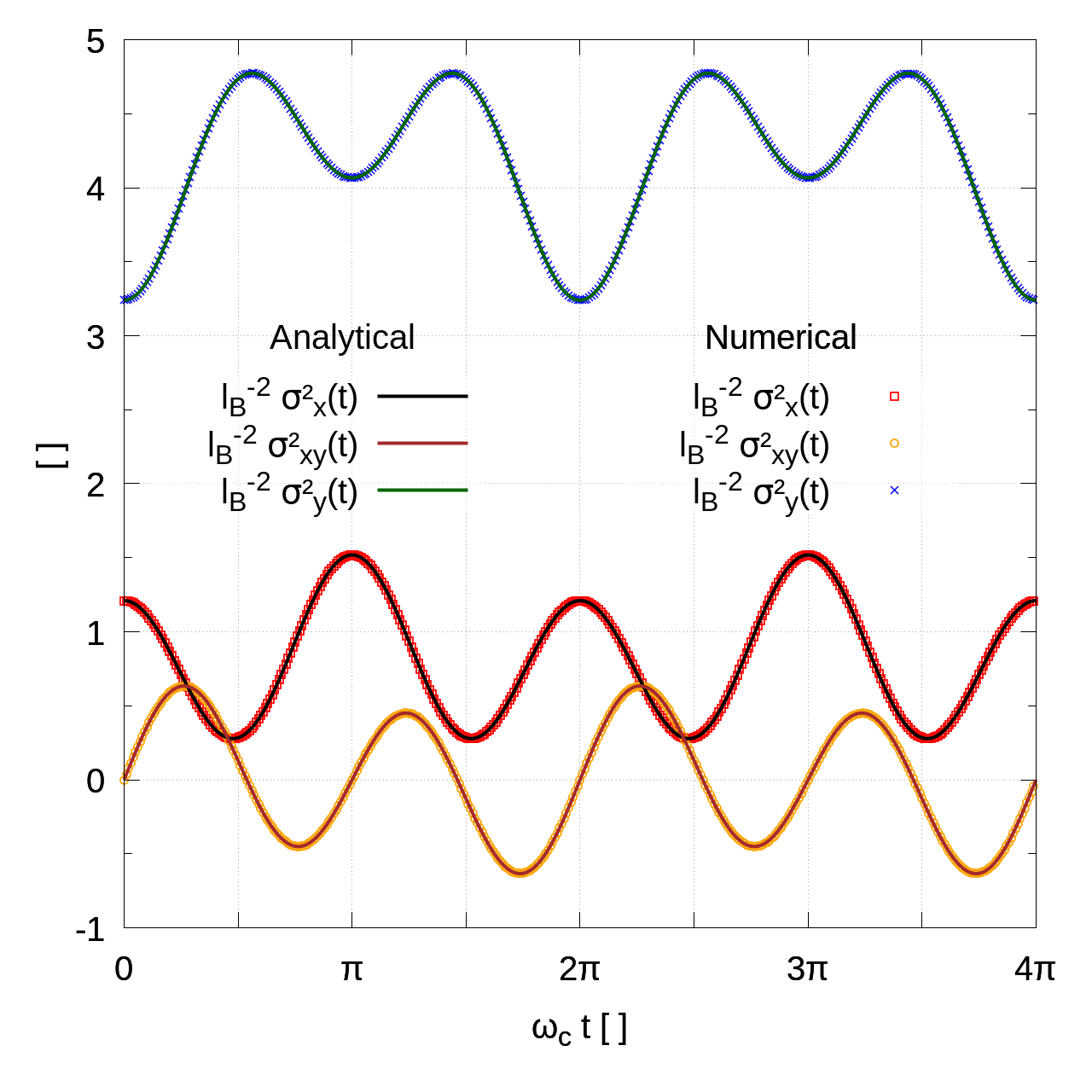}
	\end{minipage}%
	\begin{minipage}{0.5\textwidth}
		\centering
		\includegraphics[width=1.0\textwidth]{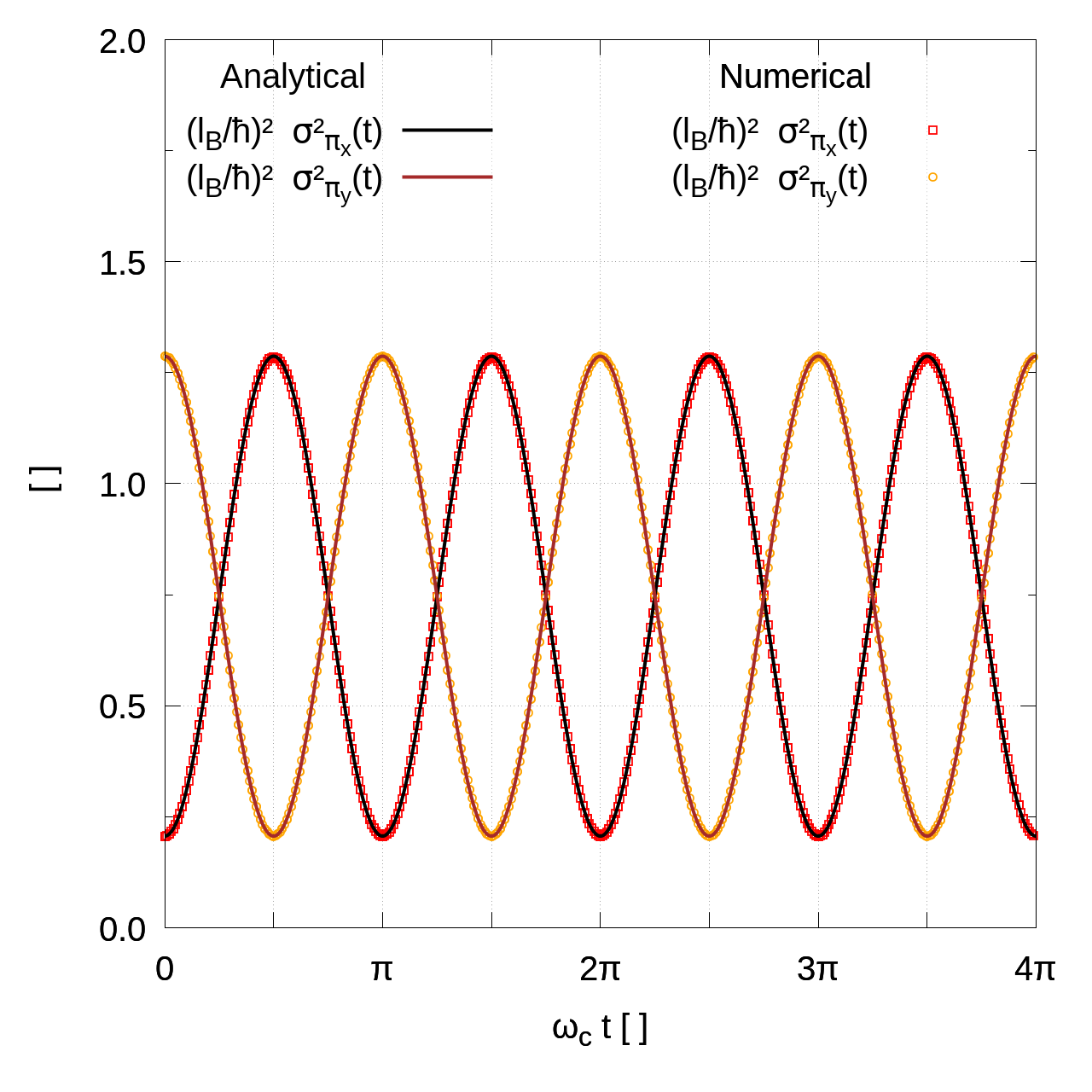}
	\end{minipage}    
	\caption[The LOF caption]{The panel on the left hand side shows moments which are second order in the coordinates (i.e. $\sigma_{x_ix_j}^2=\braket{x_ix_j}-\braket{x_i}\braket{x_j}$), comparing the numerical results with the theoretical predictions. On the right hand side the second order moments in the kinetic momenta are plotted instead.}
	\label{fig:SecondOrderMoments}
\end{figure}

\noindent In general the variances of $\pi_x$ and $\pi_y$ will obey equations of the form eq. \ref{eq:pix2_pixpiy_piy2_expectation}. Since $\braket{s_{\pi_x,\pi_y}}_0-2\braket{\pi_x}_0\braket{\pi_y}_0=0$, the simpler eq. \ref{eq:sigmas} can be used.
The widths of the packet instead obey eq. \ref{eq:SpatialVariances}.
The numerical simulations results are compared with the analytical ones in Fig. \ref{fig:SecondOrderMoments}. The committed errors are not shown, being exactly analogous to those plotted above in the case of the first order moments (Fig. \ref{fig:FirstOrderMoments}).
As discussed in the appendix we can see that as $\sigma_{\pi_x}$ decreases $\sigma_x$ increases, and vice-versa, and analogously for the other pair of conjugated variables $\{\sigma_{\pi_y}, \sigma_{y}\}$ and $\{\sigma_{\pi_y}, \sigma_{\pi_x}\}$; the Gaussian, while following the classical trajectory, \virgolette{pulsates} and tilts periodically (with the same period $\omega_c T = 2\pi$ of the cyclotron orbit). 
Notice that interestingly the Gaussian packet widths at $\omega_c t=\pi$ are not the same as the initial ones; a semiclassical heuristic explanation for this seemingly odd phenomenon is given at the end of the Appendix \ref{app:GaussianMoments}.

\begin{figure}[htp!]
	\begin{minipage}{.5\textwidth}
		\centering
		\includegraphics[width=1.0\textwidth]{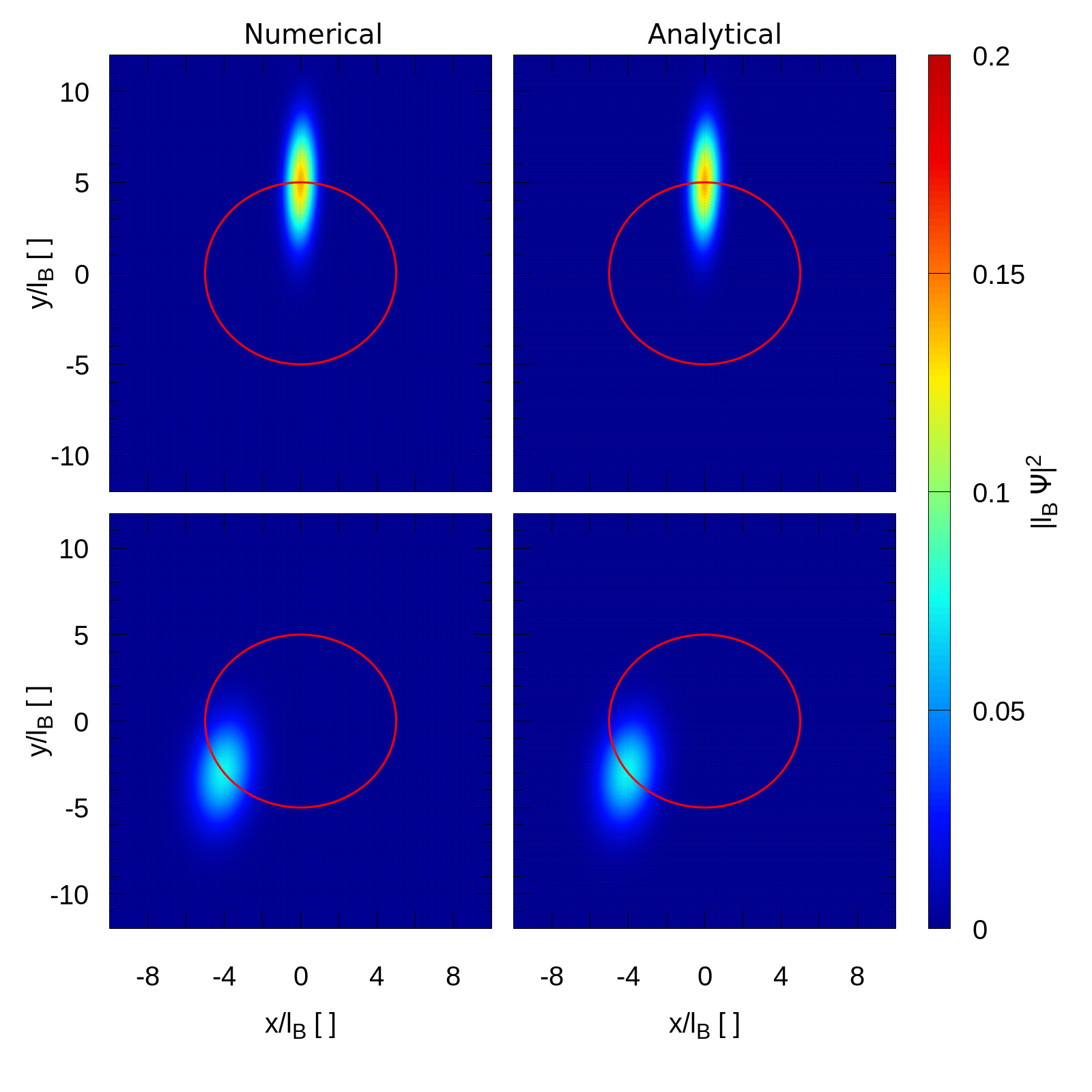}
	\end{minipage}%
	\begin{minipage}{0.5\textwidth}
		\centering
		\includegraphics[width=1.0\textwidth]{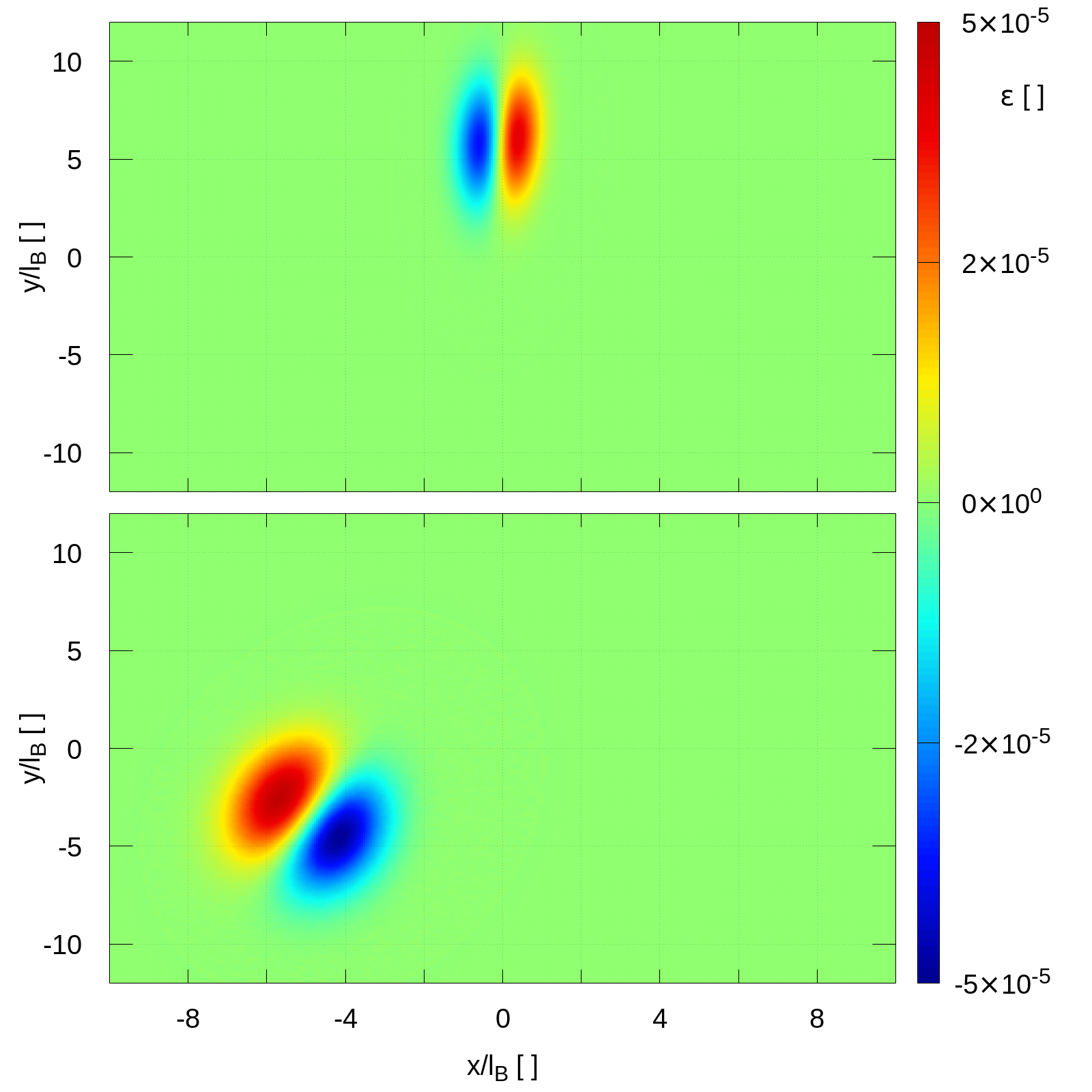}
	\end{minipage}    
	\caption[The LOF caption]{The images in the left hand side panel qualitatively compare the square modulus of the state $\psi(x,y;t)$ obtained by numerically evolving (using eq. \ref{eq:ADI}) the initial state $\psi(x,y;t_0)$ (eq. \ref{eq:InitialStateGaussian}) with the analytical expression eq. \ref{eq:wavepacket_time_evo}. 
	Numerical data are shown in the leftmost panels at two different time instants, $\omega_c t=\frac{\pi}{2}$ (top) and $\omega_c t=\frac{6}{5}\pi$ (bottom). The analytical expression evaluated at the same time instants is plotted alongside.
	\newline In the right hand side panel the differences between the numerical and analytical data is plotted instead.}
	\label{fig:WFcomparison} 
\end{figure}

A \virgolette{full} comparison between the square modulus of the numerically time-evolved wavepacket and the analytical solution given in the appendix (eq. \ref{eq:wavepacket_time_evo}) is shown in Fig. \ref{fig:WFcomparison} in the left hand side image. The other panel shows the difference between the analytical and numerical results, making clear that the agreement between the two results is \virgolette{good}. 
\newline As a final note, the wavefunction normalization has been directly checked at every time-step, and no significant variation of the norm has been observed. 

\begin{figure}[htp!]
	\centering
	\includegraphics[width=1.0\textwidth]{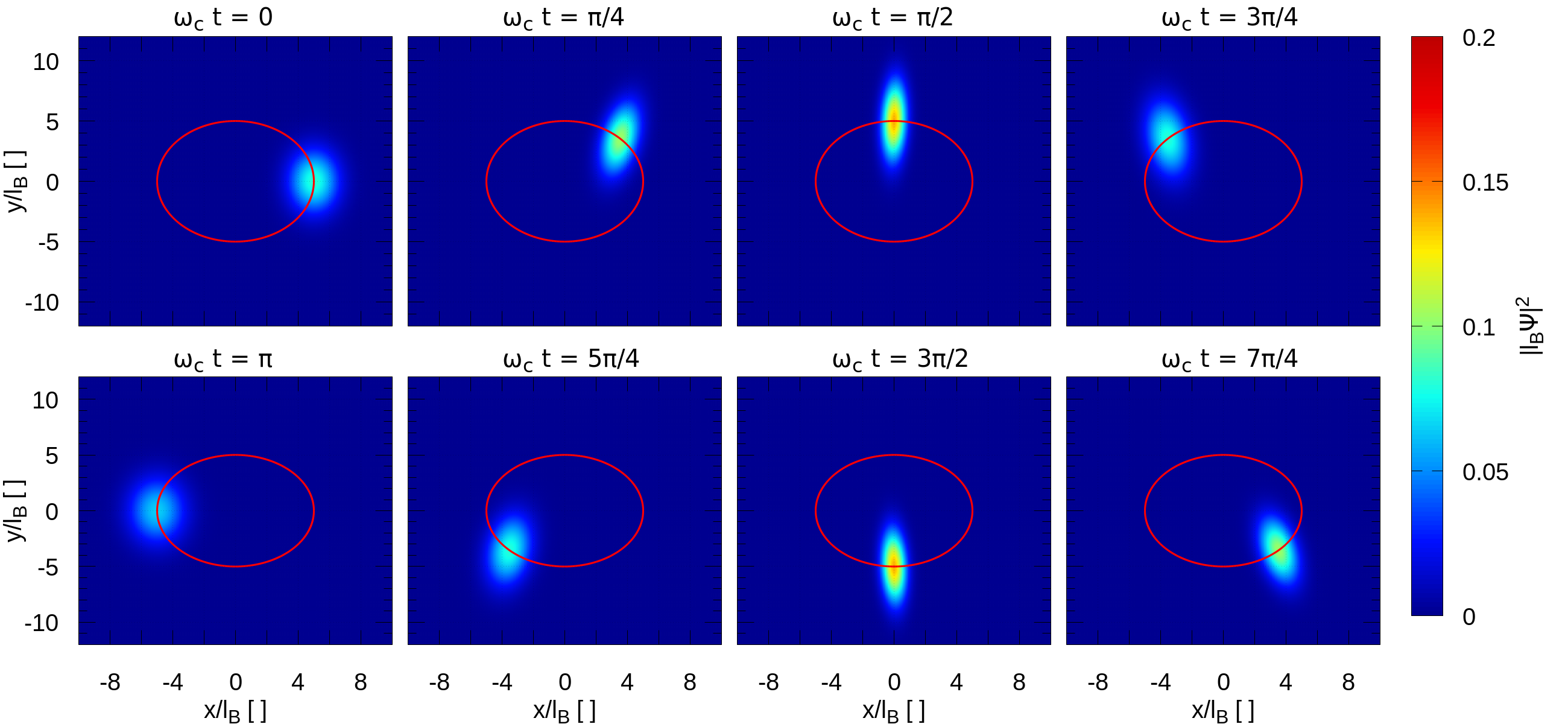}
	\caption[The LOF caption]{The images show the numerically computed wavepacket square-modulus at different times, increasing from left to right and from top to bottom (the snapshots have been taken at regular intervals $\omega_c \Delta t=\frac{\pi}{4}$, from $\omega_c t=0$ to $\omega_c t=2\pi$. The $\omega_c t=2\pi$ heat map is actually not shown, being the same as the one at $\omega_c t=0$).}
	\label{fig:gaussian_time_evo}
\end{figure}

\noindent For the sake of completeness, several snapshots taken during the numerical time evolution are shown in Fig. \ref{fig:gaussian_time_evo}.
Notice that by reflecting $y\rightarrow-y$ the squared modulus of the wavepacket at time $\omega_c t$ looks exactly the same as the one computed at time $2\pi-\omega_c t$. This is not a coincidence but a symmetry of the time-evolution equation together with the initial condition. The initial state eq. \ref{eq:InitialStateGaussian} is invariant under a reflection through the $y=y_0$ axis, $y'=2y_0-y$ (in the image $y_0=0$); if the same transformation is applied to the complex conjugated and time reversed $t'=-t$ Schrödinger equation
\begin{equation}
i\,\frac{\partial\,\psi^*(x,y';t')}{\partial t'}=\left[-\frac{1}{2}\frac{\partial^2}{\partial x^2}+\frac{1}{2}\,\left(x-i\frac{\partial}{\partial y}\right)^2 \right]\,\psi^*(x,y';t')
\end{equation}
we see that $\psi^*(x,y';t')$ evolves in time in exactly the same way as $\psi(x,y;t)$ does, with the same initial condition; then $\psi(x,y;t)=\psi^*(x,2y_0-y;-t)$.
This fact, together with the motion being periodic with period $\frac{2\pi}{\omega_c}$, explains the observation made. 

The algorithm proved to works as expected, at least when boundary effects can be neglected and the Hamiltonian does not depend on time. In the next two sections these aspects are investigated.

\subsection{Time evolution of an eigenstate}
An eigenstate of the system Hamiltonian undergoing free evolution has a rather trivial time-evolution given by a simple phase factor $e^{-i \omega t}$, with $\omega$ being the Bohr frequency of the state. 
Verifying whether the numerical results are able to reproduce this behaviour is a simple check for the correct algorithmic implementation of the boundary conditions as well as for the time evolution in general.
\begin{figure}[htp!]
	\begin{minipage}{.5\textwidth}
		\centering
		\includegraphics[width=1.0\textwidth]{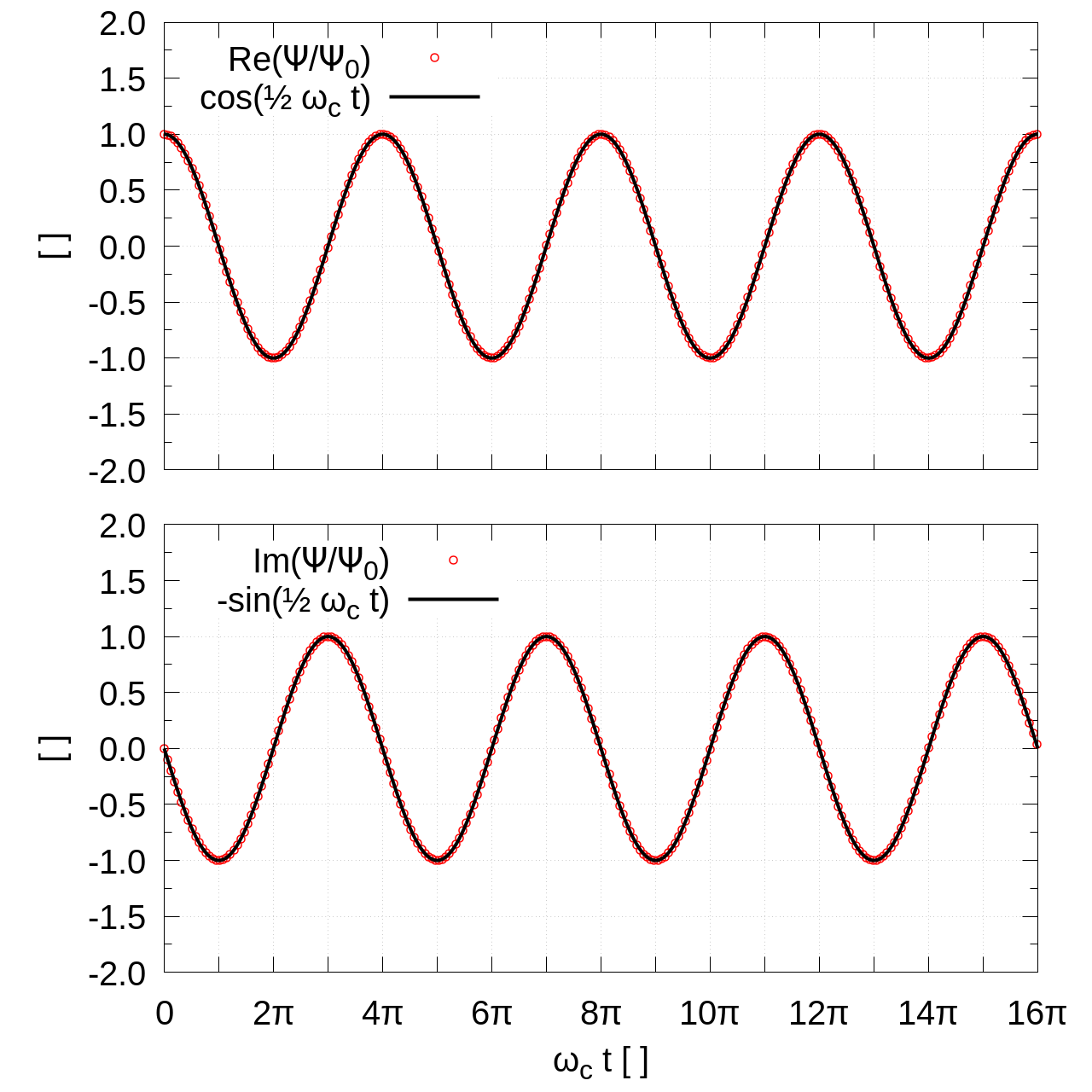}
	\end{minipage}%
	\begin{minipage}{0.5\textwidth}
		\centering
		\includegraphics[width=1.0\textwidth]{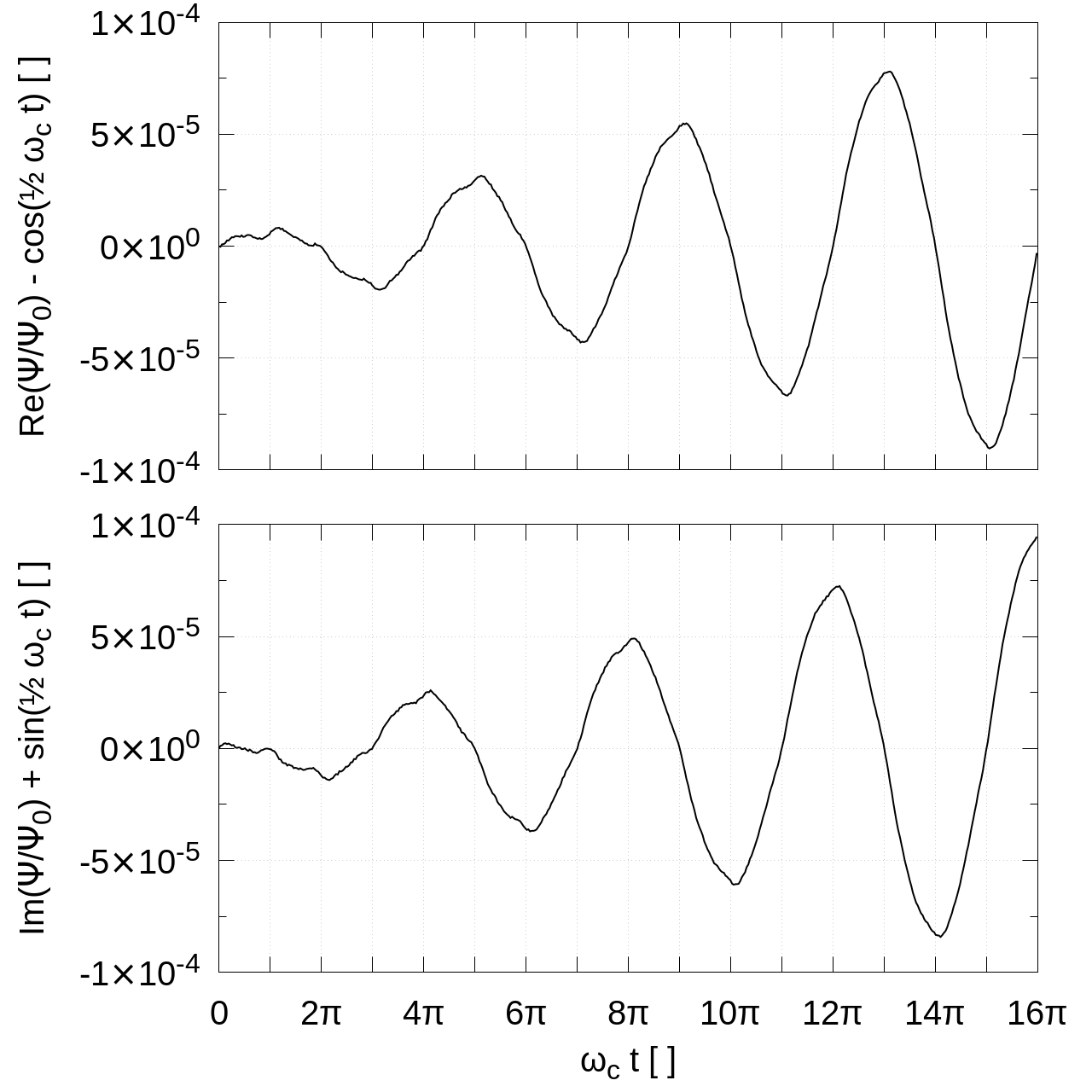}
	\end{minipage}    
	\caption[The LOF caption]{On the left hand side the numerical time evolution of $\frac{\psi(x,y; t)}{\psi(x,y; 0)}$ computed at fixed $(x,y)=(0,0)$ in the case of the bulk state $(n=0,k=0)$ is compared with the expected phase factor $e^{-i \frac{\omega_c t}{2}}$. On the right hand side the difference between these two quantities is shown instead.
	\newline $dx=dy=0.05l_B$ and $\omega_c t=0.1$ have been used.}
	\label{fig:BohrOscillations1}
\end{figure}

In order to check this behaviour we looked at $\frac{\psi(x,y; t)}{\psi(x,y; 0)}=\frac{\psi(x,y; t)}{\psi_{n,k}(x,y)}$, which should equal $e^{-i E_{n,k} t}$. The quantity $\frac{\psi(x,y; t)}{\psi(x,y; 0)}$ has been computed at a fixed point in the sample, which has been chosen so that the initial wavefunction was significatively different from zero there. 
The numerically computed eigenfunctions $\psi_{n,k}(x,y)$ have been used to this purpose.

The bulk behaviour has been checked first, by time-evolving with eq. \ref{eq:ADI} the bulk state with vanish momentum $k=0$ in the lowest Landau level $n=0$. We expect in this case a time-evolution set by half the cyclotron frequency $e^{-i \frac{\omega_c t}{2}}$. 
Left panel of Figure \ref{fig:BohrOscillations1} qualitatively compares the expectation with the numerical results, while on the right hand side a more quantitative measure of the error committed is shown.

\begin{figure}[htp!]
	\begin{minipage}{.5\textwidth}
		\centering
		\includegraphics[width=1.0\textwidth]{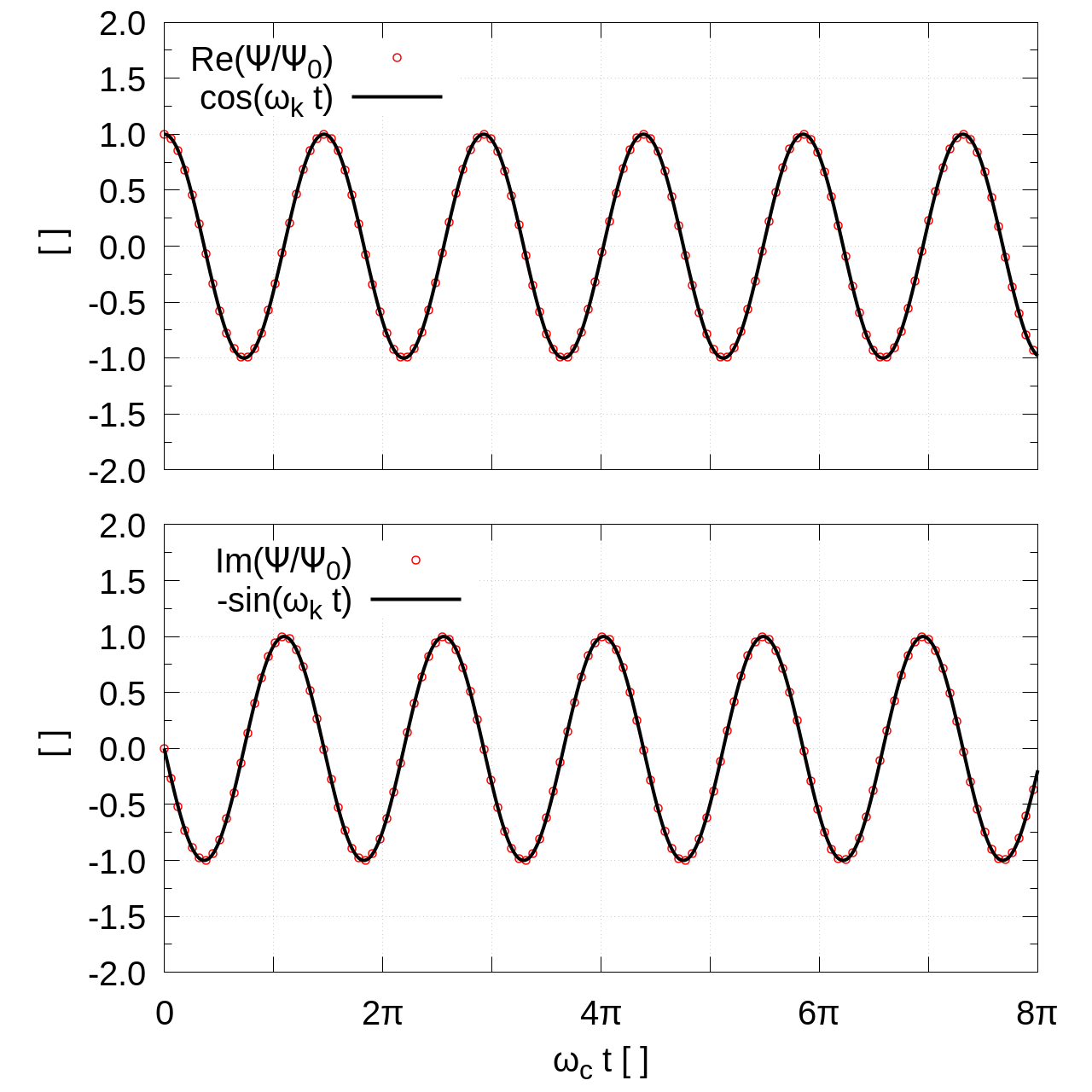}
	\end{minipage}%
	\begin{minipage}{0.5\textwidth}
		\centering
		\includegraphics[width=1.0\textwidth]{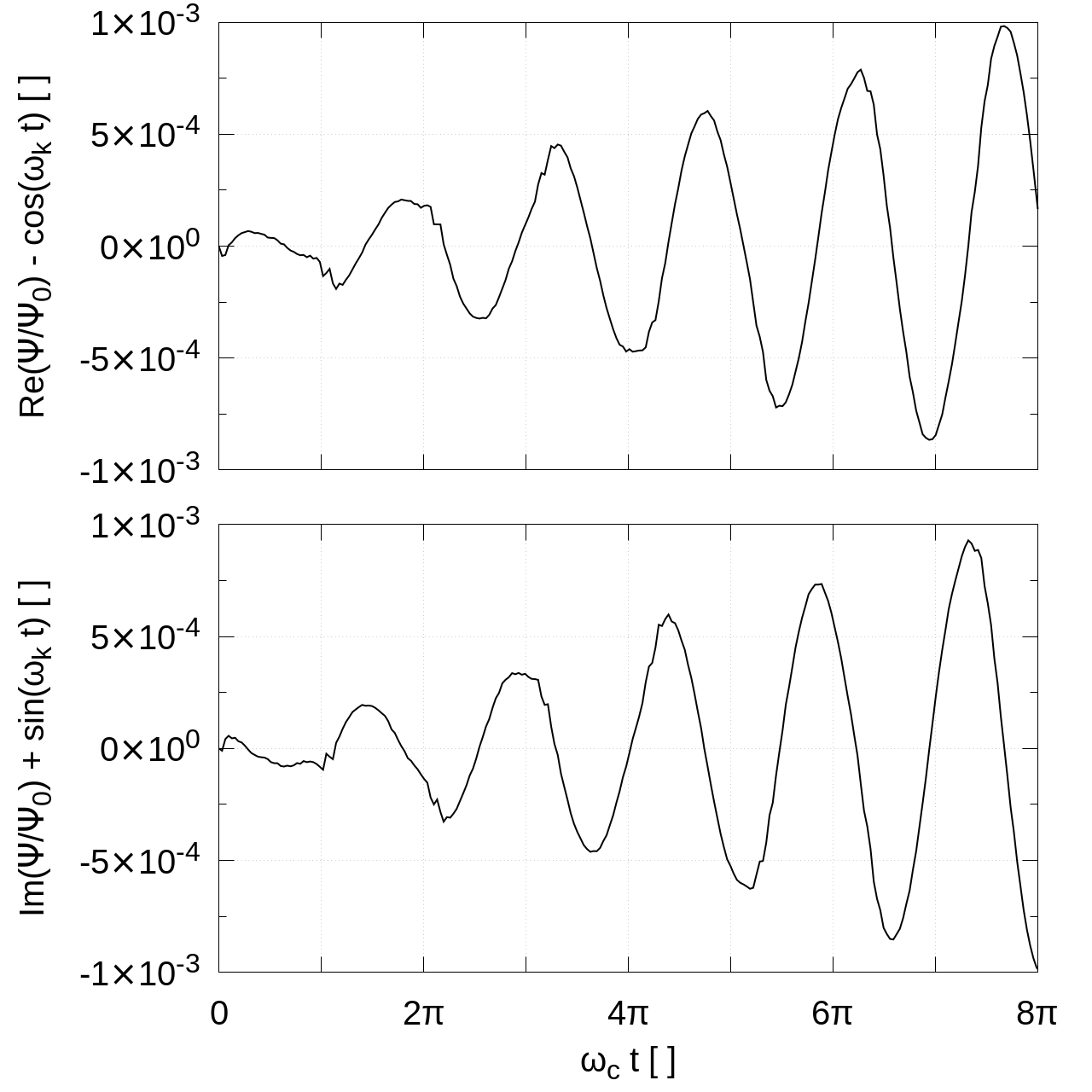}
	\end{minipage}    
	\caption[The LOF caption]{On the left hand side the numerical time evolution of $\frac{\psi(x,y; t)}{\psi(x,y; 0)}$ computed at fixed $(x,y)=(9l_B,0)$ in the case of an edge state $(n=0,k\simeq -9.84l_B^{-1})$ is compared with the expected phase factor $e^{-i E_k t}$, with $E_k\simeq 1.37\hbar\omega_c$. On the right hand side the difference between these two quantities is shown instead.
	\newline The confining potential eq. \ref{eq:confining_potential} parameters have been chosen to be $V_0=30\hbar\omega_c$, $\sigma_c=0.1l_B$ and the system length along $x$ $L_x=20l_B$.
	\newline $dx=dy=0.05l_B$ and $\omega_c t=0.1$ have been used.}
	\label{fig:BohrOscillations2}
\end{figure}
The same analysis has been carried over with a numerically computed edge state. The results are displayed in Fig. \ref{fig:BohrOscillations2}.
The error is larger than before, probably due to a faster time evolution (the energy of the state is larger than before, but the same time-step has been used); it can however be reduced by increasing the space-time grid resolution.

\subsection{Time dependent potentials}
To check whether time-dependent potentials are correctly treated, we look at how a bulk state evolves when a uniform electric field directed along $\hat{x}$ is adiabatically turned on. 
\newline The bulk Hamiltonian reads
\begin{equation}
\mathcal{H}=\frac{p_x^2}{2}+\frac{1}{2}\left(x+p_y\right)^2+\lambda(t)\,x
\end{equation}
where $\lambda(t)$ is the electric field strength which for a uniform field is a function of time alone.
\newline Suppose the initial state is an eigenstate of the unperturbed bulk Hamiltonian
\begin{equation}
\psi(x,y,t_0)=\psi_{n,k}(x,y)=\Phi_{n}(x+k)\frac{e^{iky}}{\sqrt{L_y}}.
\end{equation}
Since $e^{iky}$ is still an eigenstate of the perturbed Hamiltonian (translational invariance along $y$ is preserved) we can eliminate such a dependence by setting $\psi(x,y,t)=e^{iky}\psi_k(x,t)$, which evolves according to
\begin{equation}
i\frac{\partial \psi_k(x,t)}{\partial t}=\left(\frac{p_x^2}{2}+\frac{1}{2}\left(x+k\right)^2+\lambda(t)\,x\right)\psi_k(x,t).
\end{equation}
Introducing creation and annihilation operators for the unperturbed oscillator modes
\begin{equation}
\begin{alignedat}{2}
\begin{cases}
x+k&=\frac{1}{\sqrt{2}}\left(a^\dagger+a\right)
\\
p_x&=\frac{i}{\sqrt{2}}\left(a^\dagger-a\right)
\end{cases}
\end{alignedat}
\end{equation}
decomposing the wavefunction over the complete harmonic oscillator states
\begin{equation}
\label{eq:TDEF_FourierProjection}
\psi_k(x,t)=\sum_n C_n(t)\,\bbrakket*{x+k}{n}
\end{equation} 
and exploiting orthonormality one gets an (infinite) set of linear non-homogeneous coupled differential equations
\begin{equation}
\label{eq:TDEF_FourierComponentsTimeEvo}
i\frac{\partial C_n(t)}{\partial t}\,
=
\left(n+\frac{1}{2}-k \lambda(t)\right)C_n(t)
+\lambda(t)\left(\sqrt{\frac{n+1}{2}}\,\,C_{n+1}(t) +\sqrt{\frac{n}{2}}\,\,C_{n-1}(t) \right)
\end{equation}
which have to be supplemented by adequate initial conditions. The most obvious choice is to consider the system as being initially in the ground-state, i.e. $C_{n}(t_0)=\delta_{n0}$.
\begin{figure}[htp!]
	\begin{minipage}{.5\textwidth}
		\centering
		\includegraphics[width=1.0\textwidth]{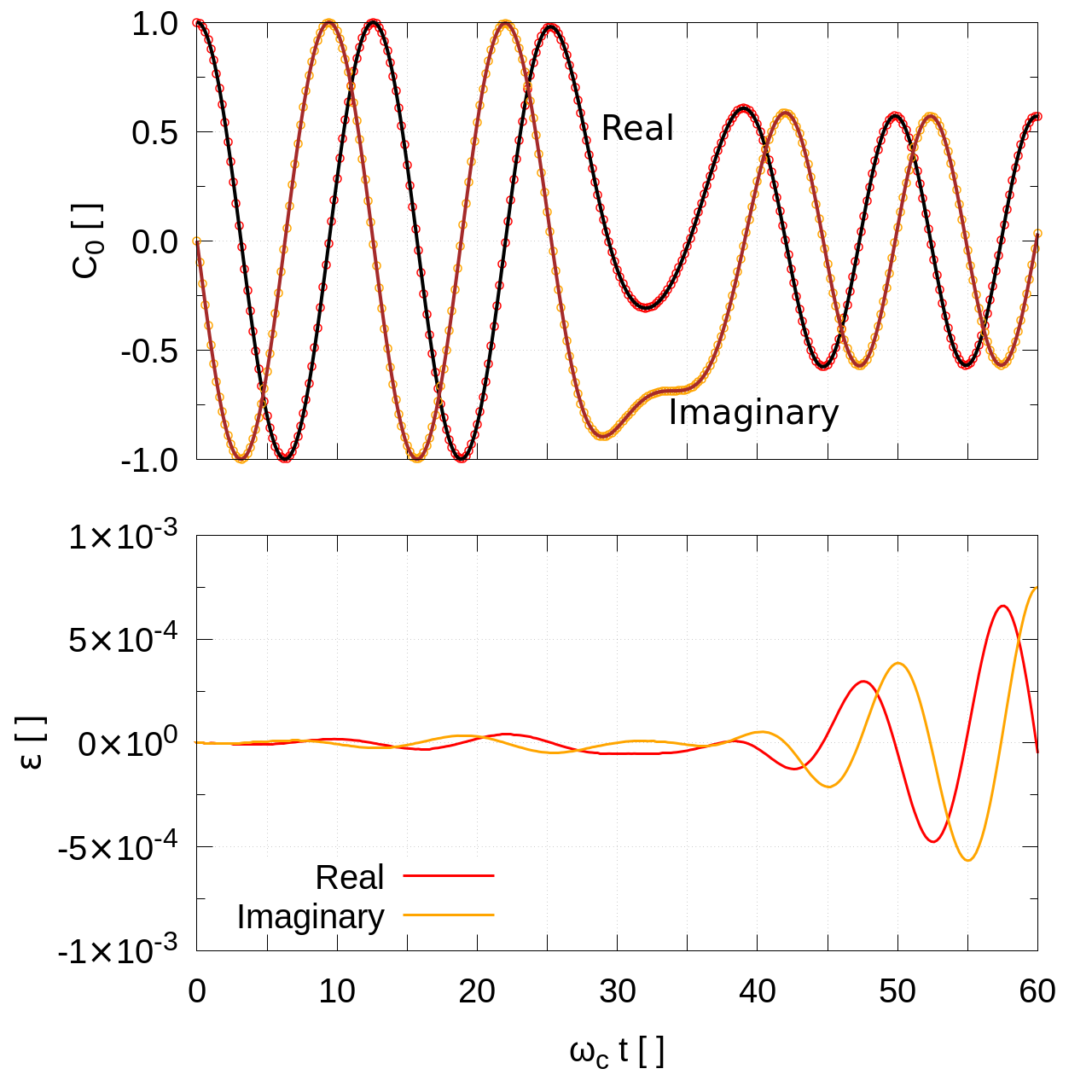}
	\end{minipage}%
	\begin{minipage}{0.5\textwidth}
		\centering
		\includegraphics[width=1.0\textwidth]{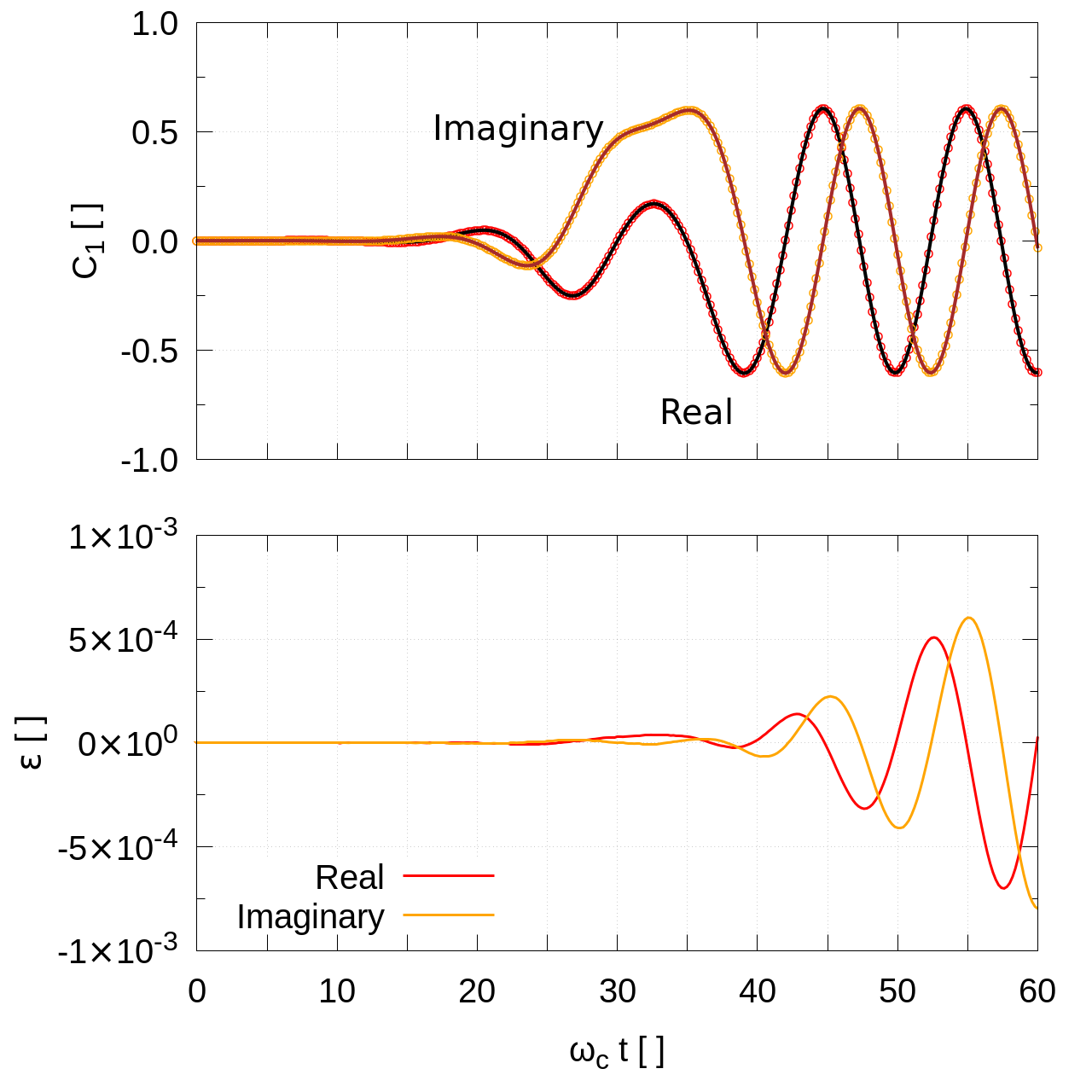}
	\end{minipage}    
	\caption[The LOF caption]{In the top-left hand side image the real and imaginary parts of the Fourier coefficient $C_{0}$ in eq. \ref{eq:TDEF_FourierProjection} obtained by using the numerical time-evolution algorithm eq. \ref{eq:ADI} (points) are directly compared with the same quantities obtained through the numerical solution of eq. \ref{eq:TDEF_FourierComponentsTimeEvo} (lines). In this last case the set of first order linear differential equations has been integrated using the Runge-Kutta 4 algorithm.
	In the bottom panel their difference (which is an estimate of numerical error) is plotted.
	\newline The right hand side panel is analogous, but the $C_1$ coefficient is shown instead.
	\newline Both the images have been obtained using $k=0$; the parameters of the time-dependent electric field in eq. \ref{eq:TDEF_potential} have been chosen to be $\eta=0.3\omega_c$, $t_0=30\omega_c^{-1}$ and $\lambda_0=1.5\hbar\omega_c$.}
	\label{fig:TDPotentialsCheck}
\end{figure}
The system of equations can be truncated at some point since oscillator modes with high enough energy will not be excited by the time-dependent excitation; it can then be solved by standard numerical methods. The explicit RK4 algorithm has been used and the results compared with those obtained from the time-evolution algorithm described above; in the latter case the Fourier coefficients have been obtained by straightforward numerical projection of the wavefunction at a given time onto the harmonic oscillator basis. 

Figure \ref{fig:TDPotentialsCheck} compares the results for $C_0(t)$ and $C_1(t)$ obtained from the two methods, with the choice $k=0$ and
\begin{equation}
\label{eq:TDEF_potential} 
\lambda(t)=\lambda_0\,\frac{e^{\eta(t-t_0)}}{e^{\eta(t-t_0)}+1}.
\end{equation}
The other coefficients exhibit a similar behaviour. The comparison has been repeated for different values of $k$; since the results are completely analogous to the ones which can be seen in Fig. \ref{fig:TDPotentialsCheck}, they will not be shown.
%The comparison has been repeated in the case $k=2l_B^{-1}$ and the results are analogous, as can be seen in Fig. \ref{fig:TDPotentialsCheck2}.
%\begin{figure}[htp!]
%	\begin{minipage}{.5\textwidth}
%		\centering
%		\includegraphics[width=1.0\textwidth]{TDPotentialsCheck2_c0.png}
%	\end{minipage}%
%	\begin{minipage}{0.5\textwidth}
%		\centering
%		\includegraphics[width=1.0\textwidth]{TDPotentialsCheck2_c1.png}
%	\end{minipage}    
%	\caption[The LOF caption]{The images are completely analogous to those in \ref{fig:TDPotentialsCheck}, but have been obtained using $k=2l_B^{-1}$.}
%	\label{fig:TDPotentialsCheck2}
%\end{figure}

\begin{figure}[htp!]
	\begin{minipage}{.5\textwidth}
		\centering		
		\includegraphics[width=1.0\textwidth]{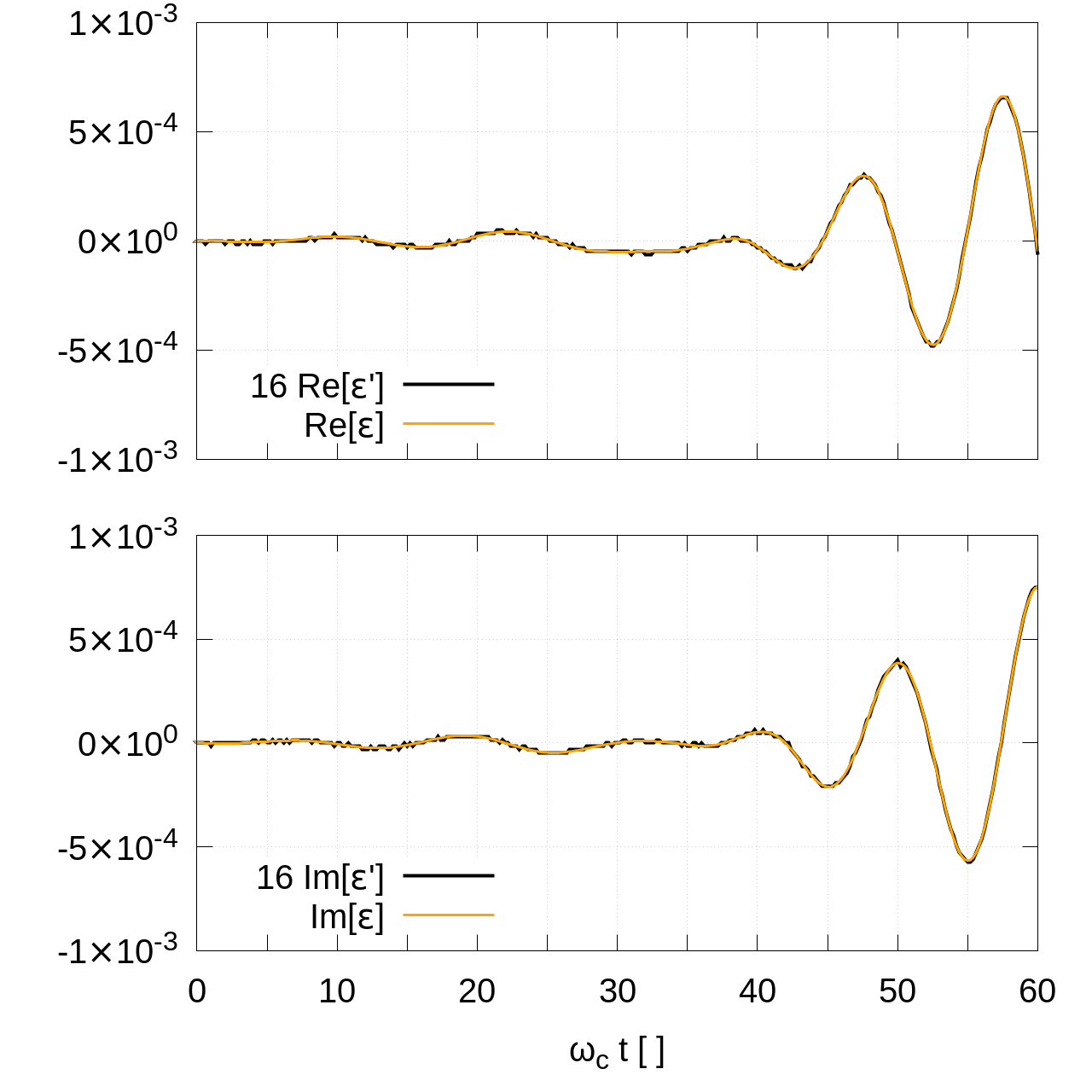}
	\end{minipage}%
	\begin{minipage}{0.5\textwidth}
		\centering
		\includegraphics[width=1.0\textwidth]{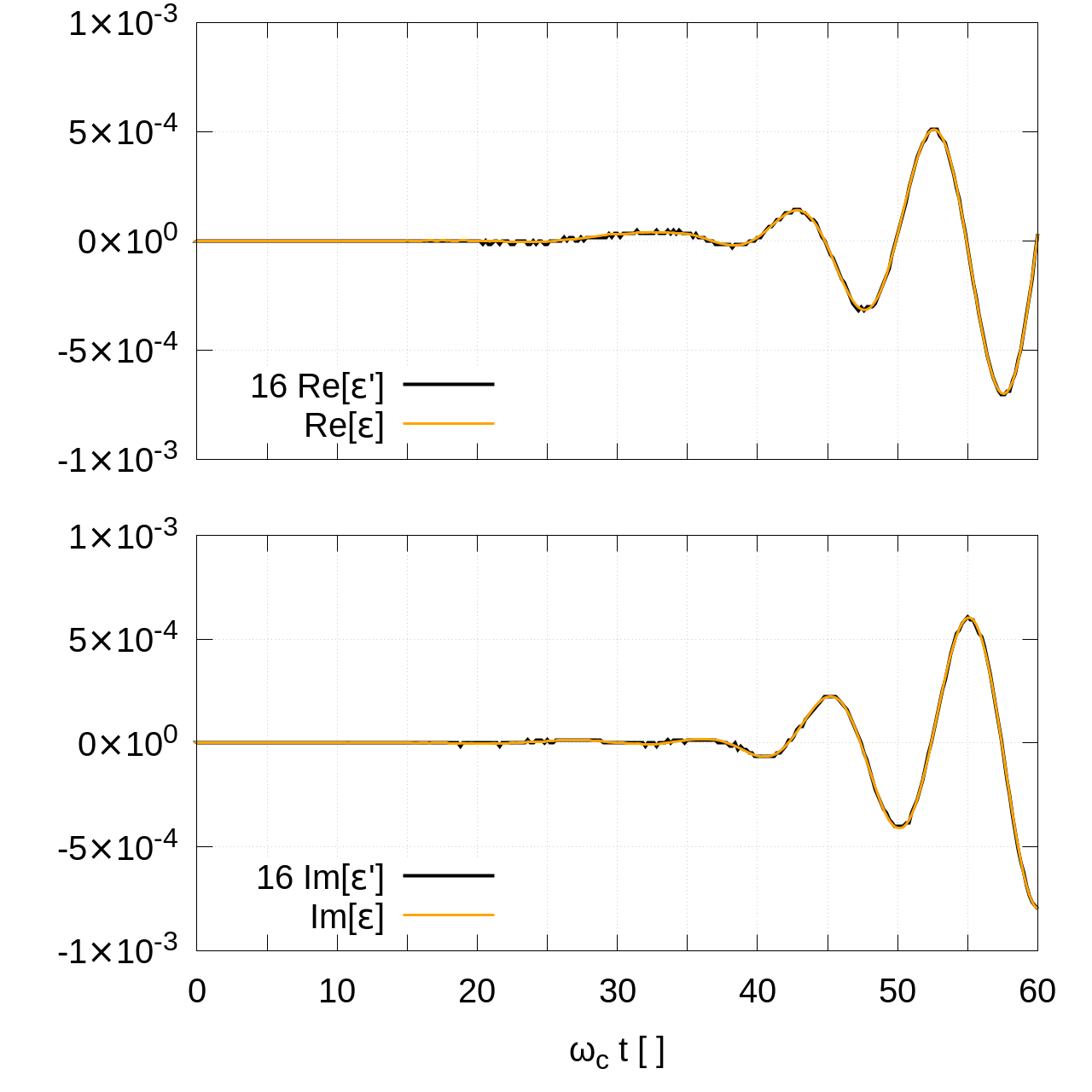}
	\end{minipage}    
	\caption[The LOF caption]{The two images show how the error on $C_0(t)$ (left panels) and $C_1(t)$ (right panels) scales when the spatial resolution is doubled and the temporal one increased by a factor of four, as in eq. \ref{eq:error_scaling2}; the finer-mesh error is consequently multiplied by a factor $16$. (The noise on $\epsilon'$ is caused by truncation error; too few digits for the purpose were saved as output at the end of the time-evolution cycle.)
}
	\label{fig:errorScalingTDP}
\end{figure}

\noindent The agreement can be seen to be quite good; notice that the error starts to grow only after $t=t_0$ due to the truncation error which causes the numerical solution computed using eq. \ref{eq:ADI} to slowly go out of phase with respect to the one obtained by solving eq. \ref{eq:TDEF_FourierComponentsTimeEvo}. It has however been checked that the expected error scaling $\mathcal{O}(dx^4,dy^4,dt^2)$ is obeyed, so that the error can be decreased by making the mesh finer.
In Fig. \ref{fig:errorScalingTDP} this scaling is tested by halving the spatial steps $dx$, $dy$ and dividing by four the temporal one $dt$ (see eq. \ref{eq:error_scaling2}). 

%\begin{figure}[htp!]
%	\begin{minipage}{.5\textwidth}
%		\centering
%		\includegraphics[width=1.0\textwidth]{TDPotentialsCheck2_c0.png}
%	\end{minipage}%
%	\begin{minipage}{0.5\textwidth}
%		\centering
%		\includegraphics[width=1.0\textwidth]{TDPotentialsCheck2_c1.png}
%	\end{minipage}    
%	\caption[The LOF caption]{The images are completely analogous to those in \ref{fig:TDPotentialsCheck}, but have been obtained using $k=2l_B^{-1}$.}
%	\label{fig:TDPotentialsCheck2}
%\end{figure}

This proves that the time evolution algorithm works as expected even in the presence of strong ($\lambda=1.5\hbar\omega_c$) time-dependent external potentials.

\section{Small numerical expedients}
Now some little expedients which have adopted to speed up the computations and to save some memory when saving the outputs of the algorithm will be briefly outlined.

\subsection{Following the slow temporal dynamics}
Suppose an electron is initially prepared in an eigenstate $\Psi_{n_0k_0}(x,y)$ of the Hamiltonian, and some sort of perturbation is adiabatically turned on. 
One is not interested in following the fast free evolution; the \virgolette{slow} dynamics determined by the slowly-varying perturbation can be isolated by setting
\begin{equation}
\Psi'(x,y;t)=e^{i E_{n_0k_0} t}\Psi(x,y;t)
\end{equation}
which obeys
\begin{equation}
i\,\frac{\partial \Psi'}{\partial t}=(\mathcal{H} - E_{n_0k_0})\Psi'.
\end{equation}
Such a substitution is very convenient since it allows one to use a coarser temporal mesh.

\subsection{Following the slow spatial dynamics}
If we are considering the semi-infinite free system the fast spatial $e^{i k_0 y}$ dependence of the eigenstates can be eliminated as well in an analogous way, by defining the slowly varying
\begin{equation}
\Psi''(x,y;t)=e^{-i k_0 y}\Psi'(x,y;t)
\end{equation}
which obeys
\begin{equation}
\begin{cases}
i\,\frac{\partial \Psi''}{\partial t}=\left(\mathcal{H}_{k_0}- E_{n_0k_0}\right)\Psi''
\\
\mathcal{H}_{k_0}=-\frac{1}{2}\,\partial_x^2+\frac{1}{2}(x+k_0-i\partial_y)^2+V_\text{c}(x)+V_\text{ext}(x,y;t).
\end{cases}
\end{equation}
Again the substitution is quite convenient: the mesh in the $y$ direction can be made coarser since $\partial_y$ now acts on slowly varying functions of $y$ (as long as the external potential $V_\text{ext}(x,y;t)$ is a long-wavelength excitation of the system of course). 

\subsection{Saving the output}
In order to save physical memory it is convenient to save the Fourier coefficients obtained by projecting $\Psi'$ over the eigenstates of the free Hamiltonian $\mathcal{H}_0$ instead of the full $\Psi'$. 
This has mainly two advantages.
First of all, saving the wavefunction over the entire mesh is quite memory expensive: saving a relatively small number of Fourier coefficients (much smaller than $N_x N_y$, $N_x$ and $N_y$ being the number of mesh points along the two spatial directions) allows to save large portions of computer memory.
Moreover, since the state will be practically zero over a large portion of the sample this would be quite a waste of space.
\newline Secondly, once the perturbation is shut down each Fourier component evolves in time as a pure phase with frequency set by the free problem eigenenergies, and thus its time evolution becomes trivial.
\newline From
\begin{equation}
\Psi''(x,y;t)=e^{i E_{n_0k_0}t}e^{-i k_0 y}\sum_{nk}b_{nk}(t)e^{-i E_{nk}t}\psi_{nk}(x,y)
\end{equation}
one straightforwardly obtains
\begin{equation}
\label{eq:FC_computation}
b_{nk}(t)=e^{-i(E_{n_0k_0}-E_{nk})t}\int\Phi_{nk}(x)\left(\int\Psi''(x,y;t)\frac{e^{-i(k-k_0)y}}{\sqrt{L_y}}dy\right)dx.
\end{equation} 
Moreover, since the externally applied potential is adiabatically turned off at large times, the $b_k$ coefficients will become time-independent; the algorithm can hence be stopped at some large time $t_f$ when the perturbation is negligibly small; the various basis vectors will then evolve freely as eigenstates of $\mathcal{H}\simeq\mathcal{H}_0$, i.e. we can use
\begin{equation}
\Psi''(x,y;t>t_f)\simeq e^{i E_{n_0k_0}t}e^{-i k_0 y}\sum_{nk}b_{nk}(t_f)e^{-i E_{nk}t}\psi_{nk}(x,y)
\end{equation}
with negligible error. This fact has been used to extend the evolution to otherwise prohibitively large times.
\newline From the numerical solution of the time evolution equations eq. \ref{eq:ADI} we get $\Psi''(x,y;t)$; this function can be used to compute the Fourier coefficients through eq. \ref{eq:FC_computation}.
In practice since the external potential is quite small when compared to the Landau level spacing $\sim\hbar\omega_C$ and has a finite width in momentum space (say $\frac{1}{\sigma_y}$, where $\sigma_y$ is the typical lengthscale over which the external potential varies), the significatively non vanishing $b_{nk}$ coefficients will be those with $n$ a few Landau levels away from $n_0$ at most and with wavevector $k$ within some order-one multiple of $\frac{1}{\sigma_y}$.

We now introduce a very useful tool which serves as an intermediate step for the computation of more involved quantities, as will become apparent in a while. Let
\begin{equation}
C_{k}(x;t)=\int\Psi''(x,y;t)\frac{e^{-i(k-k_0)y}}{\sqrt{L_y}}dy.
\end{equation}
Obviously it is possible to obtain these functions directly from the Fourier coefficient expansion of $\Psi''$.
It is easily checked indeed that %$\delta(x-y)=\sum_{n}\Phi_{nk}(x)\Phi_{nk}(y)$
\begin{equation}
C_{k}(x;t)=\sum_n e^{i(E_{n_0k_0}-E_{nk})t}\,b_{nk}(t)\,\Phi_{nk}(x).
\end{equation}

The quantities of interest can be computed by using the $C_k$ coefficients. For example suppose we are interested in the computation of the Fourier transform of the system's density. 
The density of the non-interacting many-body quantum state can be obtained by summing over the square moduli of all the single-particle wavefunctions of the constituents of the system $\rho(x,y; t)=\sum_{n_0k_0}|\Psi|^2$. 
The Fourier transform of this quantity with respect to $y$ is 
\begin{equation}
\rho_q(x,t)=\int\,e^{-iqy}\,\rho(x,y;t)\,dy 
\end{equation}
which after some simple algebra simplifies to
\begin{equation}
\label{eq:denisty_general_form}
\rho_q=\sum_{n_0k_0}\sum_k\,C_{k}^*(x;t)C_{k+q}(x;t).
\end{equation}

The (probability) current can be computed too. For a single electron initially in the state labelled by $n_0,k_0$ we have
\begin{equation}
\mathbf{J}(x,y;t)=\,\Re\left[\Psi(x,y;t)^*\boldsymbol{\pi}\,\Psi(x,y;t)\right].
\end{equation}
Its $y$ Fourier transforms 
\begin{equation}
\mathbf{J}(x,q;t)=\int\,e^{-iqy}\,\mathbf{J}(x,y;t)\,dy
\end{equation}
can easily be computed in terms of the $C_k(x;t)$
\begin{equation}
\label{eq:currents_general_form}
\begin{cases}
J_x(x,q;t)=\frac{1}{2}\sum_k \left[\,C_{k}^*(-i\partial_x)C_{k+q} + (q\rightarrow-q)^*\right]
\\
J_y(x,q;t)=\frac{1}{2}\sum_k \left[\,C_{k}^*(k+q+x)C_{k+q}+(q\rightarrow-q)^*\right].
\end{cases}
\end{equation}

Finally, the Fourier transforms can then be inverted to get the real space quantities. 
%!TEX root = ../main.tex
% Chapter Template

\graphicspath{{./pic3/}}

\chapter{One-dimensional effective theory for the edge excitations}\label{ch3}
\lhead{Chapter 4. \emph{One-dimensional effective theory for the edge excitations}}
	The aim of this chapter is to study under general terms (with some simplifying assumptions) what happens when the two-dimensional electron gas in a ideal integer quantum Hall state is slightly perturbed. %(without coupling the two edges though, i.e. assuming that there is no net electronic flow from one edge to the other one so that the two can be treated independently).
	\newline It is well known that a gas of interacting one-dimensional spinless electrons can be mapped into a so called Tomonaga-Luttinger liquid (\cite{Tomonaga}, \cite{Luttinger}, \cite{MattisLieb}, \cite{Mahan}) which is a model of non-interacting bosons. 
	In the case of the edge of a quantum Hall system the story is slightly different, since near the edge we only have right or left propagating excitations; the theory, developed by Wen through abstract algebraic techniques for a fractional quantum Hall state, is known as a chiral Luttinger liquid (\cite{Wen1990}).
	\newline In this chapter I try to recover these well-established results in the case of the one-dimensional edges of a non-interacting integer quantum Hall state described by the Hamiltonian of eq. $\ref{eq:problem_hamiltonitan}$. 
	Using a second quantisation approach, focusing on the small excitations of the system near the Fermi surface and integrating out one spatial dimension an effective chiral wave equation describing the system edge dynamics is recovered. External time-dependent potentials are included as well in the description, which appear in the effective theory as a driving term.
	\newline At the end of the chapter a brief comparison with a classical fluid is made.

\section{Linearisation of the theory near the Fermi points}
We begin by rewriting the Hamiltonian operator eq. \ref{eq:problem_hamiltonitan} in the second quantization formalism
\begin{equation}
\mathcal{H}=\sum_{n,k}E_{n,k}\,C^\dagger_{n,k}C^{\vphantom{\dagger}}_{n,k}
\end{equation}
where the creation/annihilation operators are associated to the single particle basis $\mathcal{H}\psi_{n,k}=E_{n,k}\psi_{n,k}$ obtained above.

The low-energy physics is about small excitations near the Fermi surface, so we may expand
\begin{equation}
\mathcal{H}= \sum_{n}
\left(
\sum_{k=\kappa_n-\Lambda}^{\kappa_n+\Lambda} 
+\sum_{k=-\kappa_n-\Lambda}^{-\kappa_n+\Lambda}
+\sum_{|k|<\kappa_n-\Lambda}
+\sum_{|k|>\kappa_n+\Lambda}
\right) E_{n,k}\,C^\dagger_{n,k}C^{\vphantom{\dagger}}_{n,k}
\end{equation}
where $\kappa_n$ is the n-th Fermi wavevector, and $\Lambda$ is a hard cut-off at wavevectors much larger than the typical ones at the energy scale we are considering\footnote{One expects the low-energy physics to be $\Lambda$ independent; as it will be seen the final results indeed are.}, yet much smaller than $\kappa_n$ so that we can linearise the n-th Landau level about such a point.
\newline The low-lying excitations on the zero-temperature ground state of the system will then be described by
\begin{equation}
\begin{split}
\mathcal{H}_\text{eff}=& \sum_{n}
\left(
\sum_{k=\kappa_n-\Lambda}^{\kappa_n+\Lambda} 
+\sum_{k=-\kappa_n-\Lambda}^{-\kappa_n+\Lambda}
\right) E_{n,k}\,C^\dagger_{n,k}C^{\vphantom{\dagger}}_{n,k}=\\
=&
\sum_{n}
\sum_{k=-\Lambda}^{\Lambda}
\left( E_{n,k-\kappa_n}\,C^\dagger_{n,k-\kappa_n}C^{\vphantom{\dagger}}_{n,k-\kappa_n}+E_{n,k+\kappa_n}\,C^\dagger_{n,k+\kappa_n}C^{\vphantom{\dagger}}_{n,k+\kappa_n}\right).
\end{split}
\end{equation}
%The energy spectrum can be expanded near the Fermi surfaces
The energy spectrum can be linearised near the Fermi surfaces
\begin{equation}
\begin{split}
E_{n,k\pm\kappa_n}\simeq& \,\, E_{n,\pm\kappa_n} + k \frac{\partial E_{n,k}}{\partial k}\Bigl.\Bigr|_{\pm\kappa_n}% + \frac{k^2}{2}\, \frac{\partial^2 E_{n,k}}{\partial k^2}\Bigl.\Bigr|_{\pm\kappa_n}+\mathcal{O}(k)^3=\\
=%&
\,\,E_{n,\kappa_n} \pm \hbar v_n\, k%  + \frac{\hbar^2 k^2}{2m^*_n}
%\pm \lambda_n k^3
%\frac{k^3}{6}\,\frac{\partial^3 E_{n,k}}{\partial k^3}\Bigl.\Bigr|_{\pm\kappa_n}
\end{split}
\end{equation}
where in the last equality the relation $E_{n,-k}=E_{n,k}$ was used.
\newline %Let us begin by keeping only the linear term
This is obviously valid as long as $\Delta k\frac{\partial_k^2E_{n,k}}{\partial_k E_{n,k}}\ll1$, $\Delta k$ being a typical wavevector scale for the system (set by the external potential)
\begin{equation}
\begin{split}
\mathcal{H}_\text{eff}=& 
\sum_{n}
E_{n,\kappa_n}\,\sum_{k=-\Lambda}^{\Lambda}
\left(C^\dagger_{n,k+\kappa_n}C^{\vphantom{\dagger}}_{n,k+\kappa_n}+C^\dagger_{n,k-\kappa_n}C^{\vphantom{\dagger}}_{n,k-\kappa_n}\right)+\\
+& 
\sum_{k=-\Lambda}^{\Lambda}\sum_{n}\hbar v_n \,k \left(C^\dagger_{n,k+\kappa_n}C^{\vphantom{\dagger}}_{n,k+\kappa_n}-C^\dagger_{n,k-\kappa_n}C^{\vphantom{\dagger}}_{n,k-\kappa_n}\right).
\end{split}
\end{equation}
If we can treat the number of electrons near one of the two edges in a given branch $n$ as being conserved, the first piece of the previous Hamiltonian can be neglected, since it is then just a constant energy shift. The Hamiltonian then reduces to
\begin{equation}
\begin{split}
\mathcal{H}_\text{eff}=& 
\sum_{k=-\Lambda}^{\Lambda}\sum_{n}\hbar v_n\,k \left(C^\dagger_{n,k+\kappa_n}C^{\vphantom{\dagger}}_{n,k+\kappa_n}-C^\dagger_{n,k-\kappa_n}C^{\vphantom{\dagger}}_{n,k-\kappa_n}\right).
\end{split}
\end{equation}
Carrying out the exact same calculation for the operator annihilating a particle at position $(x,y)$
\begin{equation}
\psi(x,y)=\sum_{n,k}\,\psi_{n,k}(x,y)C^{\vphantom{\dagger}}_{n,k}
\end{equation}
(where $\psi_{n,k}(x,y)=\Phi_{n,k}(x) \frac{e^{i k\,y}}{\sqrt{L_y}}$ are the eigenfunctions of the Hamiltonian \ref{eq:problem_hamiltonitan}) one obtains
\begin{equation}
\begin{split}
\psi_\text{eff}(x,y)=&\sum_{n}\underbrace{e^{i \kappa_n y}\,\sum_{k=-\Lambda}^{\Lambda}
	\frac{e^{i k y}}{\sqrt{L_y}}\Phi_{n,k+\kappa_n}(x) C_{n,k+\kappa_n}}_{\psi_{R,n}(x,y)}+\\
+&\sum_{n}\underbrace{e^{-i \kappa_n y}\,\sum_{k=-\Lambda}^{\Lambda}\frac{e^{i k y}}{\sqrt{L_y}}\Phi_{n,k-\kappa_n}(x) C_{n,k-\kappa_n}}_{\psi_{L,n}(x,y)}.
\end{split}
\end{equation}
Since the $k$ summation are restricted to $|k|<\Lambda\ll\kappa_n$, the annihilation operators $C_{n,k+\kappa_n}$ and $C_{n,k-\kappa_n}$ can be meaningfully identified to be the spectral components of a right/left moving field respectively (i.e. towards positive/negative y); these will accordingly be denoted as $R_{n,k}=C_{n,k+\kappa_n}$ and $L_{n,k}=C_{n,k-\kappa_n}$.
\newline Since $|k|\ll\Lambda\ll\kappa_n$, these new operators will inherit the following anticommutation relations
\begin{equation}
\begin{cases}
\{L_{n,k},L_{m,k'}\}=\{R_{n,k},R_{m,k'}\}=\{L_{n,k},R_{m,k'}\}=\{L_{n,k},R^\dagger_{m,k'}\}=0
\\
\{L_{n,k},L^\dagger_{m,k'}\}=\{R_{n,k},R^\dagger_{m,k'}\}=\delta_{k,k'}\delta_{m,n}
\end{cases}
\end{equation}
and the Hamiltonian can be written as
\begin{equation}
\begin{split}
\mathcal{H}_\text{eff}=& 
\sum_{k=-\Lambda}^{\Lambda}\sum_{n}\hbar v_n\,k \left(R^\dagger_{n,k}R^{\vphantom{\dagger}}_{n,k}-L^\dagger_{n,k}L^{\vphantom{\dagger}}_{n,k}\right).
\end{split}
\end{equation}

\section{An effective one-dimensional density operator}
We introduce the real space density associated to the right movers
\begin{equation}
\begin{split}
\rho_{R,n}(x,y) =& \,\psi_{R,n}^\dagger(x,y) \psi_{R,n}(x,y) =\\
%=& \sum_{k=-\Lambda}^{\Lambda}\sum_{q=-\Lambda}^{\Lambda} e^{i (k-q) y}\Phi_{n,k+\kappa_n}(x)\Phi_{n,q+\kappa_n}(x) R^\dagger_{n,q}R_{n,k}=\\
%=&\sum_{k=-\Lambda}^{\Lambda}\sum_{q=-\Lambda-k}^{\Lambda-k} e^{-i q y}\Phi_{n,k+\kappa_n}(x)\Phi_{n,q+k+\kappa_n}(x) R^\dagger_{n,k+q}R_{n,k}=\\
%=&\sum_{q} e^{-i q y}\underbrace{\sum_{k=-\Lambda}^{\Lambda}\theta(q+\Lambda+k)\theta(-q+\Lambda-k)}_{\sum'_k}\,\Phi_{n,k+\kappa_n}\Phi_{n,q+k+\kappa_n} R^\dagger_{n,q+k}R_{n,k}\\
=&\sum_{|q|\leq 2\Lambda} \frac{e^{i q y}}{L_y}
\sideset{}{'}\sum_k\,\Phi_{n,k+\kappa_n}(x)\Phi_{n,k-q+\kappa_n}^*(x) R^\dagger_{n,k-q}R_{n,k}
\end{split}
\end{equation}
where $\sum'_k=\sum_{k=-\Lambda}^{\Lambda}\theta(-q+\Lambda+k)\theta(q+\Lambda-k)=\sum_{k=\max\{-\Lambda,-\Lambda+q\}}^{\min\{\Lambda,\Lambda+q\}}$ is a a summation restricted to momenta satisfying
\begin{equation}
\begin{cases}
-\Lambda+q \leq k\leq \Lambda     \qquad q\geq 0 \\
-\Lambda   \leq k\leq \Lambda+q   \qquad q < 0.
\end{cases}
\end{equation}
Near the edge, we can expect the eigenfunctions $\Phi_{k}(x)$ of eq. \ref{eq:problem_hamiltonitan_transformed} not to depend strongly on $k$, since the only effect of increasing $k$ is to squeeze a little the wavefunction, keeping its average position in place (due to the presence of the confining potential).
As a rough approximation then, since $|q|\ll\Lambda\ll\kappa_n$, we set\footnote{Choosing properly the irrelevant global phase.}
\begin{equation}
\int \Phi_{n,k+\kappa_n}(x)\Phi_{n,k-q+\kappa_n}^*(x)\,dx\approx 1
\end{equation}
so that we can define an effective one-dimensional density by integrating out the $x$ coordinate
\begin{equation}
\rho_{R,n}(y)=\int \rho_{R,n}(x,y)\,dx\approx \sum_{q} \frac{e^{i q y}}{L_y}
\sideset{}{'}\sum_k\, R^\dagger_{n,k-q}R_{n,k}
\end{equation}
The same procedure may be carried out for the left-movers density.
This approximation is consistent with the linearisation of the energy band, i.e. is a good approximation as long as the excitations of the system ground state are long-wavelength ones. 
The heuristic argument exposed above can be numerically confirmed, as shown in Fig. \ref{fig:EnergyBands_TaylorExpansion} of the next chapter.
\newline We may then read from the previous expression the Fourier components of the density operator
\begin{equation}
\label{eq:rhoQeff}
\rho_{R,n}(q)=\sideset{}{'}\sum_k R^\dagger_{n,k-q}R^{\vphantom{\dagger}}_{n,k}
\end{equation}
which gives
\begin{equation}
\begin{cases}
\rho_{R,n}(y)= \frac{1}{L_y}\sum_{q} e^{i q y} \rho_{R,n}(q)
\\
\rho_{R,n}(q)=\int e^{-i q y} \rho_{R,n}(y)dy
\end{cases}
\end{equation}
and analogously for the left counterpart. %The $L_y^{-1}$ factor has been included to make the (expectation value of the) $\rho(q)$ operators an intensive quantity as the real-space density, i.e. a sample-size independent quantity.
\newline  Notice that
\begin{equation}
\rho_{R,n}^\dagger(q) = \rho_{R,n}(-q)
\end{equation}
i.e. creating a density excitation with wavevector $q$ is the same as annihilating one with wavevector $-q$.
\newline The commutation relations between these operators turn out to be very important. We begin by calculating $[\rho_{R,n}(q),\rho_{R,m}(-q')]$, with both $q, q'>0$
\begin{equation}
[\rho_{R,n}(q),\rho_{R,m}(-q')]=
\sideset{}{'} \sum_{k,k'}[R^\dagger_{n,k-q}R^{\vphantom{\dagger}}_{n,k}\,,R^\dagger_{m,k'+q'}R^{\vphantom{\dagger}}_{m,k'}]
\end{equation}
here the allowed values for $k$ are $-\Lambda +q \leq k \leq \Lambda$, while for $k'$ one has $-\Lambda \leq k' \leq \Lambda+q'$. After some algebra one obtains
\begin{equation}
\begin{split}
[\rho_{R,n}(q),&\rho_{R,m}(-q')]=\\
%\sideset{}{'} \sum_{k,k'}[R^\dagger_{n,k-q}R^{\vphantom{\dagger}}_{n,k}\,,R^\dagger_{m,k'+q'}R^{\vphantom{\dagger}}_{m,k'}]=\\
%=& %This is the most symmetric choice
=&\delta_{n,m}\Biggl(\sideset{}{'} \sum_{k'}R^\dagger_{n,k'+q'-q}R^{\vphantom{\dagger}}_{n,k'}
\sideset{}{'} \sum_{k}\delta_{k,k'+q'} - \sideset{}{'} \sum_{k}R^\dagger_{n,k+q'-q}R^{\vphantom{\dagger}}_{n,k}
\sideset{}{'} \sum_{k'}\delta_{k',k-q}\Biggr)\\
=&
\delta_{n,m} \Biggl(
\sum_{\max\{-\Lambda,-\Lambda-\Delta q\}}^{\Lambda-q'}
-
\sum_{-\Lambda+q}^{\min\{\Lambda,\Lambda-\Delta q\}}
\Biggr) R^\dagger_{n,k+\Delta q}R^{\vphantom{\dagger}}_{n,k}
\end{split}
\end{equation}
where $\Delta q=q'-q$. %, and the primed sum over $k$ was used for $q$, and analogously for $k'$ and $q'$
We assumed that the relevant momenta are much smaller than the cut-off scale; then we will have $|q|,|q'|\ll\Lambda$ too. Many terms, except near $\pm\Lambda$, will cancel out and the ones which survive can be replaced with their expectation value on the ground state (the occupation numbers of the system being practically unaffected at $|q|\sim\Lambda$)
\begin{equation}
\braket{R^\dagger_{n,k+\Delta q}R^{\vphantom{\dagger}}_{n,k}}_0= \theta(-k)\delta_{q,q'}.
\end{equation}
Finally then\footnote{The maximum value in the summation in the second line should actually read $-\Lambda+q-\frac{2\pi}{L_y}$. The error committed including such a point is however negligibly small in the thermodynamic limit.}$^,$\footnote{The equality is strictly valid only for $|q|\leq\Lambda$; however as already said the error committed should be small since the relevant momenta are much smaller than the cut-off one.}
\begin{equation}
\begin{split}
[\rho_{R,n}(q),\rho_{R,m}(-q')]\simeq&
 \delta_{n,m}\delta_{q,q'}\sum_{k=-\Lambda}^{-\Lambda+q} 
\rightarrow \frac{\delta_{n,m}}{L_y^2}\delta_{q,q'}\frac{L_y}{2\pi}\,\int_{-\Lambda}^{-\Lambda+q}\,dk
=\\
=&\frac{qL_y}{2\pi}\delta_{n,m}\delta_{q,q'}
\end{split}
\end{equation}
It can be checked that if $q$, $q'$ have opposite sign the commutator vanishes.
If on the other hand both $q,q'<0$
\begin{equation}
[\rho_{R,n}(q),\rho_{R,m}(-q')] = -[\rho_{R,m}(|q'|),\rho_{R,n}(-|q|)] 
=\frac{qL_y}{2\pi}\delta_{n,m}\delta_{q,q'}
\end{equation}
exactly as before.
We thus see that the operators $\rho_{R,n}(q)$ obey bosonic commutation relations.
\newline The same kind of calculation is performed for the left movers. The only difference is in the equilibrium average occupation number,
$\braket{L^\dagger_{n,k}L^{\vphantom{\dagger}}_{n,k}}_0 = \theta(k)$, which results in
\begin{equation}
[\rho_{L,n}(q),\rho_{L,m}(-q')] = - \frac{qL_y}{2\pi}\,\delta_{q,q'}\delta_{n,m}.
\end{equation}
Finally, one gets
\begin{equation}
[\rho_{L,n}(q),\rho_{R,m}(-q')] = 0.
\end{equation}
\newline The effective one-dimensional density operators at different points obey the following commutation relation
\begin{equation}
\begin{split}
[\rho_{R,n}(y), \rho_{R,n}(y')] =& \sum_q\,\frac{q}{2\pi\,L_y}\,e^{i q (y-y')}=
\\ =&\frac{-i}{2\pi\,L_y}\,\partial_y \sum_q e^{i q (y-y')}\rightarrow \,\,\frac{-i}{2\pi} \partial_y\,\delta_{\Lambda^{-1}}(y-y')
\end{split}
\end{equation}
where $\delta_{\Lambda^{-1}}(y)$ is a delta-function smeared over a region of size $\Lambda^{-1}$, owing to the presence of the cut-off in the dispersion relation. 
Notice that $\partial_y\,\delta_{\Lambda^{-1}}(y-y')$ vanishes as $y'$ approaches $y$, since $\rho_{R,n}(y)$ must commute with itself.
In the large-$\Lambda$ limit $\delta_{\Lambda^{-1}}(y)\rightarrow \delta(y)$.

The commutation relations of these density operators with the effective Hamiltonian can be computed as well
\begin{equation}
\label{eq:HrRqcomm}
\begin{split}
[\mathcal{H}_\text{eff},\rho_{R,n}(q)] = & \sideset{}{'}\sum_{k'}\sum_{k=-\Lambda}^{\Lambda}\sum_{m}\hbar v_m k[ R_{m,k}^\dagger R_{m,k}^{\vphantom{\dagger}}\,,\, R^\dagger_{n,k'-q}R^{\vphantom{\dagger}}_{n,k'}]=\\
=&
\hbar v_n\sideset{}{'}\sum_{k'}R_{n,k'-q}^{\dagger}R_{n,k'}^{\vphantom{\dagger}}\sum_{k=-\Lambda}^{\Lambda} k
\left(\delta_{k,k'-q}-\delta_{k,k'} \right) = \\
=& -\hbar v_n\,q\,\rho_{R,n}(q)
\end{split}
\end{equation}
and analogously
\begin{equation}
\label{eq:HrLqcomm}
\begin{split}
[\mathcal{H}_\text{eff},\rho_{L,n}(q)]
%&  -\sideset{}{'}\sum_{k'}\sum_{k=-\Lambda}^{\Lambda}\sum_{m}\hbar v_m k[ L_{m,k}^\dagger L_{m,k}^{\vphantom{\dagger}}\,,\, L^\dagger_{n,k'+q}L^{\vphantom{\dagger}}_{n,k'}] =\\
=& \hbar v_n\,q\,\rho_{L,n}(q).
\end{split}
\end{equation}
i.e. we see that different modes are uncoupled. Notice that this fact crucially depends on the dispersion law being linear. We also see that $\rho_{R,n}(q)$ lowers the energy of an $\mathcal{H}_\text{eff}$ eigenstate by $-\hbar v_n q$, which leads us to identify it as being an annihilation operator for our system. $\rho_{R,n}(-q)$ acts on the system eigenstates as a creation operator instead. The same considerations hold for the $L$ counterpart of these operators.
\newline From equations eq. \ref{eq:HrRqcomm} and \ref{eq:HrLqcomm} we see that the Heisenberg representation operators have the simple time evolution
\begin{equation}
\begin{cases}
\rho_{R,n}(q,t)=e^{-i v_n q (t-t_0)}\rho_{R,n}(q,t_0) 
\\
\rho_{L,n}(q,t)=e^{+i v_n q (t-t_0)}\rho_{L,n}(q,t_0).
\end{cases}
\end{equation}
The real-space one-dimensional effective density in the Heisenberg picture then reads
\begin{equation}
\begin{cases}
\rho_{R,n}(y,t)=\sum_{q} e^{+i q \left(y - v_n (t-t_0)\right)}\rho_{R,n}(q,t_0) 
\\
\rho_{L,n}(y,t)=\sum_{q} e^{+i q \left(y + v_n (t-t_0)\right)}\rho_{L,n}(q,t_0).
\end{cases}
\end{equation}
The two operators $\rho_{R,n}(y,t)$ and $\rho_{L,n}(y,t)$ depend on space-time coordinates only through $y\pm v_n t$, and thus will satisfy the chiral wave (operator) equation\footnote{This could have been derived straightforwardly by evaluating $\dot{\rho}=\frac{i}{\hbar}\,[\mathcal{H}_\text{eff},\rho(y)]$.} $\partial_t \rho \pm v \partial_x \rho=0$, as could have been expected on (semiclassical) physical grounds.
We can indeed interpret $\pm v\rho$ as being a (one-dimensional) current operator $J$ and rewrite the equation as $\partial_t \rho +\partial_x J=0$, which is just a continuity equation.
\newline Within the approximations employed, we therefore see that localized density excitations $\braket{\rho}$ will propagate undamped and rigidly at a speed set by the single-particle energy dispersion curve $E_{n,k}$, which depends on the characteristics of the confining potential.

As a final comment, notice that effective one-dimensional fermion operators can be defined through
\begin{equation}
\psi_{R,n}(y)=e^{i \kappa_n y}\sum_{|k|\leq\Lambda}\frac{e^{i k y}}{\sqrt{L_y}}\,R_{n,k}
\end{equation}
and these consistently lead to the effective one dimensional density used above (eq. \ref{eq:rhoQeff}).
Analogously can be done for the left counterpart. 
It is easy to check that in the large $\Lambda$ limit proper anticommutation relations are recovered
\begin{equation}
\begin{cases}
\{\psi^\dagger_{R,n}(y'),\psi^{\vphantom{\dagger}}_{R,n}(y)\}\rightarrow\delta(y'-y)
\\
\{\psi^{\vphantom{\dagger}}_{R,n}(y'),\psi^{\vphantom{\dagger}}_{R,n}(y)\}=\{\psi^{{\dagger}}_{R,n}(y'),\psi^{{\dagger}}_{R,n}(y)\}=0.
\end{cases}
\end{equation}
The commutation relations with the Hamiltonian are easily established as well
\begin{equation}
[\mathcal{H}_\text{eff}, \psi_{R,n}(y)] = %\sum_{|k|\leq\Lambda}\sum_{m}
-\hbar v_n\,e^{i \kappa_n y}\sum_{|k|\leq\Lambda}\frac{k\,e^{i k y}}{\sqrt{L_y}}%\left[R^\dagger_{m,k}R^{\vphantom{\dagger}}_{m,k}, \,R_{n,k'}\right].
%\delta_{k,k'}\delta_{n,m}
R^{\vphantom{\dagger}}_{n,k}
\end{equation}
which give the Heisenberg representation form of such a operator as
\begin{equation}
\psi_{R,n}(y,t) = e^{i \kappa_n y}\sum_{|k|\leq\Lambda}\frac{e^{i k (y-v_n t)}}{\sqrt{L_y}}
R^{\vphantom{\dagger}}_{n,k}
\end{equation}
Notice that it is not a function of $y-v_n t$ alone; rather there is a local phase factor $e^{i \kappa_n y}$ which however does not affect the density operator.
\newline The commutators between these fields and the density operator can be computed as well
\begin{equation}
\begin{split}
[\psi_{R,n}(y),\rho_{R,n}(q)]%=&	e^{i \kappa_n y}\sum_{|k'|\leq\Lambda}\frac{e^{i k' y}}{\sqrt{L_y}}\,\sideset{}{'}\sum_{k}\,[R_{n,k'}^{\vphantom{\dagger}},R_{n,k+q}^{\dagger}R_{n,k}]=\\
%=&e^{i q y}e^{i \kappa_n y}\sum_{k'-q=-\Lambda-q}^{\Lambda-q}\frac{e^{i (k'-q) y}}{\sqrt{L_y}}\,\sideset{}{'}\sum_{k}\,R_{n,k'-q}^{\vphantom{\dagger}} \delta_{k,k'-q}
%=&\frac{e^{-i q y}}{L_y}\,e^{i \kappa_n y}\sum_{k''=-\Lambda-q}^{\Lambda-q}\frac{e^{i k'' y}}{\sqrt{L_y}}\,R_{n,k''}^{\vphantom{\dagger}} \,\sideset{}{'}\sum_{k}\,\delta_{k,k''}\approx\\
&\simeq e^{-i q y}\, \psi_{R,n}(y)
\end{split}
\end{equation}
%where the fact that $|q|\ll\Lambda$ has been used.
which also implies that
\begin{equation}
[\psi_{R,n}(y),\rho_{R,n}(y')]=\delta(y'-y)\psi_{R,n}(y).
\end{equation} 
as expected for fermionic operators obeying anticommutation relations. The same relations can be derived for $\psi_{L,n}$ and $\rho_L$.

\section{Inclusion of an external potential}
%Before attempting to tackle the dispersion relation beyond its linear approximation, 
It is interesting to see the effects of the inclusion of some external single-particle potential.
There will be an additional piece in the Hamiltonian 
\begin{equation}
\mathcal{V}(t)=\sum_{\substack{{n,k}\\{m,q}}} \bra{n,k}V(t)\ket{m,q} C^\dagger_{n,k}C^{\vphantom{\dagger}}_{m,q}\simeq\sum_n\sum_{k,q} \bra{n,k}V\ket{n,q} C^\dagger_{n,k}C^{\vphantom{\dagger}}_{n,q}
\end{equation}
where we assumed that the potential does not couple (in any non-negligible way) different Landau levels.
\newline If the potential $V$ is localized near one edge (say the one corresponding to positive values of $k$) only the momenta whose associated wavefunction is localized in the same region will give non-vanishing matrix elements\footnote{This assumption may actually be relaxed; even if the external potential extends over the whole sample, as long as it is long-wavelength electrons will not scatter from one edge of the system to the opposite one so that the two edges are effectively not coupled.}. 
We will assume that all the relevant wavevectors lie within the linearity range $\pm\Lambda$ from the Fermi surface $\kappa_n$ of the n-th Landau level.
\newline The $k$ and $q$ dependence of the matrix element $\bra{n,k}V\ket{m,q}$ can be equivalently expressed in terms of their difference $k-q$ and average $\frac{k+q}{2}$.
As pointed out above, near the edge of the sample the wavevector dependence of the wavefunction is not that drastic, so that we can replace $\frac{k+q}{2}\approx \kappa_n$ in the n-th branch of the dispersion relation and keep only the dependence on the momentum transfer $k-q$ of the matrix element; correspondingly we write $\bra{n,k}V\ket{n,q}\approx \frac{1}{L_y} V_n(k-q;t)$, the reason for the presence of the $L_y^{-1}$ factor will become clear in a moment. 
More quantitatively
\begin{equation}
\begin{split}
\bra{n,k}V\ket{n,q} =&
\int \Phi_{n,k}^*(x) \frac{e^{-i k y}}{\sqrt{L_y}} \,V(x,y; t)\,
\Phi_{n,q}(x) \frac{e^{i q y}}{\sqrt{L_y}}\,\,dx\,dy \approx\\
\approx& \frac{1}{L_y}\int dy \, \, e^{-i (k-q) y}\,\underbrace{\int \Phi_{n,\kappa_n}^*(x) V(x,y; t) \Phi_{n,\kappa_n}(x) \,dx}_{U_n(y; t)}
\end{split}
\end{equation}
with $U_n(y)$ an effective edge potential. We then get the pair of Fourier relations
\begin{equation}
\begin{cases}
V_n(k;t)=\int \, e^{-i k y}\,U_n(y; t)\,dy\\
U_n(y; t) = \frac{1}{L_y}\sum_q\,V_n(q; t)e^{i q y}.
\end{cases}
\end{equation}
Putting together all these considerations leads to
\begin{equation}
\begin{split}
\mathcal{V}(t) \approx& 
\sum_n\frac{1}{L_y}\sum_{|k|,|q|\leq \Lambda} V_n(k-q;t) C^\dagger_{n,k+\kappa_n}C^{\vphantom{\dagger}}_{n,q+\kappa_n}=\\
%\\=&\sum_n\sum_{q =-\Lambda}^{\Lambda}\sum_{k-q=-\Lambda-q}^{\Lambda-q} V_n(k-q) R^\dagger_{n,k-q+q}R^{\vphantom{\dagger}}_{n,q}
%\sum_n\sum_{q =-\Lambda}^{\Lambda}\sum_{k=-\Lambda-q}^{\Lambda-q} V_n(k) R^\dagger_{n,k+q}R^{\vphantom{\dagger}}_{n,q}=\\
%=&\sum_n\sum_{q =-\Lambda}^{\Lambda}\sum_{k}\theta(k+\Lambda+q)\theta(-k+\Lambda-q) V_n(k) R^\dagger_{n,k+q}R^{\vphantom{\dagger}}_{n,q}=\\
%=&\sum_n\sum_{k}V_n(k) \sum_{q =\max\{-\Lambda,-\Lambda-k\}}^{\min\{\Lambda,\Lambda-k\}}  R^\dagger_{n,k+q}R^{\vphantom{\dagger}}_{n,q}
=&\sum_n\frac{1}{L_y}\sum_{k}V_n(k;t) \sideset{}{'}\sum_{q}  R^\dagger_{n,k+q}R^{\vphantom{\dagger}}_{n,q}
=\sum_n\,\frac{1}{L_y}\sum_{k}V_n(k;t) \rho_{R,n}(-k)
\end{split}
\end{equation}
Notice that this can be rewritten in coordinate space as
\begin{equation}
\label{eq:VtrhoR}
\mathcal{V}(t) = \int \,\, U_n(y)\,\rho_{R,n}(y)dy
\end{equation}
i.e. the approximated expression is nothing but a minimal coupling between the potential averaged along the x-direction and the effective one-dimensional density.
\newline The results are substantially the same if the potential is localised at the other edge of the sample; if on the other hand the potential is extended but does not directly couple the two edges (so that cross coupling terms can be neglected) one simply adds to \ref{eq:VtrhoR} an analogous expression with $\rho_{L,n}$.

The commutation relations with $\rho_{R,n}(q)$ are simple
\begin{equation}
[\mathcal{V}(t), \rho_{R,n}(q)] = -\frac{q}{2\pi}\,V_{n}(q;t)
\end{equation}
which leads to the driven evolution equation (in the Heisenberg representation)%with respect to $\mathcal{H}_\text{eff}$)
\begin{equation}
\label{eq:effective_1d_dynamics}
\dot{\rho}_{R,n}(q,t)=\frac{i}{\hbar}\,[\mathcal{H}_\text{eff}+\mathcal{V}(t),\rho_{R,n}(q)]=-i v_n\,q\,\rho_{R,n}(q, t) - i\frac{q}{2\pi\,\hbar}\, V_{n}(q;t)
\end{equation}
which shows that each mode evolves independently from one other, and is driven by the external potential term. In coordinate space
\begin{equation}
\label{eq:coordinate_spce_effective_1d_dynamics}
\begin{split}
\partial_t\rho_{R,n}(y;t)=&\frac{i}{\hbar}\,[\mathcal{H}_\text{eff}+\mathcal{V}(t),\rho_{R,n}(y,t)]=\\
=&-v_n \,\partial_y \rho_{R,n}(y;t) - \frac{1}{2\pi\,\hbar}\partial_y U_{n}(y;t).
\end{split}
\end{equation}
Which is a chiral wave equation with a source term proportional to minus the gradient of an effective one-dimensional potential. 
\newline In the next subsection a comparison between this chiral Luttinger liquid result and a classical gas is made, and the peculiarities of this equation are discussed.

\subsection*{Comparison with a classical ideal fluid}
I now open a small parenthesis to briefly compare eq. \ref{eq:coordinate_spce_effective_1d_dynamics} with the small-oscillations dynamics of a classical fluid.
\newline The small oscillations about the equilibrium configuration of a classical ideal gas can be described by the continuity and Euler equations
\begin{equation}
\label{eq:EulerEquations}
\begin{cases}
\frac{\partial\rho}{\partial t}=-\mathbf{\nabla}\cdot\rho\mathbf{v}
\\
\frac{\partial \mathbf{v}}{\partial t} + (\mathbf{v}\cdot\mathbf{\nabla})\mathbf{v}=-\frac{\mathbf{\nabla}P}{m\rho}
\end{cases}
\end{equation}
where $\rho$, $\mathbf{v}$ and $P$ are the fluid number density, velocity and pressure fields describing the state of the fluid, and $m$ is the mass of the fluid particles.
\newline Suppose we now subject the fluid to some kind of perturbation, which we describe by adding the additional term $-\frac{1}{m}\mathbf{\nabla}U$ to the right hand side of the second equation of \ref{eq:EulerEquations}. Correspondingly, we look for solutions near the equilibrium configuration and consequently expand $\rho=\rho_0+\delta\rho$, $\mathbf{v}=\delta\mathbf{v}$. 
We furthermore assume the internal pressure to depend only on the density itself $P=P(\rho)$ (we assume the other thermodynamic variables to be fixed), so that $\frac{1}{m}\mathbf{\nabla}P(\rho)=\left(\frac{1}{m}\frac{\partial P}{\partial\rho}\right)_{\rho_0}\mathbf{\nabla}\delta\rho$. Since the coefficient $\left(\frac{1}{m}\frac{\partial P}{\partial\rho}\right)_{\rho_0}$ has the dimensions of a squared velocity, we denote it by $c^2$.
\newline Linearising the two equations above (comprehensive of the source term), the non-linear term drops out and their structure simplifies considerably
\begin{equation}
\begin{cases}
\frac{\partial\delta\rho}{\partial t}=-\rho_0\,\mathbf{\nabla}\cdot\delta\mathbf{v}
\\
\rho_0\,\frac{\partial \delta\mathbf{v}}{\partial t}=-c^2\mathbf{\nabla}\delta\rho-\frac{\rho_0}{m}\,\mathbf{\nabla}U.
\end{cases}
\end{equation}
differentiating the first one with respect to the time variable and using the second one we arrive at
\begin{equation}
\label{eq:classical_fluid}
\frac{\partial^2\delta\rho}{\partial t^2}
=c^2\,\frac{\partial^2\delta\rho}{\partial y^2}+\frac{\rho_0}{m}\,\frac{\partial^2U}{\partial y^2}
\end{equation}
(where we specialised to one-spatial dimension only, replacing $\mathbf{\nabla}^2$ with $\partial_y^2$). Without the source term, this is a standard wave equation (second order both in space and time), admitting at the same time \virgolette{forward} and \virgolette{backwards} rigidly propagating solutions $\delta\rho(y\pm ct)$, opposed to the chiral equation above \ref{eq:coordinate_spce_effective_1d_dynamics}.
The source term in \ref{eq:classical_fluid} will produce ripples propagating at speed $c$ in both directions outwards of the region where the source is localised, in agreement with our classical experience. 
On the other hand the chiral equation produces only one such a ripple, propagating away from the source at speed $v$ in a single preferential direction.
\newline Notice that the structure of the source term is inherently different; in a classical gas it is the gradient of the force which produces a density variation (particles need an acceleration gradient in order to rarefy or become more dense). Eq. \ref{eq:coordinate_spce_effective_1d_dynamics} looks quite striking in this regard at first, since density variations are caused by the gradient of the potential itself. 
This is however an effect of the system not being truly one-dimensional; in the presence of a strong magnetic field indeed electrons drift along the direction which is orthogonal to the gradient of the potential (as discussed in ch \ref{ch1}). 
This transverse Hall current will introduce a density modulation in the system density which obviously reflects on its one-dimensional effective counterpart.

\graphicspath{{./pic4/}}

\chapter{Spatially periodic excitation}\label{ch4}
\lhead{Chapter 5. \emph{Spatially periodic excitation}} % Change X to a consecutive number; this is for the header on each page - perhaps a shortened title

From this chapter on the original results which have been obtained will be presented.
\newline In order to study the dynamics of the system, it has been subjected to a spatially periodic perturbation adiabatically turned on and then switched off, of the form
\begin{equation}
\label{eq:sinusoidal_excitation}
V(y,t)=\lambda\, \xi(t) \sin^2(Ky)=\lambda \,e^{-\left(\frac{t-t_0}{\tau}\right)^2} \sin^2(Ky)
\end{equation}
i.e. uniform in the $x$ direction. Notice that $K$ is an integer (or half integer) multiple of $\frac{2\pi}{L_y}$ in order to satisfy the periodic boundary conditions imposed along the $y$ direction. 
The parameters have been chosen so that at $t=0$ the perturbation was negligibly small.

\noindent The one-body density and currents obtained from the numerical algorithm developed in Chapter \ref{ch2} (Sec. \ref{sec:time_evo_alg}) in the weak coupling regime have been compared with those obtained by using time-dependent perturbation theory, from which some simple predictions can be derived. 
\newline The non-linear dynamics is both numerically investigated as well as theoretically by means of second order perturbation theory.

\noindent The advantage of a perturbative treatment resides in the possibility of taking the thermodynamic limit $L_y\rightarrow\infty$ and to get results which do not depend on the presence of periodic boundary conditions.

Throughout the whole section $\omega_c\tau=15$ and $\omega_c t_0=50$ have been used (except if otherwise specified), as well as $L_x=20l_B$. The confining potential parameters used are $V_0=30\hbar\omega_c$ and $\sigma_c=0.1l_B$, which give a steep rise of the confining potential on a lengthscale roughly of the order of the magnetic length $l_B$.
\newline The system length along $y$, the number of electrons (or equivalently the Fermi point $k_F$), the perturbation wavevector $K$ and the perturbation strength parameter $\lambda$ have been varied and will be explicitly written from time to time when looking at numerical results. 

\section{One body density}
First of all a single particle solution for the Hamiltonian evolution of an eigenstate of eq. \ref{eq:problem_hamiltonitan} under the periodic excitation eq. \ref{eq:sinusoidal_excitation} is obtained at second perturbative order; this solution is then used to study both the bulk and edge dynamics of the many-body non interacting system.
\newline Numerical results are throughout compared with the analytical ones, and the dynamics beyond the perturbative level is analysed.

\subsection{Perturbative solution}\label{subsection:sin_PT}
If $\lambda\ll\hbar\omega_C$, we can find a perturbative solution for the single particle time-dependent Schrödinger equation
\begin{equation}
\begin{cases}
i\hbar\,\frac{\partial \Psi}{\partial t}=(\mathcal{H}_0+V)\Psi\\
\Psi(t=0)=\psi_{n_0,k_0}=\Phi_{n_0,k_0}(x)\frac{e^{ik_0y}}{\sqrt{L_y}}.
\end{cases}
\end{equation}
Expanding the electron state over the eigenstates of $\mathcal{H}_0$
\begin{equation}
\Psi=\sum_{n,k}c_{n,k}(t)e^{-i\omega_{n,k}t}\psi_{n,k}
\end{equation}
and exploiting orthonormality
\begin{equation}
\begin{cases}
i\hbar\,\frac{\partial c_{n,k}(t)}{\partial t}=\sum_{m,q}c_{m,q}(t)\,e^{i\Delta\omega_{n,m}^{k,q}\,t}\int\psi_{n,k}^*V\psi_{m,q}dxdy\\
c_{n,k}(t=0)=\delta_{n,n_0}\delta_{k,k_0}
\end{cases}
\end{equation}
where $\Delta\omega_{n,m}^{k,q}=\omega_{n,k}-\omega_{m,q}$. The integral simplifies quite a bit
\begin{equation}
\int\psi_{n,k}^*V\psi_{m,q}\,dx\,dy=\lambda\, \xi(t)\,\underbrace{\int\Phi_{m,q}\Phi_{n,k}\,dx}_{d_{n,m}^{k,q}}\,\left(\frac{\delta_{q,k}}{2}-\frac{\delta_{q-k,2K}+\delta_{q-k,-2K}}{4}\right)
\end{equation}
and thus, since $d_{n,m}^{k,k}=\delta_{n,m}$ we get
\begin{equation}
\begin{split}
\frac{\partial c_{n,k}}{\partial t}=&-i\frac{\lambda\,\xi(t)}{2\hbar}\, \Biggl[\Biggr.c_{n,k}-\sum_m\Biggl(\Biggr.\frac{d_{n,m}^{k,k+2K}}{2}\,c_{m,k+2K}\,e^{i\Delta\omega_{n,m}^{k,k+2K}t}+\\&+\frac{d_{n,m}^{k,k-2K}}{2}\,c_{m,k-2K}\,e^{i\Delta\omega_{n,m}^{k,k-2K}t}\Biggl.\Biggr)\Biggl.\Biggr].
\end{split}
\end{equation}
We now perturbatively expand the Fourier coefficient to quadratic order in $\lambda$
\begin{equation}
\label{eq:perturbative_expansion_cnk}
c_{n,k}^{(n_0,k_0)}(t)=\delta_{n,n_0}\delta_{k,k_0}+\lambda f_{n,k}^{(n_0,k_0)}(t)+\lambda^2 g_{n,k}^{(n_0,k_0)}+\mathcal{O}(\lambda)^3
\end{equation}
and feed this back into the previous equation. Superscripts reminding the initial electron state have been inserted for convenience, since we will ultimately sum over all the occupied states expressions involving these variables. 
\newline Matching the coefficients of the various powers of the expansion parameter $\lambda$ provides simple differential equations for $f_{n,k}(t)$ and $g_{n,k}(t)$. 

\subsubsection*{First perturbative order}
\begin{figure}[htp!]
	\begin{minipage}{.5\textwidth}
		\centering
		\includegraphics[width=1.\textwidth]{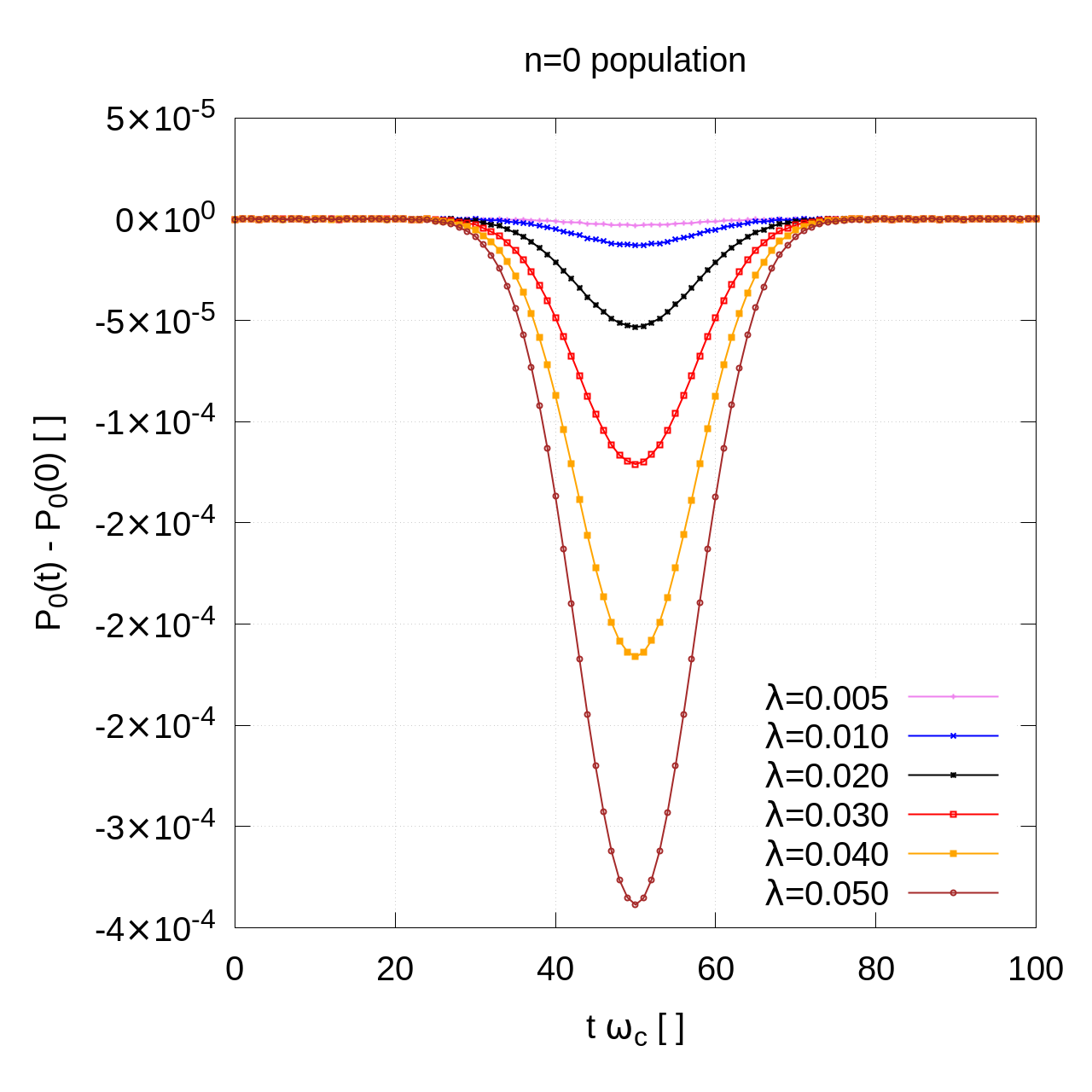}
	\end{minipage}%
	\begin{minipage}{0.5\textwidth}
		\centering
		\includegraphics[width=1.\textwidth]{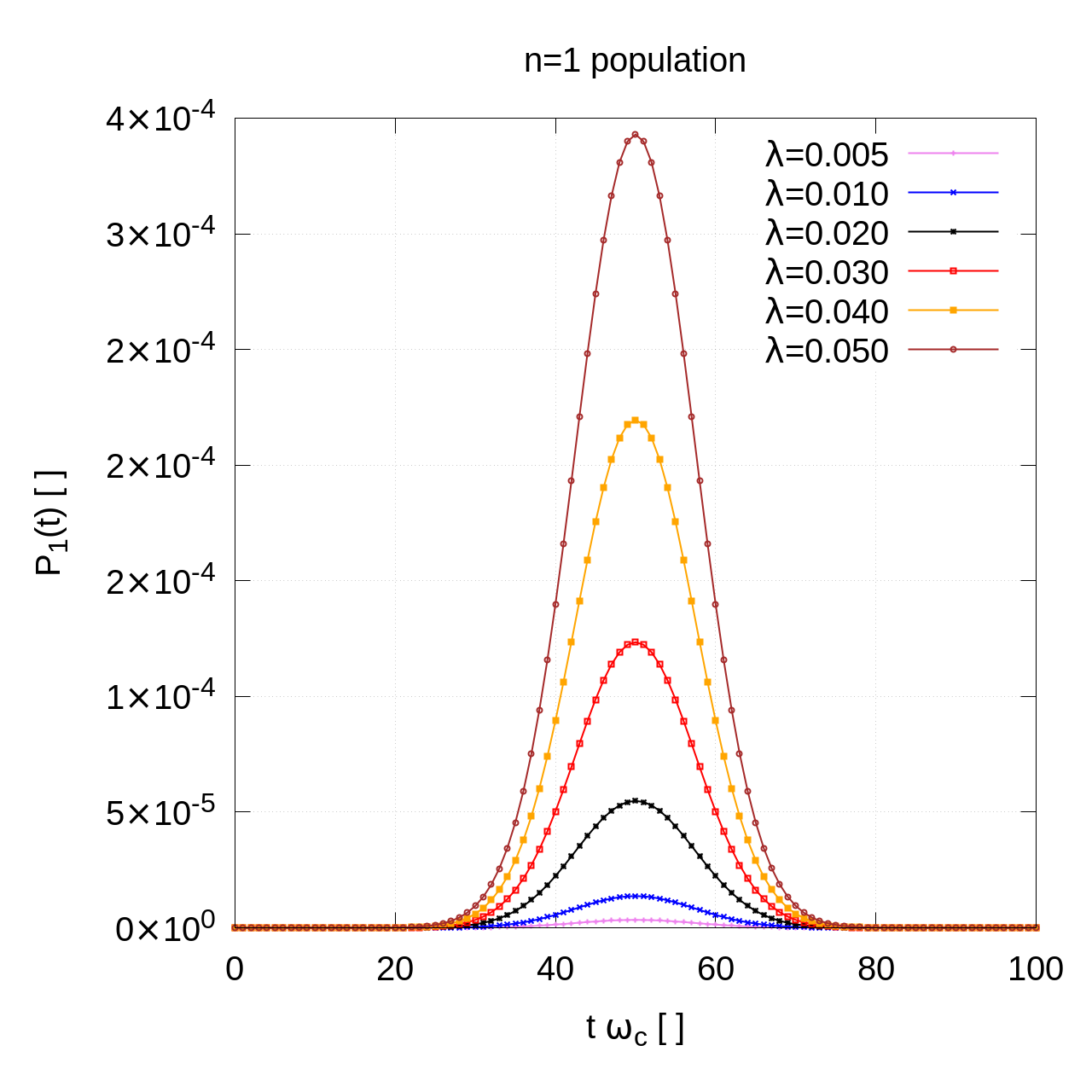}
	\end{minipage}    
	\caption[The LOF caption]{The images show the numerically computed populations of the lowest Landau level (left hand side panel) and of the first excited one (on the right). The perturbation wavevector was chosen to be $K=4\Delta k$, while the length of the sample $L_y=400l_B$; the Fermi point $k_F=546\Delta k\simeq 8.58l_B^{-1}$. Various values of the perturbation strength $\lambda$ are shown; it is apparent (and physically intuitive) that non-resonant excitations to higher Landau levels are more likely to occur as $\lambda$ gets bigger. Since the timescale over which the perturbation varies is significatively longer ($\omega_c\tau=15$) than the free evolution typical time, and $\lambda$ is much smaller than the typical spacing $\hbar\omega_c$, the populations are adiabatically following the excitation temporal profile, and long after the perturbation has been turned off no electron can be found in the excited Landau level.}
	\label{fig:populations}
\end{figure}

The equations for $f_{n,k}(t)$ read
\begin{equation}
\begin{cases}
\label{eq:pert_sin1}
\begin{alignedat}{2}
&\frac{\partial f_{n,k_0}^{(n_0,k_0)}}{\partial t} &&=-i\frac{\xi(t)}{2\hbar}\delta_{n,n_0}\\
&\frac{\partial f_{n,k_0\pm2K}^{(n_0,k_0)}}{\partial t} &&= +i \frac{\xi(t)}{4\hbar}d_{n,n_0}^{k_0\pm2K,k_0}e^{-i\Delta\omega_{n_0,n}^{k_0,k_0\pm2K}t}
\end{alignedat}
\end{cases}
\end{equation}
due to the temporally fast oscillations when $n\neq n_0$, the coupling between different Landau levels will be neglected (unless otherwise specified) since the time integration will average the coefficient to zero. 
For the moment the $n\neq n_0$ Landau level will henceforth be neglected, and the Landau level indices will consequently be dropped. 
\newline The inclusion of higher Landau levels may get relevant as long as the excitation potential \virgolette{is turned on} since they may get excited non-resonantly, as it can be seen from the numerically evaluated populations of the lowest and first excited Landau levels shown in Fig. \ref{fig:populations} (which follow adiabatically the excitation temporal profile). These transitions will be briefly discussed when we focus on the bulk dynamics.

\begin{figure}[htp!]
	\begin{minipage}{0.5\textwidth}
		\centering
		\includegraphics[width=1.\textwidth]{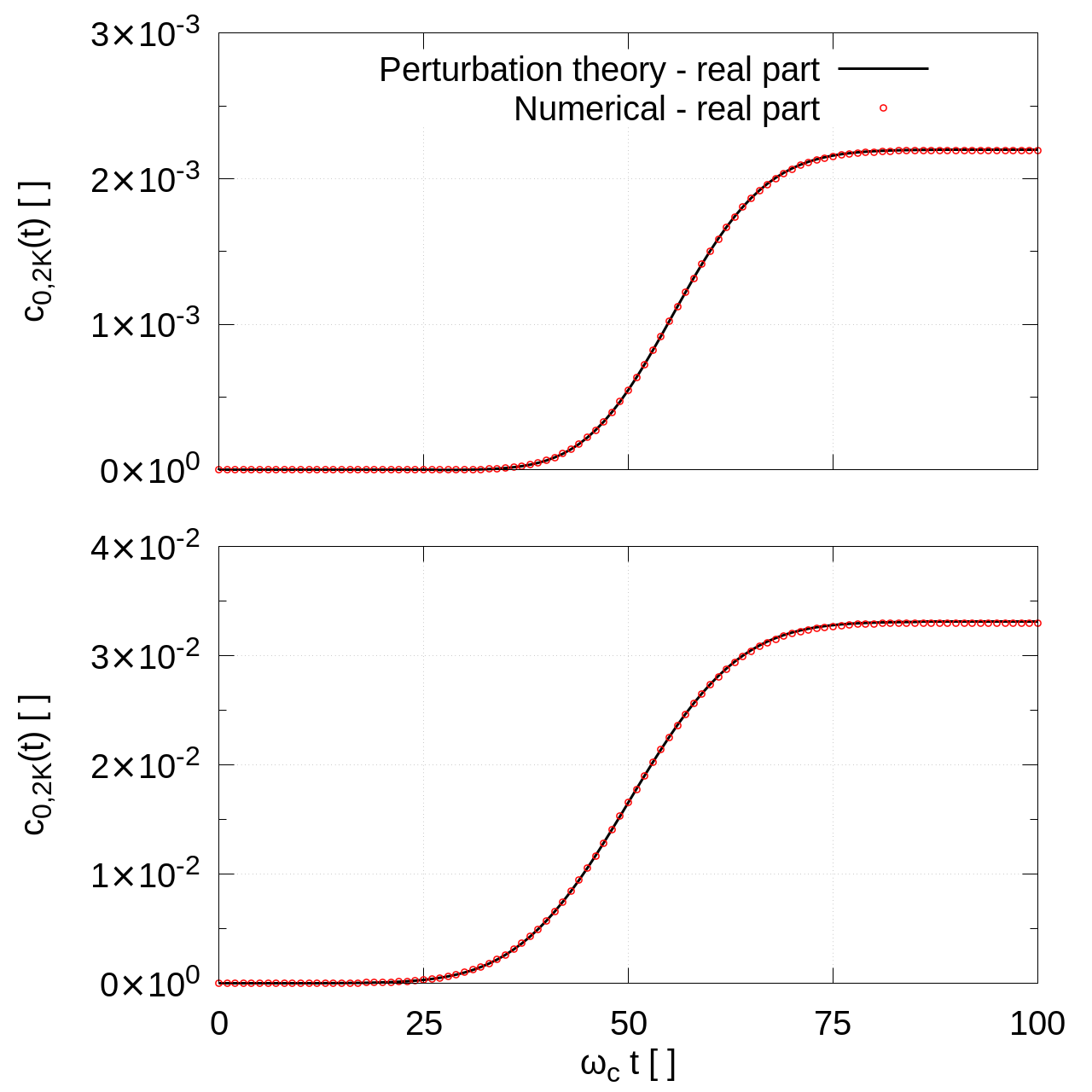}
	\end{minipage}%
	\begin{minipage}{0.5\textwidth}
		\centering
		\includegraphics[width=1.\textwidth]{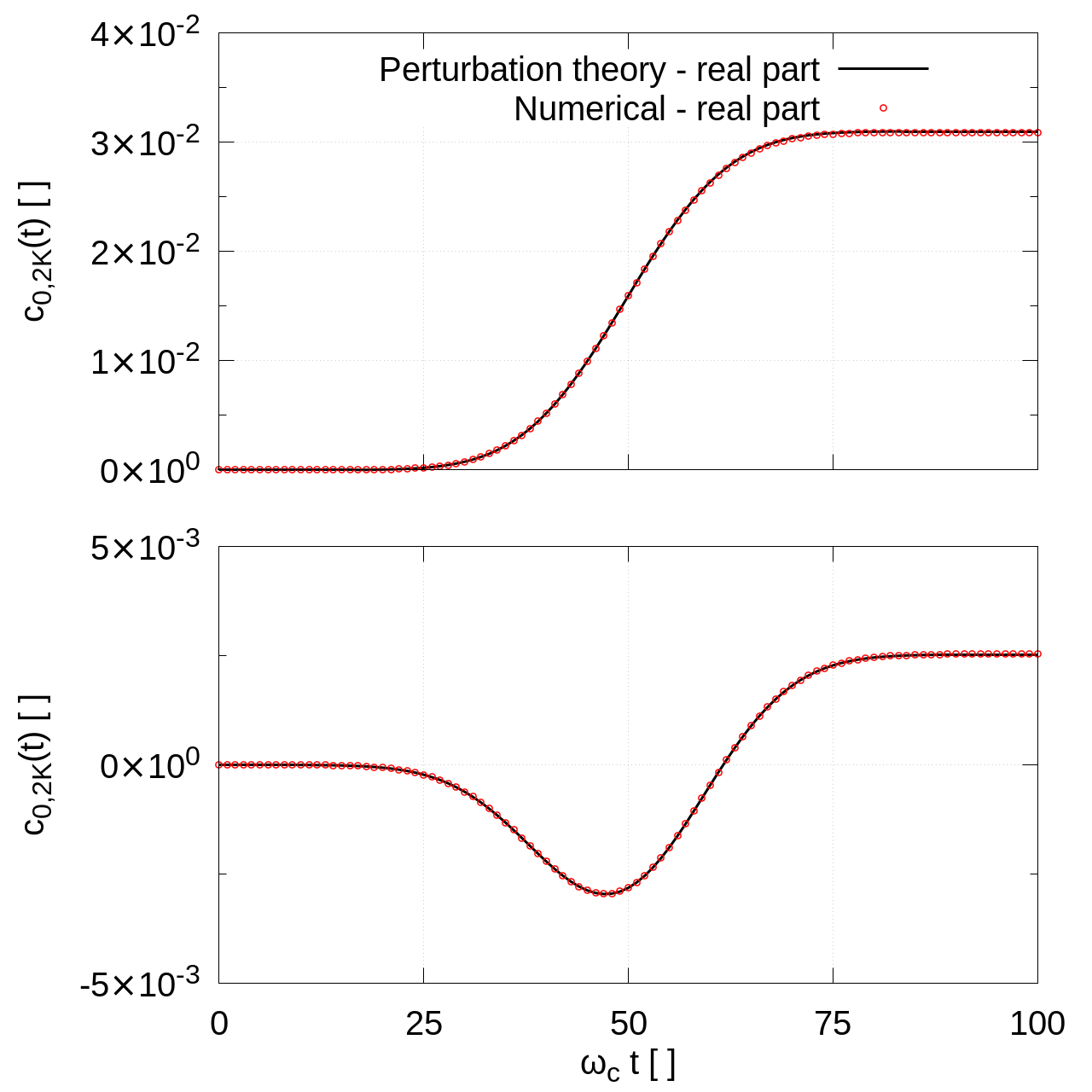}
	\end{minipage}    
	\caption[The LOF caption]{The two figures compare the results obtained using first order perturbation theory and the numerical time evolution program for a bulk state (left hand side panel, $n_0=0$, $k_0=0$) and an edge state (panel on the right, $n_0=0$, $k_0\simeq8.58l_B^{-1}$. The system length was $L_y=200l_B$, the perturbation strength $\lambda=0.005\hbar\omega_c$, the perturbation wavevector $K=4\Delta k$.)}
	\label{fig:ComparisonWithPT}
\end{figure}

The pair of equations \ref{eq:pert_sin1} can be analytically solved
\begin{equation}
\label{eq:f}
\begin{cases}
\begin{alignedat}{2}
&f_{k_0}^{(k_0)}(t)&&=-\frac{i}{2\hbar}F_t(0)\\
&f_{k_0\pm2K}^{(k_0)}(t) &&= +\frac{i}{4\hbar}d_{k_0,k_0\pm2K}F_t(\Delta\omega_{k_0,k_0\pm2K})
\end{alignedat}
\end{cases}
\end{equation}
where
\begin{equation}
\label{eq:ft}
\begin{split}
F_t(\omega)&=\int_0^t d\theta\,\xi(\theta)\,e^{-i\omega\theta}=\\&
=\frac{\sqrt{\pi}}{2}\,\tau\,e^{-\left(\frac{\omega\tau}{2}\right)^2-i\omega t_0}\,\left[\text{erf}\left(\frac{t-t_0}{\tau}+i\frac{\omega\tau}{2}\right)+\underbrace{\text{erf}\left(\frac{t_0}{\tau}-i\frac{\omega\tau}{2}\right)}_{\simeq 1}\right]
\end{split}
\end{equation}
$\text{erf}(x)=\frac{2}{\sqrt{\pi}}\int_0^x e^{-y^2}dy$ being the error function (with imaginary argument).
\newline Notice that at the linear order only the \virgolette{nearby} modes $\pm2K$ are excited by the perturbation. This is easily understood in terms of Feynman diagrams (at linear order we have a single momentum kick of magnitude $2K$).
\newline In Fig.  \ref{fig:ComparisonWithPT} numerically obtained data are compared with the perturbative expression eq. \ref{eq:f}.

The density variation can be computed at linear order
\begin{equation}
\label{eq:density_variation_general_form_real_space}
\begin{split}
\delta&\rho_{k_0}(x,y;t)=|\Psi(t)|^2-|\psi_{k_0}|^2=\\&=2\lambda\, \Re\Bigl(\Bigr.f_{k_0+2K}^{(k_0)}\,e^{i\Delta\omega_{k_0,k_0+2K}\,t}\,\Phi_{k_0}(x)\Phi_{k_0+2K}(x)\,\frac{e^{i2K y}}{L_y}+(K\rightarrow-K)\Bigl.\Bigr)
\end{split}
\end{equation}
and its Fourier transform $\delta\rho(x,q;t)=\int e^{-i q y}\delta\rho(x,y;t)dy$ reads
\begin{equation}
\begin{split}
\label{eq:pert_sin2}
\delta&\rho_{k_0}(x,q;t)=\lambda\left[(A_{k_0}(x)+B^*_{k_0}(x))\delta_{q,2K}+(A^*_{k_0}(x)+B_{k_0}(x))\delta_{q,-2K}\right]
\end{split}
\end{equation}
where
\begin{equation}
\begin{cases}
\label{eq:pert_sin2_1}
A_{k_0}(x)=f_{k_0+2K}^{(k_0)}\,e^{i\Delta\omega_{k_0,k_0+2K}\,t}\,\Phi_{k_0}(x)\Phi_{k_0+2K}(x)
\\
B_{k_0}(x)=f_{k_0-2K}^{(k_0)}\,e^{i\Delta\omega_{k_0,k_0-2K}\,t}\,\Phi_{k_0}(x)\Phi_{k_0-2K}(x).
\end{cases}
\end{equation}
Notice that at linear order $\delta\rho(q=0)$ vanishes, which ensures that $\Psi(t)$ is normalized up to higher order terms.
%With some approximations (linear dispersion) this comes to agree with the effective theory equation described in the previous chapter, as will be discussed below.
\begin{figure}[htp!]
	\begin{minipage}{.5\textwidth}
		\centering
		\includegraphics[width=1.\textwidth]{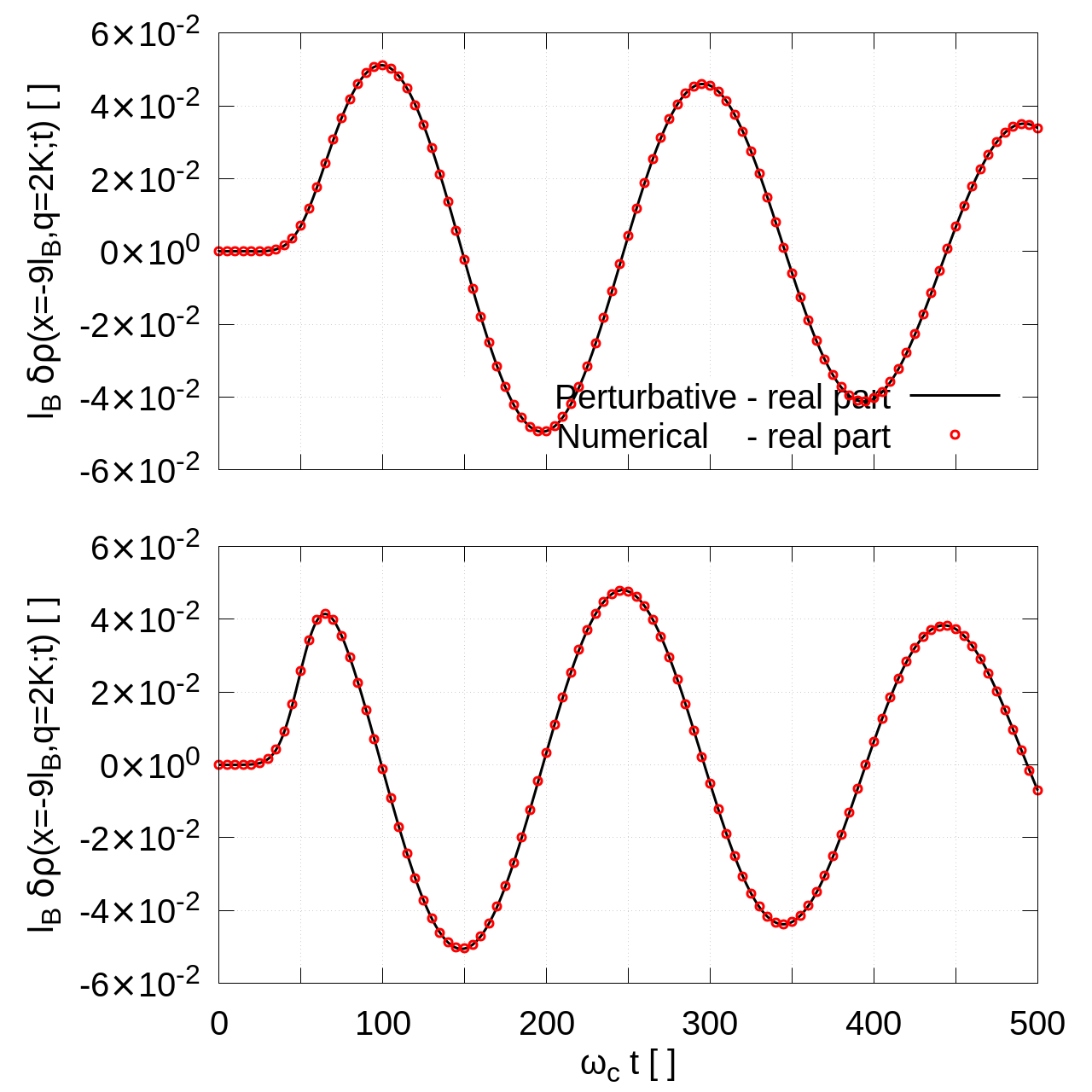}
	\end{minipage}%
	\begin{minipage}{0.5\textwidth}
		\centering
		\includegraphics[width=1.\textwidth]{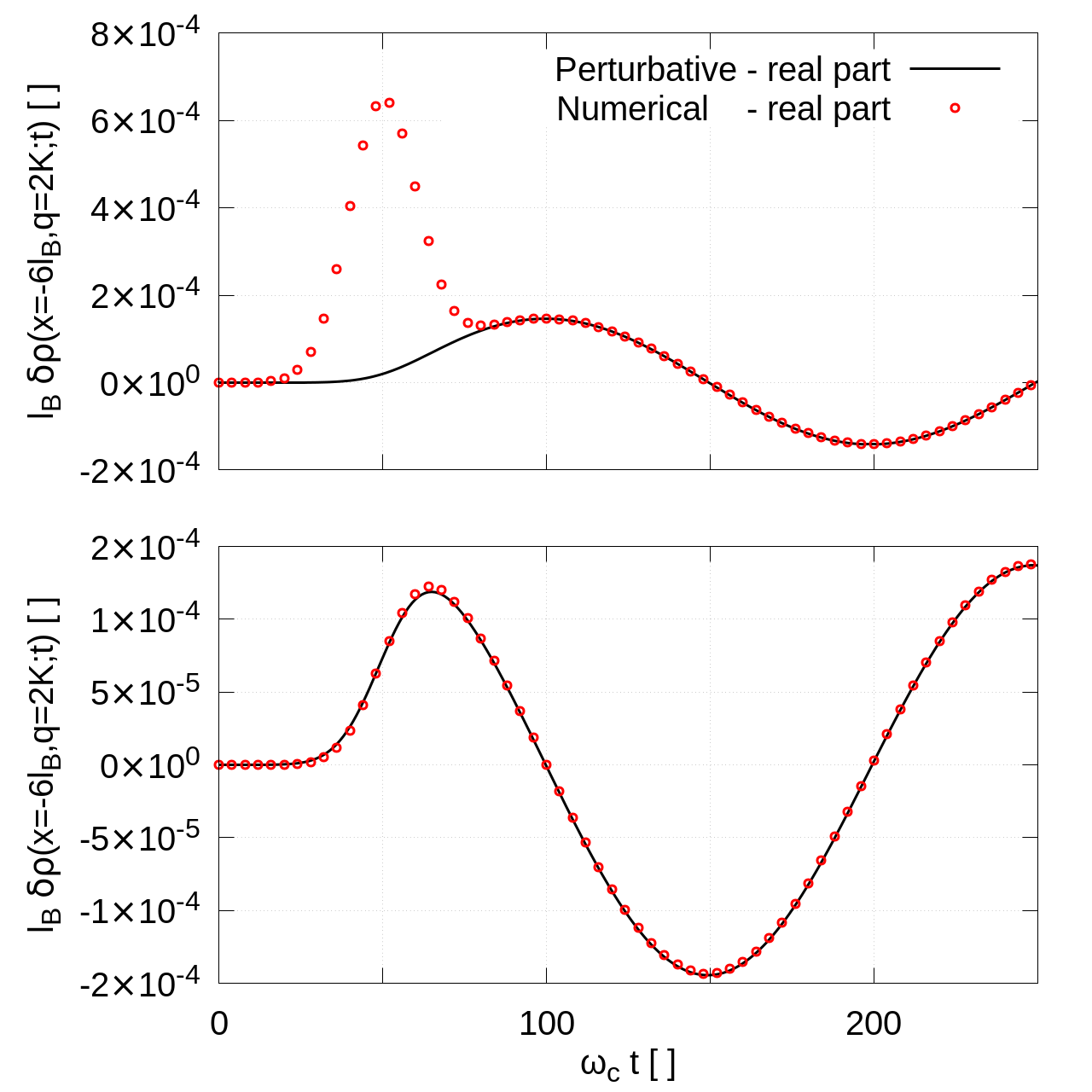}
	\end{minipage}    
	\caption[The LOF caption]{The $q=2K$ components of the Fourier transform of the one body density are here shown for two values of $x$, $x=-9l_B$ on the left hand side, $x=-6l_B$ on the right one. I used $L_y=200$, $k_F=273\Delta k\simeq 8.58l_B^{-1}$, $2K=4\Delta k$ and $\lambda=0.005\hbar\omega_c$.}
	\label{fig:ComparisonWithPT2}
\end{figure}

\noindent These single-particle results can be used for the many-body non-interacting system just by summing over all the filled states labelled by $k_0$.
\newline In Fig. \ref{fig:ComparisonWithPT2} the $q=2K$ component of the density variation at fixed $x$ is compared with the numerical results. The agreement appears to be qualitatively good at the extreme edges of the system (left hand side panel); we see however a bump emerging at times corresponding to the perturbation transient as we move towards the bulk of the system (right hand side panel). Such a bump corresponds to non-resonant excitations from the $n=0$ Landau level to the higher ones (mainly the $n=1$ level) which have been neglected in the perturbative computation performed.
\newline The fact that the bump is relevant in the bulk but it is not near the edges will be explained in more detail below; intuitively, this occurs because in the bulk all the $n=0$ Landau level is completely filled and there is no state to which an electron can jump:
if we consider these transitions only the system density will not change. The story changes with the inclusion of the $n=1$ Landau level, which allows some virtual transition to occur, giving a small contribution to the density variation $\delta\rho$ as long as the excitation potential is \virgolette{on} (the bump seen in the right hand side panel).
The discussion changes near the edges: free electron states are available above the Fermi surface; low energy transitions near the Fermi surface will affect $\delta\rho$, making the contributions arising from the $n=1$ Landau level negligible.

\subsubsection*{Second perturbative order}
The evolution equations for the $g_k$ coefficients in eq. \ref{eq:perturbative_expansion_cnk} can be written down too
\begin{equation}
\label{eq:g}
\begin{cases}
\begin{alignedat}{2}
&\frac{\partial g_{k_0}^{(k_0)}}{\partial t} &&=-i\frac{\xi(t)}{2\hbar}\left(f_{k_0}^{(k_0)}-\frac{1}{2}\left(d_{k_0,k_0-2K}e^{i\Delta\omega_{k_0,k_0-2K}t}f_{k_0-2K}^{(k_0)}+(K\rightarrow-K)\right)\right)\\
&\frac{\partial g_{k_0\pm2K}^{(k_0)}}{\partial t} &&= -i \frac{\xi(t)}{2\hbar}\left(f_{k_0\pm2K}^{(k_0)}-\frac{1}{2}d_{k_0,k_0\pm2K}e^{i\Delta\omega_{k_0\pm2K,k_0}t}f_{k_0}^{(k_0)}\right)\\
&\frac{\partial g_{k_0\pm4K}^{(k_0)}}{\partial t} &&= +i \frac{\xi(t)}{4\hbar} d_{k_0\pm4K,k_0\pm2K}e^{i\Delta\omega_{k_0\pm4K,k_0\pm2K}t}f_{k_0\pm2K}^{(k_0)}
\end{alignedat}
\end{cases}
\end{equation}
however (to the best of my knowledge) (some of\footnote{\label{footnote:g2K}Those for $g_{k_0\pm2K}$ can be reduced to the integrals encountered at first perturbative order, and thus also be analytically performed. The solution is
\[
g_{k_0\pm2K}^{(k_0)}=\frac{d_{k_0,k_0\pm2K}}{8\hbar^2}\left(\int_0^t \xi(\tau)e^{i\Delta\omega_{k_0\pm2K,k_0}t} d\tau\right)\left(\int_0^t \xi(\tau) d\tau\right)=-f^{* (k_0)}_{k_0}f_{k_0+2K}^{(k_0)}.
\] 
This result will be used later to establish that corrections to $\delta\rho_{k_0}(x,2K;t)$ given in eq. \ref{eq:pert_sin2} are $\mathcal{O}(\lambda)^3$.
}) these integrals need to be computed numerically or with some approximations, which makes it hard to go beyond the quadratic term in the perturbative expansion. 

Notice that the modes $\pm4K$ get excited at quadratic order, which can be understood in terms of Feynman diagrams as resulting from a \virgolette{double} interaction with the external potential. It is intuitive that higher harmonics will get excited at even higher orders.
\newline It is easy to check that
\begin{equation}
\sum_{n,k}|c_{n,k}^{(k_0)}|^2=1+\lambda^2\underbrace{\left(2\Re\left(g_{k_0}^{(k_0)}\right)+\left|f_{k_0+2K}^{(k_0)}\right|^2+\left|f_{k_0-2K}^{(k_0)}\right|^2+\left|f_{k_0}^{(k_0)}\right|^2\right)}_{\Theta(t)}+\mathcal{O}(\lambda^3)
\end{equation}
Normalization requires $\sum_{n,k}\left|c_{n,k}^{(k_0)}\right|^2=1$, and thus $\Theta(t)$ needs to vanish at any time. This can be easily shown to be true by noticing that $\Theta(t=0)=0$ and $\partial_t\Theta=0$.

\noindent It is easy to generalize the computation of the Fourier transform $\delta\rho(x,q;t)$ to include the quadratic terms as well. The $2K$ Fourier component of the density variation at second perturbative order reads
\begin{equation}
\label{eq:lambda2_correction_to_rho2K}
\begin{split}
\delta\rho_{k_0}&(x, 2K; t) = \left[\lambda\,f_{k_0+ 2K}^{(k_0)}+\lambda^2\left(f_{k_0}^{(k_0)*}f_{k_0+ 2K}^{(k_0)}+g_{k_0+ 2K}^{(k_0)}\right)\right]\,e^{i\Delta\omega_{k_0, k_0+2K}\,t}\Phi_{k_0}\Phi_{k_0+2K}+\\&+
\left[\lambda\,f_{k_0- 2K}^{(k_0)*}+\lambda^2\left(f_{k_0}^{(k_0)}f_{k_0- 2K}^{(k_0)*}+g_{k_0- 2K}^{(k_0)*}\right)\right]\,e^{-i\Delta\omega_{k_0, k_0-2K}\,t}\Phi_{k_0}\Phi_{k_0-2K}.
\end{split}
\end{equation}
Since\footnoteref{footnote:g2K} $g_{k_0+2K}^{(k_0)}=-f^{* (k_0)}_{k_0}f_{k_0+2K}^{(k_0)}$, the second order corrections are identically zero. 
\newline This cancellation can be easily understood as being a symmetry consequence. One can grasp it by thinking of the scattering processes in terms of Feynman diagrams. At each interaction in perturbation theory an electron gets a momentum kick equal to the perturbation wavevector, $2K$.
Only odd-order processes can then excite \virgolette{odd} modes, i.e. modes having a wavevector which is an odd multiple of the fundamental one $2K$; analogously even-order processes will excite \virgolette{even} modes, at any order in perturbation theory.
\begin{figure}[htp!]
	\centering
	\includegraphics[width=1.\textwidth]{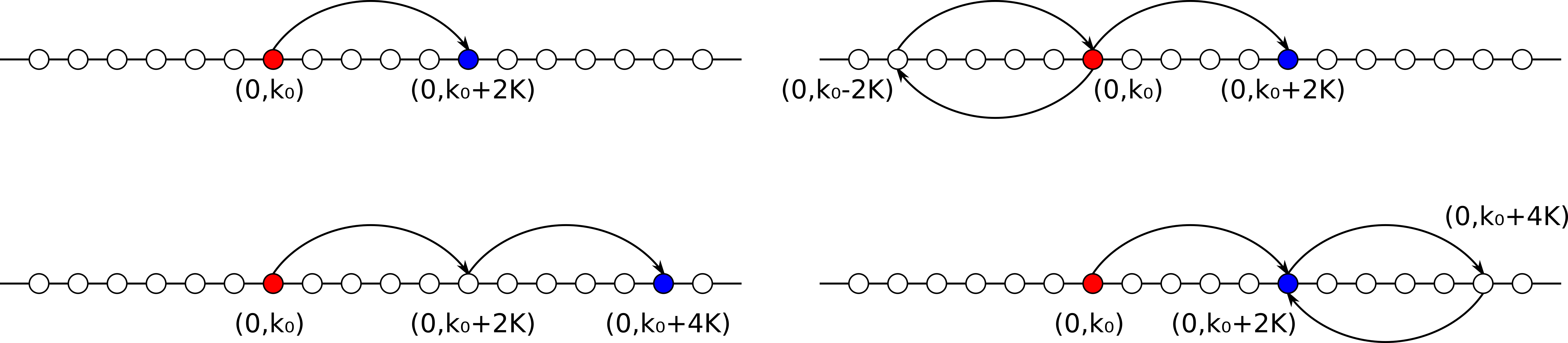}
	\caption[The LOF caption]{Some of the Feynman diagrams discussed in the main text. The initial state is depicted in red, the final one in blue. On the left hand side first and second order transitions are shown, the right panel shows examples of third order ones. Other diagrams are possible for the second and third order transitions connecting the same initial and final states, involving virtual transitions between different Landau levels. These are however not shown.}
	\label{fig:FeynmanDiagrams}
\end{figure}
The diagrams corresponding to the computed contributions are shown in Fig. \ref{fig:FeynmanDiagrams}. Some of the diagrams corresponding to third order contributions to the $2K$ mode are shown too. Diagrams of processes involving non-resonant excitations to higher Landau levels have not been drawn.

\begin{figure}[htp!]
	\begin{minipage}{.5\textwidth}
		\centering
		\includegraphics[width=1.\textwidth]{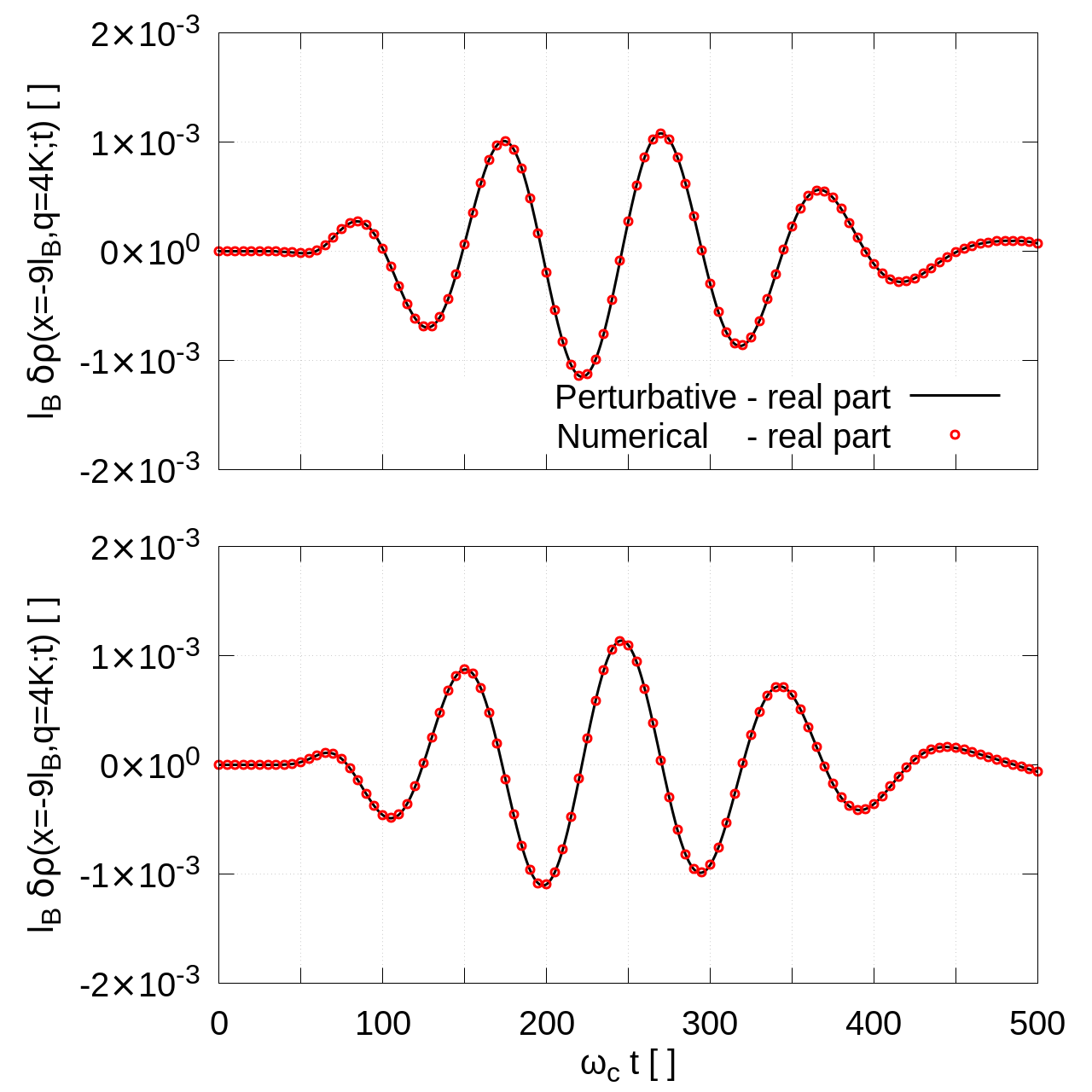}
	\end{minipage}%
	\begin{minipage}{0.5\textwidth}
		\centering
		\includegraphics[width=1.\textwidth]{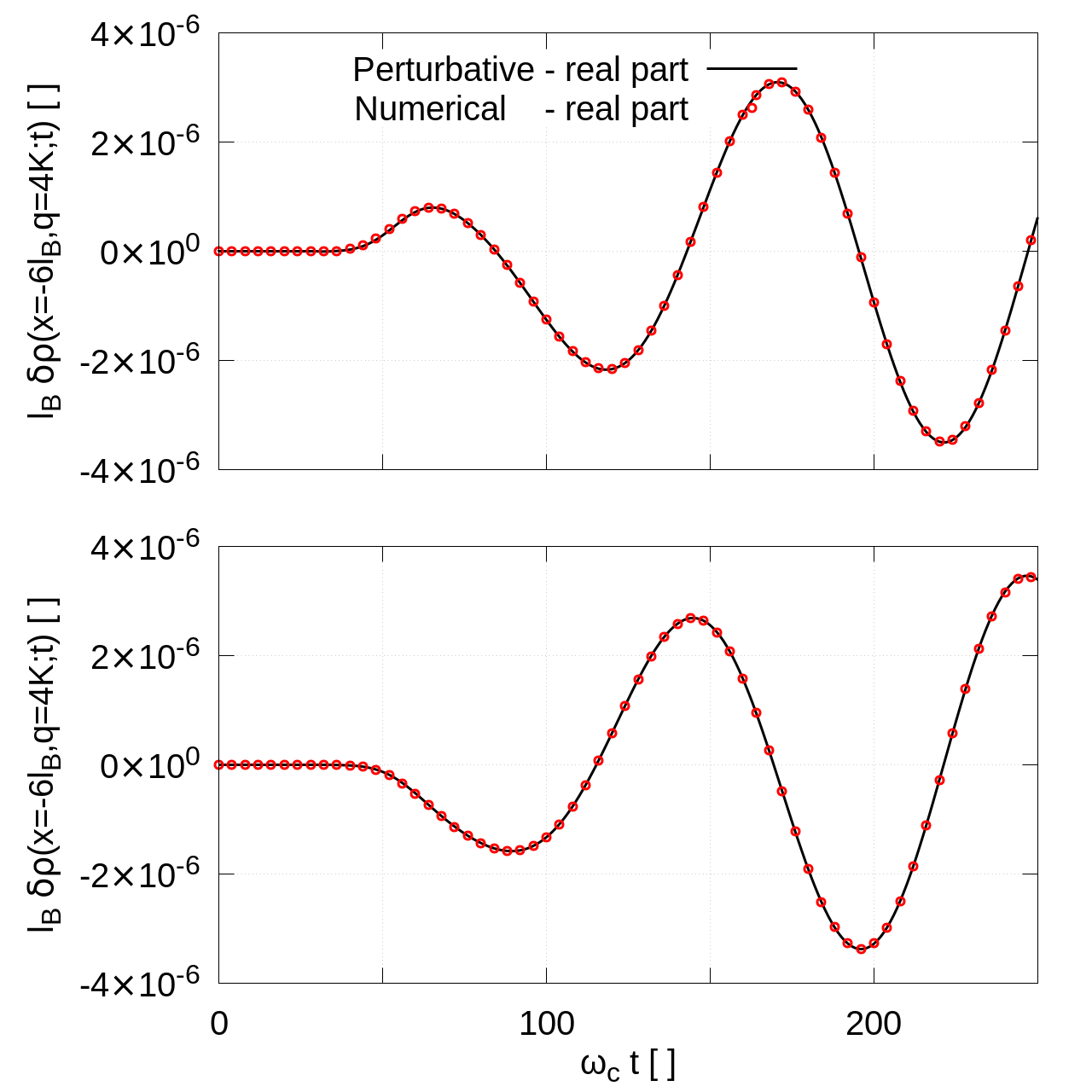}
	\end{minipage}    
	\caption[The LOF caption]{The $q=4K$ components of the Fourier transform of the one body density are here shown for two values of $x$, $x=-9l_B$ on the left hand side, $x=-6l_B$ on the right one. I used $L_y=200$, $k_F=273\Delta k\simeq 8.58l_B^{-1}$, $2K=4\Delta k$ and $\lambda=0.005\hbar\omega_c$.}
	\label{fig:ComparisonWithPT3}
\end{figure}

The other Fourier components of the density variation are not explicitly written here; the $q=4K$ one will however be used below when discussing the generation of higher harmonics.
In Fig. \ref{fig:ComparisonWithPT3} $\delta\rho(x,q=4K;t)$ is compared with the numerical data. It can be seen that the results do qualitatively agree even at the second perturbative order.

\subsection{The bulk}\label{subsection:sin_PTin_bulk}
Before diving into the edge dynamics, it is worth studying the bulk response to the periodic excitation \ref{eq:sinusoidal_excitation}.
 
Consider the $q=2K$ component of the Fourier transform, at linear order. In the bulk of the system we have degenerate Landau levels, so
\begin{equation}
\Delta \omega_{k_0+2K, k_0} \simeq \Delta \omega_{k_0,k_0-2K}\simeq 0.
\end{equation}
In the bulk the matrix element $d_{k_0,k_0+\Delta}$ depends only on the magnitude of $\Delta$. A quick computation using the $n=0$ Landau level gives $d_{k_0,k_0+\Delta}=e^{-\left(\frac{\Delta l_B}{2}\right)^2}$.
The $q=2K$ component of the density Fourier transform $\delta \rho_{k_0}(x, 2K; t)$ given in equation \ref{eq:pert_sin2} then reads
\begin{equation}
\delta \rho_{k_0}
%\simeq\lambda e^{-i2K v\,t}\left[f_{k_0+2K}\,\Phi_{k_0}^*\Phi_{k_0+2K}+f_{k_0-2K}^*\,\Phi_{k_0}\Phi_{k_0-2K}^*\right]
\simeq i\frac{\lambda}{4\hbar} F_t(0)e^{-\left(\frac{\Delta l_B}{2}\right)^2}\left(\Phi_{k_0}(x)\Phi_{k_0+2K}(x)-
\Phi_{k_0}(x)\Phi_{k_0-2K}(x)\right).
\end{equation}
It is easy to notice that when summing over all the occupied states of the ground state of the system (i.e. a fully occupied Landau level) this vanishes\footnote{Except for boundary terms which are here neglected since in any case the dispersion ceases to be flat.}, i.e.
\begin{equation}
\delta \rho(2K, t)\simeq 0
\end{equation}
which we expected on the basis of the particle-hole excitations discussed above: there is no empty state to which bulk electrons can be excited by a \virgolette{small} perturbation. The bulk of the system behaves then as an incompressible medium.
\newline On physical grounds we then expect such a cancellation to occur for the generic bulk of the system, i.e. for a general non-flat dispersion relation of bulk electrons, since the lowest lying excitations are gapped. This can be immediately checked by noticing that the contribution to the density variation given by the electron with labelled by $k_0$ are exactly counterbalanced by the electrons with wavevector $k_0\pm2K$, indeed, since $F_t(\omega)^*=F_t(-\omega)$ and $\Delta\omega_{k,q}=-\Delta\omega_{q,k}$ we have
\begin{equation}
\begin{cases}
A_{k_0}(x)=-B^*_{k_0+2K}(x)\\
B^*_{k_0}(x)=-A_{k_0-2K}(x)
\end{cases}
\end{equation}
so that they cancel out when summing over $k_0$, and $\delta\rho(2K,t)\simeq0$, as stated above.
\begin{figure}[htp!]
	\begin{minipage}{.5\textwidth}
		\centering
		\includegraphics[width=1.\textwidth]{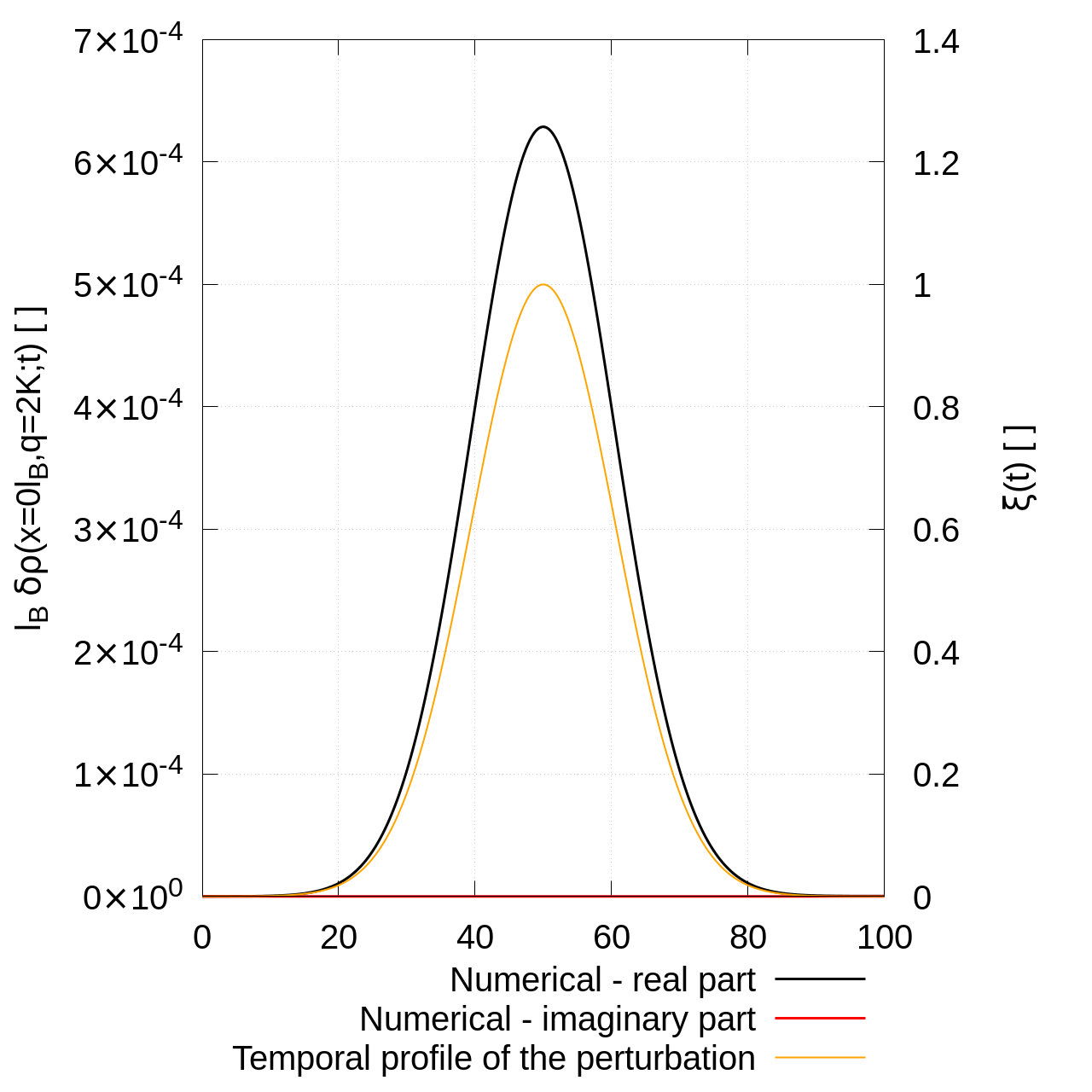}
	\end{minipage}%
	\begin{minipage}{0.5\textwidth}
		\centering
		\includegraphics[width=1.05\textwidth]{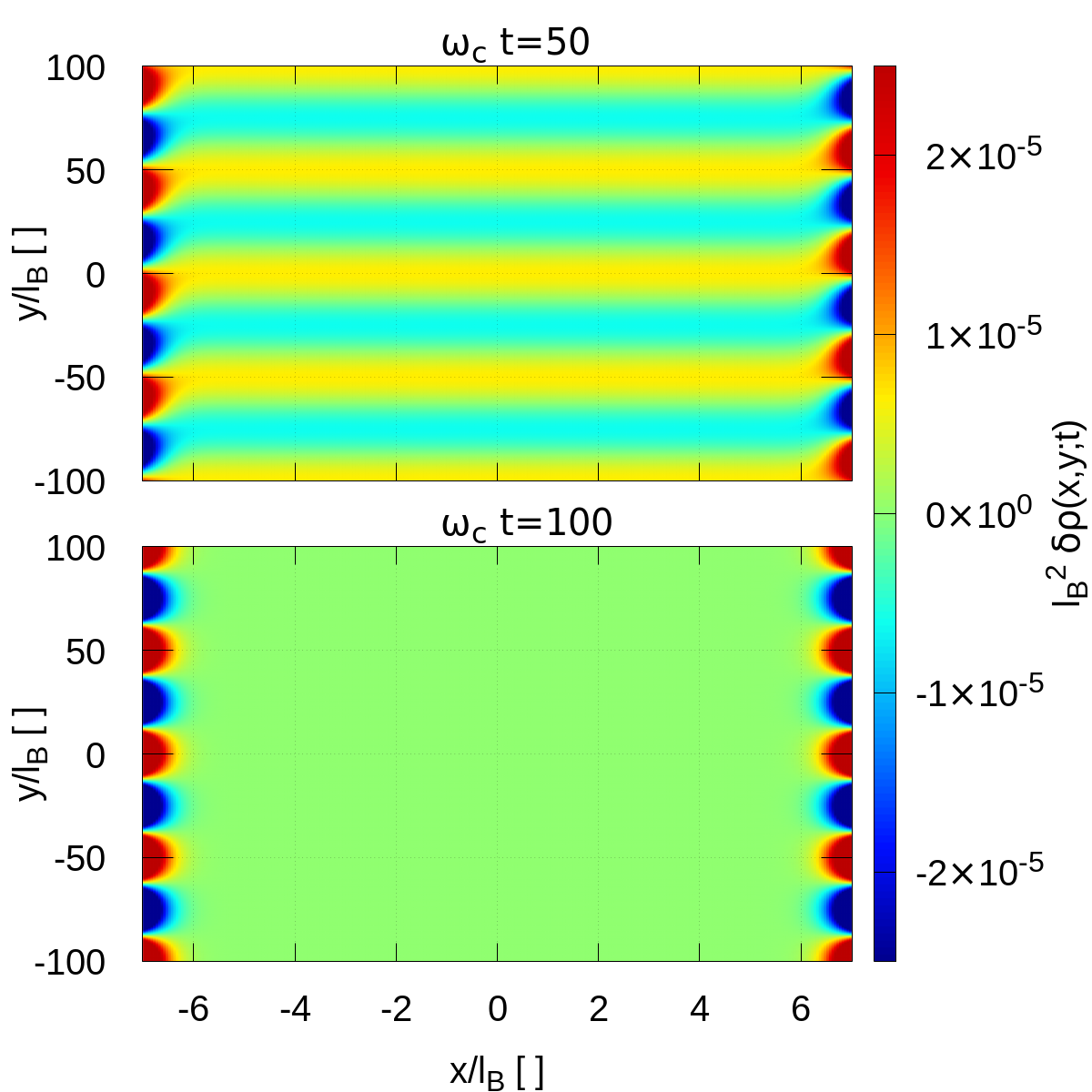}
	\end{minipage}    
	\caption[The LOF caption]{The left hand side plot shows the real (black curve) and imaginary (red) parts of the the $q=2K$ component of the Fourier transform of the one body density variation $\delta\rho$, at $x=0$, i.e. in the bulk of the system. The temporal profile of the externally applied excitation is also shown (orange curve). 
	\newline The right-hand side panel instead compares the bulk spatial density variation at two different times: the top panel shows the system density when the perturbation is still \virgolette{on}, the bottom panel after it has been turned off. The edges have been excluded from the $x$-range since they were too much off-scale.
	\newline The images have been obtained using $L_y=200$, $k_F=273\Delta k\simeq 8.58l_B^{-1}$, $2K=4\Delta k$, $\lambda=0.005\hbar\omega_c$}
	\label{fig:bulk_incompressibility}
\end{figure}

It can be checked straightforwardly that such a cancellation occurs even at the second perturbative order\footnote{In order to show that when summing over all the electrons also the $q=4K$ component of the Fourier transform vanishes, the identity $g_{k_0+2K}^{(k_0-2K)*}+f_{k_0+2K}^{(k_0)*}f_{k_0-2K}^{(k_0)}+g_{k_0-2K}^{(k_0+2K)}$ has been used. It can be easily proven by using the identity $\int_0^t f(x)dx\int_0^t g(y)dy=\int_0^tf(x)\int_0^xg(y)dy\,dx+\int_0^tg(x)\int_0^xf(y)dy\,dx$.}; although the result is interesting, the cumbersome computations are not reported here due too excessively tedious book-keeping. This however agrees with our understanding: as long as excitations to higher Landau levels are neglected, the system bulk cannot be \virgolette{deformed}; this physical motivation makes us believe that this will probably be true at every perturbative order. 

In Fig. \ref{fig:bulk_incompressibility} we see that the bulk slightly deforms though (top right panel). This occurs in the transient, since the ground state is adiabatically following the perturbation, as a result of available single electron states in higher energy Landau levels; these states can be excited non-resonantly (the perturbation being turned on very slowly compared to $\omega_c^{-1}$, and its intensity being much smaller than $\hbar\omega_c$), so after the perturbation is slowly turned off the bulk returns to its initial configuration.

\begin{figure}[htp!]
	\begin{minipage}{.5\textwidth}
		\centering
		\includegraphics[width=1.\textwidth]{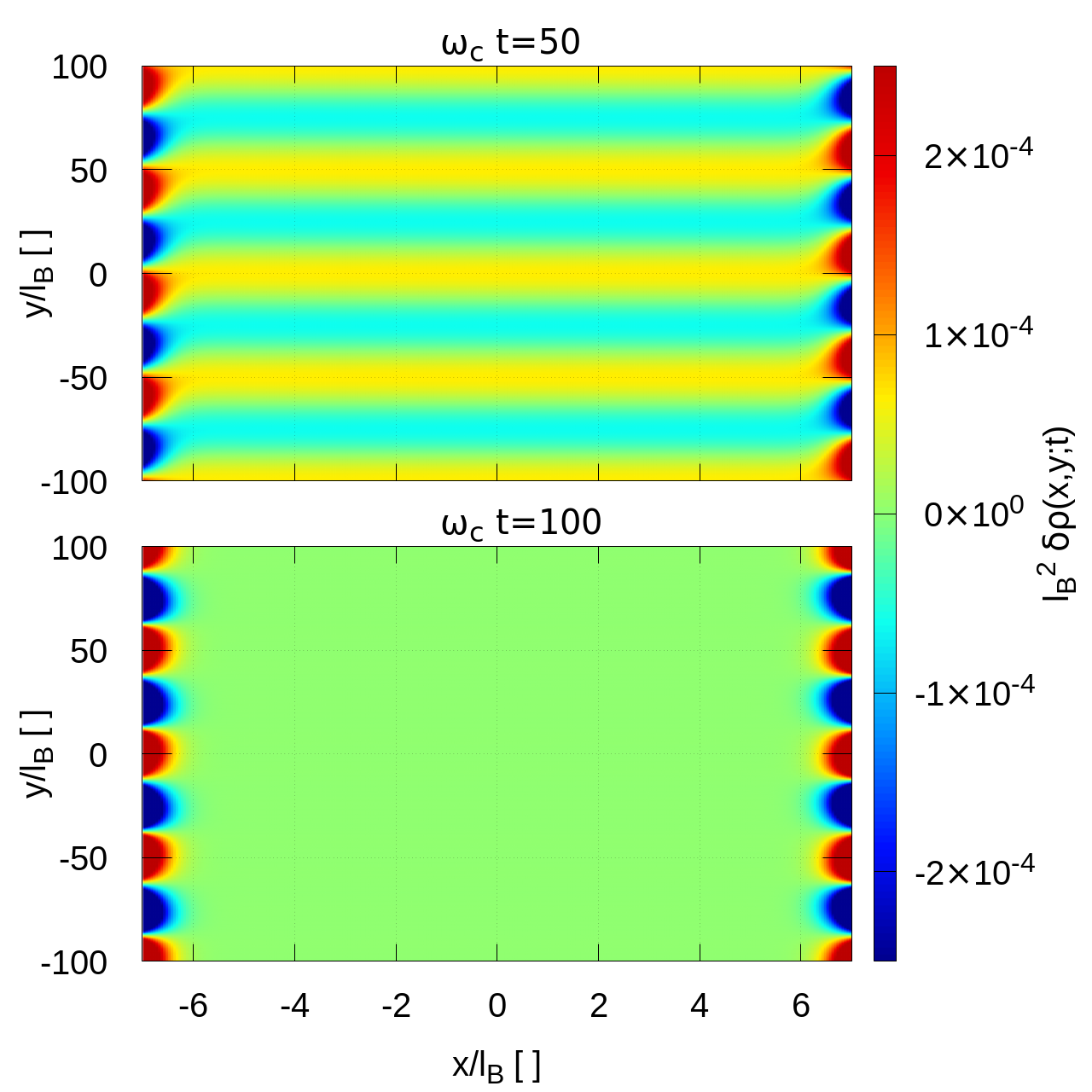}
	\end{minipage}%
	\begin{minipage}{0.5\textwidth}
		\centering
		\includegraphics[width=1.\textwidth]{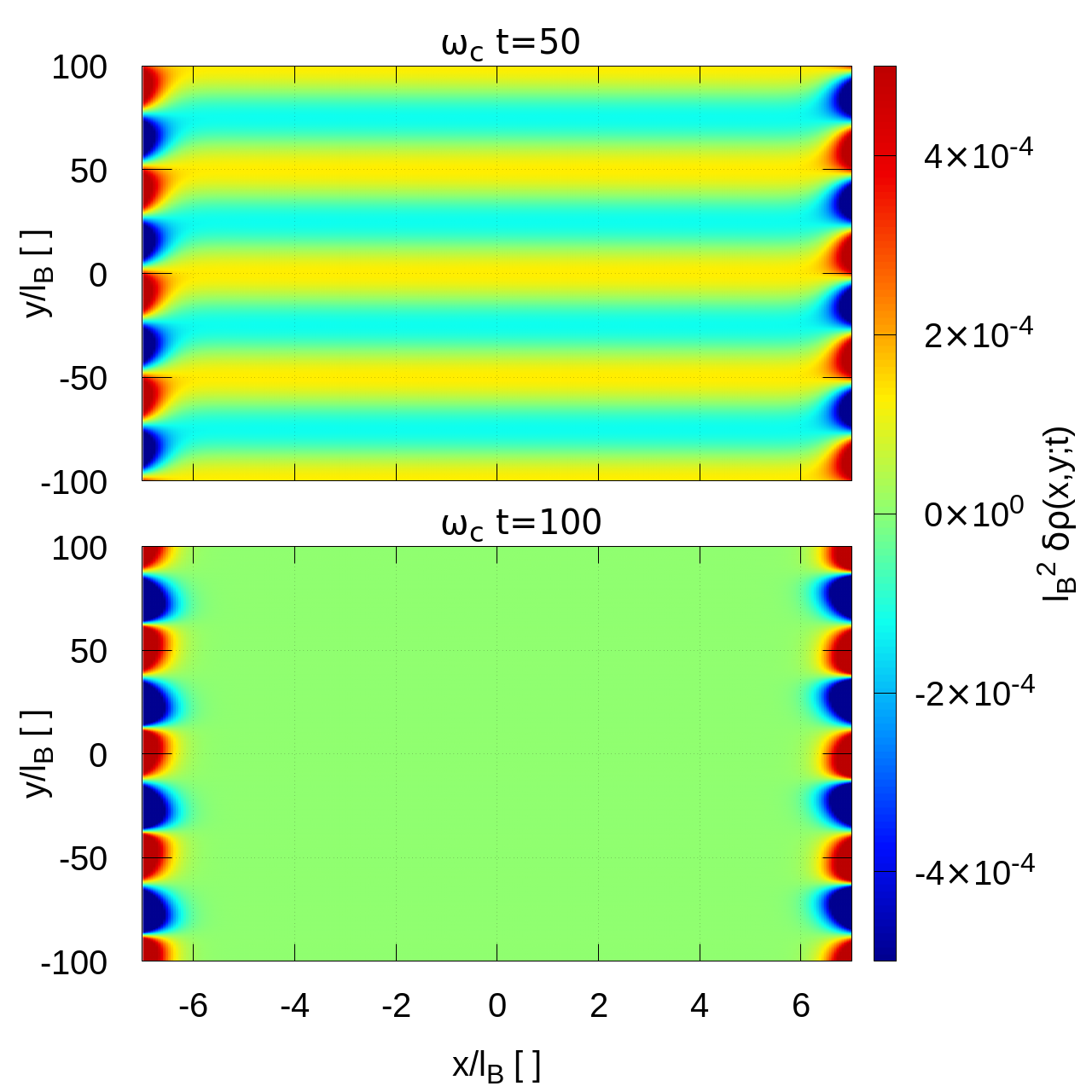}
	\end{minipage}    
	\caption[The LOF caption]{The two images are analogous to the one in the right hand side panel of Fig. \ref{fig:bulk_incompressibility}. They both compare the bulk spatial density variation at two different times: the top panel shows the system density during the perturbation transient, the bottom panel after it has been turned off. The edges have been excluded from the $x$-range since they were too much off-scale.
	\newline The images have been obtained using $L_y=200$, $k_F=273\Delta k\simeq 8.58l_B^{-1}$, $2K=4\Delta k$; the left hand side image has been obtained with $\lambda=0.05\hbar\omega_c$, the right hand side one for $\lambda=0.10\hbar\omega_c$ (i.e. excitation strengths ten and twenty times larger than the one used for Fig. \ref{fig:bulk_incompressibility}).}
	\label{fig:bulk_incompressibility2}
\end{figure}

In Fig. \ref{fig:bulk_incompressibility2} non-perturbative ($\frac{\lambda}{\hbar\omega_c}=0.05$ and $\frac{\lambda}{\hbar\omega_c}=0.10$) numerical results analogous to those plotted in the right hand side panel of Fig. \ref{fig:bulk_incompressibility} are shown; we see that the previous considerations still hold (but it can be seen that the edge contours substantially change their shape).

As a final comment, notice that the relevant matrix elements regulating transition rates to higher Landau levels get smaller as the perturbation wavevector is reduced
\begin{equation}
d^{k_0,k_0+\Delta}_{0,n}=\sqrt{\frac{1}{2^n\,n!}}\,(\Delta l_B)^n\,\exp\left[-\left(\frac{\Delta l_B}{2}\right)^2\right]\propto \Delta^n
\end{equation}
so that low order transitions to higher Landau levels not only are suppressed by the high energy gap between different Landau levels, but also by orthogonality, so that the bulk deformation becomes insignificant to any extent. 
\newline Notice however that for shorter wavelengths such a matrix element may get relevant; it is indeed maximized at wavevectors $l_B \Delta_M(n)=\sqrt{2n}$. Exciting the system with a perturbation with such a wavevector would maximize transitions to higher Landau levels during its transient.
\newline I actually hid a subtlety: the sum over all the electrons may get large. This is however not the case. Following the same steps that will be carried out in one of the last sections of this chapter (subsection \ref{subsection:Jx_transient_bulk}, in which $x$ component of the probability current $\mathbf{J}(x,2K;t)$ is computed) one can easily obtain the contribution resulting from the inclusion of the $n=1$ Landau level to the bulk density variation.
If the excitation is long wavelength (the matrix element has been series expanded) we obtain
\begin{equation}
\label{eq:bulk_n=1_density_variation}
\delta\rho(x,2K;t)\simeq\frac{\lambda}{4\hbar\omega_c}\,\frac{2K\,l_B}{\sqrt{2}}\,e^{-\left(\frac{t-t_0}{\tau}\right)^2}\,\sum_{k_0}
\Bigl[
\Phi_{0,k_0}(x)\Phi_{1,k_0+2K}(x)-\Phi_{0,k_0}(x)\Phi_{1,k_0-2K}(x)
\Bigr].
\end{equation}
Since the term in the square brackets is odd in $2K$, such a term is at least $\mathcal{O}(2K)$, making the bulk density variation\footnote{Notice the presence of the $2K$ coefficient arising from the long wavelength expansion of the matrix element $d^{0,1}_{k_0,k_0+2K}$.} $\delta\rho(x,2K;t)=\mathcal{O}(2K)^2$, as compared to the edge result which will be derived in the following section which is $\mathcal{O}(2K)$ (eq. \ref{eq:effective_density_variation_linear_dispersion}): the bulk excitation will thus be negligible compared to the edge one if the wavelength of the external perturbation is sufficiently small, as can indeed be seen in Fig. \ref{fig:bulk_incompressibility}.

\begin{figure}[htp!]
	\begin{minipage}{.5\textwidth}
		\centering
		\includegraphics[width=1.\textwidth]{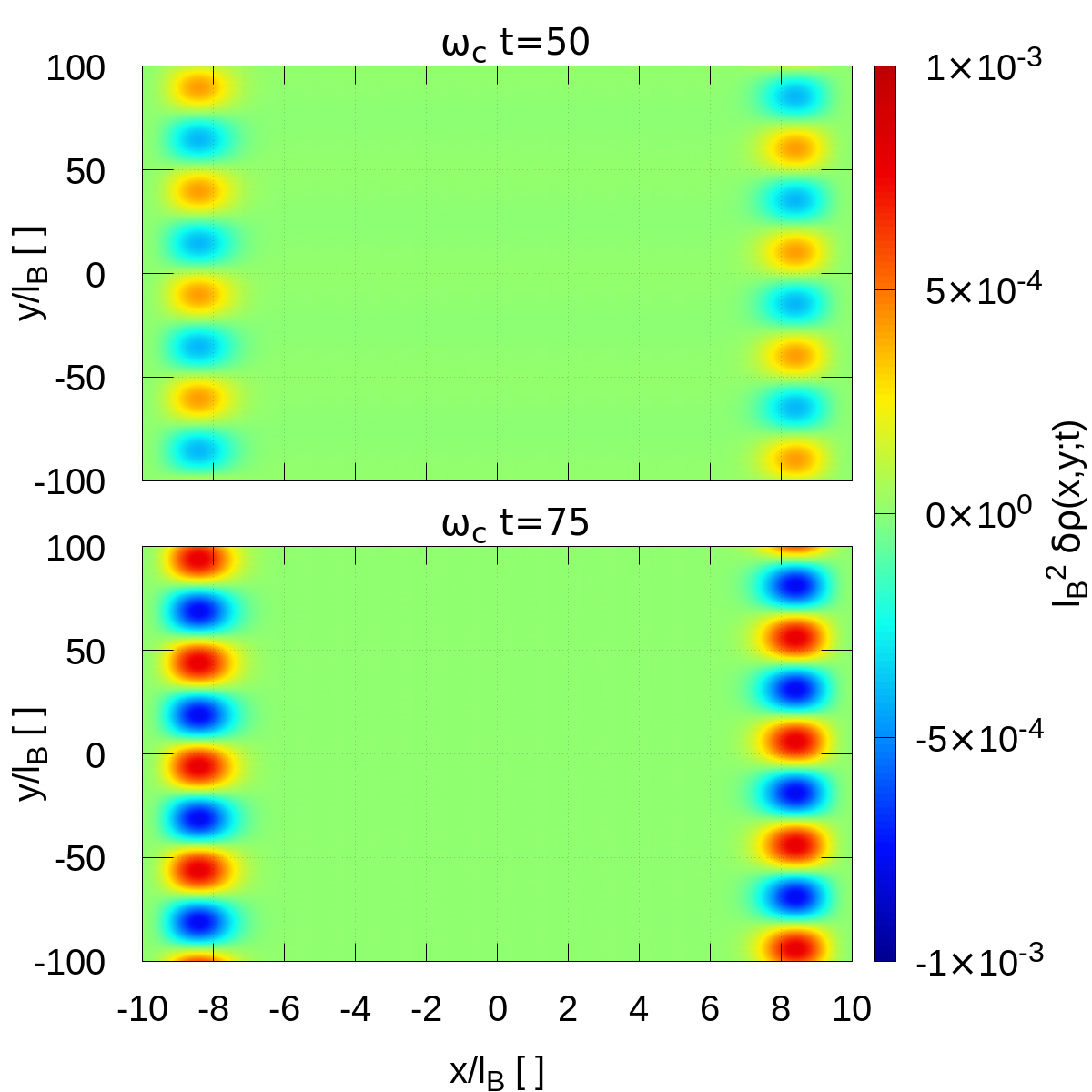}
	\end{minipage}%
	\begin{minipage}{0.5\textwidth}
		\centering
		\includegraphics[width=1.\textwidth]{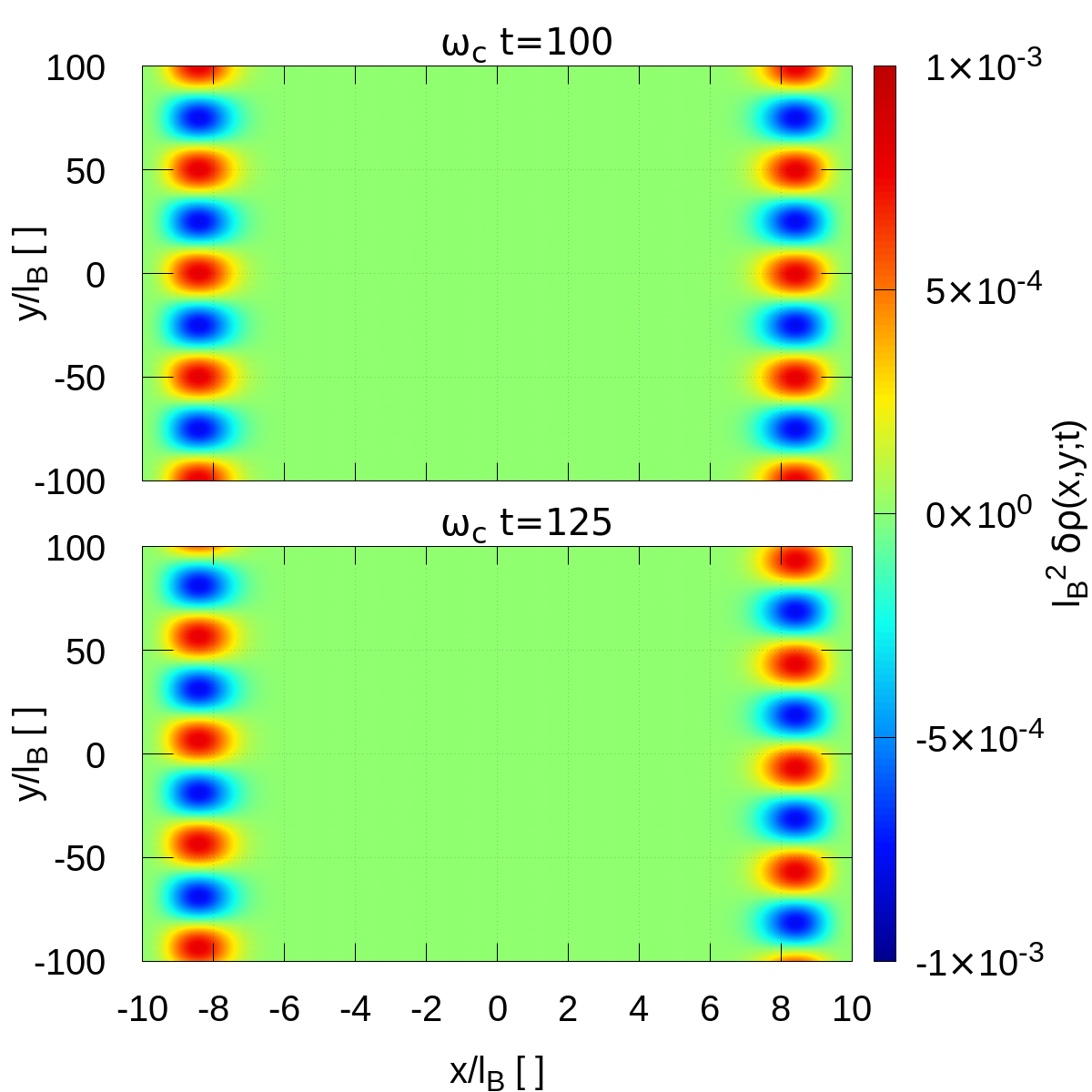}
	\end{minipage}    
	\caption[The LOF caption]{Both the images show the one-body density variation at different times.
		These have been obtained using $L_y=200$, $k_F=273\Delta k\simeq 8.58l_B^{-1}$, $2K=4\Delta k$, $\lambda=0.005\hbar\omega_c$}
	\label{fig:whole_system_heat_plots}
\end{figure}

As a final comment, notice that the cancellation between different electronic contributions can possibly occur in the bulk only; near the Fermi surface of the system it will not be complete, since there are no electrons above the Fermi point which can cancel the contributions of the ones immediately below it; density excitations will then as expected be spatially localized near the system edges, as can be seen in Fig. \ref{fig:whole_system_heat_plots}. 
In the following sections the discussion will focus on this aspect.

\subsection{Edge dynamics}
We now focus on the dynamics at the edges of the system. As before, we begin by studying the dynamics at the perturbative level using the results obtained in Sec. \ref{subsection:sin_PT}, and then see what happens as the perturbation strength is increased. 
The results obtained are also compared with the effective dynamics equation described in the previous chapter eq. \ref{eq:effective_1d_dynamics} and \ref{eq:real_space_eff_drho}; if the dispersion relation at the Fermi surface is perfectly linear, these results are recovered. We begin by studying this simpler case and then analyse what happens if the curvature of the band does matter.

\subsubsection{Linear dispersion approximation}\label{section:linear_dispersion_approximation_sin}
Near the system edges if the excitation has long enough wavelength the dispersion relation can be linearised. 
\newline Consider the $q=2K$ component of the Fourier transform, at linear perturbative order, as given in eq. \ref{eq:pert_sin2}. If the dispersion relation is approximatively linear, we have
\begin{equation}
\Delta \omega_{k_0+2K, k_0} \simeq \Delta \omega_{k_0,k_0-2K}\simeq 2K v
\end{equation}
where $v$ is the slope of the dispersion relation. Then the $q=2K$ component of the density Fourier transform $\delta \rho_{k_0}(x, 2K; t)$ from \ref{eq:pert_sin2} reads
\begin{equation}
\delta \rho_{k_0}
%\simeq\lambda e^{-i2K v\,t}\left[f_{k_0+2K}\,\Phi_{k_0}^*\Phi_{k_0+2K}+f_{k_0-2K}^*\,\Phi_{k_0}\Phi_{k_0-2K}^*\right]
\simeq i\frac{\lambda}{4\hbar} e^{-i2K v\,t}F_t(-2K v)\left(d_{k_0,k_0+2K}\,\Phi_{k_0}\Phi_{k_0+2K}-
d_{k_0,k_0-2K}\,\Phi_{k_0}\Phi_{k_0-2K}\right)
\end{equation}
i.e. every electron contributes with the same time dependence, which can be factored out of the summation which will hence be just a numerical (in general $x$-dependent) coefficient. This obviously is no longer the case once higher order terms in the dispersion relation are included, and the effects of the curvature of the level will extensively be discussed in the following sections.
\newline Summing over all the occupied states below the Fermi sphere\footnote{the one corresponding to positive Fermi momentum $k_F$ only, the generalization being straightforward; we are thus neglecting the presence of the other edge, which is however not a problem in this context since the two edges are practically decoupled.}, many terms cancel out leaving
\begin{equation}
\label{eq:sum_over_electrons_linear_edge}
\begin{split}
\delta \rho(2K, t)
%\simeq\lambda e^{-i2K v\,t}\left[f_{k_0+2K}\,\Phi_{k_0}^*\Phi_{k_0+2K}+f_{k_0-2K}^*\,\Phi_{k_0}\Phi_{k_0-2K}^*\right]
\simeq i\frac{\lambda}{4\hbar} &e^{-i2K v\,t}F_t(-2K v)\Bigl.\Bigr[d_{k_F,k_F+2K}\,\Phi_{k_F}\Phi_{k_F+2K}+\dots+\\&+d_{k_F-2K+\Delta k,k_F+\Delta k}\,\Phi_{k_F+\Delta k}\Phi_{k_F-2K+\Delta K}\Bigl.\Bigr].
\end{split}
\end{equation}

\begin{figure}[htp!]
	\begin{minipage}{.5\textwidth}
		\centering
		\includegraphics[width=1.\textwidth]{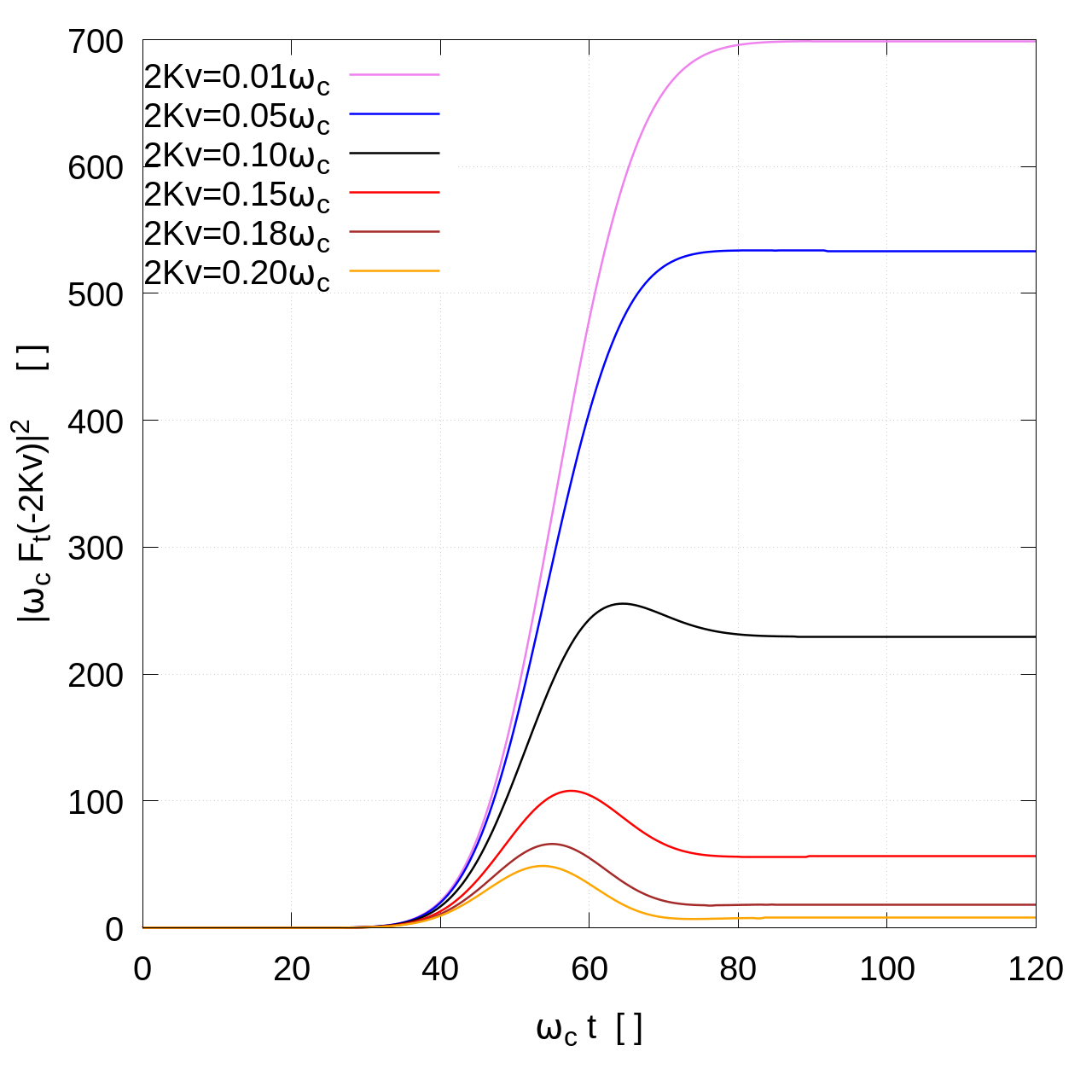}
	\end{minipage}%
	\begin{minipage}{0.5\textwidth}
		\centering
		\includegraphics[width=1.\textwidth]{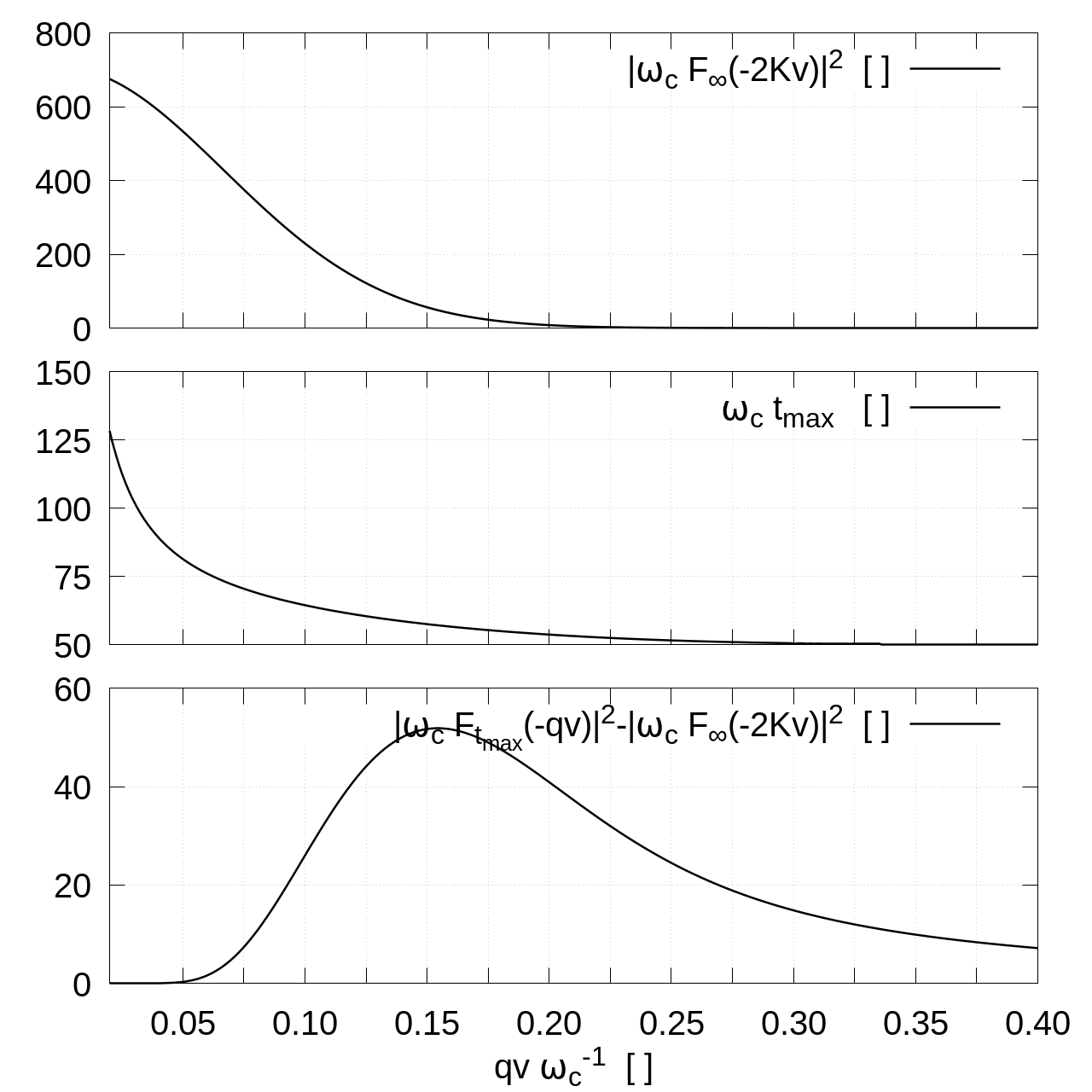}
	\end{minipage}    
	\caption[The LOF caption]{The figure on the left hand side shows $|F_t(-2Kv)|^2$ as a function of time, for different values of $2Kv$. The right hand side, the position of its maximum (top) and its height (from which the value at infinity has been subtracted off, so that \virgolette{no peak} means the difference approaches zero).}
	\label{fig:Ft(qv)_analysis}
\end{figure}

\noindent If we now integrate out the $x$ variable (in the bulk the density variation is negligibly small; far beyond the system edge it vanishes exponentially, as can be seen in Fig. \ref{fig:whole_system_heat_plots}) in order to get an effective description of the system\footnote{Notice that the two edges of the system are not coupled, i.e. we have not any net flow of charge from one edge to the other one.}
\begin{equation}
\label{eq:linear_dispersion_sin_result}
\delta \rho_{\text{eff}}(2K, t)
%\simeq\lambda e^{-i2K v\,t}\left[f_{k_0+2K}\,\Phi_{k_0}^*\Phi_{k_0+2K}+f_{k_0-2K}^*\,\Phi_{k_0}\Phi_{k_0-2K}^*\right]
\simeq i\frac{\lambda}{4\hbar} e^{-i2K v\,t}F_t(-2K v)\left[d_{k_F,k_F+2K}^2+\dots+d_{k_F-2K+\Delta k,k_F+\Delta k}^2\right]
\end{equation}
given a a large enough system so that $\Delta k\ll k_F$ and if the excitation is a long-wavelength one, i.e. $\frac{K}{\Delta k}$ is \virgolette{not too large}, each coefficient within the square brackets is approximatively equal to $1$; it is easy to realize that we are summing $\frac{2K}{\Delta k}$ terms, so one finally obtains
\begin{equation}
\label{eq:effective_density_variation_linear_dispersion}
\frac{1}{L_y}\,\delta \rho_{\text{eff}}(2K, t)\simeq i\frac{\lambda}{4\hbar} \frac{2K}{2\pi}\,e^{-i2K v\,t}F_t(-2K v)
\end{equation}
which is just the exact same solution which can be obtained by solving the effective density evolution equation derived in the previous chapter (eq. \ref{eq:effective_1d_dynamics}), in the presence of an external potential.
\newline Notice that the presence of the system length $L_y$ in eq. \ref{eq:linear_dispersion_sin_result} is just an artifact of the Fourier convention; inversion indeed gives
\begin{equation}
\delta \rho_{\text{eff}}(y, t)\simeq 2\Re \left[i\frac{\lambda}{4\hbar} \frac{2K}{2\pi}\,e^{i2K(y-v\,t)}F_t(-2K v)\right].
\end{equation}
Notice that in the large time limit $F_t(\omega)$ becomes time-independent (since $\xi(t)$ has a finite width)
\begin{equation}
F_t(-2Kv)\xrightarrow[t\rightarrow\infty]{}\sqrt{\pi}\tau\,e^{-\left(\frac{2Kv\tau}{2}\right)^2+i2Kvt_0}
\end{equation}
whose squared modulus is plotted in the top right panel of Fig. \ref{fig:Ft(qv)_analysis}.
We see that $\delta \rho_{\text{eff}}(2K, t)$ in the large time limit has a simple time dependence $e^{-i2K v\,t}$, which in turn means that $\delta\rho_{\text{eff}}(y,t)$ depends on $y$ and $t$ only through their linear combination $y-vt$. 
Notice moreover that the result is sample size independent without the need of taking the thermodynamic limit.

It is worth discussing a bit in more detail the solution we obtained (eq. \ref{eq:effective_density_variation_linear_dispersion}).
The behaviour of its square modulus $|\delta\rho_\text{eff}(2K,t)|^2$ (which is just the squared amplitude of the system response to the excitation) as a function of time is entirely regulated by $|F_t(-2Kv)|^2$, which is plotted for different values of $2Kv$ in the left hand side panel of Fig. \ref{fig:Ft(qv)_analysis}.
\newline We see that as $2Kv$ increases a peak starts to appear and its position approaches $t_0$ (mid panel of Fig. \ref{fig:Ft(qv)_analysis}). 
By inspecting the functional form of $F_t(-2Kv)$, we see that in the limit $Kv\,\tau\gg1$ (see the top right hand side panel of Fig. \ref{fig:Ft(qv)_analysis}), the edge excitation is adiabatically following the perturbation
\begin{equation}
%F_t(\omega)\xrightarrow[Kv\,\tau\gg1]{} i\,\frac{1}{\omega}e^{- i \omega t}\exp\left[-\left(\frac{t-t_0}{\tau}\right)^2\right]
F_t(-2Kv)\xrightarrow[Kv\,\tau\gg1]{} -\frac{i}{2Kv}e^{ i 2Kv t}\exp\left[-\left(\frac{t-t_0}{\tau}\right)^2\right]
\end{equation}
which gives
\begin{equation}
\frac{1}{L_y}\,\delta \rho_{\text{eff}}(2K, t)
\xrightarrow[Kv\,\tau\gg1]{} 
\frac{\lambda}{4\hbar} \frac{1}{2\pi\,v} \,\exp\left[-\left(\frac{t-t_0}{\tau}\right)^2\right]
\end{equation}
and hence no propagating \virgolette{ripples} will be found in the system edge after the perturbation has been turned off.
We can easily understand such a result: $(qv)^{-1}$ is the typical time regulating the edge dynamics.
If the perturbation varies slowly compared to such a parameter, i.e. $Kv\tau\gg1$, we expect the evolution of edge observables to adiabatically follow the excitation.
\begin{figure}[htp!]
	\begin{minipage}{.5\textwidth}
		\centering
		\includegraphics[width=1.\textwidth]{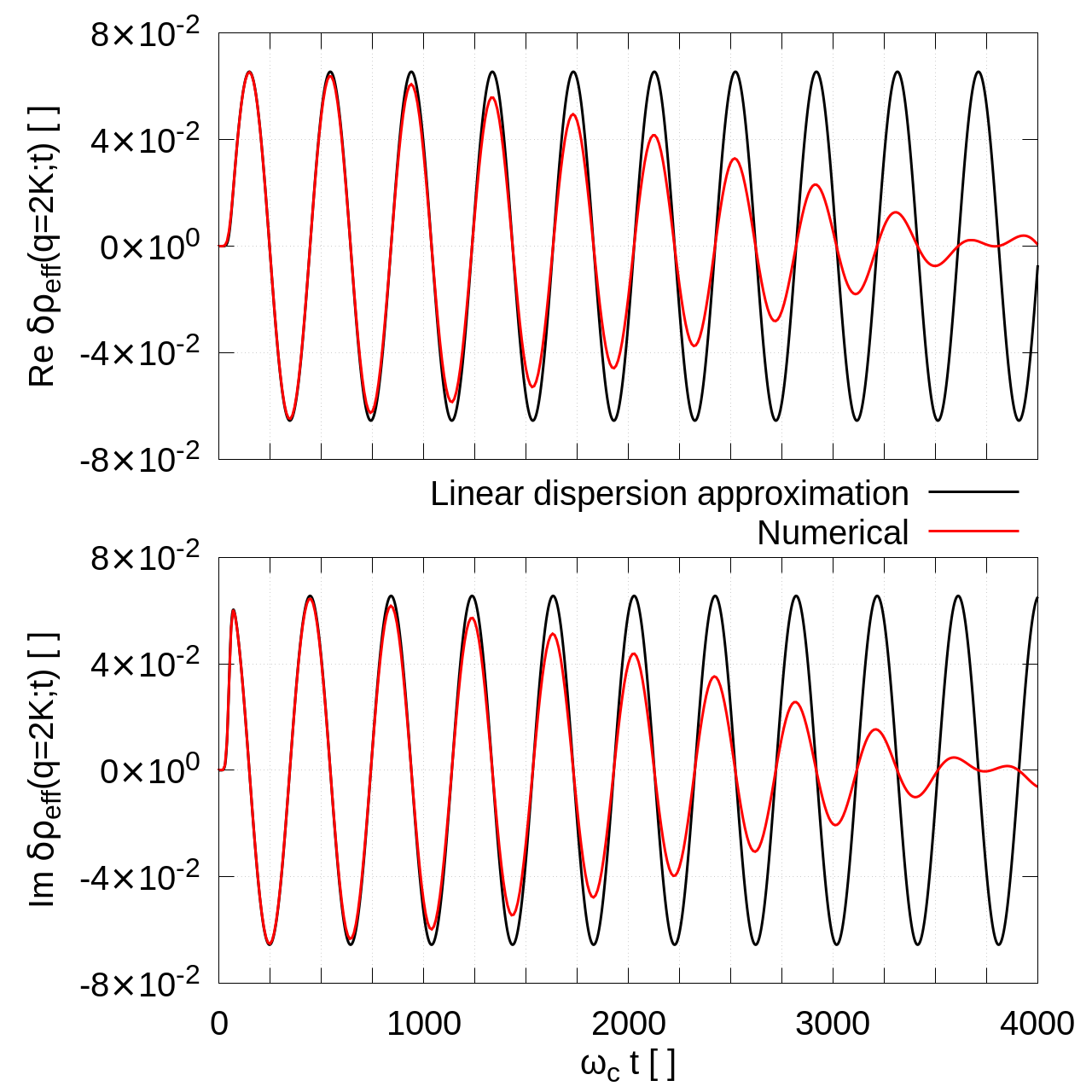}
	\end{minipage}%
	\begin{minipage}{0.5\textwidth}
		\centering
		\includegraphics[width=1.\textwidth]{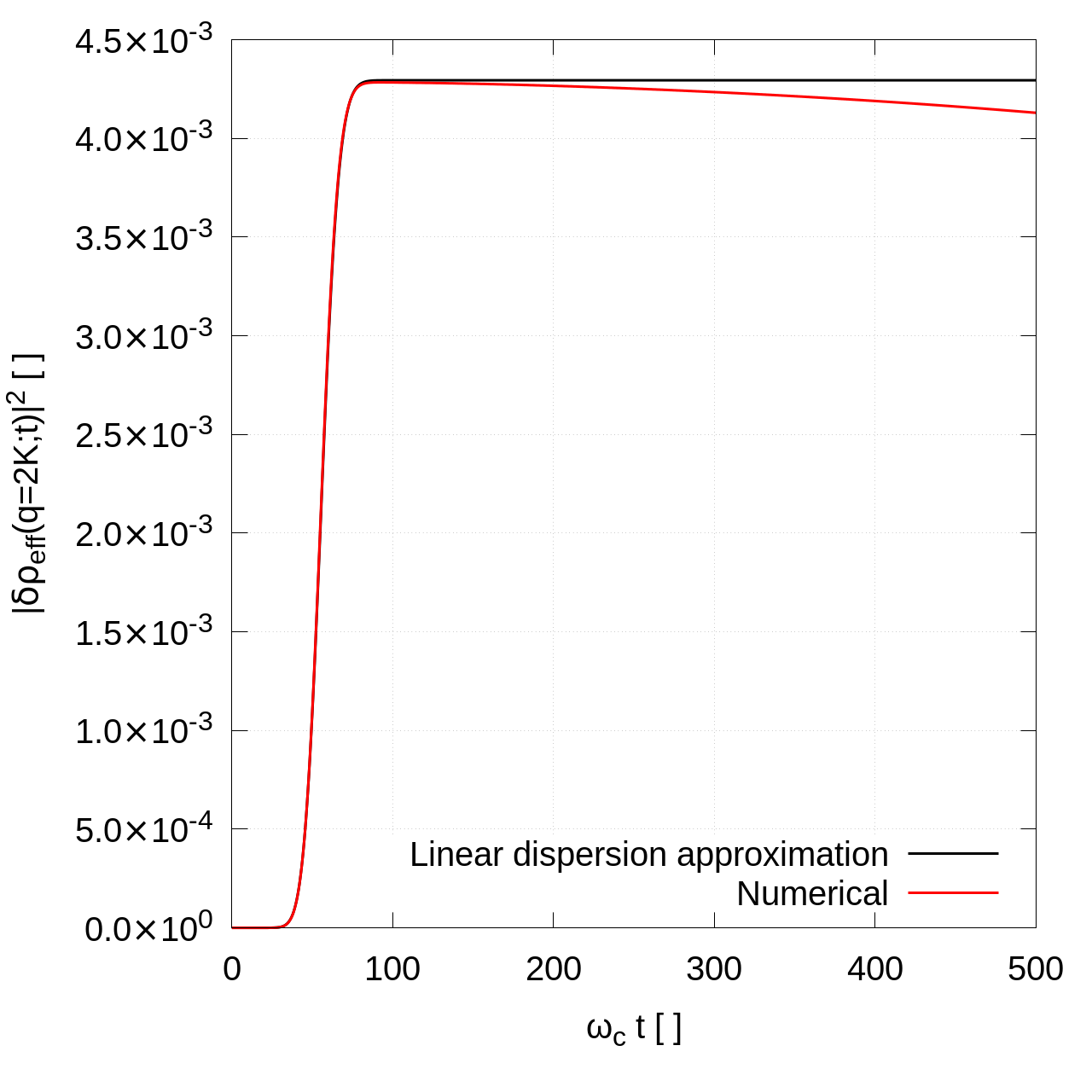}
	\end{minipage}    
	\caption[The LOF caption]{The two images show a comparison between the numerical data and the linear-band result for the effective one-body density. The real part is shown in the top left panel, the imaginary part just below this one and the square modulus in the right panel. The simulation parameters were $L_y=200$, $k_F=273\Delta k\simeq 8.58l_B^{-1}$, $2K=2\Delta k$, $\lambda=0.005\hbar\omega_c$.}
	\label{fig:ComparisonWithPT_linear_dispersion}
\end{figure}

On the other hand when $Kv\tau$ is of order unity (or smaller) $F_t$ will approach at large times a value which is significatively different from zero (again, see the top right hand side panel of Fig. \ref{fig:Ft(qv)_analysis}), so that we successfully managed to excite rigidly travelling modes.
At large times indeed we have
\begin{equation}
\label{eq:maximum_amplitude_sin_linear_approx}
\begin{split}
\left|\frac{1}{L_y}\,\delta \rho_{\text{eff}}(2K, t)\right|^2
&\xrightarrow[t-t_0\gg\tau]{} \frac{\lambda^2}{16\hbar^2} \left(\frac{2K}{2\pi}\right)^2\,
\pi\,\tau^2\,e^{-\frac{1}{2}\left(2Kv\tau\right)^2}.
\end{split}
\end{equation}
As a function of $2K$ the behaviour is not monotonic, rather it is maximized when $2K=(v\tau)^{-1}$; however notice that as the excitation wavevector increases the behaviour predicted by eq. \ref{eq:maximum_amplitude_sin_linear_approx} will break up, since the dispersion relation can not be (in general) safely linearised anymore. 
In order for it to provide quantitatively correct results we need the timescale set by curvature terms to be much longer than that set by the Fermi velocity.

\noindent The results derived assuming the dispersion relation to be perfectly linear will be accurate at short times, but will inevitably fail at larger times though when any small deviation from linearity will start to matter.
This can be seen at a glance from Fig. \ref{fig:ComparisonWithPT_linear_dispersion}. It will be discussed in the next section that curvature terms in the dispersion relation inevitably lead to a decay of the excited mode.

\subsubsection{What if the dispersion relation at the Fermi point is not linear?}\label{section:sin_curvature}
\begin{figure}[htp!]
	\begin{minipage}{.5\textwidth}
		\centering
		\includegraphics[width=1.\textwidth]{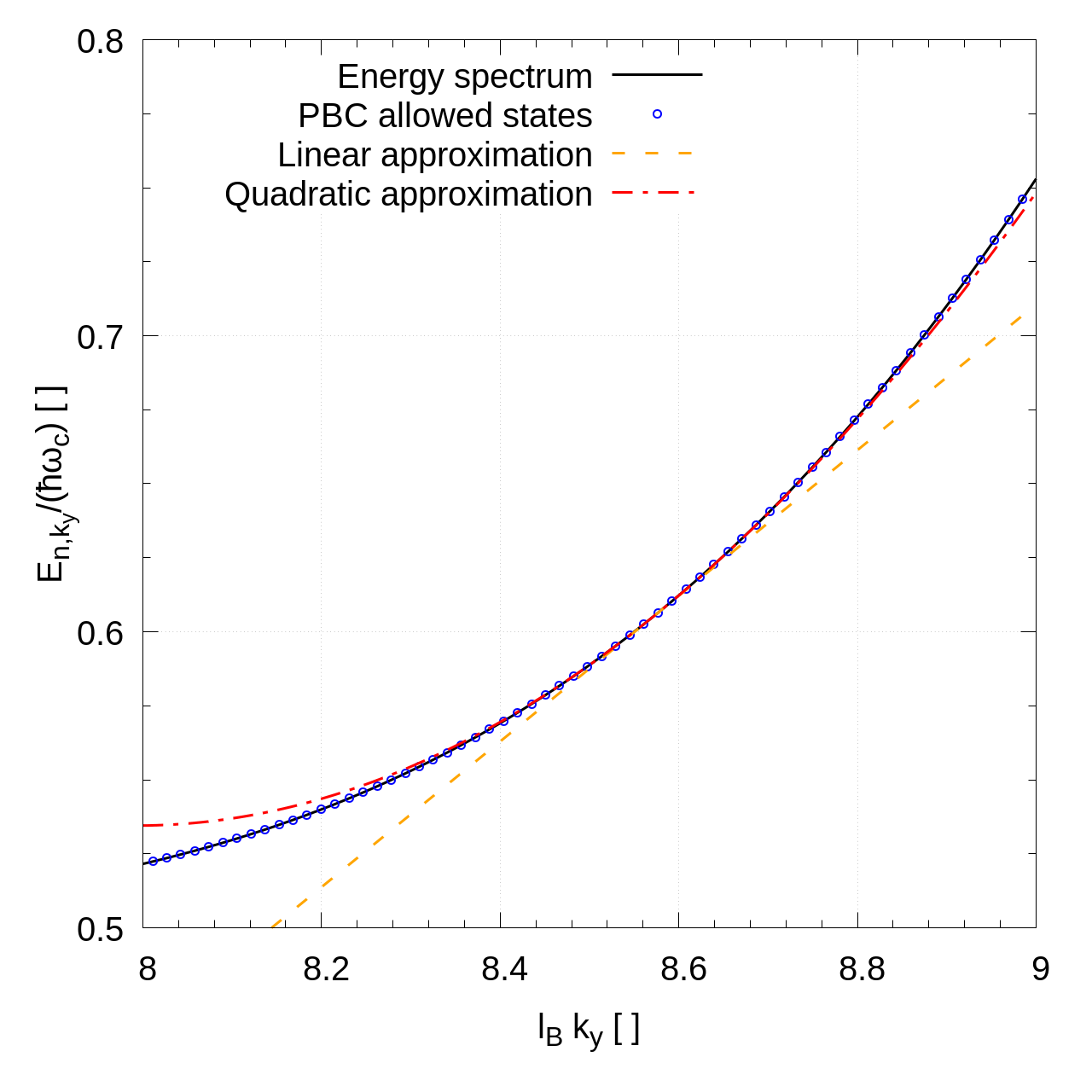}
	\end{minipage}%
	\begin{minipage}{0.5\textwidth}
		\centering
		\includegraphics[width=1.\textwidth]{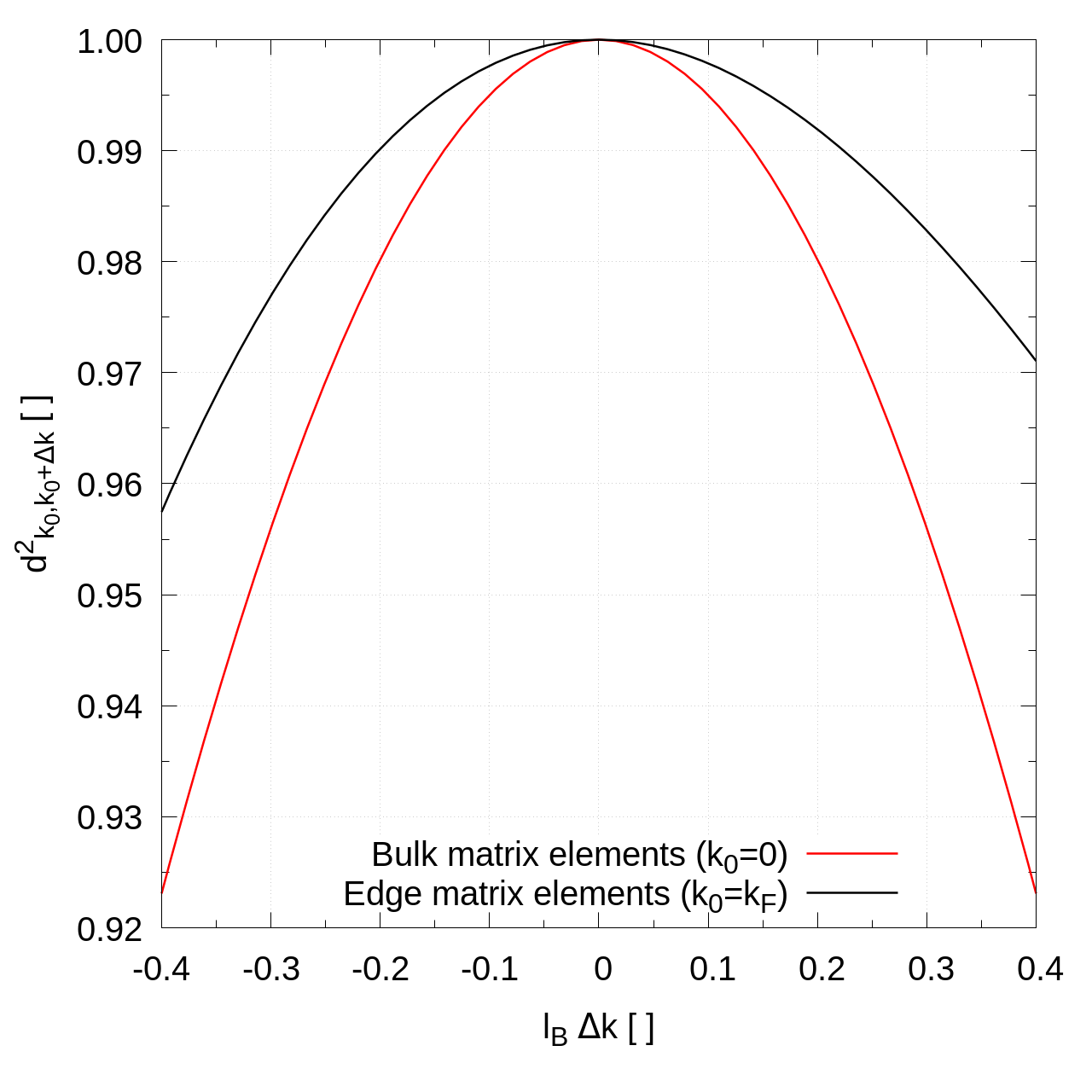}
	\end{minipage}    
	\caption[The LOF caption]{The left hand side plot shows the spectrum of the unperturbed Hamiltonian $\mathcal{H}_0$ near the chosen Fermi point $k_F$, as well as the linear and quadratic Taylor expansions of the energy band about such a point.
	\newline The right hand panel compares the matrix elements $d^2_{k_0,k_0+\Delta k}$ for two different choices of $k_0$, one corresponding to a bulk state and the other one to an edge state.
	\newline For realising the images, $k_F=273\Delta k\simeq 8.58l_B^{-1}$ and $L_y=400l_B$ were used.}
	\label{fig:EnergyBands_TaylorExpansion}
\end{figure}
We have seen in Fig. \ref{fig:ComparisonWithPT_linear_dispersion} that approximating the dispersion as being perfectly linear does not get things quite right at large times; in this section the role of the curvature of the dispersion relation at the Fermi point is highlighted by slightly generalising the calculations made in the previous section.

We want to retrace the steps which lead us to eq. \ref{eq:sum_over_electrons_linear_edge} for the density variation $\delta\rho$ in the linear dispersion case. We saw there that, differently from the bulk case, the cancellation between the contributions arising from the various electrons is not complete near the edges, since in the ground state no electron can be found above the Fermi surface.
We want however to be more general and therefore we do not (for the moment) make any assumption on the structure of the Landau levels.
\newline We start from eq. \ref{eq:pert_sin2} and sum it over all the system electrons; with a little effort one realises that only $\frac{2K}{\Delta k}$ terms do survive the cancellation\footnote{We here focus on just one of the two edges, which is valid under the assumption that the two Fermi surfaces are not directly coupled by the perturbation and can henceforth be treated separately.}, but contrary to the linear dispersion case they will evolve with different frequencies. 
It is only a matter of book-keeping to obtain the following expression for the $q=2K$ component of the $y$-Fourier transform of the density variation $\delta\rho(x,y;t)$
\begin{equation}
\label{eq:pert_sin_general_single_edge_1bd_variation}
\delta \rho(x, 2K; t)=i\,\frac{\lambda}{4\hbar}\sum_{j=0}^{\frac{2K}{\Delta k}-1}\,d_{k_j, k_j+2K}\,F_t(\Delta \omega_{k_j,k_j+2K})e^{i\Delta \omega_{k_j,k_j+2K}\,t}\,\Phi_{k_j}(x)\Phi_{k_j+2K}(x)
\end{equation}
where $k_j=k_F-j\Delta k$. Notice that the $\propto\lambda^2$ correction to this expression vanishes identically, as explained when discussing \ref{eq:lambda2_correction_to_rho2K}. 
\newline If we now integrate along the $x$ direction\footnote{The integral extends from the bulk, where the density variation vanishes, to far beyond one of the edges, where it vanishes too.} and approximate \mbox{$d_{k_j, k_j+2K}^2\simeq 1$} as above (from Fig. \ref{fig:EnergyBands_TaylorExpansion} it is apparent that the approximation works better near the system edges rather than in the bulk, the physical reasons have already been discussed in the previous chapter, and that it stays good as long as the perturbation is long-wavelength), we get
\begin{equation}
\label{eq:pert_sin_general_single_edge_1bd_variation_app1}
\delta \rho_\text{eff}(2K, t)=i\,\frac{\lambda}{4\hbar}\,\sum_{j=0}^{\frac{2K}{\Delta k}-1}\,\,F_t(\Delta \omega_{k_j,k_j+2K})e^{i\Delta \omega_{k_j,k_j+2K}\,t}.
\end{equation}
Notice that for $\frac{K}{\Delta k}=\frac{1}{2}$ and $\frac{K}{\Delta k}=1$ we have a single and two terms in the sum respectively, so that the square modulus $|\delta \rho_\text{eff}(2K, t)|^2$ in the large time region $t\gg t_0$ is time independent in the first case and oscillates as a pure sine-wave in the second, with frequency $\Delta \omega_{k_F,k_F+2\Delta k}-\omega_{k_F-\Delta k, k_F+\Delta k}\sim2\Delta k^2\,\partial_k^2\,\omega|_{k_F}$.  

It is worth looking at the large-time free evolution, where\footnote{Since we also have $t_0\gg\tau$.} (see \ref{eq:ft})
\begin{equation}
\label{eq:LargeTime_ft}
F_{t\gg t_0}(\omega)\simeq\sqrt{\pi}\,\tau\,e^{-\left(\frac{\omega\tau}{2}\right)^2-i\omega t_0}
\end{equation}
since we are then left with a sum of oscillating functions, which will periodically constructively or destructively interfere.
\newline If the dispersion relation can be expanded at quadratic order about the Fermi point\footnote{This will hold if the perturbation is long-wavelength (or \virgolette{not-too-large} $\frac{K}{\Delta k}$).} the summation in \eqref{eq:pert_sin_general_single_edge_1bd_variation_app1} can be nicely written in a closed formula from which many interesting quantities can be computed
\begin{equation}
\omega_k\simeq\omega_{k_F}+v(k-k_F)+\frac{c}{2}(k-k_F)^2+\mathcal{O}(k-k_F)^3
\end{equation}
\begin{figure}[htp!]
	\begin{minipage}{.5\textwidth}
		\centering
		\includegraphics[width=1.\textwidth]{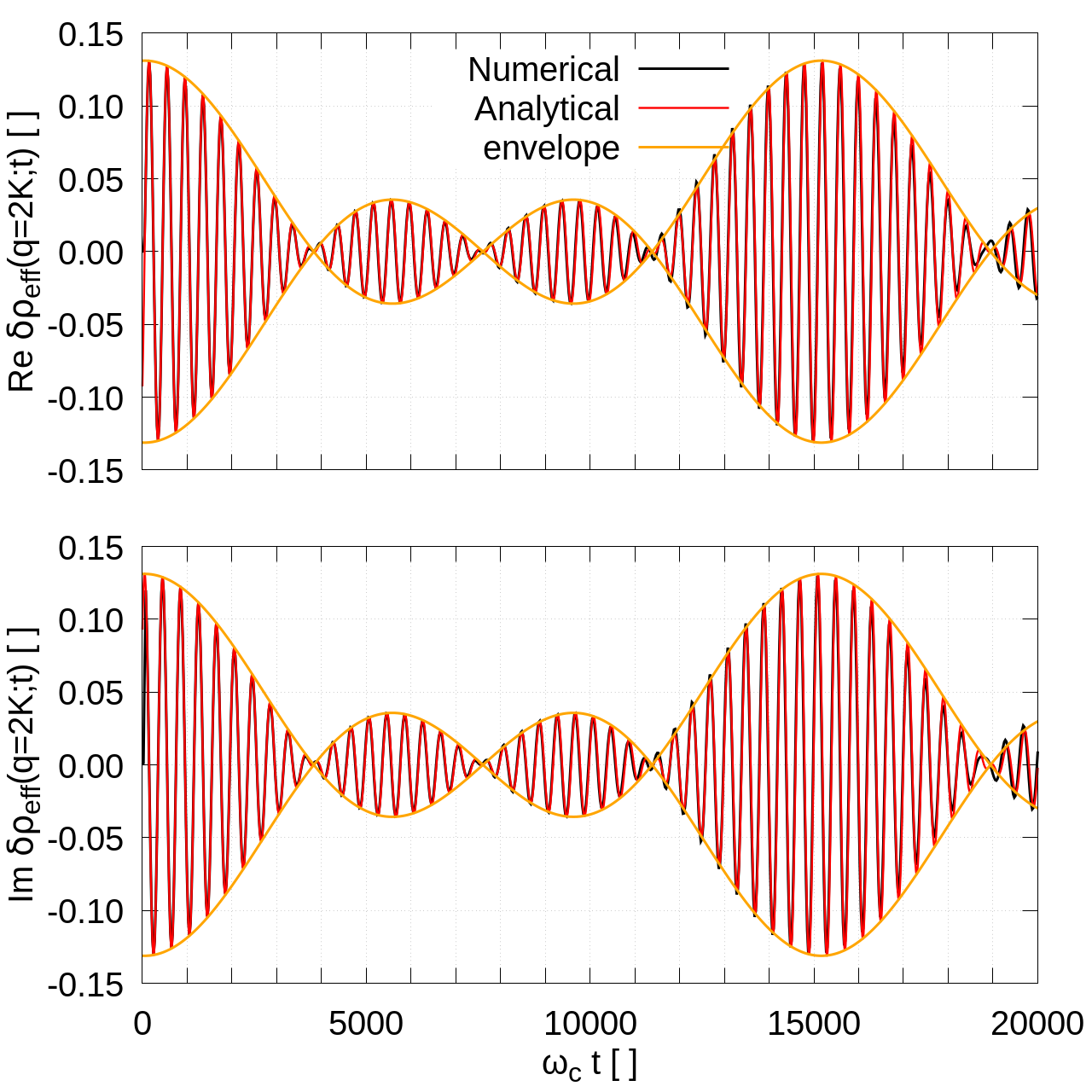}
	\end{minipage}%
	\begin{minipage}{0.5\textwidth}
		\centering
		\includegraphics[width=1.\textwidth]{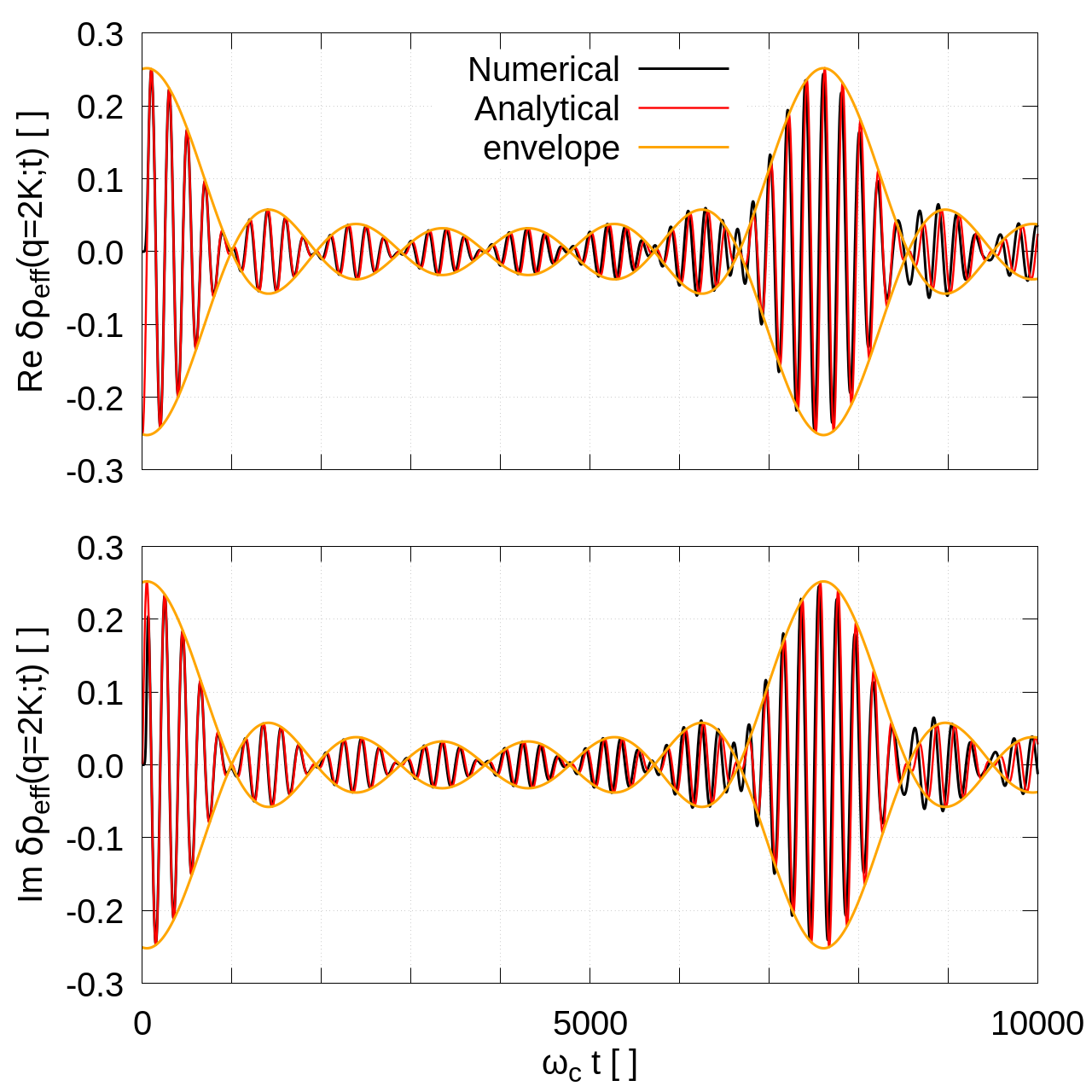}
	\end{minipage}    
	\caption[The LOF caption]{The two figures compare the numerical results for $\delta \rho_\text{eff}(2K, t)$ and the analytical expression in eq. \ref{eq:pert_sin_general_single_edge_1bd_variation_app2}. The real parts are compared in the top panels, the imaginary parts in the bottom ones. 
		\newline I used $L_y=400l_B$, $k_F=546\Delta k\simeq 8.58l_B^{-1}$ and $\lambda=0.005\hbar\omega_c$ in both cases. The left hand side plot compares the results obtained using $2K=4\Delta k=\frac{\pi}{50} l_B^{-1}$; on the right we have $2K=8\Delta k=\frac{\pi}{25} l_B^{-1}$.}
	\label{fig:quadratic_energy_band_approximation_1}
\end{figure}

After some algebra (and neglecting higher order terms when squaring $\Delta \omega$) one finds that
\begin{equation}
\label{eq:pert_sin_general_single_edge_1bd_variation_app2}
\delta \rho_\text{eff}(2K, t\gg t_0)\simeq i\,\frac{\lambda}{4\hbar}\sqrt{\pi}\,\tau\,
e^{-\zeta(t)}
e^{i\left(\frac{2K}{\Delta k}-1\right)\eta(t)}\,\frac{\sin\left(\frac{2K}{\Delta k} \eta(t)\right)}{\sin\left(\eta(t)\right)}
\end{equation}
where
\begin{comment}
\begin{equation}
\begin{cases}
\begin{split}
\zeta(t)&=\frac{1}{4}(2K\, v\,\tau)^2\,\left(1+\frac{c}{v}\,2K\right) + i(2Kv)\left(1+\frac{c}{2 v}\, 2K\right)(t-t_0)
\\&\approx\frac{1}{4}(2K\, v\,\tau)^2 + i(2Kv)\left(1+\frac{c}{2 v}\, 2K\right)(t-t_0)
\end{split}
\\
\begin{split}
\eta(t)&=\frac{c}{2}\,2K\, \Delta k\,(t-t_0)-i\,\frac{1}{4}\Delta k\,\frac{c}{v}\left(2K\,v\,\tau\right)^2
\\&\approx \frac{c}{2}\,2K\, \Delta k\,(t-t_0).
\end{split}
\end{cases}
\end{equation}
\end{comment}
\begin{equation}
\begin{cases}
\zeta(t)=\frac{1}{4}(2K\, v\,\tau)^2 + i(2Kv)\left(1+\frac{c}{2 v}\, 2K\right)(t-t_0)
\\
\eta(t)= \frac{c}{2}\,2K\, \Delta k\,(t-t_0).
\end{cases}
\end{equation}
The ratio between sines defines the (temporal) slowly-varying envelope for this Fourier component of the effective density difference; the complex exponentials on the other hand represent \virgolette{free evolution}, at frequency
\begin{equation}
\omega'_{2K}=2Kv\left(1+\frac{c}{2 v}\, 2K\right) - \frac{c}{2}\,2K\, \Delta k \left(\frac{2K}{\Delta k}-1\right) = 2Kv+\mathcal{O}(\Delta k).
\end{equation}
We see that the frequency at which $\delta\rho_\text{eff}(2K,t)$ oscillates (neglecting the envelope function) is (in the thermodynamics limit $\Delta k\rightarrow 0$) the same we met in the linear dispersion case (see eq. \ref{eq:effective_density_variation_linear_dispersion}). It is curious that the Landau level curvature does not introduce any shift $\propto c$ in this frequency.
\newline Notice that the typical frequency of the envelope is set by $\Omega_2=\frac{c}{2}(2K)^2$, which one could have guessed from scratch since it is just the second order term in the Taylor expansion of the dispersion relation at the Fermi point, using the wavevector characterising the excitation $2K$ in place of $k-k_F$ as an order of magnitude excitation.
\begin{figure}[htp!]
	\begin{minipage}{.5\textwidth}
		\centering
		\includegraphics[width=1.\textwidth]{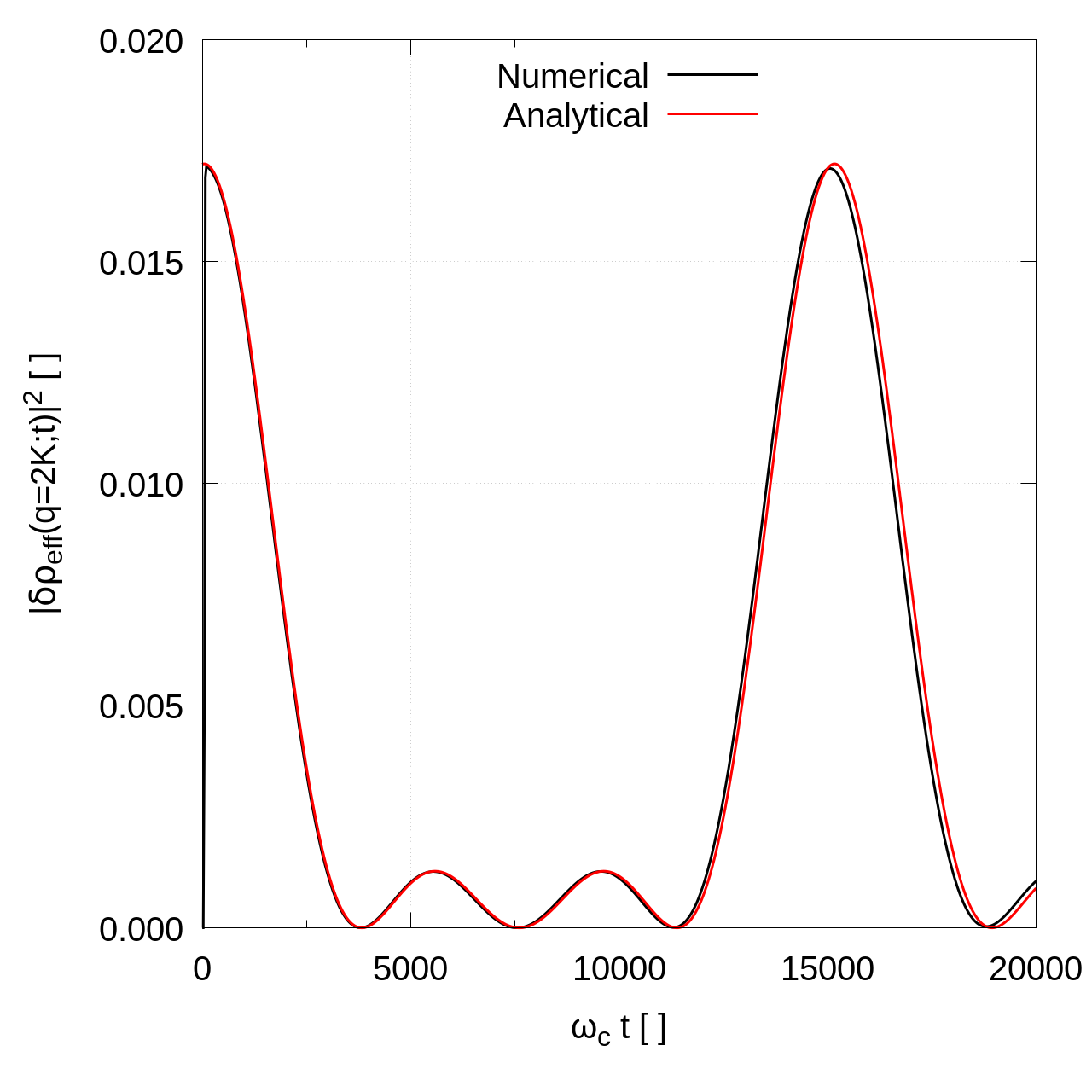}
	\end{minipage}%
	\begin{minipage}{0.5\textwidth}
		\centering
		\includegraphics[width=1.\textwidth]{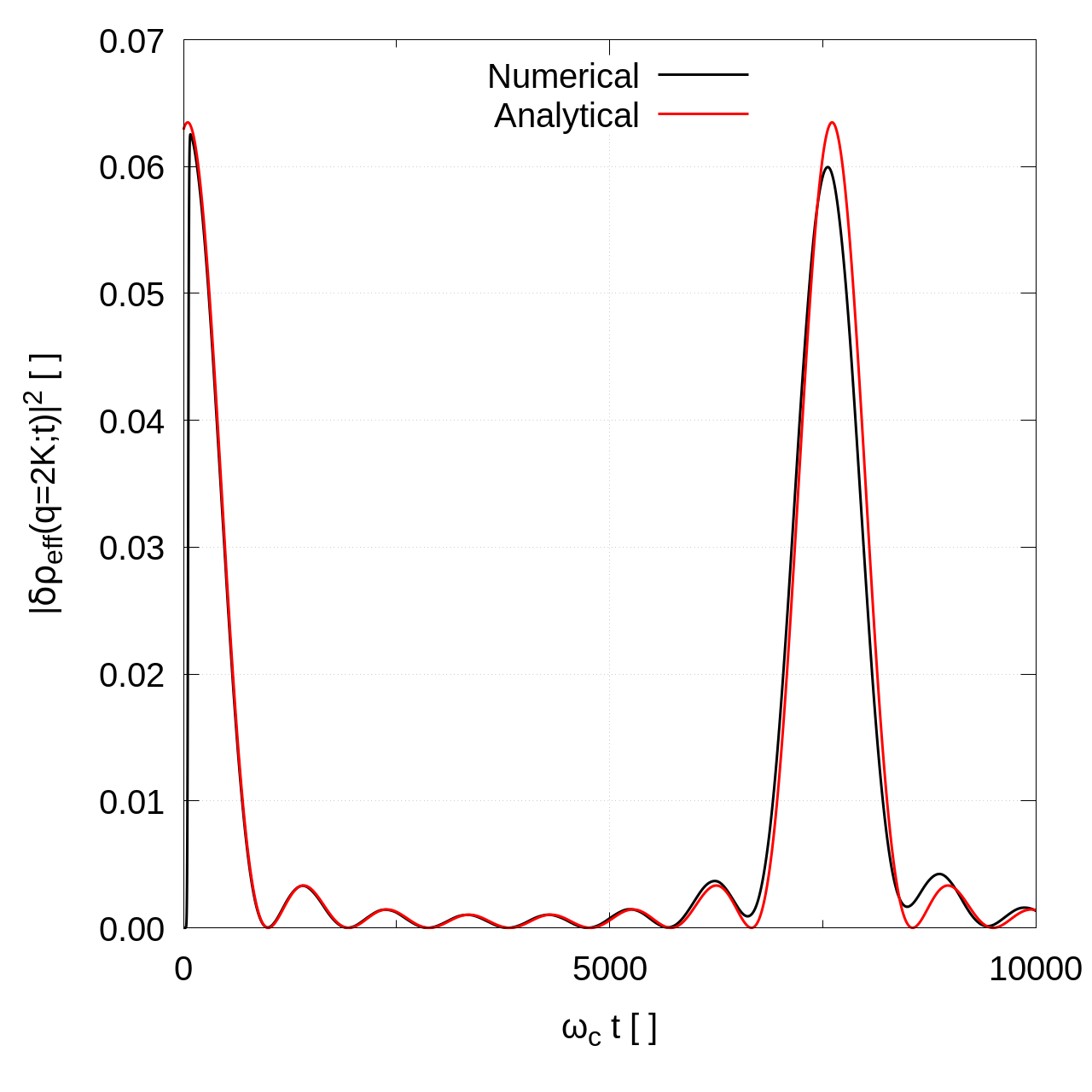}
	\end{minipage}    
	\caption[The LOF caption]{The two figures compare the numerical results for $|\delta \rho_\text{eff}(2K, t)|^2$ and the analytical expression in eq. \ref{eq:pert_sin_general_single_edge_1bd_variation_app2}. \newline I used $L_y=400l_B$, $k_F=546\Delta k\simeq 8.58l_B^{-1}$ and $\lambda=0.005\hbar\omega_c$ in both cases. The left hand side plot compares the results obtained using $2K=4\Delta k=\frac{\pi}{50} l_B^{-1}$; on the right we have $2K=8\Delta k=\frac{\pi}{25} l_B^{-1}$.}
	\label{fig:quadratic_energy_band_approximation_2}
\end{figure}

In Fig. \ref{fig:quadratic_energy_band_approximation_1}
the slowly varying function $\propto\frac{\sin\left(\frac{c}{2}(2K)^2\Delta t\right)}{\sin\left(\frac{c}{2}2K\Delta k\Delta t\right)}$ enveloping the faster \virgolette{zero curvature} oscillation $e^{-i 2K v t}$ is shown in various cases and compared to the numerical results. In Fig. \ref{fig:quadratic_energy_band_approximation_2} the numerical and theoretical results for the square modulus of $\delta\rho_\text{eff}$ are compared instead.
From these figures (\ref{fig:quadratic_energy_band_approximation_1} and  \ref{fig:quadratic_energy_band_approximation_2}) we see that the approximation works better for smaller perturbation wavelengths, i.e. when cubic terms in the dispersion relation are less relevant and thus will not scramble things up at larger and larger times. 
Without much a surprise it can be seen that at short times (when the perturbation is still \virgolette{on}) the analytical expression (eq. \ref{eq:pert_sin_general_single_edge_1bd_variation_app2}) fails to predict the correct response of the system, simply because it was derived by assuming the perturbation was already turned off.

\noindent If we take the square modulus of the relevant Fourier component eq. \ref{eq:pert_sin_general_single_edge_1bd_variation_app2} only the slowly varying envelope survives%\footnote{Compatibly with the approximations made above, one needs to replace $\cosh^2\left[2K\,\frac{c}{v}\,(Kv\tau)^2\right]\simeq1$ and $\sinh^2\left[2K\,\frac{c}{v}\,(Kv\tau)^2\right]\simeq0$ apart for higher order corrections.}
\begin{equation}
|\delta \rho_\text{eff}(2K, t\gg t_0)|^2\simeq \frac{\lambda^2}{16\hbar^2}\pi\,\tau^2\,
e^{-\frac{1}{2}(2K v \tau)^2}
\frac{\sin^2\left[\frac{c}{2}\,(2K)^2(t-t_0)\right]}{\sin^2\left[\frac{c}{2}\,2K\,\Delta k\,(t-t_0)\right]}.
\end{equation}
\begin{comment}
\begin{equation}
\delta \rho_\text{eff}(2K, t\gg t_0)\simeq i\,\frac{\lambda}{4\hbar}\sqrt{\pi}\,\tau\,
e^{-\zeta(t)}
e^{i \frac{\eta(t)}{2}\,\left(\frac{2K}{\Delta k}-1\right)}\,\frac{\sin\left(\frac{\eta(t)\frac{2K}{\Delta k}}{2}\right)}{\sin\left(\frac{\eta(t)}{2}\right)}
\end{equation}
where
\begin{equation}
\begin{cases}
\zeta(t)=\left(K v\tau\right)^2\left(1+\frac{c}{v}\,2K\right)+i\, 2Kv \left(1+\frac{c}{2 v}\, 2K\right)(t-t_0)
\\
\eta(t)=c\, 2K\,\Delta k\,(t-t_0)-i\, 2\Delta k\,\frac{c}{v}\left(Kv\tau\right)^2.
\end{cases}
\end{equation}
If we take the square modulus of the previous expression we can simplify things a bit\footnote{Compatibly with the approximations made above, one needs to replace $\cosh^2\left[2K\,\frac{c}{v}\,(Kv\tau)^2\right]\simeq1$ and $\sinh^2\left[2K\,\frac{c}{v}\,(Kv\tau)^2\right]\simeq0$ apart for higher order corrections.}
\begin{equation}
|\delta \rho_\text{eff}(2K, t\gg t_0)|^2\simeq \frac{\lambda^2}{16\hbar^2}\pi\,\tau^2\,
e^{-2(K v \tau)^2\left(1+\frac{c}{v}\,\Delta k \right)}\frac{1}{L_y^2}
\frac{\sin^2\left[\frac{c}{2}\,(2K)^2(t-t_0)\right]}{\sin^2\left[\frac{c}{2}\,2K\,\Delta k\,(t-t_0)\right]}.
\end{equation}
\end{comment}
From this linear-order expression many things can be predicted.

\begin{figure}[htp!]
	\begin{minipage}{.5\textwidth}
		\centering
		\includegraphics[width=1.\textwidth]{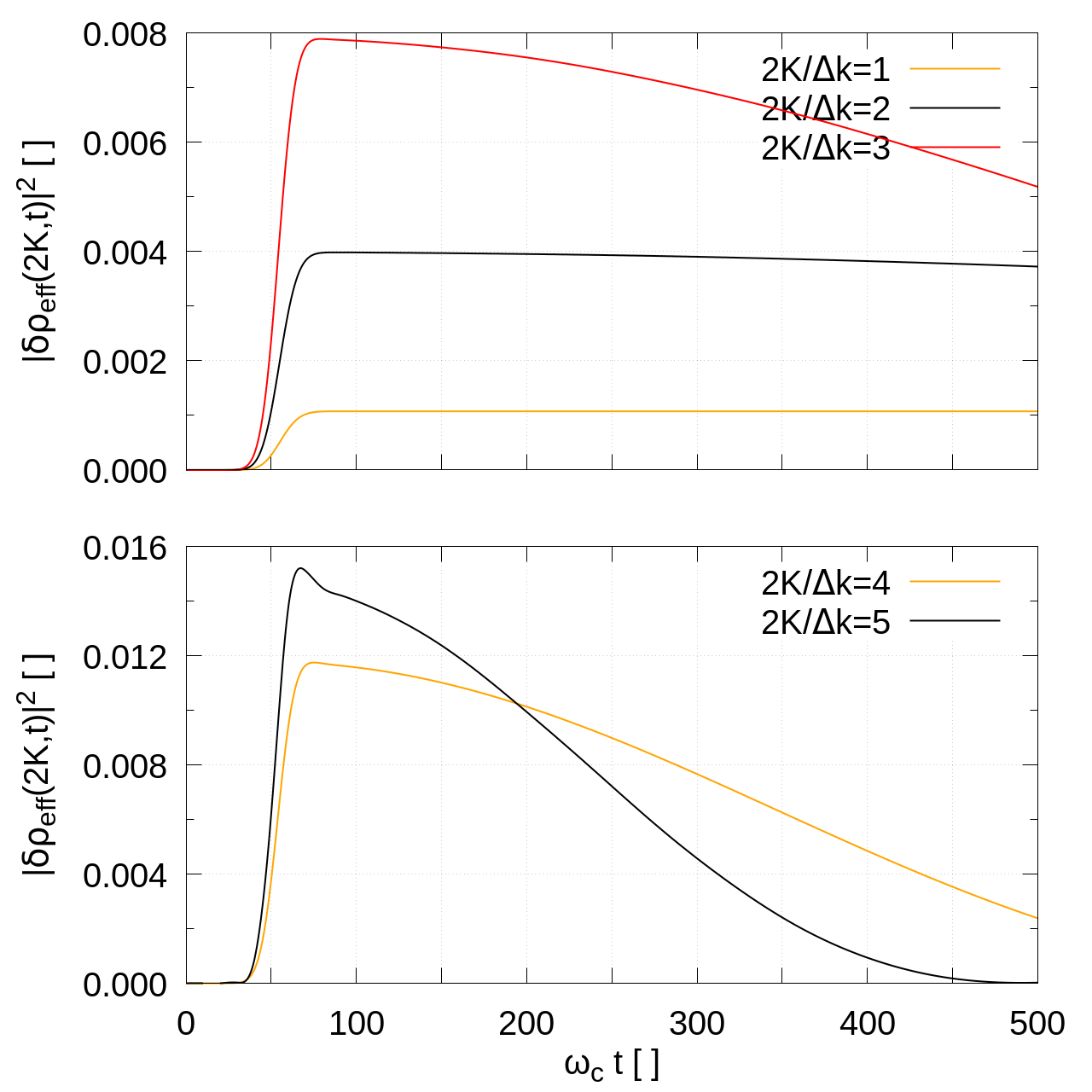}
	\end{minipage}%
	\begin{minipage}{0.5\textwidth}
		\centering
		\includegraphics[width=1.\textwidth]{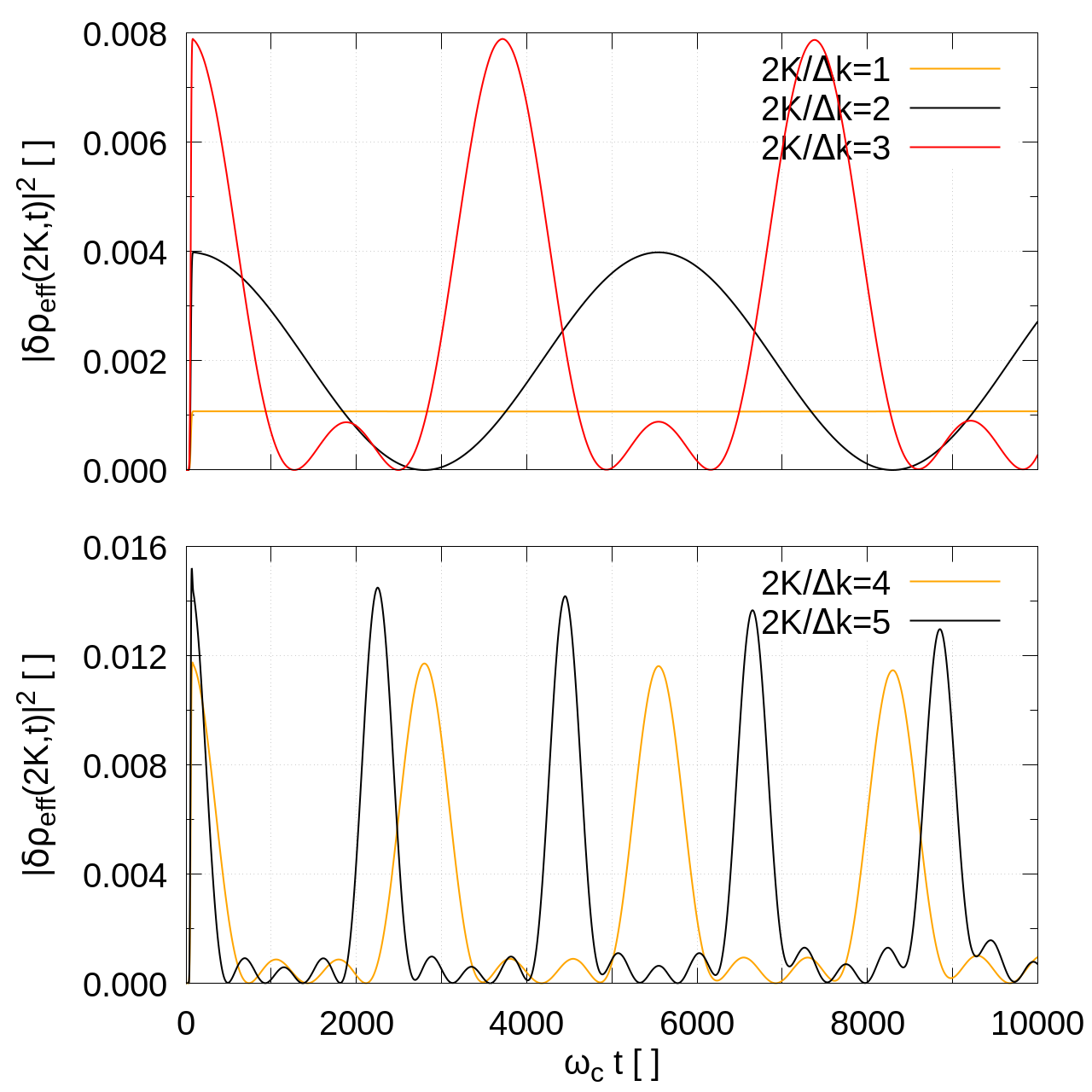}
	\end{minipage}    
	\caption[The LOF caption]{Both the images show the numerically computed square modulus of the effective density $|\delta\rho_\text{eff}(2K,t)|^2$ as a function of time, for different values of the excitation wavevector. The plot on the left zooms onto the time evolution at short times, while the one on the left looks to longer times. $L_y=200l_B$, $k_F=287\Delta k\simeq 9.02l_B^{-1}$ and $\lambda=0.005\hbar\omega_c$ were used.}
	\label{fig:comparison_different_k}
\end{figure}

First of all, if $K=\frac{1}{2}\Delta k$, $|\delta \rho_\text{eff}(2K, t\gg t_0)|^2$ will not depend on time at large time, but approaches a constant value $\frac{\lambda^2}{16\hbar^2}\pi\,\tau^2\,\frac{1}{L_y^2}\,
e^{-\frac{1}{4}(2K v \tau)^2}$ instead. For $K=\Delta k$, we will have a purely harmonic time dependence, as the ratio between squared sines simplifies to $4\cos^2\left[c\,\Delta k^2(t-t_0)\right]$. These results were indeed expected from just looking at eq. \ref{eq:pert_sin_general_single_edge_1bd_variation_app1}.
\newline Interesting things start to occur for greater $\frac{K}{\Delta k}$ ratios, since more electrons start to interfere.
The denominator oscillates at a frequency which is $\frac{\Delta k}{2K}$ times smaller than the one at which the numerator oscillates. Since the ratio becomes large when the denominator approaches zero, we will periodically see some higher spikes (constructive interference) interspersed with smaller side lobes (destructive); see Fig. \ref{fig:comparison_different_k}.

\begin{figure}[htp!]
	\begin{minipage}{.5\textwidth}
		\centering
		\includegraphics[width=1.\textwidth]{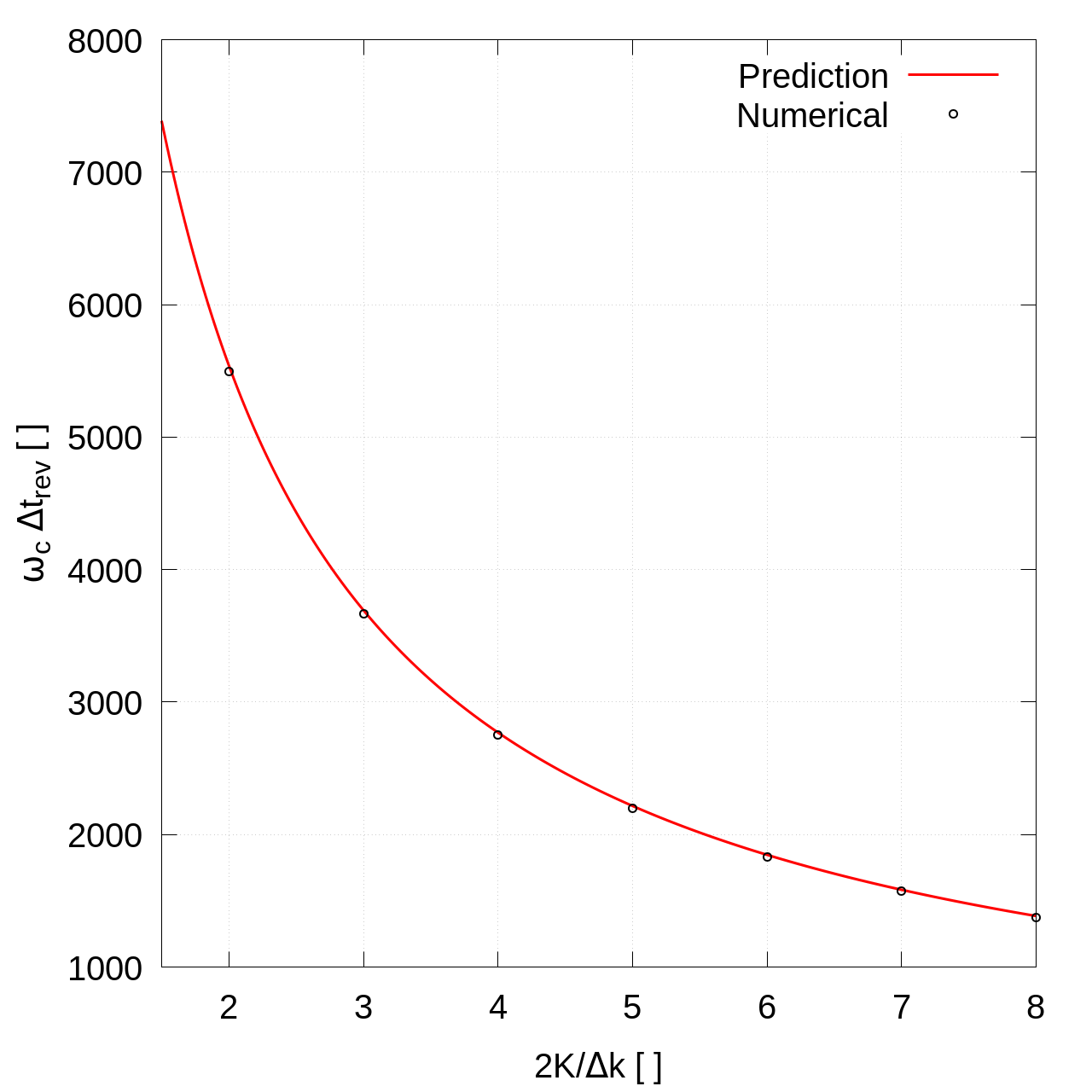}
	\end{minipage}%
	\begin{minipage}{0.5\textwidth}
		\centering
		\includegraphics[width=1.\textwidth]{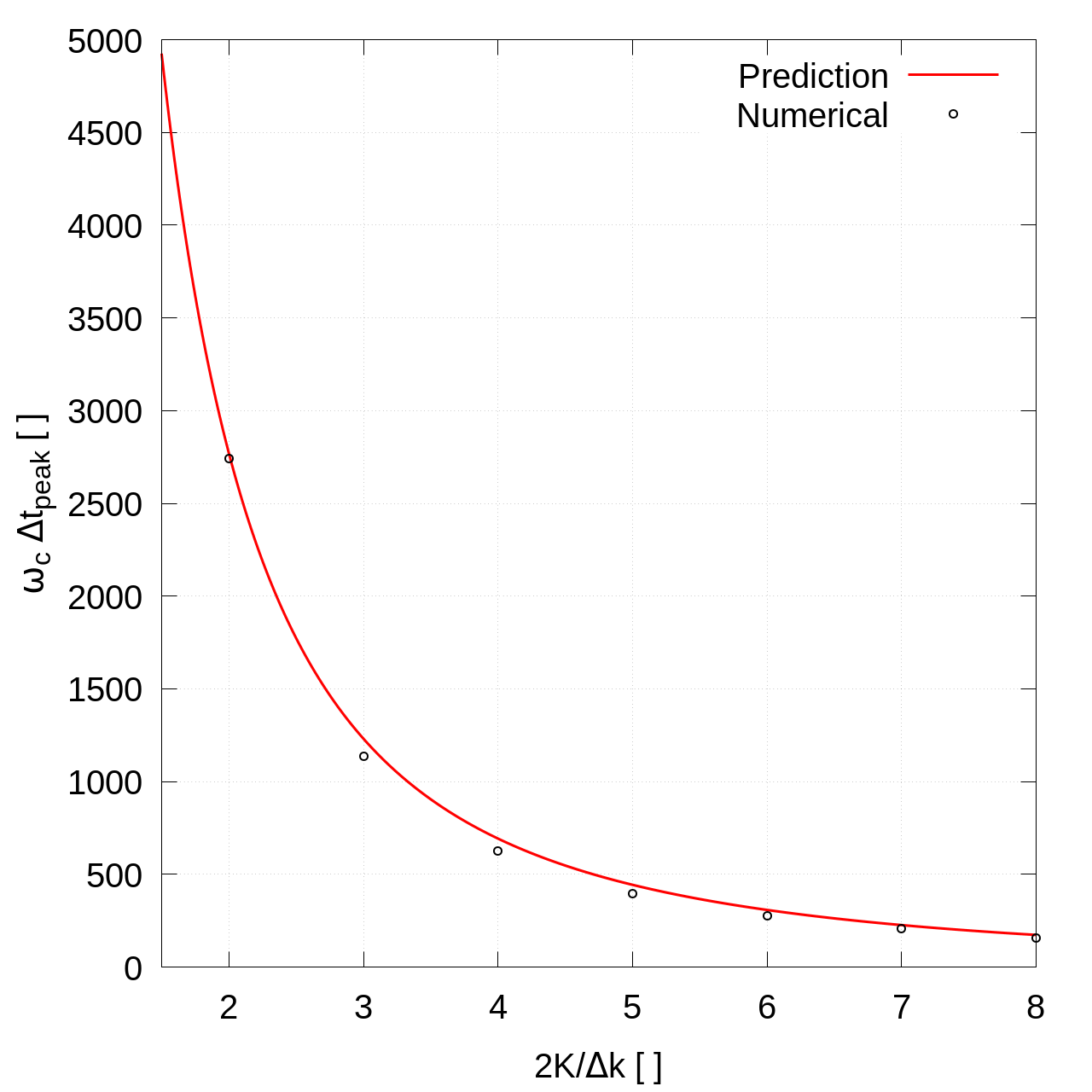}
	\end{minipage}    
	\caption[The LOF caption]{The two images compare the numerically computed revival time (on the left) and the full width at half maximum of the first \virgolette{revived} peak (right) with the theoretical expressions eq. \ref{eq:sin_deltaT_rev} and eq. \ref{eq:sin_deltaT_peak}.
		\newline $L_y=200$, $k_F=287\Delta k\simeq9.02l_B^{-1}$ and $\lambda=0.005\hbar\omega_c$ have been used.}
	\label{fig:decay_and_revival}
\end{figure}

The revival time can easily be computed since two consecutive spikes occur when the denominator gains a $\pi$ phase difference
\begin{equation}
\label{eq:sin_deltaT_rev}
\Delta t_\text{rev}=\frac{2\pi}{c\,2K\Delta k}\propto\frac{1}{K}.
\end{equation}
$\Delta t_\text{rev}$ has been numerically computed for various value of the excitation wavevector and compared to the above equation in Fig. \ref{fig:decay_and_revival} (left hand side panel).
\newline The width of the spikes can be estimated as well. $h(t)=\frac{\sin^2(M \omega t)}{\sin^2(\omega t)}$ has period $\frac{\pi}{\omega}$, and its integral over the period evaluates to $\int_{-\frac{\pi}{2\omega}}^{\frac{\pi}{2\omega}}\,h(t)=M\,\frac{\pi}{\omega}$. Since the maximum of such a function within the period is $M^2$, and the largest part of the area is concentrated below such a spike, the width of the peak will be of the order of\footnote{This is actually an overestimation since the area below the side-lobes is not fully negligible; this approximation however gives an order of magnitude estimation.} $\frac{\pi}{M \omega}$. In our case
\begin{equation}
\label{eq:sin_deltaT_peak}
\Delta t_\text{peak} \lesssim \frac{2\pi}{c\,\left(2K\right)^2}\propto\frac{1}{K^2}.
\end{equation}
This behaviour is indeed correct as can be seen from Fig. \ref{fig:decay_and_revival}.

\begin{figure}[htp!]
	\begin{minipage}{.5\textwidth}
		\centering
		\includegraphics[width=1.\textwidth]{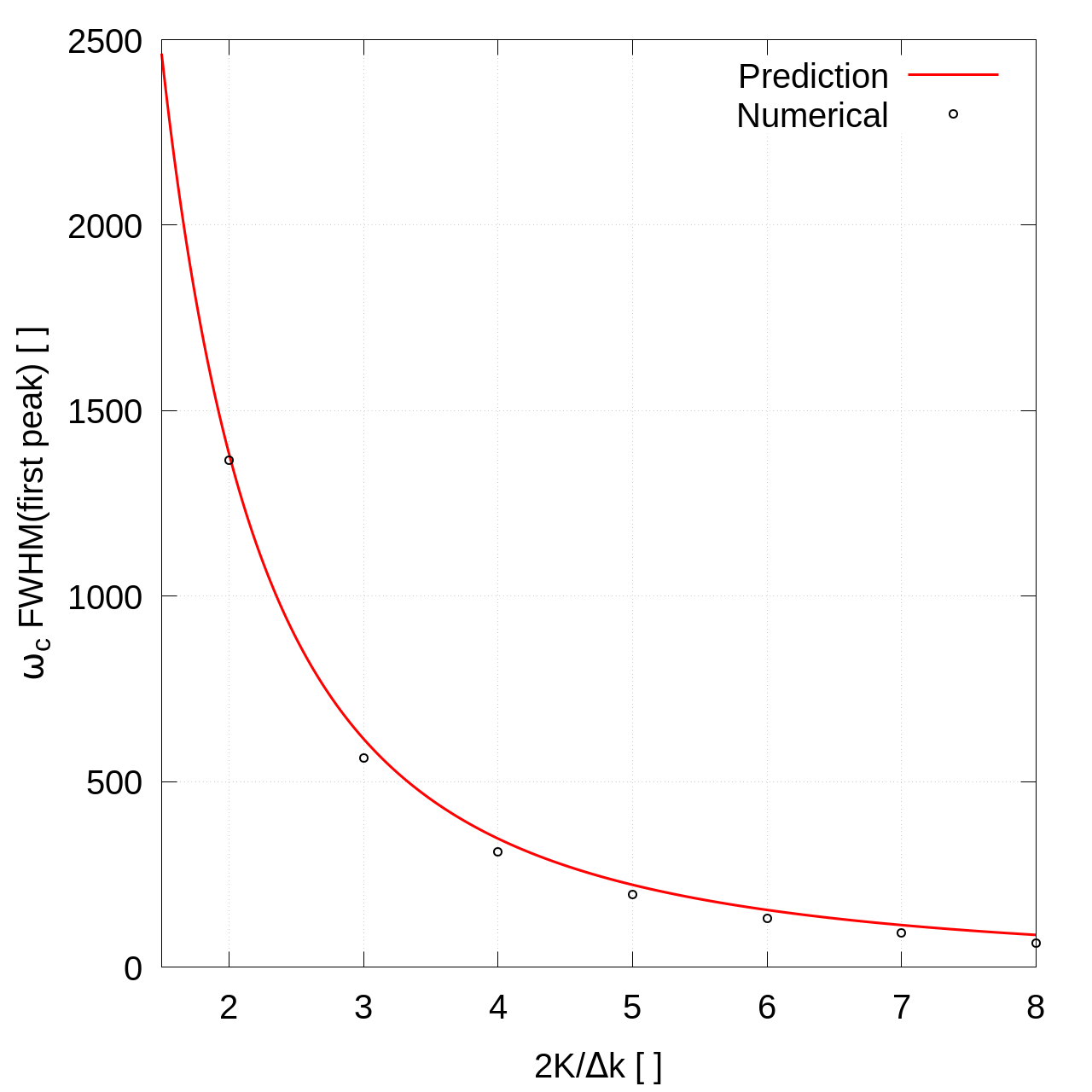}
	\end{minipage}%
	\begin{minipage}{0.5\textwidth}
		\centering
		\includegraphics[width=1.\textwidth]{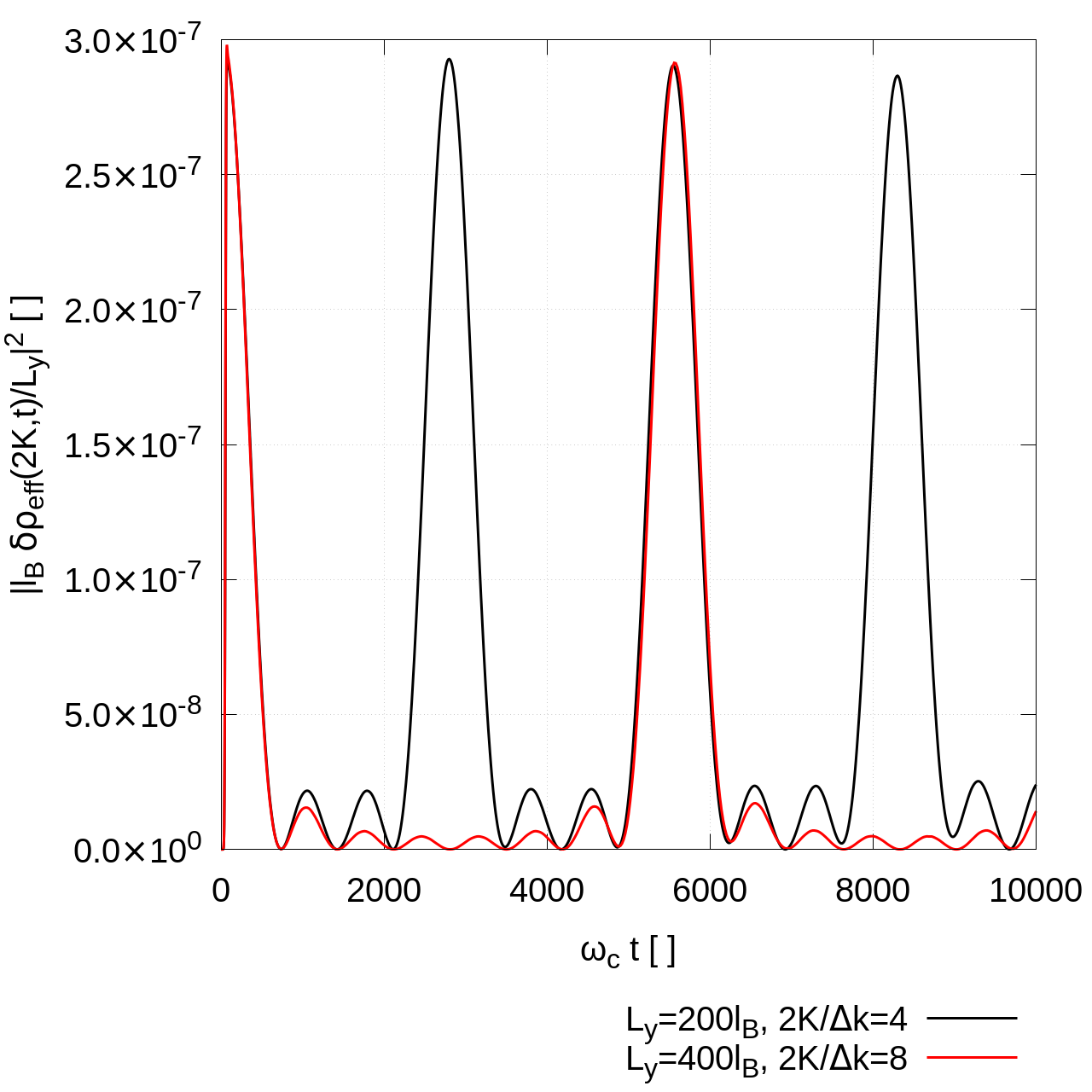}
	\end{minipage}    
	\caption[The LOF caption]{The left hand side image compares the numerically computed full width at half maximum of the first peak with the order of magnitude estimation $\Delta t_\text{decay}\sim\frac{1}{2}\Delta t_\text{peak}\sim\frac{\pi}{c(2K)^2}$. $L_y=200$, $k_F=287\Delta k\simeq9.02l_B^{-1}$ and $\lambda=0.005\hbar\omega_c$ have been used.
		\newline On the right $|\delta\rho_\text{eff}(2K,t)|^2$ is plotted as a function of time, for different system size lengths $L_y$ but fixed perturbation wavevector $2K=\frac{\pi}{25} l_B^{-1}$.}
	\label{fig:thermodynamic_limit}
\end{figure}

Finally, the number of lobes between two spikes can be computed. The period of the numerator of $h(t)$ is $\frac{2\pi}{M\omega}$, the one of the denominator $\frac{2\pi}{\omega}$. The denominator period exactly fits $M$ numerator periods in; the number of maxima within a period will thus be $M$, but two of these do correspond to the spikes. There will consequently be $M-2$ lobes. In our case
\begin{equation}
N_\text{l} = \frac{2K}{\Delta k}-2
\end{equation}
as can be seen in Fig. \ref{fig:comparison_different_k} and from the right hand side panel of Fig. \ref{fig:thermodynamic_limit}.

It is also interesting to see what happens if we keep the perturbation wavevector $K$ fixed, taking the large sample-size $L_y\rightarrow\infty$. If such a limit is taken, the revival time (eq. \ref{eq:sin_deltaT_rev}) blows up, while the spike width  (eq. \ref{eq:sin_deltaT_peak})  is independent of the sample size. For times $t_0\ll t\ll \frac{1}{c\,2K\,\Delta k}\propto L_y$, we may approximate eq. \ref{eq:pert_sin_general_single_edge_1bd_variation_app2}\footnote{The same result could have more straightforwardly been obtained by directly passing in the thermodynamic limit $L_y\rightarrow\infty$ which allows to replace the summation with an integral $\frac{1}{L_y}\sum_{j=0}^{\frac{2K}{\Delta k}-1}G(k_j)\rightarrow-\frac{1}{2\pi}\int_{k_F}^{k_F-2K}G(k')dk'$.}
\begin{equation}
\frac{1}{L_y}\delta \rho_\text{eff}(2K, t\gg t_0)\simeq i\,\frac{\lambda\tau}{8\sqrt{\pi}\hbar}\,
2K\,e^{-\frac{1}{4}(2K\, v\,\tau)^2}
e^{-i 2K v \Delta t}\,\text{sinc}\left(\frac{c}{2}\,(2K)^2\,\Delta t\right)
\end{equation}
so we see that if curvature effects are not negligible the excited mode decays slowly in time $\propto \frac{1}{(t-t_0)}$.
The square modulus
\begin{equation}
\left|\frac{1}{L_y}\delta \rho_\text{eff}(2K, t\gg t_0)\right|^2\simeq \frac{\lambda^2\tau^2}{64\pi\hbar^2}\,
(2K)^2e^{-\frac{1}{2}(2K v \tau)^2}
\text{sinc}^2\left(\frac{c}{2}\,(2K)^2(t-t_0)\right)
\end{equation}
Notice that in the limit $c\rightarrow0$ (when the band curvature vanishes identically) we get back the previous results since $\text{sinc}\left(\frac{c}{2}\,(2K)^2(t-t_0)\right)\rightarrow1$: the excited mode propagates undamped. 
Notice that the curvature parameter $c$ only appears in the argument of the $\text{sinc}$ function, so that it only stretches it out without otherwise affecting its shape and functional behaviour.

The spike width computed in eq. \ref{eq:sin_deltaT_peak} gives a rough order of magnitude estimation for the decay time
\begin{equation}
\label{eq:decay_time}
\Delta t_\text{decay}\sim\frac{1}{2}\Delta t_\text{peak}\sim\frac{\pi}{c(2K)^2}.
\end{equation}
In the left hand side panel of Fig. \ref{fig:thermodynamic_limit} we qualitatively recognize the predicted behaviour. 

\subsubsection*{Beyond the quadratic order}
\begin{figure}[htp!]
	\centering
	\includegraphics[width=.8\textwidth]{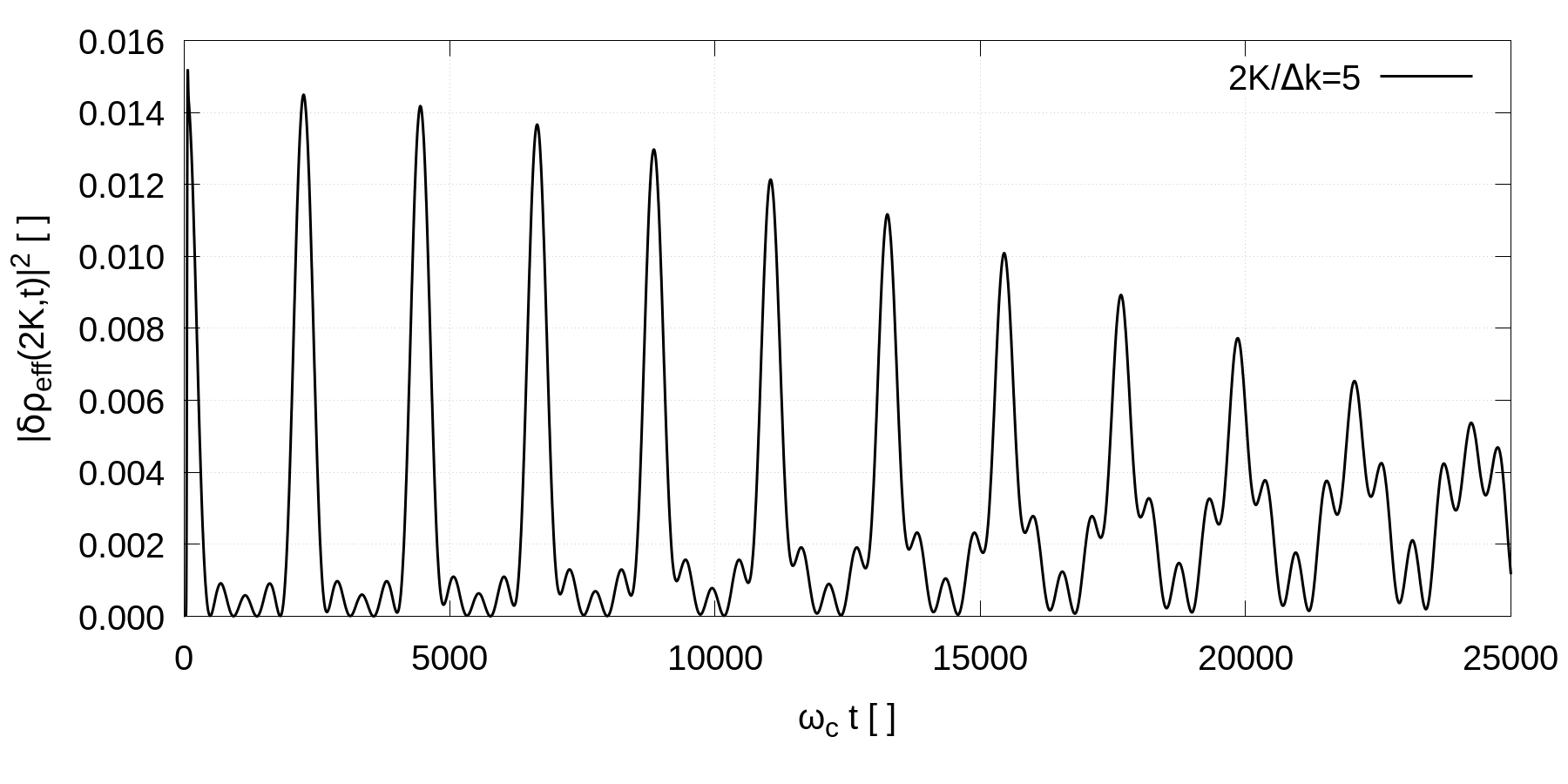}
	\caption[The LOF caption]{The image plots $|\delta\rho_\text{eff}(2K,t)|^2$ as a function of time and makes manifest the importance of correction terms in the dispersion relation beyond the quadratic one on the very long time evolution.  This is not a big deal though, as long as we are interested in taking the thermodynamic limit and look at the initial region alone, since terms beyond the quadratic order will not have much influence over such a short timescale.
		\newline For the image, $L_y=200l_B$, $k_F=287\Delta k\simeq 9.02l_B^{-1}$, $2K=5\Delta k$ and $\lambda=0.005\hbar\omega_c$ were used.
		\newline Using the numerical data, the evolution given by the curvature term can be estimated; its typical evolution frequency is set by $\Omega_2=\frac{c}{2}(2K)^2\approx7\times10^{-3}\omega_c$, much shorter than the time evolution set by third derivative terms, which have the typical frequency $\Omega_3=\frac{\partial_k^3\omega_{k_F}}{3}(2K)^3\approx1.9\times10^{-4}\omega_c$. The timescales $\Omega_2^{-1}$ and $\Omega_3^{-1}$ can easily be recognized in the image.}
	\label{fig:long_time_mess}
\end{figure}

The relevance of the dispersion relation coefficients at the Fermi point beyond the quadratic order will now be qualitatively discussed.
\newline Notice that by doubling the system length $L_y$ (and keeping the perturbation wavevector fixed) the revival time should exactly double, since $\Delta t_\text{rev}\propto L_y$. This is not perfectly the case, as can indeed be seen in the right hand side panel of Fig. \ref{fig:thermodynamic_limit}. The error is not numerical; rather it resides in having neglected terms beyond the quadratic one in the dispersion relation in order to derive the above expressions.
If terms beyond the quadratic one are indeed relevant, at some large time everything will eventually mess up (de-phase), as can be seen in Fig. \ref{fig:long_time_mess}. 

\noindent The typical time over which these effects will take place can easily be estimated by generalising the discussion about $\Omega_2$ made above: we expect such a timescale to be given by $\Omega_3^{-1}$, with $\Omega_3=\frac{\partial_k^3\omega_{k_F}}{3}(2K)^3$ and $\partial_k^3\omega_{k_F}$ being the third order derivative of the energy dispersion evaluated at the Fermi point.
\newline We are ultimately interested in taking the thermodynamic limit though, so that these effects will not affect the discussion made above, at least when the wavevector of the perturbation is small enough so that $\Omega_3^{-1}\gg\Omega_2^{-1}$.

As a final comment, notice that at the lowest perturbative order the only excited mode is the one with the same wavevector as the external perturbation. In this case the real space one-body density $\rho(y)$ while evolving in time will not change its shape but rather only decay while propagating, since at linear order higher modes are not excited.
If non-linear corrections become relevant (i.e. higher harmonic terms get excited) this obviously ceases to be true: during the decay the sinusoidal shape will deform.
The non-linear response $\delta\rho_\text{eff}(4K,t)$ will be the argument of section \ref{section:sin_hhg}; in the next section the non-linear corrections to $\delta\rho_\text{eff}(2K,t)$ are numerically discussed instead.

\subsection{Increasing the excitation strength: numerical results beyond perturbation theory}
In this section the non-linear effects on the $q=2K$ component of the Fourier transform of the effective density variation are briefly discussed by looking at the numerical results obtained with the time evolution algorithm \ref{eq:ADI}. Recall that these corrections are $\propto\lambda^3$.

\begin{figure}[htp!]
	\begin{minipage}{.5\textwidth}
		\centering
		\includegraphics[width=1.\textwidth]{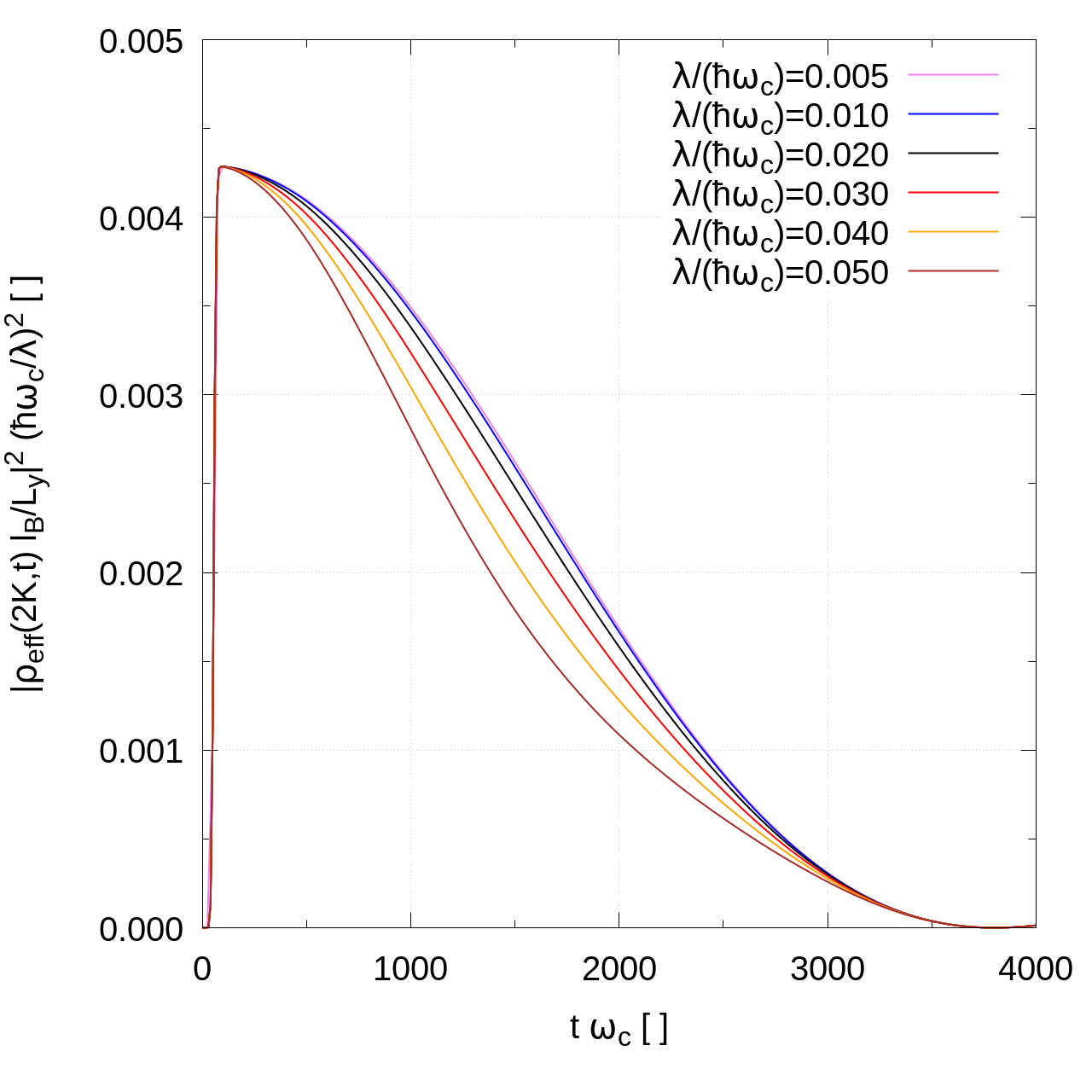}
	\end{minipage}%
	\begin{minipage}{0.5\textwidth}
		\centering
		\includegraphics[width=1.\textwidth]{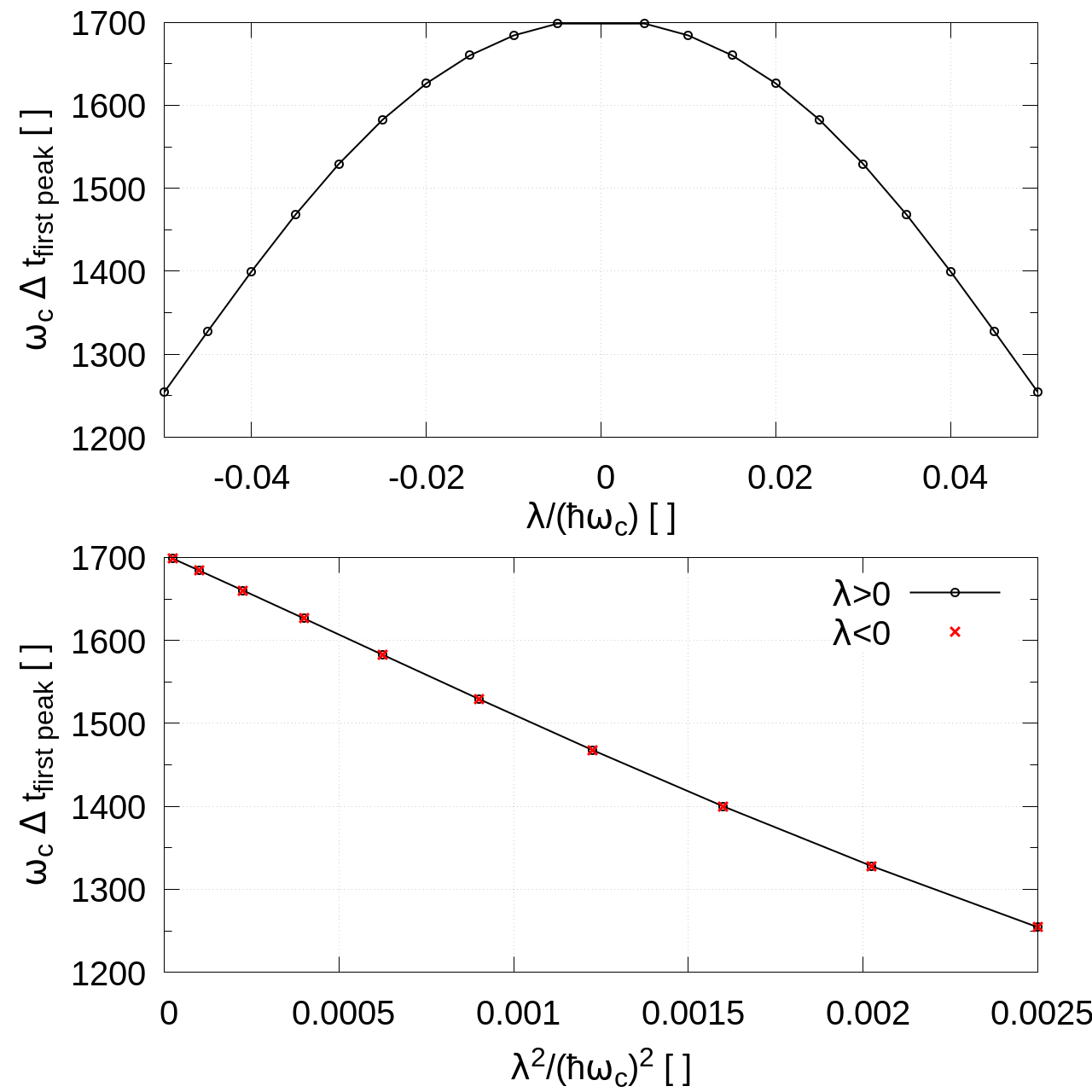}
	\end{minipage}    
	\caption[The LOF caption]{On the left I show the short time evolution (the excitation decay) of $|\delta\rho_\text{eff}(2K,t)|^2$ divided by $\lambda^2$ for different values of the excitation strength $\lambda$. The image on the right compares the full width at half maximum of such a peak, as a function of $\lambda$ (top) and $\lambda^2$ (below), both for positive and negative values of $\lambda$.
		\newline $L_y=400l_B$, $k_F=546\Delta k\simeq8.58 l_B^{-1}$ and $2K=4\Delta k$ were used.}
	\label{fig:decay_lambda}
\end{figure}

Since higher order transitions are more likely to occur for bigger values of $\lambda$, as this parameter is increased more terms oscillating at different frequencies need to be added up; these terms will inevitably interfere so we expect a faster decay of the excitation, as can indeed be seen in the left-hand side panel of Fig. \ref{fig:decay_lambda}.
\newline The behaviour of the decay time as a function of $\lambda$ can be easily predicted with some heuristic argument. Since the perturbation is periodic if we change $\lambda$ with $-\lambda$ the $y$-dependent part of the perturbation $\propto \cos(2K y)$ physically represents the same kind of excitation. If it was not for the $y$ independent part of the potential, $\mathcal{H}(\lambda)$ and $\mathcal{H}(-\lambda)$ would indeed be related by a simple translation along $y$. Since the constant part of the potential only couples different Landau levels we expect the typical width over which the excitation induced in the system decays to zero to be an even function of $\lambda$, and thus to decrease $\propto \lambda^2$ as $\lambda$ increases. This is approximatively the case, as can be seen in the image in the panel on the right of Fig. \ref{fig:decay_lambda}; one could notice a small deviation from a straight line, which may arise from higher order $\propto \lambda^4$ correction (as can be seen from the image the fact that the width is an even function of $\lambda$ has been directly numerically checked with good accuracy).
\newline Notice that we could have also expected the corrections to the first peak width to vary $\propto\lambda^2$ from the fact that corrections to the linear order expression for $\delta\rho_\text{eff}(2K,t)$ given in eq. \ref{eq:pert_sin_general_single_edge_1bd_variation} are $\propto\lambda^3$, as discussed above.

\begin{figure}[htp!]
	\begin{minipage}{.5\textwidth}
		\centering
		\includegraphics[width=1.\textwidth]{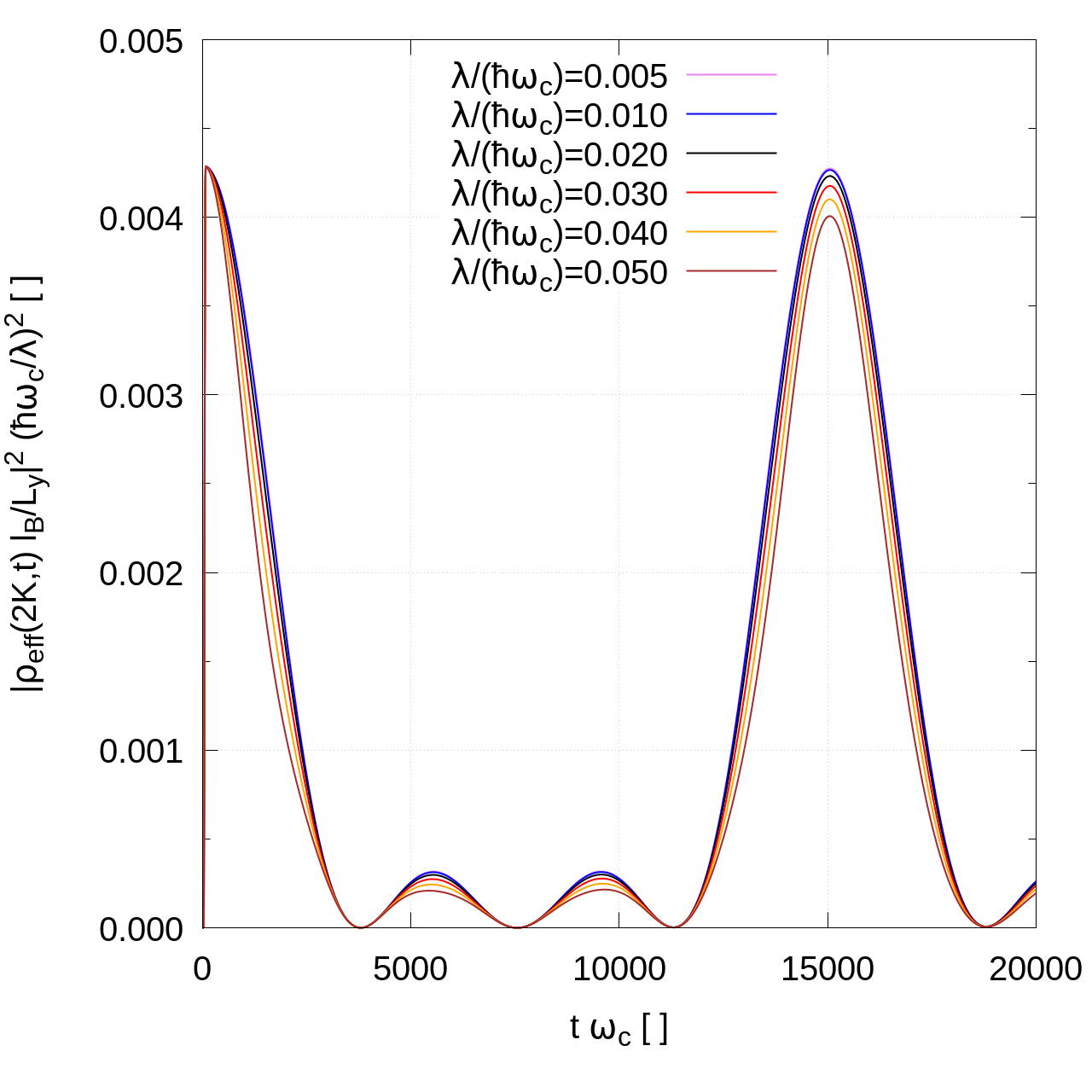}
	\end{minipage}%
	\begin{minipage}{0.5\textwidth}
		\centering
		\includegraphics[width=1.\textwidth]{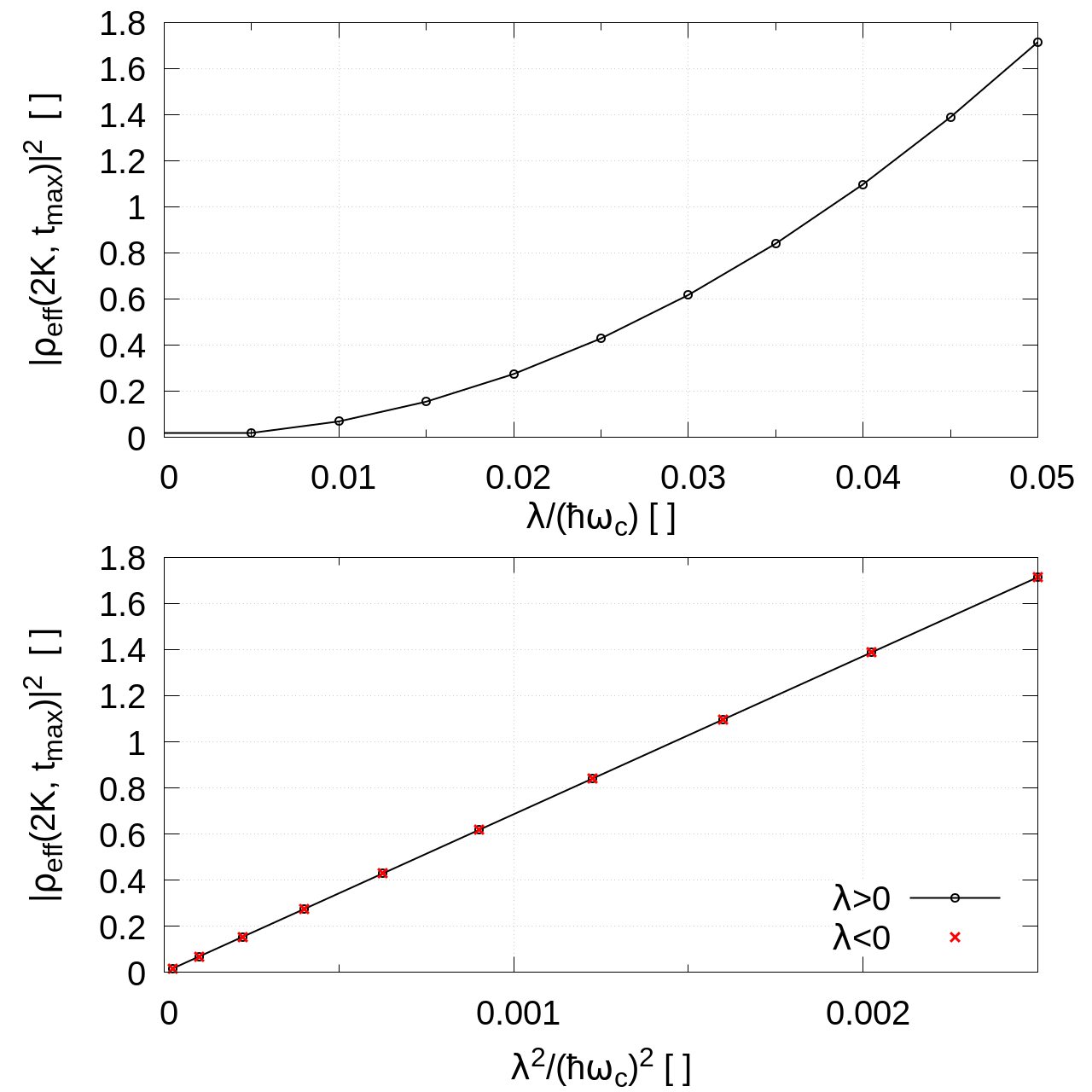}
	\end{minipage} 
	\caption[The LOF caption]{In the left hand side panel the time evolution of $|\delta\rho_\text{eff}(2K,t)|^2$ divided by $\lambda^2$ (for different values of the excitation strength $\lambda$) is shown over a larger time interval than the one shown in Fig. \ref{fig:decay_lambda}. 
	\newline The right hand side image shows the height of the first peak, i.e. $|\delta\rho_\text{eff}(2K,t_\text{max})|^2$, as a function of the excitation strength $\lambda$, plotted versus $\lambda$ (top panel) and $\lambda^2$ (bottom panel). 
	\newline $L_y=400l_B$, $k_F=546\Delta k\simeq8.58 l_B^{-1}$ and $2K=4\Delta k$ were used.}
	\label{fig:first_peak_height}
\end{figure}

From the image in the left hand side panel of Fig. \ref{fig:decay_lambda} we see that the height of the first peak increases $\appropto\lambda^2$, i.e. at short times the linear order perturbation theory result is scarcely affected as the excitation strength increases (at least for the values which have been tested). 
This is seen in the right hand side panel of Fig. \ref{fig:first_peak_height}, where the height of such a peak is plotted against $\lambda$ and $\lambda^2$.
In the left hand side panel of Fig. \ref{fig:first_peak_height} the square modulus of $\delta\rho_\text{eff}(2K,t)$ is plotted again (it was also plotted in the left hand side panel of Fig. \ref{fig:decay_lambda}), over a larger time interval though. 
We can see how increasing the excitation strength parameter $\lambda$ affects more strongly the evolution at longer times.

%The height of the first peak can be computed as well, and can be compared with the above derived expression eq. \ref{eq:maximum_amplitude_sin_linear_approx}, since $2Kv\tau\approx 0.23\leq1$. 

\subsection{Higher harmonics generation}\label{section:sin_hhg}
In this section we first of all study the non-linear dynamics by using the second-order perturbation theory results derived in section \ref{subsection:sin_PT}; we recall that at quadratic order the correction to the $q=2K$ density mode vanishes identically. At this order however the $q=4K$ component gets non-vanishing contributions from second-order transitions; otherwise stated we should in principle be able to dynamically generate higher harmonics.
From the edge effective theory (eq. \ref{eq:effective_1d_dynamics}) we however expect that these processes can possibly occur only if the curvature of the Landau level does not vanish, otherwise each mode would evolve independently from the other ones. This is precisely what it is found;
not only the curvature plays the role of making the $\delta\rho_\text{eff}(2K,t)$ decay in time, it also regulates how different modes \virgolette{interact} among each other.
\newline Numerical results are finally shown and discussed.

The $q=4K$ component of the density variation looks a bit scary, but it is worth analysing it. Putting together the results of section \ref{subsection:sin_PT} one obtains the contribution to $\delta\rho(x, 4K; t)$ arising from a single electron as
 \begin{equation}
 \begin{split}
 \delta\rho_{k_0}&(x, 4K; t) = \lambda^2
 \Biggl.\Biggr[
 f_{k_0+ 2K}^{(k_0)}f_{k_0- 2K}^{(k_0)*}e^{i\Delta\omega_{k_0-2K, k_0+2K}\,t}\Phi_{k_0+2K}(x)\Phi_{k_0-2K}(x)
 +\\&+
 g_{k_0+ 4K}^{(k_0)}e^{i\Delta\omega_{k_0, k_0+4K}\,t}\Phi_{k_0+4K}(x)\Phi_{k_0}(x)
 +
 g_{k_0- 4K}^{(k_0)*}e^{-i\Delta\omega_{k_0, k_0-4K}\,t}\Phi_{k_0-4K}(x)\Phi_{k_0}(x)
 \Biggl.\Biggr].
 \end{split}
 \end{equation}
 As already stated above, when we sum over all the occupied electron states in the bulk of the system we get terms which look like $g_{k_0+2K}^{(k_0-2K)*}+f_{k_0+2K}^{(k_0)*}f_{k_0-2K}^{(k_0)}+g_{k_0-2K}^{(k_0+2K)}$, and it can easily be checked that this sum vanishes identically, as indeed expected from our considerations.
 \newline Near the system edges, this is no longer the case since there are available levels above the Fermi surface to which electrons can jump to. It is easy to show that only finitely many terms survive; we get (considering only the positive momentum $+k_F$ Fermi point)
 \begin{equation}
 \begin{split}
 \delta\rho&(x, 4K; t)=
 \lambda^2\sum_{j=0}^{\frac{2K}{\Delta k}-1}\,f_{k_j+ 2K}^{(k_j)}f_{k_j- 2K}^{(k_j)*}e^{i\Delta\omega_{k_j-2K, k_j+2K}\,t}\Phi_{k_0+2K}\Phi_{k_0-2K}
 +\\&+
 \lambda^2\sum_{j=0}^{\frac{4K}{\Delta k}-1}\,g_{k_j+ 4K}^{(k_j)}e^{i\Delta\omega_{k_0, k_0+4K}\,t}\Phi_{k_0+4K}\Phi_{k_0}
 \end{split}
 \end{equation}
 where as before we set $k_j=k_F-j\Delta k$. Plugging in the results for the coefficients $f$ and $g$ (equations \ref{eq:f} and \ref{eq:g}), integrating out the $x$ variable and performing the usual long-wavelength approximation $d^2_{k_1,k_2}\approx1$ we get
 \begin{equation}
 \delta\rho_\text{eff}(4K; t)=S_1+S_2
 \end{equation}
 where
 \begin{equation}
 \begin{cases}
 S_1=+\left(\frac{\lambda}{4\hbar}\right)^2\,\sum_{j=0}^{\frac{2K}{\Delta k}-1}e^{i\Delta\omega_{k_j-2K, k_j+2K}\,t}
 F_t(\Delta\omega_{k_j,k_j+2K})F_t(\Delta\omega_{k_j-2K,k_j})
 \\
 S_2=-\left(\frac{\lambda}{4\hbar}\right)^2\,\sum_{j=0}^{\frac{4K}{\Delta k}-1}
 e^{i\Delta\omega_{k_j, k_j+4K}t}\int_0^t \xi(t')e^{i\Delta\omega_{k_j+4K,k_j+2K}t'}F_{t'}(\Delta\omega_{k_j,k_j+2K})dt'
 \end{cases}
 \end{equation}
 \newline From our discussion in chapter \ref{ch3}, we would expect this to vanish in the case of a perfectly linear dispersion relation $\omega_q=vq$, since under this condition the $\rho_q$ modes are decoupled one from the other, prohibiting the generation of higher harmonics. This is indeed easily checked. One obtains
 \begin{equation}
 S_1=-S_2=\left(\frac{\lambda}{4\hbar}\right)^2 F_t(-2Kv)^2\,e^{i4Kvt}\,\frac{2K}{\Delta k}
 \end{equation}
 and thus they cancel out when summed.
 \begin{figure}[htp!]
 	\centering
 	\includegraphics[width=.8\textwidth]{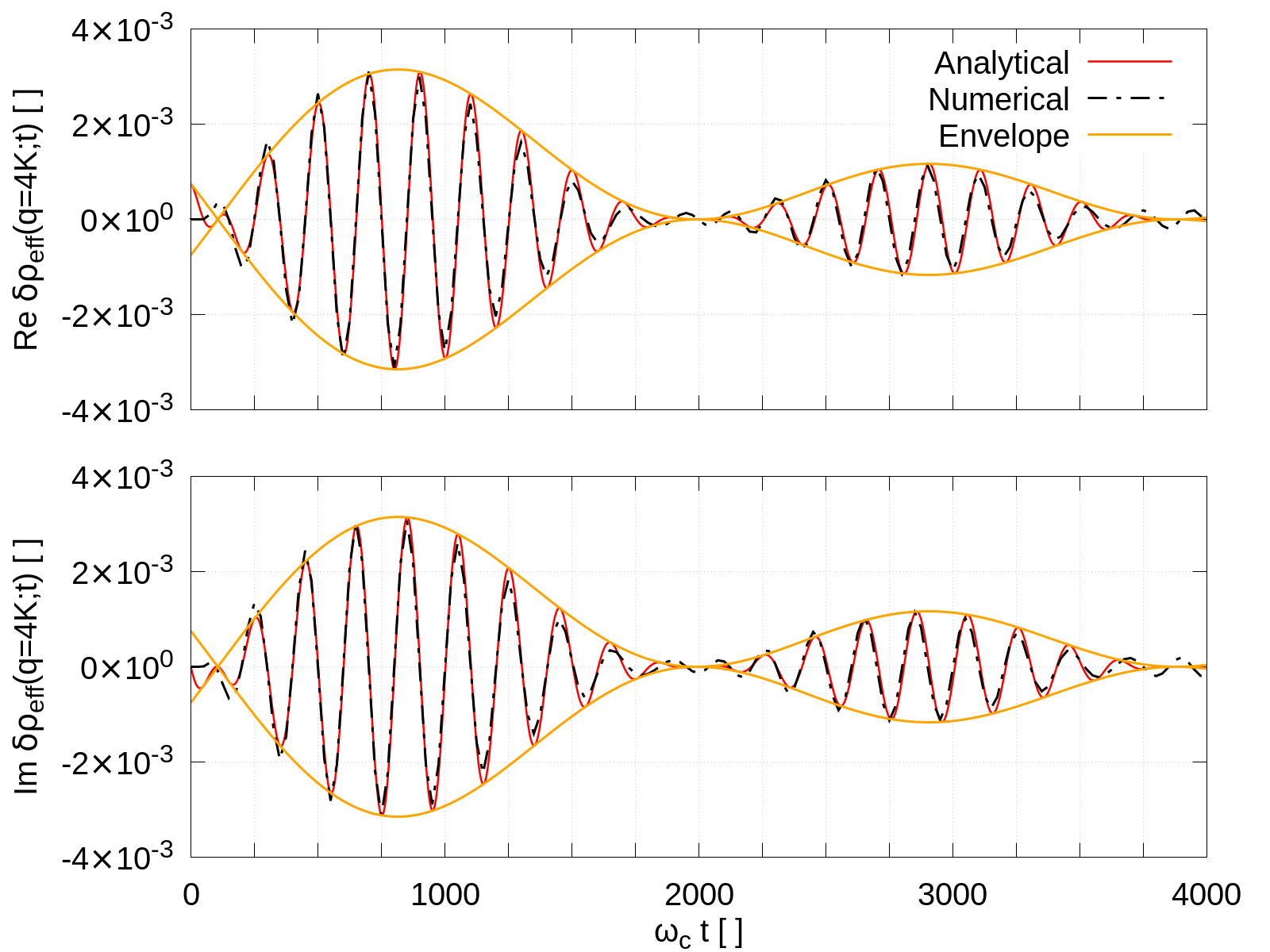}
 	\caption[The LOF caption]{The image compares the numerical results for $\delta \rho_\text{eff}(4K, t)$ with the analytical expression in eq. \ref{eq:higher_harmonic_generation1}. Both the real and imaginary parts are shown, as well as its enveloping function.  \newline I used $L_y=400l_B$, $k_F=546\Delta k\simeq 8.58l_B^{-1}$, $\lambda=0.005\hbar\omega_c$ and $2K=4\Delta k=\frac{\pi}{50} l_B^{-1}$.}
 	\label{fig:higher_harmonics_generation}
 \end{figure}
 
 By expanding the dispersion relation at quadratic order, we get a non-vanishing correction for this result, i.e. the previously uncoupled modes become \virgolette{interacting}. As before, we can obtain a nice closed formula by considering the time-evolution long after the perturbation has been turned off and neglecting higher order correction terms when squaring $\Delta\omega$. Using the identity $\sum_{j=0}^{n-1}e^{ijx}=e^{i(n-1)\frac{x}{2}}\frac{\sin\left(\frac{nx}{2}\right)}{\sin\left(\frac{x}{2}\right)}$ we get
 \begin{equation}
 \begin{cases}
 %S_1\simeq \left(\frac{1}{4\hbar}\right)^2\,\pi\tau^2\,e^{-2\left(\frac{2Kv\tau}{2}\right)^2}\,\sum_{j=0}^{\frac{2K}{\Delta k}-1}e^{i\Delta\omega_{k_j-2K, k_j+2K}\,(t-t_0)}
 S_1\simeq+\pi\left(\frac{\lambda\tau}{4\hbar}\right)^2\,e^{-2\left(\frac{2Kv\tau}{2}\right)^2}\,
 e^{-i\,4Kv\,\Delta t}\,
 e^{i\frac{c}{4}[(4K)^2- 8K \Delta k]\,\Delta t}\,
 \frac{\sin\left(\frac{c}{4} (4K)^2 \,\Delta t\right)}{\sin\left(\frac{c}{2}4K\Delta k\,\Delta t\right)}
 \\
 S_2\simeq-\frac{\pi}{2}\left(\frac{\lambda\tau}{4\hbar}\right)^2
 e^{-2\left(\frac{2Kv\tau}{2}\right)^2}\left(1+i\,\alpha\right)\,
 e^{-i\,4Kv\,\Delta t}%\\&
 e^{-i4K\frac{c}{2}\Delta t\Delta k}\,
 \frac{\sin\left(\frac{c}{2} (4K)^2 \,\Delta t\right)}{\sin\left(\frac{c}{2}4K\Delta k\,\Delta t\right)}
 \end{cases}
 \end{equation}
 where $\Delta t=t-t_0$ and $\alpha=\frac{c(2K)^2\tau}{\sqrt{2\pi}}$. In order to compute $S_2$, both the integration limits were pushed to infinity (since we have both $t_0,\,t-t_0\gg\tau$) and with a little effort\footnote{
 	Changing the integration variables and exploiting the analyticity of the integrand function, one is left with an integral of the form $I(\epsilon)=\int_{-\infty}^{\infty}e^{-(z-\epsilon)^2}\text{erf}(z)dz$. Evidently $I(0)=0$ by symmetry. Integrating by parts the derivative with respect to $\epsilon$ one finds $I'(\epsilon)=\sqrt{2}\,e^{-\frac{\epsilon^2}{2}}$, and thus $I(\epsilon)=\sqrt{\pi}\,\text{erf}\left(\frac{\epsilon}{\sqrt{2}}\right)$.} it can be shown that 
 \[
 \begin{split}
 \int_{-\infty}^\infty \exp\left[-\left(\frac{t'-t_0}{\tau}\right)^2\right]&\,e^{-i\omega_1 t'}\text{erf}\left(\frac{t'-t_0}{\tau}+i\frac{\omega_2\tau}{2}\right)dt'=\\&e^{-i\omega_1 t_0}e^{-\left(\frac{\omega_1\tau}{2}\right)^2}\sqrt{\pi}\tau\,\text{erf}\left[i\frac{(\omega_2-\omega_1)\tau}{2\sqrt{2}}\right].
 \end{split}
 \] 
 \newline Compatibly with the above approximations we replaced $\text{erf}\left(i\frac{c(2K)^2\tau}{2\sqrt{2}}\right)\simeq i\frac{c(2K)^2\tau}{\sqrt{2\pi}}=i\alpha$.
 \newline Adding the two terms and simplifying by using some elementary trigonometric identity we end up with
 \begin{equation}
 \label{eq:higher_harmonic_generation1}
 \begin{split}
 \delta\rho_\text{eff}(4K; t)=&
 i\pi\left(\frac{\lambda\tau}{4\hbar}\right)^2\,e^{-2\left(\frac{2Kv\tau}{2}\right)^2}\,e^{-i\,4Kv\,\Delta t}\,e^{-i4K \frac{c}{2} \Delta t \Delta k}\\&
 \frac{\sin^2\left(\frac{c}{4} (4K)^2 \,\Delta t\right)%\sin\left(\frac{c}{4}(4K)^2\,\Delta t\right)
 	-\frac{\alpha}{2} \sin\left(\frac{c}{2} (4K)^2 \,\Delta t\right)
 }{\sin\left(\frac{c}{2}4K\Delta k\,\Delta t\right)}
 \end{split}
 \end{equation}
 It is easy to check that this result vanishes as $c\rightarrow0$.
 \newline In Fig. \ref{fig:higher_harmonics_generation} we see that the theoretical expression \ref{eq:higher_harmonic_generation1} shows a qualitatively good agreement with the numerical data, which is not perfect though. This probably arises from the various approximations introduced for the derivation of this expression.
 \newline We notice that $\delta\rho_\text{eff}(4K; t)$ mainly evolves with frequency $4Kv$, which is the expected result in the linear dispersion case, but has an envelope slowly evolving with frequency $\frac{c}{4} (4K)^2$, as indeed expected.
 Deviations from our predictions will occur on timescales of the order of $\Omega_3^{-1}$, with $\Omega_3=\frac{\partial^3_k\omega_{k_F}}{6}\,(4K)^3$ i.e. set by terms beyond the quadratic one in the Taylor expansion of the Landau level.
 
 It is easy to take the thermodynamic limit $L\rightarrow\infty$ in \ref{eq:higher_harmonic_generation1}

 %\begin{equation}
 %\begin{cases}
 %S_1\simeq \left(\frac{1}{4\hbar}\right)^2\,\pi\tau^2\,e^{-2\left(\frac{2Kv\tau}{2}\right)^2}\,\sum_{j=0}^{\frac{2K}{\Delta k}-1}e^{i\Delta\omega_{k_j-2K, k_j+2K}\,(t-t_0)}
 %\frac{1}{L}S_1\simeq \left(\frac{\lambda}{4\hbar}\right)^2\,\frac{\tau^2}{2}\,e^{-2\left(\frac{2Kv\tau}{2}\right)^2}\,
 %e^{-i\,4Kv\,\Delta t}
 %e^{i\frac{c}{4}(4K)^2\,\Delta t}
 %\frac{\sin\left(\frac{c}{4} (4K)^2 \,\Delta t\right)}{\frac{c}{2}4K\,\Delta t}
 %\\
 %\frac{1}{L}S_2\simeq-\frac{1}{2}\left(\frac{\lambda}{4\hbar}\right)^2\pi\frac{\tau^2}{2}
 %\left(1+i\,\text{erfi}\left(\frac{c(2K)^2\tau}{2\sqrt{2}}\right)\right)
 %e^{-2\left(\frac{2Kv\tau}{2}\right)^2}
 %e^{-i\,4Kv\,\Delta t}
 %\frac{\sin\left(\frac{c}{2} (4K)^2 \,\Delta t\right)}{\frac{c}{2}4K\Delta k\,\Delta t}
 %\end{cases}
 %\end{equation}
 \begin{equation}
 \frac{1}{L_y}\delta\rho_\text{eff}(4K; t)=
 i\,4K\,\left(\frac{\lambda\tau}{4\hbar}\right)^2
 e^{-2\left(\frac{2Kv\tau}{2}\right)^2}e^{-i\,4Kv\,\Delta t}\,
 G\left(\frac{c}{4} (4K)^2 \,\Delta t\right)
 \end{equation}
 where 
 \begin{equation}
 G(x)%=\frac{\sin\left(x\right)\,e^{ix}-\frac{1}{2}\left(1+i\,\alpha\right)\sin\left(2x\right)}{4x}
 =\frac{\sin^2\left(x\right)-\frac{\alpha}{2}\sin\left(2x\right)}{4x}.
 \end{equation}
 \newline As it was the case for $\delta\rho_\text{eff}(2K; t)$, we see that also the $q=4K$ mode decays in time $\propto \Delta t^{-1}$. (In the finite system case, the excitation will revive after an initial decay).
 \newline $G(x)$ is an envelope function whose dependence on the curvature parameter $c$ is \virgolette{small}. If we neglect the $\propto\alpha$ term\footnote{Its presence may be easily included by performing a series expansion. For non zero $\alpha$, we find that at linear order the maximum position shifts at $x(\alpha)\simeq x_0+\frac{\alpha}{2}\left(1-\frac{\tan(2x_0)}{2x_0}\right)$.}, its stationary points occur when $2x \cos(x)-\sin(x)=0$ or $\sin(x)=0$. The first maximum needs to be numerically determined; it turns out to be located at $x_0\simeq1.166$, with $G(x_0)\simeq0.181$.
 Then, if $\frac{c}{4}(4K)^2 \tau \ll 1$ such a maximum occurs in a region satisfying the requirement $\Delta t\gg\tau$; i.e. where our results are expected to hold. The time at which the induced non-linear density response is maximised is set by
 \begin{equation}
 \label{eq:tM_hhg}
 t_M-t_0\simeq \frac{4x_0}{c(4K)^2}
 \end{equation}
 with amplitude
 \begin{equation}
 \label{eq:amplitude_hhg}
 \left|\frac{1}{L_y}\delta\rho_\text{eff}(4K; t_M)\right|\simeq
 G\left(x_0\right)\left(\frac{\lambda\tau}{4\hbar}\right)^2
 \,4K\,
 e^{-2\left(\frac{2Kv\tau}{2}\right)^2}.
 \end{equation}
 Notice that the non-linear response amplitude is independent of the Landau level curvature, which only tunes \textit{when} its maximum will occur.
 
 \begin{figure}[htp!]
 	\begin{minipage}{.5\textwidth}
 		\centering
 		\includegraphics[width=1.\textwidth]{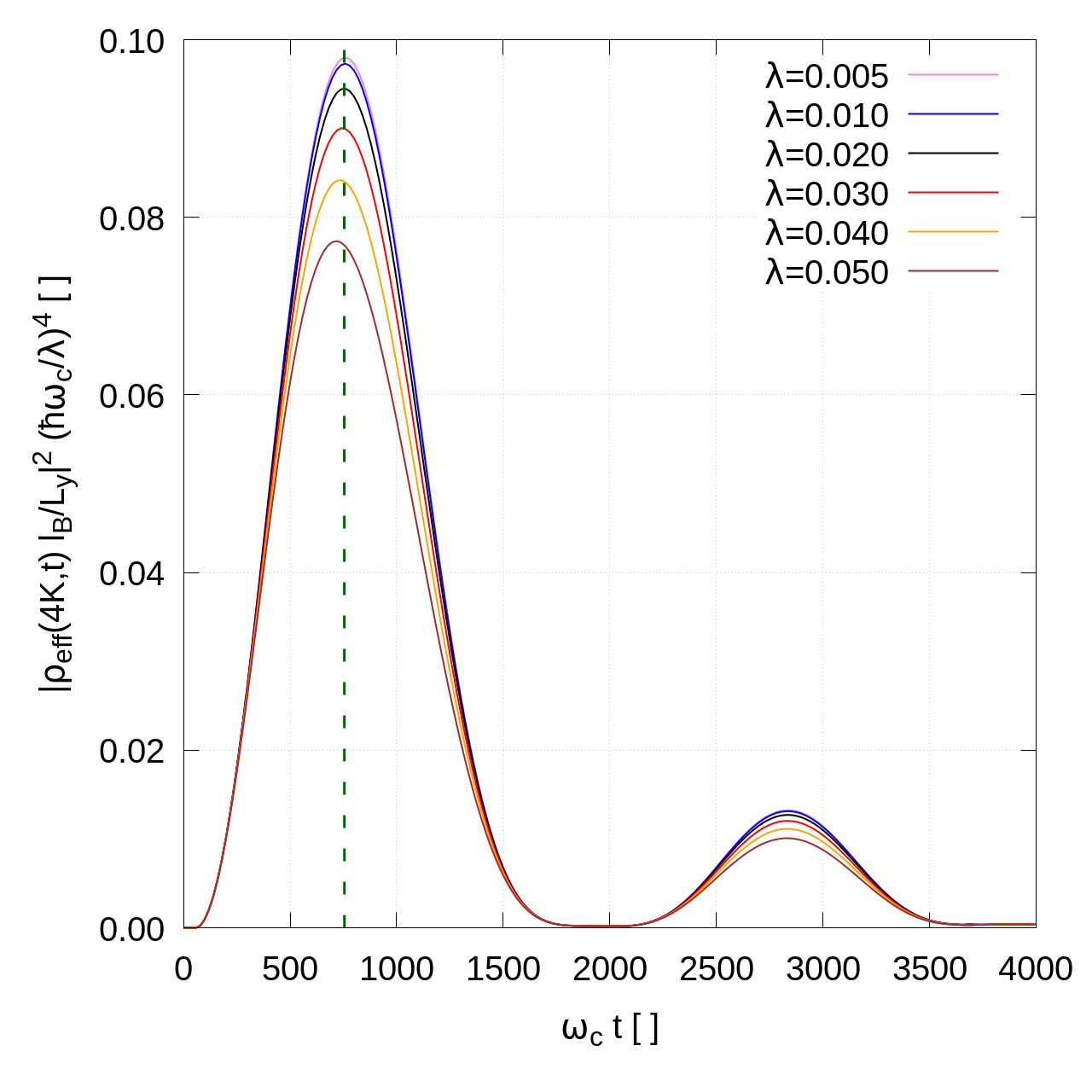}
 	\end{minipage}%
 	\begin{minipage}{0.5\textwidth}
 		\centering
 		\includegraphics[width=1.\textwidth]{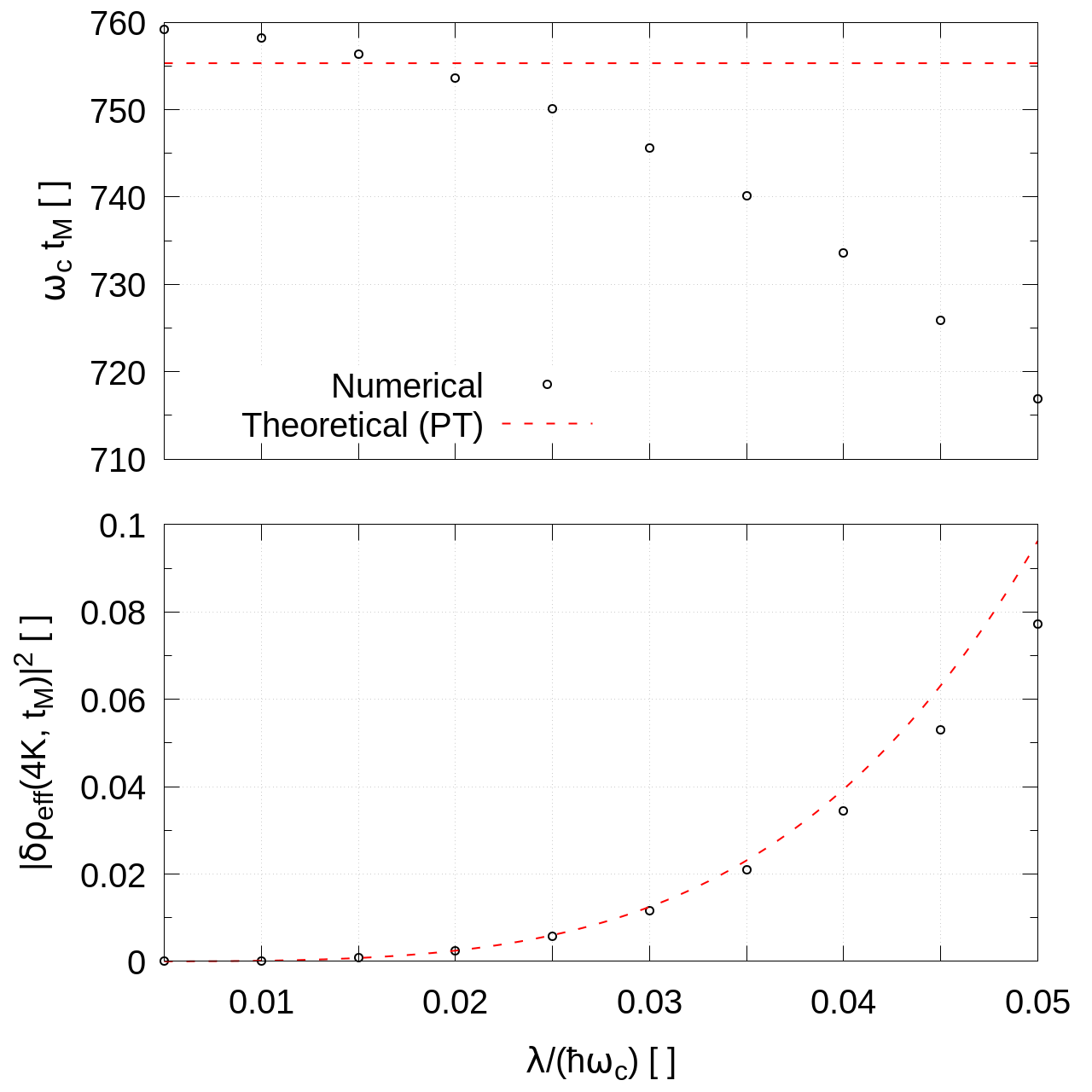}
 	\end{minipage}    
 	\caption[The LOF caption]{The left panel image shows $|\delta\rho_\text{eff}(4K,t)|^2$ divided by $\lambda^4$, when the perturbation strength $\lambda$ is varied. In the right panel the time at which the peak occurs (top) and its height (bottom). The parameters I used in order to generate these images were $L_y=400$, $2K=4\Delta k$ and $k_F=273\Delta k\simeq 8.58l_B^{-1}$.}
 	\label{fig:HigherHarmonics2}
 \end{figure}
 
 The images in Fig. \ref{fig:HigherHarmonics2} show how the discussed behaviour varies with $\lambda$.
 \newline We also see that $\delta\rho_\text{eff}(4K; t_M)\propto \lambda^2$ (as predicted by eq. \ref{eq:amplitude_hhg}) is accurate for small $\lambda$, but it picks up some corrections for large $\lambda$ which make the non-linear response smaller than expected, as can be seen from the image on the left and the one at the bottom on the right hand side.
 The image at the top of the right panel compares the numerically determined $t_M$ with the expression given in eq. \ref{eq:tM_hhg}; the agreement is qualitatively good (the relative error is less than a percent) but not perfect. The deviation may arise for the causes already discussed above.
 
 In real space the main consequence of the non-vanishing higher harmonic $\delta\rho_\text{eff}(4K,t)$ is the deformation of the spatial density profile as it propagates; in this case the system response ceases to be a pure sine wave with the same spatial modulation as the external excitation. The numerical obtained density profiles are discussed in the next section.
 
\subsection{Non-linear dynamics in real space}
\begin{figure}[htp!]
 	\begin{minipage}{.5\textwidth}
 		\centering
 		\includegraphics[width=1.\textwidth]{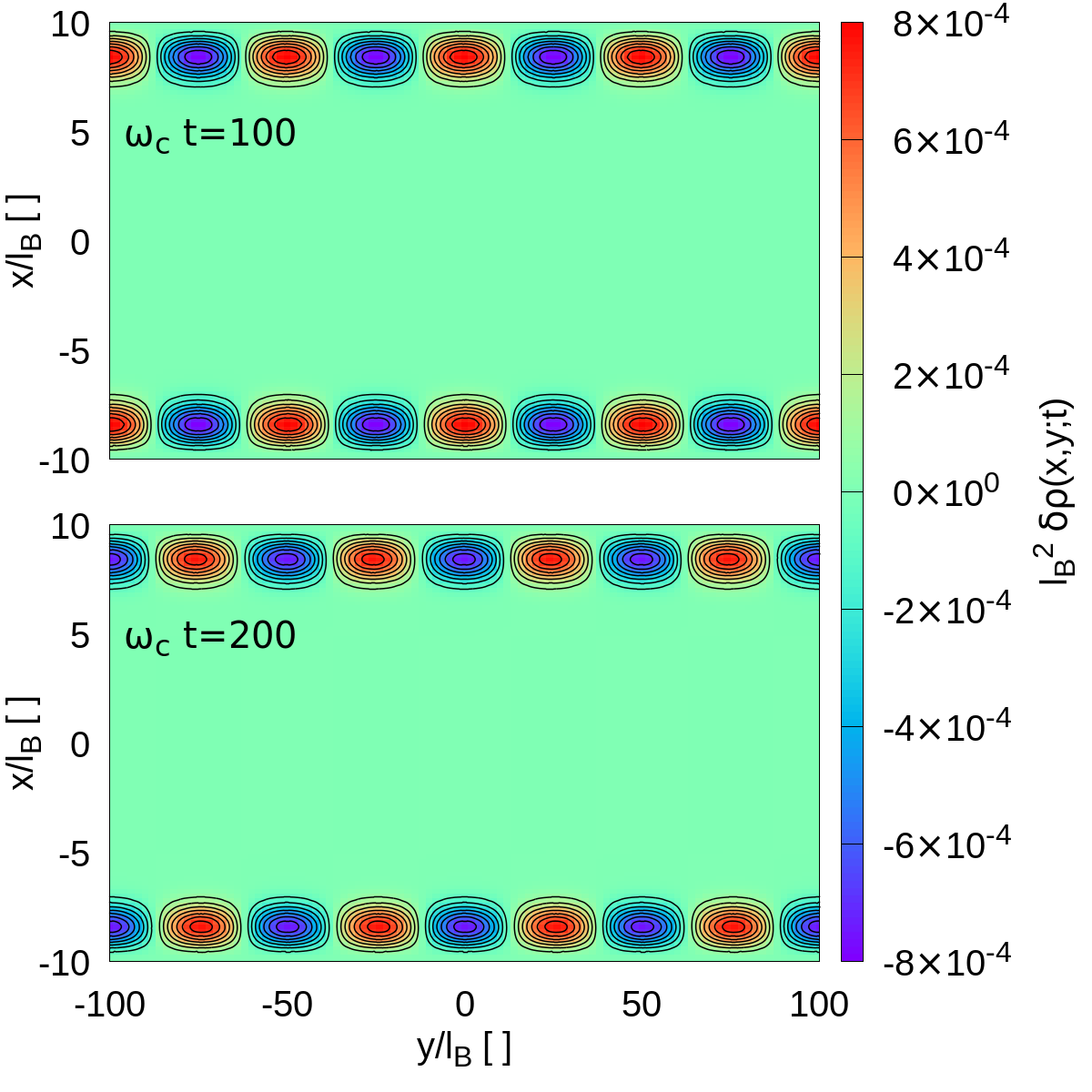}
 	\end{minipage}%
 	\begin{minipage}{0.5\textwidth}
 		\centering
 		\includegraphics[width=1.\textwidth]{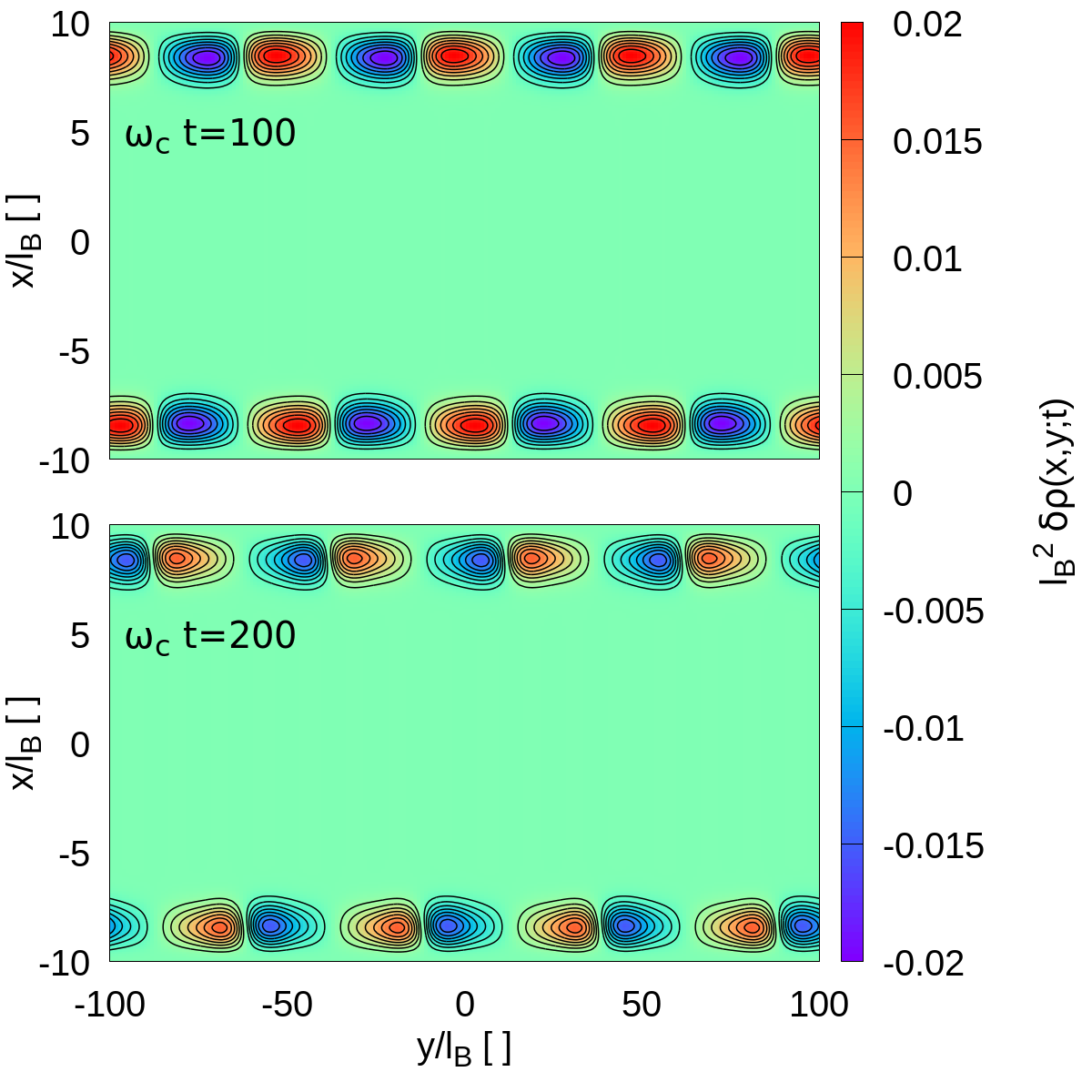}
 	\end{minipage}    
 	\caption[The LOF caption]{On the left hand side the system one-body density response when excited by eq. \ref{eq:sinusoidal_excitation} with strength $\lambda=0.005\hbar\omega_c$ is shown at $\omega_c t=100$ (top) and $\omega_c t=200$ (bottom) as a heat-map. The black lines are isocontours. On the right hand side we have $\lambda=0.1\hbar\omega_c$. \newline In both cases $L_x=20l_B$, $L_y=200l_B$, $2K=4\Delta k$ and $k_F=273\Delta k\simeq8.58l_B^{-1}$ have been used.}
 	\label{fig:sin_changing_profile}
 \end{figure}

 As discussed above for large enough values of the perturbation strength $\lambda$ the spatial profile of the density excitation will deform while propagating since higher order transitions become relevant (and curvature effects start to matter). This behaviour can be recognised when comparing the left and right images in Fig. \ref{fig:sin_changing_profile}, which show the full density variation $\delta\rho(x,y)$ at two different times as a colour-map in the cases $\lambda=0.005\hbar\omega_c$ and $\lambda=0.1\hbar\omega_c$.
 On the left hand side the only qualitative change between the top and bottom panels is a rigid translation of the \virgolette{density packets} by $v\Delta t\approx25l_B$ (with different signs on opposite sides of the sample); on the right hand side on the other hand there is a significant deformation of the shape of the excitation.
 
\begin{figure}[htp!]
 	\begin{minipage}{.5\textwidth}
 		\centering
 		\includegraphics[width=1.\textwidth]{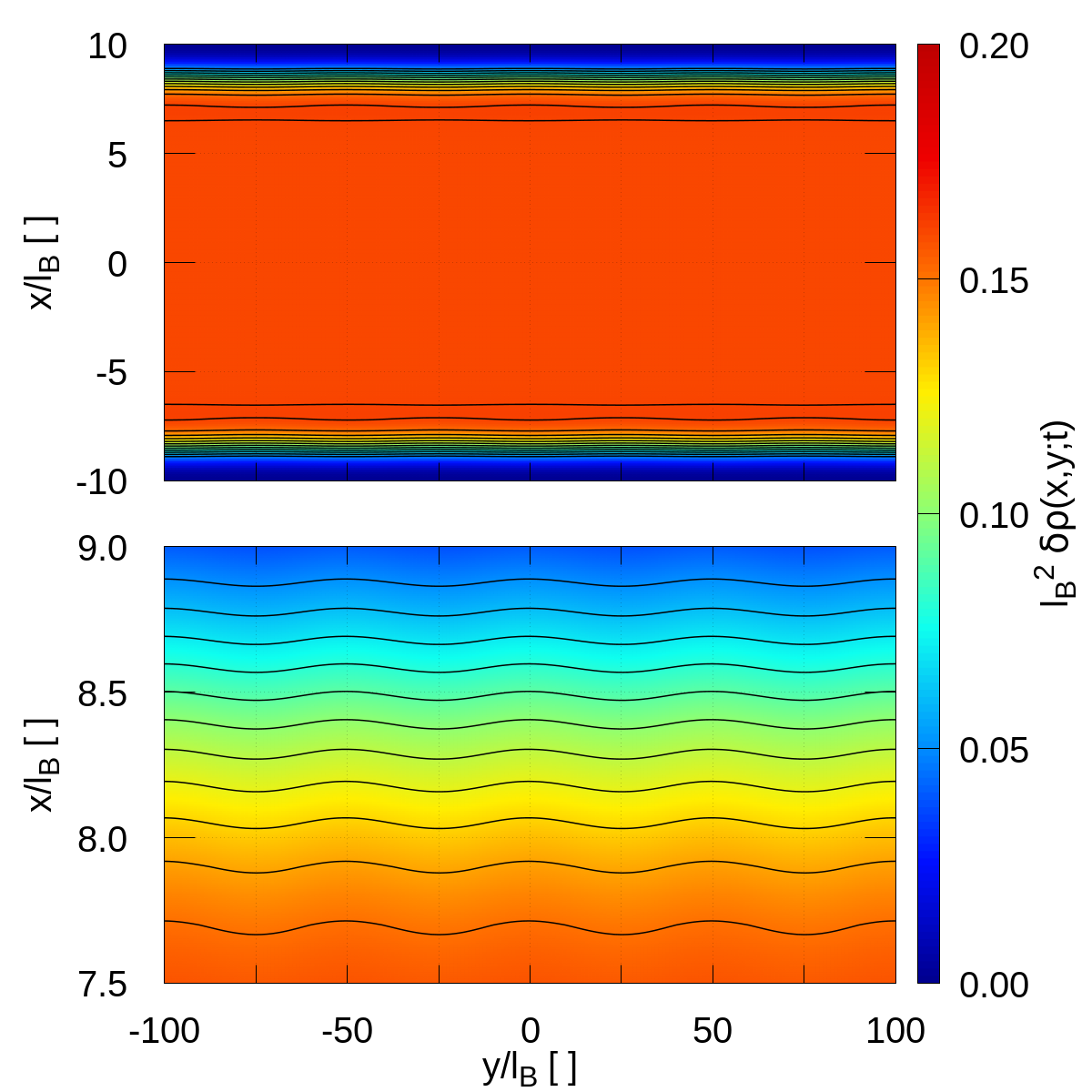}
 	\end{minipage}%
 	\begin{minipage}{0.5\textwidth}
 		\centering
 		\includegraphics[width=1.\textwidth]{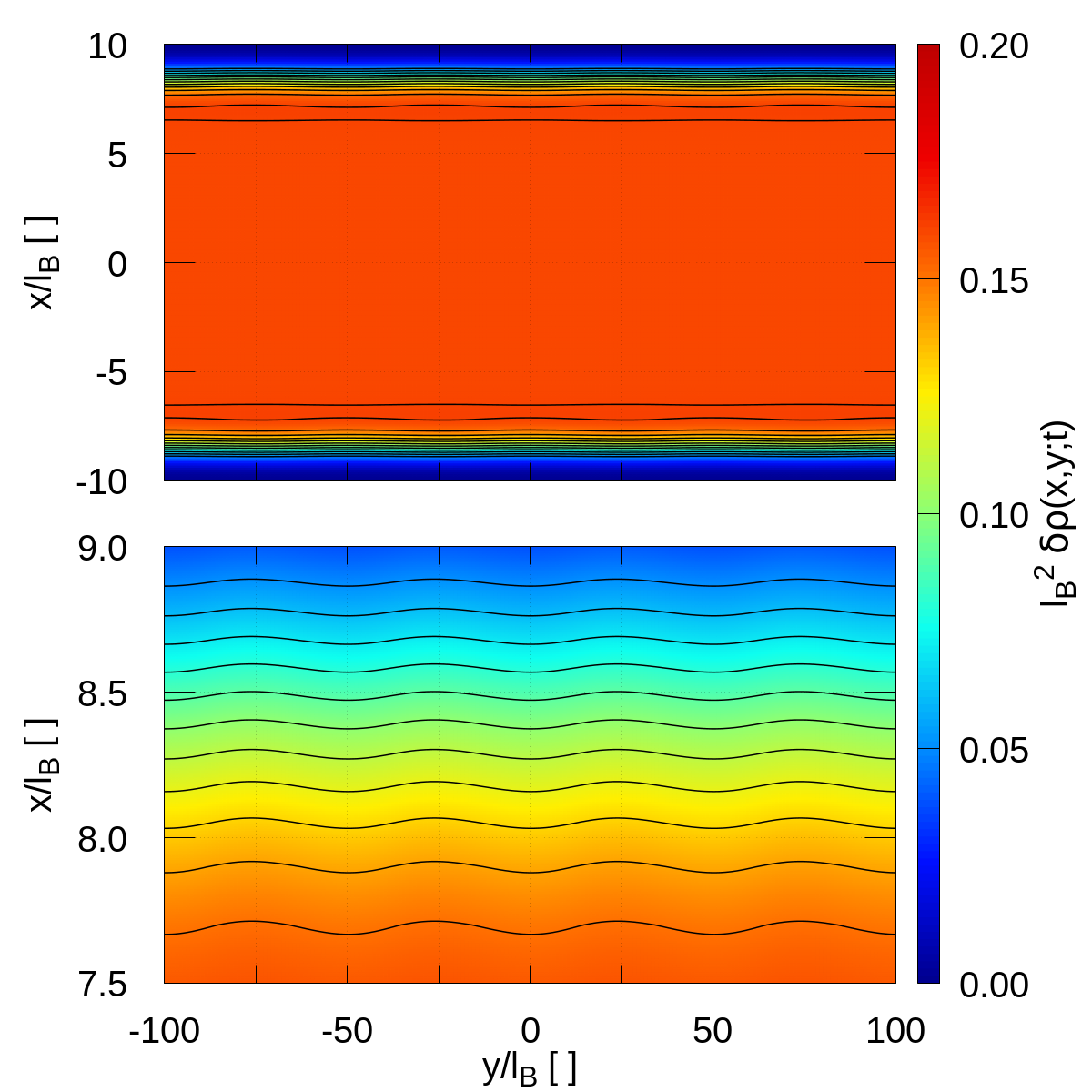}
 	\end{minipage}    
 	\caption[The LOF caption]{The two images show the \textit{full} numerically computed system density $\rho(x,y;t)$ for $\lambda=0.01\hbar\omega_c$, at $\omega_c t=100$ (left hand side) and at $\omega_c t=200$ (right hand side).
 		On top the system is shown between $\pm\frac{L_x}{2}$; the bottom images on the other hand are zoomed onto the system $x>0$ edge.
 		\newline In both cases $L_y=200l_B$, $2K=4\Delta k$ and $k_F=273\Delta k\simeq8.58l_B^{-1}$ have been used.}
 	\label{fig:edge_density0}
 \end{figure}

\begin{figure}[htp!]
	\begin{minipage}{.5\textwidth}
		\centering
		\includegraphics[width=1.\textwidth]{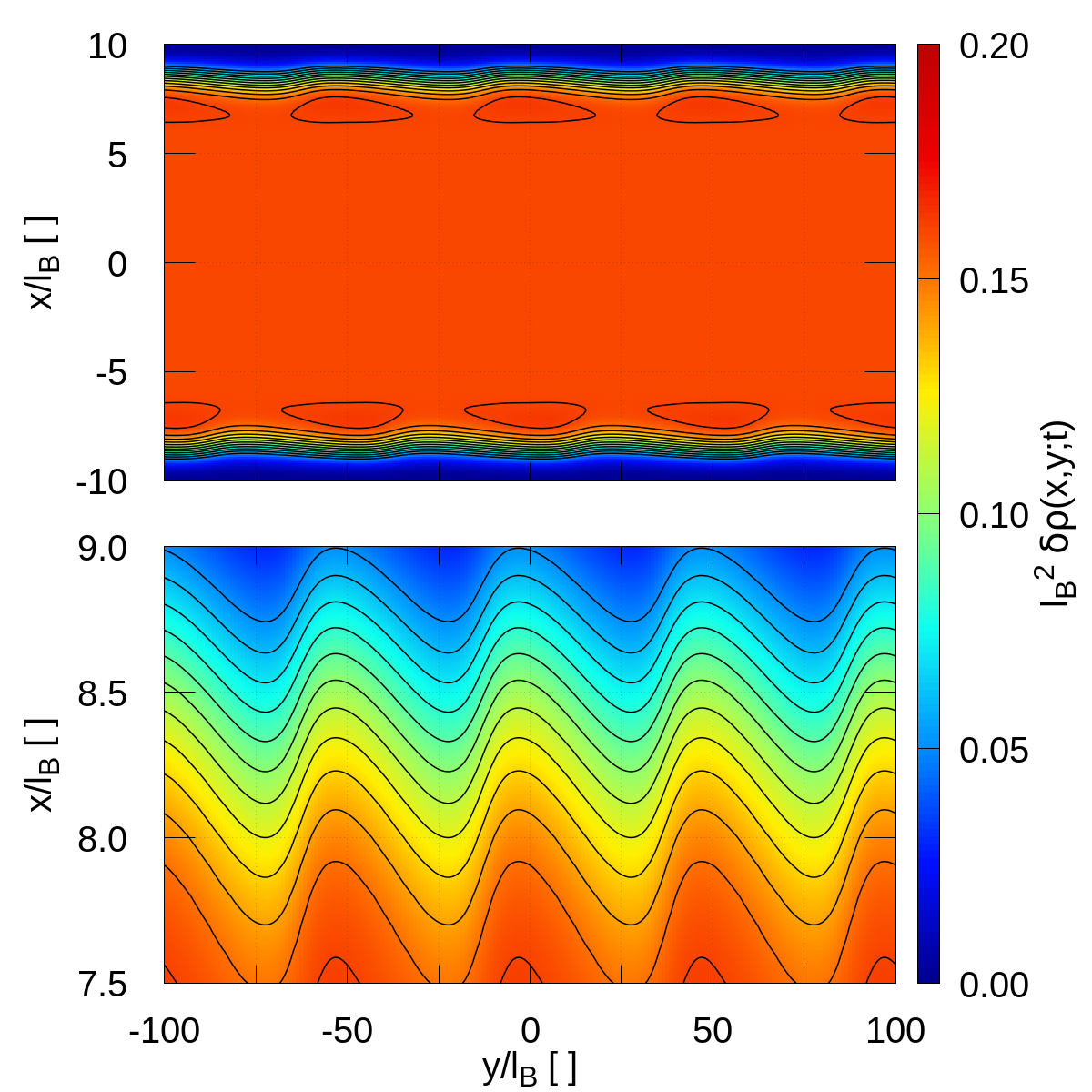}
	\end{minipage}%
	\begin{minipage}{0.5\textwidth}
		\centering
		\includegraphics[width=1.\textwidth]{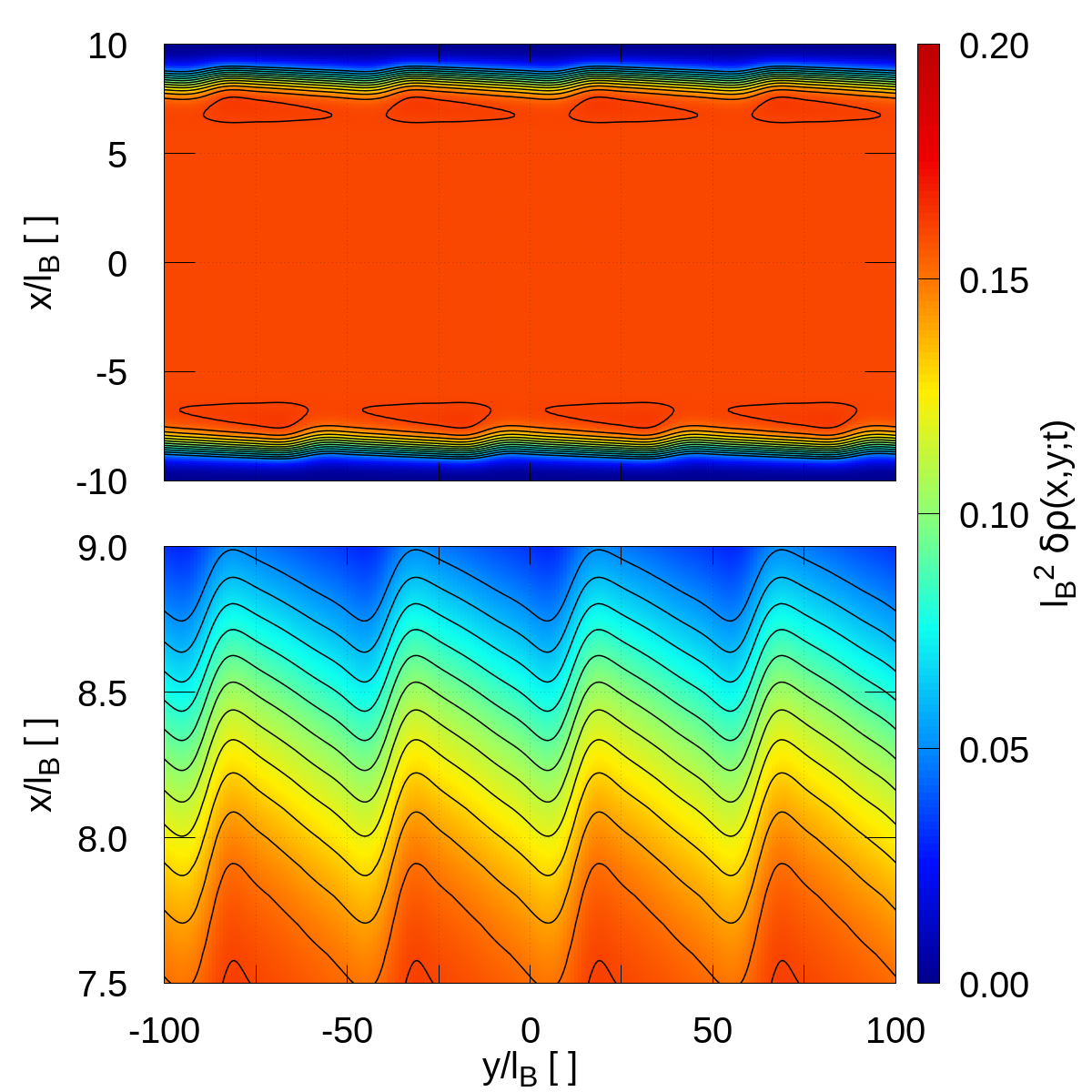}
	\end{minipage}    
	\caption[The LOF caption]{The two images show the \textit{full} numerically computed system density $\rho(x,y;t)$ for $\lambda=0.1\hbar\omega_c$, at $\omega_c t=100$ (left hand side) and at $\omega_c t=200$ (right hand side).
		On top the system is shown between $\pm\frac{L_x}{2}$; the bottom images on the other hand are zoomed onto the system $x>0$ edge.
		\newline In both cases $L_y=200l_B$, $2K=4\Delta k$ and $k_F=273\Delta k\simeq8.58l_B^{-1}$ have been used.}
	\label{fig:edge_density1}
\end{figure}

 Figures \ref{fig:edge_density0} and \ref{fig:edge_density1} show the full system density $\rho(x,y;t)$ for two different excitation strengths ($\lambda=0.01\hbar\omega_c$ and $\lambda=0.10\hbar\omega_c$ respectively). 
 The black lines again represent isocontours; by looking at these lines (or at the underlying colour map) it is apparent that in the first case (Fig. \ref{fig:edge_density0}), when effects beyond the leading perturbative order are negligible, the density \virgolette{ripples} which are propagating in the $y$ direction have a sinusoidal shape with the same wavevector as the externally applied excitation.
 \newline Notice that the propagation occurs in the negative $y$ direction in the zoomed plot, since the $x>0$ edge corresponds to the $k<0$ one, and thus negative $\partial_k E$. As above, the sinusoid shifts in this direction an amount set by $v\Delta t\approx25l_B$.
On the other hand when higher-harmonics are not negligible (higher $\lambda$, relevant Landau level curvature so that the coupling between different density modes is not inhibited) these ripples, which are obviously bigger in magnitude, significatively deviate from the pure sinusoidal shape seen in the previous case (Fig. \ref{fig:edge_density1}).
\newline At relatively short times after the perturbation has been turned off the ripples tilt in a sawtooth wave with the sharp side facing the direction of the sound propagation.
At later times the same mode decay discussed above for the linear case (eq. \ref{eq:decay_time}) causes the excited density wave to disappear.

The non-linear dynamics in real space will be discussed in more detail in the next chapters in the case of localised density excitations, since it can be better visualised in this latter case.
 
\section{Probability current}
It is interesting to also analyse the probability current density. In this section this quantity will be related to the density variation $\delta\rho$.
\newline The same computations carried out above can be performed again, but will here be omitted for brevity, being exactly analogous to the derivation of eq. \ref{eq:pert_sin_general_single_edge_1bd_variation}.

\subsection{y-component of the current}
\begin{figure}[htp!]
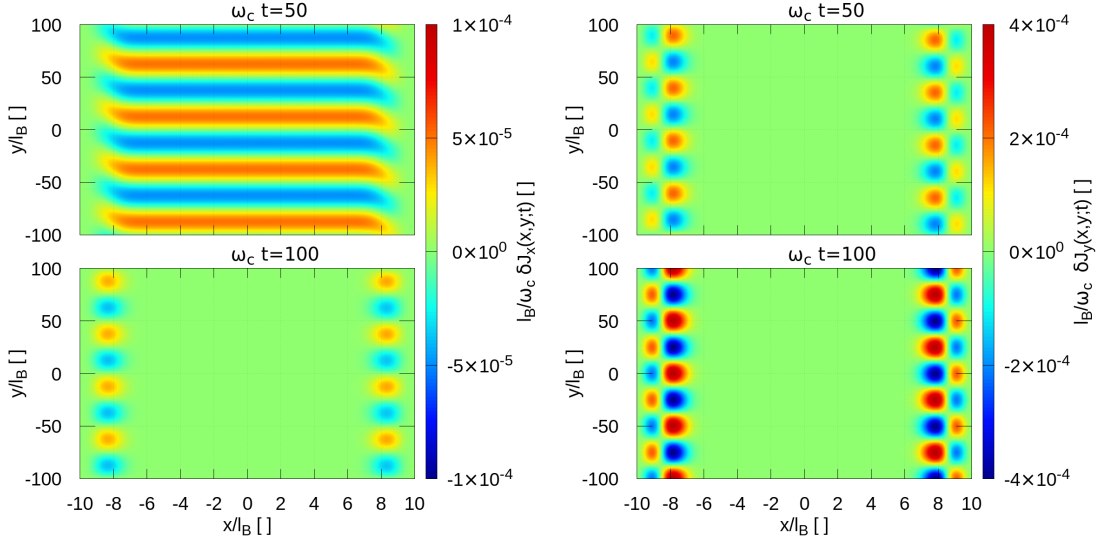

	\begin{minipage}{.5\textwidth}
		\centering
		\includegraphics[width=1.\textwidth]{jx_t=50,100.png}
	\end{minipage}%
	\begin{minipage}{0.5\textwidth}
		\centering
		\includegraphics[width=1.\textwidth]{jy_t=50,100.png}
	\end{minipage}    
	\caption[The LOF caption]{The two images show the numerically computed real space current components $\delta J_x$ (left hand side) and $\delta J_y$ (right) at $\omega_c t=50$ (top) and $\omega_c t=100$ (bottom); $L_y=200l_B$, $k_F=273\Delta k\simeq8.58l_B^{-1}$, $2K=4\Delta k$ and $\lambda=0.005\hbar\omega_c$ are the parameters used to generate the images.}
	\label{fig:bulkJx_Jy}
\end{figure}

We shall begin by discussing the $y$ component of the current density $\delta J_y=\delta\mathbf{J}\cdot\hat{y}$ (from which the unperturbed system current has been subtracted off), without making any approximation on the dispersion relation. Using eq. \ref{eq:currents_general_form}, as well as the general perturbation theory results obtained in section \ref{subsection:sin_PT}, we obtain that at linear perturbative order in the excitation strength $\lambda$ the Fourier transform $\delta J_y(x,q=2K;t)$ can be written as
\begin{equation}
\label{eq:current_jy_1} 
\begin{split}
\delta J_y(x,2K;t)=&%\frac{\hbar}{m\,l_B}\,\lambda\sum_{k_0}
%\Biggl[\Biggr.
%\left(l_B(k_0+K)+\frac{x}{l_B}\right)C^{k_0\,*}_{(k_0)}C^{(k_0)}_{k_0+2K}+\\&
%+\left(l_B(k_0-K)+\frac{x}{l_B}\right)C_{k_0}^{(k_0)}C^{(k_0)\,*}_{k_0-2K}
%\Biggl.\Biggr]
\frac{\hbar\lambda}{m\,l_B}\sum_{k_0}
\Biggl[\Biggr.
f_{k_0+2K}^{(k_0)}e^{i\Delta\omega_{k_0,k_0+2K}t}\left(l_B(k_0+K)+\frac{x}{l_B}\right)\Phi_{k_0}(x)\Phi_{k_0+2K}(x)+\\&
+
f_{k_0-2K}^{(k_0)\,*}e^{-i\Delta\omega_{k_0,k_0-2K}t}\left(l_B(k_0-K)+\frac{x}{l_B}\right)\Phi_{k_0}(x)\Phi_{k_0-2K}(x)
\Biggl.\Biggr]
\end{split}
\end{equation}
and it is easy to check that within the bulk the contributions arising from different electronic states do cancel out, as can also be seen in the right hand side panel of Fig. \ref{fig:bulkJx_Jy}, both in the perturbation transient (top right image) and after it has been switched off (bottom right). ($J_x$ is plotted as well, though it will be discussed later below). This is also the behaviour we do expect by classical reasoning: since the local electric field along $x$ vanishes in the bulk, there will not be any current (variation actually) directed along $y$ in such a region.
The same thing does not however occur near the edges, where instead we are left with a finite number of contributions
\begin{equation}
\begin{split}
\delta J_y(x,2K;t)=i\frac{\lambda}{4m\,l_B}\,\sum_{j=0}^{\frac{2K}{\Delta k}-1}&\,d_{k_j,k_j+2K}\,F_t(\Delta\omega_{k_j,k_j+2K})\left(l_B(k_j+K)+\frac{x}{l_B}\right)\\& e^{i\Delta\omega_{k_j,k_j+2K}t}\Phi_{k_j+2K}(x)\Phi_{k_j}(x).
\end{split}
\end{equation}
%\begin{comment}
%We now employ the linear-dispersion approximation, which allows to take out the energy differences of the sum, as $\Delta\omega_{k_j,k_j+2K}=-2Kv$, i.e. $k_j$ independent.
%\begin{equation}
%\begin{split}
%\delta J_y(x,2K;t)=i\frac{\lambda}{4m\,l_B}\,F_t(-2Kv)\,&e^{-i\,2Kv\,t}\,\sum_{j=0}^{\frac{2K}{\Delta k}-1}\,d_{k_j,k_j+2K}\,\\&\left(l_B(k_j+K)+\frac{x}{l_B}\right) \Phi_{k_j+2K}\Phi_{k_j}.
%\end{split}
%\end{equation}
%As usual, we integrate over $x$ to get an effective edge description
%\begin{equation}
%\begin{split}
%\delta J_{\text{eff},\,y}(2K,t)=i\frac{\lambda}{4m\,l_B}\,F_t(-2Kv)\,&e^{-i\,2Kv\,t}\,\sum_{j=0}^{\frac{2K}{\Delta k}-1}\,d_{k_j,k_j+2K}\,\\&\int\Phi_{k_j+2K}\left(l_B(k_j+K)+\frac{x}{l_B}\right) \Phi_{k_j}dx.
%\end{split}
%\end{equation}
%\end{comment}
As usual, we integrate over $x$ to get an effective edge description
\begin{equation}
\label{eq:current_y_app1}
\begin{split}
\delta J_{\text{eff},\,y}(2K,t)=i\frac{\lambda}{4m\,l_B}\,\sum_{j=0}^{\frac{2K}{\Delta k}-1}&\,d_{k_j,k_j+2K}\,F_t(\Delta\omega_{k_j,k_j+2K})e^{i\Delta\omega_{k_j,k_j+2K}t}\\& \int\Phi_{k_j+2K}(x)\left(l_B(k_j+K)+\frac{x}{l_B}\right) \Phi_{k_j}(x)dx.
\end{split}
\end{equation}
We expect the integral to roughly be proportional to the Fermi velocity. Indeed, using the bra-ket notation for simplicity we have
\begin{equation}
\begin{cases}
\begin{alignedat}{2}
&E_{q+\delta q}&&=\frac{\bra{q+\delta q}\widetilde{\mathcal{H}}_{q+\delta q}\ket{q}}{\braket{q+\delta q|q}}
\\
&E_{q} &&=\frac{\bra{q+\delta q}\widetilde{\mathcal{H}}_{q}\ket{q}}{\braket{q+\delta q|q}}
\end{alignedat}
\end{cases}
\end{equation}
where the Landau level index has been omitted. Here $\widetilde{\mathcal{H}}_q$ is the transformed Hamiltonian in eq. \ref{eq:problem_hamiltonitan_transformed}, $\ket{q}=\ket{0,q}$ its lowest lying eigenstate (i.e. $\braket{x|q}=\Phi_q(x)$) and thus $\braket{q+\delta q|q}=d_{q+\delta q,q}$.  
\newline Subtracting these two equations, using eq. \ref{eq:problem_hamiltonitan_transformed} to simplify the difference $\mathcal{H}_{q+\delta q}-\mathcal{H}_q$, and dividing both sides by $\delta q$, we get (the result is not an approximation)
\begin{equation}
\label{eq:velocity_relation1}
\frac{E_{q+\delta q}-E_q}{\delta q}=l_B\,\hbar\omega_c\frac{\bra{q+\delta q}\frac{x}{l_B}+l_B\left(q+\frac{\delta q}{2}\right)\ket{q}}{\braket{q+\delta q|q}}.
\end{equation}
\begin{figure}[htp!]
		\centering
		\includegraphics[width=1.\textwidth]{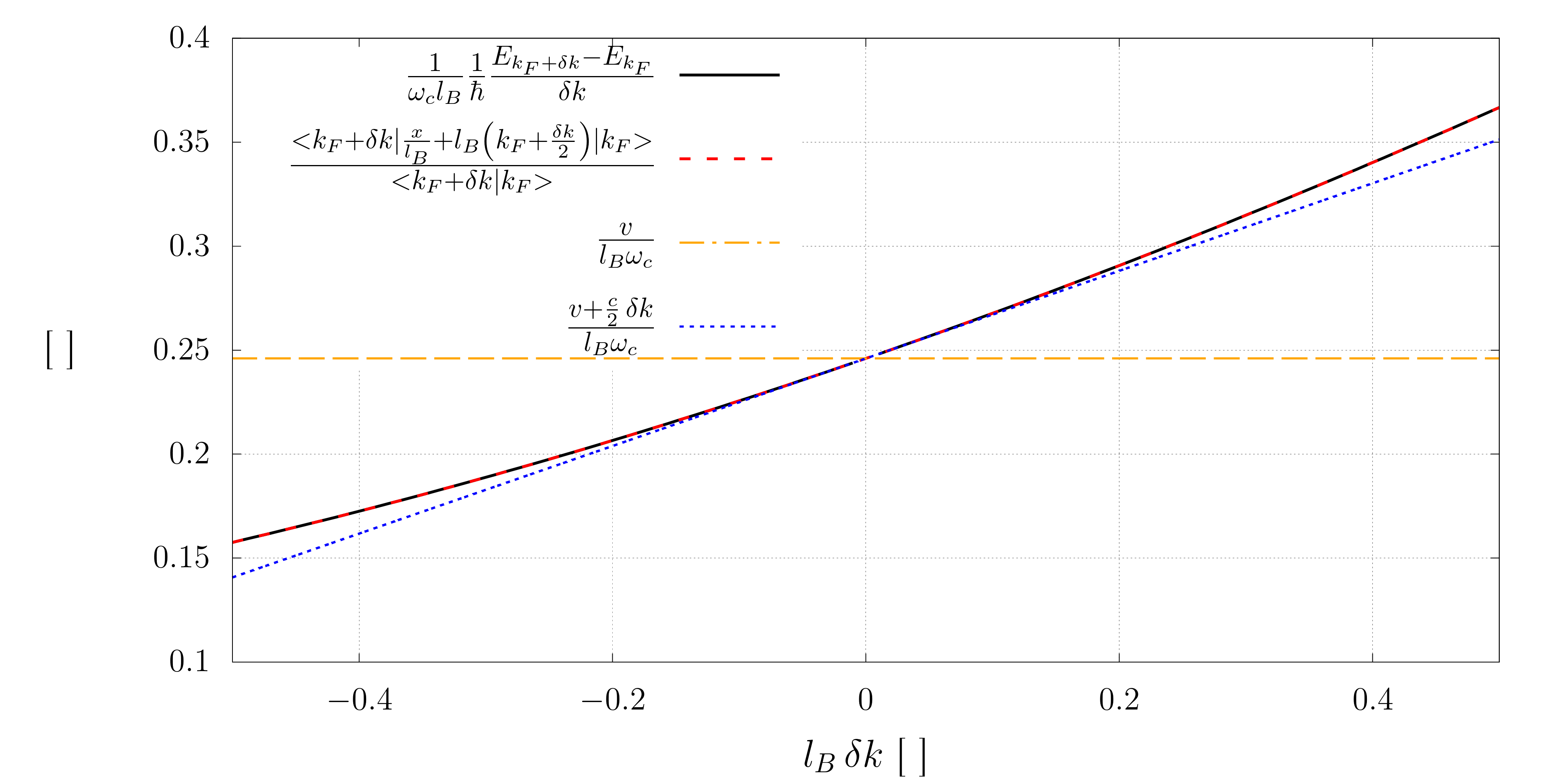}
	\caption[The LOF caption]{In the image we plot both the right hand side and the left hand side of eq. \ref{eq:velocity_relation1} to prove its validity; this curve is compared with the lowest order approximation in eq. \ref{eq:fermi_velocity_and_x_expectation_value}, and it is apparent that the two curves meet as $\delta k\rightarrow 0$. It is also shown that, as expected by inspection of \ref{eq:velocity_relation1}, the curvature of the energy band provides the linear order correction to the approximation \ref{eq:fermi_velocity_and_x_expectation_value}.
	\newline The relations are plotted at fixed $q=k_F\simeq 8.58l_B^{-1}$ as a function of the momentum difference $\delta k$. }
	\label{fig:velocity_relation}
\end{figure}
Given a linear dispersion relation, the left-hand side is just the Fermi velocity (multiplied by $\hbar$). More generally however we are allowed to replace the left hand side with the Fermi velocity only in the long-wavelength limit $\delta q\rightarrow0$, the committed error being of order $\delta q$ and proportional to the curvature of the Landau level at the Fermi surface. The validity of \ref{eq:velocity_relation1} is numerically checked and exhibited in Fig. \ref{fig:velocity_relation}.
We can therefore approximate, as long as $K$ corresponds to a long-wavelength excitation of the system, the integral appearing in the last line of eq. \ref{eq:current_y_app1} as
\begin{equation}
\label{eq:fermi_velocity_and_x_expectation_value}
\int\Phi_{k_j+2K}\left(l_B(k_j+K)+\frac{x}{l_B}\right) \Phi_{k_j}dx \simeq \frac{v}{l_B\omega_c} \,d_{k_j,k_j+2K}.
\end{equation}
Using eq. \ref{eq:fermi_velocity_and_x_expectation_value} into eq. \ref{eq:current_y_app1}
\begin{equation}
\frac{1}{L_y}\,\delta J_{\text{eff},\,y}(2K,t)=v\underbrace{ i\frac{\lambda}{4\,\hbar}\,\sum_{j=0}^{\frac{2K}{\Delta k}-1}\,d_{k_j,k_j+2K}^2\,F_t(\Delta\omega_{k_j,k_j+2K})e^{i\Delta\omega_{k_j,k_j+2K}t}}_{\frac{1}{L_y} \delta\rho_\text{eff}(2K,t)}
\end{equation}
and thus, as expected\footnote{It is easy to check that the $q=0$ mode does not contribute.}
\begin{equation}
\label{eq:JY_rho_linked}
\delta J_{\text{eff},\,y}(y,t)=v\,\delta\rho_\text{eff}(y,t).
\end{equation}
This result can be interpreted as being the consequence of a one-dimensional continuity equation; indeed comparing the previous result with the real space effective equation eq. \ref{eq:coordinate_spce_effective_1d_dynamics}
\begin{equation}
\partial_t \delta\rho_\text{eff}(y,t)=-\partial_y\left(v\delta\rho_\text{eff}(y,t)\right)=-\partial_y\delta J_{\text{eff},\,y}(y,t).
\end{equation}

\begin{figure}[htp!]
	\begin{minipage}{.5\textwidth}
		\centering
		\includegraphics[width=1.\textwidth]{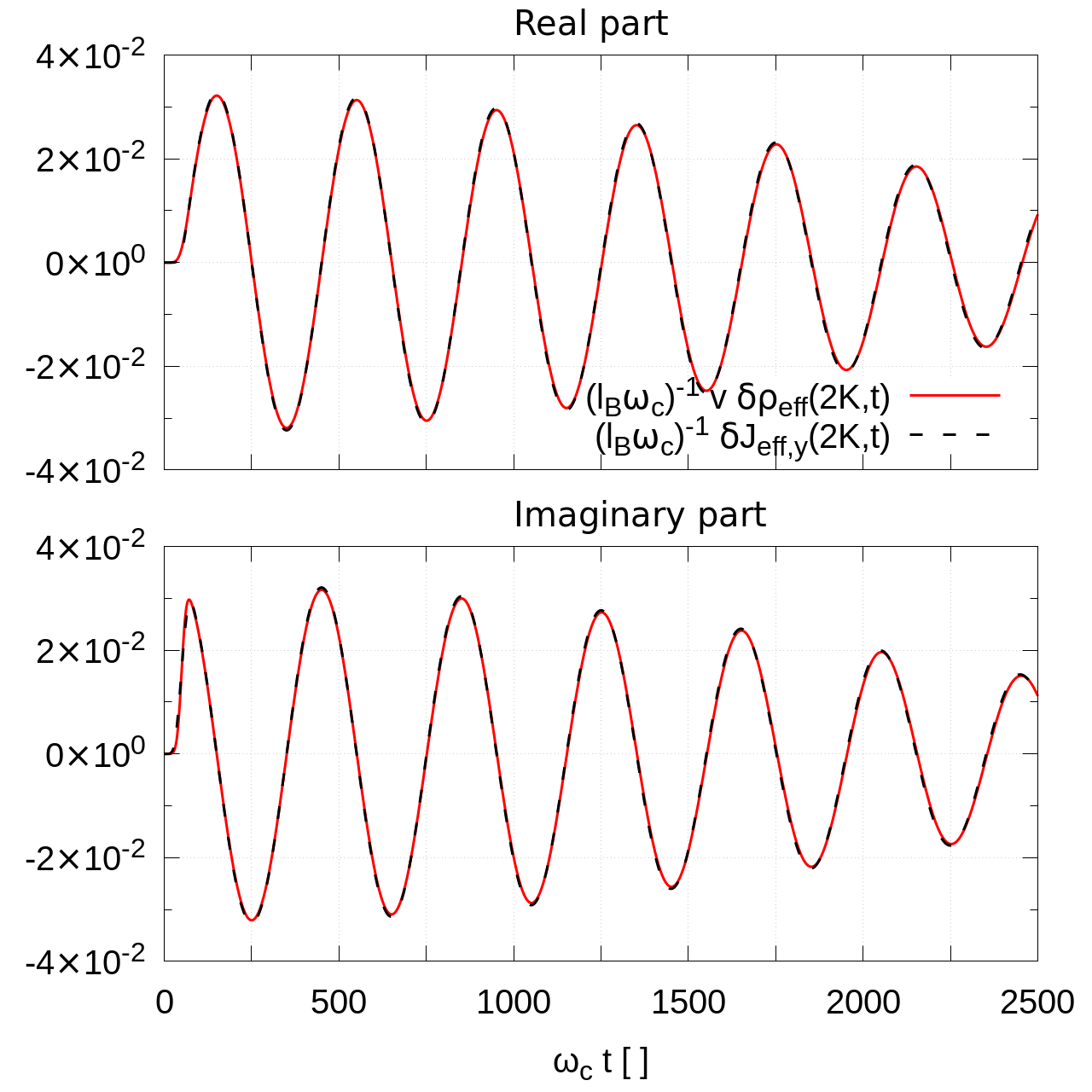}
	\end{minipage}%
	\begin{minipage}{0.5\textwidth}
		\centering
		\includegraphics[width=1.\textwidth]{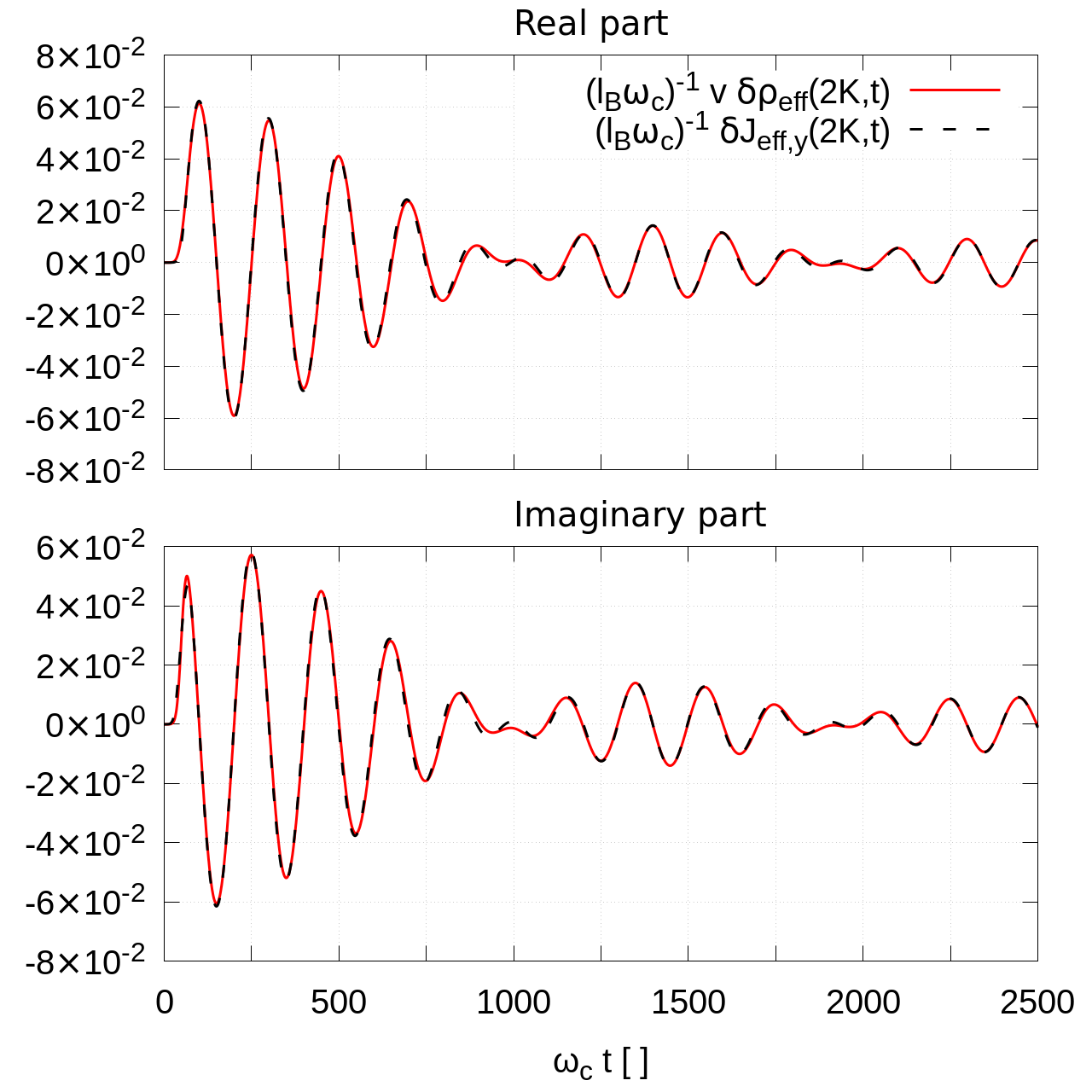}
	\end{minipage}    
	\caption[The LOF caption]{The two images show the real and imaginary parts of $\delta J_{\text{eff},\,y}(2K,t)$ as a function of time, compared with $v\,\delta\rho_\text{eff}(2K,t)$. On the left hand side $2K=\frac{\pi}{50} l_B^{-1}$ was used, $2K=\frac{\pi}{25} l_B^{-1}$ for the other one. The system length was set to be $L_y=400l_B$, the Fermi wavevector $k_F\simeq 8.58 l_B^{-1}$. $\lambda=0.005\hbar\omega_c$ has been used.}
	\label{fig:Jy}
\end{figure}

\subsection{x-component of the current}
Now we discuss the x component of the current density $\delta J_x=\delta\mathbf{J}\cdot\hat{x}$. 
From eq. \ref{eq:currents_general_form} and the perturbation theory results obtained in section \ref{subsection:sin_PT}, one obtains an expression for $J_x$ analogous to eq. \ref{eq:current_jy_1} for $J_y$ but a bit more lengthy. Higher Landau level contributions are included from scratch, since their contribution will turn out to be relevant during the transient in the system's bulk.
\begin{equation}
\label{eq:current_jx_1} 
\begin{split}
\delta J_x(x,2K;t)=&%\frac{1}{2}\sum_{k_0}\left[C_{k_0}^{(k_0)*}(-i\partial_x)C_{k_0+2K}^{(k_0)}+C_{k_0}^{(k_0-2K)*}(-i\partial_x)C_{k_0}^{(k_0)}+(2K\rightarrow-2K)^*\right]=
\frac{\hbar\lambda}{2m}\sum_{k_0}\sum_n\Biggl[\Biggr.
f^{(n_0,k_0)}_{n,k_0+2K}e^{-i\Delta\omega_{n,n_0}^{k_0+2K,k_0}t}\Phi_{n_0,k_0}(-i\partial_x)\Phi_{n,k_0+2K}
+\\&+
f^{(n_0,k_0)*}_{n,k_0-2K}e^{i\Delta\omega_{n,n_0}^{k_0-2K,k_0}t}\Phi_{n,k_0-2K}(-i\partial_x)\Phi_{n_0,k_0}
+(2K\rightarrow-2K)^*\Biggl.\Biggr]
\end{split}
\end{equation}
It is easy to check that if only $n=0$ is considered, within the bulk the contributions arising from different electrons do cancel out as expected. As can be seen from the left hand side panel in Fig. \ref{fig:bulkJx_Jy} this is indeed the case once the perturbation is off, but during the transient higher Landau levels need to be included, their presence being far from negligible. %\footnote{This behaviour has been checked numerically, both by including the first excited Landau level in the perturbative computation as well as using only the lowest Landau level in the \virgolette{exact} time evolution program, and cross-checking the results.}. 
The behaviour can be understood quite easily since the electrons drift orthogonally to the local electric field they feel, as expected from the discussion made when in section \ref{section:classical_electric_field}.
%qualitatively be understood on purely classical grounds: the electrons locally respond to the excitation as if an electric field (whose direction is purely the y axis) was turned on: electrons will roll down the hills of the potential $\propto-\cos(2Ky)$ while being subjected to the uniform orthogonal magnetic field which makes them curve towards the positive or negative verse of the $x$-axis, i.e. they move orthogonally to the local electric field they feel, as expected from the discussion made when in section \ref{section:classical_electric_field}. 
A more quantitative discussion for what concerns the bulk will be made in the upcoming subsection \ref{subsection:Jx_transient_bulk}.

\begin{figure}[htp!]
	\centering
	\includegraphics[width=1.\textwidth]{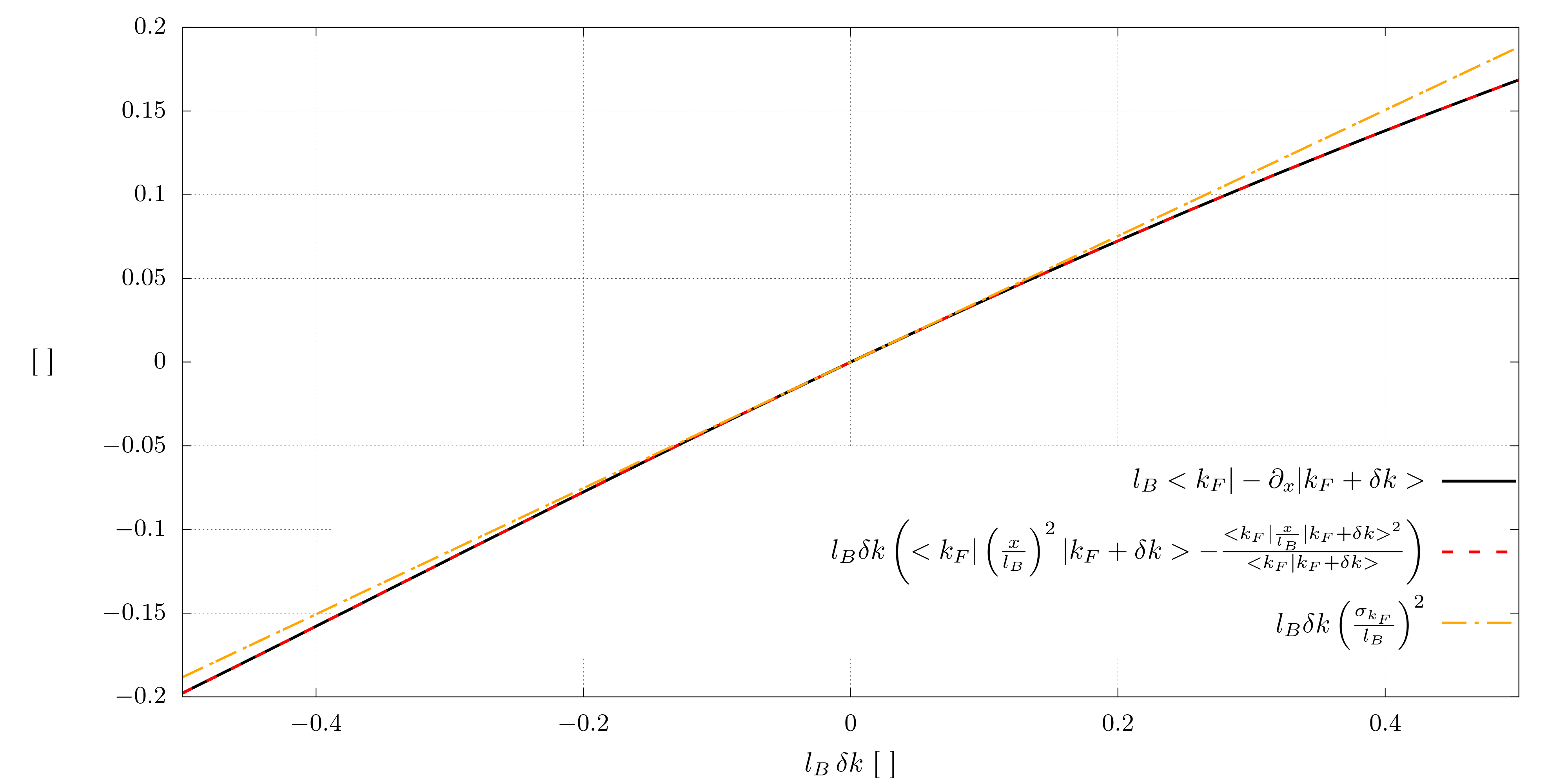}
	\caption[The LOF caption]{In order to numerically prove its validity of eq. \ref{eq:momentum_relation1}, both its right hand side and its left hand side are plotted in this image as a function of the momentum difference $\delta k$, at fixed $q=k_F\simeq 8.58l_B^{-1}$. 
	\newline The curve is also compared with the lowest order approximation in eq. \ref{eq:pdx_sigmax}.}
	\label{fig:momentum_relation}
\end{figure}

We now focus on the edges, considering the presence of the lowest Landau level only.
As usual in such a region the cancellation is not complete and we get that the $q=2K$ Fourier component of $J_x$ can be written as\footnote{As usual only one edge of the sample, the one corresponding to positive momentum, has been considered.} (at linear perturbative order in the excitation strength $\lambda$)
\begin{equation}
\begin{split}
\delta J_x(x,2K;t) =& i\frac{\lambda}{8m}\,
\sum_{j=0}^{\frac{2K}{\Delta k}-1} 
d_{k_0,k_0+2K}F_t(\Delta\omega_{k_0,k_0+2K})e^{i\Delta\omega_{k_0,k_0+2K}t}\\&
\left[
\Phi_{k_j}(x)(-i\partial_x\Phi_{k_j+2K}(x))+
\Phi_{k_j+2K}(x)(+i\partial_x\Phi_{k_j}(x))
\right].
\end{split}
\end{equation}
If we integrate over x we get
\begin{equation}
\label{eq:x_integrated_edge_Jx}
\begin{split}
\delta J_{\text{eff},\,x}(2K,t) =&
i\frac{\lambda}{4m}\,
\sum_{j=0}^{\frac{2K}{\Delta k}-1} 
d_{k_0,k_0+2K}F_t(\Delta\omega_{k_0,k_0+2K})e^{i\Delta\omega_{k_0,k_0+2K}t}\\&
\int \Phi_{k_j}(x)(-i\partial_x\Phi_{k_j+2K}(x))\,dx.
\end{split}
\end{equation}
%\begin{comment}
%If we approximate the dispersion relation as being linear and integrate over x we get
%\begin{equation}
%\begin{split}
%\delta J_{\text{eff},\,x}(2K,t) =
%i\frac{\lambda}{4m}\,F_t(-2Kv)e^{-i2Kvt}
%\sum_{j=0}^{\frac{2K}{\Delta k}-1} 
%d_{k_0,k_0+2K}
%\int \Phi_{k_j}(-i\partial_x\Phi_{k_j+2K})\,dx.
%\end{split}
%\end{equation}
%\end{comment}
In the long wavelength limit, the integral in the last line vanishes linearly with $K$, as can easily be inferred by a series expansion. The proportionality constant will now be related to the squared width $\braket{(x-\braket{x})^2}_{k_j}$ of the state $k_j$. Since $[p_x, x]=-i\hbar $, we can write
\begin{equation}
\begin{split}
p_x=&i\frac{m}{\hbar}\left[\frac{p_x^2}{2m}, x\right]
=i\frac{m}{\hbar}\left[\widetilde{\mathcal{H}}_k, x\right]
=i\frac{m}{\hbar}\left(\widetilde{\mathcal{H}}_k\,x-x\widetilde{\mathcal{H}}_{k+\delta k}+x(\widetilde{\mathcal{H}}_{k+\delta k}-\widetilde{\mathcal{H}}_{k})\right)=\\&
=i\,\frac{\hbar}{l_B}\left(\frac{\widetilde{\mathcal{H}}_k}{\hbar\omega_c}\,\frac{x}{l_B}-\frac{x}{l_B}\frac{\widetilde{\mathcal{H}}_{k+\delta k}}{\hbar\omega_c}+\frac{x}{l_B}\,l_B\delta k\left(\frac{x}{l_B}+l_B\left(k+\frac{\delta k}{2}\right)\right)\right)
\end{split}
\end{equation}
where $\widetilde{\mathcal{H}}_k$ is the transformed Hamiltonian in eq. \ref{eq:problem_hamiltonitan_transformed}.
Taking the matrix element between the eigenstates $\ket{0,k}=\ket{k}$ of $\widetilde{\mathcal{H}}_k$ and $\ket{0,k+\delta k}$ of $\widetilde{\mathcal{H}}_{k+\delta k}$ we get an expression for the integral appearing in the last line of eq. \ref{eq:x_integrated_edge_Jx}.
\begin{equation}
\label{eq:momentum_relation1}
\begin{split}
\bra{k}p_x\ket{k+\delta k}=&
i\frac{\hbar}{l_B}\,l_B\delta k\,
\Biggl[\Biggr.
	 \frac{1}{l_B^2}\bra{k}x^2\ket{k+\delta k}
	-\\&-\underbrace{\left(
	 \frac{1}{\hbar\omega_c}\frac{E_{k+\delta k}-E_k}{l_B\delta k} -l_B\left(k+\frac{\delta k}{2}\right)
	 \right)}_{\frac{1}{l_B}\frac{\bra{k}x\ket{k+\delta k}}{\braket{k|k+\delta k}}}\frac{1}{l_B}
\bra{k}x\ket{k+\delta k}
\Biggl.\Biggr]\\=&
i\,\hbar\,\delta k\,
\frac{1}{l_B^2}\Biggl[\Biggr.
\bra{k}x^2\ket{k+\delta k}
-\frac{\bra{k}x\ket{k+\delta k}^2}{\braket{k|k+\delta k}}
\Biggl.\Biggr]
\end{split}
\end{equation}
where the relation between the Fermi velocity and the expectation value of the position operators (eq. \ref{eq:fermi_velocity_and_x_expectation_value}) has been used (under the big bracket). In the long wavelength limit $\delta k\rightarrow0$ we can replace $\ket{k+\delta k}$ with $\ket{k}$ up to higher order corrections, so
\begin{equation}
\label{eq:pdx_sigmax}
\begin{split}
\bra{k}-i\partial_x\ket{k+\delta k}&\simeq
i\,\delta k\,
\frac{\bra{k}x^2\ket{k}
	-\bra{k}x\ket{k}^2}{l_B^2}+\mathcal{O}(\delta k^2)
\\& 
=i\,\delta k\,
\left(\frac{\sigma_k}{l_B}\right)^2+\mathcal{O}(\delta k^2)
\end{split}
\end{equation}
where $\sigma_k^2=\bra{k}x^2\ket{k}-\bra{k}x\ket{k}^2$.
\newline The validity of eq. \ref{eq:momentum_relation1} is numerically tested in Fig. \ref{fig:momentum_relation}, as well as the validity of the approximation \ref{eq:pdx_sigmax}.

\begin{figure}[htp!]
	\begin{minipage}{.5\textwidth}
		\centering
		\includegraphics[width=1.\textwidth]{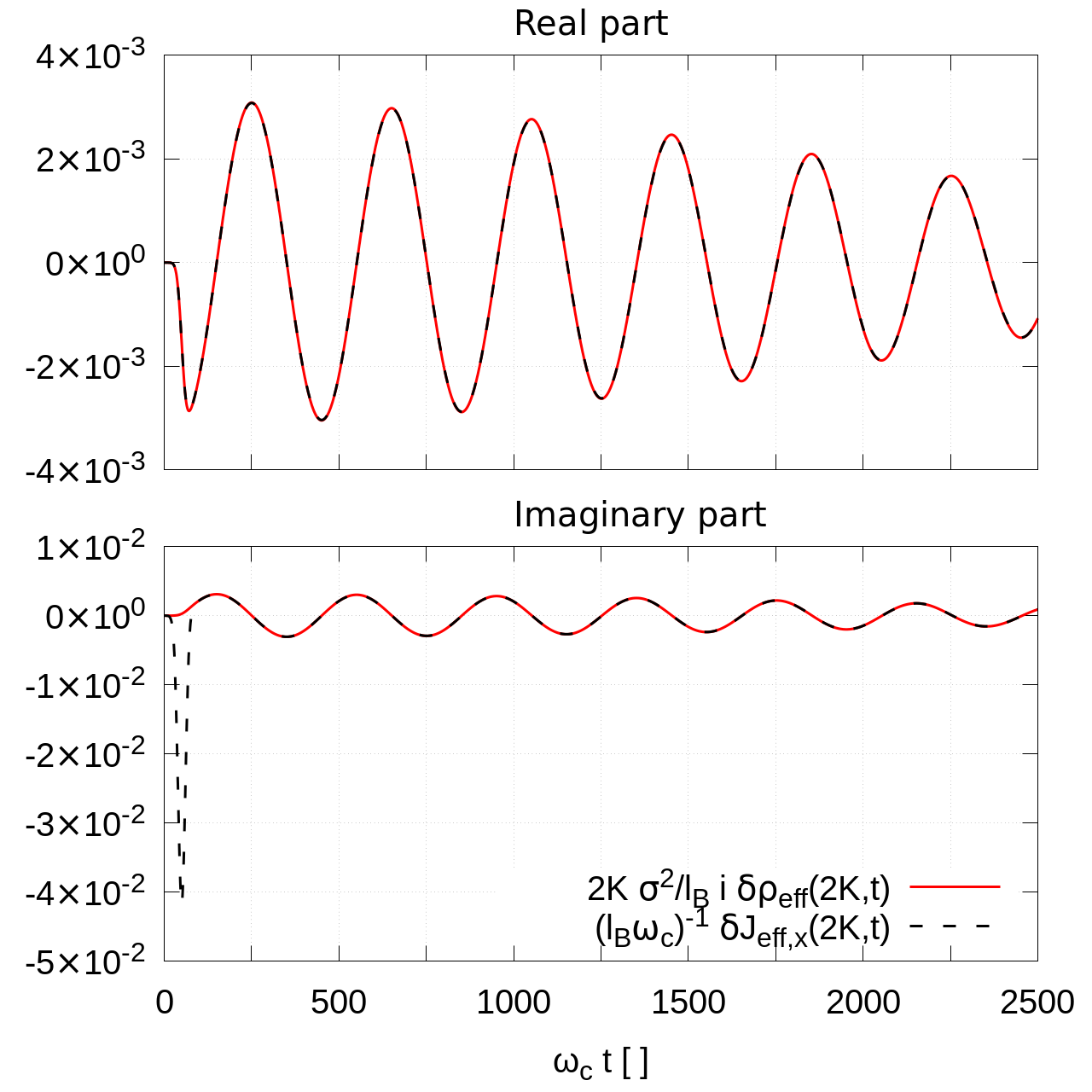}
	\end{minipage}%
	\begin{minipage}{0.5\textwidth}
		\centering
		\includegraphics[width=1.\textwidth]{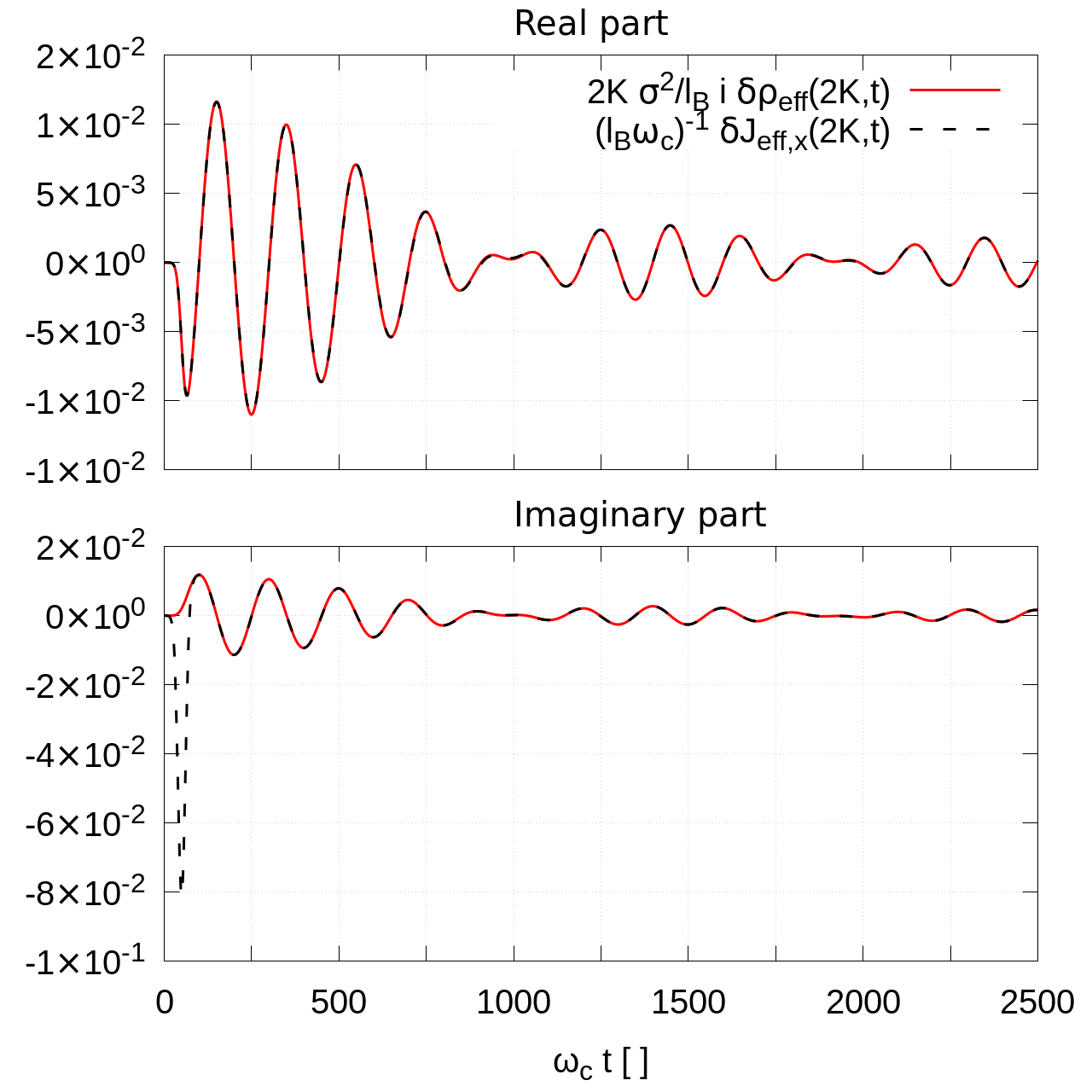}
	\end{minipage}    
	\caption[The LOF caption]{The two images show the real and imaginary parts of $\delta J_{\text{eff},\,x}(2K,t)$ as a function of time, compared with $i\, 2K\,\frac{\hbar}{m}\,\left(\frac{\sigma_{k_F}}{l_B}\right)^2 \delta\rho_\text{eff}(2K,t)$. 
	On the left hand side $2K=\frac{\pi}{50} l_B^{-1}$ was used, $2K=\frac{\pi}{25} l_B^{-1}$ for the other one. The system length was set to be $L_y=400l_B$, the Fermi wavevector $k_F\Delta k\simeq 8.58 l_B^{-1}$. $\lambda=0.005\hbar\omega_c$ has been used.}
	\label{fig:Jx}
\end{figure}

Finally, plugging eq. \ref{eq:pdx_sigmax} into eq. \ref{eq:x_integrated_edge_Jx}, and approximating $\sigma_{k_j}^2\approx \sigma_{k_F}$, we get
\begin{equation}
\label{eq:JxLinked}
\begin{split}
\frac{1}{L_y}\delta J_{\text{eff},\,x}(2K,t) \approx
i\, 2K\,\frac{\hbar}{m}\,\left(\frac{\sigma_{k_F}}{l_B}\right)^2 
\left(\frac{1}{L_y} \delta\rho_\text{eff}(2K,t)\right)
\end{split}
\end{equation}
so we see that $\delta J_{\text{eff},\,x}$ is proportional to $\delta\rho_\text{eff}$ and they have a $\frac{\pi}{2}$ phase difference. %A physical interpretation of this result will be given at the end of this section.
\newline From the comparison between numerical simulations and perturbation theory in Fig. \ref{fig:Jx} we that the prediction for $J_x$ fails at short times, when the perturbation is still \virgolette{on}, as indeed expected from the above discussion.

We see that $J_{\text{eff},x}$ does not vanish even if the band curvature $\partial^2_kE_k$ vanishes identically. From equations \ref{eq:JY_rho_linked} and \ref{eq:JX_rho_linked} we rather obtain
\begin{equation}
\label{eq:JX_rho_linked}
\left|\frac{\delta J_{\text{eff},\,x}(2K,t)}{\delta J_{\text{eff},\,y}(2K,t)}\right|\approx \frac{v}{2K\,\frac{\hbar}{m}\,\left(\frac{\sigma_{k_F}}{l_B}\right)^2 }
\end{equation}
i.e. $\delta J_{\text{eff},\,x}(2K,t)$ becomes negligible with respect to $\delta J_{\text{eff},\,y}(2K,t)$ if the excitation is long wavelength. This is not in contrast with the effective continuity equation \mbox{$\partial_t \rho_\text{eff}=-\partial_yJ_{\text{eff},y}$} we derived above; 
if $J_x$ vanishes both in the bulk and at the edge of the sample we have $\int \partial_x J_x dx=0$, so that once we integrate along x the \virgolette{exact} continuity equation $\partial_t\rho=-\boldsymbol{\nabla}\cdot\mathbf{J}$ we precisely get the previous one, even if $J_{\text{eff},x}\neq0$.

\subsection{Increasing the excitation strength}
\begin{figure}[htp!]
	\begin{minipage}{.5\textwidth}
		\centering
		\includegraphics[width=1.\textwidth]{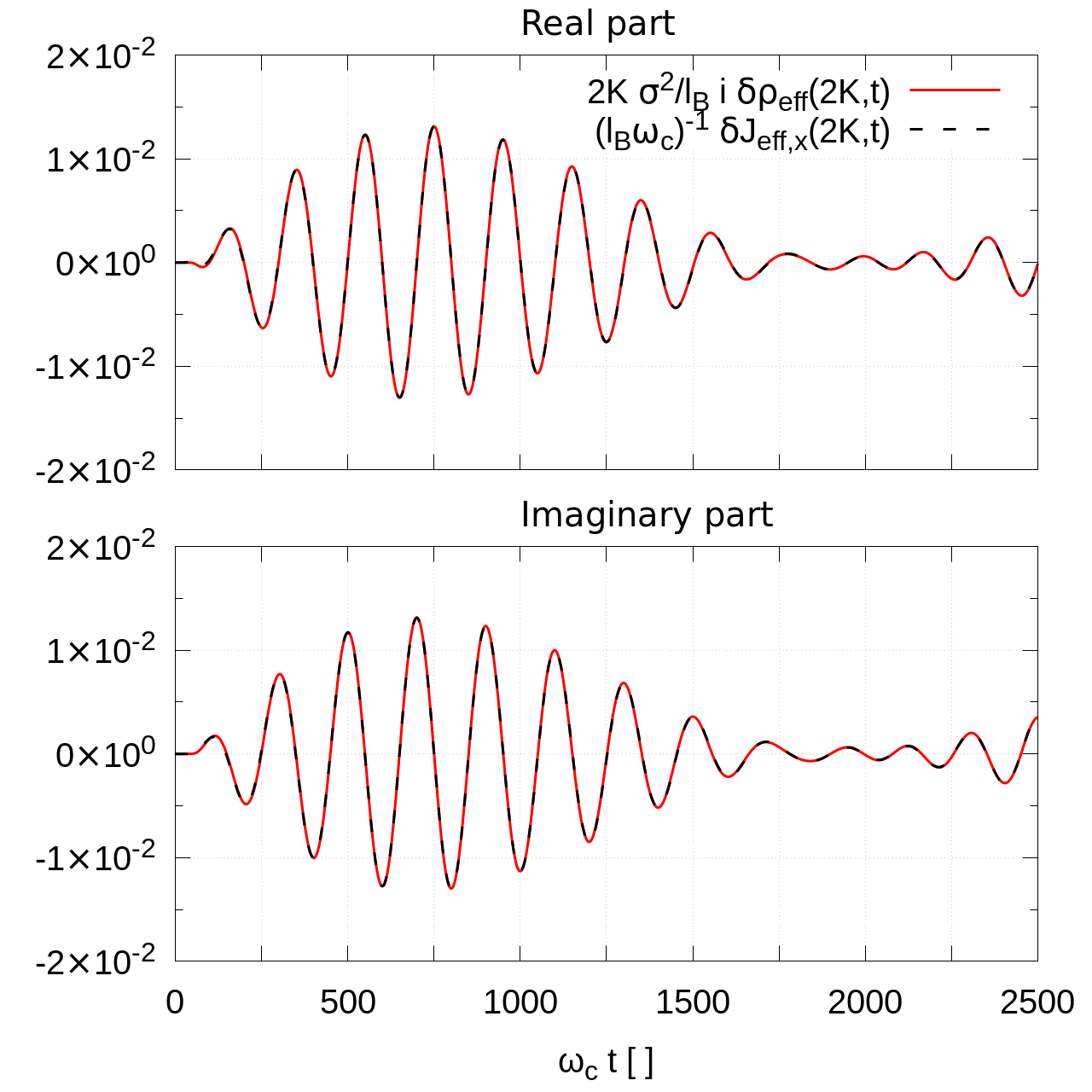}
	\end{minipage}%
	\begin{minipage}{0.5\textwidth}
		\centering
		\includegraphics[width=1.\textwidth]{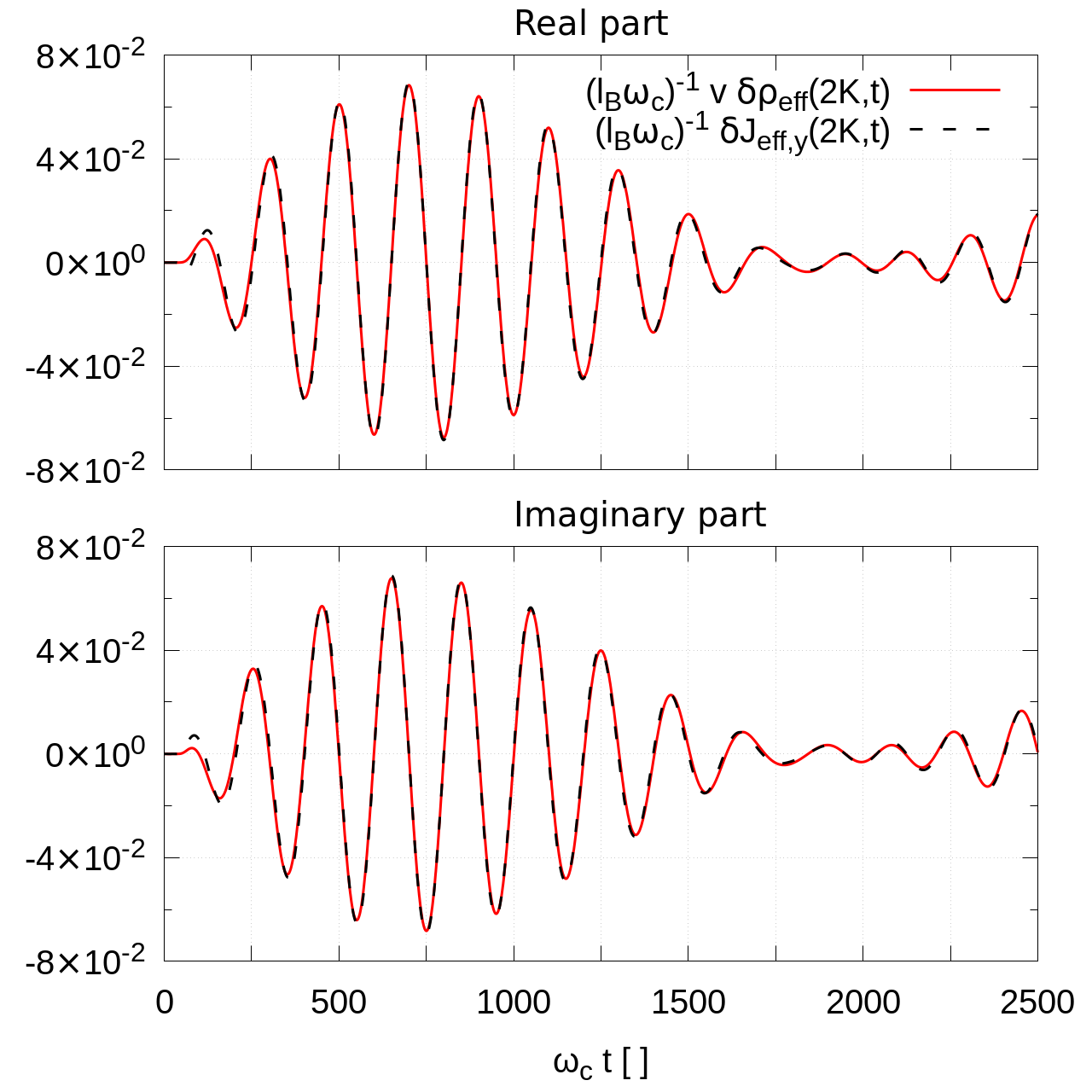}
	\end{minipage}    
	\caption[The LOF caption]{The two images compare the numerically computed real and imaginary parts of $\delta J_{\text{eff},\,x}(2K,t)$ (left hand side panel) and of $\delta J_{\text{eff},\,y}(2K,t)$ (right hand side panel) with the expressions in eq. \ref{eq:JxLinked} and \ref{eq:JY_rho_linked} respectively, in a high-excitation strength regime.	
	\newline	Both images have been obtained by using $2K=\frac{\pi}{50} l_B^{-1}$, $L_y=400l_B$, $k_F\simeq 8.58 l_B^{-1}$ and $\lambda=0.05\hbar\omega_c$.}
	\label{fig:JxHyNonPerturbative}
\end{figure}

The behaviour of the current modes is briefly discussed as the excitation strength is increased, based on the numerically obtained data.
\begin{figure}[htp!]
	\centering
	\includegraphics[width=.7\textwidth]{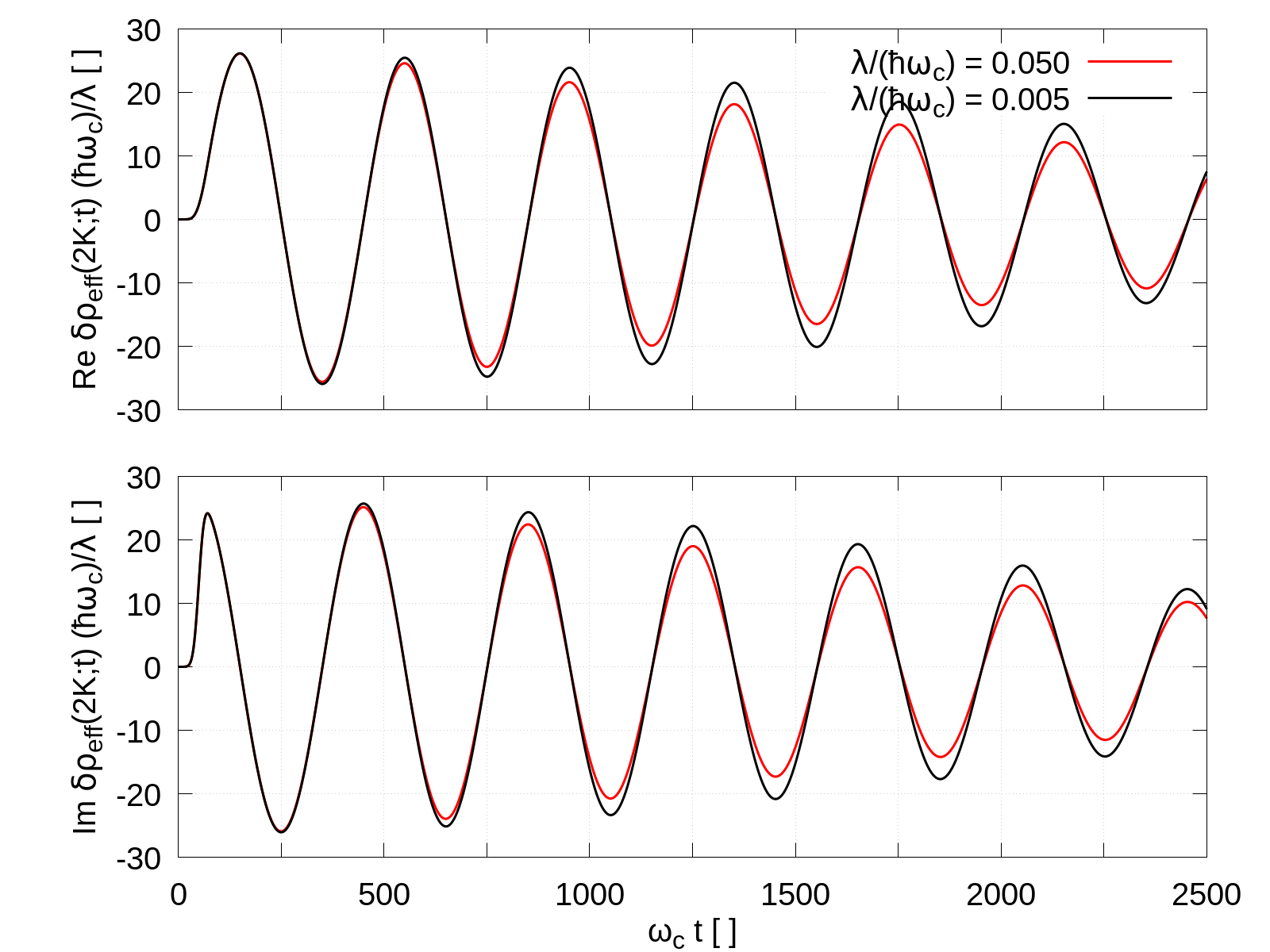}
	\caption[The LOF caption]{In the image the effective density variation $\delta\rho_\text{eff}(2K,t)$ real and imaginary parts are compared as a function of time in the cases $\lambda=0.005\hbar\omega_c$ and  $\lambda=0.050\hbar\omega_c$. The curves have been obtained with $2K=\frac{\pi}{50} l_B^{-1}$, $L_y=400l_B$, $k_F\simeq 8.58 l_B^{-1}$.}
	\label{fig:0005vs0050}
\end{figure}
As the excitation strength parameter $\lambda$ is increased, despite $\delta\rho_\text{eff}(2K,t)$ taking up non-negligible non-linear contributions (see Fig. \ref{fig:0005vs0050}) we can see in Fig. \ref{fig:JxHyNonPerturbative} that the relations between $\delta J_{\text{eff},\,x}$, $\delta J_{\text{eff},\,y}$ and $\delta\rho_\text{eff}$ derived above by using linear order perturbation theory (equations \ref{eq:JY_rho_linked} and \ref{eq:JxLinked}) still exhibit a good agreement, which makes us think that these expressions still retain their validity even beyond the linear perturbative order; this is reasonable, at least for $\delta J_{\text{eff},\,y}$, given its physical interpretation as being the effective current associated to the one-dimensional chiral edge mode.

\subsubsection{The bulk in the transient}\label{subsection:Jx_transient_bulk}
It is interesting to see how a large current towards $x$ can build up in the bulk as long as the perturbation is still on (as seen in the right hand side panel of Fig. \ref{fig:bulkJx_Jy}).

As seen above, excitations to higher Landau levels in the bulk cannot be neglected and they produce a sinusoidally modulated current. This can be shown more quantitatively. The $n=0$ contribution in eq. \ref{eq:current_jx_1} is identically vanishing; the first non-zero correction arises from the coupling to the $n=1$ Landau level. 
\newline Keeping only the $n=1$ level in eq. \ref{eq:current_jx_1} we get (in the bulk the energy differences $\Delta\omega$ equal plus or minus the cyclotron frequency $\omega_c$).
\begin{equation}
\label{eq:dJx_n=1}
\begin{split}
\delta J_x(x,2K;t)=&%\frac{1}{2}\sum_{k_0}\left[C_{k_0}^{(k_0)*}(-i\partial_x)C_{k_0+2K}^{(k_0)}+C_{k_0}^{(k_0-2K)*}(-i\partial_x)C_{k_0}^{(k_0)}+(2K\rightarrow-2K)^*\right]=
i\frac{\hbar\lambda}{2m}\sum_{k_0}\Biggl[\Biggr.
f^{(0,k_0)}_{1,k_0+2K}e^{-i\omega_c t}
\Bigl[
\Phi_{1,k_0+2K}(x)\partial_x\Phi_{0,k_0}(x)-\Phi_{0,k_0}(x)\partial_x\Phi_{1,k_0+2K}(x)
\Bigr]
+\\&+
f^{(0,k_0)*}_{1,k_0-2K}e^{i\omega_c t}
\Bigl[ 
\Phi_{0,k_0}(x)\partial_x\Phi_{1,k_0-2K}(x)-\Phi_{1,k_0-2K}(x)\partial_x\Phi_{0,k_0}(x)
\Bigr]
\Biggl.\Biggr].
\end{split}
\end{equation}
The matrix element $d^{k_0,k_0\pm2K}_{0,1}$ contained in $f^{(0,k_0)}_{1,k_0\pm2K}$ is odd in $2K$ and can be series-expanded in the long wavelength limit
\begin{equation}
d^{k_0,k_0+2K}_{0,1} = \frac{2K l_B\,e^{-\left(\frac{2Kl_B}{2}\right)^2}}{\sqrt{2}}\simeq\frac{2Kl_B}{\sqrt{2}} +\mathcal{O}(2K)^3.
\end{equation}
Looking at eq. \ref{eq:f}  and eq. \ref{eq:ft} we can notice that $f^{(0,k_0)*}_{1,k_0-2K}e^{i\omega_c t}=-\left(f^{(0,k_0)}_{1,k_0+2K}e^{-i\omega_c t}\right)^*$. 
It suffices therefore to analyse only one of the two, e.g. $f^{(0,k_0)}_{1,k_0+2K}e^{-i\omega_c t}$, which more explicitly reads (using $t_0\gg\tau$)
\begin{equation}
\begin{split}
f&^{(0,k_0)}_{1,k_0+2K}e^{-i\omega_c t}=
%\frac{i}{4\hbar}d^{k_0,k_0+2K}_{0,1}F_t(-\omega_c)e^{-i\omega_c t}=
\\&
\frac{i}{4\hbar}d^{k_0,k_0+2K}_{0,1}e^{-i\omega_c (t-t_0)}
\frac{\sqrt{\pi}}{2}
\,\tau\,
e^{-\left(\frac{\omega_c\tau}{2}\right)^2}
\,\left[\text{erf}\left(\frac{t-t_0}{\tau}-i\frac{\omega_c\tau}{2}\right)+1\right]
\end{split}
\end{equation}
Since the perturbation is adiabatic with respect to the microscopic dynamics of the system we have $\omega_c\tau\gg 1$; we can then series expand the error function and neglect exponentially small terms
\begin{equation}
f^{(0,k_0)}_{1,k_0+2K}e^{-i\omega_c t}=\frac{1}{4\hbar\omega_c}d^{k_0,k_0+2K}_{0,1}\,e^{-\left(\frac{t-t_0}{\tau}\right)^2}.
\end{equation}
eq. \ref{eq:dJx_n=1} then becomes
\begin{equation}
\begin{split}
\delta J_x(x,2K;t)\simeq
i\frac{\lambda}{2}&\,
\frac{1}{4m\omega_c}\,\frac{2Kl_B}{\sqrt{2}}\,e^{-\left(\frac{t-t_0}{\tau}\right)^2}
\\&\sum_{k_0}\Biggl[\Biggr.
\Phi_{1,k_0+2K}(x)\partial_x\Phi_{0,k_0}(x)-\Phi_{0,k_0}(x)\partial_x\Phi_{1,k_0+2K}(x)
-\\&-
\Phi_{0,k_0}(x)\partial_x\Phi_{1,k_0-2K}(x)+\Phi_{1,k_0-2K}(x)\partial_x\Phi_{0,k_0}(x)
\Biggl.\Biggr].
\end{split}
\end{equation}
The term within the big brackets is even in $2K$, and the lowest order term in the series expansion is therefore non-vanishing in the long-wavelength excitation limit\footnote{As opposed to what happens in the case of $\delta\rho$ (as seen in eq. \ref{eq:bulk_n=1_density_variation})}.
\newline Up to higher order correction terms in $2K$ (which are $\mathcal{O}(2K)^3$) then
\begin{equation}
\label{eq:dJx_n=1_2}
\begin{split}
\delta J_x(x,2K;t)&\simeq
i\frac{\lambda}{4m\omega_c}\,\frac{2Kl_B}{\sqrt{2}}\,e^{-\left(\frac{t-t_0}{\tau}\right)^2}
\sum_{k_0}\Bigl[
\Phi_{1,k_0}(x)\partial_x\Phi_{0,k_0}(x)-\Phi_{0,k_0}(x)\partial_x\Phi_{1,k_0}(x)\Bigr].
\end{split}
\end{equation}
Using the Harmonic oscillator wavefunctions it is a matter of straightforward algebra to obtain
\begin{equation}
\Phi_{1,k_0}(x)\partial_x\Phi_{0,k_0}(x)-\Phi_{0,k_0}(x)\partial_x\Phi_{1,k_0}(x)=-\frac{1}{l_B^2}\,\sqrt{\frac{2}{\pi}}\,e^{-\left(\frac{x}{l_B}+k_0 l_B\right)^2}.
\end{equation}
\begin{figure}[htp!]
	\centering
	\includegraphics[width=.8\textwidth]{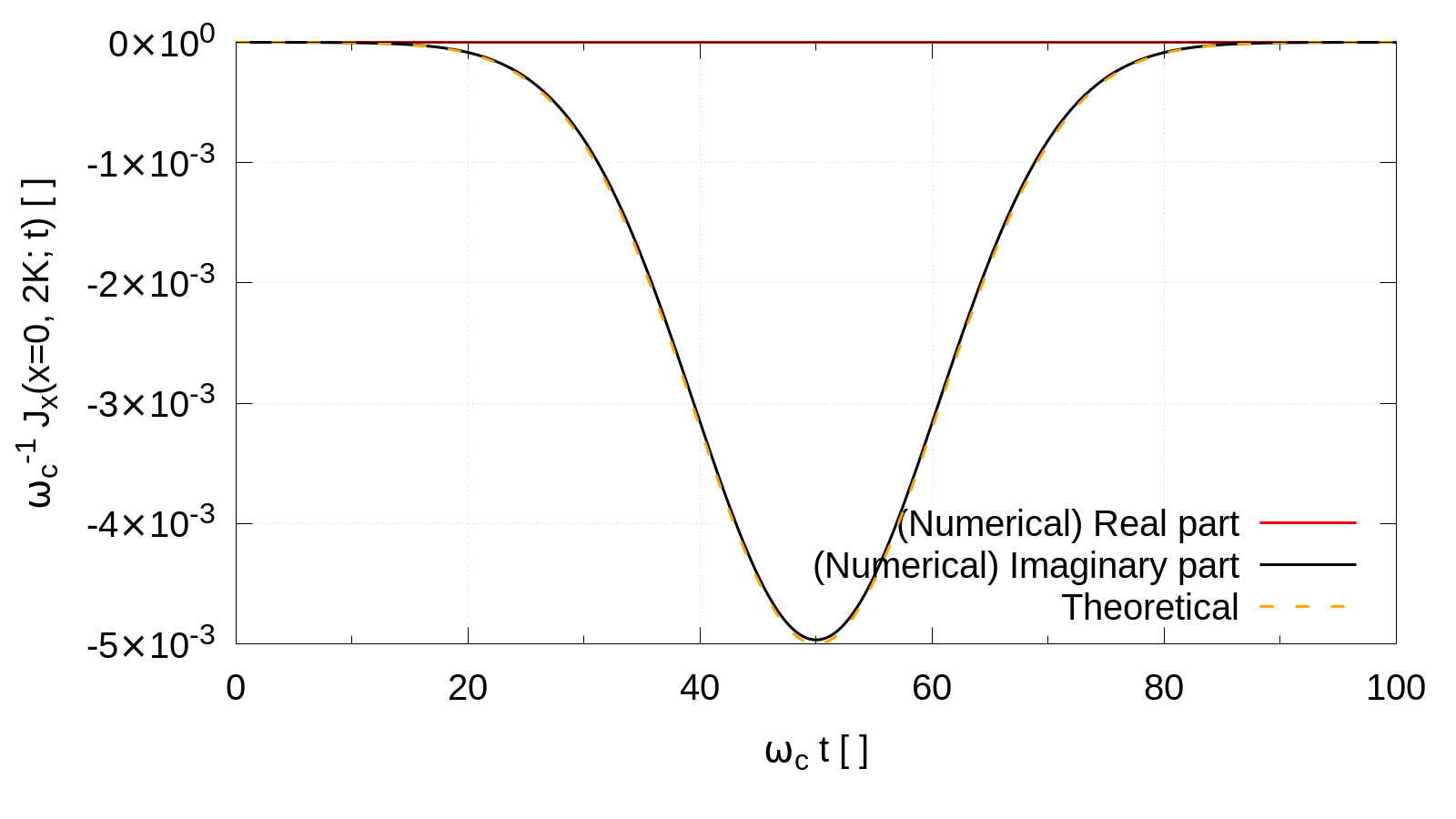}
	\caption[The LOF caption]{The expression \ref{eq:bulk_jx} for the bulk Fourier transform of the $x$-component of the probability current is tested against the numerically computed data, obtained by performing a simulation with $L_y=200l_B$, $k_F=273\Delta k\simeq 8.58l_B^{-1}$, $\lambda=0.005\hbar\omega_c$ and $2K=4\Delta k$.}
	\label{fig:bulk_jx}
\end{figure}
In the thermodynamic limit the summation is $x$-independent and becomes quite large, effectively compensating for the small matrix element $d^{k_0,k_0+2K}_{0,1}$
\begin{equation}
\sum_{k_0}
\Phi_{1,k_0}(x)\partial_x\Phi_{0,k_0}(x)-\Phi_{0,k_0}(x)\partial_x\Phi_{1,k_0}(x)\rightarrow
-\frac{\sqrt{2}}{l_B^3\,\Delta k}.
\end{equation}
Plugging this result back into eq. \ref{eq:dJx_n=1_2} we finally obtain
\begin{equation}
\label{eq:bulk_jx}
\begin{split}
\frac{1}{L_y}\delta J_x(x,2K;t)&\simeq
-i\frac{\lambda}{4\hbar}\,\,\frac{2K}{2\pi}\,e^{-\left(\frac{t-t_0}{\tau}\right)^2}
\end{split}
\end{equation}
and thus
\begin{equation}
\begin{split}
\delta J_x(x,y;t)\simeq
\frac{\lambda}{2\hbar}\,\frac{2K}{2\pi}\,e^{-\left(\frac{t-t_0}{\tau}\right)^2}\sin(2K y)
\end{split}
\end{equation}
i.e. the $x$ component of the probability current adiabatically follows the external perturbation, out of phase by $\frac{\pi}{2}$ though, as expected from the classical picture of electrons drifting orthogonally to the local electric field, with velocity proportional to its gradient (as seen in Sec. \ref{section:classical_electric_field}); notice that $\lambda\,2K\,e^{-\left(\frac{t-t_0}{\tau}\right)^2}\sin(2K y)$ is the $y$-derivative of eq. \ref{eq:sinusoidal_excitation} indeed.
\newline A comparison between $\delta J_x(x,2K;t)$ evaluated at $x=0$ from the numerical data and the one obtained in eq. \ref{eq:bulk_jx} is shown if Fig. \ref{fig:bulk_jx}. 
%!TEX root = ../main.tex
% Chapter Template

\graphicspath{{./pic5/}}

\chapter{Gaussian excitation}\label{ch5}
\lhead{Chapter 6. \emph{Gaussian excitation}} % Change X to a consecutive number; this is for the header on each page - perhaps a shortened title

After having studied in the previous chapter the system response to a spatially periodic excitation, we now turn to the case of a spatially localised excitation generated by a Gaussian external potential
\begin{equation}
\label{eq:gaussian_external_exc}
V(y,t)=-\lambda\, \xi(t) e^{-\left(\frac{y}{\sigma}\right)^2} =-\lambda \,e^{-\left(\frac{t-t_0}{\tau}\right)^2} e^{-\left(\frac{y}{\sigma}\right)^2}
\end{equation}
uniform in the $x$ direction.
This will help us to develop some understanding on the dynamics (both linear and non-linear) of a localised perturbation on top of the ground state density of a $\nu=1$ integer quantum Hall state, in particular the interplay between the band curvature and the non-linear effects.

The parameters have been chosen so that both at $t=0$ and at $y=\pm\frac{L_y}{2}$ the perturbation is negligibly small (notice however that by symmetry the periodic boundary conditions are exactly satisfied).
The results have been compared with those obtained by using time-dependent perturbation theory, which as in the previous chapter allow to make some simple considerations (valid in the weak excitation regime) as well as to get rid of the dependence on the choice of the boundary conditions along the y direction by taking the thermodynamic limit.
\newline The non-linear dynamics generated by non-perturbative excitation strength is then studied.

\noindent Throughout the whole section $\omega_c\tau=15$ and $\omega_c t_0=50$ have been used (except if otherwise specified), as well as $L_x=20l_B$. The confining potential parameters used are $V_0=30\hbar\omega_c$ and $\sigma_c=0.1l_B$, which give a steep rise of the confining potential on a lengthscale roughly of the order of the magnetic length $l_B$.
\newline The system length along $y$ has been fixed at $L_y=800l_B$, in order to make as more irrelevant as possible the presence of the periodic boundary conditions. The spatial width of the excitation has been fixed at $\sigma=20l_B$.
The Fermi wavevector has been fixed at $k_F=1148\Delta k\simeq 9.02l_B$, or equivalently $N=2297$ electrons occupying the lowest Landau level.
\newline The strength parameter $\lambda$ of the perturbation on the other hand has been varied and its values will therefore be explicitly written from time to time when looking at numerical results. 

\section{Semiclassical considerations}\label{section:semiclassical_considerations_gaussian}
In this short section the shape of the density at the edge is understood in semiclassical terms by analysing the electron current flow generated by the Gaussian excitation eq. \ref{eq:gaussian_external_exc}. 
The system response when a more complicated potential (localised near the edges) is used is briefly analysed and the numerical results compared with the ones simpler case (eq. \ref{eq:gaussian_external_exc}) treated in this chapter.
\begin{figure}[htp!]
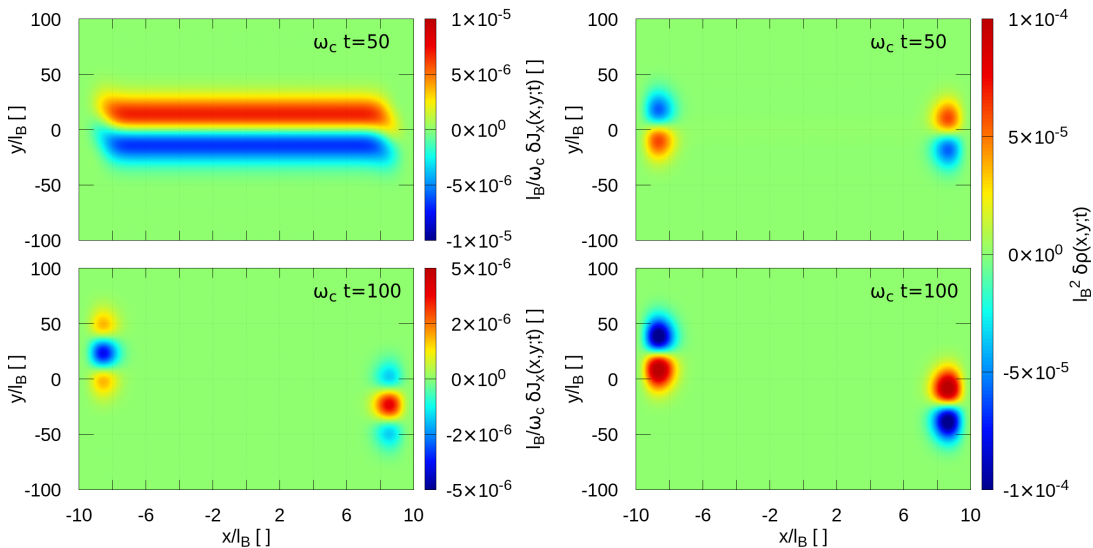

	\begin{minipage}{.5\textwidth}
		\centering
		\includegraphics[width=1.\textwidth]{jx_t=50,100.png}
	\end{minipage}%
	\begin{minipage}{0.5\textwidth}
		\centering
		\includegraphics[width=1.\textwidth]{density_50_100.png}
	\end{minipage}    
	\caption[The LOF caption]{On the left hand side the numerically computed $x$-component of the probability current density is shown in the case $\frac{\lambda}{\hbar\omega_c}=0.001$, at times $\omega_c t=50$ and $\omega_c t=100$. The picture on the right hand side shows the variation of the density with respect to the ground state one, under the same circumstances.
		\newline Notice that the plots have been restricted to a subdomain of the full $y$ length $L_y=800l_B$.}
	\label{fig:transient_current_and_density}
\end{figure}
\noindent During the transient we classically expect that currents will start to flow orthogonally to the local electric field $\propto \partial_y V(y,t) \,\hat{y}$. Some electrons will flow from one side of the sample to the opposite one, symmetrically creating a density dip and a bump localised near the edges.
\newline The qualitatively described behaviour\footnote{The resulting density variation could as well have been guessed based on the effective one-dimensional dynamics equation eq. \ref{eq:coordinate_spce_effective_1d_dynamics}, since the time derivative of the one-dimensional effective density is proportional to (minus) the gradient of $V(y,t)$.} can be recognised in Fig. \ref{fig:transient_current_and_density}.
\begin{figure}[htp!]
	\begin{minipage}{.5\textwidth}
		\centering
		\includegraphics[width=1.\textwidth]{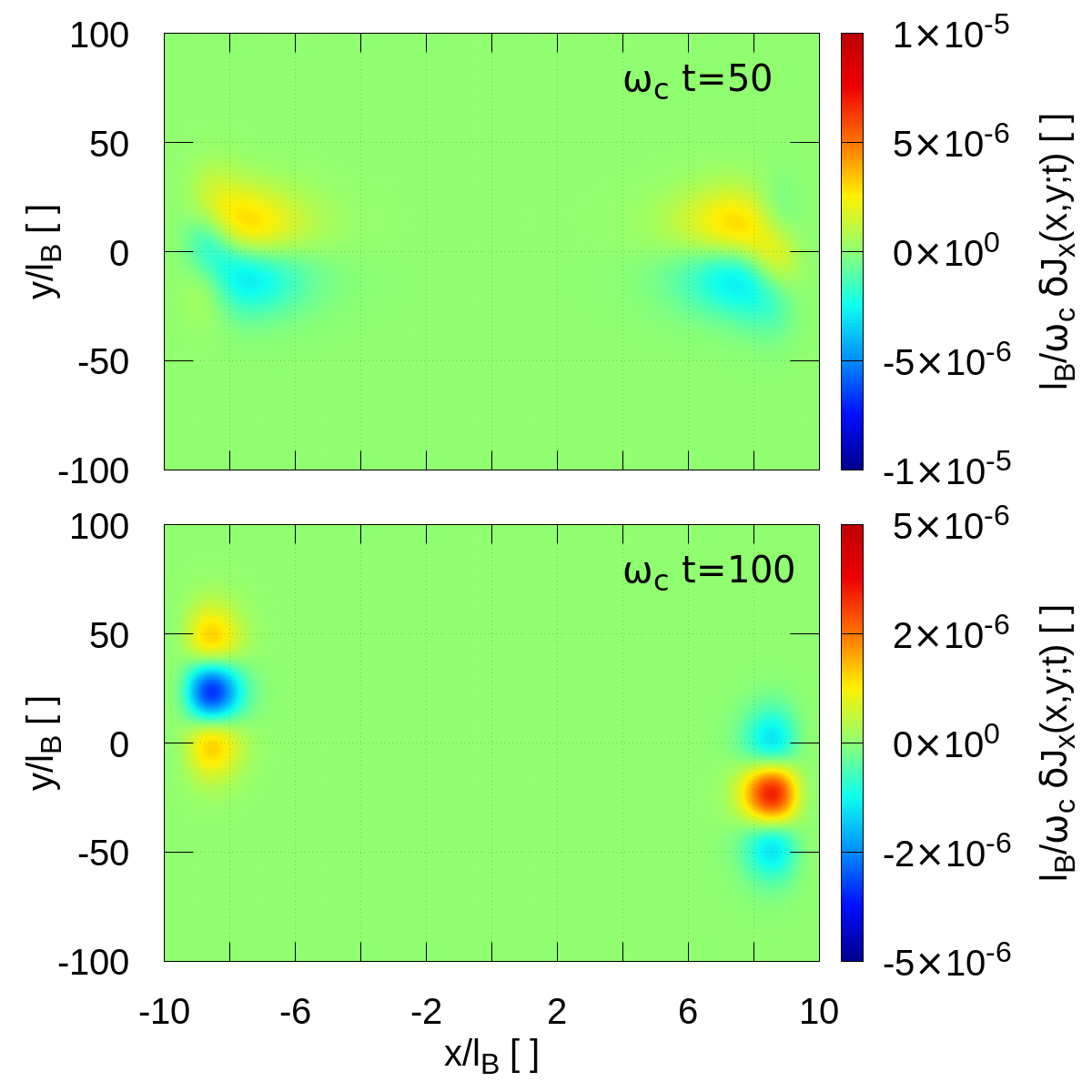}
	\end{minipage}%
	\begin{minipage}{0.5\textwidth}
		\centering
		\includegraphics[width=1.\textwidth]{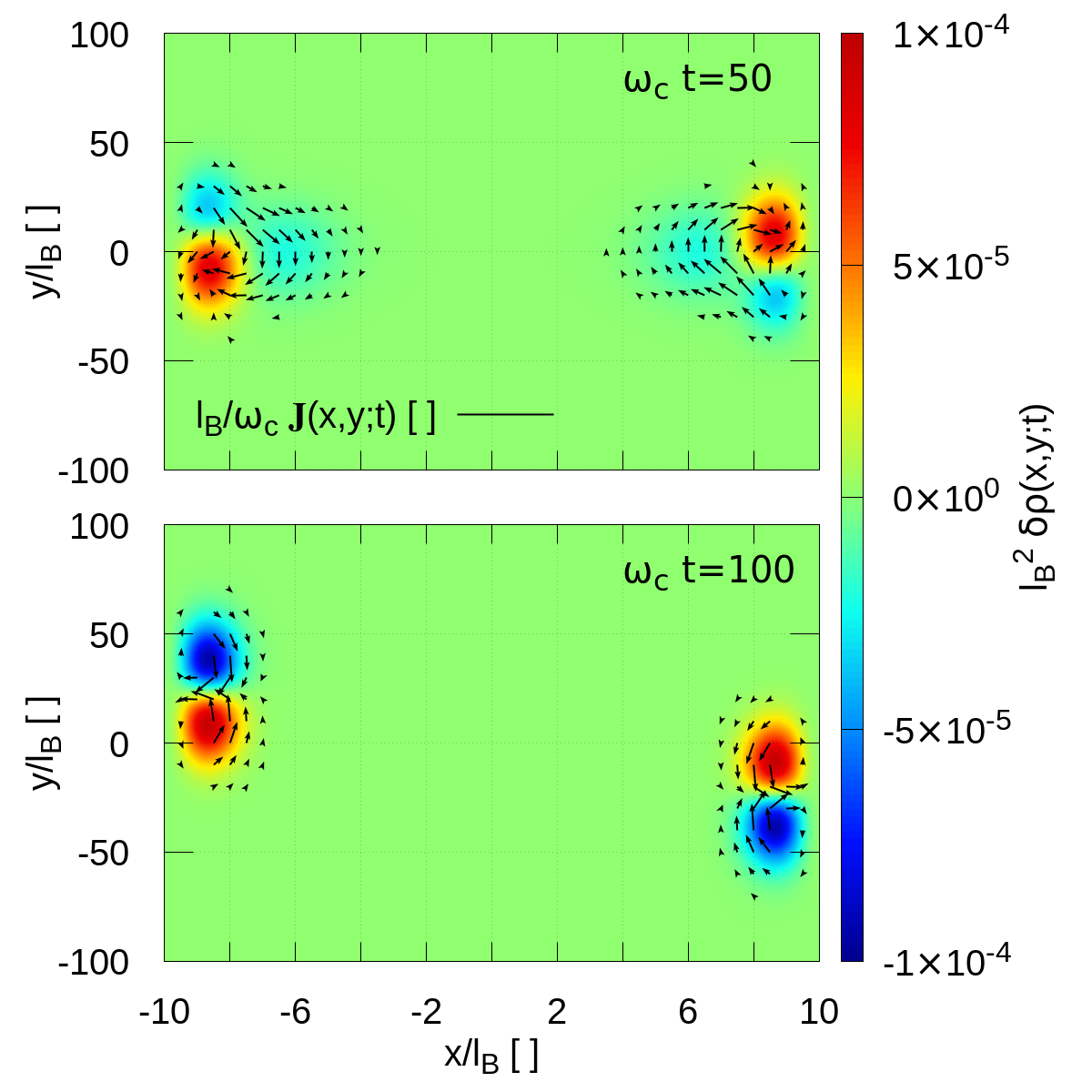}
	\end{minipage}    
	\caption[The LOF caption]{The two pictures have been generated with an excitation analogous to the one used in the chapter (eq. \ref{eq:gaussian_external_exc}), but localised at the system edges and vanishing in the system's bulk. The model potential has been chosen of the form
	\[
		V'(x,y;t)=\frac{V(y,t)}{1+\exp\left[-\frac{x^2-\left(\frac{L_\text{e}}{2}\right)^2}{\sigma_\text{e}L_\text{e}}\right]}
	\]
	with $L_\text{e}=15l_B$ and $\sigma_\text{e}=0.5l_B$.
	\newline On the left hand side the $x$-component of the probability current density is shown in the case $\frac{\lambda}{\hbar\omega_c}=0.001$, at times $\omega_c t=50$ and $\omega_c t=100$. The picture on the right hand side shows the variation of the density with respect to the ground state one, under the same circumstances. The probability current density field has been plotted as black arrows with length proportional to the field magnitude. Too short arrows are not shown. 
	\newline The two pictures serve as a comparison between the \virgolette{extended} version of the potential and should be compared with those in Fig. \ref{fig:transient_current_and_density}.}
	\label{fig:external_potential_x_localised}
\end{figure}
As already noted in the previous section (fig. \ref{fig:bulkJx_Jy}) during the transient the $x$-component of the probability current extends over the whole sample (from $\sim-\frac{L_x}{2}$ to $\sim\frac{L_x}{2}$). This is a consequence of the externally applied potential being $x$-independent; if it vanishes in the bulk, $J_x$ is localised near the edge also during the transient, as can be seen in Fig. \ref{fig:external_potential_x_localised}, which have been generated by using a potential localised at the system edges. 
\newline For completeness also the $y$ component of the probability current is compared in Fig. \ref{fig:JyComparison} for the two cases. One immediately notices that the qualitative differences are substantial as long as the excitation is still \virgolette{turned on}, but the profiles are qualitatively very similar once it has been turned off. Although the fine details will depend on the excitation to which the system has been subjected, the considerations which can be obtained by using the substantially simpler eq. \ref{eq:gaussian_external_exc} are expected to hold even in these more complicated cases.

\begin{figure}[htp!]
	\begin{minipage}{.5\textwidth}
		\centering
		\includegraphics[width=1.\textwidth]{jy_t=50,100.png}
	\end{minipage}%
	\begin{minipage}{0.5\textwidth}
		\centering
		\includegraphics[width=1.\textwidth]{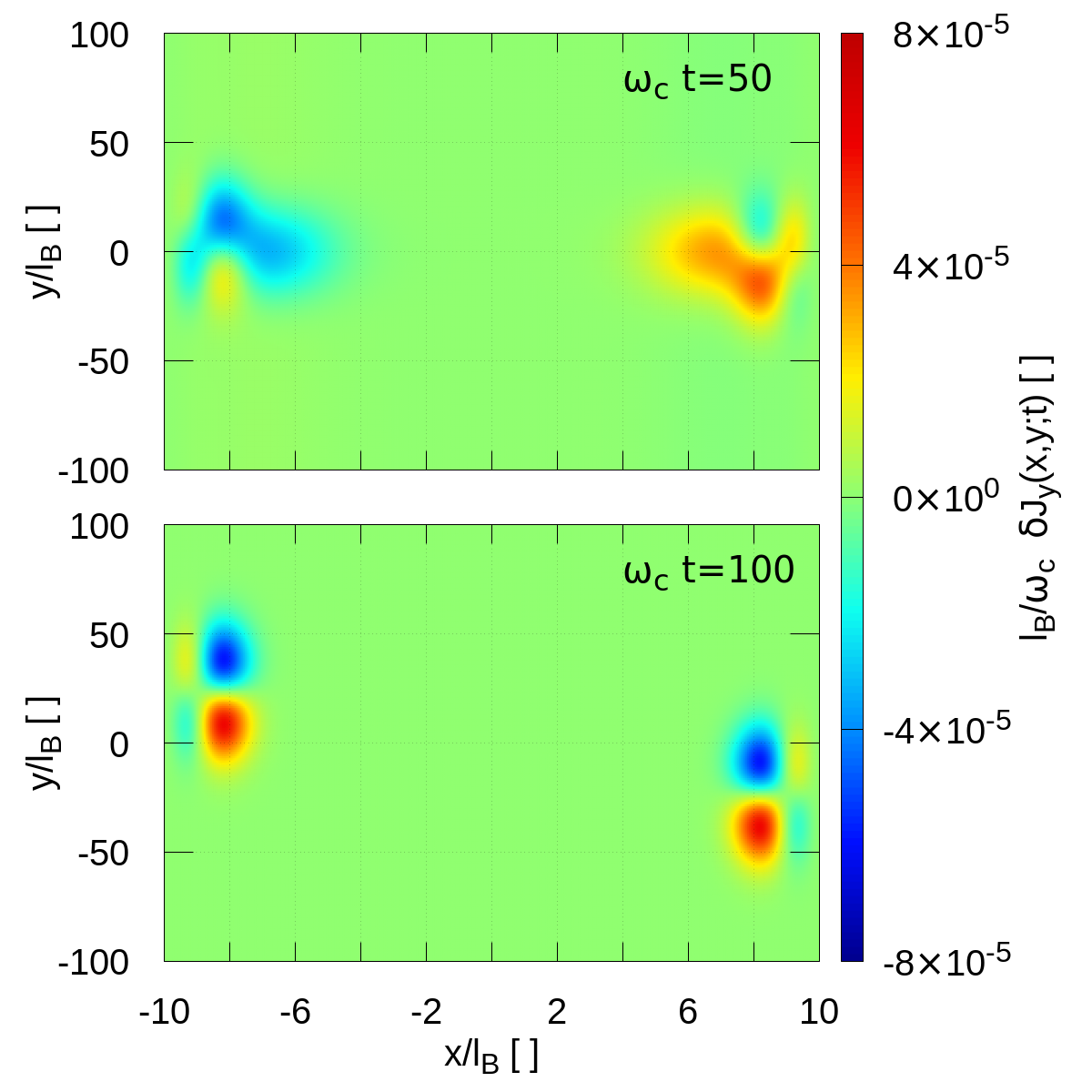}
	\end{minipage}    
	\caption[The LOF caption]{The two pictures compare the $y$ component of the probability current density (from which the ground state one has been subtracted off) in the two cases discussed above: when the perturbation extends into the system's bulk (left hand side picture) and when it is localised at the edges (on the right), at times  $\omega_c t=50$ (excitation transient) and $\omega_c t=100$ (the excitation has been turned off). $\frac{\lambda}{\hbar\omega_c}=0.001$ has been used.
	\newline For the right hand side image the excitation potential is the same as the one described in \ref{fig:external_potential_x_localised}.}
	\label{fig:JyComparison}
\end{figure}

\section{Perturbative computation of the one-body density}
The diagonal part of the one-body density matrix is perturbatively computed at linear order (much of the details will be skipped though, being exactly analogous to the calculations performed in Chapter \ref{ch4}) and used to study the edge dynamics. 
The obtained expression is further simplified by making some approximations on the structure of the levels at the Fermi point. From the closed formula which is obtained some simple predictions can in turn be made.

\begin{figure}[htp!]
	\centering
	\includegraphics[width=1.\textwidth]{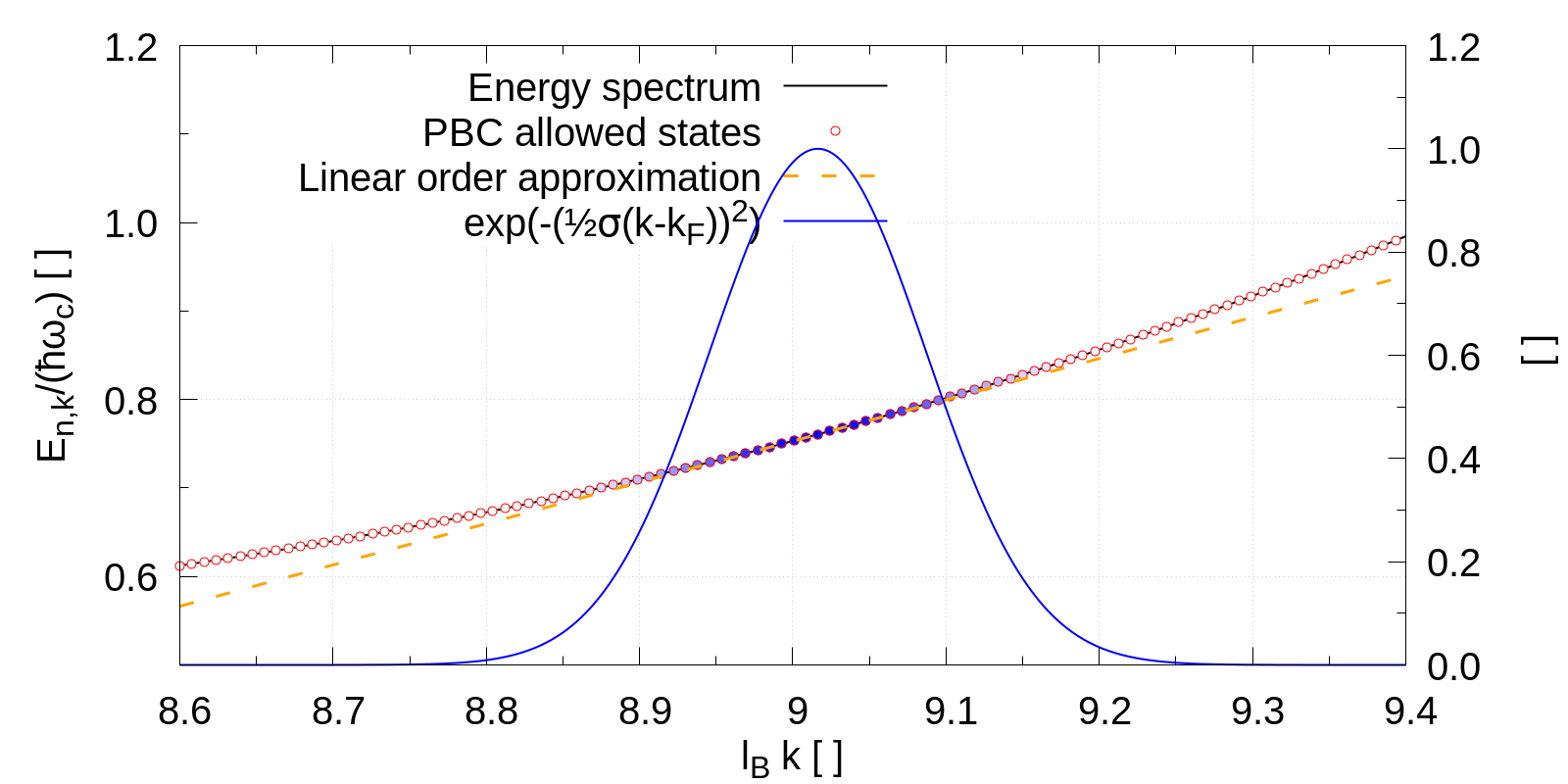}	
	\caption[The LOF caption]{On the main y-axis the allowed electronic levels are shown near the Fermi point at positive momentum. On the second axis $\exp\left[-\left(\frac{(k-k_F)\sigma}{2}\right)^2\right]$, which is the Fourier transform of the external excitation's spatial profile $e^{-\frac{y^2}{\sigma^2}}$ (apart for a constant which is irrelevant for the discussion) is shown instead, centred at the Fermi wavevector. The allowed electronic levels are coloured according to the magnitude of such a Fourier transform.}
	\label{fig:EnergySpectrumAndGaussian}
\end{figure}

Using the same procedure which has been explicitly carried out in sec. \ref{subsection:sin_PT}, one obtains a set of first order linear differential equations for the Fourier coefficients of the expansion of the system wavefunction over unperturbed Landau levels
\begin{equation}
\label{eq:diffeq_foucoeff_gauss}
\frac{\partial c_{n,k}}{\partial t}=i\,\xi(t)\,\frac{\lambda}{\hbar}\,\frac{\sigma}{L_y}\sum_{m,q} \sqrt{\pi}\,e^{-\left(\frac{\sigma}{2}(q-k)\right)^2}\,d_{n,m}^{k,q}e^{i\Delta\omega^{k,q}_{n,m}t}\,c_{m,q}(t)
\end{equation}
where the $d_{n,m}^{k,q}$ are the matrix elements between two different eigenstates of the unperturbed Hamiltonian and $\Delta\omega^{k,q}_{n,m}$ the difference between their eigenfrequencies.
Notice that the Fourier transform $\propto e^{-\left(\frac{\sigma}{2}(q-k)\right)^2}$ of the spatial part of the external potential (eq. \ref{eq:gaussian_external_exc}) makes irrelevant contributions to the summation at momenta which are greater than some $\sim \sigma^{-1}$.
In Fig. \ref{fig:EnergySpectrumAndGaussian} the allowed electronic levels are shown near the Fermi point at positive momentum, and are coloured according to the magnitude of the Fourier transform of the external potential spatial profile centred at the Fermi wavevector.

The integration of eq. \ref{eq:diffeq_foucoeff_gauss} with initial condition $c_{n,k}(t_0)=\delta_{k,k_0}\delta_{n,n_0}$ is straightforward at the lowest perturbative order. Setting $c_{n,k}(t)=c_{n,k}^{(n_0,k_0)}=c_{n,k}^{(n_0,k_0)}(t_0)+\lambda f_{n,k}^{(n_0,k_0)}(t)$ with $f_{n,k}^{(n_0,k_0)}(t_0)=0$ we get
\begin{equation}
f_{n,k}^{(n_0,k_0)}(t)=\frac{i}{\hbar}\,\frac{\sigma}{L_y} \sqrt{\pi}\,e^{-\left(\frac{\sigma}{2}(k-k_0)\right)^2}\,d^{k_0,k}_{n_0,n} \,F_t\left(\Delta\omega^{k_0,k}_{n_0,n}\right).
\end{equation}
where $F_t(\omega)$ is the same function defined in eq. \ref{eq:ft}. 
We again compute the density variation (referred to the one of the unperturbed system)
\begin{equation}
\delta\rho_{n_0,k_0}(x,y;t)=2\lambda \sum_{m,q}\Re\left[f_{m,q}^{(n_0,k_0)}(t)\,e^{i\Delta\omega_{n_0,m}^{k_0,q}t}\,\Phi_{n_0,k_0}(x)\Phi_{m,q}(x)\frac{e^{i(q-k_0)y}}{L_y}\right]
\end{equation}
(which is exactly analogous to eq. \ref{eq:density_variation_general_form_real_space}).
Following the same procedure used in section \ref{subsection:sin_PT}, we compute the $y$-Fourier transform of the density variation associated to an electron initially in the quantum state labelled by $(n_0,k_0)$
\begin{equation}
\delta\rho_{n_0,k_0}(x,q;t)= 2\lambda \sum_m \left[
f_{m,k_0+q}^{(n_0,k_0)}(t)\,e^{i\Delta\omega_{n_0,m}^{k_0,k_0+q}t}\,\Phi_{n_0,k_0}(x)\Phi_{m,k_0+q}(x) 
+
(q\rightarrow-q)^*\right].
\end{equation}
In order to obtain the non-interacting many-body state we just sum over the indices labelling the initially occupied electronic states. As before for simplicity and in order to obtain analytically closed formulae we consider a single Landau level (and omit the relative label). 

In the bulk the same cancellations discussed in section \ref{subsection:sin_PTin_bulk} do occur, owing to its incompressibility; near the Fermi points however some terms do survive. Considering just a single edge we get, analogously to \ref{eq:pert_sin_general_single_edge_1bd_variation}
\begin{equation}
\label{eq:drho_gauss_general}
\delta\rho(x,q;t)
=i\,\frac{\lambda\sigma}{2\sqrt{\pi}\hbar}\,\,e^{-\left(\frac{\sigma q}{2}\right)^2}
\,\frac{2\pi}{L_y}\sum_{j=0}^{\frac{q}{\Delta k}-1}F_t(\Delta \omega_{k_j,k_j+q})e^{i\Delta \omega_{k_j,k_j+q}\,t}\, d_{k_j,k_j+q} \Phi_{k_j}(x)\Phi_{k_j+q}(x)
\end{equation}
where $k_j=k_F-j \Delta k$. 
As expected we see that for long wavelength excitations (large enough $\sigma$) only electrons with $k\sim k_F$ are relevant. The expression is exactly analogous to that obtained in the previous chapter, the only difference being the $y$ Fourier transform of the external potential. The discussion of the result is analogous and much will henceforth be omitted.

\subsubsection*{Linear dispersion approximation}
If we can linearise the dispersion relation at the Fermi surface $\omega_k\sim\omega_{k_F}+v(k-k_F)$, integrating out the $x$ variable and making the usual long wavelength approximation $d_{k_F,k_F+q}\sim1$ (i.e. following the same steps as those explicitly performed in section \ref{section:linear_dispersion_approximation_sin}) we obtain 
\begin{equation}
\label{eq:momentum_space_eff_drho}
\delta\rho_\text{eff}(q;t)=
i \,\frac{\lambda\sigma}{\hbar}\,\frac{q}{2\sqrt{\pi}}\,e^{-\left(\frac{\sigma q}{2}\right)^2}\,e^{-i q v t}\,F_t(-q v)
\end{equation}
which is exactly what one gets by integrating eq. \ref{eq:effective_1d_dynamics}, i.e. the same solution which can be derived from the effective density equation derived in the third chapter, which predicts that once the perturbation has been turned off the excited density variation propagates rigidly.
\newline Notice that this is vanishes as $q\rightarrow 0$, as required by the conservation of the number of the electrons in the half system when no net electronic transfer occurs between the two edges.

In the thermodynamic limit the real space effective density variation can be analytically computed at any time by inverting eq. \ref{eq:momentum_space_eff_drho}. This general and quite lengthy computation does not provide much insight though; the solution at large times is simpler and more interesting
\begin{equation}
\label{eq:real_space_eff_drho}
\delta\rho_\text{eff}(y;t)=
-\frac{\lambda\sigma\tau}{8\sqrt{\pi}\hbar}
\frac{\Delta}{\alpha^3}\,\exp\left[-\left(\frac{\Delta}{2\alpha}\right)^2\right]
\end{equation}
where $\Delta=y-v (t-t_0)$ and $\alpha^2=\left(\frac{\sigma}{2}\right)^2+\left(\frac{v \tau}{2}\right)^2$; as expected the result depends only on $\Delta$, and it is antisymmetric in $\Delta$. 
\begin{figure}[htp!]
	\begin{minipage}{.5\textwidth}
		\centering
		\includegraphics[width=1.\textwidth]{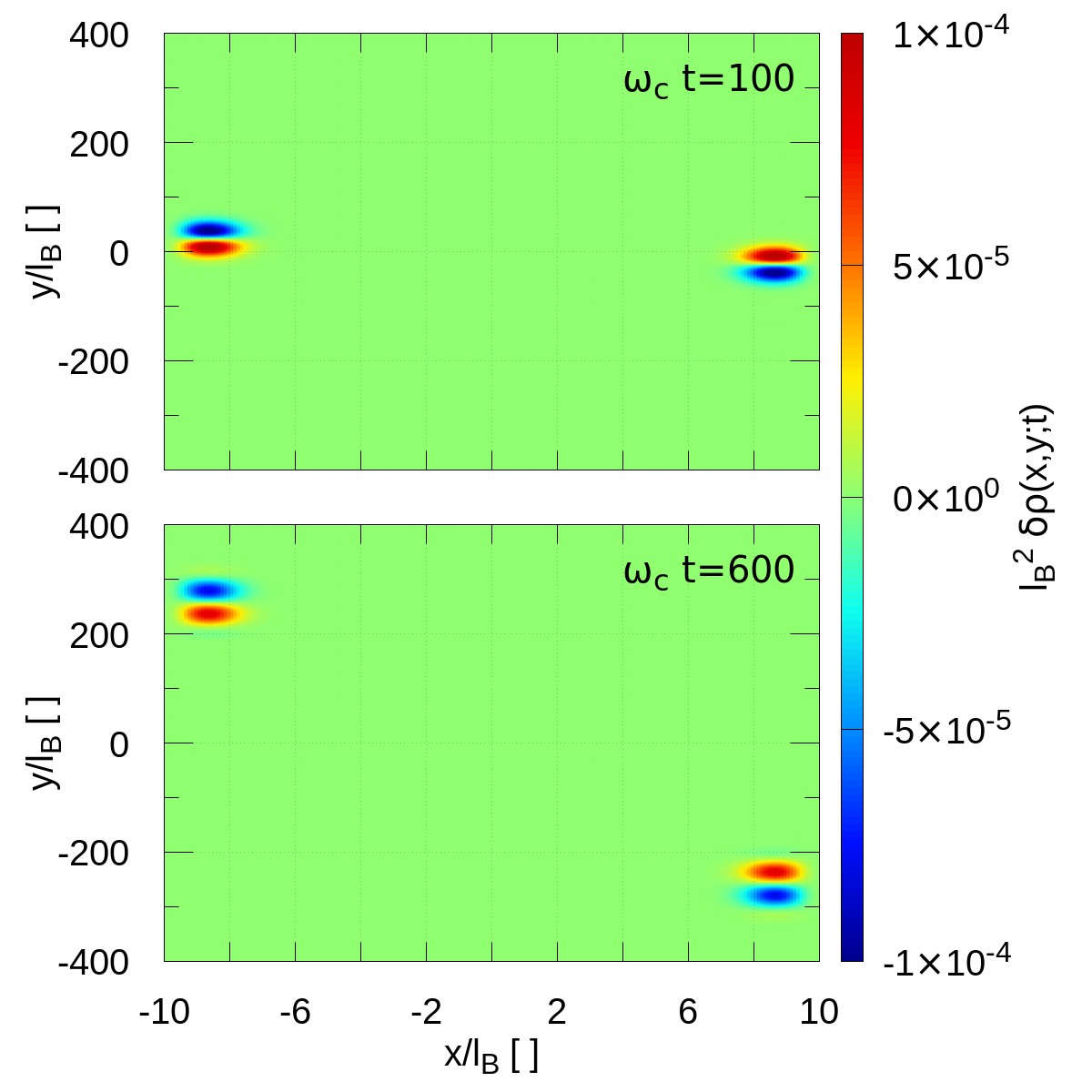}
	\end{minipage}%
	\begin{minipage}{0.5\textwidth}
		\centering
		\includegraphics[width=1.\textwidth]{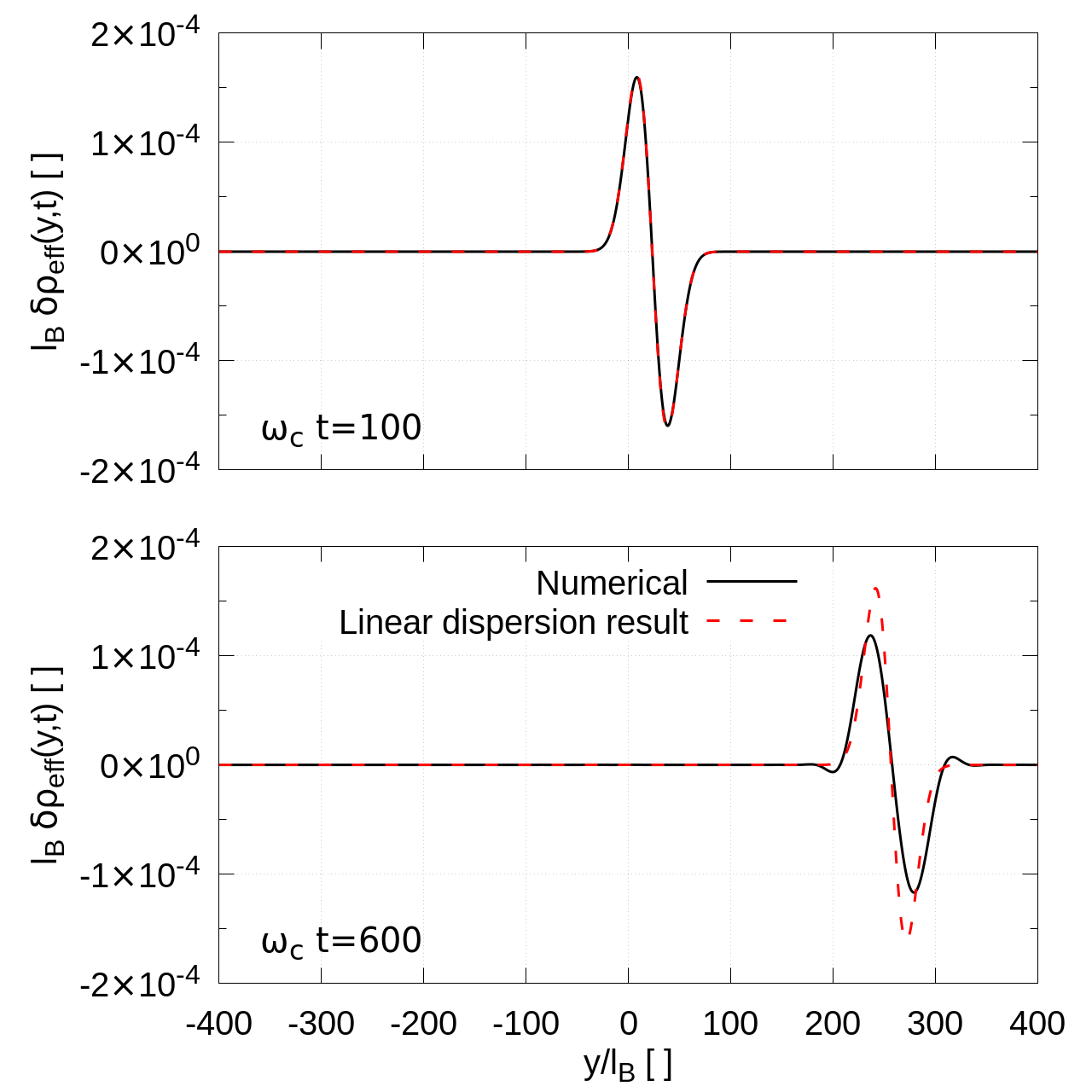}
	\end{minipage}    
	\caption[The LOF caption]{On the left hand side the density variation $\delta\rho(x,y;t)$ induced by the Gaussian perturbation is plotted as a heat-map at $\omega_c t=100$ and $\omega_c t=600$. On the right hand side the effective density computed numerically is compared with the linear dispersion result eq. \ref{eq:width_linear_band}. $\lambda=0.001\hbar\omega_c$ has been used.}
	\label{fig:comparison_linear_gauss}
\end{figure}

The width can be estimated by measuring the distance between the two lobes
\begin{equation}
\label{eq:width_linear_band}
W=2\sqrt{2}\,\sqrt{\left(\frac{\sigma}{2}\right)^2+\left(\frac{v \tau}{2}\right)^2}
\end{equation}
obviously of order $\alpha$. Its physical interpretation is quite simple.
The $\sigma$ contribution evidently arises from the width of the excitation, and it is a \virgolette{static} contribution (the system deforms following the excitation); the $v \tau$ contribution is a \virgolette{dynamical} one instead. It arises from the generated density excitation at $y\sim 0$ within $\sigma$ travelling chirally outside this region at speed $v$ for a time given by the duration $\tau$ of the excitation. 
The overall effect is a competition between these two physically different processes.
\newline From the discussion made in Chapter \ref{ch4} we expect the above result to be quite accurate as long as the Landau level curvature can be neglected. 

A comparison between the numerical results and \ref{eq:real_space_eff_drho} is shown in the right hand side panel of Fig. \ref{fig:comparison_linear_gauss}. The image on the left shows the full density variation $\delta\rho(x,y;t)$. 
It is apparent that as the time passes the effect of the curvature of the Landau level becomes relevant; the effects of the curvature will now be analysed.

\subsubsection*{Beyond the linear dispersion}
If the dispersion relation is Taylor expanded at quadratic order (and making the usual long wavelength assumptions), we again obtain that each Fourier component of the effective density has the same large time evolution (i.e. long after the excitation has been turned off) described in the previous chapter $\propto \frac{\sin\left[\frac{c}{2}\,q^2(t-t_0)\right]}{\sin\left[\frac{c}{2}\,q\,\Delta k\,(t-t_0)\right]}$ (see eq. \ref{eq:pert_sin_general_single_edge_1bd_variation_app2})
\begin{equation}
\label{eq:finite_size_rhoQ_gauss}
\begin{split}
\delta\rho_{\text{eff}}(q;t)
=
i\,\frac{\lambda\sigma\tau}{2\hbar}\frac{2\pi}{L_y}\,e^{-q^2\left(\left(\frac{\sigma}{2}\right)^2+\left(\frac{v \tau}{2}\right)^2\right)}e^{-i vq\,\Delta t}\frac{\sin\left(\frac{cq^2}{2}\Delta t\right)}{\sin\left(\frac{c q\Delta k}{2}\Delta t\right)}.
\end{split}
\end{equation}
%\begin{figure}[htp!]
%	\centering
%	\includegraphics[width=1.\textwidth]{F_spectrum.png}
%	\caption[The LOF caption]{The image compares some of the numerically computed Fourier components of the $x$-integrated density variation of the system with the perturbative result \ref{eq:finite_size_rhoQ_gauss}. The excitation strength was set to $\frac{\lambda}{\hbar\omega_c}=0.001$.}
%	\label{fig:F_spectrum}
%\end{figure}
%As a function of time this result is compared (for different values of $q$) with the numerical results in Fig. \ref{fig:F_spectrum}.

We notice two things.
\newline First of all even at linear perturbative order (with respect to the expansion parameter $\lambda\ll\hbar\omega_c$) the propagating packet will not only be damped in time but will also change its shape; this was not the case for the harmonic excitation analysed in Chapter \ref{ch4}, where only the density mode with the same wavevector of the excitation is significantly excited when the perturbation strength parameter is small enough.
\newline Additionally in the large-sample limit, each mode decays in time as
\begin{equation}
\label{eq:sinc_curvature_large_sample}
\propto \frac{\sin\left[\frac{c}{2}\,q^2\,\Delta t\right]}{\sin\left[\frac{c}{2}\,q\,\Delta k\,\Delta t\right]}\xrightarrow[L_y\rightarrow\infty]{} \frac{\sin\left[\frac{c}{2}\,q^2(t-t_0)\right]}{\frac{c}{2}\,q\,\Delta k\,(t-t_0)}
\end{equation}
i.e. it becomes irrelevant in a time interval $\Delta t_\text{decay}$ which depends on the mode wavevector and is set by eq. \ref{eq:decay_time}
\begin{equation}
\Delta t_\text{decay}\sim\frac{\pi}{c(2K)^2}.
\end{equation}
At a given time only the modes with wavevector satisfying the condition $\frac{\Delta t}{\Delta  t_\text{decay}}\lesssim1$ will then be significatively excited. Equivalently, the momentum space width of the excitation at a given time will roughly be of order
\begin{equation}
\widetilde{W} \approx \sqrt{\frac{\pi}{c\Delta t}}.
\end{equation} 
In the real space this means that the wavepacket will spread as
\begin{equation}
\label{eq:packet_width}
W\sim \widetilde{W}^{-1}=\sqrt{\frac{c\Delta t}{\pi}}\propto \sqrt{c\Delta t}.
\end{equation}
This  behaviour is different from the one of a localised quantum-mechanical free single electron, whose spatial spread is asymptotically linearly increasing in time $\sim \frac{c\Delta t}{\sigma}$.

We can be a bit more quantitative than this. If we let $L_y\rightarrow\infty$ in eq. \ref{eq:finite_size_rhoQ_gauss}, we get
\begin{equation}
%\begin{split}
\delta\rho_{\text{eff}}(q;t)\rightarrow
%=&i\,\frac{\lambda\sigma\tau}{2\hbar}\,e^{-q^2\left(\left(\frac{\sigma}{2}\right)^2+\left(\frac{v \tau}{2}\right)^2\right)}e^{-i vq\,\Delta t}e^{-i \frac{cq^2}{2}\,\Delta t}
%\,\int_{0}^{q}e^{i cqk\,\Delta t}dk=\\=&
i\,\frac{\lambda\sigma\tau}{\hbar}\,e^{-q^2\left(\left(\frac{\sigma}{2}\right)^2+\left(\frac{v \tau}{2}\right)^2\right)}e^{-i vq\,\Delta t}\frac{\sin\left(\frac{cq^2}{2}\Delta t\right)}{cq\Delta t}
%\end{split}
\end{equation}
where $\Delta t=t-t_0$. This Fourier transform can be inverted (the integration is not obvious, but quite standard; the whole procedure is however quite lengthy and will therefore be carried out in Appendix \ref{appendix:gaussian_pert_computation}; many details which are here quoted without explicit proof are also discussed there)
\begin{equation}
\label{eq:curvature_gaussian_eff_density}
\delta\rho_{\text{eff}}(y;t)= \frac{\lambda\sigma\tau}{2\hbar c}\,\frac{1}{\Delta t}\,\,\Im\left[\text{erf}\left(\frac{y-v\Delta t}{2\sqrt{\left(\frac{\sigma}{2}\right)^2+\left(\frac{v\tau}{2}\right)^2+i \frac{c\Delta t}{2}}}\right)\right].
\end{equation}
We notice that the excitation exactly vanishes at $y=v\Delta t$ at any time, and that it is an odd function about such a point, as in the linear dispersion case (see eq. \ref{eq:real_space_eff_drho}). 
In the appendix (eq. \ref{eq:vanishing_curvature_limit}) it is also shown that this result agrees with eq. \ref{eq:real_space_eff_drho} in the limit of vanishingly small curvature $c\rightarrow 0$.
\newline As $\Delta t$ increases, the original packet changes its shape and side-lobes start to appear symmetrically about $y=v\Delta t$ (the behaviour becomes oscillatory); this is evidently an effect of the dispersion relation at the Fermi point being non-linear, with the consequence that different modes are coupled and different points of the profile propagate at different velocities.
\begin{figure}[htp!]
	\centering
	\includegraphics[width=1.\textwidth]{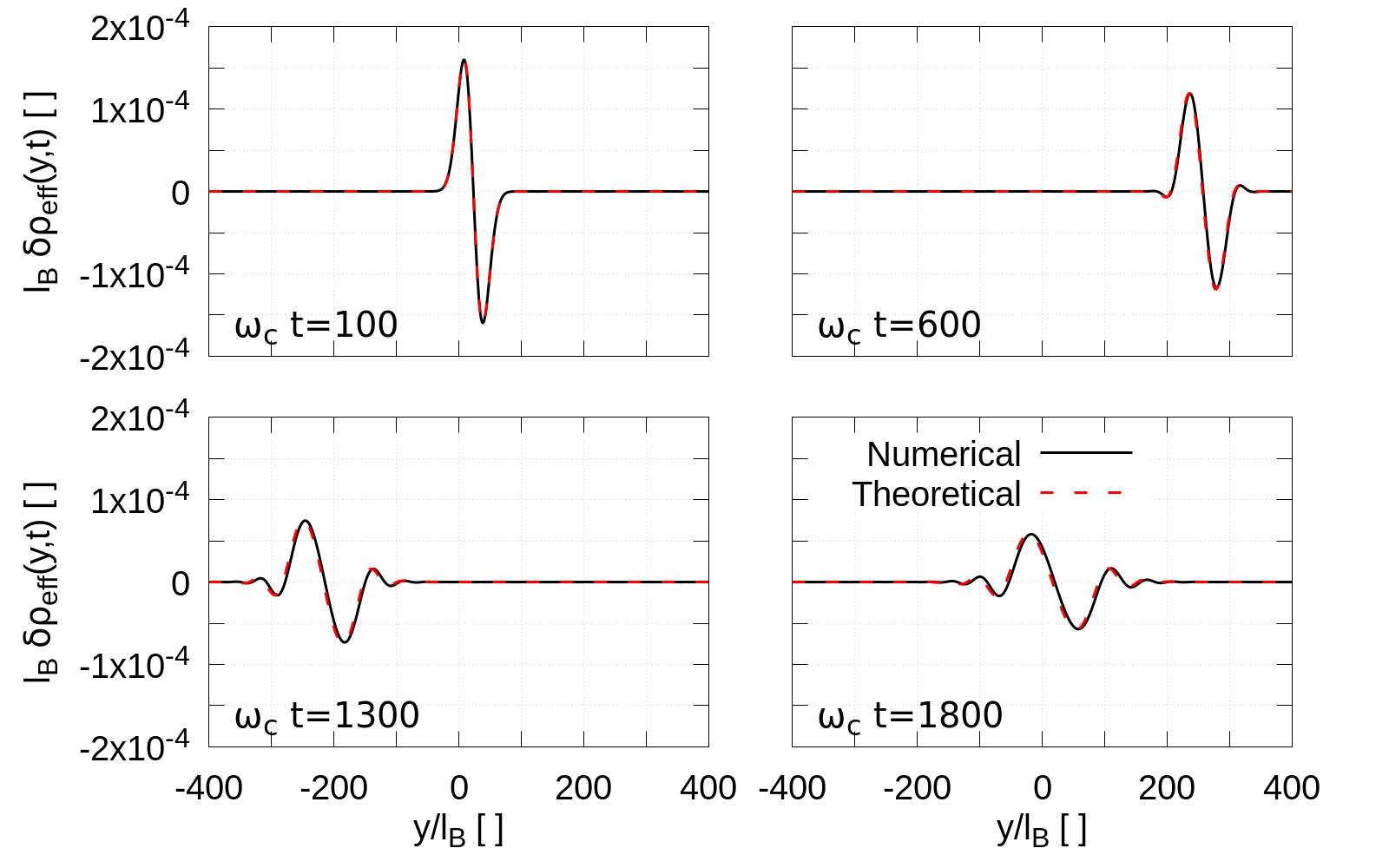}
	\caption[The LOF caption]{The various panels compare (at different times) the numerically computed effective density variation in real space $\delta\rho_\text{eff}(y,t)$ with the analytical expression in eq. \ref{eq:curvature_gaussian_eff_density}.
	The excitation strength was set to $\frac{\lambda}{\hbar\omega_c}=0.001$.}
	\label{fig:curvature_dispersion_gaussian}
\end{figure}

In Fig. \ref{fig:curvature_dispersion_gaussian} the effective density is compared with eq. \ref{eq:curvature_gaussian_eff_density} at different times. It can be seen that lobes (symmetrically about $y=v\Delta t$) appear as $\Delta t$ increases. 
We notice also that the packet height decreases while its width grows.
The agreement is not perfect but still quite good.

\begin{figure}[htp!]
	\begin{minipage}{.5\textwidth}
		\centering
		\includegraphics[width=1.\textwidth]{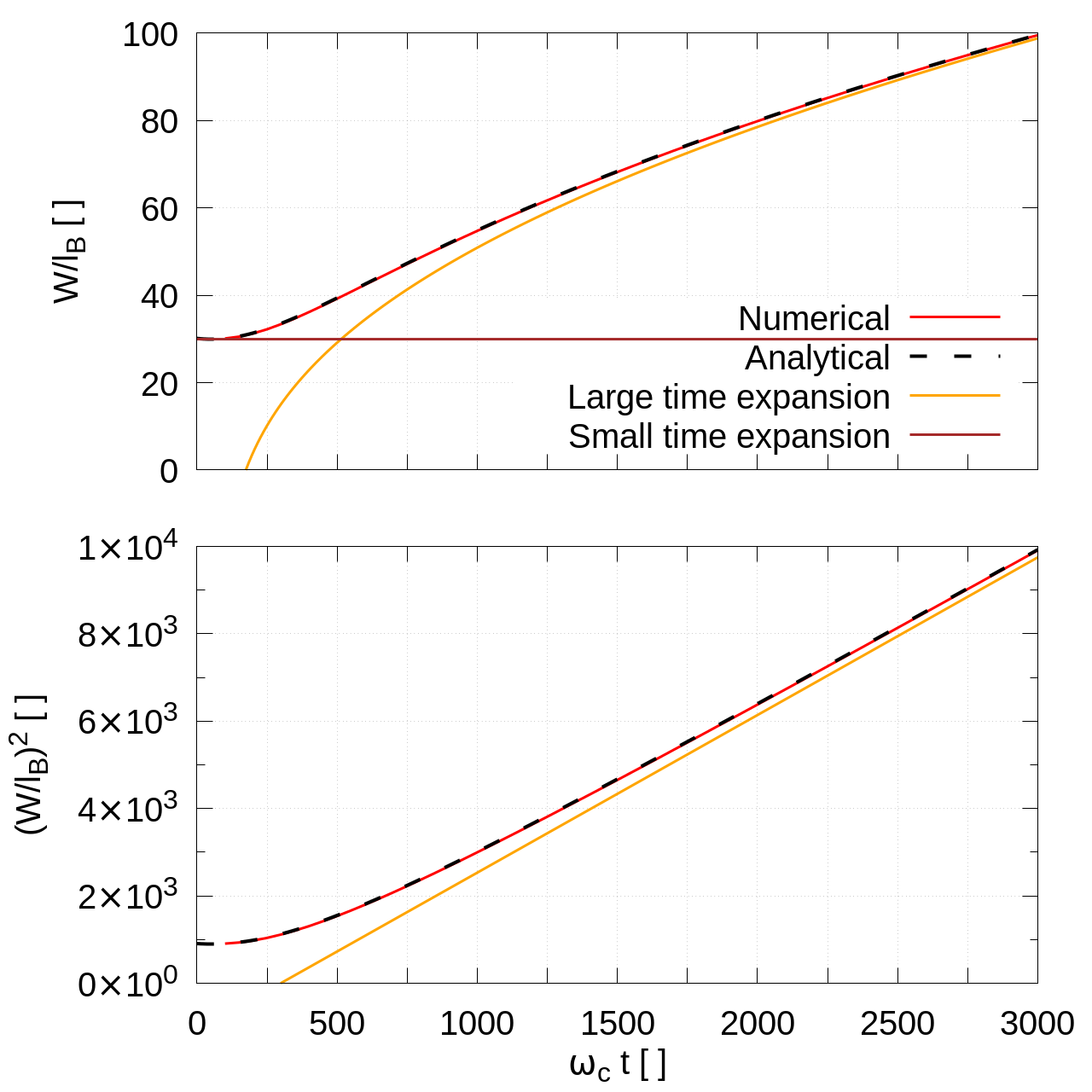}
	\end{minipage}%
	\begin{minipage}{0.5\textwidth}
		\centering
		\includegraphics[width=1.\textwidth]{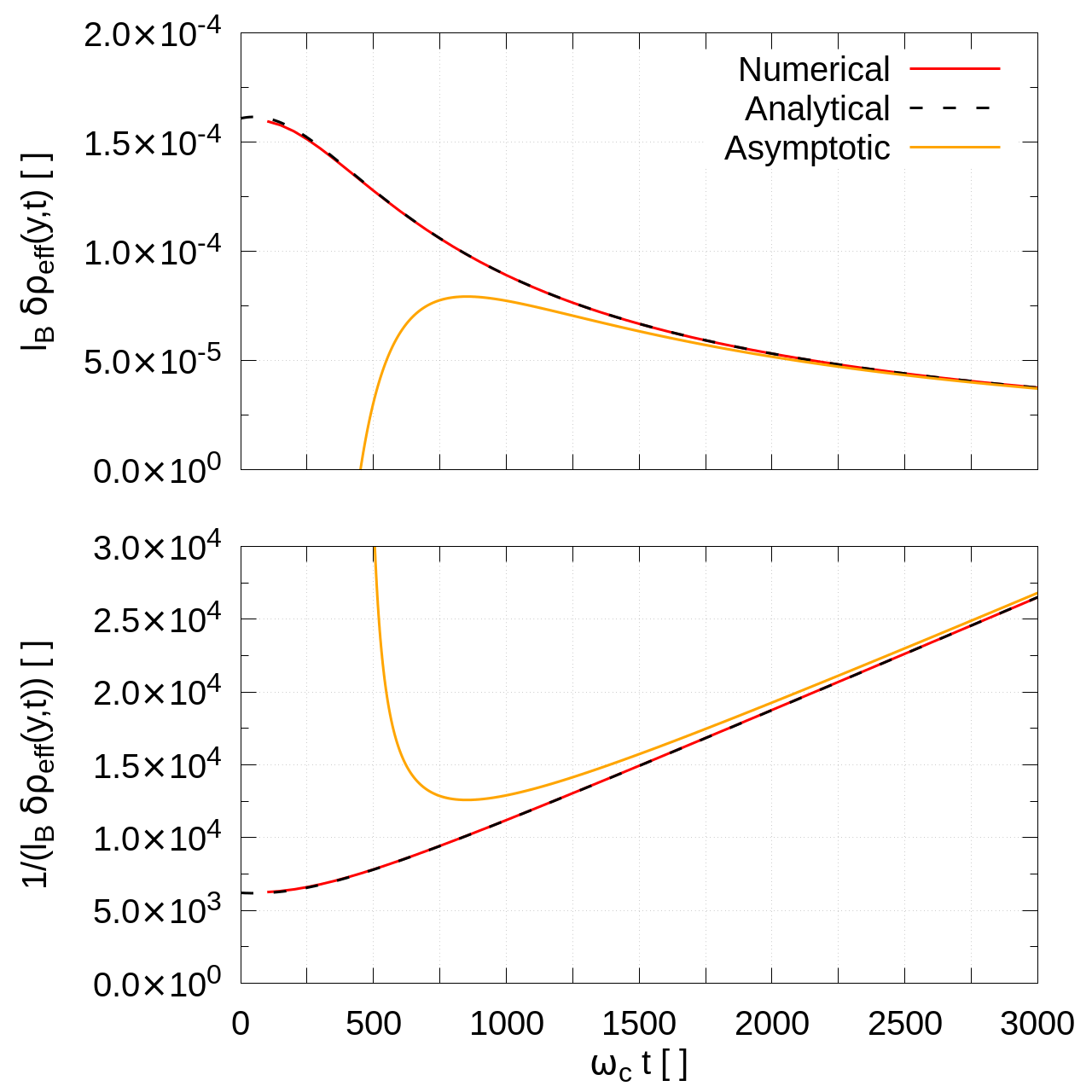}
	\end{minipage}    
	\caption[The LOF caption]{In the left hand panel images the numerically computed distance between the absolute maximum and minimum of the packet is compared with the theoretical expression \ref{eq:Gwidth1}. The asymptotic regimes \ref{eq:large_time_width} and \ref{eq:large_time_squared_width} are also shown.
		\newline On the right hand side the height of the absolute maximum of the packet is plotted; the numerical data are compared to eq. \ref{eq:curvature_gaussian_eff_density} evaluated at the first lobe position (given in eq. \ref{eq:Globes}). 
		The asymptotic expression eq. \ref{eq:decay_gaussian_exc} is also shown.
		\newline In both cases $\lambda=0.001\hbar\omega_c$ has been used.}
	\label{fig:gauss_width_and_height}
\end{figure}

The typical length over which the packet is non-vanishing increases with time. According to equations \ref{eq:gaussian_exc_width} and \ref{eq:delta_alpha_beta}
\begin{equation}
\label{eq:Gwidth1}
W=4\,\sqrt{\,\frac{
\left[\left(\frac{\sigma}{2}\right)^2+\left(\frac{v\tau}{2}\right)^2\right]^2+\left(\frac{c\Delta t}{2}\right)^2	
}{c\Delta t}\,\arctan\left(\frac{\frac{c\Delta t}{2}}{
\left(\frac{\sigma}{2}\right)^2+\left(\frac{v\tau}{2}\right)^2
}\right)}.
\end{equation}
We see a competition between the width which was computed by neglecting the curvature term (eq. \ref{eq:width_linear_band}) and the mode-decay contribution heuristically derived in eq. \ref{eq:packet_width}.
When curvature effects are negligible, $c\Delta t\ll \sigma^2+\left(v\tau\right)^2$
\begin{equation}
\label{eq:small_time_width}
W\sim 2\sqrt{2}\sqrt{\left(\frac{\sigma}{2}\right)^2+\left(\frac{v\tau}{2}\right)^2}.
\end{equation}
and we get precisely eq. \ref{eq:width_linear_band}.
\begin{figure}[htp!]
	\centering
	\includegraphics[width=1.\textwidth]{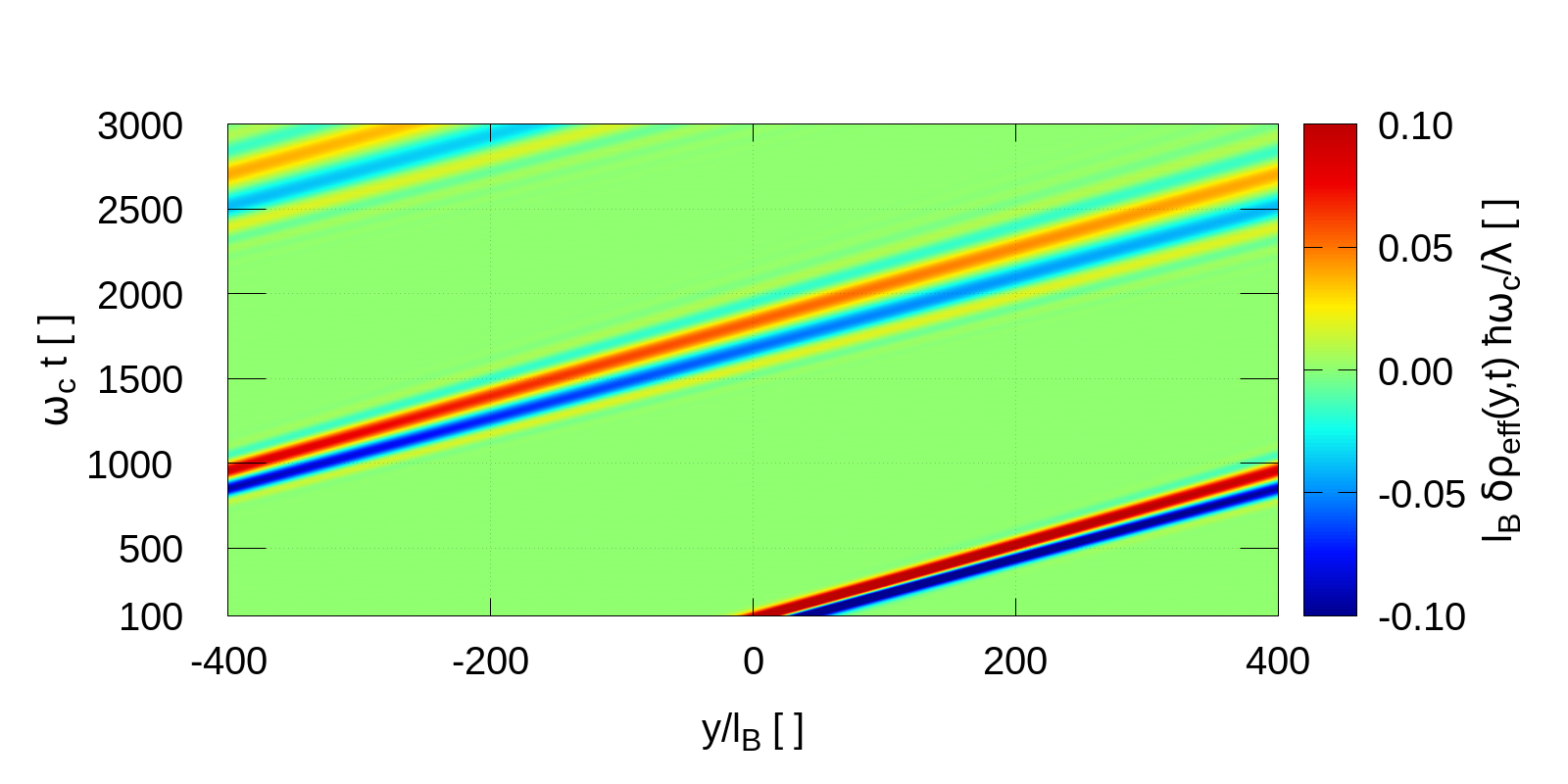}
	\caption[The LOF caption]{The image shows $\delta\rho_\text{eff}(y,t)$ (divided by the excitation strength $\lambda$) as a heat-map in the $t-y$ plane. The \virgolette{zigzagging} behaviour is caused by the presence of the periodic boundary conditions. $\lambda=0.01\hbar\omega_c$ has been used.}
	\label{fig:gauss_ty_plane}
\end{figure}
\newline In the opposite limit $c\Delta t\gg \sigma^2+(v\tau)^2$
\begin{equation}
\label{eq:large_time_width}
\begin{split}
W\sim& \sqrt{2\pi c\Delta t} - 2\sqrt{\frac{2}{\pi\,c\Delta t}}\,\left[\left(\frac{\sigma}{2}\right)^2+\left(\frac{v\tau}{2}\right)^2\right]\\
&\rightarrow \sqrt{2\pi c\Delta t}
\end{split}
\end{equation}
as expected from eq. \ref{eq:packet_width}. The correction to this result $\propto \frac{1}{\sqrt{c\Delta t}}$ does actually matter since it becomes negligible quite slowly. The squared width will be linearly increasing at large time
\begin{equation}
\label{eq:large_time_squared_width}
W^2\sim 2\pi c\Delta t -8\,\left[\left(\frac{\sigma}{2}\right)^2+\left(\frac{v\tau}{2}\right)^2\right]+\mathcal{O}\left(\frac{1}{c\Delta t}\right)
\end{equation}
as can indeed be seen in Fig. \ref{fig:gauss_width_and_height}.
\newline We notice that the lobes are getting further away from $y=c\Delta t$, even in the direction which is opposite to the Fermi velocity; I think this point deserves a little discussion, since we expect the edge modes to be unidirectionally propagating. 
In the limits of validity of the approximations introduced above in order to derive the results we are discussing, it has been (numerically) checked that $\frac{1}{v}\frac{dW}{dt}$ is always safely smaller than the unit value (this is apparent in the very large-time limit), so that the chirality of the propagating modes is safe and well defined.
In Fig. \ref{fig:gauss_ty_plane} the effective density $\delta\rho_{\text{eff}}$ is plotted as a heat map in the $t-y$ plane. It is apparent that the propagation is unidirectional.

Finally, in the limit  $c\Delta t\gg \sigma^2+(v\tau)^2$ the packet height decreases as (eq. \ref{eq:decay_gaussian_exc}) 
\begin{equation}
\label{eq:packet_height_analytical_large_times}
\delta\rho_{\text{eff}}\Bigl(y;c\Delta t\gg \sigma^2+(v\tau)^2\Bigr) \propto \frac{1}{\Delta t}+\mathcal{O}(\Delta t^{-2}).
\end{equation}
This behaviour can be recognized in Fig. \ref{fig:gauss_width_and_height}.

\section{Non-linear phenomena}
In the following section the non-linear dynamics is explored by increasing the modulus of the excitation strength parameter $\lambda$; the results for a given $\lambda$ are also compared with those obtained with\footnote{Notice that positive and negative excitation strength parameters $\lambda$ correspond to physically different Hamiltonians, since the two cannot be related by some symmetry transformation. In general then $\delta\rho_\text{eff}$ will change if the sign of the excitation strength is reversed.} $-\lambda$. Important differences are found.

\begin{figure}[htp!]
	\begin{minipage}{.5\textwidth}
		\centering
		\includegraphics[width=1.\textwidth]{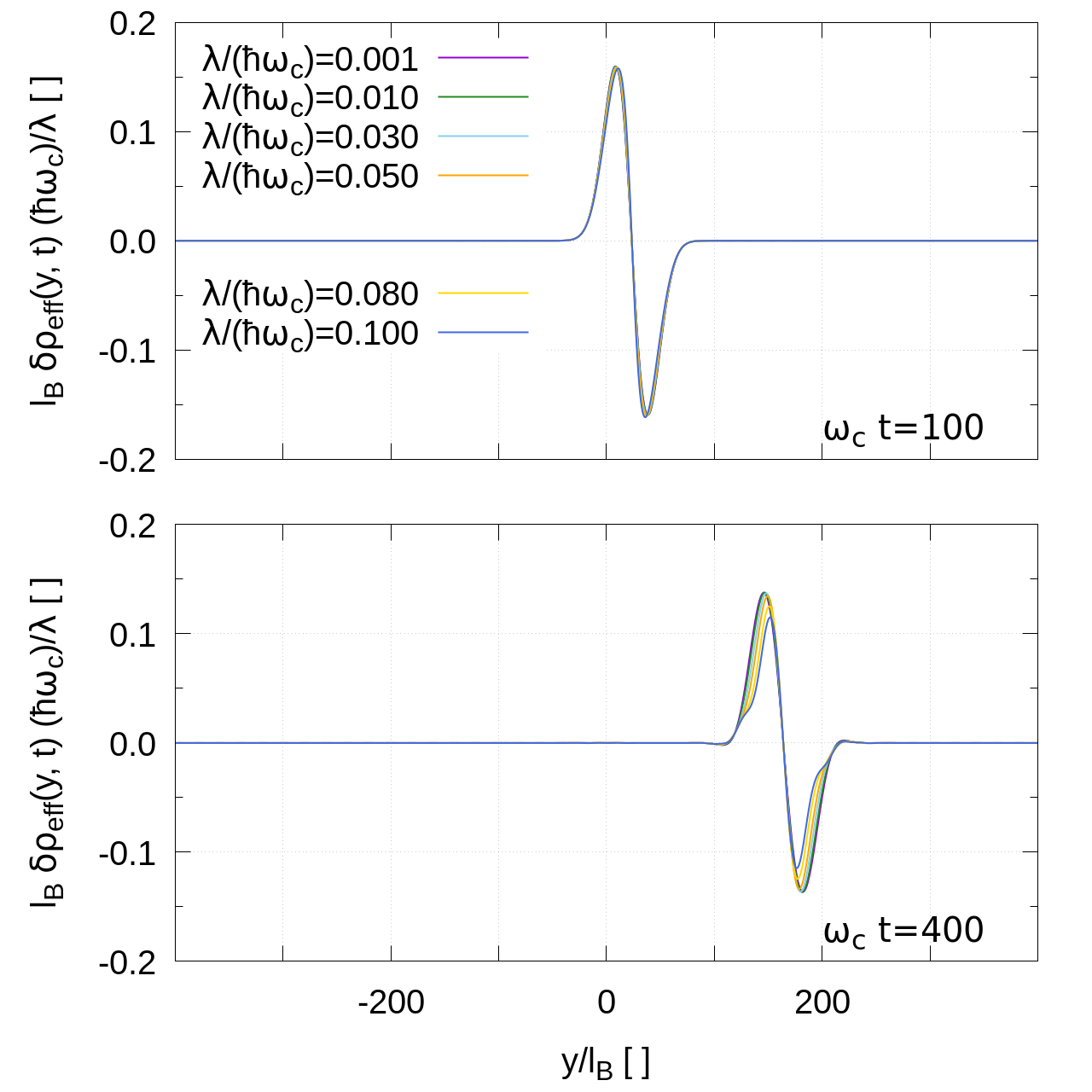}
	\end{minipage}%
	\begin{minipage}{0.5\textwidth}
		\centering
		\includegraphics[width=1.\textwidth]{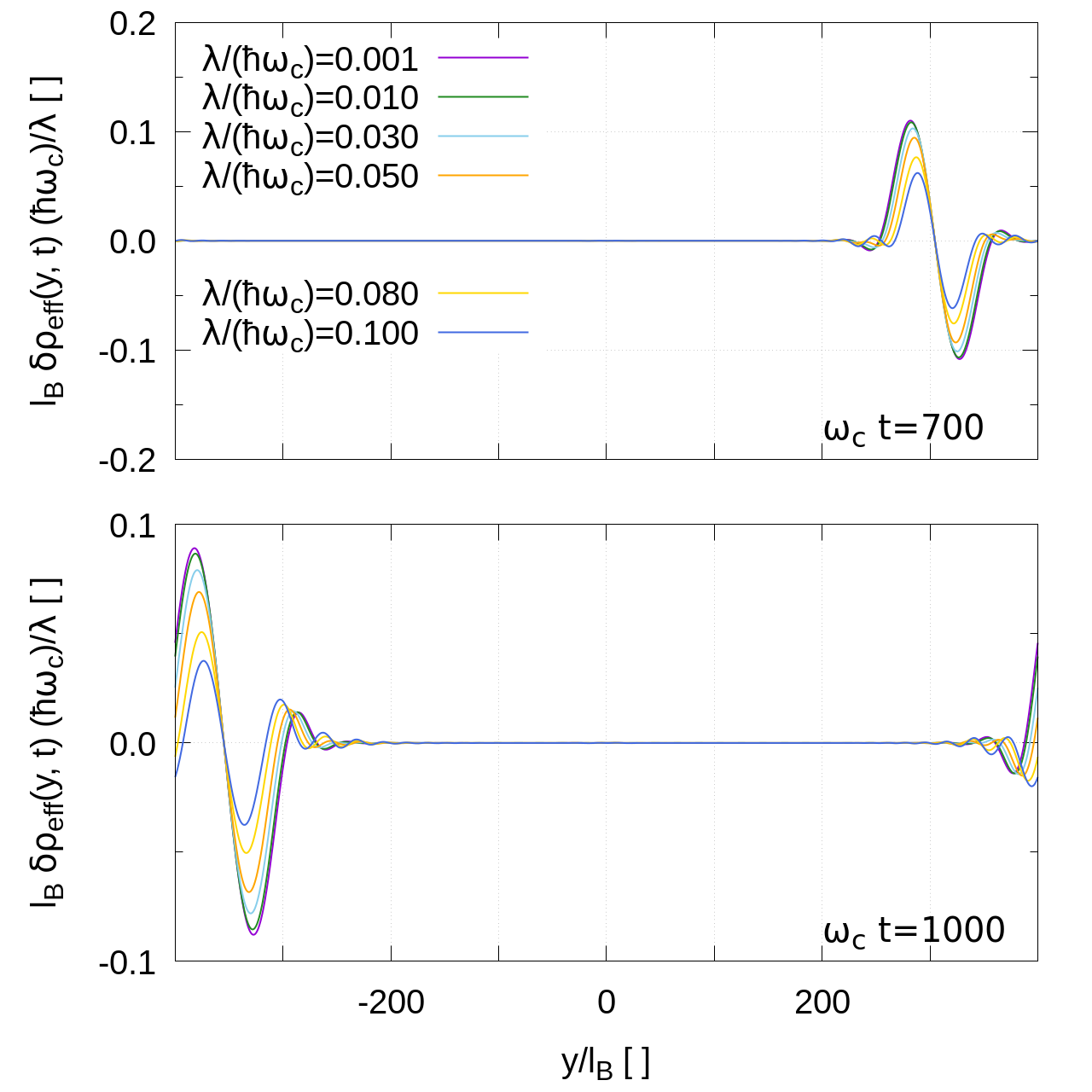}
	\end{minipage}    
	\caption[The LOF caption]{The two images show the integrated density variation $\delta\rho_\text{eff}$ (divided by the excitation strength $\lambda$) at different times, for various values of $\lambda$.}
	\label{fig:DifferentLambdaGaussComparison}
\end{figure}

\begin{figure}[htp!]
	\begin{minipage}{.5\textwidth}
		\centering
		\includegraphics[width=1.\textwidth]{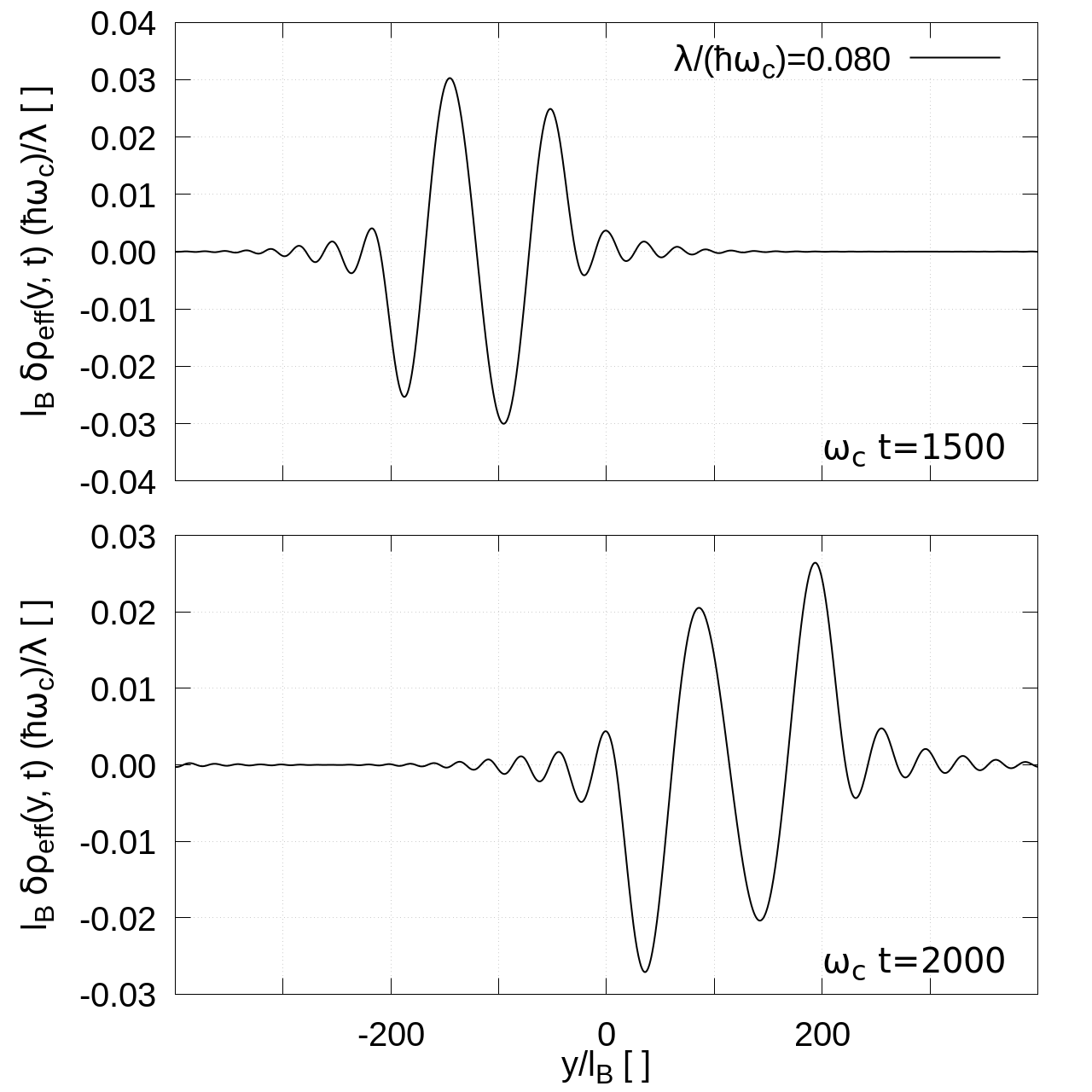}
	\end{minipage}%
	\begin{minipage}{0.5\textwidth}
		\centering
		\includegraphics[width=1.\textwidth]{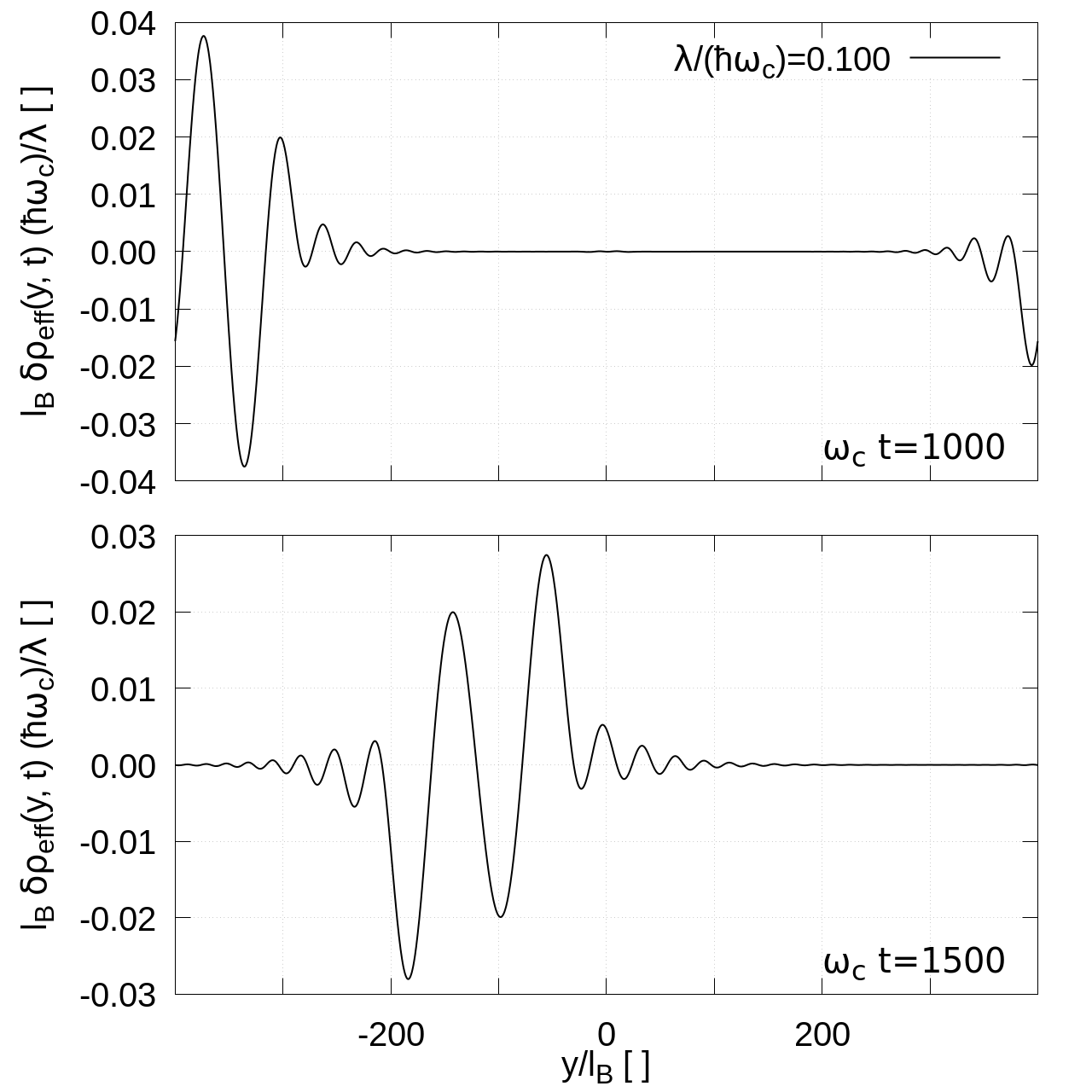}
	\end{minipage}    
	\caption[The LOF caption]{Both the images show the secondary lobes becoming more important than the first ones; the integrated densities $\delta\rho_\text{eff}$ (divided by the excitation strength parameter $\lambda$) are to this purpose compared at two different times (top and bottom panels) in the cases $\frac{\lambda}{\hbar\omega_c}=0.08$ (left hand panels) and $\frac{\lambda}{\hbar\omega_c}=0.10$ (right hand ones).}
	\label{fig:side_lobes_becoming_more_important}
\end{figure}

In Fig. \ref{fig:DifferentLambdaGaussComparison} the density variation integrated along the $x$ direction over the half system (and divided by the excitation strength $\lambda$) is shown at different times when the excitation strength increases.
Despite the density profiles being very similar short after the excitation potential has been turned off (top left panel) the differences amplify during the time-evolution. It is apparent that the packet changes its shape more rapidly as the excitation strength $\lambda$ increases: side-lobes appear sooner, and it decays faster. 
\newline Notice that interestingly the \virgolette{centre} of the packet stays at $y=v\Delta t$ regardless of the value of $\lambda$; the packet itself is moreover close to be odd about such a point.
%It will be seen in the next Chapter \ref{ch6} that this is the result of the packet being odd with respect to such a point rather than a general fact.
\newline We can qualitatively see that the distance between the absolute maximum and minimum becomes smaller as $\lambda$ increases; the presence of the side lobes makes the extension of the packets approximatively the same in the various cases though.
\begin{figure}[htp!]
	\begin{minipage}{.5\textwidth}
		\centering
		\includegraphics[width=1.\textwidth]{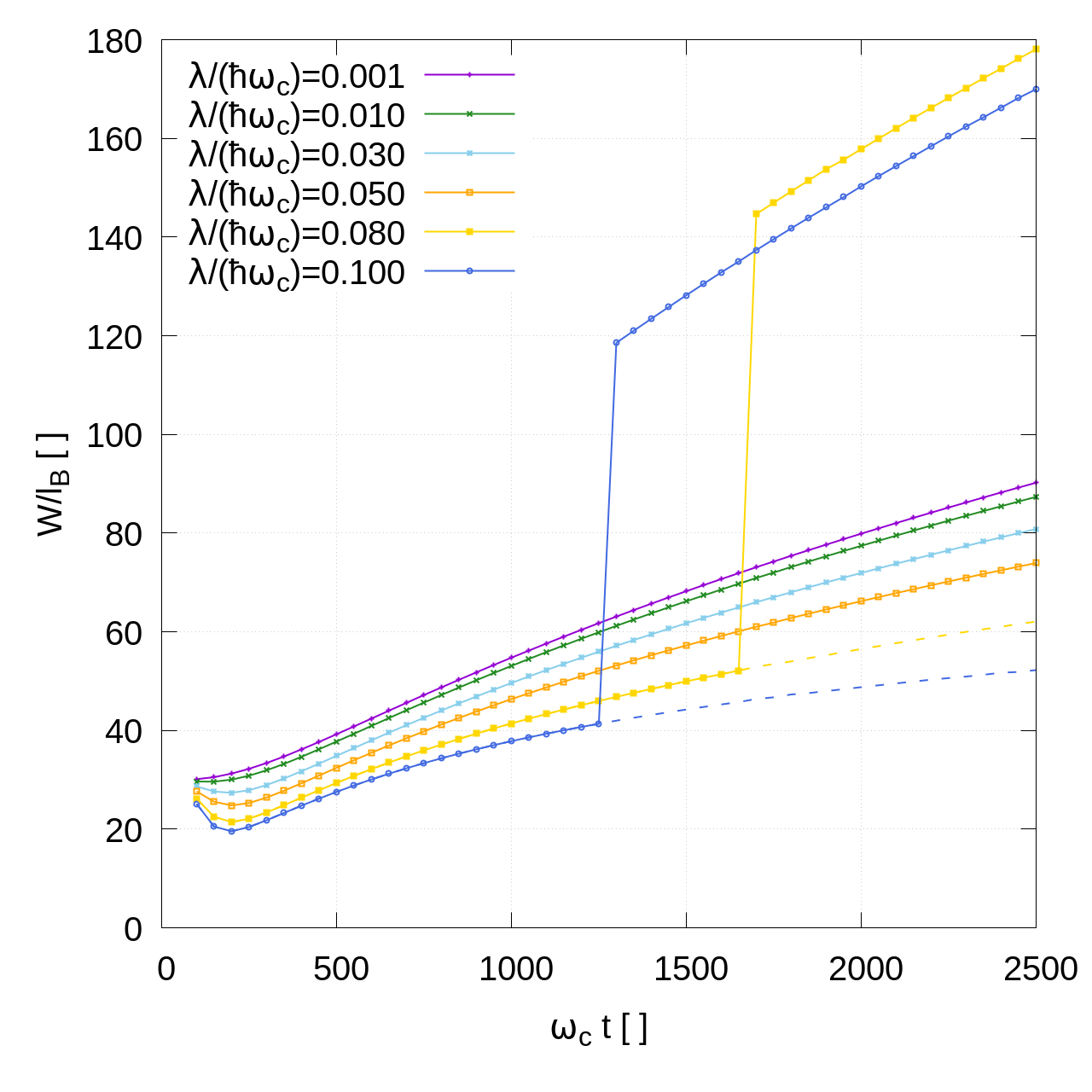}
	\end{minipage}%
	\begin{minipage}{0.5\textwidth}
		\centering
		\includegraphics[width=1.\textwidth]{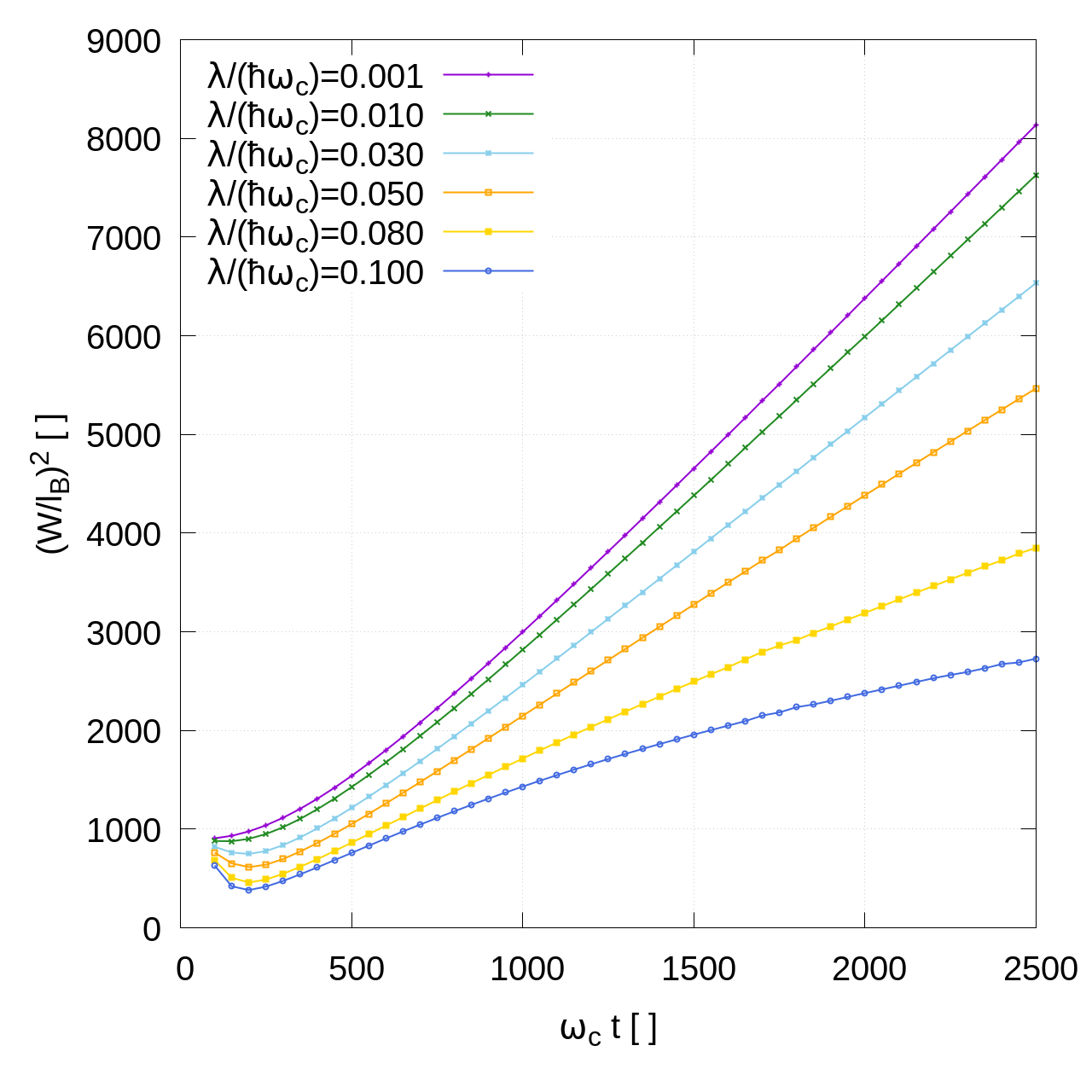}
	\end{minipage}
	\caption[The LOF caption]{In the left panel the distance between the absolute maximum and minimum of the density profiles plotted in Fig. \ref{fig:DifferentLambdaGaussComparison} is shown for different values of the excitation strength $\lambda$. The dashed lines correspond to the distance between the lobes nearest to the packet centre at $y=v\Delta t$.
	\newline On the right hand side the squared distance between these first two lobes is plotted instead.}
	\label{fig:WidthDifferentLambdaGaussComparison}
\end{figure}
\begin{figure}[htp!]
	\begin{minipage}{.5\textwidth}
		\centering
		\includegraphics[width=1.\textwidth]{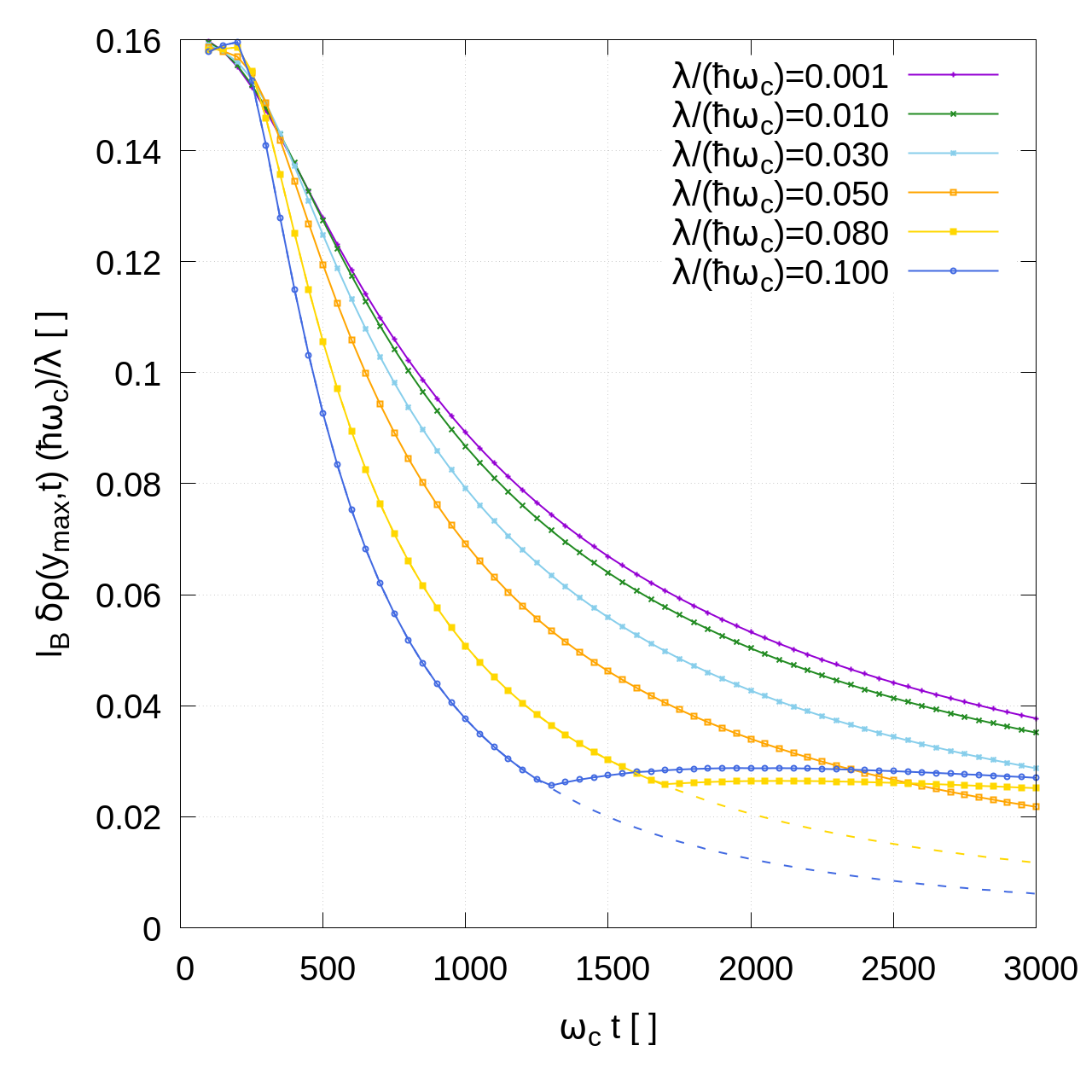}
	\end{minipage}%
	\begin{minipage}{0.5\textwidth}
		\centering
		\includegraphics[width=1.\textwidth]{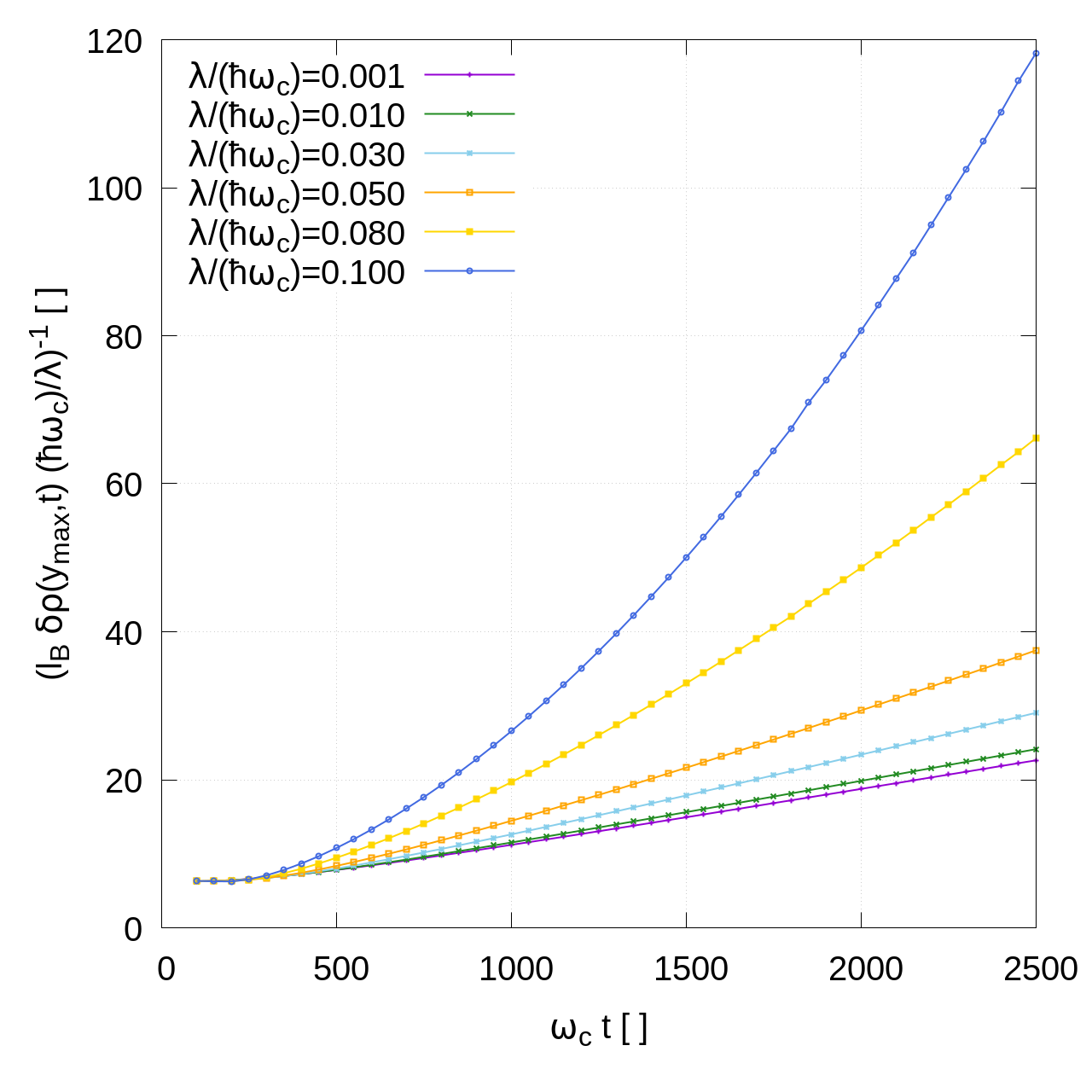}
	\end{minipage}
	\caption[The LOF caption]{The image on the left hand side shows the height of the absolute maximum of the density profiles plotted in Fig. \ref{fig:DifferentLambdaGaussComparison}, for different values of the excitation strength $\lambda$. 
		The dashed lines correspond height of the \virgolette{first} lobe.
		\newline On the right hand side the reciprocal of the first lobe height is plotted instead.}
	\label{fig:HeightDifferentLambdaGaussComparison}
\end{figure}
In Fig. \ref{fig:WidthDifferentLambdaGaussComparison} the distance between the absolute maximum and minimum of the packet is shown in the left hand panel; on the right hand side the same quantity squared is plotted. 
For small values of $\lambda$ we see that at large times this distance increases roughly $\propto\sqrt{c \Delta t}$, as predicted at linear order in eq. \ref{eq:large_time_width}. However as the excitation strength is large and positive significant deviations from this result are observed: the squared distance between the first two lobes increases less than linearly indeed.
\newline The discontinuity of the curves in the left hand side image of Fig. \ref{fig:WidthDifferentLambdaGaussComparison} is a consequence of the side-lobes (the secondary ripples dynamically appearing during the propagation) becoming more important than the lobes nearest to $y=v\Delta t$ (as can be seen in Fig. \ref{fig:side_lobes_becoming_more_important}); this aspect will be further discussed in a while.

\noindent In the left hand side panel of Fig. \ref{fig:HeightDifferentLambdaGaussComparison} the height of the absolute maximum is shown; on the right hand side the reciprocal of the height of the first maximum is plotted instead. For large and positive values of the excitation strength we see that the decay initially occurs faster than $\propto \Delta t^{-1}$ (compare with eq. \ref{eq:packet_height_analytical_large_times}) but at large enough times the behaviour eventually seems to be the same as the one predicted at linear perturbative order.
\newline The \virgolette{bouncing} phenomenon occurring at high $\lambda$ is again a consequence of the secondary ripples becoming larger than the two main lobes (those which are nearest to $y=v\Delta t$).

\begin{figure}[htp!]
	\begin{minipage}{.5\textwidth}
		\centering
		\includegraphics[width=1.\textwidth]{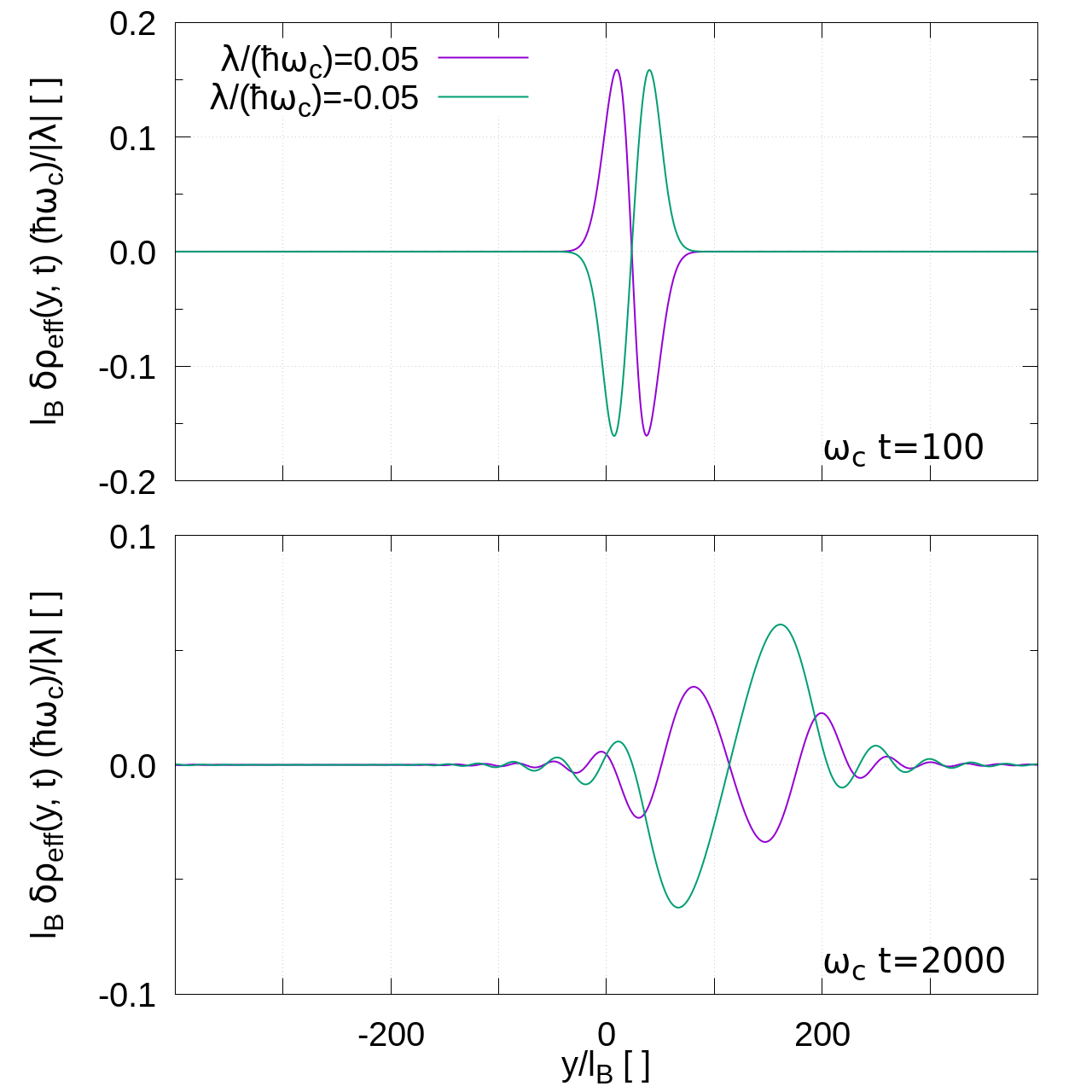}
	\end{minipage}%
	\begin{minipage}{0.5\textwidth}
		\centering
		\includegraphics[width=1.\textwidth]{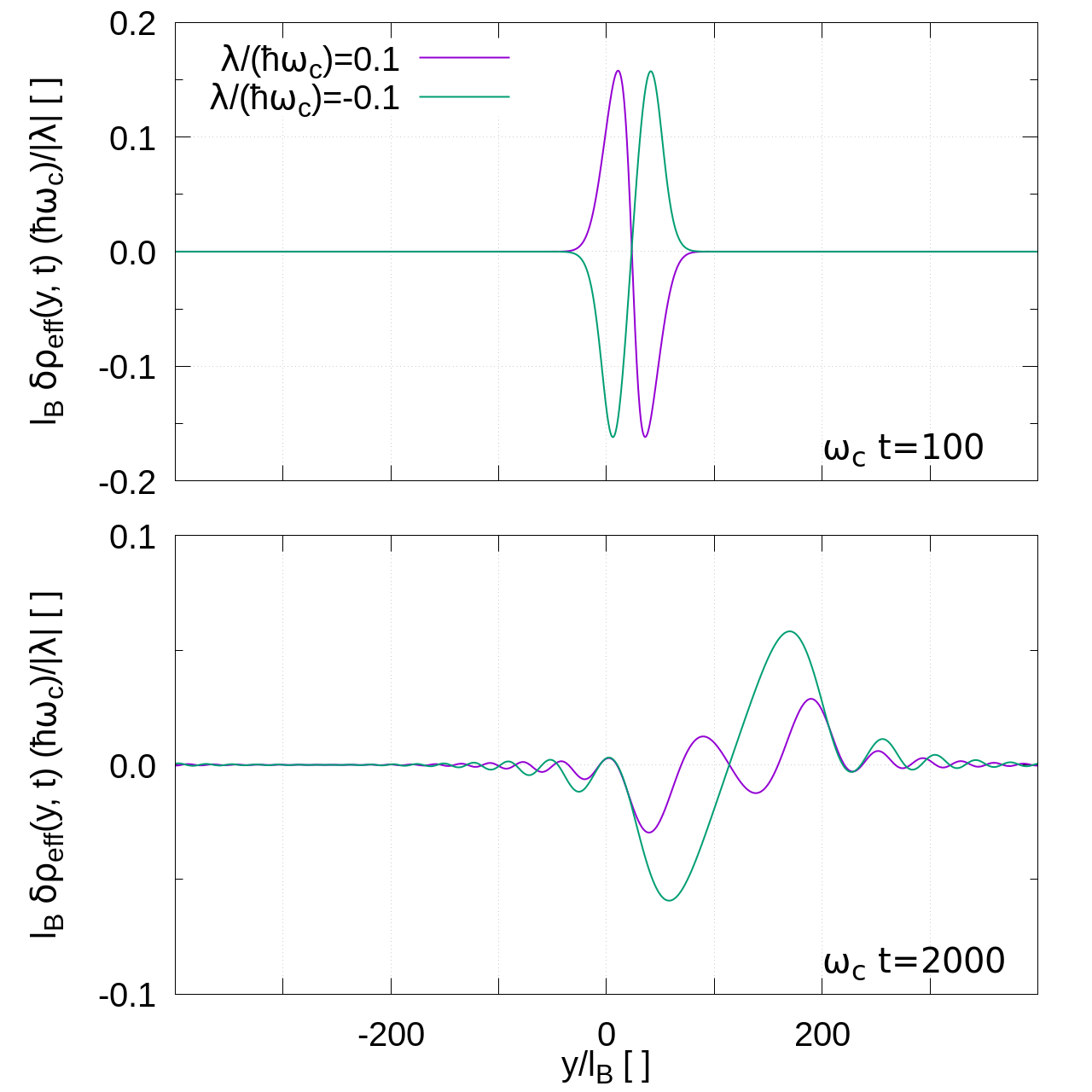}
	\end{minipage}    
	\caption[The LOF caption]{Both the images compare $\delta\rho_\text{eff}$ (divided by $|\lambda|$) for positive and negative values of the excitation strength parameter $\lambda$ (the modulus being the same), at different times.  In the left panel $\frac{\lambda}{\hbar\omega_c}=0.05$ is used; in the right one $\frac{\lambda}{\hbar\omega_c}=0.10$.}
	\label{fig:PositiveNegativeLambdaComparison}
\end{figure}

\noindent These \virgolette{jumps} and \virgolette{bouncings} are apparently an artifact of the computation: as already stated they both are caused by the absolute maximum and minimum shifting from the first lobes to the second ones. The phenomenon is more explicitly shown in Fig. \ref{fig:side_lobes_becoming_more_important}. They however signal something interesting occurring.
As discussed in the appendix \ref{appendix:gaussian_pert_computation}, the linear order expression derived above eq. \ref{eq:curvature_gaussian_eff_density} does not predict this effect: the lobes which are the closest to $y=v\Delta t$ should always be the greatest ones. 
We can moreover easily exclude the possibility that this phenomenon is a linear effect but related to terms in the dispersion relation at the Fermi point beyond the quadratic one (which have been dropped in the derivation of eq. \ref{eq:curvature_gaussian_eff_density}), indeed if this were the case by flipping the sign of the external excitation ($\lambda\rightarrow-\lambda$) the excited density packets would only be \virgolette{reversed}.
In Fig. \ref{fig:PositiveNegativeLambdaComparison} curves analogous to those plotted in Fig. \ref{fig:side_lobes_becoming_more_important} in the case of $\lambda>0$ are compared to those which have been obtained with $\lambda'=-\lambda<0$.
That there are radical differences in the two cases can be seen at a glance. This is the hallmark of non-linear physics occurring.

\noindent We note that the configuration in which the front\footnote{By front we mean the part facing the propagation direction, which in Fig. \ref{fig:PositiveNegativeLambdaComparison} occurs towards positive $y$ values.} of the $\delta\rho_\text{eff}$ packet is initially (i.e. as soon as the excitation has been shut down) a density bump ($\delta\rho_\text{eff}>0$) is much more \virgolette{stable} than the configuration in which we have a density dip in the packet front, in the sense that in the second case the packet shape gets scrambled up quite a lot, with the result that during the evolution the two lobes eventually \virgolette{flip}, and decays faster (see the right-bottom image in Fig. \ref{fig:PositiveNegativeLambdaComparison}, where the height of the density excitation in the case $\lambda=0.1\hbar\omega_c$ is approximatively halved with respect to the one computed in the case $\lambda=-0.1\hbar\omega_c$).
In the first case the packet deforms only a little instead: even at large times its shape is resemblant of the linear order expression eq. \ref{eq:curvature_gaussian_eff_density} (non-linear corrections however give substantial deviations from this result).

\noindent The behaviour will be better understood from the results of the upcoming Chapter \ref{ch6}. 
Long story short: by studying the dynamics of an (initially) symmetric packet about $y=v\Delta t$ (bell shaped) we will see (from the numerical results) that if we have a region of increased density large ripples are created on the side facing the propagation direction; in the tail fewer and much smaller ripples appear instead. 
\newline Vice-versa is observed if we have a region of decreased density (fewer and smaller ripples on the front side). (See Fig. \ref{fig:erfx_nonlinear_shapes_comparison}.)
\newline From these simple considerations and looking again at the top panels of Fig. \ref{fig:PositiveNegativeLambdaComparison}, we see that if $\lambda<0$ we expect ripples to be mainly created \virgolette{on the outside} of the packet, with the result that its original shape does not get drastically scrambled up in the \virgolette{inside}, near $y=v\Delta t$. 
Vice-versa for $\lambda>0$; in this case the deformation will mainly occur in the inner region, messing up with the structure of the packet during its propagation.

\begin{figure}[htp!]
	\begin{minipage}{.5\textwidth}
		\centering
		\includegraphics[width=1.\textwidth]{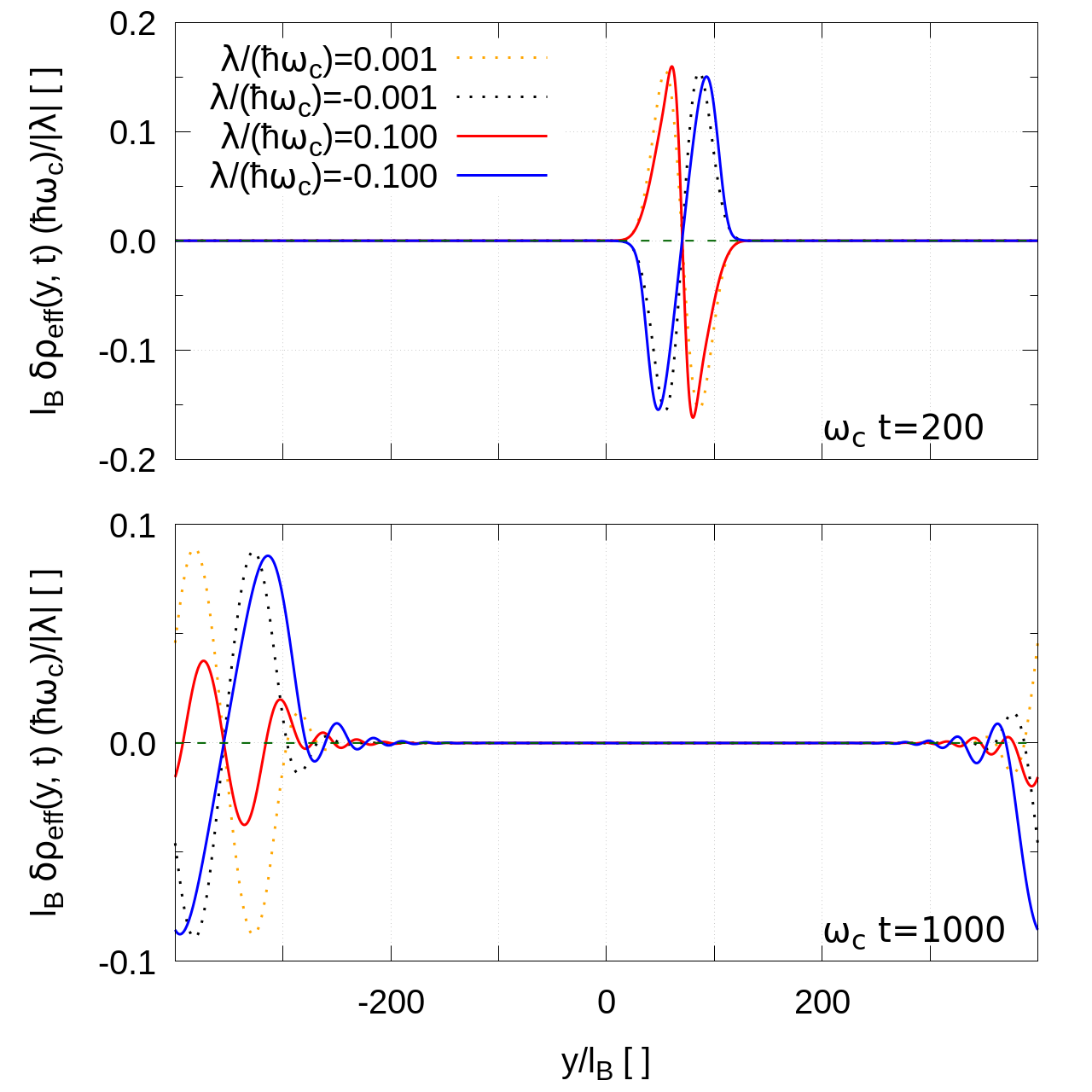}
	\end{minipage}%
	\begin{minipage}{0.5\textwidth}
		\centering
		\includegraphics[width=1.\textwidth]{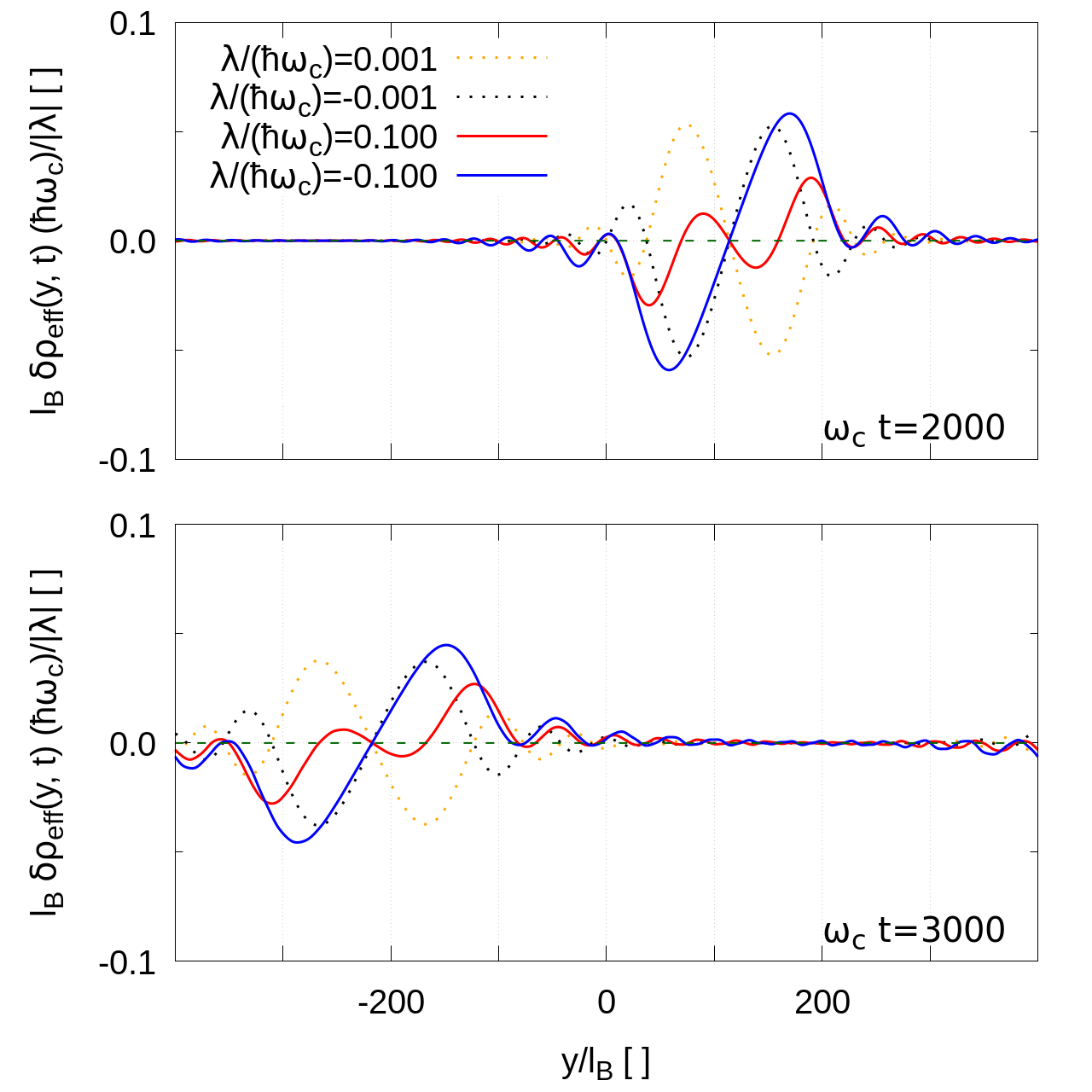}
	\end{minipage}   
	\caption[The LOF caption]{The two figures show $\delta\rho_\text{eff}$ (divided by the excitation strength modulus $|\lambda|$), at different times (long after the excitation has been turned off). The numerically obtained results in the cases $\lambda=\pm0.100\hbar\omega_c$ (continuous lines) are compared with those obtained in the perturbative regime with  $\lambda=\pm0.001\hbar\omega_c$ (dotted lines).}
	\label{fig:vFDensityDependence}
\end{figure}

Finally, we can notice that portions of the packet for which $\delta\rho_\text{eff}>0$ travel slightly faster in the non-linear regime than in the perturbative case,
while those with $\delta\rho_\text{eff}<0$ travel slightly slower, as can be seen by inspecting Fig. \ref{fig:vFDensityDependence}:
first of all, we see that at short times a hole in the system density \virgolette{tilts} backwards with respect to the direction of the motion when compared to the linear order result; 
vice-versa for the density bump.
Secondly, at larger times we see that in the $y<v\Delta t$ region we mainly have ripples which are holes in the system density, and vice-versa in the $y>v\Delta t$ region.

This behaviour could phenomenologically be described by making the propagation speed increase/decrease with the density (higher compression points travelling faster, deeper holes in the ground state density moving slower).
This is resemblant of high-amplitude non-linear classical hydrodynamics waves.

\begin{figure}[htp!]
	\begin{minipage}{.5\textwidth}
		\centering
		\includegraphics[width=1.\textwidth]{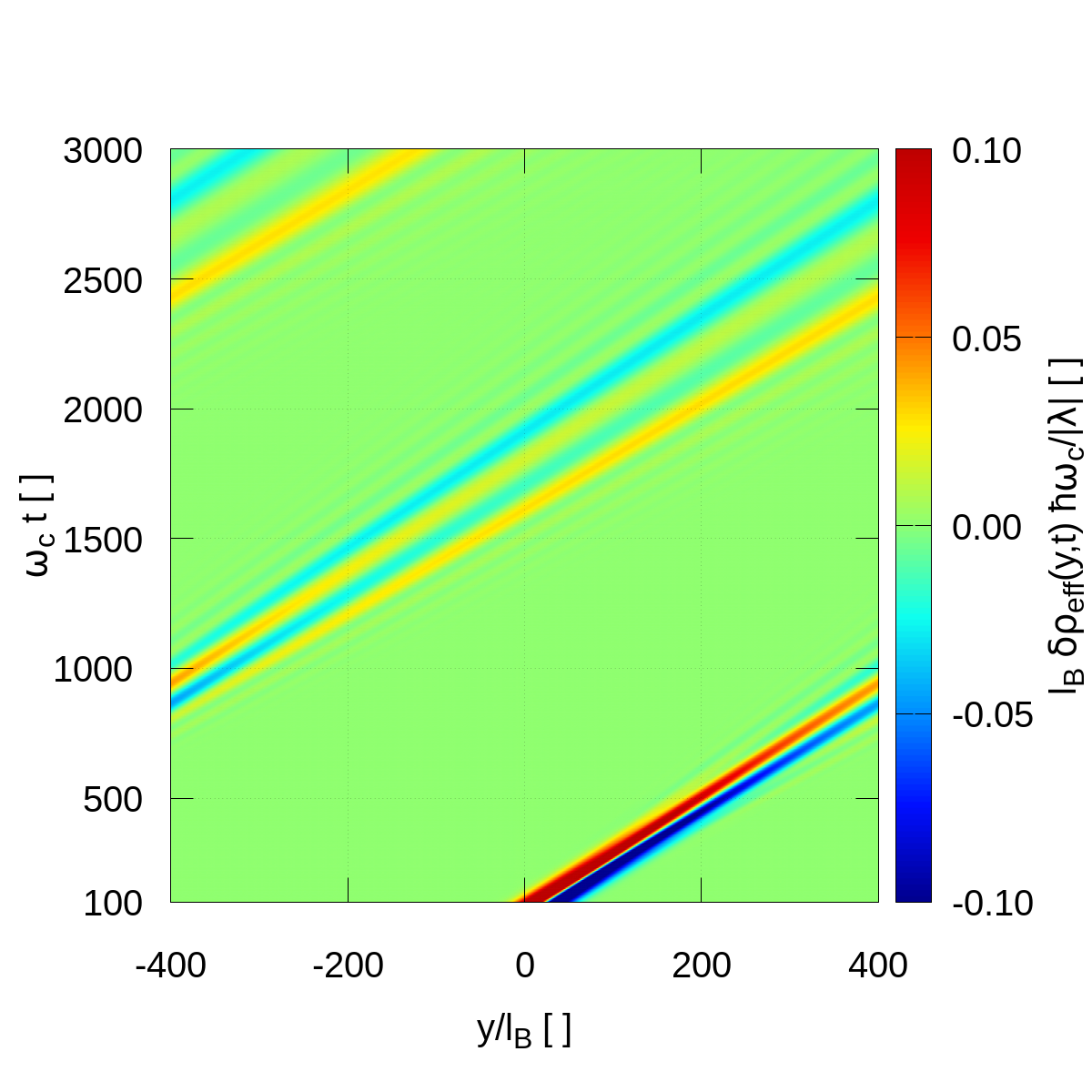}
	\end{minipage}%
	\begin{minipage}{0.5\textwidth}
		\centering
		\includegraphics[width=1.\textwidth]{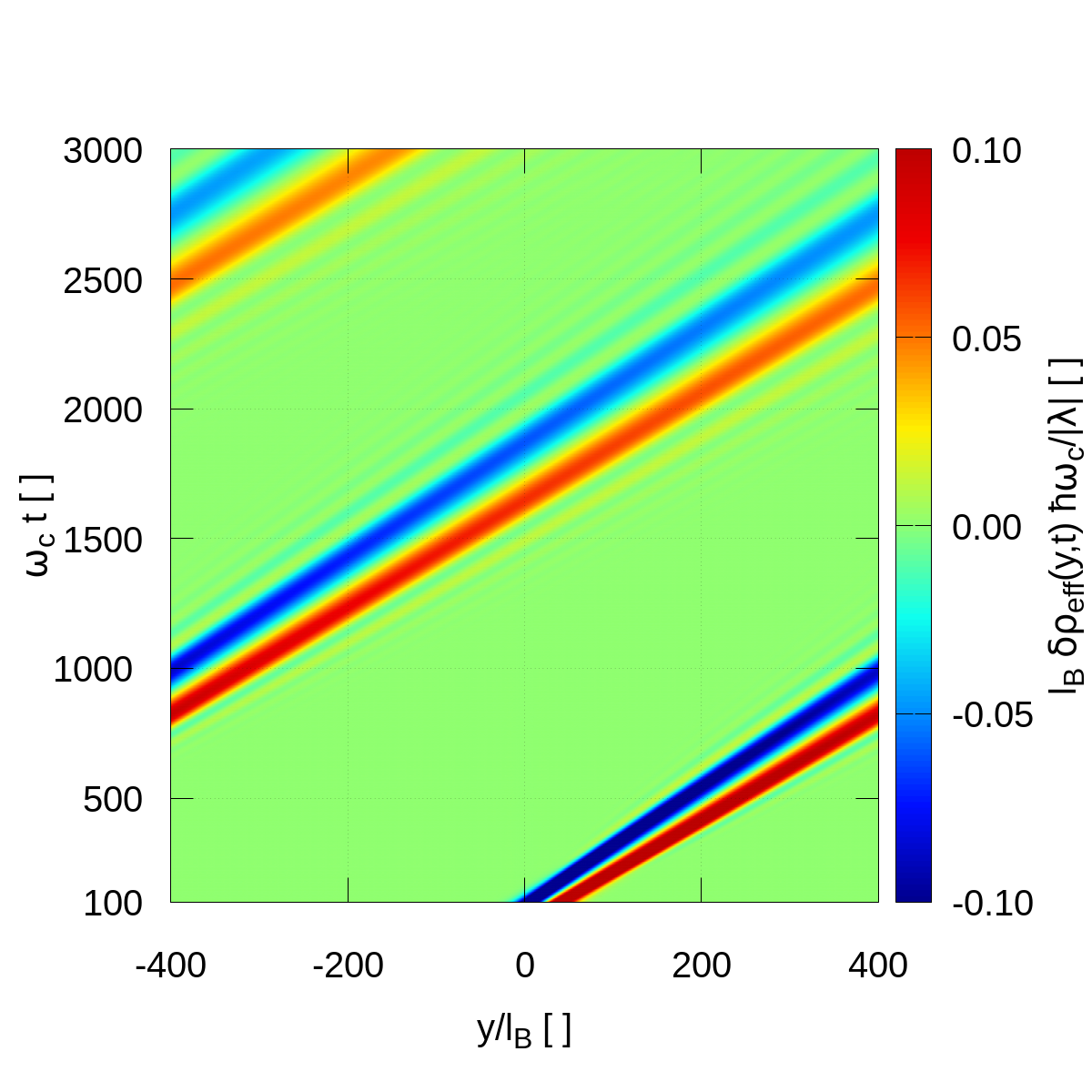}
	\end{minipage}   
	\caption[The LOF caption]{The images show $\delta\rho_\text{eff}(y,t)$ (divided by the absolute value of the excitation strength $|\lambda|$) as a heat-map in the $t-y$ plane. On the left hand side the numerical data have been generated using $\lambda=0.1\hbar\omega_c$; on the right hand side $\lambda=-0.1\hbar\omega_c$ instead.}
	\label{fig:fig_gauss_ty_plane2}
\end{figure}

In Fig. \ref{fig:fig_gauss_ty_plane2} $\delta\rho_\text{eff}(y,t)$ is plotted as a heat map, again in the cases $\lambda=\pm0.1\hbar\omega_c$. The chiral propagation is apparent, as well as the features already discussed above.

\begin{figure}[htp!]
	\begin{minipage}{.5\textwidth}
		\centering
		\includegraphics[width=1.\textwidth]{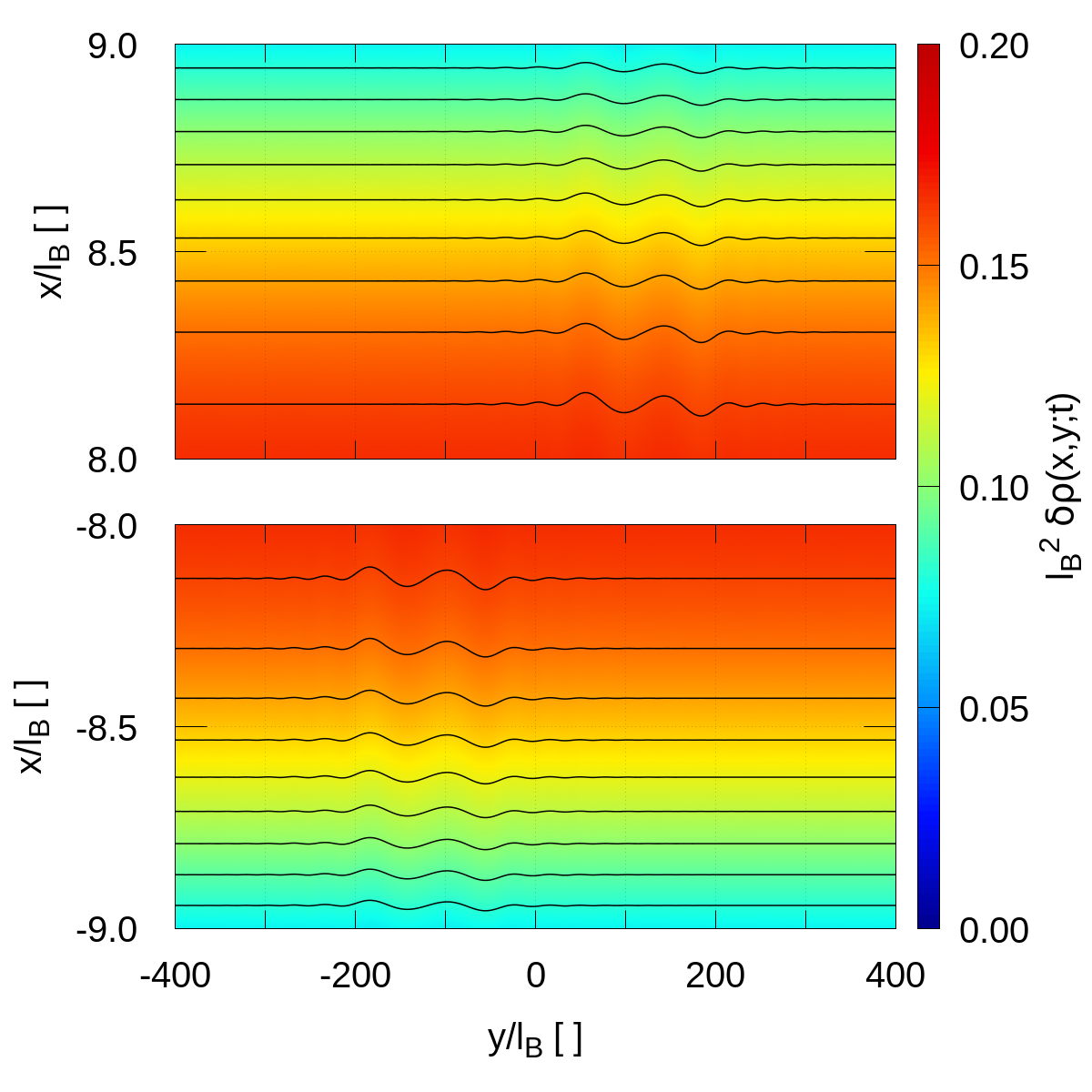}
	\end{minipage}%
	\begin{minipage}{0.5\textwidth}
		\centering
		\includegraphics[width=1.\textwidth]{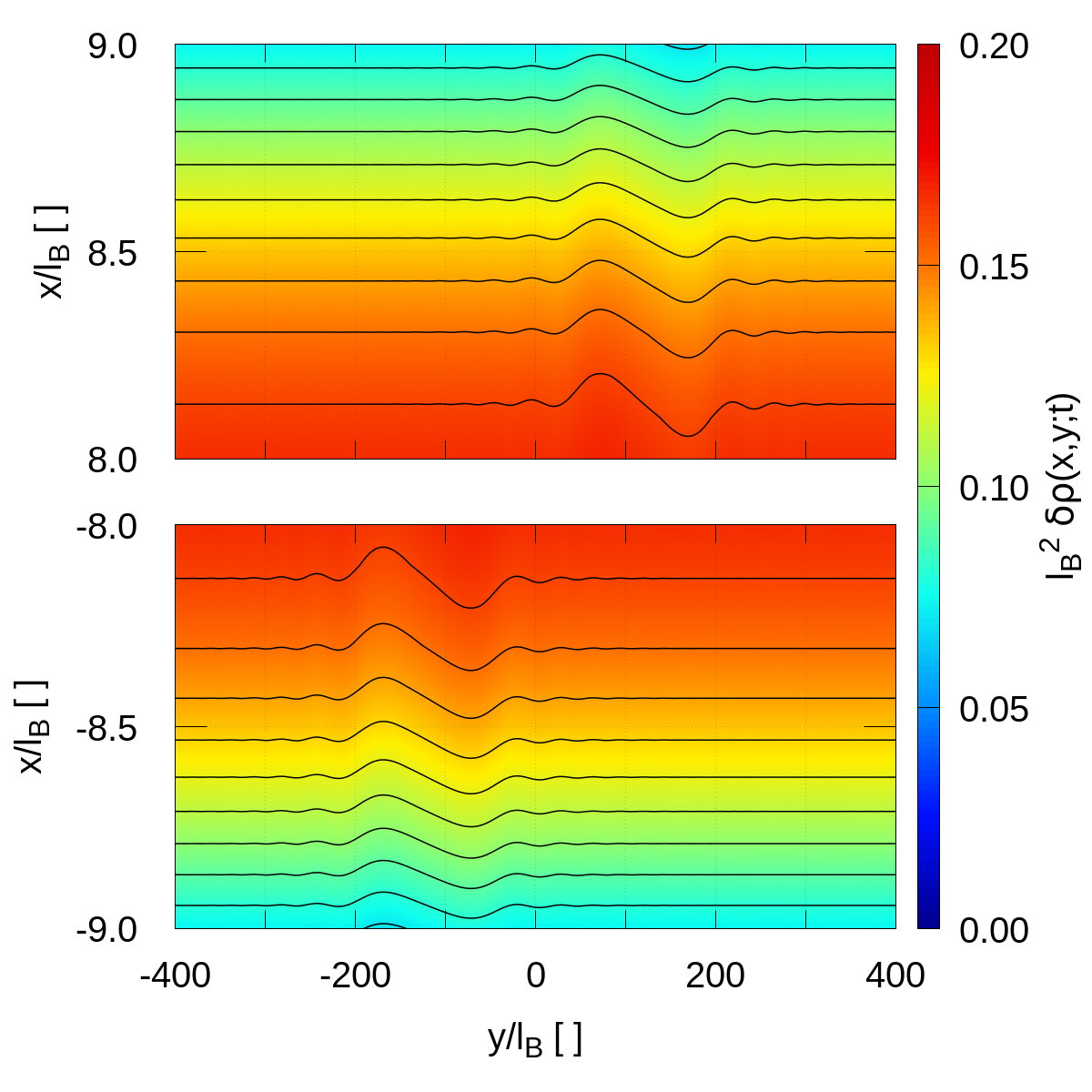}
	\end{minipage}    
	\caption[The LOF caption]{The two images show the \textit{full} system density $\rho(x,y;t)$ at time $\omega_c t=1500$, for $\lambda=+0.1\hbar\omega_c$ (left panel) and  $\lambda=-0.1\hbar\omega_c$ (right one). In the top panel the right ($x>0$) system edge is shown; in the bottom panel the left one ($x<0$). The black lines are isocontours.}
	\label{fig:gauss_edge_density0}
\end{figure}

In Fig. \ref{fig:gauss_edge_density0} two snapshots of the full system density $\rho(x,y;t)$ (zoomed into the two edges) with some isocontour are plotted in the case $\lambda=0.1\hbar\omega_c$ (in the left panel) and $\lambda=-0.1\hbar\omega_c$. 

As a final comment, notice that the non-linear phenomena just discussed are allowed by a non-vanishing curvature at the Fermi point; if it were perfectly linear we would expect that (from the discussion in chapters \ref{ch3} and \ref{ch4}) no higher order effect could possibly be observed.

\begin{comment}
\subsubsection*{Numerical}
From the numerical viewpoint, since the Gaussian term $e^{-\left(\frac{\sigma q}{2}\right)^2}$ chops the integration at $q\sim\sigma^{-1}$, by Taylor expanding both the matrix elements $d^2_{k,q}$ and the dispersion relation in a sufficiently large neighbourhood of the Fermi point $k_F$ makes the $k$-integration straightforward; by inverting the Fourier transform relating $\delta\rho_{\text{eff}}(q;t)$ and $\delta\rho_{\text{eff}}(y;t)$ the real space density has then be calculated
\begin{equation}
\begin{split}
\delta\rho_{\text{eff}}(y;t)=\frac{1}{L_y}\sum_q e^{iqy}\,\delta\rho_{\text{eff}}(q;t)&\rightarrow\,
\frac{1}{2\pi}\int dq\,e^{iqy}\,\delta\rho_{\text{eff}}(q;t)=\\&
=\frac{1}{\pi}\,\int_0^{\infty}dq\,\Re\Bigl[e^{iqy}\,\delta\rho_{\text{eff}}(q;t)\Bigr]
\end{split}
\end{equation}
(the momentum integral can be truncated at momenta of the order of $\sigma^{-1}$ without affecting the final result precision).
\end{comment}
%!TEX root = ../main.tex
% Chapter Template

\graphicspath{{./pic6/}}

\chapter{Sigmoid excitation}\label{ch6}
\lhead{Chapter 7. \emph{Sigmoid excitation}} % Change X to a consecutive number; this is for the header on each page - perhaps a shortened title

Given the peculiar response of the chiral liquid to an external excitation, we have seen that the Gaussian potential studied in the previous chapter gives rise to a two-lobed density pattern. In order to obtain an (initially) bell shaped density variation the system response to a sigmoid-like potential along the system edges is numerically studied in this final chapter.

\noindent Some considerations on the potential used are first of all made, followed by a few semiclassical expectations on the system density response. The system parameters used throughout are reported in the following section.

Both the linear and non-linear edge dynamics are then studied numerically.

\section{The excitation profile and some initial considerations}
Suppose we aim to excite a symmetric (bell-like) density variation (at least before Landau level curvature effects start to be relevant) at the system edge moving along the one-dimensional boundary. 
\newline From the (real space) effective edge dynamics equation (eq. \ref{eq:coordinate_spce_effective_1d_dynamics}) we expect that a sigmoid potential will serve the purpose, since density excitations are driven by the gradient of the external potential.
In a finite-sized system the periodic conditions at the boundary cannot however be satisfied by a purely sigmoid shaped potential;
%Note that in a finite-sized system if a density dip is created somewhere, it must increase away from it (and vice-versa); a pure sigmoid will however only drain electron from/to the region where its slope is the greatest. This is also reflected in the periodic conditions at the boundary not being met.
the problem can be easily circumvented though by superimposing to it a linear (or more generally an odd) potential. %so that the electronic density is uniformly redistributed away from the dip.
\newline We chose to use an error function. Requiring periodicity at the boundary one obtains
\begin{equation}
\label{eq:SigmoidPotential}
V(y; t)=\lambda \,\xi(t)\,\left(
\text{erf}\left(\frac{y}{\sigma}\right)
-
\frac{2y}{L_y}\,\text{erf}\left(\frac{L_y}{2\sigma}\right)
\,\right)
\end{equation}
where $\xi(t)$ is the same Gaussian temporal profile previously used (eq. \ref{eq:sinusoidal_excitation}).
One immediately notices the rather artificial presence of the system length $L_y$. This however drops out in the thermodynamic limit; its effect will become apparent in a while.
\newline Notice moreover that once again the potential has been chosen to be $x$ independent since the general features of the propagating edge mode should not be heavily modified, as qualitatively discussed in Sec. \ref{section:semiclassical_considerations_gaussian}.

Throughout the whole section $\omega_c\tau=15$ and $\omega_c t_0=50$ have been used, as well as $L_x=20l_B$. The confining potential parameters used are $V_0=30\hbar\omega_c$ and $\sigma_c=0.1l_B$, which give a steep rise of the confining potential on a lengthscale roughly of the order of the magnetic length $l_B$.
\newline The system length along $y$ has been fixed at $L_y=800l_B$, in order to make as more irrelevant as possible the presence of the periodic boundary conditions, even in the large time limit as the excited density profile broadens out. 
The typical width over which the sigmoid potential rises has been fixed at $\sigma=20l_B$.
The Fermi wavevector has been fixed at $k_F=1148\Delta k\simeq 9.02l_B$, or equivalently $N=2297$ electrons occupying the lowest Landau level.
\newline The strength parameter $\lambda$ of the perturbation on the other hand has been varied and its values will therefore be explicitly written from time to time when looking at numerical results. 

\begin{figure}[htp!]
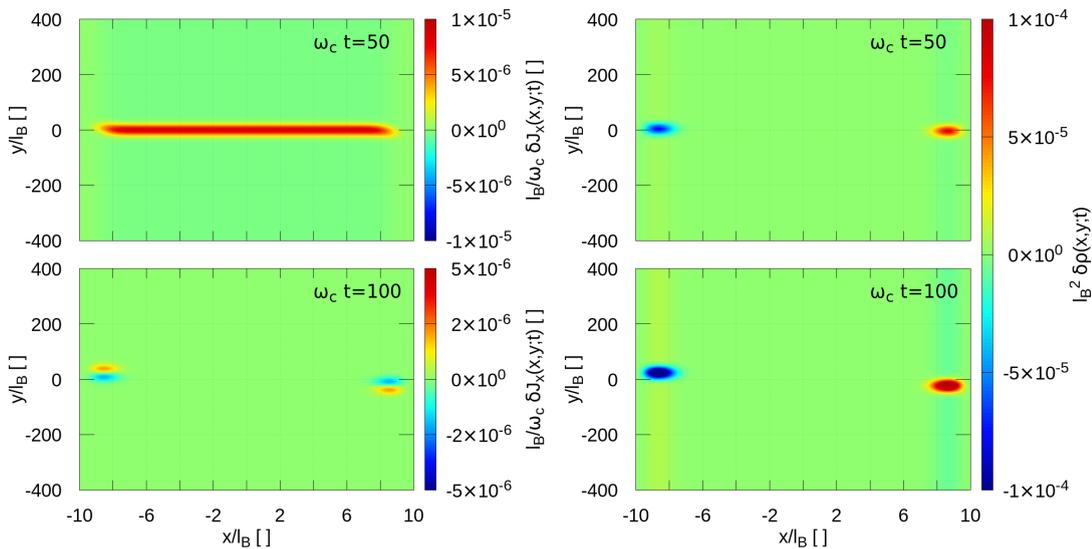

	\begin{minipage}{.5\textwidth}
		\centering
		\includegraphics[width=1.\textwidth]{jx_t=50,100.png}
	\end{minipage}%
	\begin{minipage}{0.5\textwidth}
		\centering
		\includegraphics[width=1.\textwidth]{density_50_100.png}
	\end{minipage}    
	\caption[The LOF caption]{On the left hand side the numerically computed $x$-component of the probability current density is shown in the case $\frac{\lambda}{\hbar\omega_c}=0.001$, at times $\omega_c t=50$ and $\omega_c t=100$. The picture on the right hand side shows the variation of the density with respect to the ground state one, under the same circumstances.}
	\label{fig:transient_current_and_density_erf}
\end{figure}

\begin{figure}[htp!]
	\centering
	\includegraphics[width=.8\textwidth]{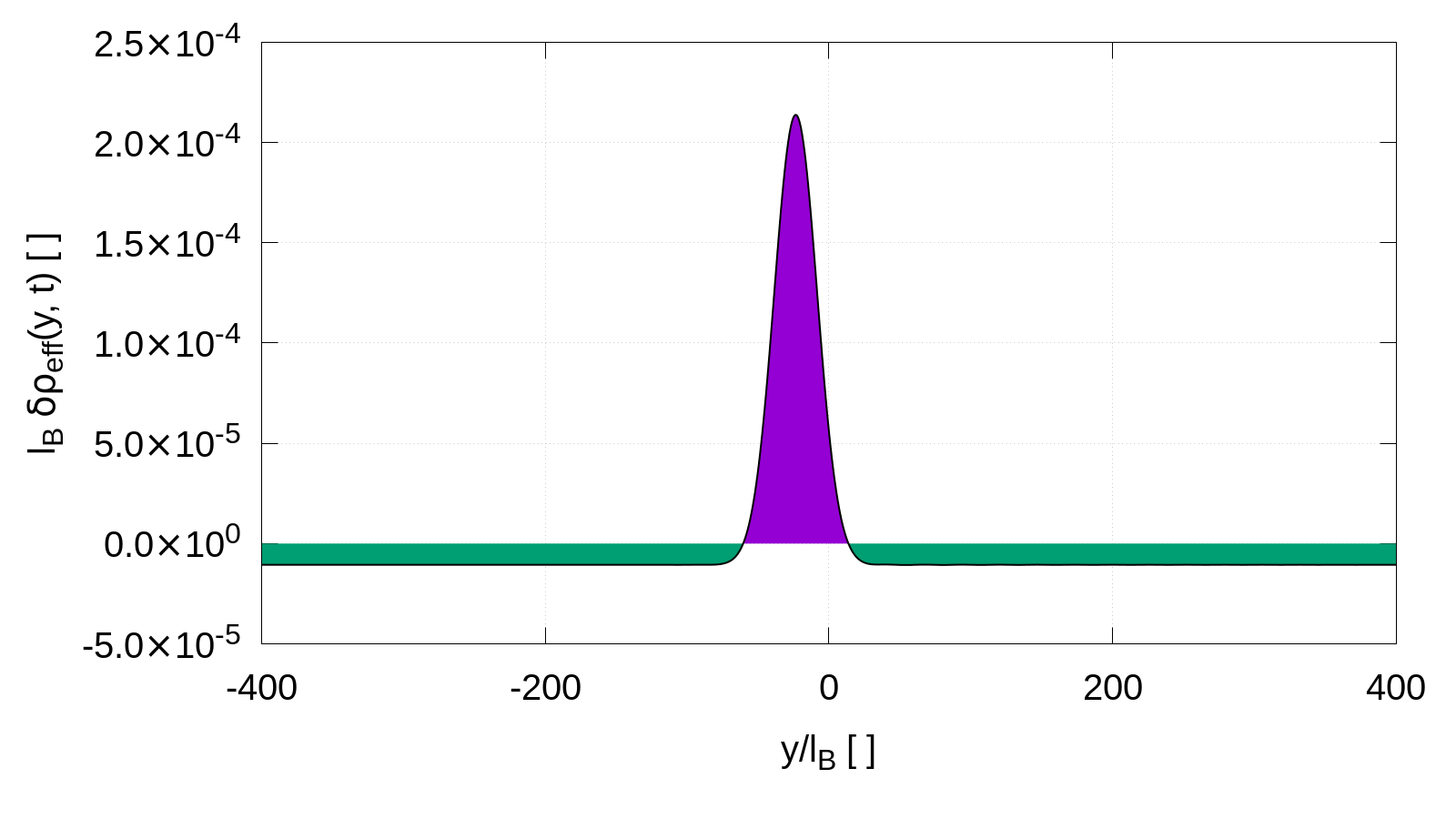}
	\caption[The LOF caption]{The image shows the system density variation integrated along the $x$ direction over the half $x>0$ half-sample, at $\omega_c t=100$. The $\delta\rho_\text{eff}>0$ and $\delta\rho_\text{eff}<0$ regions have been dashed to make the point of the discussion more clear. 
	Its integral has been numerically compute and found to vanish ($\int dy\delta\rho_\text{eff}\sim5\times10^{-12}$, which can be assumed to be purely numerical error).
	\newline The data have been obtained with $\lambda=0.001\hbar\omega_c$.}
	\label{fig:density_redistribution}
\end{figure}

As already discussed in section \ref{section:semiclassical_considerations_gaussian}, during the transient we classically expect that currents will start to flow orthogonally to the local electric field $\propto \partial_y V(y,t) \,\hat{y}$, as can be seen in the left hand side panel of \ref{fig:transient_current_and_density_erf}.
We can see that the effect of the sigmoid potential is to \virgolette{suck} electrons from one edge, pumping them into the other one, symmetrically creating a density dip and bump on opposite sides of the sample.
The linear potential has the same effect, but the flow occurs uniformly (since it has a constant gradient) in the opposite direction, as can be seen in Fig. \ref{fig:transient_current_and_density_erf} and Fig. \ref{fig:density_redistribution}. 
\newline With this particular choice we expect the number of right edge and left edge electrons to be separately conserved (see the caption of Fig. \ref{fig:density_redistribution}). 
This fact can indeed be seen perturbatively, at least at the lowest perturbative order; the computations are however not reported.
It is also possible to deduce the fact in a somewhat heuristic way. By integrating over $y$ the effective dynamics equation (eq. \ref{eq:coordinate_spce_effective_1d_dynamics}) and noting that $\int \partial_y V(y,t) dy=0$ due to the periodic conditions at the boundary, we get that $\partial_t\int \delta\rho_\text{eff}(y, t)=0$.
\noindent Notice that in the thermodynamic limit, the linear part $\propto y$ of $V(y;t)$ vanishes, which means that there will be a net charge flow from one edge to the other one; in this case the number of left edge electrons cannot be the same as that of right edge electrons.

\section{Numerical results}
In the following section the obtained numerical results are presented, analysed and qualitatively discussed.
Notice that the behaviours of the left and right hand system edges are different\footnote{Differently to the cases analysed in chapters \ref{ch4} and \ref{ch5}, where the single electron Hamiltonian had inversion ($x\rightarrow-x$, $y\rightarrow-y$) symmetry; in these cases the two edges were thus simply related by a parity transformation.}; the results will consequently be presented for both the left ($x<0$) and right ($x>0$) system edges. 
Throughout the whole section the effective density variation obtained by integrating $\rho(x,y;t)-\rho(x,y;t=0)$ over the $x>0$ half sample will be dubbed $\delta\rho_\text{eff}^{(+)}$; analogously, the quantity obtained by integration over the $x<0$ half sample $\delta\rho_\text{eff}^{(-)}$. 

We shall begin by discussing the linear physics; the non-linear effects will be discussed afterwards.

\subsection{Linear effects}
\begin{figure}[htp!]
	\begin{minipage}{.5\textwidth}
		\centering
		\includegraphics[width=1.\textwidth]{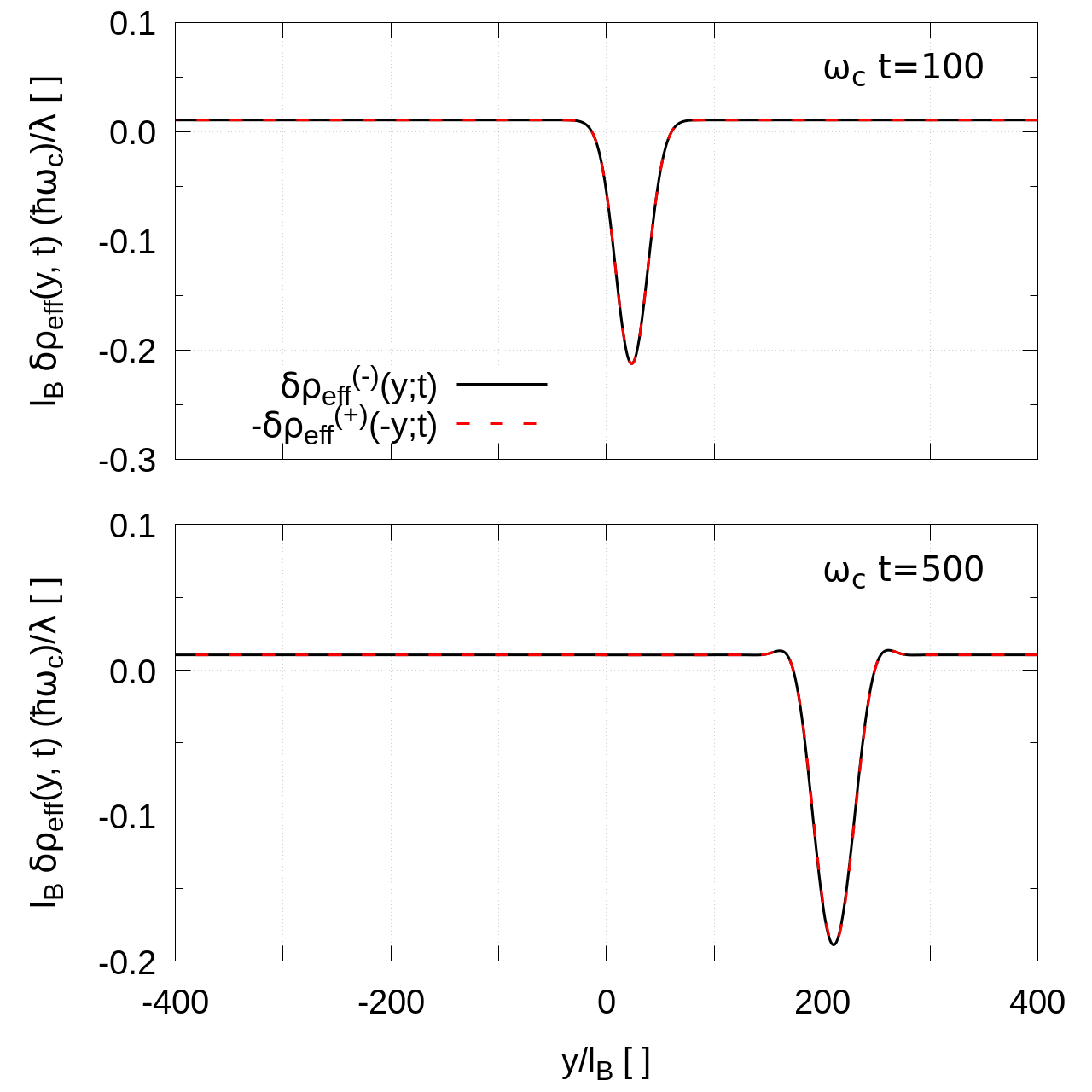}
	\end{minipage}%
	\begin{minipage}{0.5\textwidth}
		\centering
		\includegraphics[width=1.\textwidth]{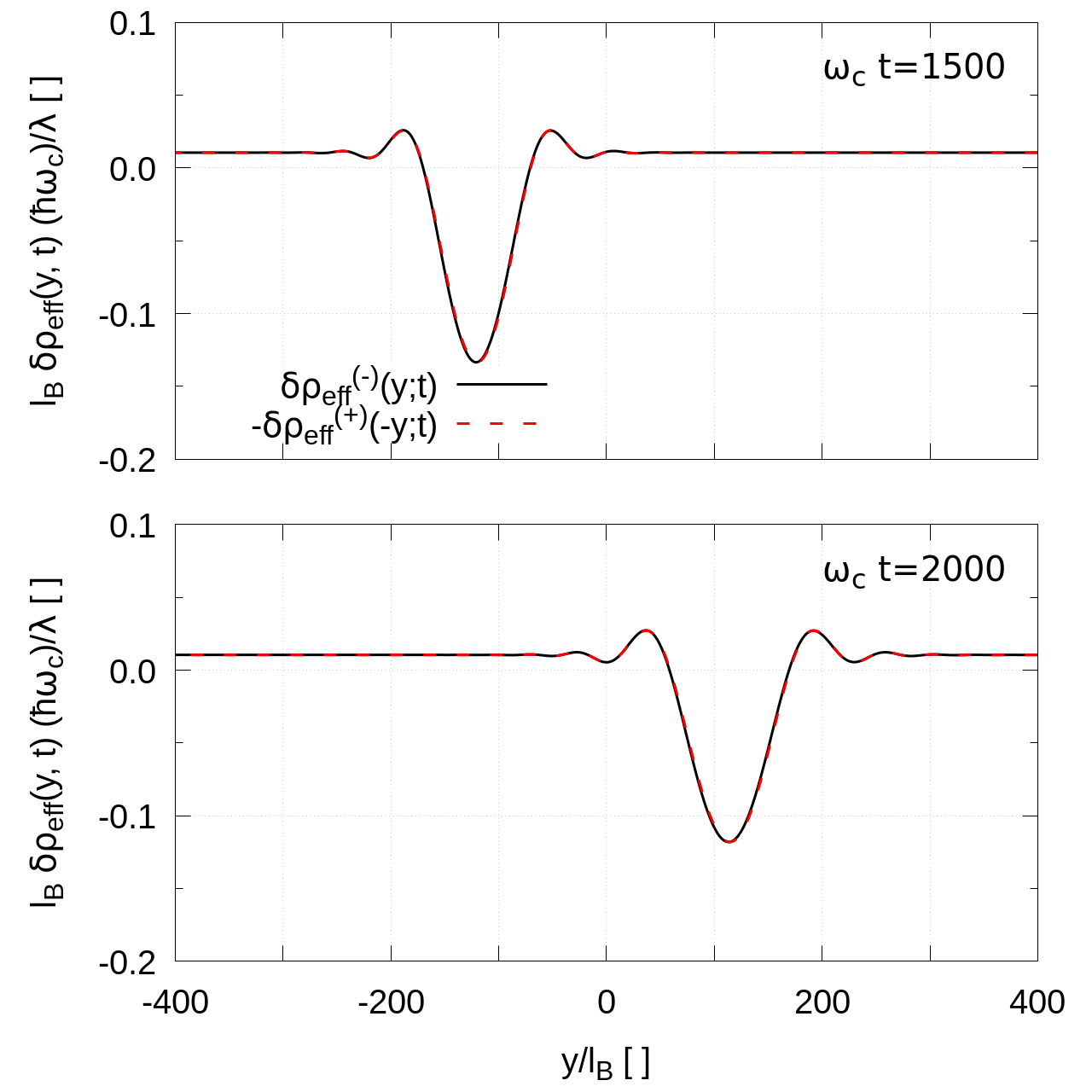}
	\end{minipage}    
	\caption[The LOF caption]{The two images show some snapshots of the time evolution of $\delta\rho_\text{eff}^{(-)}(y;t)$ (black curves) and $-\delta\rho_\text{eff}^{(+)}(-y;t)$ (red dashed ones), both divided by the excitation strength $\lambda$. 
	The data have been obtained from numerical simulations performed in the small excitation limit, with $\lambda=0.001\hbar\omega_c$.}
	\label{fig:sigmoid_linear_physicsRL}
\end{figure}

In Fig. \ref{fig:sigmoid_linear_physicsRL} the integrated density variations  $\delta\rho_\text{eff}^{(-)}(y;t)$ and $-\delta\rho_\text{eff}^{(+)}(-y;t)$ are plotted at different times in the small excitation strength limit. 
First of all we notice that the curves do overlap very well; this is a consequence of the excitation being perturbative (inversion $x\rightarrow-x$, $y\rightarrow-y$ is not a symmetry in the presence of the sigmoid potential eq. \ref{eq:SigmoidPotential}).
\newline We see that during the propagation lobes symmetrically appear about the packet peak, which drifts at the Fermi velocity; 
the appearance of these lobes is an effect of the Landau level at the Fermi point not being a straight line, as extensively discussed in the previous chapters.

\begin{figure}[htp!]
	\begin{minipage}{.5\textwidth}
		\centering
		\includegraphics[width=1.\textwidth]{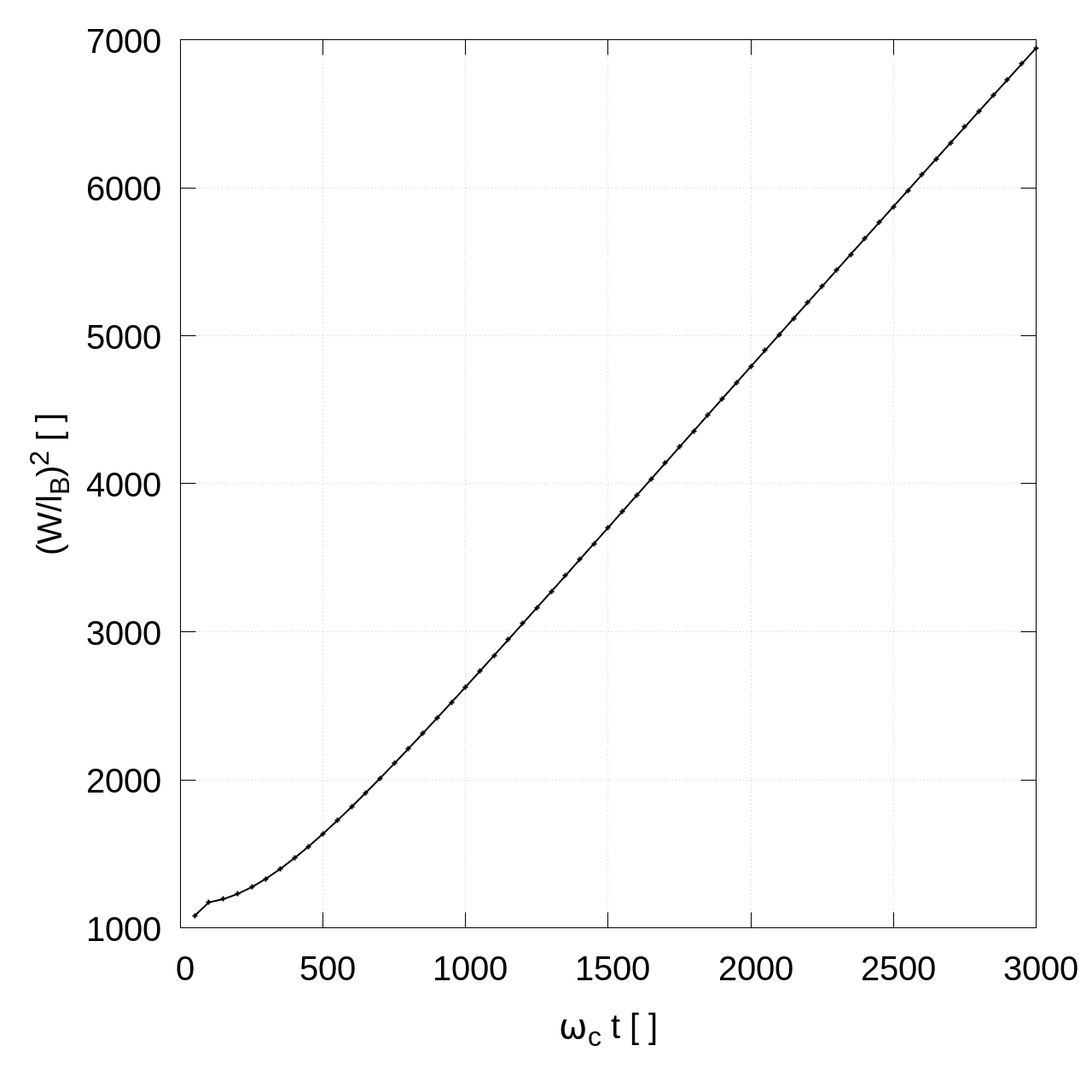}
	\end{minipage}%
	\begin{minipage}{0.5\textwidth}
		\centering
		\includegraphics[width=1.\textwidth]{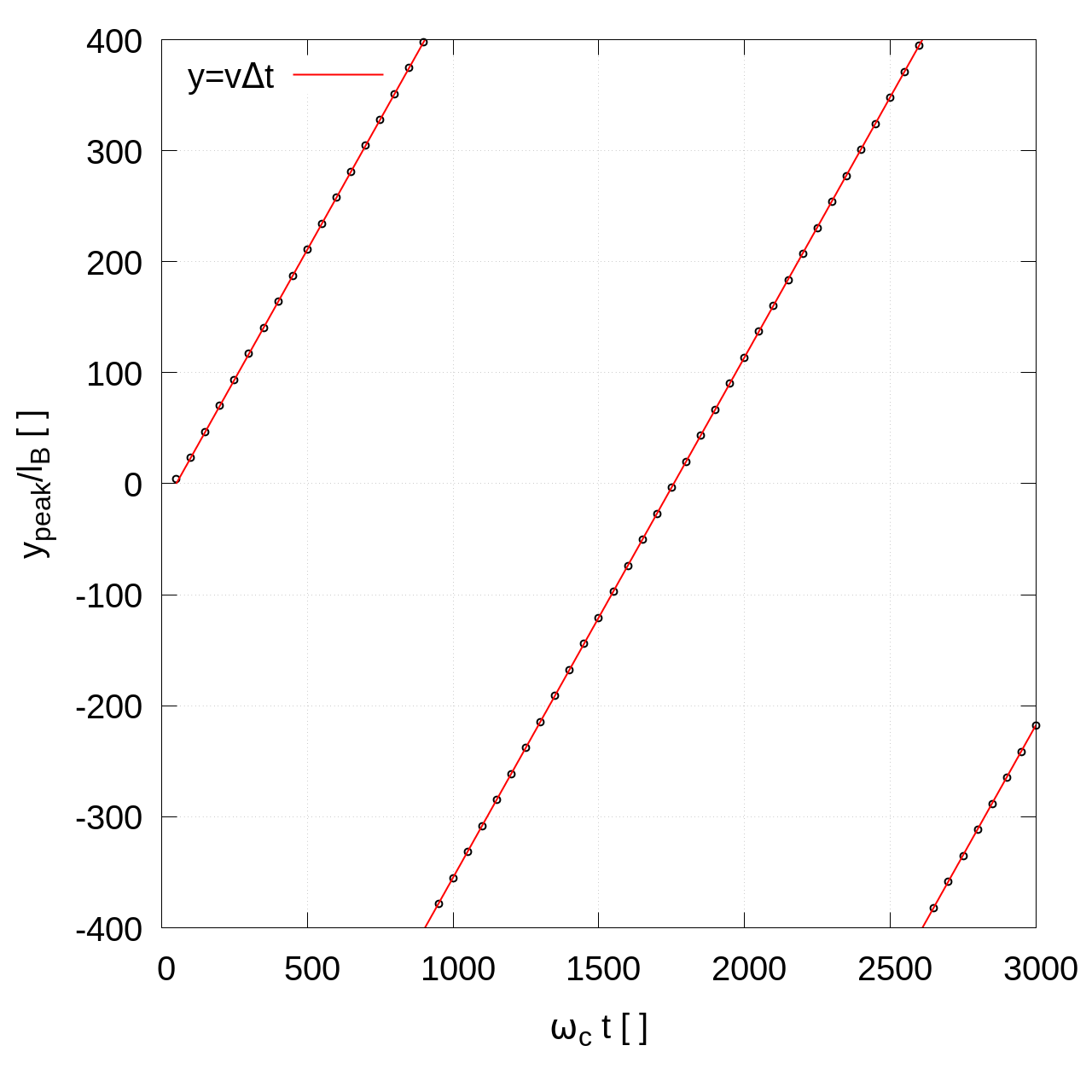}
	\end{minipage}    
	\caption[The LOF caption]{In the left hand panel the squared width is plotted; it has been computed by finding the full width at half maximum of the packet, the height being measured with respect to the $\delta\rho_\text{eff}^{(-)}(y;t)=0$ level.
	\newline In the right hand side panel the temporal evolution of the peak position is shown and compared with a straight line with slope given by the Fermi velocity $v\sim0.467l_B\omega_c$. The discontinuous jumps from $+\frac{L_y}{2}$ to $-\frac{L_y}{2}$ are caused by the presence of periodic boundary conditions along the corresponding edge.}
	\label{fig:WidthSpeedHeight}
\end{figure}

In the left hand side panel of Fig. \ref{fig:WidthSpeedHeight} the square of the width of the $\delta\rho_\text{eff}^{(-)}(y;t)$ packet is plotted. 
We see that the perturbative result derived in Chapter \ref{ch5} still holds; the reasoning which lead to eq. \ref{eq:packet_width} is indeed still valid.
In the right hand side panel of the same image, the peak position is plotted versus time and compared with $v\Delta t$ (the discontinuous jumps are evidently caused by the presence of the periodic conditions at the boundary), showing that the peak position does indeed move at the Fermi velocity.

\subsection{Non-linear effects}
\begin{figure}[htp!]
	\begin{minipage}{.5\textwidth}
		\centering
		\includegraphics[width=1.\textwidth]{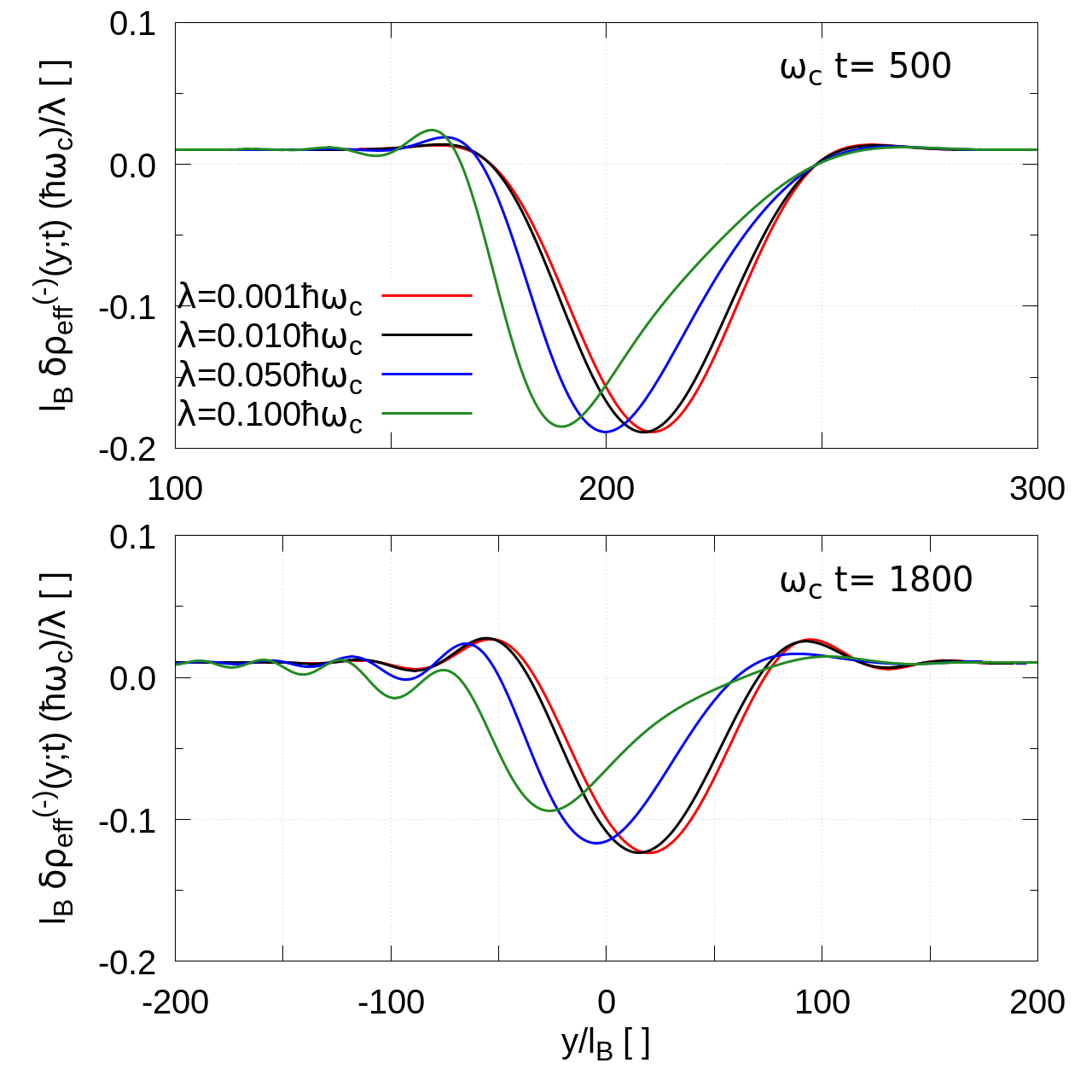}
	\end{minipage}%
	\begin{minipage}{0.5\textwidth}
		\centering
		\includegraphics[width=1.\textwidth]{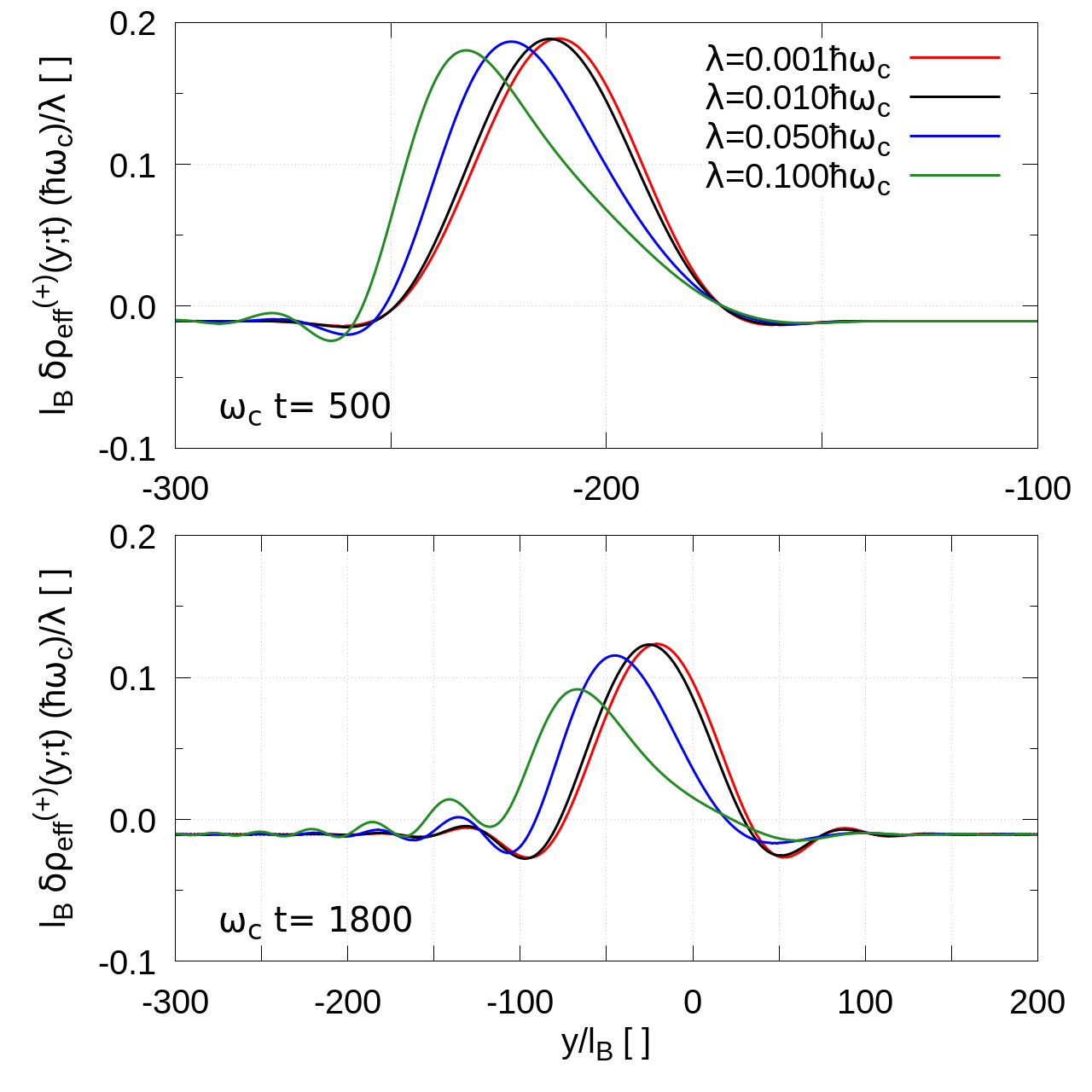}
	\end{minipage}    
	\caption[The LOF caption]{In the left hand side panels $\delta\rho_\text{eff}^{(-)}(y;t)$ is plotted at $\omega_c t=500$ (top panel) and $\omega_c t=1800$ (bottom panel) for different values of $\lambda$. The images in the right hand side are analogous, but $\delta\rho_\text{eff}^{(+)}(y;t)$ is shown instead. The density variations have been divided by the excitation strength $\lambda$ to make the comparison more clear.}
	\label{fig:increasing_excitation_comparison}
\end{figure}

In Fig. \ref{fig:increasing_excitation_comparison} the $\delta\rho_\text{eff}^{(-)}(y;t)$ and $\delta\rho_\text{eff}^{(+)}(y;t)$ packets (in the left and right panels respectively) are shown at fixed times as a function of $y$ for different values of the excitation strength $\lambda$.
It is apparent that important non-linear corrections are present, which will now be discussed.
\newline Notice first of all that the packets loose their symmetry about the $y=v \Delta t$ axis, producing small ripples preferentially on one side.

\begin{figure}[htp!]
	\begin{minipage}{.5\textwidth}
		\centering
		\includegraphics[width=1.\textwidth]{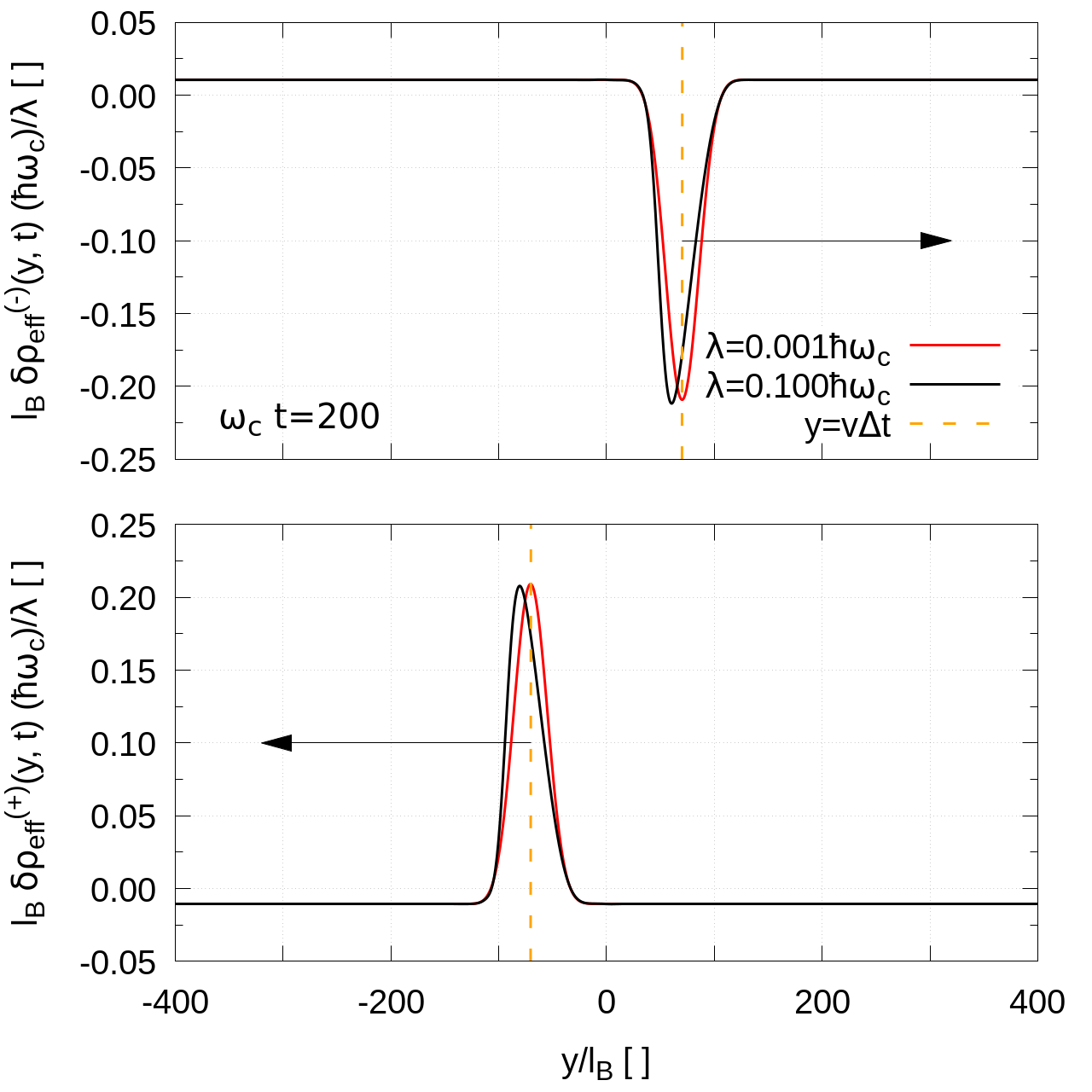}
	\end{minipage}%
	\begin{minipage}{0.5\textwidth}
		\centering
		\includegraphics[width=1.\textwidth]{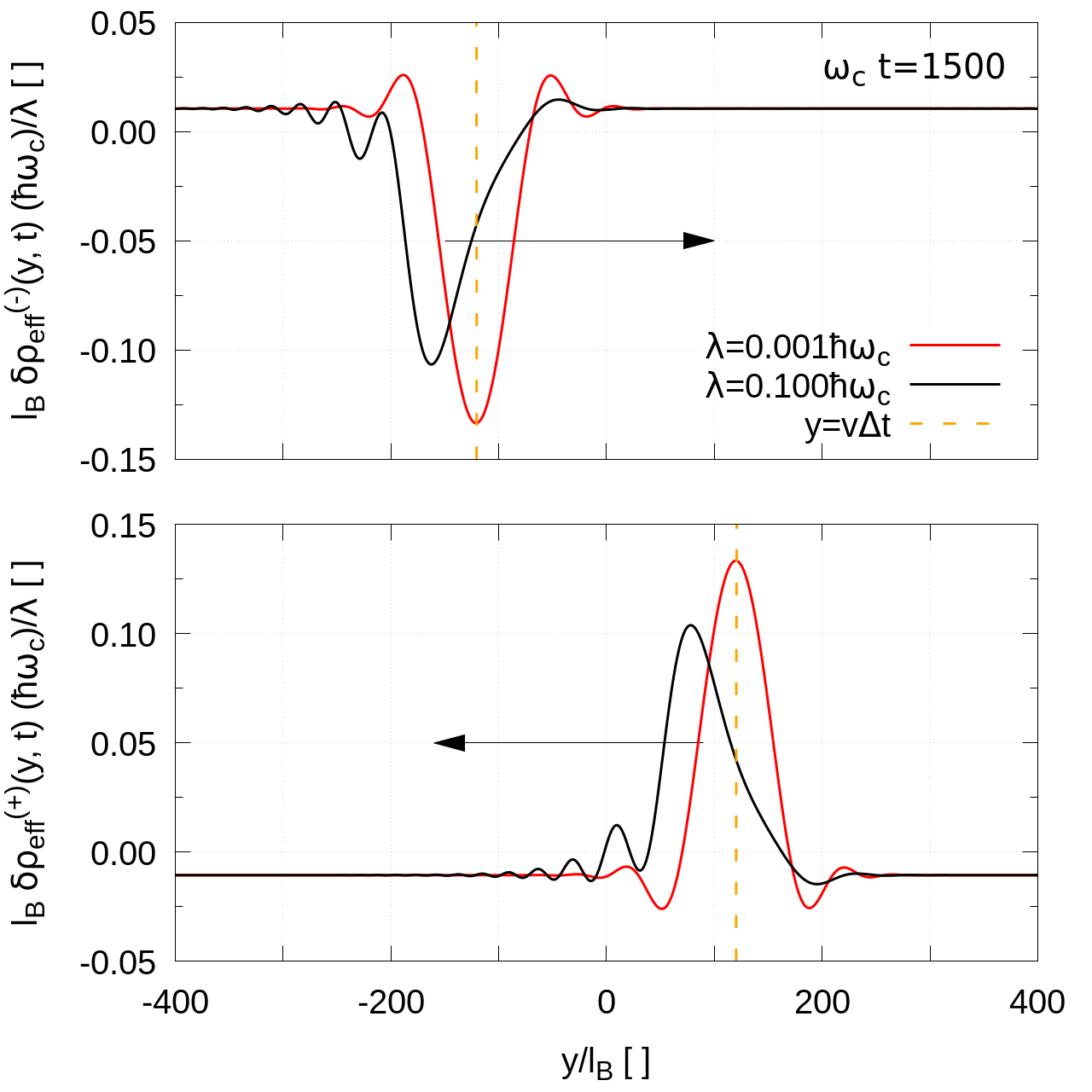}
	\end{minipage} 
	\caption[The LOF caption]{The images compare the different behaviour of $\delta\rho_\text{eff}^{(-)}(y;t)$ (top panels) and $\delta\rho_\text{eff}^{(+)}(y;t)$ (bottom ones), divided by the excitation strength $\lambda$, at $\omega_c t=200$ and $\omega_c t=1500$ (left and right hand side panels respectively).
	\newline The data have been generated with an excitation strength parameter $\lambda=0.100\hbar\omega_c$ (black curves) and  $\lambda=0.001\hbar\omega_c$ (red curves). 
	\newline Arrows showing the propagation direction are schematically shown, and the orange dashed line corresponds to $y=v\Delta t$ (opportunely accommodated for the presence of the periodic conditions at the boundary).}
	\label{fig:erfx_nonlinear_shapes_comparison}
\end{figure}

In Fig. \ref{fig:erfx_nonlinear_shapes_comparison} $\delta\rho_\text{eff}^{(-)}(y;t)$ and  $\delta\rho_\text{eff}^{(+)}(y;t)$ are plotted as a function of $y$ in the top and bottom panels respectively, at $\omega_c t=200$ (left panels) and $\omega_c t=1500$ (panels on the right). 
The arrows serve as a reminder for the propagation direction of the chiral mode. The small strength excitation limit profiles are shown as a comparison.
\newline We see that $\delta\rho_\text{eff}^{(-)}(y;t)$ (plotted in the top panels) is a density dip (compare also with the right hand side panel of Fig. \ref{fig:transient_current_and_density_erf}) travelling towards positive $y$ values. In the non-perturbative regime (black curve), as soon as the excitation has been turned off the peak starts to \virgolette{bend} backwards with respect to the direction of the motion, making the tail of the dip steeper than the front.
Lobes interestingly appear during the propagation more prominently in the tail of the dip, while its front gets flattened out:
we can indeed see that the tail lobes are magnified with respect to those appearing in the weak coupling regime; the front lobes instead are smaller.
\newline Conversely  $\delta\rho_\text{eff}^{(+)}(y;t)$ (plotted in the bottom panels) represents a density \virgolette{bump} chirally drifting towards negative $y$ values. 
After the excitation has been turned off, the packet bends in the forward motion direction making this side steeper. 
Greater lobes are this time magnified in the packet front and flattened out in the tail.
\newline In both cases the ripples are generated mainly on the steeper side of the dip/bump, and it can be seen that the peak travels at a speed which slightly smaller or higher than the Fermi velocity, i.e. the propagation speed appears to be density dependent.

Finally, notice that by changing $\lambda$ with $-\lambda$ we would have the density dip propagating towards the negative $y$ axis direction and the bump towards positive $y$. 
However if we perform a parity transformation, letting $x\rightarrow-x$ and $y\rightarrow-y$, we see that the perturbed ground state Hamiltonian $\mathcal{H}+V(y;t)$ (see equations \ref{eq:problem_hamiltonitan} and \ref{eq:SigmoidPotential}) stays the same, which means that\footnote{This is the reason why numerical simulations have not been performed in the case $\lambda<0$ as in Chapter \ref{ch5}.} $\rho_\lambda(x,y;t)=\rho_{-\lambda}(-x,-y)$;
this in turn implies e.g. that a density bump localised on the $x<0$ edge and chirally propagating in the $y>0$ direction will deform in the exact same way as the bump shown in Fig. \ref{fig:erfx_nonlinear_shapes_comparison}, i.e. by bending towards the direction of the motion and \virgolette{producing} ripples on this side.

We notice that the behaviour is resemblant of non-linear hydrodynamics dissipationless shock-waves: a propagating packet develops a steep front due to different propagation velocity of its parts; this front will eventually become unstable and the density hole/bump will decay into many oscillating features with smaller amplitude.

\begin{figure}[htp!]
	\begin{minipage}{.5\textwidth}
		\centering
		\includegraphics[width=1.\textwidth]{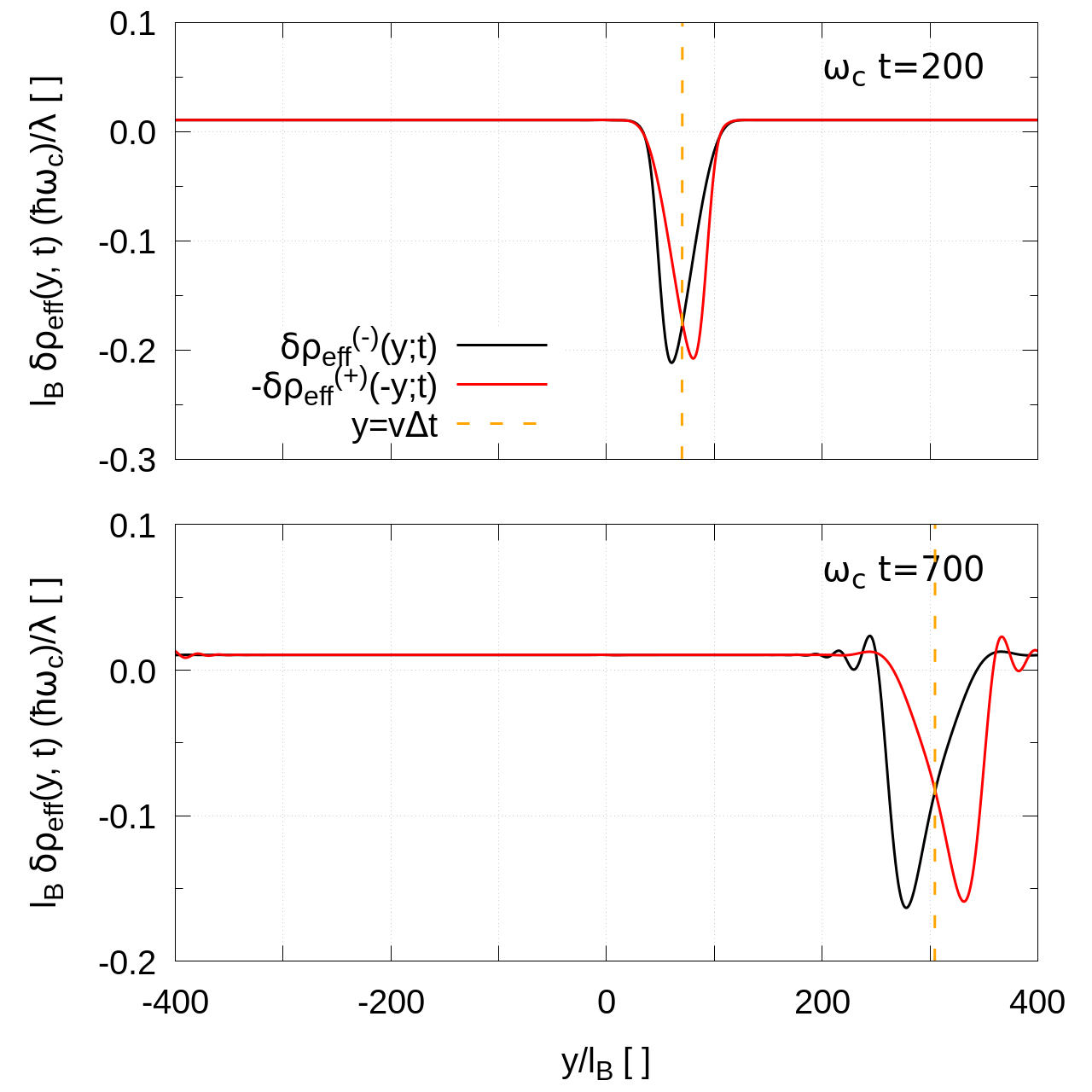}
	\end{minipage}%
	\begin{minipage}{0.5\textwidth}
		\centering
		\includegraphics[width=1.\textwidth]{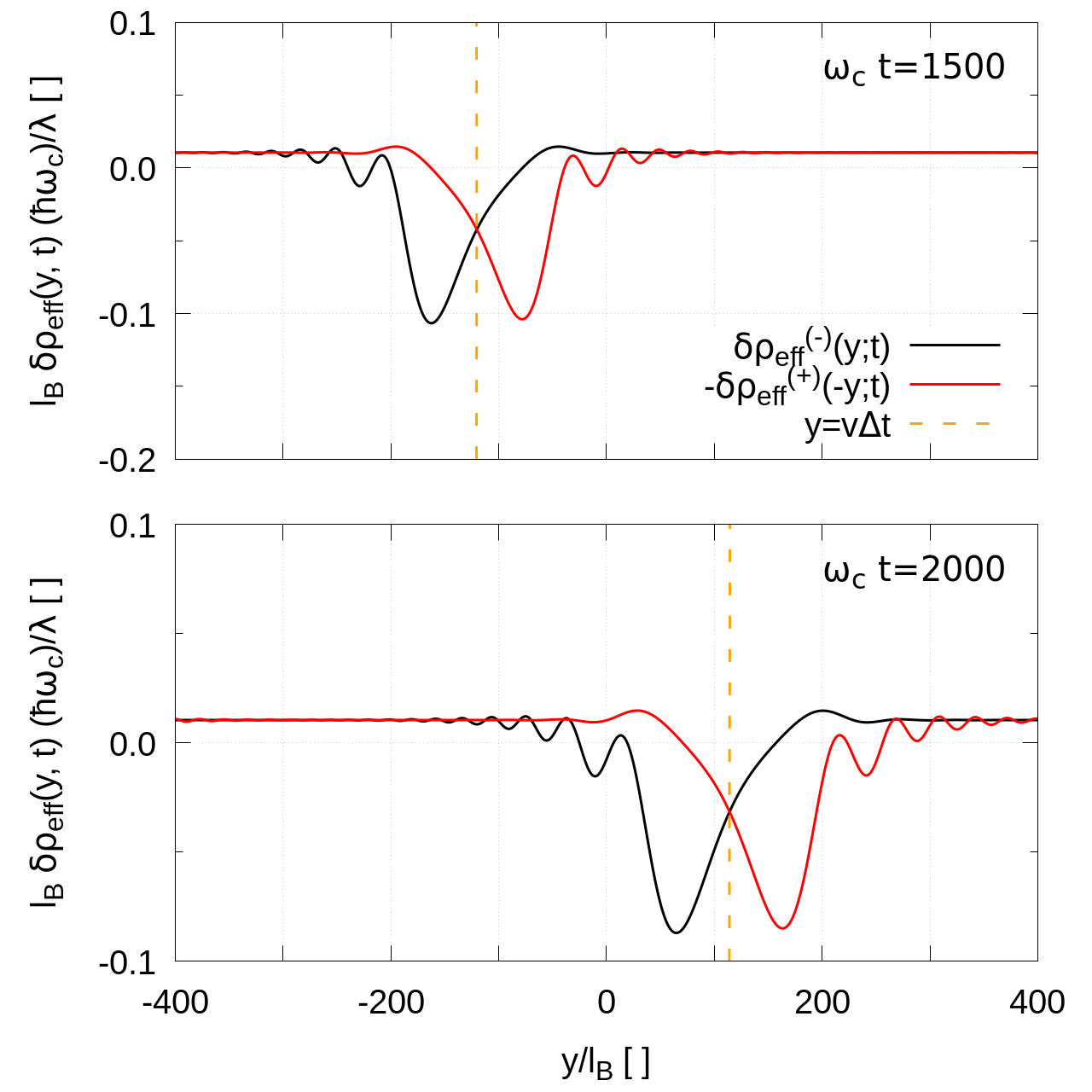}
	\end{minipage}    
	\caption[The LOF caption]{The two images show some snapshots of the time evolution of $\delta\rho_\text{eff}^{(-)}(y;t)$ (black curves) and $-\delta\rho_\text{eff}^{(+)}(-y;t)$ (red dashed ones), both divided by the excitation strength $\lambda$. 
		\newline The data have been obtained from numerical simulations performed in the non-perturbative excitation limit, with $\lambda=0.1\hbar\omega_c$.
		\newline The dashed orange line is a straight line through $y=v\Delta t$.}
	\label{fig:NonLinearTimeEvolution}
\end{figure}

\begin{figure}[htp!]
	\begin{minipage}{.5\textwidth}
		\centering
		\includegraphics[width=1.\textwidth]{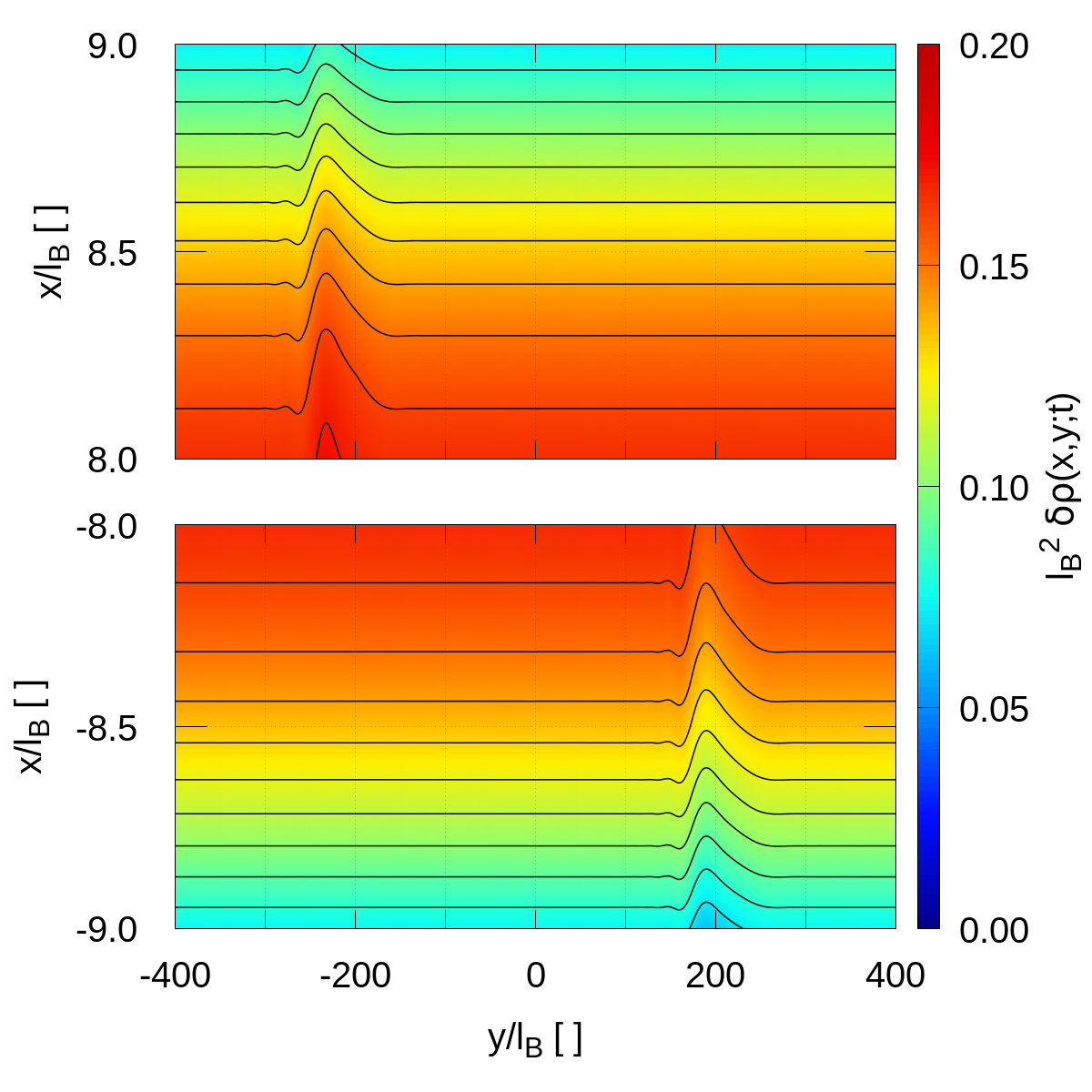}
	\end{minipage}%
	\begin{minipage}{0.5\textwidth}
		\centering
		\includegraphics[width=1.\textwidth]{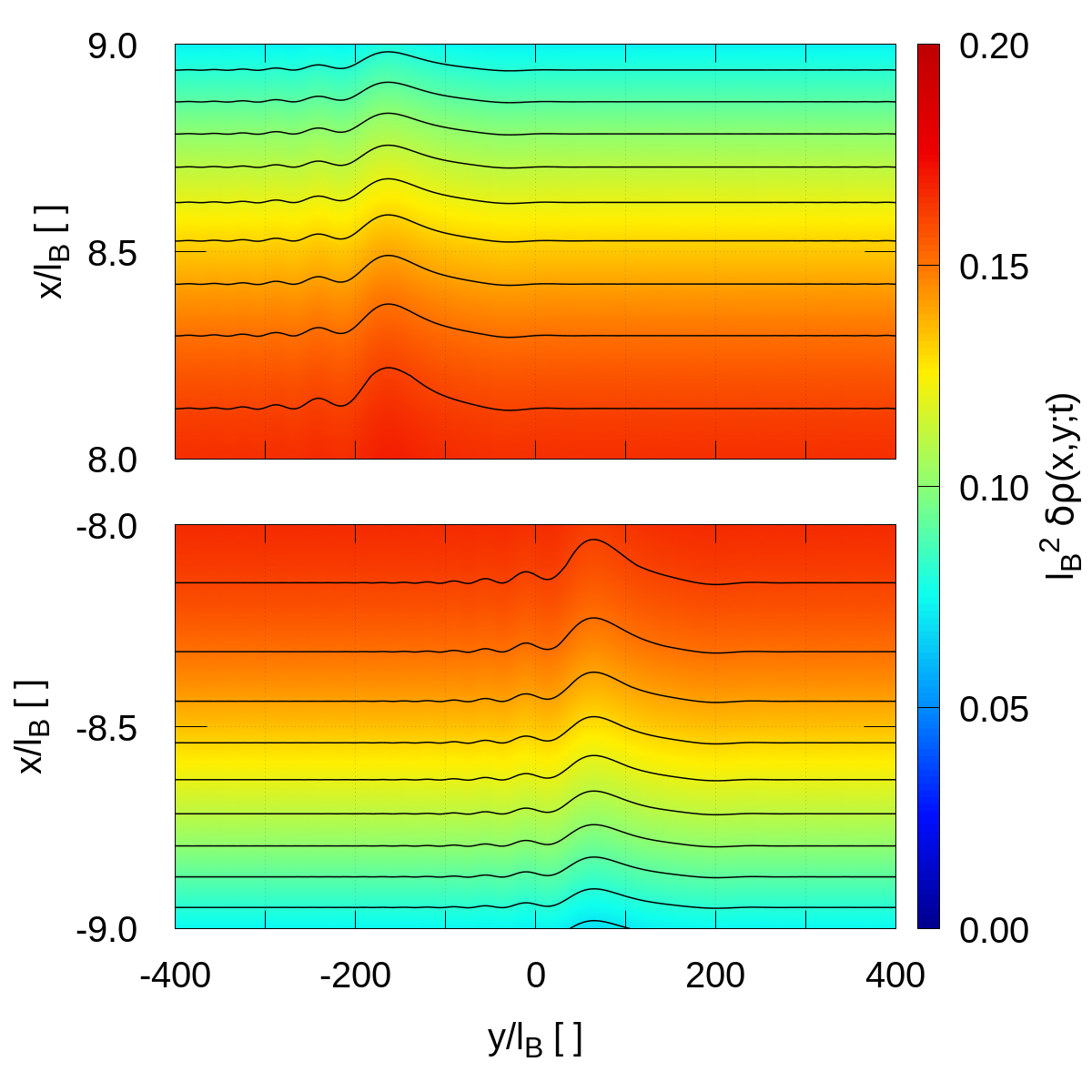}
	\end{minipage}
	\caption[The LOF caption]{The two images show the \textit{full} system density $\rho(x,y;t)$ excited by the potential term described in eq. \ref{eq:SigmoidPotential} with $\lambda=0.100\hbar\omega_c$, when $\omega_c t=500$ and $\omega_c t=2000$, respectively in the left and right hand side panels. 
		The top and bottom panels are enlarged pictures of the $x>0$ and $x<0$ system edges.}
	\label{fig:full_rho_erf_snapshots}
\end{figure}

Another intriguing fact may be observed. 
As the excitation strength is increased, there is no reason why $\delta\rho_\text{eff}^{(-)}(y;t)$ and $-\delta\rho_\text{eff}^{(+)}(-y;t)$ should be equal; this is not the case indeed, as can be seen in the direct comparisons showed both in the left and right panels of Fig. \ref{fig:NonLinearTimeEvolution}.
Notice however that for some to me yet not-understood reason the curves $\delta\rho_\text{eff}^{(-)}(y;t)$ and $-\delta\rho_\text{eff}^{(+)}(-y;t)$ shown are very close to being mirror images through the $y=v\Delta t$ axis, i.e. there is some sort of symmetry linking the dynamic evolution of density holes and bumps. This insight will be useful for devising an effective non-linear hydrodynamic picture of the edge evolution.
	
In Fig. \ref{fig:full_rho_erf_snapshots} two snapshots of the full system density $\rho(x,y;t)$ are finally shown.

%Incidentally this would also explain why the density profiles analysed in Chapter \ref{ch5} were close to be odd about $y=v\Delta t$ regardless of the excitation strength. 
%!TEX root = ../main.tex
% Chapter Template

\chapter{Conclusions and future perspectives}\label{ch7}
\lhead{Chapter 8. Conclusions and future perspectives} 

In this final chapter the work and the main results which have been found are briefly recapped.
Some perspectives of future work are also made.

%\subsection*{Results}
The edge dynamics of an integer Quantum Hall state has been studied in the case of a simple semi-infinite Hall bar geometry, confined along one spatial direction by means of a simple model potential; along the other direction periodic boundary conditions have been used.
This choice allowed us to exactly study the linear but more importantly the non-linear edge dynamics without renouncing to a microscopic viewpoint nor introducing severe approximations which are otherwise mandatory in order to treat many-body interacting fermion systems.

\noindent The single electron Hamiltonian has been numerically diagonalised with the purpose of building the many-body non-interacting ground state; in order to numerically investigate the dynamics an alternate direction implicit Crank-Nicolson time evolution scheme has been implemented, which has thoroughly been tested.
\newline The dynamics of the $\nu=1$ integer quantum Hall state has been studied by subjecting the system to several time dependent excitations of different strength and modulated along the edge direction.
In particular, the density response to spatially periodic, Gaussian and sigmoid shaped excitations has been considered.
In the weak excitation regime the problem has been tackled by means of perturbation theory too, and some simple results beyond the edge linear dispersion paradigm have been derived; the decay of the induced density excitations as well as the broadening of a localised density bump or hole (owing to dispersive phenomena) have been given a simple characterisation in terms of the Landau level curvature at the system Fermi surface. 
In the case of a spatially periodic excitation, the role the curvature parameter plays in the generation of higher harmonics in the density response has been investigated through second order perturbation theory.
\newline The non-linear dynamics has been investigated by means of the developed algorithm.
The deviations from the simple linear order predictions have been numerically investigated, and important differences between the dynamical behaviour of a density hole and bump have been observed, as well as an intriguing non-linear dynamic behaviour.
In particular, the bottom of a hole in the system density is found to travel slightly slower than the Fermi velocity; this has the effect of producing a steeper tail, where large ripples are produced, and a smoother front.
On the other hand, the peak of a density heap travels slighlty faster than the Fermi velocity; the effect is the packet-front steepening, where lobes facing the propagation direction grow.
\newline This simple phenomenological characterization is more manifestly observed in the case of a sigmoid shaped potential (producing an initially bell shaped hole or bump) and allows to qualitatively understand the non-linear response to both the spatially periodic and Gaussian excitation; in particular it explains the strongly asymmetric behaviour when the sign of the Gaussian potential is reversed.

\noindent The non-linear effects have been qualitatively discussed and characterised based on the numerical simulations. It will be the subject of future work to restate the problem in terms of some non-linear effective chiral wave equation for the 1D edge density, possibly derived from the microscopic theory of the system; 
the observed behaviour is indeed somehow resemblant of dissipationless shock-wave phenomena in the realm of non-linear hydrodynamics.
The study of a (non-linear) hydrodynamic effective description of the system edge is indeed a current research topic in the study of the boundary modes of a fractional quantum Hall system, and will surely be the subject of future study and research, since
fractional quantum Hall systems offer a way richer physics, were electron-electron correlations play a prominent role and fractionally charged excitations come into play.
These phenomena are being investigated in the literature through very abstract techniques; our future work will rather try to focus into developing a more transparent understanding of the underlying physics, both in continuous geometries as well as discrete lattice systems with a non-trivial band topology.
This will be particularly interesting in the context of highly tunable and clean quantum Hall analogues, for instance confined cold atoms subjected to synthetic gauge fields or topological photonic systems.

%\subsection*{Future perspectives} 

%----------------------------------------------------------------------------------------
%	THESIS CONTENT - APPENDICES
%----------------------------------------------------------------------------------------

\addtocontents{toc}{\vspace{2em}} % Add a gap in the Contents, for aesthetics

\appendix % Cue to tell LaTeX that the following 'chapters' are Appendices

% Include the appendices of the thesis as separate files from the Appendices folder
% Uncomment the lines as you write the Appendices

\chapter{First and second order moments for a Gaussian wavepacket in a magnetic field} % Main appendix title

\graphicspath{{./picAppA/}}

\label{app:GaussianMoments} % Change X to a consecutive letter; for referencing this appendix elsewhere, use \ref{AppendixX}

\lhead{Appendix A. Gaussian wavepacket in a magnetic field} % Change X to a consecutive letter; this is for the header on each page - perhaps a shortened title

\subsection*{Relevant commutation relations}

The system evolves under the bulk Hamiltonian
\begin{equation}
\mathcal{H}=\frac{\pi_x^2+\pi_y^2}{2}.
\end{equation}
\virgolette{Natural} units are used in this section.
Some useful commutation relations between the kinetic momenta and the coordinates can be immediately derived
\begin{equation}
\label{eq:commutation_relations1}
\begin{cases}
\left[x,\pi_x\right] = \left[y,\pi_y\right] = \left[\pi_y,\pi_x\right] = i
\\
\left[x,x\right]=\left[y,y\right]=\left[x,y\right]=\left[y,\pi_x\right]=[x,\pi_y]=0
\end{cases}
\end{equation}
and from these ones those with the Hamiltonian follow immediately
\begin{equation}
\label{eq:commutation_relations2}
\begin{alignedat}{2}
\begin{cases}
\left[x,\mathcal{H}\right] = \left[\pi_y,\mathcal{H}\right] &= i\,\pi_x
\\
\left[y,\mathcal{H}\right] = - \left[\pi_x,\mathcal{H}\right] &= i\,\pi_y.
\end{cases}
\end{alignedat}
\end{equation}

\subsection*{Time evolution of first order moments}
The time evolution equations for the expectation values of the kinetic momenta immediately follow from the commutation relations \ref{eq:commutation_relations2}
\begin{equation}
\begin{cases}
\dot{\braket{\pi_x}} = \frac{1}{i}\braket{\left[\pi_x,\mathcal{H}\right]} = -\braket{\pi_y}
\\
\dot{\braket{\pi_y}} = \frac{1}{i}\braket{\left[\pi_y,\mathcal{H}\right]} = \braket{\pi_x}
\end{cases}
\end{equation}
whose solution is (choosing $t_0=0$)
\begin{equation}
\label{eq:pix_piy_expectation}
\left(
\begin{array}{c}
\braket{\pi_x}_t\\
\braket{\pi_y}_t 
\end{array}
\right)
=
\left(
\begin{array}{cc}
\cos(t) & -\sin(t) \\
\sin(t) & \phantom{+}\cos(t)
\end{array}
\right)
\left(
\begin{array}{c}
\braket{\pi_x}_0\\
\braket{\pi_y}_0
\end{array}
\right)
\end{equation}
Integration with respect to time gives the time-evolution of expectation value of the coordinates
\begin{equation}
\label{eq:x_y_expectation}
\left(
\begin{array}{c}
\braket{x}_t-\braket{x}_0\\
\braket{y}_t-\braket{y}_0
\end{array}
\right)
=
\left(
\begin{array}{cc}
\sin(t) & \cos(t)-1 \\
1-\cos(t) & \sin(t)
\end{array}
\right)
\left(
\begin{array}{c}
\braket{\pi_x}_0\\
\braket{\pi_y}_0
\end{array}
\right)
\end{equation}
i.e. the average position of the wavepacket follows the classical cyclotron orbit, as expected, centred at $\mathcal{C}=(\braket{x}_0-\braket{\pi_y}_0,\braket{y}_0-\braket{\pi_x}_0)$ and radius $\sqrt{\braket{\pi_x}_0^2+\braket{\pi_y}_0^2}\propto \sqrt{\braket{\mathcal{H}}}$.
\newline Notice that it was possible to compute the first order moments because the system of differential equations we obtained is closed and higher order moments do not appear in the differential equations. This is true at any order owing to the quadratic nature of the Hamiltonian.

\subsection*{Time evolution of second order moments}
\subsubsection*{· Quadratic in the momenta}
The quadratic terms in the kinetic momenta obey the following homogeneous system of differential equations
\begin{equation}
\begin{alignedat}{2}
\begin{cases}
\frac{d}{dt}\,\braket{\pi_x^2} &=-\braket{\pi_x \pi_y +\pi_y \pi_x}
\\
\frac{d}{dt}\,\braket{\pi_x\,\pi_y+\pi_y\pi_x} &= 2\braket{\pi_x^2 -\pi_y^2}
\\
\frac{d}{dt}\,\braket{\pi_y^2} &= +\braket{\pi_x\pi_y+\pi_y\pi_x}
\end{cases}
\end{alignedat}
\end{equation}
whose solution is straightforwardly obtained by matrix exponentiation. Defining for convenience the symmetrized product of two operators as e.g. $s_{\pi_x,\pi_y}=\pi_x\,\pi_y+\pi_y\pi_x$, the solution reads
\begin{equation}
\label{eq:pix2_pixpiy_piy2_expectation}
\left(
\begin{array}{c}
\braket{\pi_x^2}_t
\\
\braket{s_{\pi_x,\pi_y}}_t
\\
\braket{\pi_y^2}_t
\end{array}
\right)
=
\left(
\begin{array}{ccc}
\cos^2(t)&-\frac{1}{2}\sin(2t)&\phantom{+}\sin^2(t)	\\
\sin(2t)&\phantom{+}\cos(2t)&-\sin(2t)	\\
\sin^2(t)&\phantom{+}\frac{1}{2}\sin(2t)&\phantom{+}\cos^2(t)	\\
\end{array}
\right)
\left(
\begin{array}{c}
\braket{\pi_x^2}_0
\\
\braket{s_{\pi_x,\pi_y}}_0
\\
\braket{\pi_y^2}_0
\end{array}
\right).
\end{equation}
It is easy to check that $\braket{\pi_x^2}_t+\braket{\pi_y^2}_t = \braket{\pi_x^2}_0 + \braket{\pi_y^2}_0$, as required by energy conservation; however notice that they are not (in general) separately constants of motion, rather energy \virgolette{bounces} between the two contributions.
\newline If $\braket{s_{\pi_x,\pi_y}}_0=0$ and $\braket{\pi_x^2}_0=\braket{\pi_y^2}_0$ there is no time-evolution at all for these variables, which can be seen as a symmetry consequence (The Hamiltonian being invariant under $\pi_x \leftrightarrow \pi_y$.).
\newline Finally, notice that the variances do obey the same equations (with the $t=0$ coefficients substituted by adequate terms), as can be easily seen by directly computing for example $\sigma_{\pi_x}^2(t)=\braket{\pi_x^2}_t-\braket{\pi_x}_t^2$. %By choosing $\sigma_{\pi_x}^2(t=0)=\sigma_{\pi_y}^2(t=0)$ and $\braket{s_{\pi_x,\pi_y}}_0-2\braket{\pi_x}_0\braket{\pi_y}_0=0$, the variances will not evolve in time.
As a consequence, the sum $\sigma_{\pi_x}^2+\sigma_{\pi_y}^2$ is a constant of motion, but each term is a function of time (the discussion being completely analogous to the one above).
\newline Consider the simple case $\braket{s_{\pi_x,\pi_y}}_0-2\braket{\pi_x}_0\braket{\pi_y}_0=0$. In this case
\begin{equation}
\label{eq:sigmas}
\begin{cases}
\sigma_{\pi_x}^2(t) = \frac{\sigma_{\pi_x}^2(0)+\sigma_{\pi_y}^2(0)}{2} + \frac{\sigma_{\pi_x}^2(0)-\sigma_{\pi_y}^2(0)}{2}\,\cos(2t)
\\
\sigma_{\pi_y}^2(t) = \frac{\sigma_{\pi_x}^2(0)+\sigma_{\pi_y}^2(0)}{2} - \frac{\sigma_{\pi_x}^2(0)-\sigma_{\pi_y}^2(0)}{2}\,\cos(2t)
\end{cases}
\end{equation}
and we see that the variances have an average value $\frac{\sigma_{\pi_x}^2(0)+\sigma_{\pi_y}^2(0)}{2}$ but oscillate with  amplitude $\frac{\sigma_{\pi_x}^2(0)-\sigma_{\pi_y}^2(0)}{2}$ at twice the cyclotron frequency.
\newline Notice that when $\sigma_{\pi_x}^2(t)$ increases when $\sigma_{\pi_y}^2(t)$ decreases (in order to keep their sum constant), and vice-versa. This can be understood also as being a consequence of the Heisenberg uncertainty principle, since from \ref{eq:commutation_relations1} we see that $\{\pi_x,\pi_y\}$ are a pair of conjugated variables. 

\noindent This kind of reasoning can be generalised. Since $[x,\pi_x]=i$, the Heisenberg uncertainty theorem gives $\sigma_x\sigma_{\pi_x}\geq\frac{1}{2}$. Heuristically we therefore expect the width of the packet to increase when $\sigma_{\pi_x}$ decreases. For example,
if at the initial instant $\sigma_{\pi_x}^2$ is larger than $\sigma_{\pi_y}^2$, $\sigma_{\pi_x}^2(t)$ will initially decrease in time as a consequence of eq. \ref{eq:sigmas}; correspondingly,  $\sigma_x^2$ will initially increase.
\newline In exactly the same way these results can be extended to the pair $\{y,\pi_y\}$.

\noindent In conclusion we can thus understand the spatial \virgolette{pulsation} of the wavepacket as its centre of mass rotates following a cyclotron orbit as being a consequence of the Heisenberg uncertainty principle (together with some symmetry of the bulk Hamiltonian).

\subsubsection*{· Mixed terms in the momenta and the coordinates}
The equations for $s_{x,\pi_x}$ and $s_{x,\pi_y}$ are coupled and contain source terms
\begin{equation}
\label{eq:x_pix__x_piy}
\begin{alignedat}{2}
\begin{cases}
\frac{d}{dt}\,\braket{x\pi_x+\pi_x x} &=-\braket{x \pi_y +\pi_y x} + 2\braket{\pi_x^2}
\\
\frac{d}{dt}\,\braket{x \pi_y +\pi_y x} &= + \braket{x\pi_x+\pi_x x} + \braket{\pi_x\pi_y + \pi_y\pi_x}.
\end{cases}
\end{alignedat}
\end{equation}
The general solution is easily found. By using the following ansatz
\begin{equation}
\begin{cases}
\braket{s_{x,\pi_x}}_t= \left[\braket{s_{x,\pi_x}}_0 + F(t)\right]\,\cos(t) - \left[\braket{s_{x,\pi_y}}_0 + G(t)\right]\,\sin(t)
\\
\braket{s_{x,\pi_y}}_t= \left[\braket{s_{x,\pi_x}}_0 + F(t)\right]\,\sin(t) + \left[\braket{s_{x,\pi_y}}_0 + G(t)\right]\,\cos(t)
\end{cases}
\end{equation}
and performing some straightforward algebra one finds the functions $F(t)$ and $G(t)$ to be given by
\begin{equation}
\begin{cases}
F(t) = \int_0^t\,d\tau \left(+2\braket{\pi_x^2}_\tau\,\cos(\tau) + \braket{s_{\pi_x\pi_y}}_\tau\sin(\tau)\right)
\\
G(t) = \int_0^t\,d\tau\left(-2\braket{\pi_x^2}_\tau\,\sin(\tau)+\braket{s_{\pi_x\pi_y}}_\tau\cos(\tau)\right).
\end{cases}
\end{equation}

The equations for $s_{y,\pi_x}$ and $s_{y,\pi_y}$ are completely analogous (symmetric) to the ones above for $s_{x,\pi_x}$ and $s_{x,\pi_y}$ (eq. \ref{eq:x_pix__x_piy})
\begin{equation}
\label{eq:y_piy__y_pix}
\begin{alignedat}{2}
\begin{cases}
\frac{d}{dt}\,\braket{y\pi_x+\pi_x y} &=-\braket{y \pi_y +\pi_y y} + \braket{\pi_x\pi_y + \pi_y\pi_x}
\\
\frac{d}{dt}\,\braket{y \pi_y +\pi_y y} &= + \braket{y\pi_x+\pi_x y} + 2\braket{\pi_y^2}
\end{cases}
\end{alignedat}
\end{equation}
and the solution is the straightforward generalization of the one written above
\begin{equation}
\begin{cases}
\braket{s_{y,\pi_y}}_t = +\left[\braket{s_{y,\pi_y}}_0 + F'(t)\right]\,\cos(t) + \left[\braket{s_{y,\pi_x}}_0 + G'(t)\right]\,\sin(t)
\\
\braket{s_{y,\pi_x}}_t= -\left[\braket{s_{y,\pi_y}}_0 + F'(t)\right]\,\sin(t) + \left[\braket{s_{y,\pi_x}}_0 + G'(t)\right]\,\cos(t).
\end{cases}
\end{equation}
where
\begin{equation}
\begin{cases}
F'(t) = \int_0^t\,d\tau \left(2\braket{\pi_y^2}_\tau\,\cos(\tau) - \braket{s_{\pi_x\pi_y}}_\tau\sin(\tau)\right)
\\
G'(t) = \int_0^t\,d\tau\left(2\braket{\pi_y^2}_\tau\,\sin(\tau)+\braket{s_{\pi_x\pi_y}}_\tau\cos(\tau)\right).
\end{cases}
\end{equation}

\subsubsection*{· Quadratic in the positions}
Finally the quadratic terms in the coordinates obey
\begin{equation}
\begin{alignedat}{2}
\begin{cases}
\frac{d}{dt}\,\braket{x^2} &= \braket{x\,\pi_x +\pi_x x}
\\
\frac{d}{dt}\,\braket{xy+yx} &= \braket{x \pi_y+\pi_y x} + \braket{y\pi_x+\pi_x y}
\\
\frac{d}{dt}\,\braket{y^2} &= \braket{y\,\pi_y +\pi_y y}
\end{cases}
\end{alignedat}
\end{equation}
and the solution is obtained by rather straightforward time-integration of the above equations (\ref{eq:x_pix__x_piy} and \ref{eq:y_piy__y_pix}).
\newline Notice that in general these moments will evolve in time differently than the squared first order moments, leading to a pulsation of the widths of a wavepacket moving along a cyclotron orbit, as indeed observed above.

\subsection*{Time evolution of a Gaussian packet}
The Landau gauge will now be used. With a Gaussian initial state of the form
\begin{equation}
\psi(x,y; t_0) = \left(\frac{1}{4\pi^2 \sigma_x^2\sigma_y^2}\right)^\frac{1}{4}\exp\left(-\frac{(x-x_0)^2}{4\sigma_x^2}\right)\exp\left(-\frac{(y-y_0)^2}{4\sigma_y^2}\right)
\end{equation}
the initial conditions are easily computed. Those for the first order moments read
\begin{equation}
\label{eq:InitialConditions1}
\begin{cases}
\braket{x}_0=x_0
\\
\braket{y}_0=y_0
\end{cases}
\quad
\begin{cases}
\braket{\pi_x}_0=0
\\
\braket{\pi_y}_0=x_0
\end{cases}
\end{equation}
and the ones for the second order moments
\begin{equation}
\label{eq:InitialConditions2}
\begin{cases}
\braket{\pi_x^2}_0=\frac{1}{4\sigma_x^2}
\\
\braket{s_{\pi_x,\pi_y}}_0=0
\\
\braket{\pi_y^2}_0=\frac{1}{4\sigma_y^2}+x_0^2+\sigma_x^2
\end{cases}
\quad
\begin{cases}
\braket{x^2}_0=x_0^2+\sigma_x^2
\\
\braket{xy}_0=x_0 y_0
\\
\braket{y^2}_0=y_0^2+\sigma_y^2
\end{cases}
\quad
\begin{cases}
\braket{s_{x,\pi_x}}_0 = \braket{s_{y,\pi_x}}_0 = 0
\\
\braket{s_{x,\pi_y}}_0 = 2\left(x_0^2+\sigma_x^2\right)
\\
\braket{s_{y,\pi_y}}_0 = 2x_0y_0.
\\
\end{cases}
\end{equation}
After some algebra one obtains the time evolution of the elements of the covariance-matrix
\begin{equation}
\label{eq:SpatialVariances}
\left\{
\begin{array}{lll}
\Delta_{xx}(t) &= \sigma_x^2
&+\frac{1}{\sigma_y^2}\,\sin^2\left(\frac{t}{2}\right)
-\left(\sigma_x^2-\frac{1}{4\sigma_x^2}+\frac{1}{4\sigma_y^2}\right)\,\sin^2(t)
\\
\Delta_{xy}(t) &= 
&+\left(\frac{1}{4\sigma_x^2}-\frac{1}{4\sigma_y^2}\right)\sin(t)
+\frac{1}{2}\left(\sigma_x^2-\frac{1}{4\sigma_x^2}+\frac{1}{4\sigma_y^2}\right)\sin(2t)
\\
\Delta_{yy}(t) &=
\sigma_y^2
&+\frac{1}{\sigma_x^2}\sin^2\left(\frac{t}{2}\right)
+\left(\sigma_x^2-\frac{1}{4\sigma_x^2}+\frac{1}{4\sigma_y^2}\right)\,\sin^2(t)
\end{array}
\right.
\end{equation}
where $\Delta_{xx}(t)=\braket{x^2}_t-\braket{x}_t^2$, $\Delta_{yy}(t)=\braket{y^2}_t-\braket{y}_t^2$ and $\Delta_{xy}(t)=\braket{xy}_t-\braket{x}_t\braket{y}_t$. 
\newline Notice that at $t=0$ $\Delta_{xx}=\sigma_x^2$ and $\sigma_{\pi_x}^2=\frac{1}{4\sigma_x^2}$, thus their product equals $\frac{1}{4}$, i.e. the Heisenberg inequality associated with the non-commuting observables $\pi_x$ and $x$ is saturated; during the time evolution the state is no longer a state of minimum uncertainty though.

The first order moments
\begin{equation}
\label{eq:first_order_moments}
\begin{cases}
\braket{x}_t = x_0 \cos(t)
\\
\braket{y}_t = y_0+x_0 \sin(t)
\end{cases}
\end{equation}
together with the spatial variances in eq. \ref{eq:SpatialVariances} uniquely determine the time-evolution of the squared modulus of the wavefunction (an additional position-dependent phase factor may be present, which however does not bother us too much if we are not interested in the study of some interference phenomena)
\begin{equation}
\label{eq:wavepacket_time_evo}
|\psi(x,y,;t)|^2=\sqrt{\frac{1}{4\pi^2\,\det(\Sigma)}}\,\exp\left(-\frac{1}{2}\,r_i\Sigma^{-1}_{ij} r_j\right)
\end{equation}
where
\begin{equation}
\Sigma=
\left(
\begin{array}{cc}
\Delta_{xx}(t)	& \Delta_{xy}(t) \\
\Delta_{xy}(t)	& \Delta_{yy}(t) \\
\end{array}
\right),
\qquad\quad
\mathbf{r}=
\left(
\begin{array}{c}
x-\braket{x}_t\\
y-\braket{y}_t
\end{array}\right).
\end{equation}

\vspace{1cm}
As a final comment, notice that (from eq. \ref{eq:SpatialVariances}) at $t=\pi$ the widths measure
\begin{equation}
\label{eq:variances_at_pi}
\begin{cases}
\Delta_{xx}(\pi)=\sigma_x^2+\frac{1}{\sigma_y^2}
\\
\Delta_{yy}(\pi)=\sigma_y^2+\frac{1}{\sigma_x^2}.
\end{cases}
\end{equation}
These are not equal to the initial ones, i.e. once the packet has completed half of its cyclotron motion its shape is deformed, which may seem quite odd at a first glance.

\begin{figure}[htp!]
	\centering
	\includegraphics[width=0.7\textwidth]{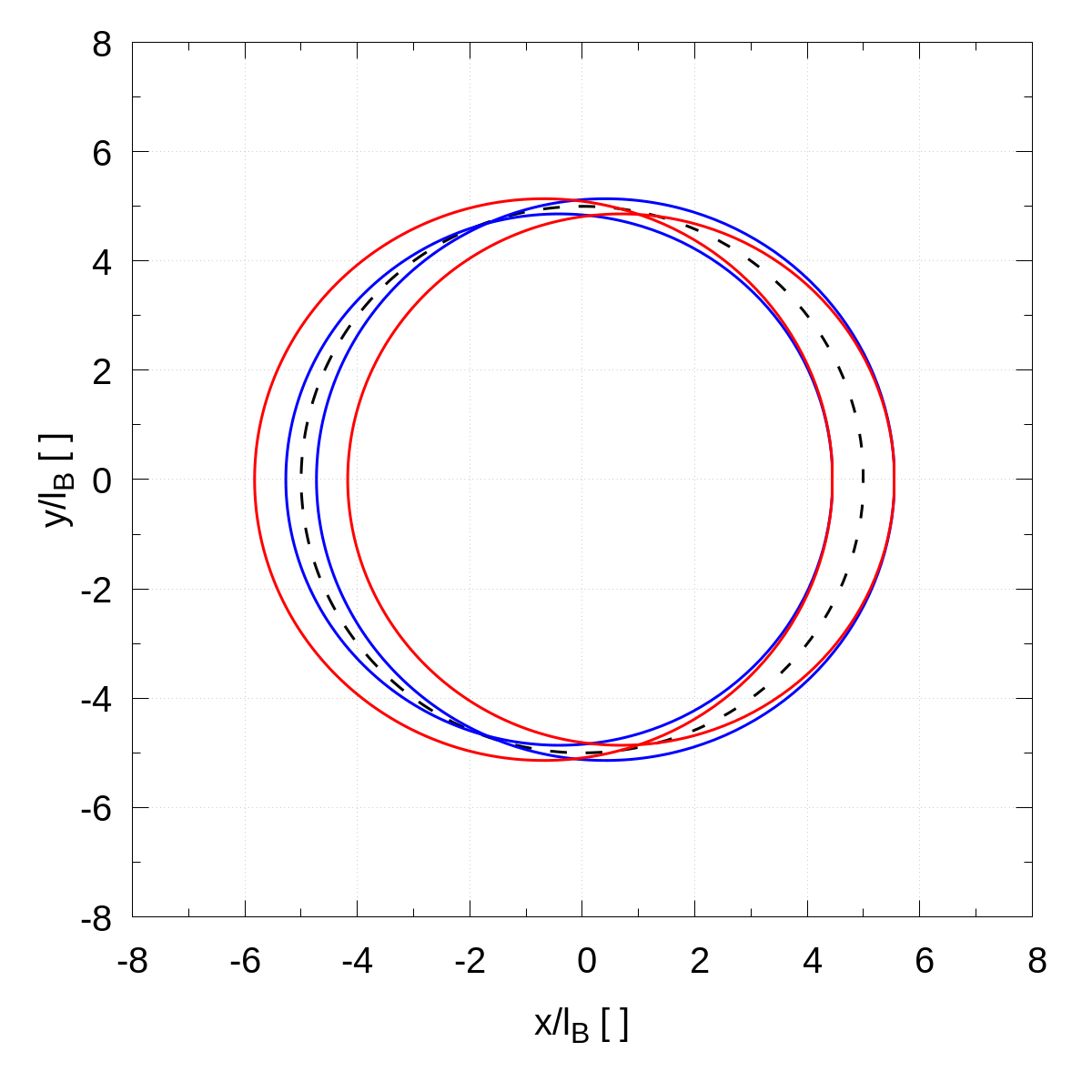}   
	\caption[The LOF caption]{The image shows the classical cyclotron trajectories used in the main text for the discussion of $\Delta_{xx}$ as blue and red circles. The black dashed circle is the true (quantum average) cyclotron orbit.}
	\label{fig:ClassicalTrajectories}
\end{figure}
Some heuristic understanding on $\Delta_{xx}$ can be can be gained from semiclassical considerations.
If a classical electron was located at the centre of the quantum wavepacket both in real and in momentum space it would simply follow the expected classical cyclotron orbit; however quantum packets represent probability distributions, so we can think of many interfering classical cyclotron orbits.
The initial probability distribution has widths $\sigma_x,\sigma_y$ in real space and $\frac{1}{2\sigma_x},\frac{1}{2\sigma_y}$ in momentum space. 
\newline Consider now a classical electron initially located at $(x_0+\frac{\sigma_x}{2},0)$, with kinetic momentum $(0,x_0+\frac{1}{4\sigma_y})$ (i.e. \virgolette{displaced} both in real and momentum space with respect to the average position and momentum).
The corresponding classical cyclotron orbit has radius $r_0=x_0+\frac{1}{4\sigma_y}$; since at $t=\pi$ the electron is in the diametrically opposite position, it will end up at $(-x_0+\frac{\sigma_x}{2}-\frac{1}{2\sigma_y},0)$.
\newline If the same considerations are applied to an electron initially at $(x_0-\frac{\sigma_x}{2},0)$ with momentum $(0,x_0-\frac{1}{4\sigma_y},0)$, one finds that it will end up at $(x',y')=(-x_0-\frac{\sigma_x}{2}+\frac{1}{2\sigma_y},0)$.
\newline At $t=\pi$ the distance between these two electrons is $\Delta=\sigma_x-\frac{1}{\sigma_y}$.
\newline The situation is graphically represented in Fig. \ref{fig:ClassicalTrajectories} (the blue curves are the classical trajectories for the two cases just discussed).
The same reasoning can be iterated by exchanging the momenta of the two electrons (red curves in Fig. \ref{fig:ClassicalTrajectories}), yielding $\Delta'=\sigma_x+\frac{1}{\sigma_y}$. If we now take the root mean square average of the two results, we get
\begin{equation}
\overline{\Delta}^2=\sigma_x^2+\frac{1}{\sigma_y^2}
\end{equation}
which is precisely the $\Delta_{xx}$ result in eq. \ref{eq:variances_at_pi}.

Despite the reasoning being quite artificial, and in spite of not being obvious how the reasoning could be extended to the case of $\Delta_{yy}(\pi)$ (since the cyclotron orbits shown in Fig. \ref{fig:ClassicalTrajectories} do not give an estimation of the packet width in the $y$ direction) it gives a rough idea of how this phenomenon emerges.

\begin{comment}
\begin{equation}
\begin{alignedat}{2}
\begin{cases}
\braket{x^2}_t &= \left(x_0^2+\frac{\sigma_x^2}{2}\right)\,\cos^2(t)+\frac{2}{\sigma_y^2}\,\sin^2\left(\frac{t}{2}\right)
+\frac{1}{2}\,\left(\frac{1}{\sigma_x^2}-\frac{1}{\sigma_y^2}\right)\,\sin^2(t)
\\
\braket{xy}_t &= \left[\left(\frac{1}{\sigma_x^2}-\frac{1}{\sigma_y^2}\right)\sin(t)-2x_0y_0\right]\,\sin^2\left(\frac{t}{2}\right)+\frac{1}{2}\left(x_0^2+\frac{\sigma_x^2}{2}\right)\,\sin(2t)
\\
\braket{y^2}_t &=
y_0^2+\frac{\sigma_y^2}{2}+\frac{2}{\sigma_x^2}\sin^2\left(\frac{t}{2}\right)
+2x_0 y_0 \sin(t)
+\left(x_0^2+\frac{\sigma_x^2}{2}\right)\,\sin(t)^2-\frac{1}{2}\left(\frac{1}{\sigma_x^2}-\frac{1}{\sigma_y^2}\right)\,\sin^2(t)
\end{cases}
\end{alignedat}
\end{equation}
\end{comment}
\chapter{Real space effective density response to a Gaussian excitation} % Main appendix title

\graphicspath{{./picAppB/}}

\label{appendix:gaussian_pert_computation} % Change X to a consecutive letter; for referencing this appendix elsewhere, use \ref{AppendixX}

\lhead{Appendix B. Real space effective density response to a Gaussian excitation} % Change X to a consecutive letter; this is for the header on each page - perhaps a shortened title

In order to compute the real-space effective ($x$-integrated) one body density we need to compute the following Fourier inversion integral 
\begin{equation}
\begin{split}
\delta\rho_{\text{eff}}(y;t)=
\frac{i}{2\pi}\frac{\lambda\sigma\tau}{\hbar c\,\Delta t}\,\int_{-\infty}^{\infty} 
\,e^{iq(y-v\Delta t)}\,e^{-q^2\,\left(\left(\frac{\sigma}{2}\right)^2+\left(\frac{v \tau}{2}\right)^2\right)}\frac{\sin\left(\frac{cq^2}{2}\Delta t\right)}{q}\,dq
\end{split}
\end{equation}
Defining
\begin{equation}
\label{eq:delta_alpha_beta}
\begin{cases}
\Delta=y-v\Delta t
\\
\alpha^2=\left(\frac{\sigma}{2}\right)^2+\left(\frac{v \tau}{2}\right)^2
\\
\beta^2=\frac{c\Delta t}{2}
\end{cases}
\end{equation}
the integral we need to compute can compactly be written as
\begin{equation}
I(\Delta,\alpha,\beta)=
\int_{-\infty}^{\infty} 
\,e^{iq\Delta}\,e^{-(q \alpha)^2}\,\frac{\sin\left[(q\beta)^2\right]}{q}\,dq.
\end{equation}
The integral is easily performed by converting it to a differential equation.
\begin{equation}
\label{eq:FirstOrderDiffEq}
\begin{cases}
\partial_\Delta I(\Delta,\alpha,\beta)=
i\int_{-\infty}^{\infty} 
\,e^{iq\Delta}\,e^{-(q \alpha)^2}\,\sin\left[(q\beta)^2\right]\,dq\\
I(0,\alpha,\beta)=0
\end{cases}
\end{equation}
where the initial condition $I(0,\alpha,\beta)=0$ is a symmetry consequence (the integrand function being odd in the integration variable when $\Delta=0$). 
We notice that $I'(\Delta)$ in eq. \ref{eq:FirstOrderDiffEq} is purely imaginary, again because of symmetry considerations.
\begin{equation}
\begin{split}
\partial_\Delta I(\Delta,\alpha,\beta)=&
i\int_{-\infty}^{\infty} 
\,e^{-(q \alpha)^2}\,\cos(q\Delta)\sin\left[(q\beta)^2\right]\,dq\\
=&
i\int_{-\infty}^{\infty} 
\,e^{-(q \alpha)^2}\,\Bigl(
\cos(q\Delta)\sin\left[(q\beta)^2\right]-\sin(q\Delta)\cos\left[(q\beta)^2\right]
\Bigr)\,dq\\
=&i\int_{-\infty}^{\infty} 
\,e^{-(q \alpha)^2}\,
\sin\left[(q\beta)^2-q\Delta\right]\,dq.
\end{split}
\end{equation}
In the second line an odd term has been added; when integrated, it thus vanishes and the identity holds. In the last line some elementary trigonometry has been used.
\newline $I'(\Delta)$ is finally written as the imaginary part of a convergent Gaussian integral
\begin{equation}
\begin{split}
\partial_\Delta I(\Delta,\alpha,\beta)=
-i \,\Im\left(
\int_{-\infty}^{\infty} 
	\,e^{iq\Delta}\,e^{-q^2(\alpha^2 +i\beta^2)}\right)\equiv -i\,\Im(J(\Delta,\alpha,\beta)).
\end{split}
\end{equation}
The integration is straightforward
\begin{equation}
J(\Delta,\alpha,\beta) = \sqrt{\frac{\pi}{\alpha^2+i\beta^2}}\,\,\exp\left[-\frac{\Delta^2}{4(\alpha^2+i\beta^2)}\right]
\end{equation}
and the differential equation is now easily solved in terms of the complex error function
\begin{equation}
I(\Delta,\alpha,\beta)=-i\,\Im\left(\int_0^\Delta J(\Delta',\alpha,\beta)\,d\Delta'\right)
=-i\,\pi\,\Im\left(\text{erf}\left[\frac{\Delta}{2\sqrt{\alpha^2+i\beta^2}}\right]\right)
\end{equation}
or
\begin{equation}
\label{eq:density_variation_gaussian_exc_result}
\begin{split}
\delta\rho_{\text{eff}}(y;t)=
\frac{\lambda\sigma\tau}{2\hbar c\,\Delta t}\,
\,\Im\left(\text{erf}\left[\frac{y-v\Delta t}{2\sqrt{\left(\frac{\sigma}{2}\right)^2+\left(\frac{v \tau}{2}\right)^2+i\frac{c\Delta t}{2}}}\right]\right)
\end{split}
\end{equation}
which is the result quoted in the thesis.

\begin{figure}[htp!]
	\begin{minipage}{.5\textwidth}
		\centering
		\includegraphics[width=1.\textwidth]{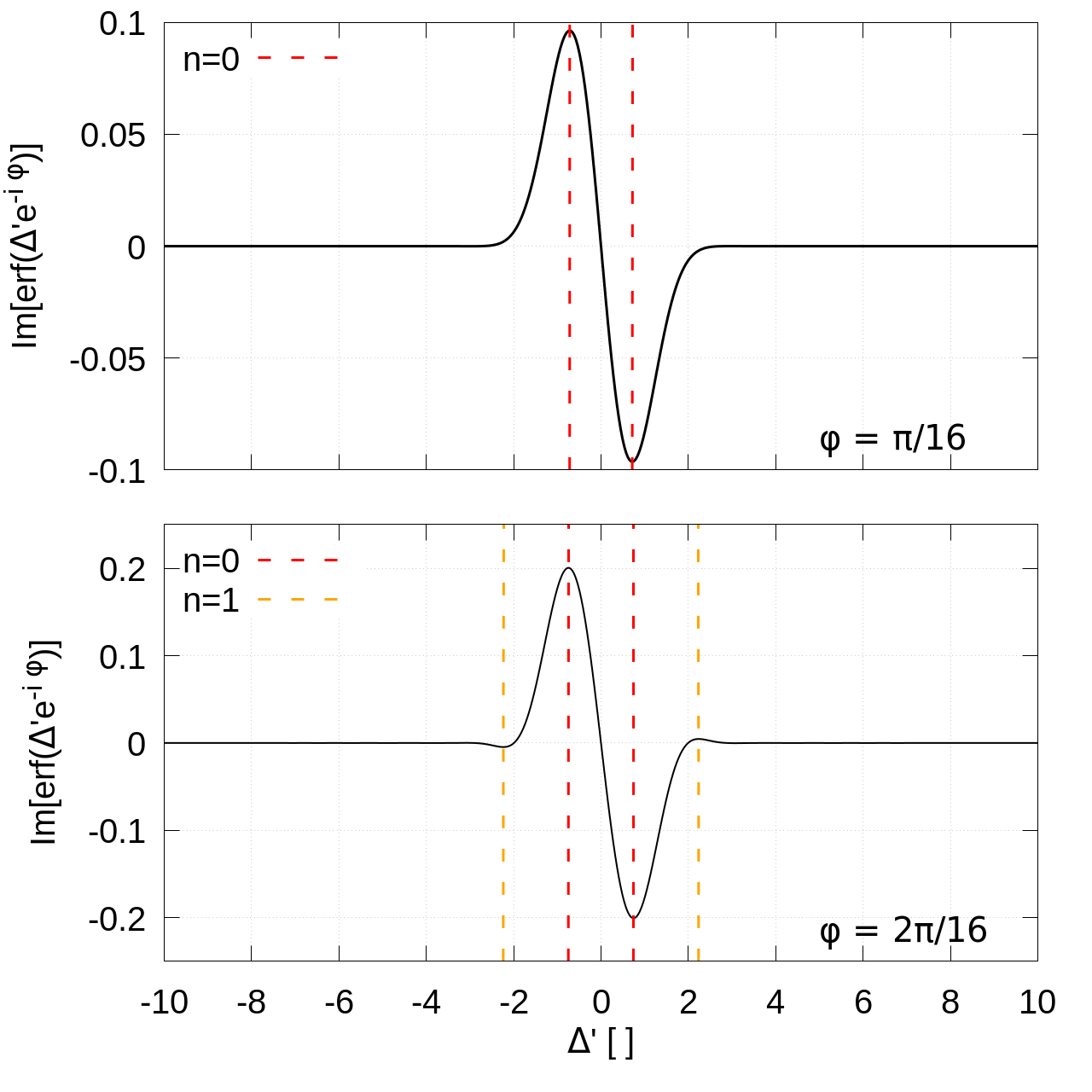}
	\end{minipage}%
	\begin{minipage}{0.5\textwidth}
		\centering
		\includegraphics[width=1.\textwidth]{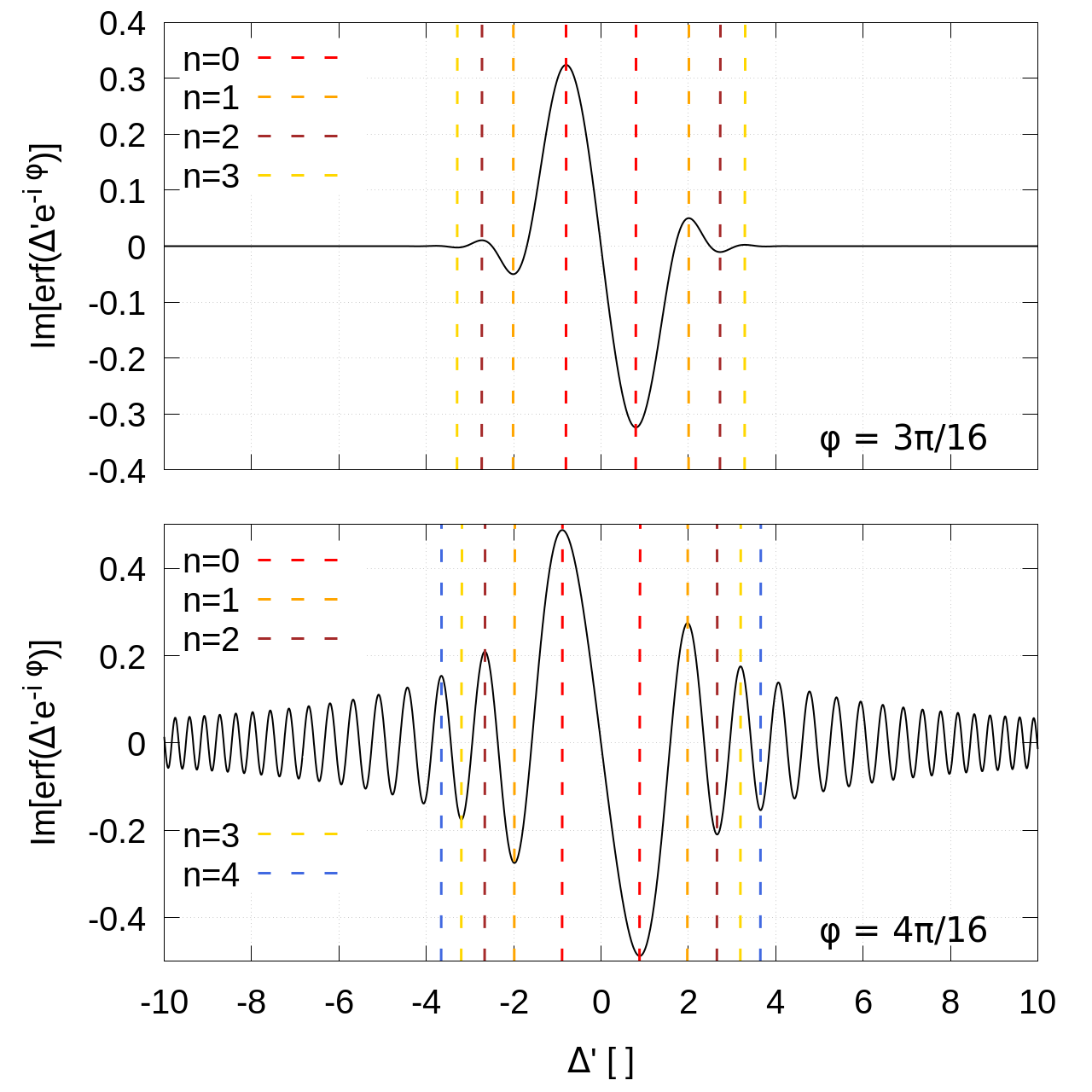}
	\end{minipage}    
	\caption[The LOF caption]{The two images show $\Im\text{erf}\left(\Delta'e^{-i \phi}\right)$ for different values of $\phi$ (i.e. at different times) as a function of $\Delta'\propto y-v\Delta t$. The stationary points of the packet (eq. \ref{eq:Globes}) are highlighted as vertical dashed lines.}
	\label{fig:appB_WFstationarypoints}
\end{figure}

Now we want to study the properties of such a function. We define
\begin{equation}
\begin{cases}
\Delta'=\frac{\Delta}{2(\alpha^4+\beta^4)^\frac{1}{4}}
\\
\phi=\frac{1}{2}\,\arctan\left(\frac{\beta^2}{\alpha^2}\right).
\end{cases}
\end{equation}
Notice first of all that $0\leq\phi\leq\frac{\pi}{4}$ and that $\phi\rightarrow\frac{\pi}{4}$ as $\Delta t\rightarrow\infty$.
To study the asymptotic behaviour of $\delta\rho_\text{eff}(y;t)$ as $y$ approaches $\pm\infty$ we use the following integral representation for the imaginary part of the error function with complex argument
\begin{equation}
\Im\,\text{erf}\left(\Delta'\,e^{-i\phi}\right) = -\frac{2}{\sqrt{\pi}}\,e^{-\Delta'^2\cos^2\phi}\int_0^{\Delta'\sin\phi} e^{u^2}\cos(2u\,\Delta'\cos\phi)\,du
\end{equation}
for which an upper bound can be easily found
\begin{equation}
\label{eq:im_erf_asymptotics}
\left|\Im\,\text{erf}\left(\Delta'\,e^{-i\phi}\right)\right|\leq e^{-\Delta'\,^2\cos^2\phi}\,\text{erfi}(|\Delta'\sin\phi|)\xrightarrow[y\rightarrow\pm\infty]{} \frac{1}{\sqrt{\pi}}\frac{\exp\left[-\cos(2\phi)\Delta'\,^2\right]}{\Delta'\sin\phi}
\end{equation}
which approaches $0$ as $y$ becomes large, since $0\leq\cos(2\phi)\leq1$.
\newline From the integral representation of the function it is also apparent that if the imaginary part of the argument $\Delta'\sin\phi$ does not vanish the behaviour of $\delta\rho_\text{eff}(y;t)$ has an oscillating component as well, i.e. we have a modulated oscillation.
\newline The characteristic length over which $\delta\rho_\text{eff}(y;t)$ is non-vanishing at a given time can be estimated by looking at the $y$ derivative
\begin{equation}
\frac{\partial}{\partial\Delta'}\Im\,\text{erf}\left(\Delta'\,e^{-i\phi}\right) = \frac{2}{\sqrt{\pi}}\, e^{-\Delta'^2\cos(2\phi)}\,\sin\Bigl(\Delta'\,^2\sin(2\phi)-\phi\Bigr)
\end{equation}
which vanishes when
\begin{equation}
\label{eq:Globes}
\Delta_n'^{(\pm)} = \frac{\Delta_n^{(\pm)}}{2(\alpha^4+\beta^4)^\frac{1}{4}} = \pm\sqrt{\frac{\phi+n\pi}{\sin(2\phi)}}
\end{equation}
where $n$ is a positive integer labelling the \virgolette{number} of the zero, i.e. its \virgolette{distance} from the symmetry point. 
Notice that if we choose the second zeroes by setting $n=\pm1$, as $\Delta t\rightarrow0$ these two points move further away, as expected (the side-lobes appear later on during the packet propagation).
If $n=0$ this is not the case though since $\frac{\phi}{\sin(2\phi)}$ is finite.
\newline In Fig. \ref{fig:appB_WFstationarypoints} $\Im\text{erf}\left(\Delta'e^{-i \phi}\right)$ is plotted at different times and the stationary points of eq. \ref{eq:Globes} are highlighted. Notice that when $\phi=\frac{\pi}{4}$ (or equivalently $\Delta t=\infty$) the function decays to zero quite slowly, $\propto \frac{1}{\Delta'}$ as can be seen in eq. \ref{eq:im_erf_asymptotics}; this is however not a problem since at $\Delta t=\infty$ the excited density packet $\delta\rho$ has already decayed to zero (since $\delta\rho$ decreases $\propto\Delta t^{-1}$, as will be shown explicitly below).
From the images it can be seen that $|\Im\,\text{erf}\left(\Delta_n'\,^{(\pm)}\,e^{-i\phi}\right)|$ is a monotonically decreasing sequence in $n$, i.e. the the lobes nearest to $y=v\Delta t$ are always larger in magnitude than those which form during the propagation of the packet.

\noindent The packet typical width (at finite time) can be roughly estimated by taking the distance between the first two stationary points (i.e. the distance between the first two lobes), $W=\Delta_{0}^{(+)}-\Delta_{0}^{(-)}$
\begin{equation}
\label{eq:gaussian_exc_width}
W=\sqrt{8\,\frac{\alpha^4+\beta^4}{\beta^2}\,\arctan\left(\frac{\beta^2}{\alpha^2}\right)}.
\end{equation}
It is worth analysing some asymptotic form of this expression.
\newline As long as the curvature effect is negligible we can approximate $\beta^2\ll\alpha^2$, and then
\begin{equation}
\label{eq:gaussian_exc_width_small_time_limit}
W\xrightarrow[\beta^2\ll\alpha^2]{} \sqrt{8}\,\alpha.
\end{equation}
In the opposite regime we instead are dominated by the effect of the curvature and
\begin{equation}
\label{eq:gaussian_exc_width_large_time_limit}
W\xrightarrow[\beta^2\gg\alpha^2]{} 2\sqrt{\pi}\,\beta\propto \sqrt{c\Delta t}.
\end{equation}

It is also easy to show that the density at the peaks decreases with time.
As $c\Delta t\gg \sigma^2+(v\tau)^2$ we have that $\phi$ approaches $\frac{\pi}{4}$. We can Taylor expand to obtain
\begin{equation}
\Im\,\text{erf}\left(\Delta_0'\,^{(\pm)}\,e^{-i\phi}\right) \xrightarrow[c\Delta t\gg \sigma^2+(v\tau)^2]{} 
\pm\Im\,\text{erf}\left(e^{-i\frac{\pi}{4}}\sqrt{\frac{\pi}{4}}\,\right)\pm\frac{1}{2\frac{\beta^2}{\alpha^2}}.
\end{equation}
the first term is just a constant $\Im\,\text{erf}\left(e^{-i\frac{\pi}{4}}\sqrt{\frac{\pi}{4}}\,\right)\approx-0.488$. The density at these peaks will thus decreases in time
\begin{equation}
\label{eq:decay_gaussian_exc}
\begin{split}
\delta\rho_{\text{eff}}(\Delta_0'\,^{(\pm)})&\sim
\pm\frac{\lambda\sigma\tau}{2\hbar c\,\Delta t}\,
\left(\Im\,\text{erf}\left(e^{-i\frac{\pi}{4}}\sqrt{\frac{\pi}{4}}\,\right)+\frac{\alpha^2}{2\beta^2}\right)\\
 &\propto \frac{1}{\Delta t}+\mathcal{O}(\Delta t^{-2})
\end{split}
\end{equation}

It is also interesting to look at the $c\rightarrow0$ limit, which is equivalent to $\phi\rightarrow0$. We can then approximate $\int_0^{\Delta'\sin\phi} e^{u^2}\cos(2xu)\,du \rightarrow \Delta'\phi$ with vanishing error. Then
\begin{equation}
\Im\,\text{erf}\left(\Delta'\,e^{i\phi}\right) \rightarrow -\frac{2}{\sqrt{\pi}}\,e^{-\Delta'\,^2}\Delta'\,\phi
\end{equation}
so
\begin{equation}
\label{eq:vanishing_curvature_limit}
\begin{split}
\delta\rho_{\text{eff}}(y;t)\rightarrow
-\frac{\lambda\sigma\tau}{8\sqrt{\pi}\hbar}\,
\,e^{-\left(\frac{\Delta}{2\alpha}\right)^2}
\frac{\Delta}{\alpha^3}
\end{split}
\end{equation}
which is precisely the result obtained by considering a straight-line dispersion relation (eq. \ref{eq:real_space_eff_drho}).

\addtocontents{toc}{\vspace{2em}} % Add a gap in the Contents, for aesthetics

\backmatter

%----------------------------------------------------------------------------------------
%	BIBLIOGRAPHY
%----------------------------------------------------------------------------------------

\label{Bibliography}

\lhead{\emph{Bibliography}} % Change the page header to say "Bibliography"

\bibliographystyle{unsrtnat} % Use the "unsrtnat" BibTeX style for formatting the Bibliography

%!TEX root = ../main.tex

\emph{}
% \bibliography{Bibliography} % The references (bibliography) information are stored in the file named "Bibliography.bib"

\clearpage
%----------------------------------------------------------------------------------------
%	ACKNOWLEDGEMENTS
%----------------------------------------------------------------------------------------

\setstretch{1.4} % Reset the line-spacing to 1.3 for body text (if it has changed)

%\acknowledgements{\addtocontents{toc}{\vspace{1em}} % Add a gap in the Contents, for aesthetics

%The acknowledgements and the people to thank go here, don't forget to include your project advisor\ldots
%}
%\clearpage % Start a new page

\end{document}